\RequirePackage{graphicx}
\RequirePackage{xspace}

\documentclass[%
superscriptaddress,
reprint,
showpacs,preprintnumbers,
nofootinbib,
amsmath,amssymb,
aps,
prd,
floatfix,
]{revtex4-1}

\usepackage{color} 
\usepackage[colorlinks=true,linkcolor=blue,citecolor=blue,allcolors=blue]{hyperref}%
\usepackage[capitalize]{cleveref}
\usepackage{siunitx}
\usepackage[utf8]{inputenc}
\usepackage{subfigure}
\usepackage{lineno}
\usepackage{multirow}
\usepackage[percent]{overpic}

\newcommand{\deltacp}{\ensuremath{\delta_\text{CP}}\xspace}
\newcommand{\cczeropi}{\ensuremath{\text{CC-}0\pi}\xspace}

\newcommand{\cconepiplus}{\ensuremath{\text{CC-}1\pi^+}\xspace}
\newcommand{\cconepiminus}{\ensuremath{\text{CC-}1\pi^-}\xspace}
\newcommand{\ccother}{\ensuremath{\text{CC-Other}}\xspace}
\newcommand{\nc}{\ensuremath{\text{NC}}\xspace}
\newcommand{\oofv}{\ensuremath{\text{out FV}}\xspace}
\newcommand{\numu}{\ensuremath{\nu_\mu}\xspace}
\newcommand{\barnumu}{\ensuremath{\bar{\nu}_\mu}\xspace}

\newcommand{\barnu}{\ensuremath{\bar{\nu}}\xspace}
\begin{document}


\title{First combined measurement of the muon neutrino and antineutrino charged-current cross section without pions in the final state at T2K}


\newcommand{\INSTHD}{\affiliation{University Autonoma Madrid, Department of Theoretical Physics, 28049 Madrid, Spain}}
\newcommand{\INSTEE}{\affiliation{University of Bern, Albert Einstein Center for Fundamental Physics, Laboratory for High Energy Physics (LHEP), Bern, Switzerland}}
\newcommand{\INSTFE}{\affiliation{Boston University, Department of Physics, Boston, Massachusetts, U.S.A.}}
\newcommand{\INSTD}{\affiliation{University of British Columbia, Department of Physics and Astronomy, Vancouver, British Columbia, Canada}}
\newcommand{\INSTGA}{\affiliation{University of California, Irvine, Department of Physics and Astronomy, Irvine, California, U.S.A.}}
\newcommand{\INSTI}{\affiliation{IRFU, CEA Saclay, Gif-sur-Yvette, France}}
\newcommand{\INSTGB}{\affiliation{University of Colorado at Boulder, Department of Physics, Boulder, Colorado, U.S.A.}}
\newcommand{\INSTFG}{\affiliation{Colorado State University, Department of Physics, Fort Collins, Colorado, U.S.A.}}
\newcommand{\INSTFH}{\affiliation{Duke University, Department of Physics, Durham, North Carolina, U.S.A.}}
\newcommand{\INSTBA}{\affiliation{Ecole Polytechnique, IN2P3-CNRS, Laboratoire Leprince-Ringuet, Palaiseau, France }}
\newcommand{\INSTEF}{\affiliation{ETH Zurich, Institute for Particle Physics and Astrophysics, Zurich, Switzerland}}
\newcommand{\INSTIE}{\affiliation{CERN European Organization for Nuclear Research, CH-1211 Genève 23, Switzerland}}
\newcommand{\INSTEG}{\affiliation{University of Geneva, Section de Physique, DPNC, Geneva, Switzerland}}
\newcommand{\INSTHJ}{\affiliation{University of Glasgow, School of Physics and Astronomy, Glasgow, United Kingdom}}
\newcommand{\INSTDG}{\affiliation{H. Niewodniczanski Institute of Nuclear Physics PAN, Cracow, Poland}}
\newcommand{\INSTCB}{\affiliation{High Energy Accelerator Research Organization (KEK), Tsukuba, Ibaraki, Japan}}
\newcommand{\INSTIB}{\affiliation{University of Houston, Department of Physics, Houston, Texas, U.S.A.}}
\newcommand{\INSTED}{\affiliation{Institut de Fisica d'Altes Energies (IFAE), The Barcelona Institute of Science and Technology, Campus UAB, Bellaterra (Barcelona) Spain}}
\newcommand{\INSTEC}{\affiliation{IFIC (CSIC \& University of Valencia), Valencia, Spain}}
\newcommand{\INSTHH}{\affiliation{Institute For Interdisciplinary Research in Science and Education (IFIRSE), ICISE, Quy Nhon, Vietnam}}
\newcommand{\INSTEI}{\affiliation{Imperial College London, Department of Physics, London, United Kingdom}}
\newcommand{\INSTGF}{\affiliation{INFN Sezione di Bari and Universit\`a e Politecnico di Bari, Dipartimento Interuniversitario di Fisica, Bari, Italy}}
\newcommand{\INSTBE}{\affiliation{INFN Sezione di Napoli and Universit\`a di Napoli, Dipartimento di Fisica, Napoli, Italy}}
\newcommand{\INSTBF}{\affiliation{INFN Sezione di Padova and Universit\`a di Padova, Dipartimento di Fisica, Padova, Italy}}
\newcommand{\INSTBD}{\affiliation{INFN Sezione di Roma and Universit\`a di Roma ``La Sapienza'', Roma, Italy}}
\newcommand{\INSTEB}{\affiliation{Institute for Nuclear Research of the Russian Academy of Sciences, Moscow, Russia}}
\newcommand{\INSTHI}{\affiliation{International Centre of Physics, Institute of Physics (IOP), Vietnam Academy of Science and Technology (VAST), 10 Dao Tan, Ba Dinh, Hanoi, Vietnam}}
\newcommand{\INSTHA}{\affiliation{Kavli Institute for the Physics and Mathematics of the Universe (WPI), The University of Tokyo Institutes for Advanced Study, University of Tokyo, Kashiwa, Chiba, Japan}}
\newcommand{\INSTID}{\affiliation{Keio University, Department of Physics, Kanagawa, Japan}}
\newcommand{\INSTIF}{\affiliation{King's College London, Department of Physics, Strand, London WC2R 2LS, United Kingdom}}
\newcommand{\INSTCC}{\affiliation{Kobe University, Kobe, Japan}}
\newcommand{\INSTCD}{\affiliation{Kyoto University, Department of Physics, Kyoto, Japan}}
\newcommand{\INSTEJ}{\affiliation{Lancaster University, Physics Department, Lancaster, United Kingdom}}
\newcommand{\INSTFC}{\affiliation{University of Liverpool, Department of Physics, Liverpool, United Kingdom}}
\newcommand{\INSTFI}{\affiliation{Louisiana State University, Department of Physics and Astronomy, Baton Rouge, Louisiana, U.S.A.}}
\newcommand{\INSTHB}{\affiliation{Michigan State University, Department of Physics and Astronomy,  East Lansing, Michigan, U.S.A.}}
\newcommand{\INSTCE}{\affiliation{Miyagi University of Education, Department of Physics, Sendai, Japan}}
\newcommand{\INSTDF}{\affiliation{National Centre for Nuclear Research, Warsaw, Poland}}
\newcommand{\INSTFJ}{\affiliation{State University of New York at Stony Brook, Department of Physics and Astronomy, Stony Brook, New York, U.S.A.}}
\newcommand{\INSTGJ}{\affiliation{Okayama University, Department of Physics, Okayama, Japan}}
\newcommand{\INSTCF}{\affiliation{Osaka City University, Department of Physics, Osaka, Japan}}
\newcommand{\INSTGG}{\affiliation{Oxford University, Department of Physics, Oxford, United Kingdom}}
\newcommand{\INSTIC}{\affiliation{University of Pennsylvania, Department of Physics and Astronomy,  Philadelphia, PA, 19104, USA.}}
\newcommand{\INSTGC}{\affiliation{University of Pittsburgh, Department of Physics and Astronomy, Pittsburgh, Pennsylvania, U.S.A.}}
\newcommand{\INSTFA}{\affiliation{Queen Mary University of London, School of Physics and Astronomy, London, United Kingdom}}
\newcommand{\INSTE}{\affiliation{University of Regina, Department of Physics, Regina, Saskatchewan, Canada}}
\newcommand{\INSTGD}{\affiliation{University of Rochester, Department of Physics and Astronomy, Rochester, New York, U.S.A.}}
\newcommand{\INSTHC}{\affiliation{Royal Holloway University of London, Department of Physics, Egham, Surrey, United Kingdom}}
\newcommand{\INSTBC}{\affiliation{RWTH Aachen University, III. Physikalisches Institut, Aachen, Germany}}
\newcommand{\INSTFB}{\affiliation{University of Sheffield, Department of Physics and Astronomy, Sheffield, United Kingdom}}
\newcommand{\INSTDI}{\affiliation{University of Silesia, Institute of Physics, Katowice, Poland}}
\newcommand{\INSTIA}{\affiliation{SLAC National Accelerator Laboratory, Stanford University, Menlo Park, California, USA}}
\newcommand{\INSTBB}{\affiliation{Sorbonne Universit\'e, Universit\'e Paris Diderot, CNRS/IN2P3, Laboratoire de Physique Nucl\'eaire et de Hautes Energies (LPNHE), Paris, France}}
\newcommand{\INSTEH}{\affiliation{STFC, Rutherford Appleton Laboratory, Harwell Oxford,  and  Daresbury Laboratory, Warrington, United Kingdom}}
\newcommand{\INSTCH}{\affiliation{University of Tokyo, Department of Physics, Tokyo, Japan}}
\newcommand{\INSTBJ}{\affiliation{University of Tokyo, Institute for Cosmic Ray Research, Kamioka Observatory, Kamioka, Japan}}
\newcommand{\INSTCG}{\affiliation{University of Tokyo, Institute for Cosmic Ray Research, Research Center for Cosmic Neutrinos, Kashiwa, Japan}}
\newcommand{\INSTHF}{\affiliation{Tokyo Institute of Technology, Department of Physics, Tokyo, Japan}}
\newcommand{\INSTGI}{\affiliation{Tokyo Metropolitan University, Department of Physics, Tokyo, Japan}}
\newcommand{\INSTHG}{\affiliation{Tokyo University of Science, Faculty of Science and Technology, Department of Physics, Noda, Chiba, Japan}}
\newcommand{\INSTF}{\affiliation{University of Toronto, Department of Physics, Toronto, Ontario, Canada}}
\newcommand{\INSTB}{\affiliation{TRIUMF, Vancouver, British Columbia, Canada}}
\newcommand{\INSTG}{\affiliation{University of Victoria, Department of Physics and Astronomy, Victoria, British Columbia, Canada}}
\newcommand{\INSTDJ}{\affiliation{University of Warsaw, Faculty of Physics, Warsaw, Poland}}
\newcommand{\INSTDH}{\affiliation{Warsaw University of Technology, Institute of Radioelectronics and Multimedia Technology, Warsaw, Poland}}
\newcommand{\INSTFD}{\affiliation{University of Warwick, Department of Physics, Coventry, United Kingdom}}
\newcommand{\INSTGH}{\affiliation{University of Winnipeg, Department of Physics, Winnipeg, Manitoba, Canada}}
\newcommand{\INSTEA}{\affiliation{Wroclaw University, Faculty of Physics and Astronomy, Wroclaw, Poland}}
\newcommand{\INSTHE}{\affiliation{Yokohama National University, Department of Physics, Yokohama, Japan}}
\newcommand{\INSTH}{\affiliation{York University, Department of Physics and Astronomy, Toronto, Ontario, Canada}}

\INSTHD
\INSTEE
\INSTFE
\INSTD
\INSTGA
\INSTI
\INSTGB
\INSTFG
\INSTFH
\INSTBA
\INSTEF
\INSTIE
\INSTEG
\INSTHJ
\INSTDG
\INSTCB
\INSTIB
\INSTED
\INSTEC
\INSTHH
\INSTEI
\INSTGF
\INSTBE
\INSTBF
\INSTBD
\INSTEB
\INSTHI
\INSTHA
\INSTID
\INSTIF
\INSTCC
\INSTCD
\INSTEJ
\INSTFC
\INSTFI
\INSTHB
\INSTCE
\INSTDF
\INSTFJ
\INSTGJ
\INSTCF
\INSTGG
\INSTIC
\INSTGC
\INSTFA
\INSTE
\INSTGD
\INSTHC
\INSTBC
\INSTFB
\INSTDI
\INSTIA
\INSTBB
\INSTEH
\INSTCH
\INSTBJ
\INSTCG
\INSTHF
\INSTGI
\INSTHG
\INSTF
\INSTB
\INSTG
\INSTDJ
\INSTDH
\INSTFD
\INSTGH
\INSTEA
\INSTHE
\INSTH

\author{K.\,Abe}\INSTBJ
\author{N.\,Akhlaq}\INSTFA
\author{R.\,Akutsu}\INSTCG
\author{A.\,Ali}\INSTCD
\author{C.\,Alt}\INSTEF
\author{C.\,Andreopoulos}\INSTEH\INSTFC
\author{L.\,Anthony}\INSTEI
\author{M.\,Antonova}\INSTEC
\author{S.\,Aoki}\INSTCC
\author{A.\,Ariga}\INSTEE
% ...

\collaboration{The T2K Collaboration}\noaffiliation




\date{\today}

\begin{abstract}
	This paper presents the first combined measurement of the double-differential muon neutrino and antineutrino charged-current cross sections with no pions in the final state on hydrocarbon at the off-axis near detector of the T2K experiment. The data analyzed in this work comprise 5.8$\times$10$^{20}$ and 6.3$\times$10$^{20}$ protons on target in neutrino and antineutrino mode respectively, at a beam energy peak of 0.6 GeV. Using the two measured cross sections, the sum, difference and asymmetry were calculated with the aim of better understanding the nuclear effects involved in such interactions. The extracted measurements have been compared with the prediction from different Monte Carlo generators and theoretical models showing that the difference between the two cross sections have interesting sensitivity to nuclear effects.
\end{abstract}


\maketitle


\section{Introduction}
\label{sec:Intro}

Current and future long-baseline neutrino oscillation experiments have as primary goals the measurements of the Charge-Parity (CP) violating phase (\deltacp), the neutrino mass ordering and the octant determination of the mixing angle $\theta_{23}$~\cite{Abe:2011ks,Adamson:2016tbqf,Abe:2014oxa,Abi:2018dnh}. To this end, the associated systematic error must be minimized. At present, the limited knowledge of (anti)neutrino-nucleus interactions dominates the uncertainties~\cite{Abe:2018wpn,Adamson:2017zcg}. The main obstacles behind a better understanding of such interactions are a result of limited modeling of the nuclear dynamics and the difficulties in measuring its effect on the cross section. Despite theoretical and experimental efforts in investigating the (anti)neutrino-nucleus cross section during the last decade, a comprehensive picture has not yet emerged~\cite{Katori:2016yel,Alvarez-Ruso:2014bla}.

Measured values of the muon neutrino and antineutrino charged-current quasi elastic scattering (CCQE) cross sections at K2K~\cite{Gran:2006jn}, MiniBooNE~\cite{AguilarArevalo:2010zc,Aguilar-Arevalo:2013dva}, MINOS~\cite{Adamson:2014pgc} and SciBooNE~\cite{AlcarazAunion:2009ku}, and more recently by T2K~\cite{Abe:2014iza,Abe:2016tmq,Abe:2017rfw,Abe:2018pwo} and MINERVA~\cite{Fiorentini:2013ezn,Fields:2013zhk,Walton:2014esl,Rodrigues:2015hik,Betancourt:2017uso,Patrick:2018gvi,Gran:2018fxa,Lu:2018stk,Ruterbories:2018gub} were found to be higher than predictions obtained using the Relativistic Fermi Gas (RFG) nuclear model. The results favored a higher value of the nucleon axial mass (M$_A^{QE}$) than those previously measured in bubble chamber experiments using deuterium as targets and pion electroproduction data~\cite{Kuzmin:2017bzt,Bodek:2007ym,Bernard:2001rs}. Furthermore, the CCQE muon neutrino and antineutrino cross sections measured by the NOMAD Collaboration at energies above 3 GeV are in agreement with a value of M$_A^{QE}$ around 1 GeV/c$^2$~\cite{Lyubushkin:2008pe}. This discrepancy highlighted the need for a more detailed description of the (anti)neutrino-nucleus scattering in the few-GeV energy region. 

In a muon neutrino CCQE interaction a negatively charged muon and proton are produced via \textit{W} exchange with a neutron, while in an antineutrino interaction of the same type a positively charged muon and neutron are produced via \textit{W} exchange with a proton:

\begin{eqnarray}
\centering
&\numu& + n \rightarrow \mu^- + p \nonumber\\
&\barnumu& + p \rightarrow \mu^+ + n.
\end{eqnarray}
As modern long-baseline neutrino experiments use relatively heavy nuclei as targets, nuclear dynamics plays an important role in the interpretation of the (anti)neutrino oscillations. Several theoretical models have proposed that these effects may explain the observed anomalies between data and Monte Carlo (MC) simulations~\cite{Martini:2009uj,Nieves:2011yp,Benhar:2015ula,Ankowski:2012ei,Butkevich:2009cp,Bodek:2016abf,Leitner:2008ue,Maieron:2003df,Meucci:2003cv,Lovato:2013cua,Pandey:2014tza,Amaro:2010sd}. If nuclear effects are considered, the particles produced in the (anti)neutrino-nucleus interaction can interact with other nucleons before exiting the nucleus. These, so-called Final State Interactions (FSI), can alter the type, number and kinematics of particles that exit the nucleus after such interactions. For example, in a muon neutrino resonant pion production interaction a pion, a proton and a muon are produced. The pion could interact with the nuclear media producing other nucleons, with the result that only the muon and the nucleons exit the nucleus. If only the muon and the proton are above the detection threshold, this would be indistinguishable from a CCQE interaction. Anyway, the observed discrepancies between data and models cannot be explained by FSI alone. Martini \textit{et al.}~\cite{Martini:2009uj} indicates that further contributions to CCQE-like processes arise from two (or more) interacting nucleons, referred to as 2p2h excitations or multi-nucleon knockout. Such interactions eject low-energy nucleons (200-500 MeV), which cannot be easily detected. Multi-nucleon knockout is expected to be less significant for antineutrinos relative to neutrinos; in particular it has a different role in the vector-axial interference term which differs by a sign for the neutrino and antineutrino cross section~\cite{Martini:2010ex}. Studying differences between CCQE-like cross sections for neutrino and antineutrino interactions could provide information about the role of multi-nucleon knockout in (anti)neutrino-nucleus interactions. The sum and the difference of the neutrino and antineutrino CCQE-like cross sections could yield this information. In the sum, the axial-vector interference term is eliminated whereas the difference isolates this term~\cite{Bernard:2001rs}. In Ref.~\cite{Ericson:2015cva}, the predicted sum and difference of the neutrino and antineutrino CCQE-like cross-sections are compared with the equivalent values computed using the CCQE-like double-differential cross sections obtained by the MiniBooNE experiment. The analysis found that additional nuclear effects, other than FSI, would be needed to explain the discrepancy between the observed and predicted values of the sum and difference. However, the analysis in Ref.~\cite{Ericson:2015cva} was limited as the neutrino and antineutrino beams peaked at different energies and the two cross sections were measured independently, implying that correlations between the two data sets were not taken into account.

A more rigorous analysis can be performed at the T2K near detector complex. Data has been taken with neutrino and antineutrino beams, both of which peak at the same energy. Combining the two data sets can exploit the correlation between them leading to a more precise cross-section measurement. This paper reports the first combined measurement of the double differential neutrino and antineutrino charged current cross sections on hydrocarbon without pions in the final state. This CCQE-like cross section will include contributions from CCQE events as well as events in which a pion was produced and then reabsorbed by the nucleus and multi-nucleon knockout events. The neutrino and antineutrino cross sections were used to compute the sum, difference and asymmetry. The neutrino-antineutrino cross-section asymmetry, which is the ratio between the difference and the sum, is a crucial quantity to control in order to avoid biases in the search for CP violation in neutrino oscillation. All these quantities have been compared with predictions made using several MC generators and models, which are discussed in this paper.

The paper is organized as follows. The T2K experiment is described in \cref{sec:T2Kexp}. The data and MC samples used in this analysis are reported in \cref{sec:datamc}. \cref{sec:anaStrategy} describes the analysis procedure, including the event selection and the cross-section extraction method. Finally the results and their interpretation are discussed in \cref{sec:resultscomp}, followed by conclusions in \cref{sec:conclusions}.

\section{The T2K experiment}
\label{sec:T2Kexp}

The Tokai-to-Kamioka (T2K) experiment~\cite{Abe:2011ks} is a long-baseline experiment that studies neutrino oscillations in an accelerator-produced $\nu_{\mu}$ ($\bar{\nu}_{\mu}$) beam. The neutrino beam, produced by the J-PARC facility, utilizes a 30\,GeV proton beam. A proton spill consisting of 8 bunches with 580\,ns spacing is produced every 2.48\,s. At a beam power of 430\,kW, this spill and repetition rate correspond to $2.25\times 10^{14}$ protons on target (p.o.t) per spill. Secondary hadrons, mainly pions and kaons, are produced when the proton beam interacts with a graphite target. Three magnetic horns are used to perform focusing and charge selection of the pions and kaons. The polarity of the magnetic horns can be changed to select positively (forward horn current) or negatively (reverse horn current) charged pions and kaons to produce a beam that is predominantly made of $\nu_{\mu}$ in the forward horn current case or $\bar{\nu}_{\mu}$ for the reverse horn current. The selected hadrons decay in a 96\,m long decay volume, to produce a (anti)neutrino beam whose direction is parallel to that of the initial proton beam. 

\begin{figure}[ht!]
	\includegraphics[width = 0.4\textwidth]{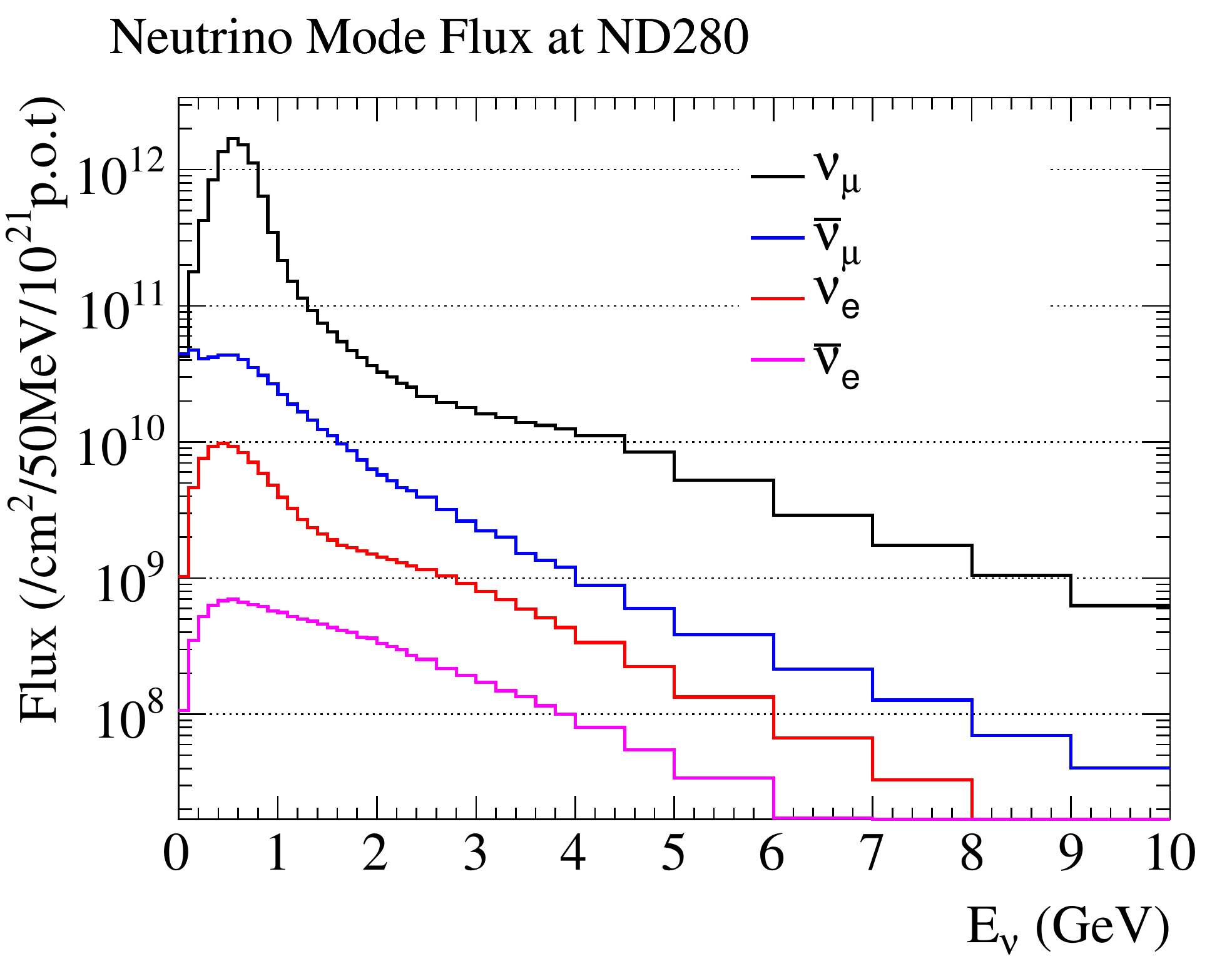}
	\includegraphics[width = 0.4\textwidth]{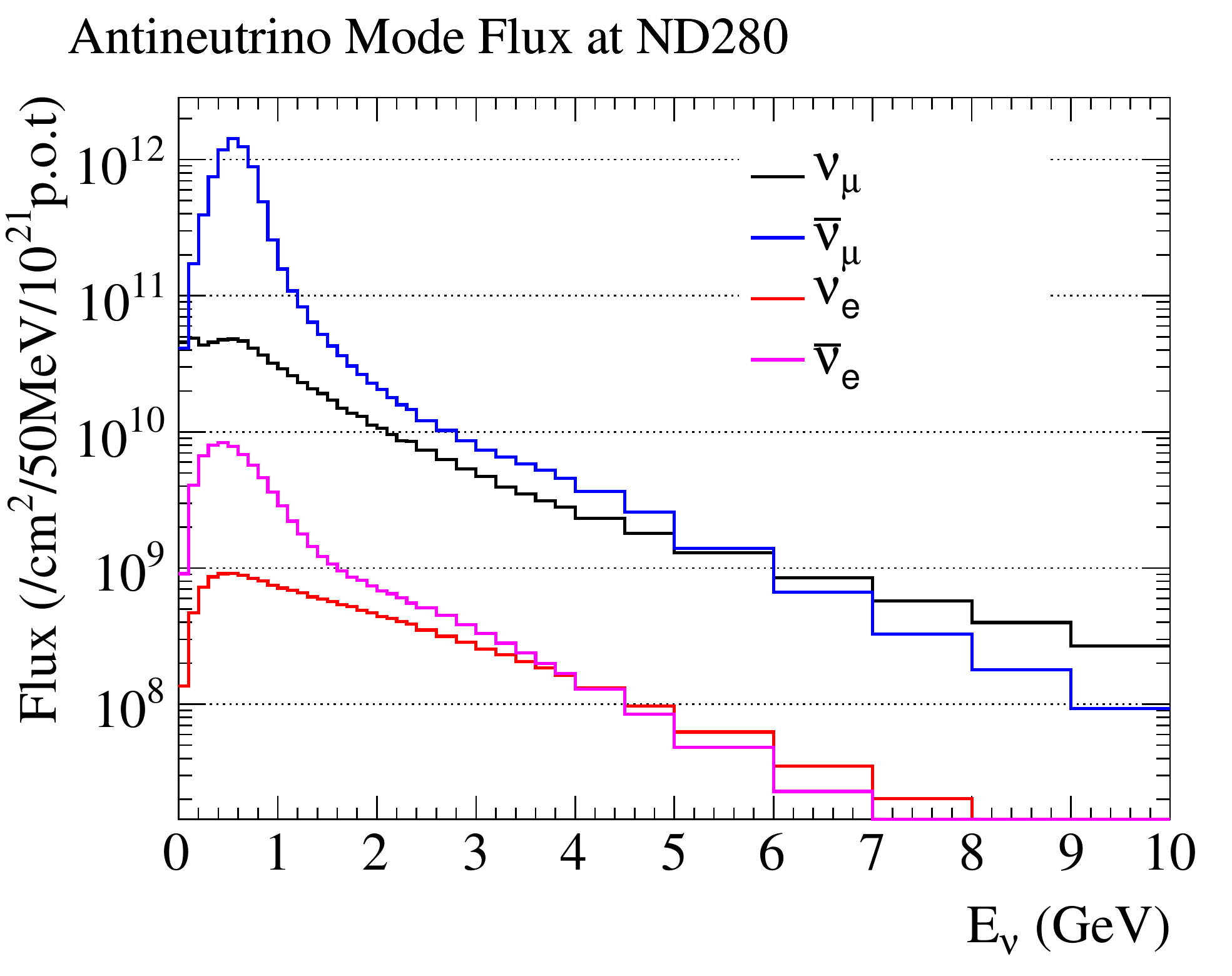}
	\caption{The predicted flux as a function of energy at the ND280 detector, for the neutrino beam (forward horn current) on top and antineutrino beam (reverse horn current) on bottom. In each case, the $\nu_{\mu}, \bar{\nu}_{\mu}, \nu_e$ and $\bar{\nu}_{e}$ components of the beam are shown.}
	\label{beam}
\end{figure}

Both the neutrino and antineutrino beam consist of a mixture of $\nu_{\mu}, \bar{\nu}_{\mu}, \nu_e$ and $\bar{\nu}_{e}$. The compositions of the neutrino and antineutrino beams are shown in \cref{beam}. In the neutrino beam mode, the ``right-sign'' $\nu_{\mu}$ (and $\nu_e$) flux is around 15\% higher around the flux peak when compared with the right-sign $\bar{\nu}_{\mu}$ (and $\bar{\nu}_e$) flux in the antineutrino mode. The background antineutrino flux is also lower in the neutrino mode compared with the neutrino flux in the antineutrino mode, especially at high energy. These differences can be attributed to the higher production multiplicities of positively, rather than negatively, charged parent particles.

The Super-Kamiokande far detector is located $2.5^{\circ}$ off the beam axis, at a distance of 295 km from the production point. The near detector complex, located 280\,m downstream from the production target, contains two sets of detectors: INGRID and ND280. INGRID~\cite{Abe:2011xv} is on-axis and monitors the flux and direction of the neutrino beam. The ND280 detector is positioned $2.5^{\circ}$ off-axis and is used to study the unoscillated beam. At an off-axis angle of $2.5^{\circ}$, the energy spectrum of the beam is narrowed and centered around 600\,MeV, which corresponds to the oscillation maximum for a baseline of 295\,km. In addition, this narrow energy spectrum suppresses the intrinsic $\nu_e$ ($\bar{\nu}_e$) and non-quasi-elastic interactions, leading to lower intrinsic backgrounds to the $\nu_e$ ($\bar{\nu}_e$) appearance search at the far detector. 

\begin{figure}[ht!]
	\includegraphics[width = 0.4\textwidth]{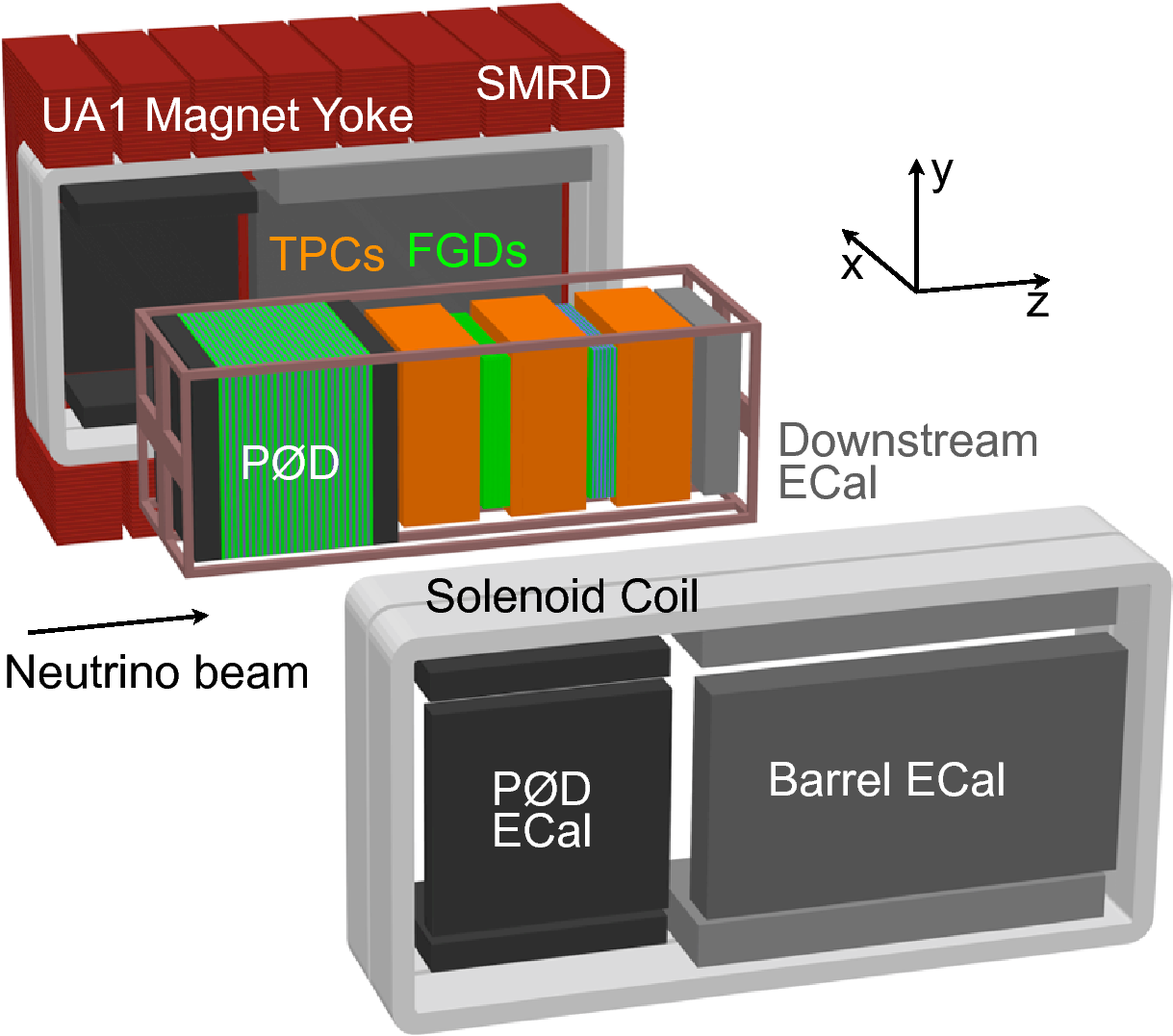}
	\caption{Schematic showing an exploded view of the ND280 off-axis detector. Each subdetector is labelled using the acronyms given in the text.}
	\label{nd280}
\end{figure}

This work has been performed using the off-axis near detector, ND280. \cref{nd280} shows a schematic of such detector. The ND280 detector is formed from five sub-detectors; an upstream $\pi^0$ detector (P$\emptyset$D)~\cite{Assylbekov:2011sh}, two Fine-Grained Detectors (FGDs)~\cite{Amaudruz:2012agx}, three Time Projection Chambers (TPCs)~\cite{Abgrall:2010hi}, Electromagnetic Calorimeter (ECal)~\cite{Allan:2013ofa} and a Side Muon Range Detector (SMRD)~\cite{Aoki:2012mf}. The P$\emptyset$D, FGDs, TPCs and ECal are contained within a magnet that provides a 0.2 T field, whilst the SMRD is embedded in the magnet. 

The measurements reported in this paper used the FGDs, TPCs, ECal and SMRD to select charged-current $\nu_{\mu}$ and $\bar{\nu}_{\mu}$ interactions. The most upstream FGD (FGD1) is formed from layers constructed from polystyrene scintillator bars. The scintillator layers are perpendicular to the beam's direction and alternating layers are orientated orthogonal to each other. The FGD is composed of 86.1\% carbon, 7.4\% hydrogen, 3.7\% oxygen, 1.7\% titanium, 1\% silicon and 0.1\% nitrogen by mass. The active region of FGD1 consists of scintillator layers only, whereas the downstream FGD (FGD2) has alternating layers of scintillator and water. The drift gas mixture used in the TPCs is Ar:CF$_4$:\textit{i}C$_4$H$_{10}$ (95:3:2). The TPCs (TPC1 the most upstream, TPC2 the central and TPC3 the most downstream) provide excellent particle identification and accurate measurement of momentum. Together the TPCs and FGDs form the tracker region of ND280. The ECal surround the tracker and consists of 13 modules made up of plastic scintillator bars alternating with lead sheets. SMRD consists of 440 modules of plastic scintillator counters.  

\section{Data and Monte Carlo samples}
\label{sec:datamc}

The studies reported in this paper use $5.80\times10^{20}$ p.o.t forward horn-current ($\nu$-mode) data and $6.27\times 10^{20}$ p.o.t of reverse horn-current ($\bar{\nu}$-mode) data broken into run periods shown in \cref{tab:runs-pot}.

\begin{table}[ht!]
	\caption{T2K neutrino and antineutrino mode runs and their associated p.o.t, filtered for spills where all ND280 detectors were flagged with good data quality.}
	\label{tab:runs-pot}
	\begin{ruledtabular}
		\begin{tabular}{ l c c c }
			Run & Dates & \(\nu\)-mode p.o.t & \(\bar{\nu}\)-mode p.o.t \\ 
			Period & & (\(\times10^{20}\)) & (\(\times10^{20}\))   \\ \hline
			Run 2 & Nov. 2010 - Mar. 2011 & 0.79 & -- \\
			Run 3 & Mar. 2012 - Jun. 2012 & 1.58 & -- \\
			Run 4 & Oct. 2012 - May 2013 & 3.42 & -- \\
			Run 5 & Jun. 2014 & -- & 0.43 \\
			Run 6 & Nov. 2014 - Apr. 2015 & -- &  3.40\\
			Run 7 & Feb. 2016 - May 2016  & -- &  2.44\\		
			\hline
			Total & Nov. 2010 - May 2016 & 5.80 &  6.27 
		\end{tabular}
	\end{ruledtabular}
\end{table}

The MC simulation used for this analysis consist of a sample corresponding to ten times the data p.o.t. It is performed generating (anti)neutrino interactions according with the flux predicted at ND280. The simulation of the $\nu_{\mu}$ and $\bar{\nu}_{\mu}$ fluxes reaching the near detector are described in detail in Ref.~\cite{Abe:2012av}. The neutrino and antineutrino interactions in the ND280 sub-detectors, as well as events inside the magnet yoke and in the rock surrounding the ND280 pit, were simulated using the \textsc{Neut} MC generator version~\texttt{5.3.2}~\cite{Hayato:2002sd}. The CCQE neutrino-nucleon cross section is simulated according to the Llewellyn-Smith formalism~\cite{LlewellynSmith:1971uhs} with a dipole axial form factor and BBBA05 vector form factors~\cite{Bradford:2006yz}. The nuclear model uses a spectral function (SF), developed in Ref.~\cite{Benhar:1994hw} with an axial mass $M_A^{QE} = 1.21$ GeV/c$^2$ based on the K2K measurement of the \numu CCQE cross section~\cite{Gran:2006jn}. It utilizes the multi-nucleon interaction model (2p2h) from Nieves \textit{et al.}~\cite{Nieves:2011pp} to simulate interactions with nucleon pairs. The model for resonant pion production (RES) is based on the Rein-Sehgal model~\cite{Berger:2007rq} with updated nucleon form factors~\cite{Graczyk:2007bc} and an invariant hadronic mass W $\leq$ 2 GeV. The DIS interaction is calculated for W $>$ 1.3 GeV, using GRV98 parton distribution functions~\cite{Gluck:1998xa} with Bodek-Yang corrections~\cite{Bodek:2003wc}. Single pion production through DIS is suppressed for W $\leq$ 2 GeV to avoid double counting with RES and it uses a custom hadronization model. For values of the invariant hadronic mass W $>$ 2 GeV, \textsc{Pythia/JetSet}~\cite{Sjostrand:1993yb} is used for hadronization. FSI, i.e. interactions of the hadrons produced by neutrino interactions with the other nucleons before leaving the nuclear environment, are simulated using a semiclassical intranuclear cascade model~\cite{Bertini:1972vz,Oset:1986sy}. 

The propagation of the final state particles through the ND280 sub-detectors is simulated using the package \textsc{Geant4} version~\texttt{4.9.4}~\cite{Agostinelli:2002hh} as detailed in Ref.~\cite{Abe:2011ks} employing the following physics lists: \texttt{QGSP\_BERT} for the hadronic physics, \texttt{emstandard\_opt3} for the electromagnetic physics and \texttt{G4DecayPhysics} for the particle decays.

\section{Analysis strategy}
\label{sec:anaStrategy}

A joint measurement of neutrino and antineutrino cross-sections, fully accounting for correlations in the systematic uncertainties, has been performed. Given the relatively large background of neutrino interactions in the antineutrino sample, such a joint analysis is mandatory for a robust antineutrino cross-section measurement. Indeed, since the neutrino and antineutrino cross-sections are largely driven by the same underlying physics, it would be inconsistent to assume to know the former while measuring the latter. A further advantage of a joint measurement, is that it exploits the full, high-statistics, neutrino sample minimizing the correlated detector and flux systematic uncertainties and thus resulting in a more precise antineutrino measurement. Finally, a joint analysis enables interesting measurements, as explained in \cref{sec:Intro}. 

An unregularized binned likelihood fit with control sample to constrain the background is performed as in Ref.~\cite{Abe:2016tmq, Abe:2018pwo,Abe:2018uhf}. This analysis method guarantees a negligible dependence on the signal model used in the simulation for the correction of detector effects, provided that a too coarse binning is not used. A simultaneous fit is applied to the antineutrino sample and the neutrino sample, the former being further sub-divided in different signal and background samples depending on the direction of the outgoing muon, while the latter depending on the kinematics of the outgoing muon and proton. The number of selected events in each bin of reconstructed kinematics ($j$) for each signal and background sample ($s$) is computed as

\begin{align}
	N^s_j  = \sum_i^\text{true bins}\Bigg[ & c{_i}{^{\numu}} \left(N^\text{MC \numu \cczeropi}_i  \prod_x w(x)_{i}^{\numu \cczeropi} \right) + \nonumber \\ 
	& c{_i}{^{\barnumu}} \left(N^\text{MC \barnumu \cczeropi}_i  \prod_x w(x)_{i}^{\barnumu\cczeropi} \right) + \nonumber \\
	& \sum_k^\text{bkg reactions} N^\text{MC bkg $k$}_{i} \prod_b w(b)_{i}^{k} \Bigg]\times \nonumber \\
	& t_{ij}^\text{det} d_j \sum_n^{\text{E}_{\numu}\:_\text{or}\:_{\barnumu}} w_n^i f_n
	\label{eq:Nreco}
\end{align}
where $N_i^\text{MC}$ is the true number of events in MC with the superscript indicating which interaction type they correspond to. The index $i$ runs over the bins of the ``true" muon kinematics prior to detector smearing effects, $k$ runs over the possible background reactions and $n$ runs over the neutrino or antineutrino energy bins. Both $c{_i}{^{\numu}}$ and $c{_i}{^{\barnumu}}$ are the parameters of interest which adjust the \numu and \barnumu \cczeropi number of events in MC, in order to match the observed number of events in data. The transfer matrix $t_{ij}^\text{det}$, relates the true ($i$) and reconstructed ($j$) muon kinematics bins and $d_j$ represents the nuisance parameters in the fit describing the detector systematics which are constrained by a prior covariance matrix. The flux parameters $f_n$ and weights $w^i_n$, describe the neutrino energy distribution for each bin of $p^{true}_\mu,\cos\theta^{true}_\mu$. The $f_n$ are nuisance parameters in the fit constrained by a prior covariance matrix. The product $\prod_{x,b}$ runs over the systematics related to the theoretical modeling of the interaction channels contributing to the signal ($x$) or the background ($b$). Each $w(x)_{i}^{\numu \cczeropi}$ and $w(b)^{k}_{i}$ term is a weighting function describing how the generated muon kinematics change (in bins $i$ for each signal and background process) as a function of the value of a particular theoretical parameter. All the parameters $x$ and $b$ are nuisance parameters in the fit and are constrained by a prior covariance matrix. Signal modeling parameters $x$ are not fitted to avoid model dependence but they must be included to account for their effect on the uncertainty of the efficiency corrections. The parameters of interest $c{_i}{^{\numu}}$ reweights the neutrino signal in neutrino mode and neutrino background in the antineutrino mode, whereas $c{_i}{^{\barnumu}}$ reweights the antineutrino background in neutrino mode and antineutrino signal in antineutrino mode. The nuisance parameters may be different in each sample and their correlations between samples are fully taken into account.

The best fit parameters are those that minimize the following log-likelihood:

\begin{align}
	\chi^2 & = \chi^2_\text{stat,$\nu$} + \chi^2_\text{stat,$\bar{\nu}$} + \chi^2_\text{syst} \nonumber\\
	& = \sum_j^\text{reco bins} 2\left(N^{\numu}_j-N_j^\text{\numu obs}+N_j^\text{\numu obs} \ln\frac{N_j^\text{\numu obs}}{N^{\numu}_j}\right) \nonumber\\
	& + \sum_j^\text{reco bins} 2\left(N^{\barnumu}_j-N_j^\text{\barnumu obs}+N_j^\text{\barnumu obs} \ln\frac{N_j^\text{\barnumu obs}}{N^{\barnumu}_j}\right) \nonumber\\
	& + \;\;\;\sum_p\left(\vec{p}-\vec{p}_\text{prior}\right)\left(V^\text{syst}_\text{cov}\right)^{-1}\left(\vec{p}-\vec{p}_\text{prior}\right)
	\label{eq:chi2}  
\end{align}
where $N_j^{\nu_\mu}$ ($N_j^{\bar{\nu}_\mu}$) is the expected total number of events in the neutrino (antineutrino) sample and $N_j^\text{\numu obs}$ ($N_j^\text{\barnumu obs}$) is the observed number of events. $\chi^2_\text{syst}$ is a penalty term for the systematics, where $\vec{p}$ are the parameters that describe the effect of nuisance parameters, $\vec{p}_\text{prior}$ are the prior values of these systematic parameters and $V^\text{syst}_\text{cov}$ is their covariance matrix which describes the confidence in the nominal parameter values, as well as, correlations between them. 

To minimize the dependence of the results on the signal model used in the simulation, two-dimensional differential cross-sections are extracted as a function of muon momentum and angle. Those are kinematic quantities directly observable in the detector and they represent all the relevant variables to characterize the detector acceptance and efficiency. The signal is defined by the absence of pions in the final state, avoiding model-dependent corrections for pion re-absorption in the nucleus.

The flux-integrated cross-sections are evaluated per nucleon and for each bin $i$ of detector unsmeared muon momentum and angle:
\begin{equation}
\begin{aligned}
\label{eq:numuxsec}
\frac{\text{d}\sigma_{\numu}}{\text{d}p_\mu \text{d}\cos\theta_\mu}&=& \frac{N^{\numu \text{\cczeropi}}_i}{\epsilon^{\numu}_i \Phi^{\numu} N^\text{FV}_\text{nucleons}} \times \frac{1}{\Delta p_\mu \Delta \cos\theta_\mu}\\
\frac{\text{d}\sigma_{\barnumu}}{\text{d}p_\mu \text{d}\cos\theta_\mu}&=& \frac{N^{\barnumu \text{\cczeropi}}_i}{\epsilon^{\barnumu}_i \Phi^{\barnumu} N^\text{FV}_\text{nucleons}} \times \frac{1}{\Delta p_\mu \Delta \cos\theta_\mu}
\end{aligned}
\end{equation}
where $N^{\numu \text{\cczeropi}}_i$ and $N^{\barnumu \text{\cczeropi}}_i$ are the number of neutrino and antineutrino \cczeropi events respectively evaluated by the fit, $\epsilon^{\numu}_i$ and $\epsilon^{\barnumu}_i$ are the efficiency evaluated from MC, $N^\text{FV}_\text{nucleons}$ is the number of target nucleons in the fiducial volume, $\Phi^{\numu}$ and $\Phi^{\barnumu}$ are the integrated fluxes for neutrino and antineutrino, $\Delta p_\mu$ and $\Delta \cos\theta_\mu$ are the bin widths of the muon momentum and cosine of the muon scattering angle w.r.t. the bean direction. The number of nucleons, computed using the areal density of the different elements composing the fiducial volume~\cite{Amaudruz:2012agx}, is equal to 5.9$\times$10$^{29}$ and it is used to extract both cross sections. The cross sections are normalized in all bins of muon kinematics with the same integrated flux to avoid a model-dependent mapping of such bins into energy intervals of the incoming neutrino.

The binning of the true muon kinematics has been optimized to reduce the bin-by-bin fluctuation derived by the extrapolation of an unsmeared quantity, as the cross-section is, and also to ensure that the systematic uncertainty are smaller than the statistical uncertainty. If the binning is too coarse, the results do not give much information about the shape of the cross section, while on the other hand if the binning is too fine, some bins could be empty causing problems with the minimization algorithm. The best binning lies in between these extreme cases and requires that the bin width is always greater than the resolution of the muon kinematics.
A MC sample simulated using the version of \textsc{Neut} described in \cref{sec:datamc} has been used as the prior of the fitting algorithm. This choice does not introduce model dependencies as extensively demonstrated in previous analyses~\cite{Abe:2018pwo,Abe:2018uhf}. The stability of the results has been confirmed by using alternative models in the fitting framework. To this end, a set of mock data samples has been created by modifying the amount of 2p2h interactions, the nuclear or the background model, and the input MC. Through them, it has been verified that the extracted cross section is always in agreement, within the uncertainties, with the mock data set predicted cross section and also produces a small $\chi^2$ computed considering the final cross section covariance matrix.

\subsection{Event selections}
\label{sec:selections}

\begin{figure*}[th!]
	\centering
	\includegraphics[width=1.\linewidth,trim={0 19cm 0 0},clip]{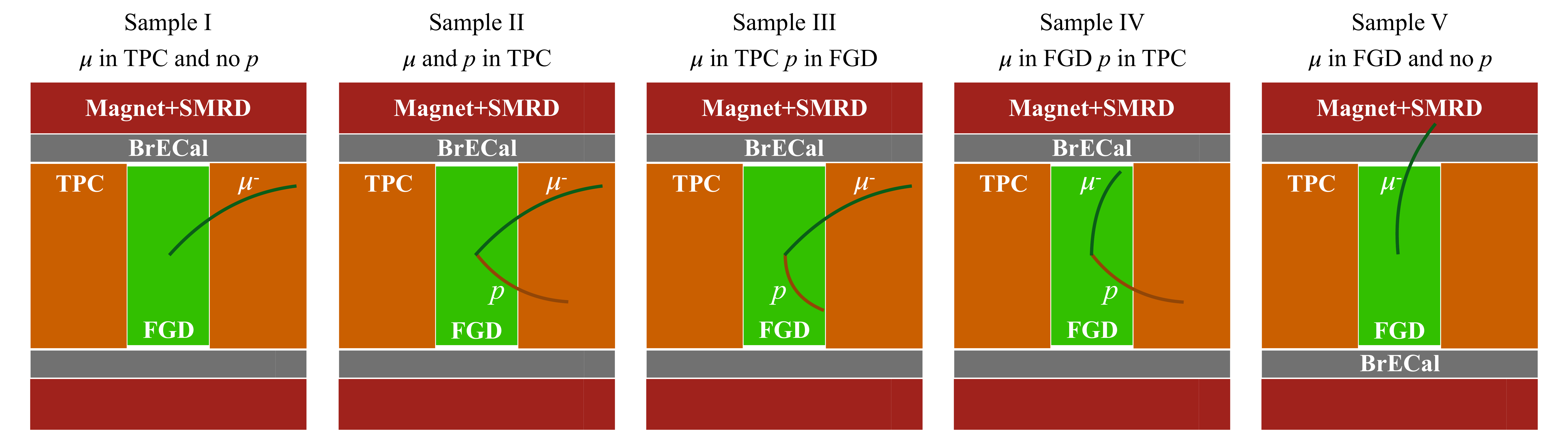}
	\caption{Schematic representation of the different \numu CC signal samples. In each drawing a \numu enters from the left and interacts in FGD1. The sub-detectors of ND280 are shown in their side view.}
	\label{fig:numusignalsamples}
\end{figure*}

The event selections developed for this analysis aim to select \numu and \barnumu \cczeropi interactions in the FGD1 and to provide appropriate control samples to constrain the main background sources. 

In previous analyses, the selection criteria were optimized to select forward going muons (with respect to the beam direction) originating from FGDs~\cite{Abe:2016tmq,Abe:2013jth,Abe:2016aoo,Abe:2019arf,Abe:2014iza}. For this analysis, the phase space of the muon kinematics was enlarged, including also high-angle and backward-going tracks. The acceptance has been increased using all the ND280 sub-detectors and the time of flight (ToF) of the particles between different sub-detectors which gives information about the direction of the track, i.e. if it is forward or backward with respect to the beam direction, following the same strategy described in Ref.~\cite{Abe:2018uhf}.

In addition to the common goal of enlarging the acceptance, the event selections have several common features:
\begin{itemize}
	\item The selection criteria have been optimized by employing a MC sample simulated using the version of \textsc{Neut} described in Sec.~\ref{sec:datamc};
	\item Particles that enter the TPCs or are fully contained in FGD1 are identified through the TPC or FGD particle identification (PID), based on d$E$/d$x$ measurements;
	\item ECal PID is performed if there is an associated ECal segment, which reduces the shower-like contamination (mostly $\pi^0$);
	\item The ratio between the track length and the electromagnetic energy associated with the track is used to reduce the proton contamination;
	\item Particles stopping in the SMRD are identified as muons, since most likely this is the only particle that will reach this detector. 
\end{itemize}

Each selection applies a set of cuts which have been optimized to give the best signal efficiency and purity. Two requirements are common to both selections:

\begin{itemize}
	\item Events must occur within the time window of one of the eight beam bunches of the spill structure of the beam and when all ND280 sub-detectors are functioning correctly;
	\item The interaction vertex, defined as the starting position of the muon candidate, must be inside the FGD1 fiducial volume (FV). Compared with the previous analyses where both a true and a reconstructed vertex in the first two scintillator layers were rejected~\cite{Abe:2016tmq,Abe:2013jth}, in this analysis the full span has been taken as the FV. Depending on the direction of the muon, the events with a reconstructed vertex in the first (forward-going muon) or the last (backward-going muon) layer have been rejected.
\end{itemize}

In the following sections, the selection strategy is discussed in detail.

\subsubsection{\numu CC event selection}
\label{sec:numuevtsel}

\begin{figure*}[th!]
	\centering
	
	\includegraphics[width=0.38\linewidth]{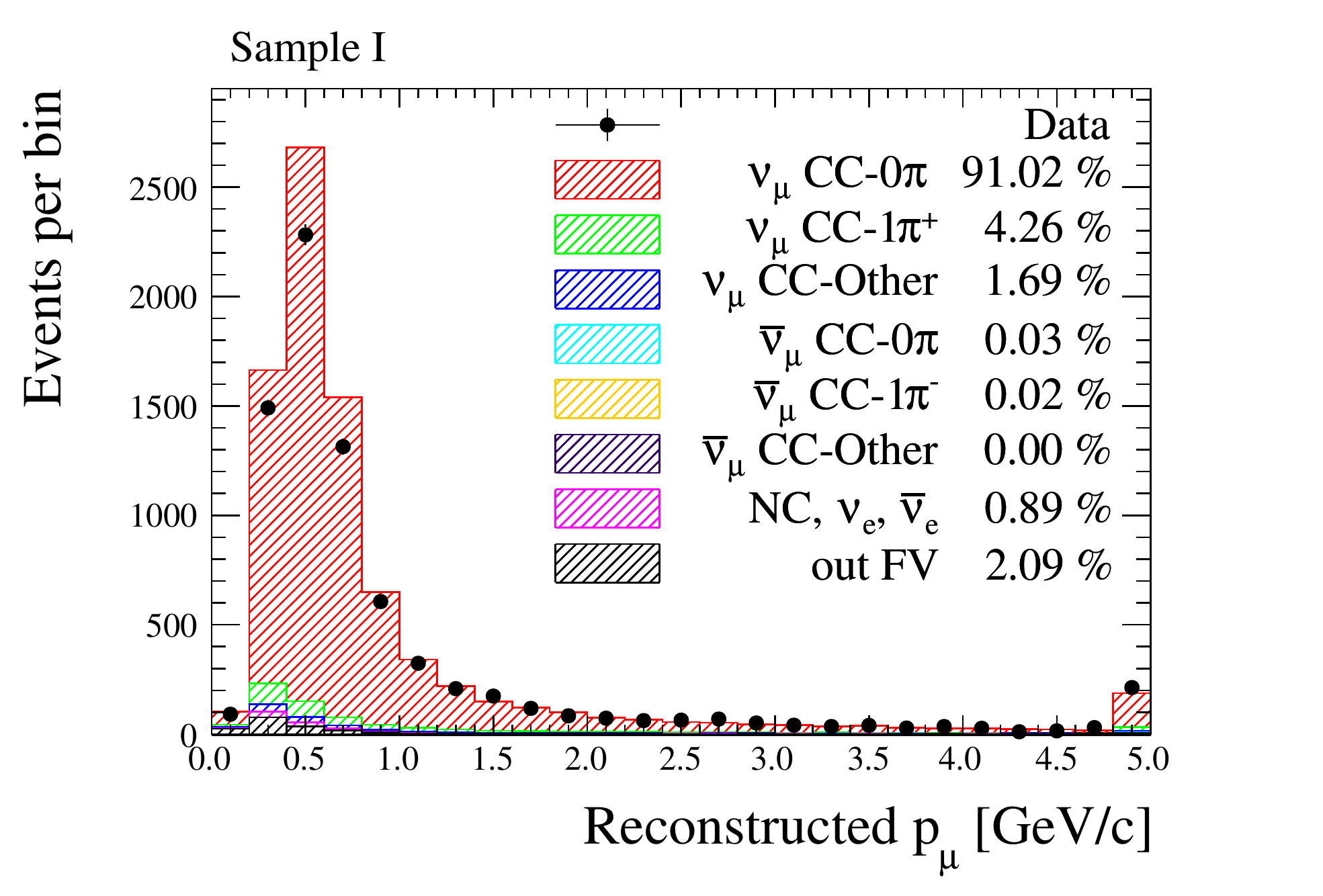}
	\includegraphics[width=0.38\linewidth]{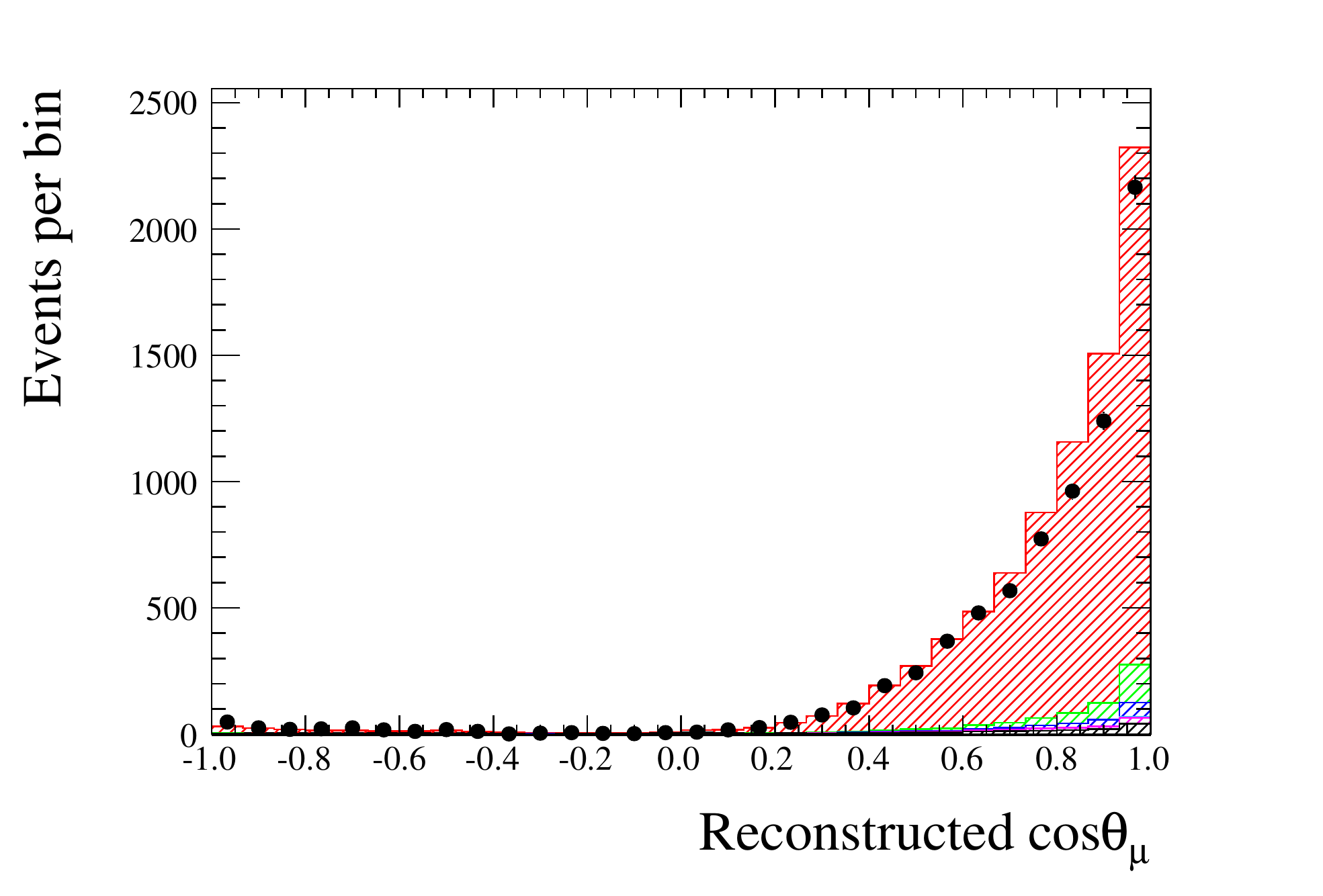}
	
	\includegraphics[width=0.38\linewidth]{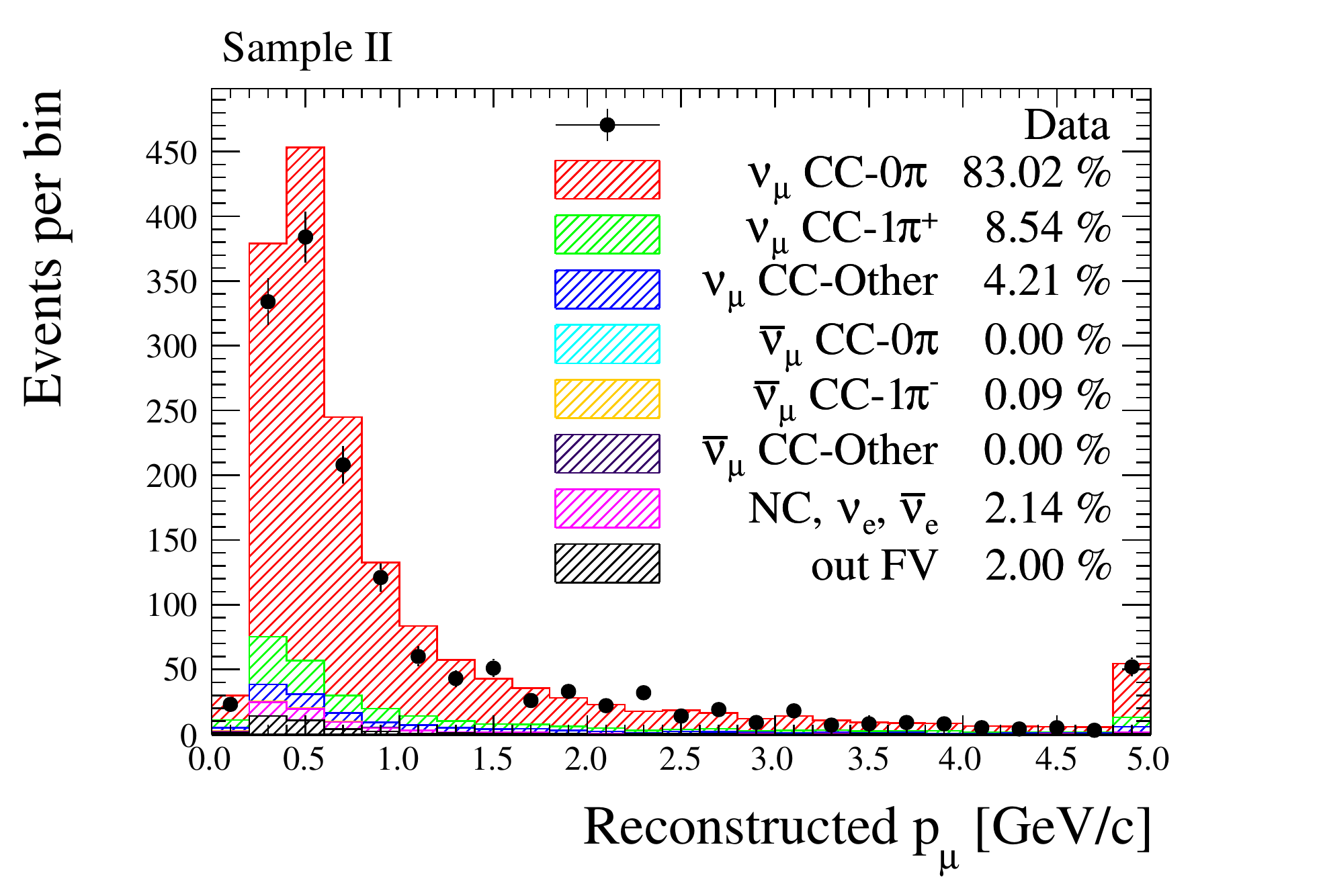}
	\includegraphics[width=0.38\linewidth]{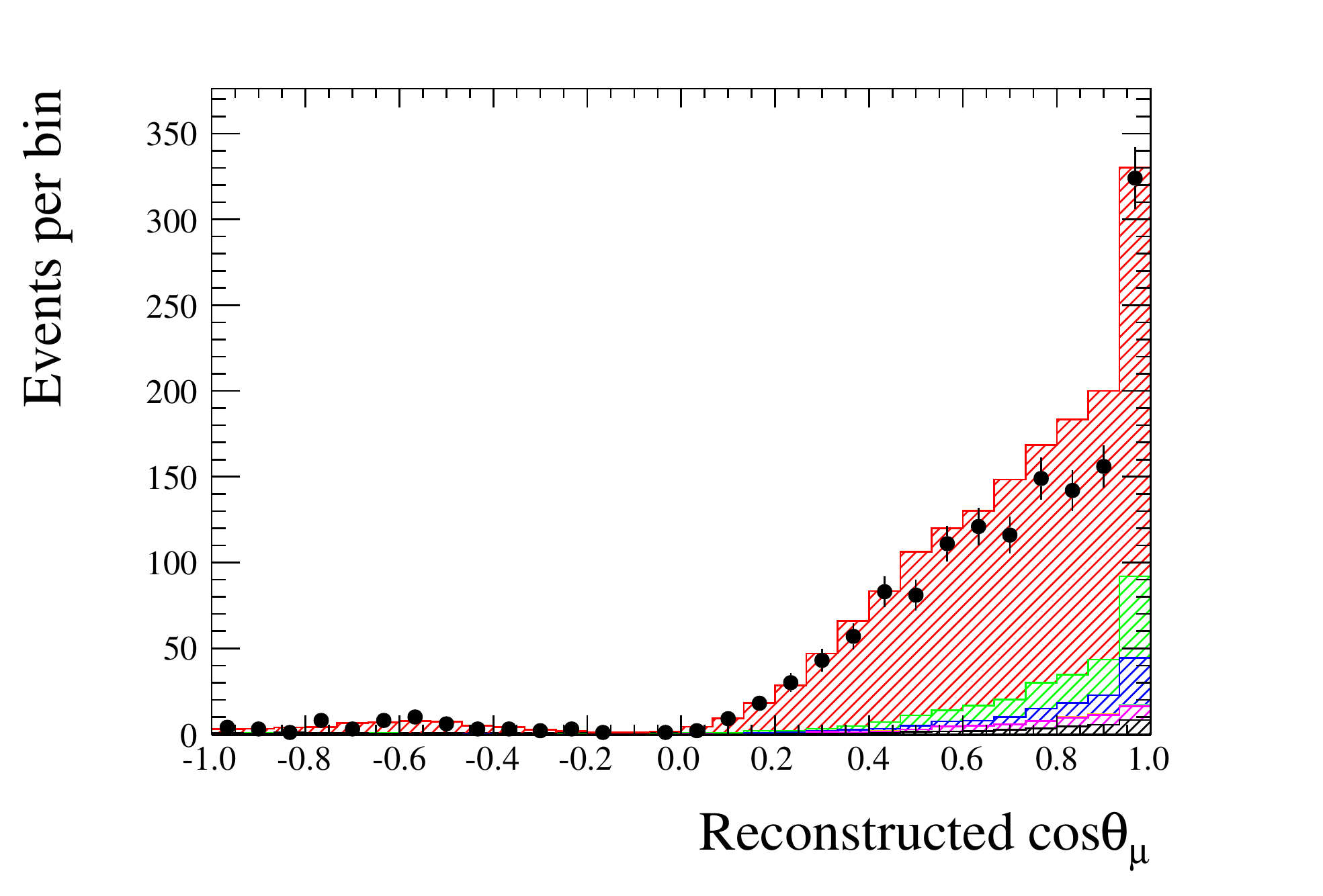}
	
	\includegraphics[width=0.38\linewidth]{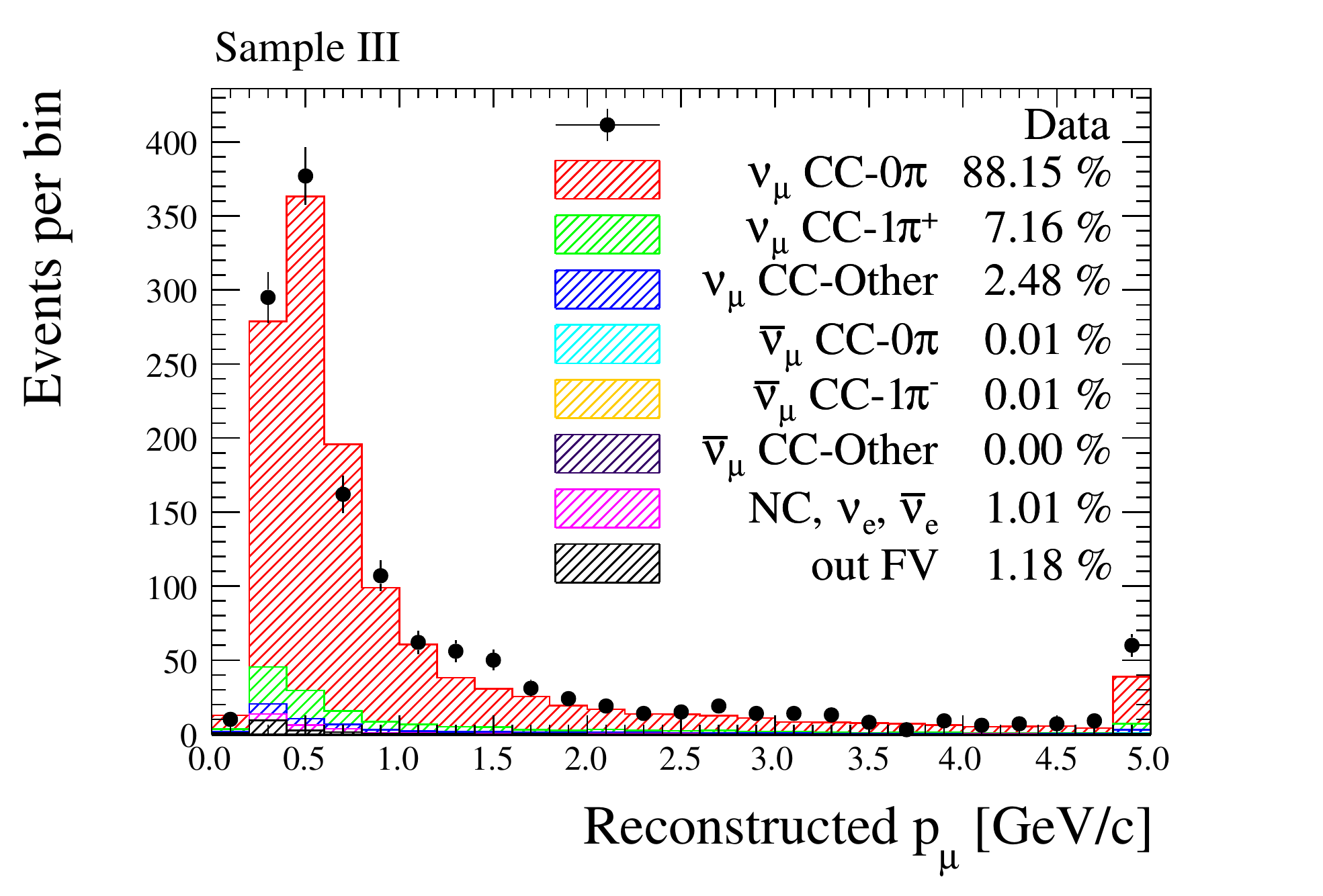}
	\includegraphics[width=0.38\linewidth]{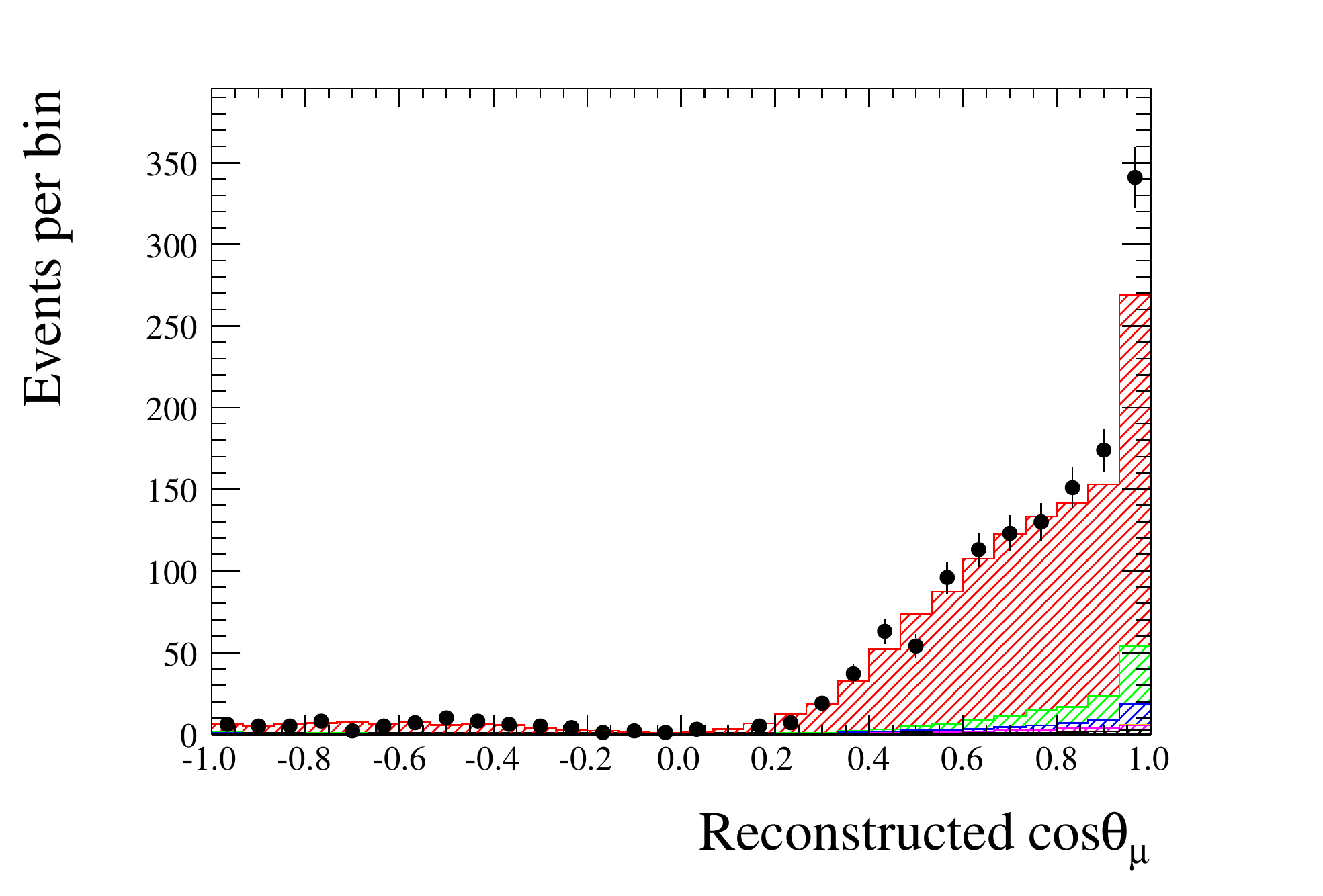}	
	
	\includegraphics[width=0.38\linewidth]{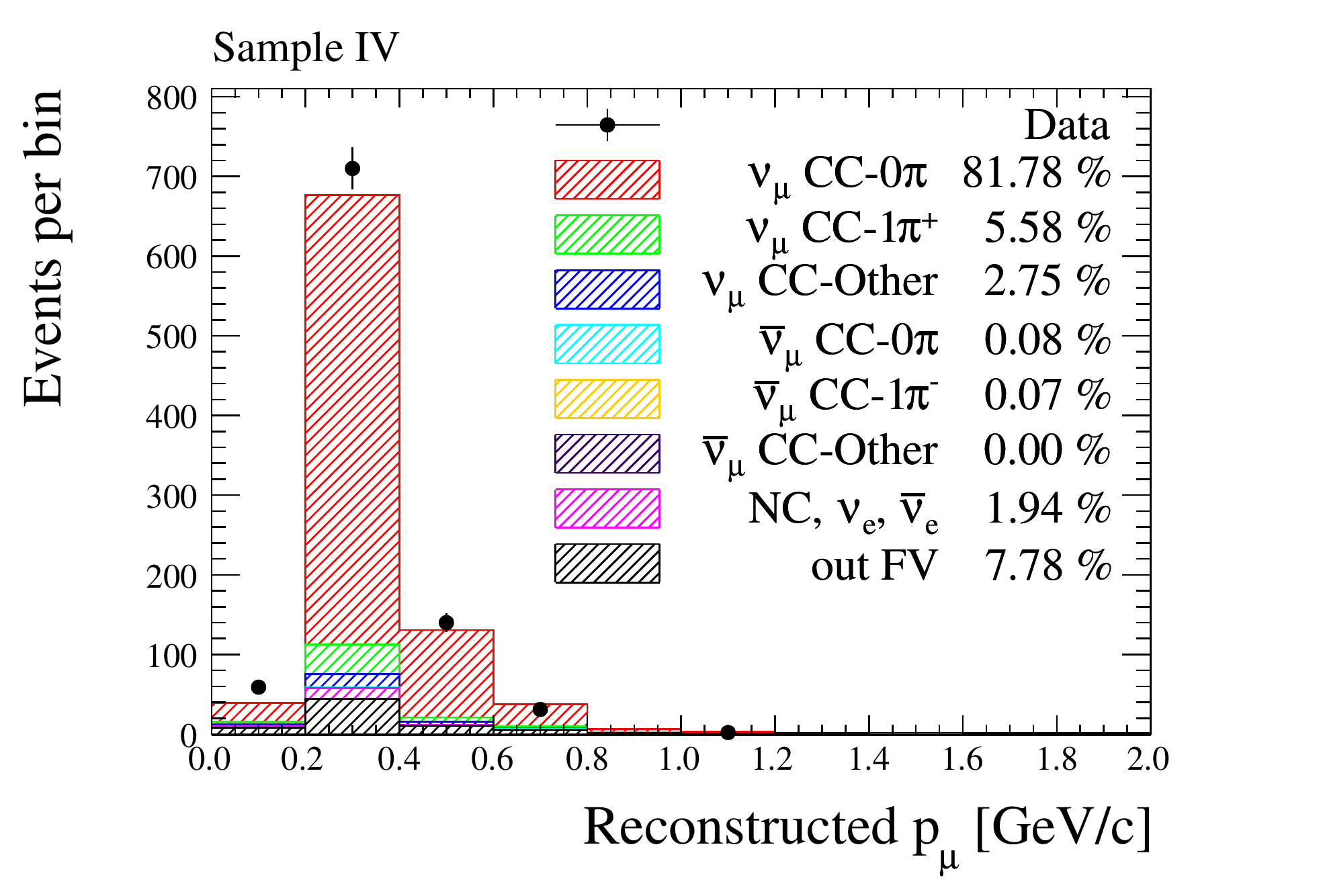}
	\includegraphics[width=0.38\linewidth]{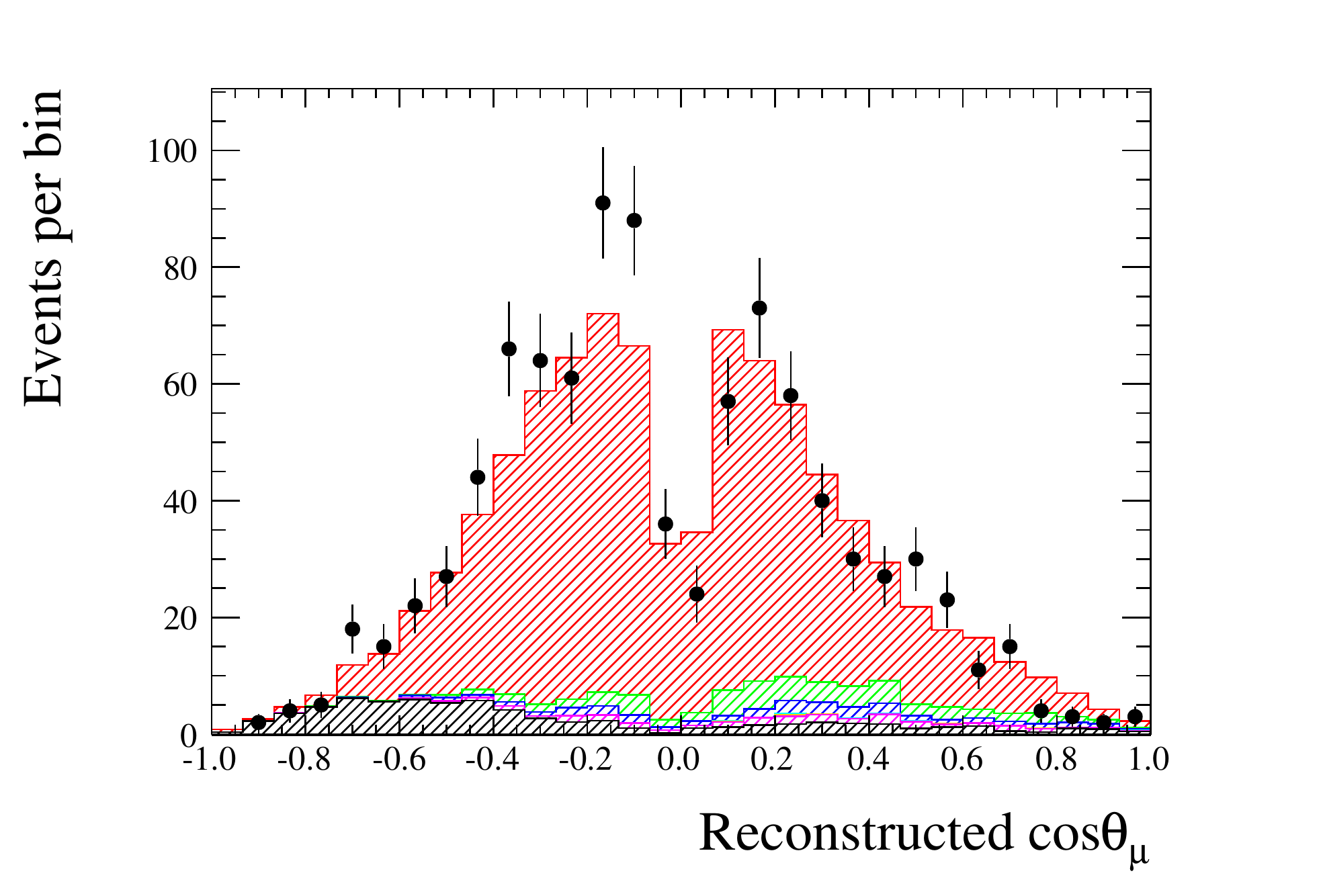}
	
	\includegraphics[width=0.38\linewidth]{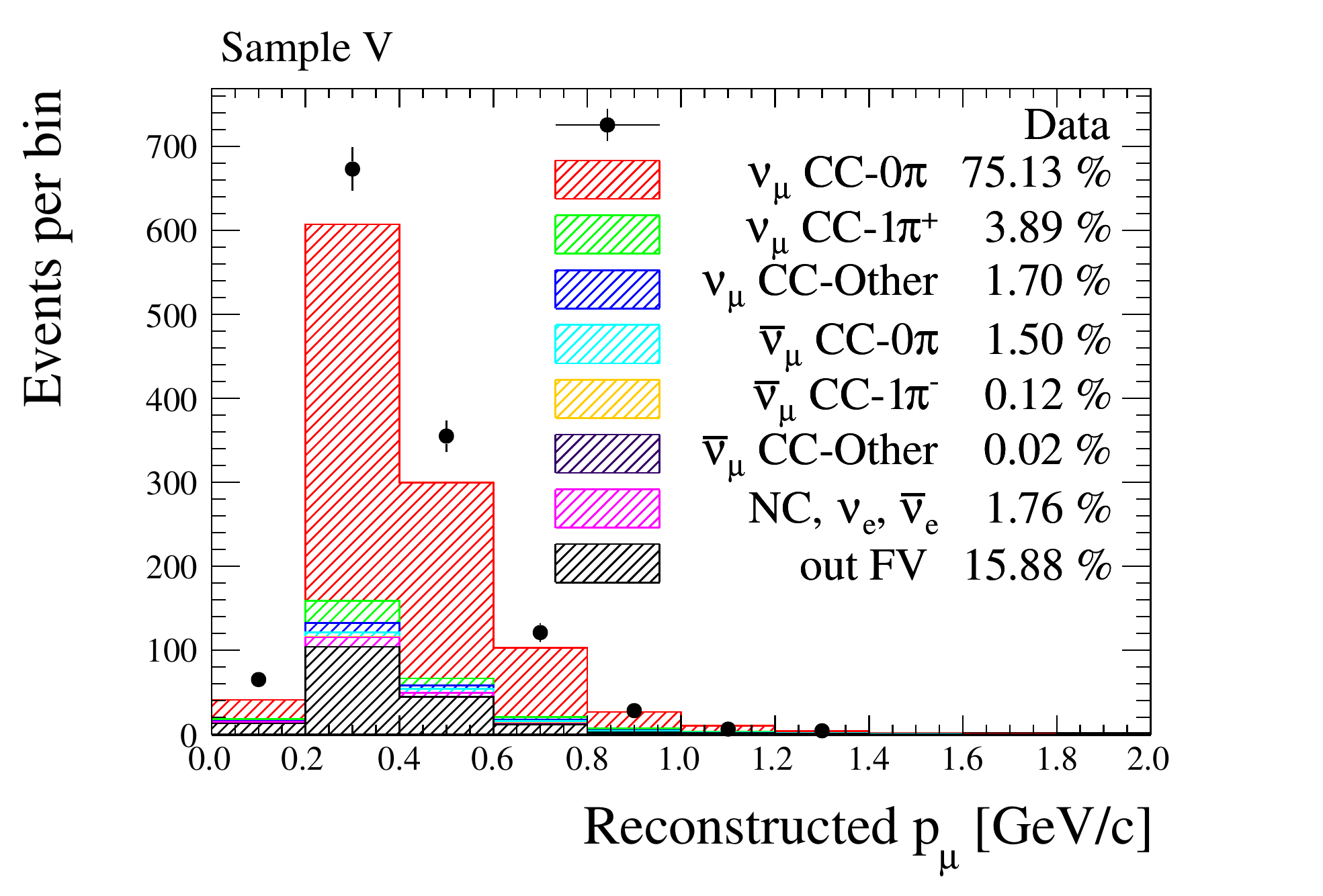}
	\includegraphics[width=0.38\linewidth]{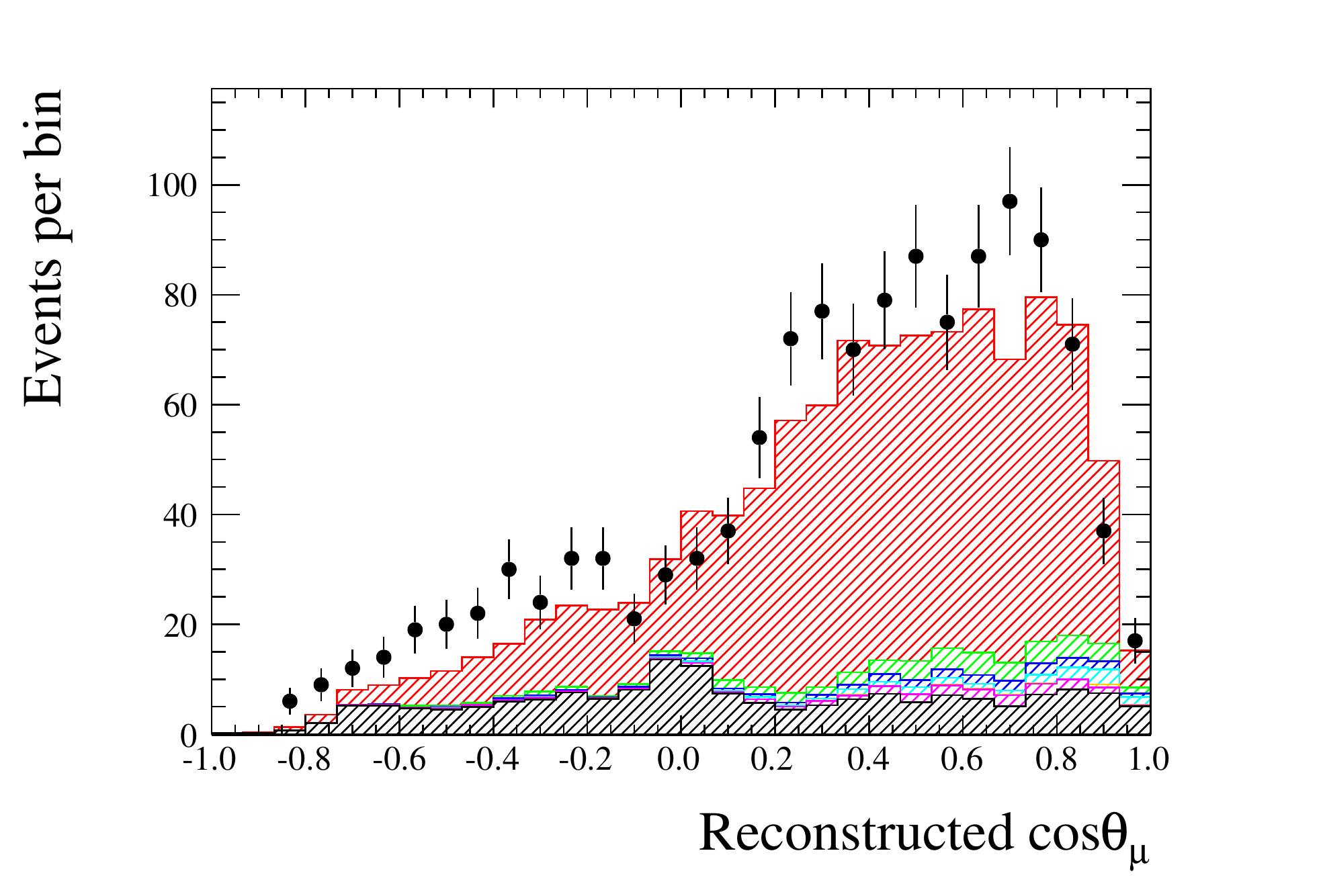}
	
	\caption{Distribution of events in the different signal samples for the neutrino sample. In the left column the number of events are plotted against the reconstructed muon momentum, while in the right column against the reconstructed muon $\cos\theta_\mu$. Histograms are stacked in true topologies. The last bin of the reconstructed muon momentum distributions contains all the events with momentum greater than 5 GeV/c for the first three samples, and 2 GeV/c for the last two. The MC has been normalized to 5.80 $\times 10^{20}$ p.o.t, the number of p.o.t in data. The legends show also the fraction for each component.}
	\label{fig:eventsDistributionsNuMu}
\end{figure*}

The selection described in this section is an improved version of the one used in Analysis I in Ref.~\cite{Abe:2016tmq}, and similar to that detailed in Ref.~\cite{Abe:2018pwo} where it has been extensively described. 

The target for \numu interactions is FGD1. This is used also as a tracker with TPC1, TPC2, ECal and SMRD. After the first requirements on the data quality and the position of the vertex are fulfilled, the selection requires tracks with a TPC segment with good reconstruction quality. For such tracks, the negatively charged one with the highest momentum, and compatible with the muon hypothesis according to the TPC PID is identified as a muon candidate. Tracks fully contained in the FGD and compatible with the energy loss by a muon have also been selected.

Protons are selected by looking for a track which starts in the FGD1 FV. The track should be identified as a positively charged in a TPC, and passes both the TPC track quality cut and PID criteria. Alternatively, if the track stops within the FGD it is identified as a proton if the track is consistent with the FGD proton hypothesis. To ensure the cross section is fully inclusive in terms of numbers of protons, events without a reconstructed proton are also included. Proton selection helps in further enlarging the phase space to high-angle and backward muons, as shown in Analysis I of Ref.~\cite{Abe:2016tmq}. The selected events are divided into five signal samples:
\begin{description}
	\item [Sample I] characterized by events with only a muon candidate in one of the TPCs (TPC2 if the muon is going forward and TPC1 if it is going backward),
	\item [Sample II] a muon candidate in one of the TPCs and one proton candidate in TPC2,
	\item [Sample III] a muon candidate in one of the TPCs and a proton candidate in FGD1,
	\item [Sample IV] a muon candidate in FGD1 and one proton in TPC2;
	\item [Sample V] only a muon candidate in FGD1 that reaches the ECal or SMRD.
\end{description}
Events with a muon candidate in FGD or TPC and more than one proton in the final state, with the leading proton in TPC, have been selected as well. As these events only accounts for 0.8\%, they have been added to the signal samples II-IV, accordingly with the muon candidate position (track in FGD only or in TPC). \cref{fig:numusignalsamples} summarizes the signal samples described above.

\begin{figure*}[th!]
	\centering
	
	\includegraphics[width=0.49\linewidth]{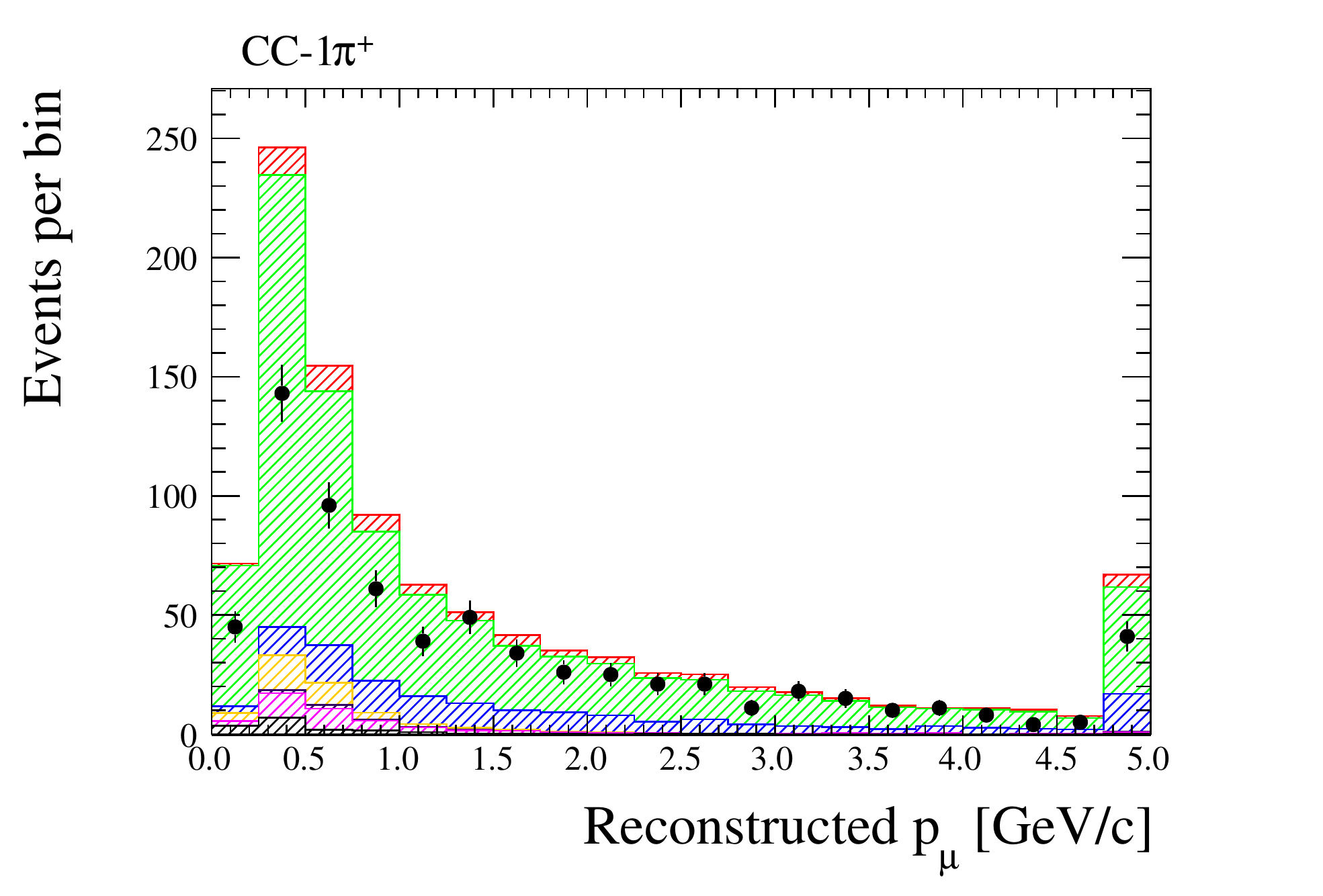}
	\includegraphics[width=0.49\linewidth]{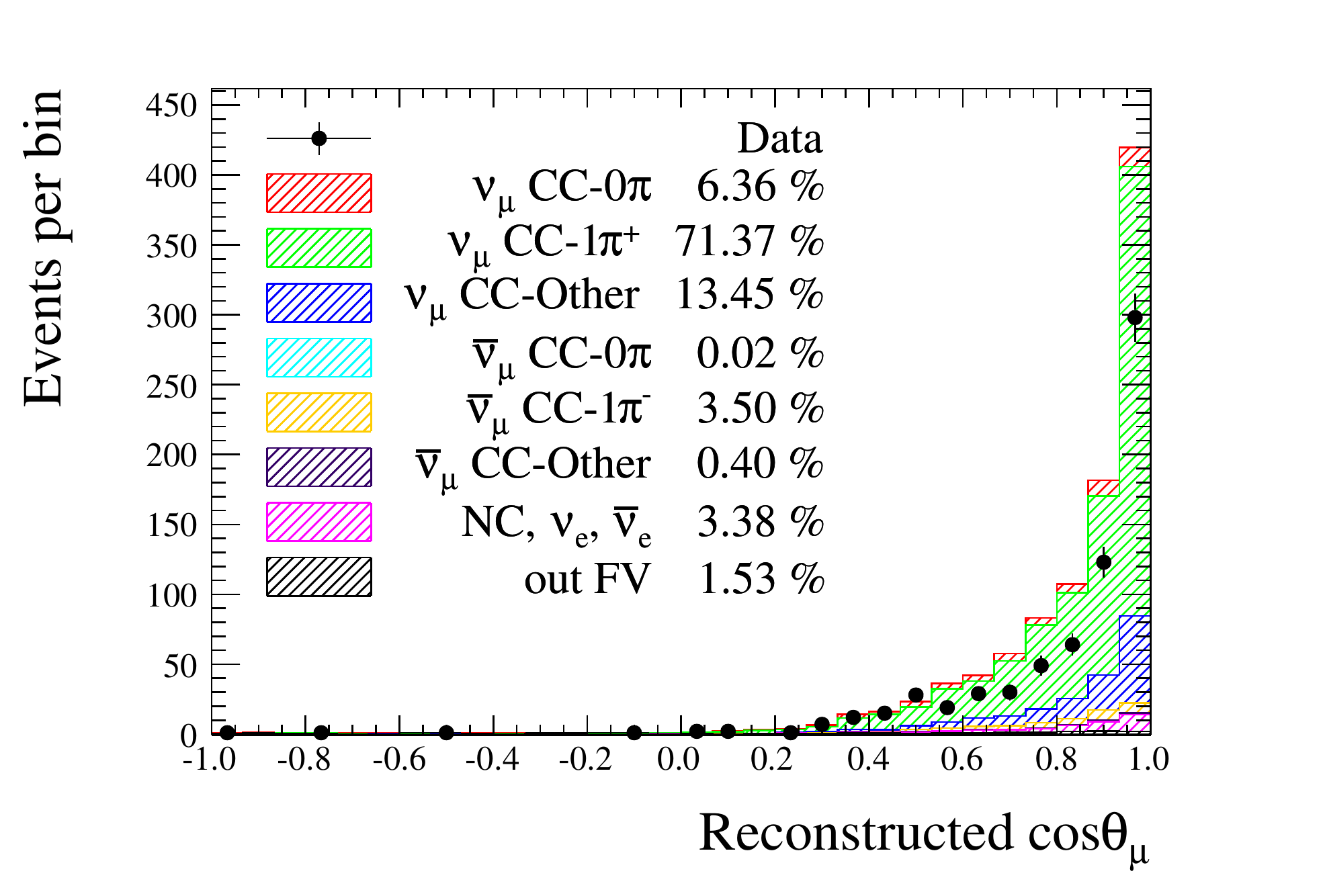}
	
	\includegraphics[width=0.49\linewidth]{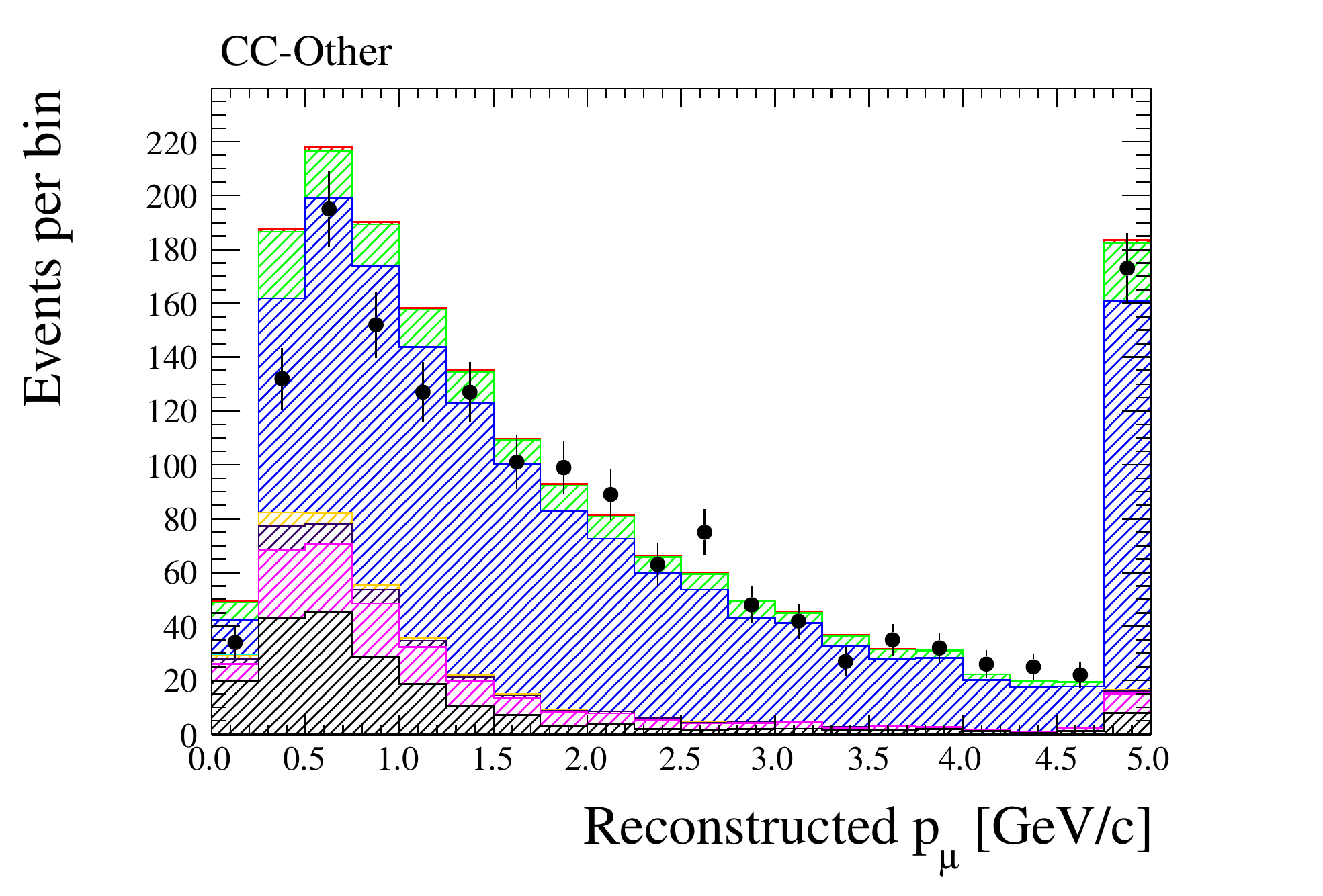}
	\includegraphics[width=0.49\linewidth]{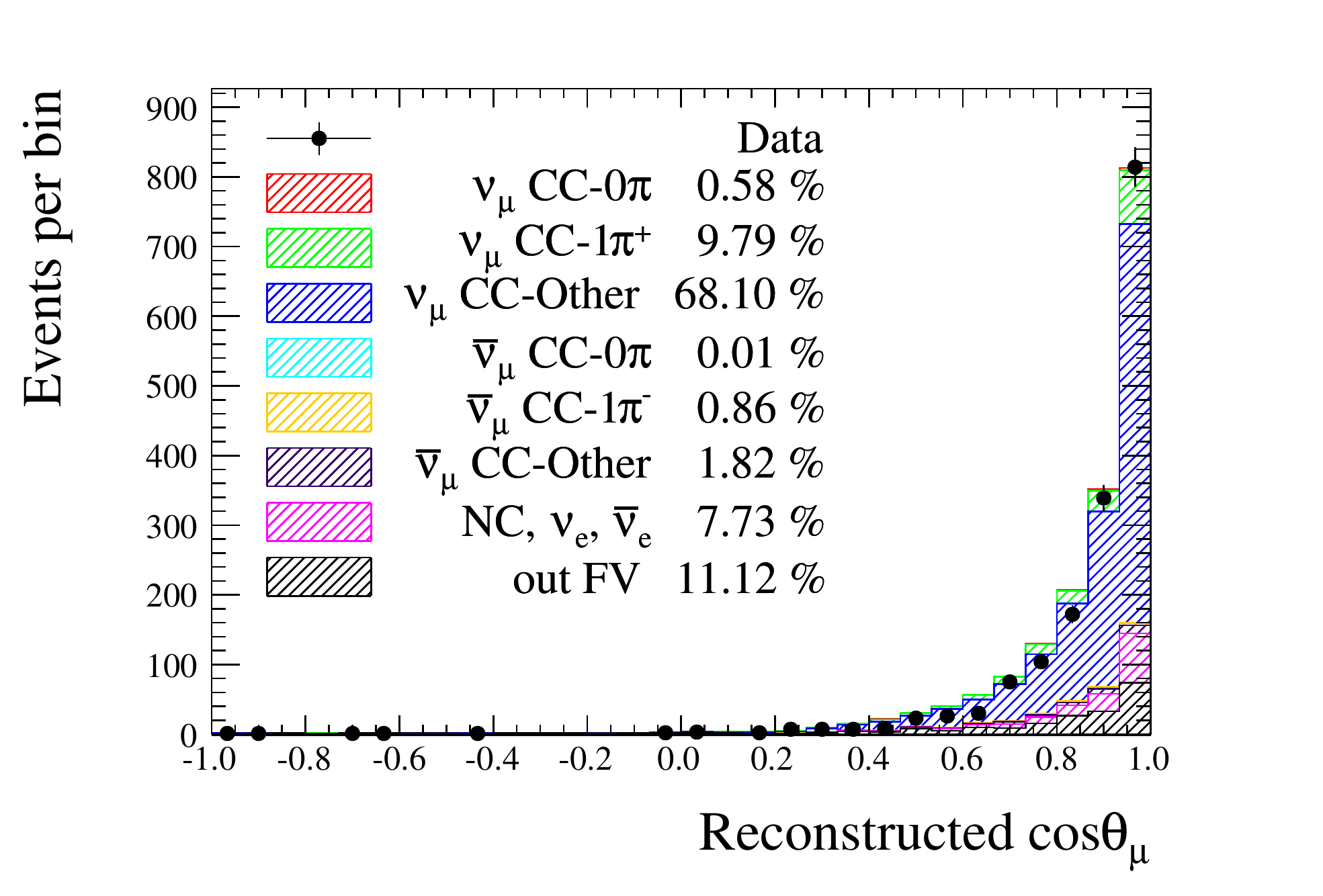}
	
	\caption{Distribution of events in the two control samples for the neutrino sample: CC-1$\pi^+$ in the top row and CC-Other in the bottom. In the left column number of events per bin are plotted against the reconstructed muon momentum, while in the right column against the reconstructed muon $\cos\theta_\mu$. The last bin of the reconstructed muon momentum distributions contains all the events with momentum greater than 5 GeV/c. Histograms are stacked in true topologies. The MC has been normalized to 5.80 $\times 10^{20}$ p.o.t, the number of p.o.t in data. The legends show also the fraction for each component.}
	\label{fig:eventsDistributionsBkgNuMu}
\end{figure*}

The kinematics of the muon candidate in each sample for the \cczeropi signal and the various backgrounds are shown in \cref{fig:eventsDistributionsNuMu} where the MC is broken down by true topologies. The selection is dominated by events with one reconstructed muon and no other tracks. The signal samples where the muon is reconstructed in the TPC have very similar momentum distributions, although events with a reconstructed proton tend to have muons at slightly larger angles. The sample with the muon in the FGD and the proton in the TPC have muons with much smaller momenta and larger angles. 

The \numu \cczeropi cross section is extracted by adding together the contributions from all the samples, but it is important to keep the events with and without protons and with the muon in different sub-detectors separated in the analysis because they are affected by different systematics and backgrounds.

The main background arises from charged-current events with one true positively charged pion (\cconepiplus), or any number of true pions (\ccother) which are misidentified or not reconstructed. Neutral current interactions (\nc) and interactions that occurred outside the FV (\oofv) but were reconstructed inside constitute a smaller background. Two control samples were selected to constrain charged current event rates with single-pion and multiple-pion production: the \cconepiplus sample is made up of events with exactly two tracks, one negatively charged muon and one positively charged pion, and the \ccother sample, made of events with more than one pion in the final state. Pions have been identified in different ways according to their charge. A $\pi^+$ can be identified by looking at the curvature of the track in the TPC and by requiring that the energy loss in this detector is consistent with a pion. $\pi^-$ are only identified by looking at the curvature of the tracks while $\pi^0$ are identified by looking for tracks in the TPC with charge depositions consistent with an electron from a $\gamma$ conversion. The kinematic distributions of the control samples are shown in \cref{fig:eventsDistributionsBkgNuMu}. The data-MC discrepancy in the kinematic distributions of the \cconepiplus control sample was already observed in previous analysis~\cite{Abe:2016tmq,Abe:2018pwo} and is corrected by the likelihood fit as shown in \cref{sec:resultscomp}.

\begin{figure*}[th!]
	\centering
	\includegraphics[width=1.\linewidth,trim={0 19cm 0 0},clip]{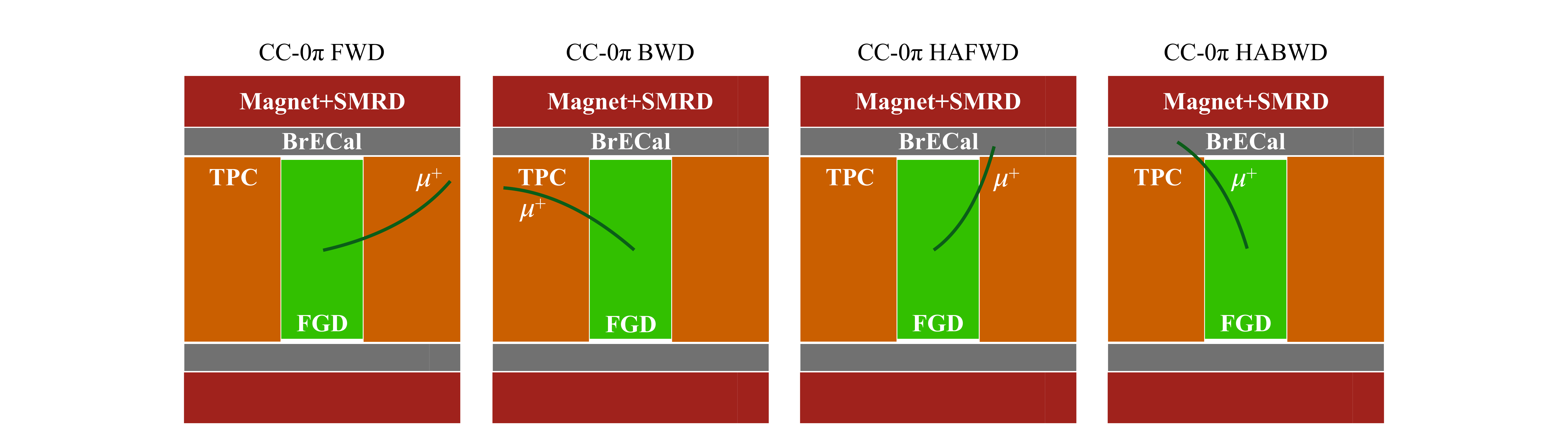}
	\caption{Schematic representation of the different \barnumu signal samples. In each drawing a \barnumu enters from the left and interacts in FGD1. The sub-detectors of ND280 are shown in their side view.}
	\label{fig:antinumusignalsamples}
\end{figure*}

\subsubsection{\barnumu CC event selection}

After the first two common requirements described at the beginning of the Sec.~\ref{sec:selections} are fulfilled, the events are divided in four samples depending on the length of the muon candidate track in the TPCs and its direction, following the same strategy described in a recent T2K publication~\cite{Abe:2018uhf}:

\begin{itemize}
	
	\item If the muon candidate travels forward w.r.t. the beam direction and the associated track has more than 18 hits in TPC2 then the event belongs to the forward (FWD) sample;
	
	\item If it travels backward and the associated track has more than 18 hits in TPC1 then the event belongs to the backward (BWD) sample;
	
	\item If the muon candidate travels forward but the track has less than 19 hits in TPC2 then the event belongs to the high angle forward (HAFWD) sample;
	
	\item if it travels backward and the associated track has less than 19 hits in TPC1 then the event belongs to the high angle backward (HABWD) sample.
	
\end{itemize}

\begin{figure*}[ht!]
	\centering
	\includegraphics[width=0.38\linewidth]{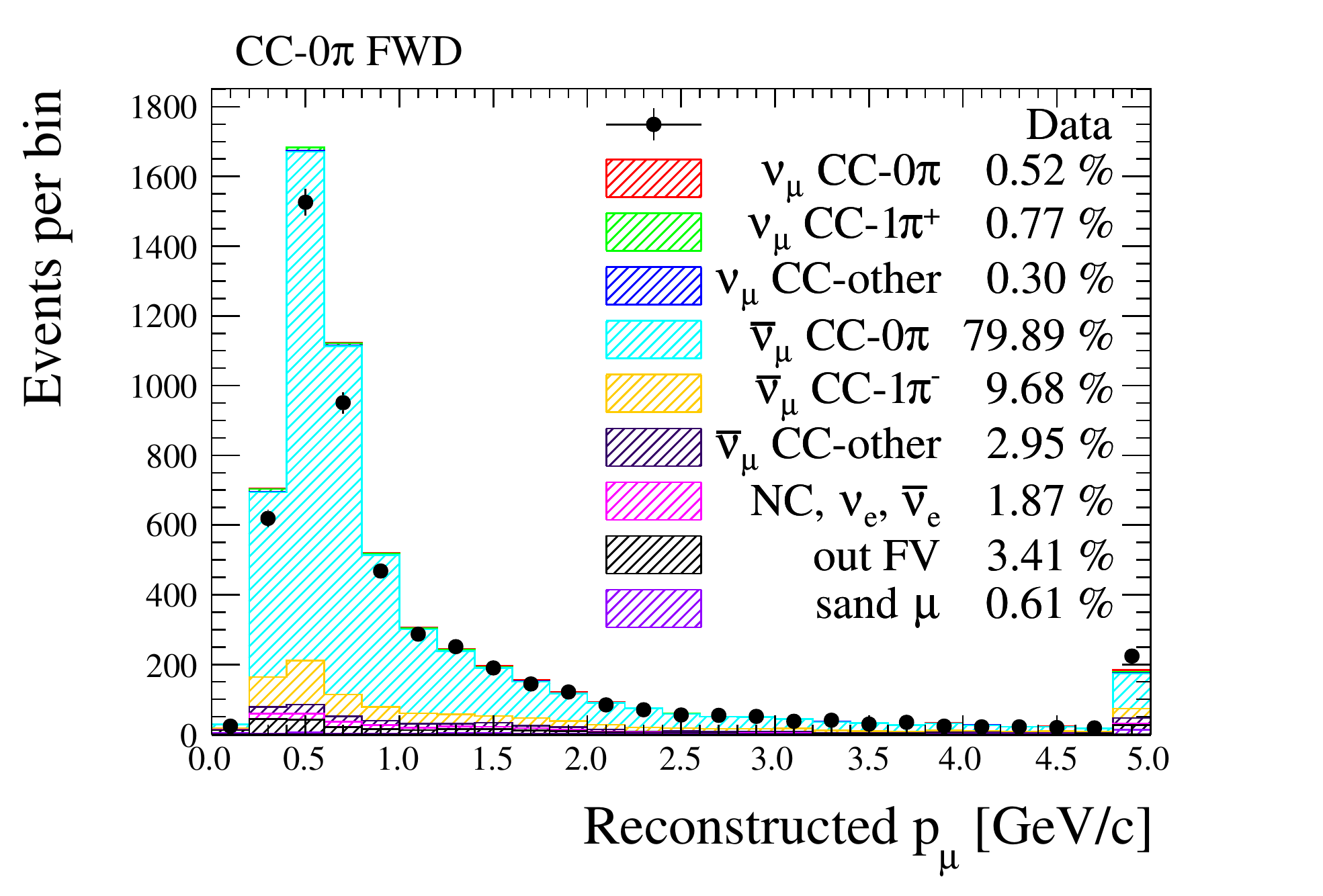}
	\includegraphics[width=0.38\linewidth]{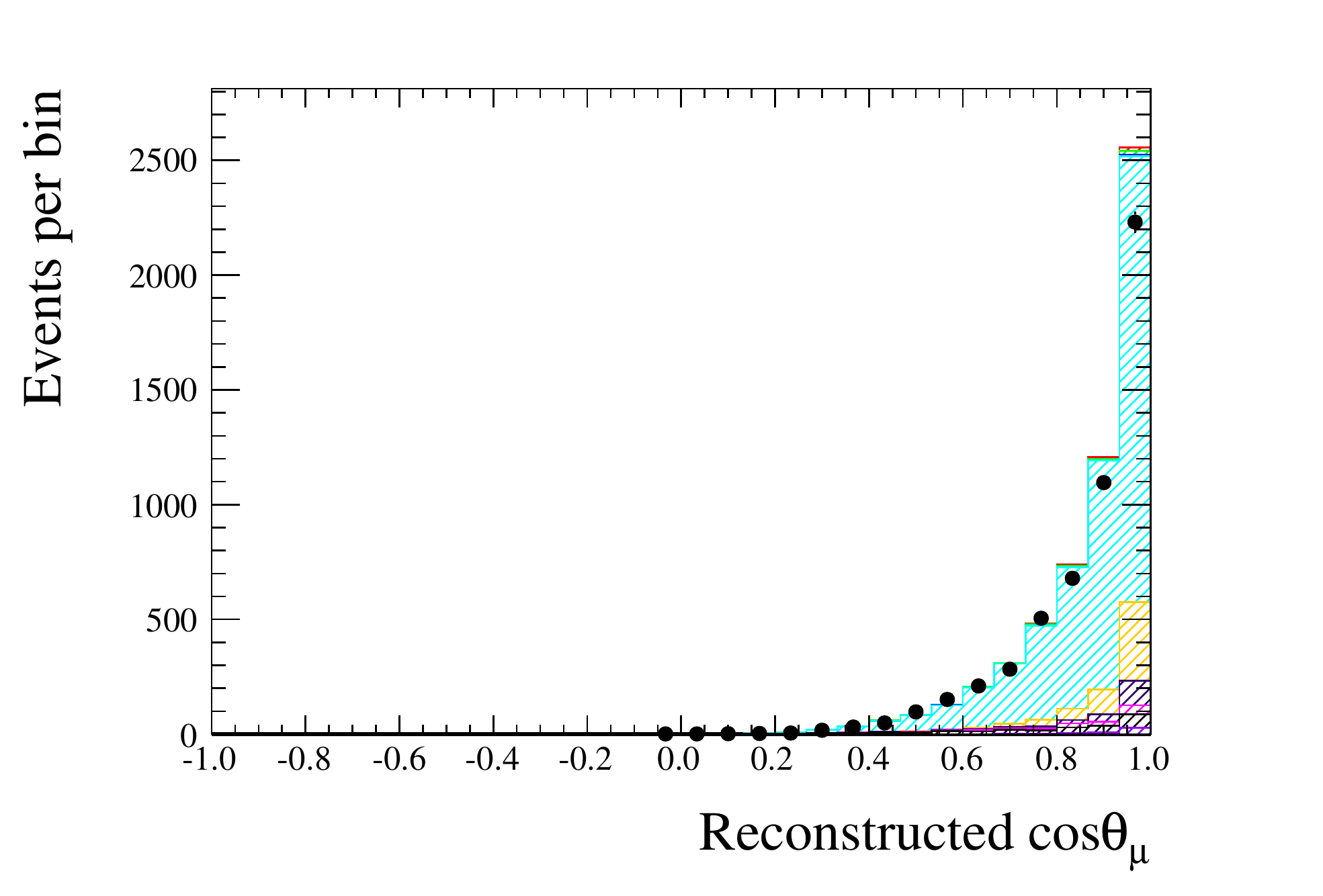}
	
	\includegraphics[width=0.38\linewidth]{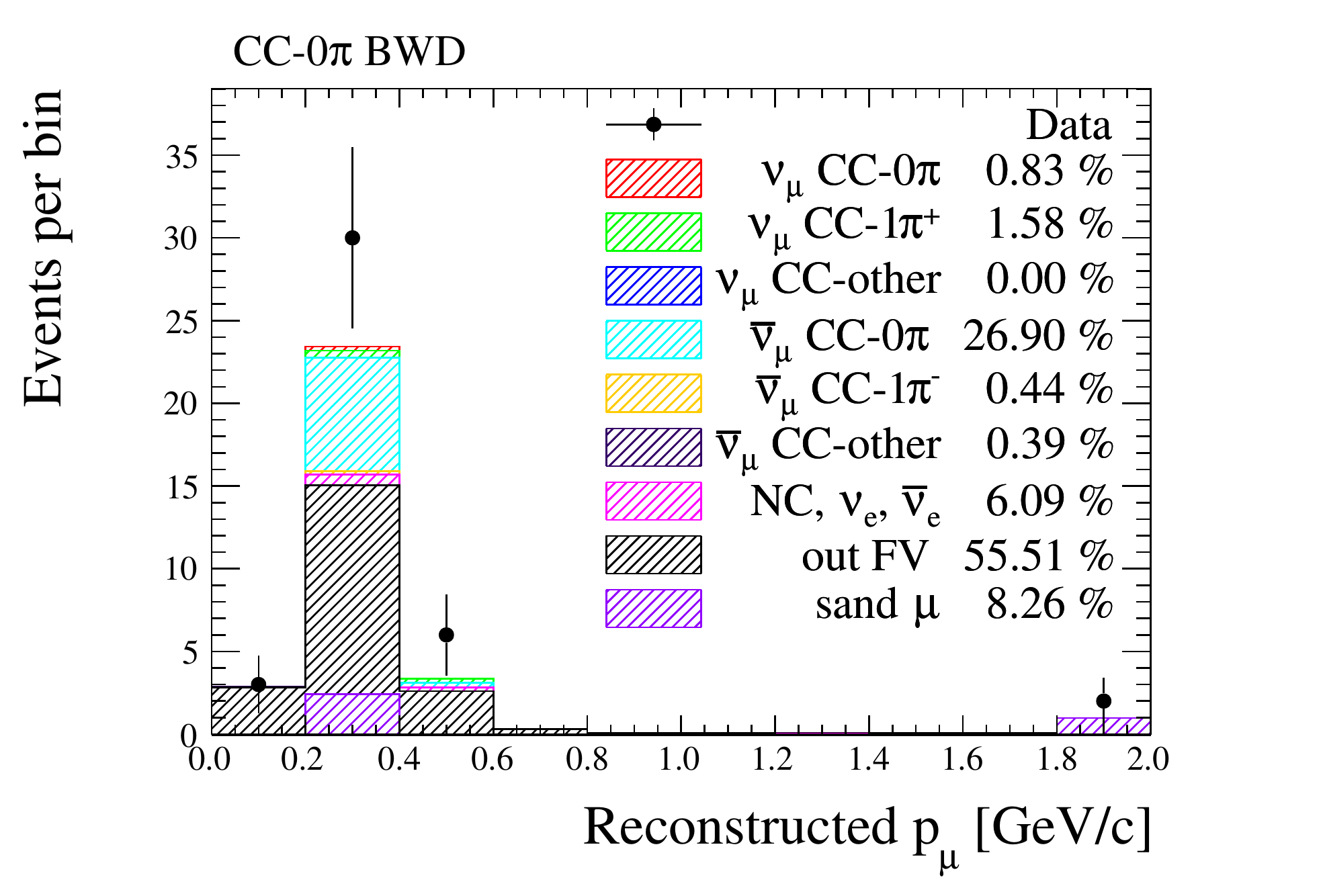}
	\includegraphics[width=0.38\linewidth]{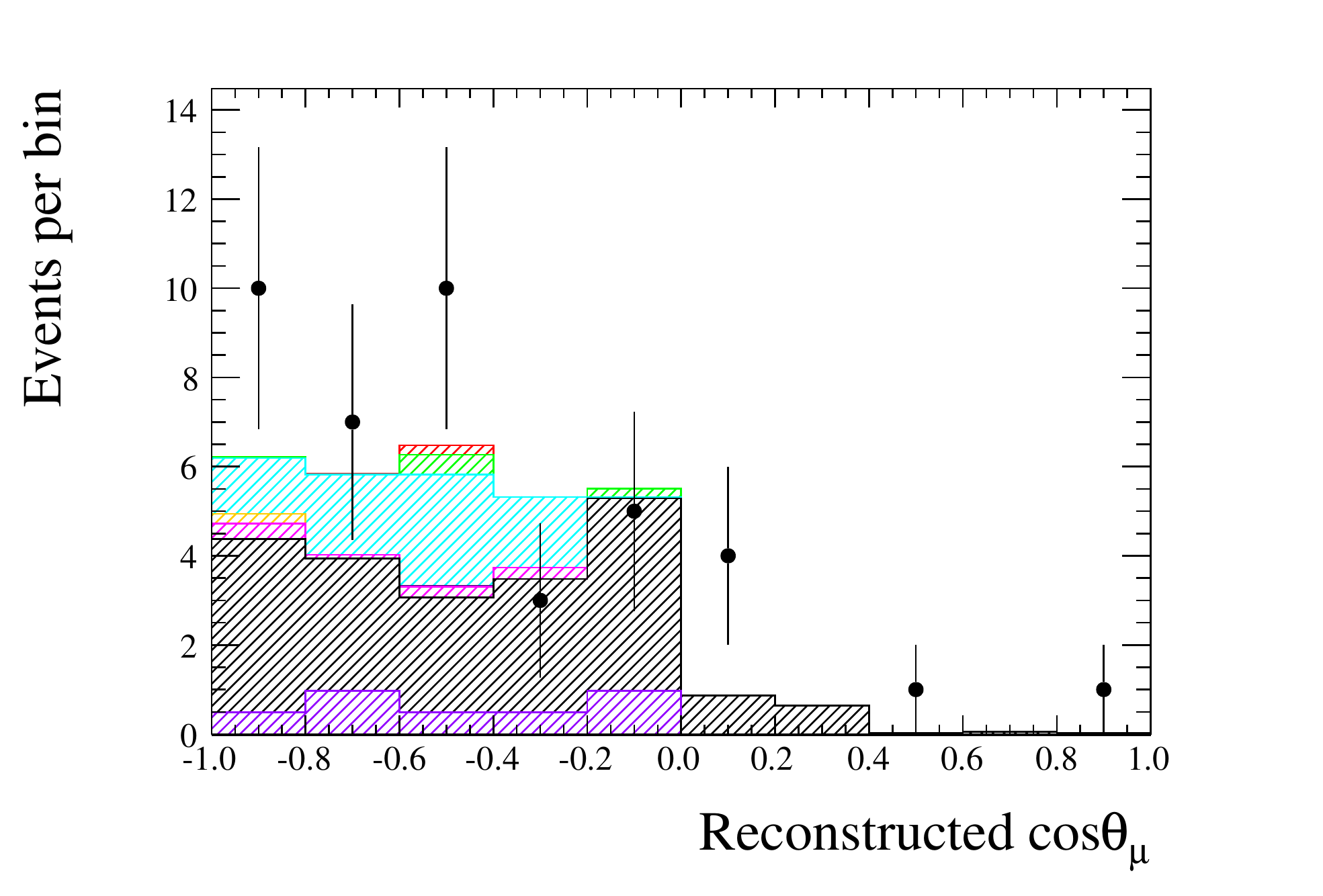}
	
	\includegraphics[width=0.38\linewidth]{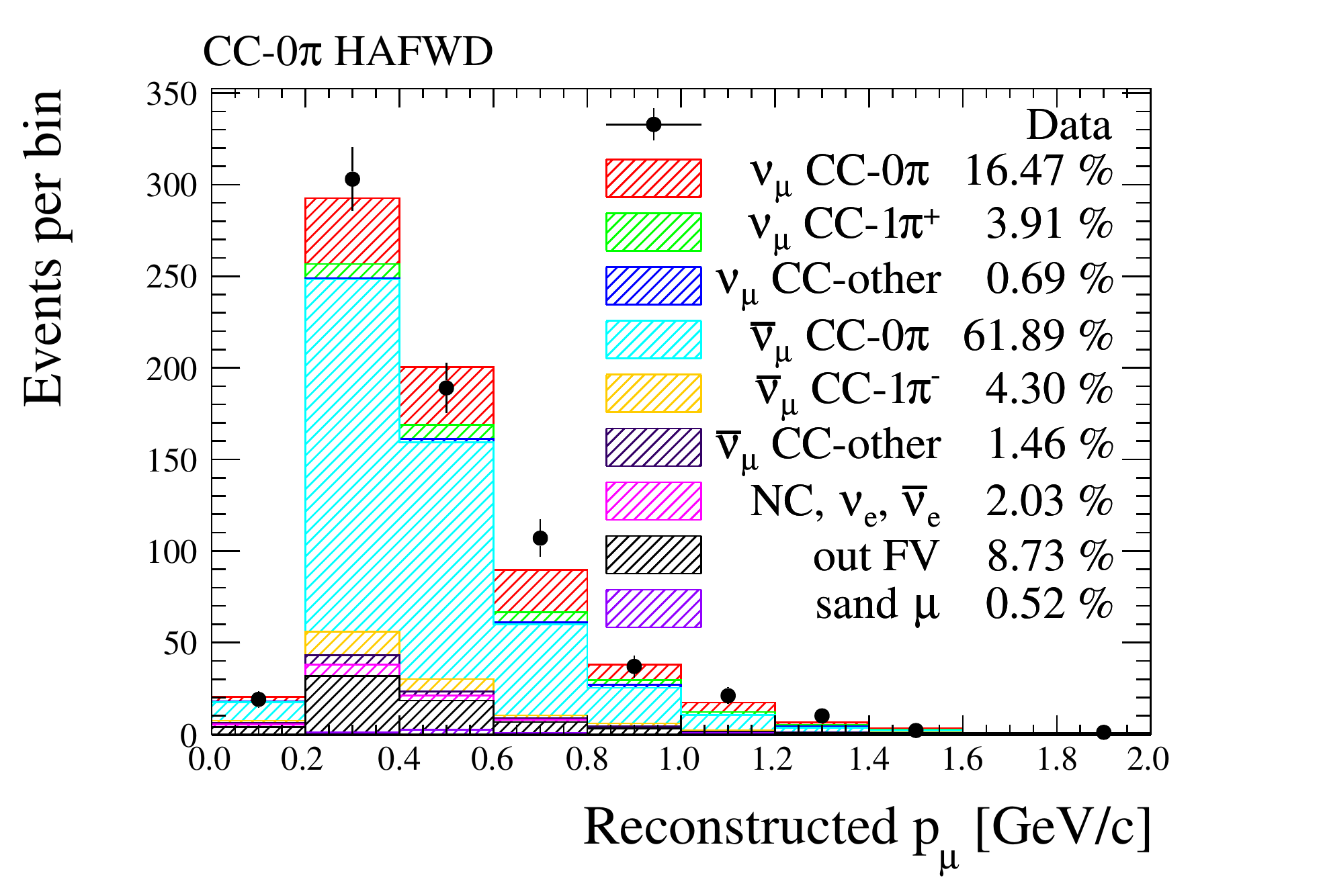}
	\includegraphics[width=0.38\linewidth]{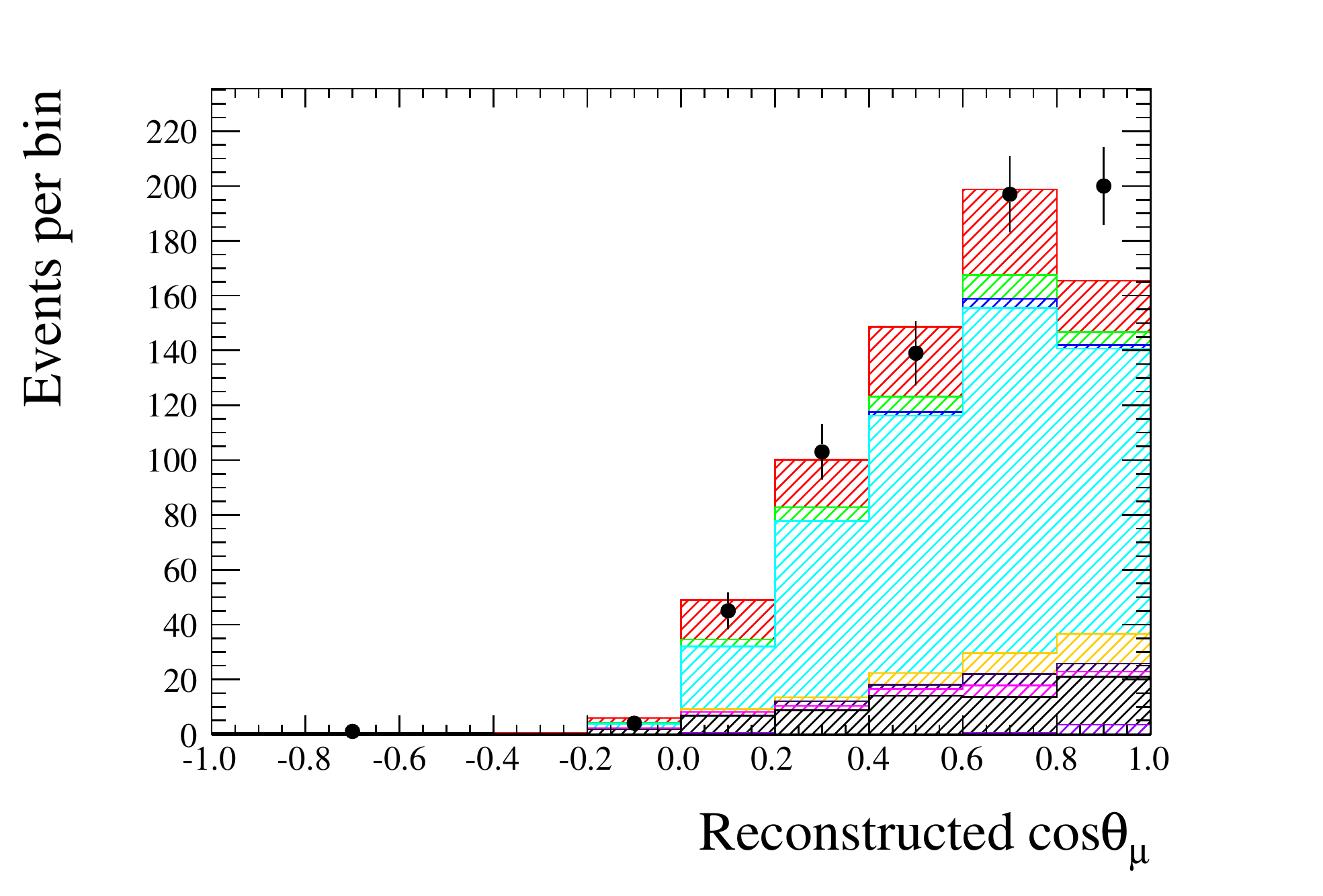}
	
	\includegraphics[width=0.38\linewidth]{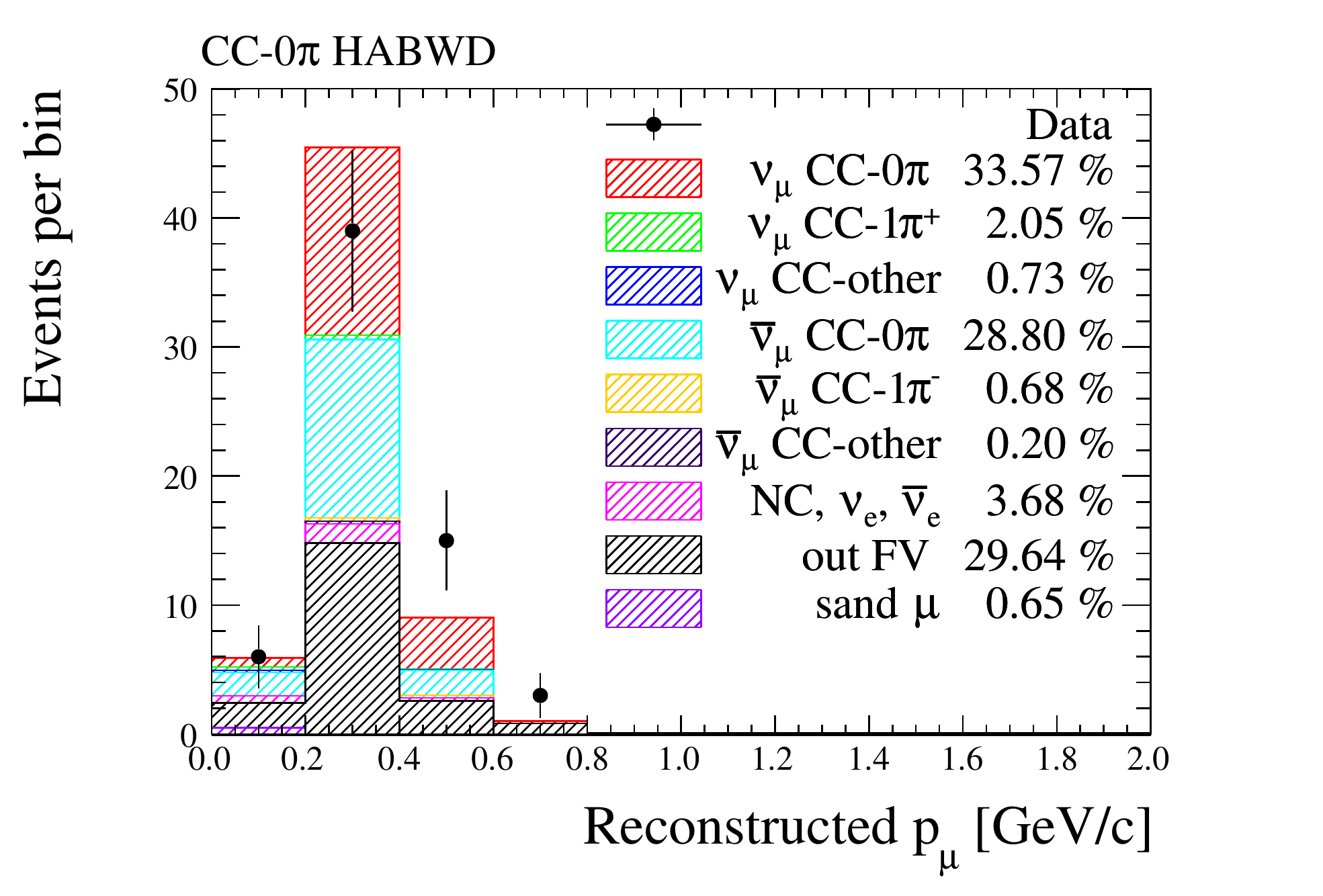}
	\includegraphics[width=0.38\linewidth]{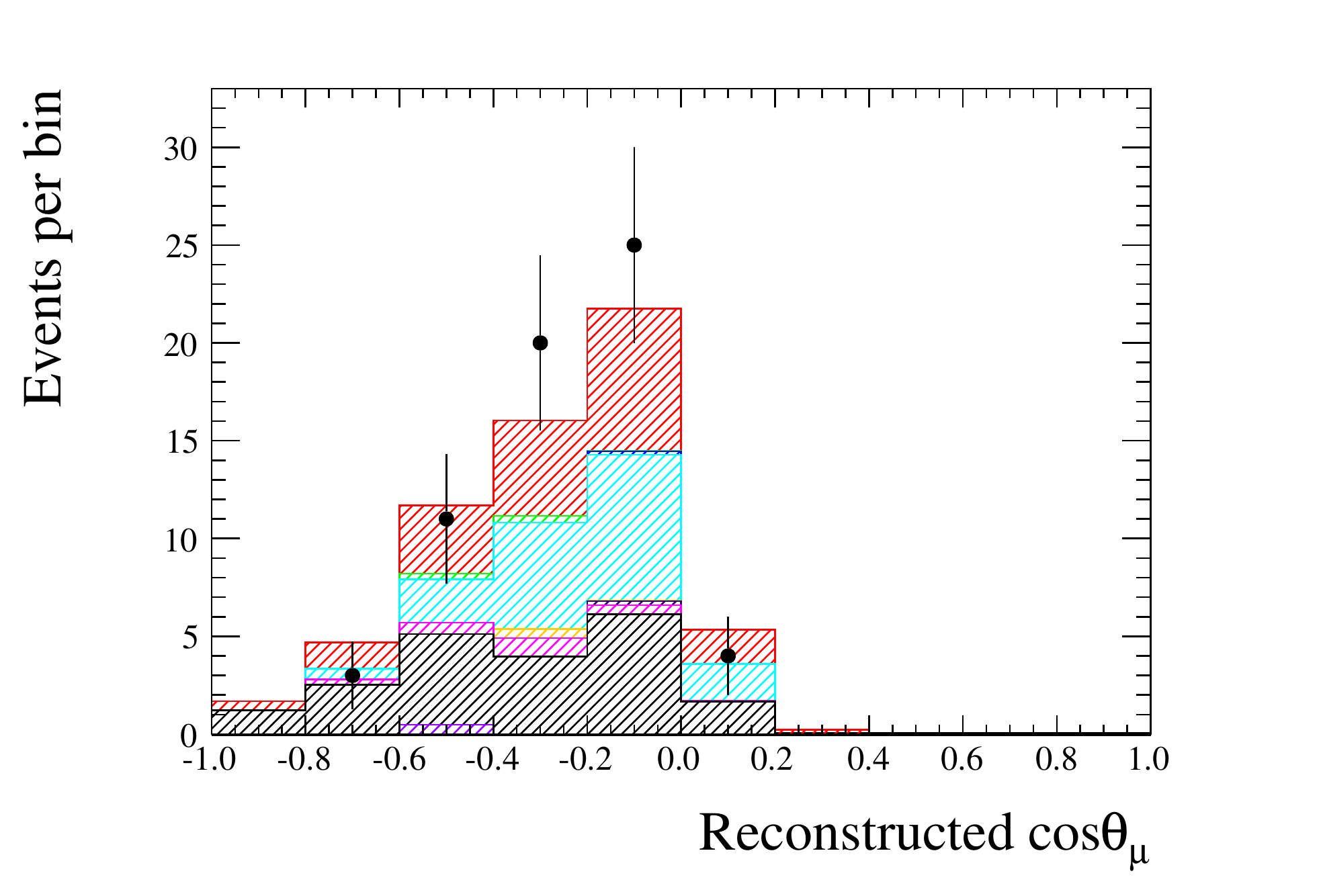}
	
	\caption{Distribution of events in the different signal samples (\barnumu \cczeropi FWD, BWD, HAFWD and HABWD). In the left column the number of events per bin are plotted against the reconstructed muon momentum, while in the right column against the reconstructed muon $\cos\theta_\mu$. The last bin of the reconstructed muon momentum distributions contains all the events with momentum greater than 5 GeV/c for the first sample, and 2 GeV/c for the others. The MC has been normalized to 6.27 $\times 10^{20}$ p.o.t, the number of p.o.t in data. Histograms are stacked in true topologies. The legends show also the fraction for each component.}	
	\label{fig:eventsDistributionsAntiNuMu}
\end{figure*}

\begin{figure*}[ht!]
	\centering
	
	\includegraphics[width=0.49\linewidth]{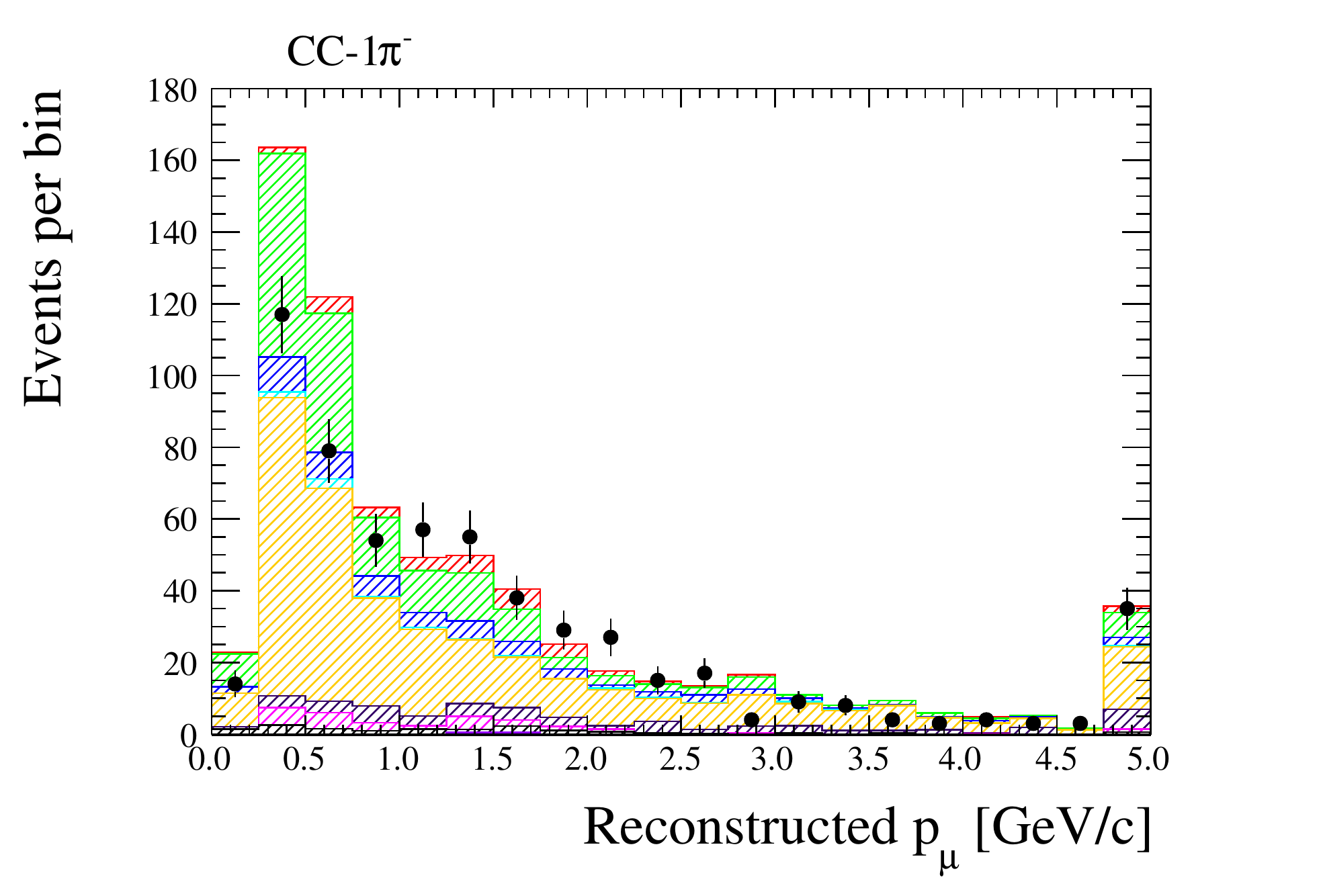}
	\includegraphics[width=0.49\linewidth]{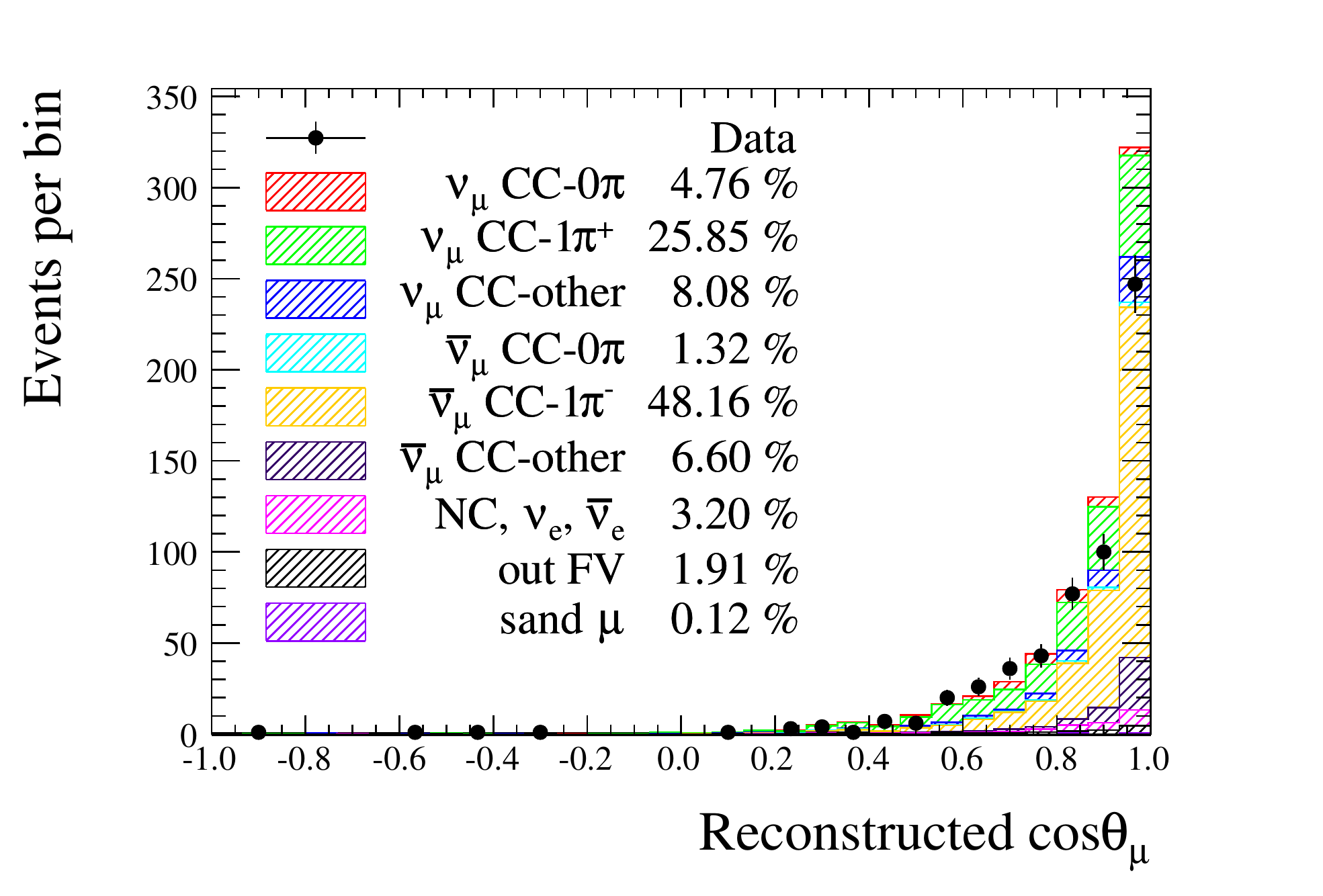}
	
	\includegraphics[width=0.49\linewidth]{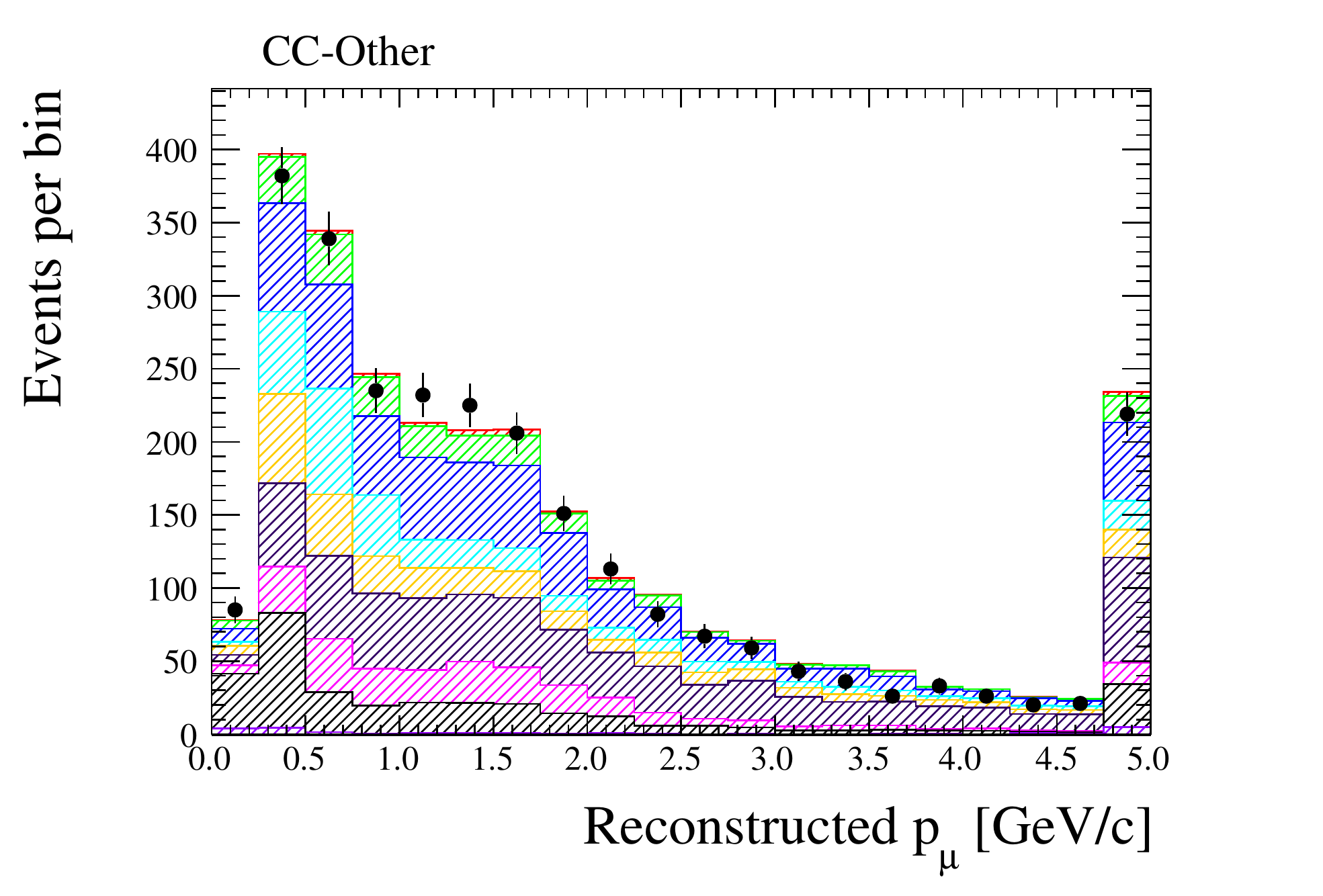}
	\includegraphics[width=0.49\linewidth]{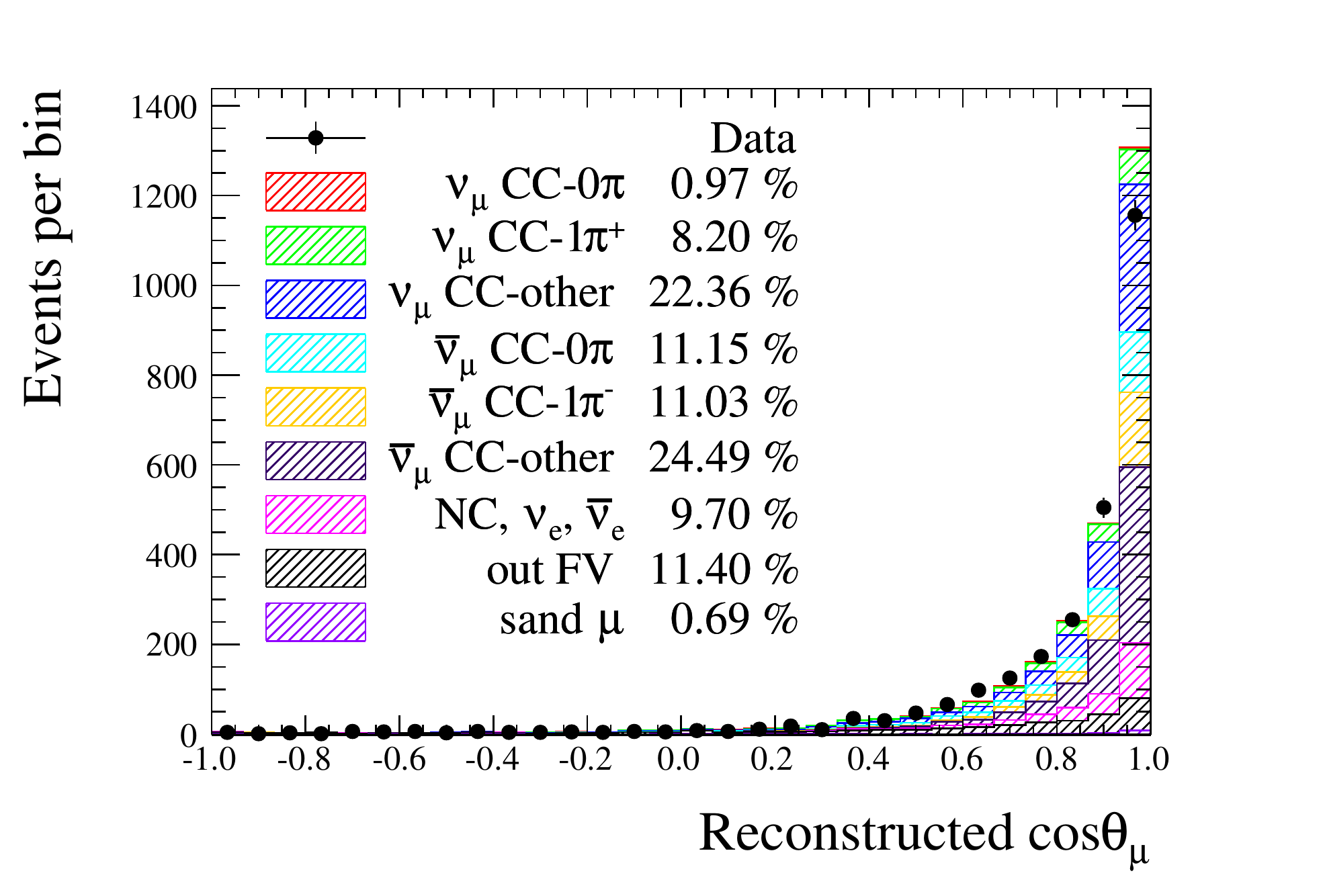}	
	
	\caption{Distribution of events in the two sidebands for the antineutrino sample: CC-1$\pi^-$ in the top row and CC-Other in the bottom. In the left column the number of events are plotted against the reconstructed muon momentum, while in the right column against the reconstructed muon $\cos\theta_\mu$. The last bin of the reconstructed muon momentum distributions contains all the events with momentum greater than 5 GeV/c. Histograms are stacked in true topologies. The MC has been normalized to 6.27 $\times 10^{20}$ p.o.t, the number of p.o.t in data. The legends show also the fraction for each component.}
	\label{fig:eventsDistributionsBkgAntiNuMu}
\end{figure*} 

For each sample, different sets of cuts have been developed to reduce the background as much as possible without decreasing the efficiency. 

One of the main backgrounds is caused by interactions that occur outside the FV (\oofv) but are incorrectly reconstructed as starting inside the FV. This can be due to a failure of the reconstruction algorithms or a scattering of the particle which can lead to two unmatched track segments, one of which may start in the FV. The ratio of the momentum of the muon candidate to the other unmatched track and also the minimal distance between the tracks are used to reduce this background. The ratio should be lower than 1 if the two segments belong to the same track. Different cut values have been chosen for the event falling in the FWD, HAFWD and HABWD samples. These cuts are not applied in the selection of the BWD sample since signal events could be rejected. Another misreconstruction pathology can break a single track into two segments, with a reconstructed vertex inside the FV and a forward-going track into the downstream TPC. This often happens near the downstream edge of the FGD and the second track is considered as the muon candidate. Therefore, events for which the start position of the track associated to the muon candidate is in one of the last two layers of FGD1 are rejected.

The muon candidate is identified as the highest momentum track that is consistent with the muon PID. If the muon candidate enters a TPC, the track must pass the TPC muon PID. If the track does not enter a TPC, the ECal portion of the reconstructed object must be consistent with the ECal muon PID. In the case where the muon candidate enters a TPC, the charge of the track will be included in the selection. For particles entering ECal the information on the charge is not available, therefore this sample of events presents a high contamination of negatively charged muons that is constrained by measuring at the same time the \numu cross section.

In the selection of the FWD sample two additional cuts have been applied to reduce the pion and proton contamination: if the muon candidate stops in FGD2 and has a momentum greater than 280 MeV/c, the candidate is most likely a pion or a proton and the event is rejected; if it reaches the ECal it must have an ECal PID compatible with the muon hypothesis. 

The described cuts select samples of muon antineutrino CC events with muons in every direction. Every sample is then split in three sub-samples according to the event pion multiplicity: events without a reconstructed pion, with one negatively charged pion, or with more than one pion in the final state. The TPC pion selection is similar to the one described previously in \cref{sec:numuevtsel}. If the pion-candidate track is contained in the FGD1, pions are identified in two ways: by requiring a charge deposition consistent with a pion, or using the delayed energy deposition in the FGD due to a decay electron coming from $\pi \rightarrow \mu$ decay. In the latter case, the pion is tagged as positively charged since the $\pi^-$ are more likely to be absorbed.

The kinematics of the muon candidate for the signal sample are shown in \cref{fig:eventsDistributionsAntiNuMu} where CC events without pions in the final state have been divided in four samples depending on the direction of the muon: \cczeropi FWD, \cczeropi BWD, \cczeropi HAFWD and \cczeropi HABWD. In the \barnumu sample, the \numu contamination is larger than the \numu contamination in the \barnumu sample. Moreover, positively charged pions (and, to a smaller extent, protons) produced in \numu interactions can be misidentified as muons constituting an irreducible background. In the high-angle selection, the charge is not reconstructed, therefore negatively charged muons are also selected (\cref{sec:anaStrategy}). The statistics of the BWD sample is limited as the antineutrino cross section is suppressed for backwards going muons. \cref{fig:antinumusignalsamples} summarizes the \barnumu signal samples. 

The background mostly arises from events with one true negatively charged pion (\cconepiminus), any number of true pions (\ccother) and \oofv, that contributes more to the BWD and HABWD samples. The \cconepiminus and \ccother samples identified through the pion tagging are employed to constrain such backgrounds. For the \oofv background there is not a dedicated control sample. The majority of them are \numu CC interactions that are constrained by the existing control samples. An uncertainty on the prediction of the \oofv interactions is taken into account as reported in \cref{sec:uncertainties}. The kinematic distributions of the control samples are shown in \cref{fig:eventsDistributionsBkgAntiNuMu}. As shown in the legend, the purity is lower than the \numu selection, at 48\% for the \cconepiminus sample and 24\% for the \ccother sample. Indeed, positively charged pions generated in \numu interactions are mis-identified as positively charged muons decreasing the purity. This difference with the \numu selection is caused by the higher \numu contamination of the antineutrino beam compared with the smaller \barnumu component in the neutrino beam. The data-MC discrepancy observed for the \cconepiminus control sample is mainly due to an overestimation of the antineutrino coherent pion production cross-section as is implemented in NEUT version~\texttt{5.3.2}~\cite{Higuera:2014azj}. Also in this case, the discrepancy is corrected by the likelihood fit (see \cref{sec:resultscomp}).

\subsection{Sources of uncertainties}
\label{sec:uncertainties}

The uncertainties can be split into the following categories: statistical uncertainty, flux uncertainty, detector systematic uncertainties, uncertainty on the modeling of signal and background processes.

\textbf{Statistical Uncertainty.} To compute the statistical uncertainty associated with the data, the nominal MC was normalized to the number of protons on target in the data. The MC was varied in each reconstructed bin according to a Poisson distribution and 1000 toy samples were generated.  A fit was performed to each toy sample. The statistical error is taken to be the width of the distribution of the cross section results for all the toys in each true bin. The difference between data and MC observed in this analysis has a negligible impact on this statistical uncertainty.

\textbf{Flux Uncertainty.} The evaluation of the uncertainties on the flux prediction are described in detail in Ref.~\cite{Abe:2012av}. It is around 10\% at the energy peak and is dominated by the hadron production model and is evaluated using data published by the NA61/SHINE experiment using a thin Carbon target~\cite{Abgrall:2011ae,Abgrall:2011ts,Abgrall:2015hmv}. The flux covariance matrix was used to generate many toy MC sets. The flux bins include separate bins for the ``right-sign" and ``wrong-sign" components of the flux in both neutrino-mode and antineutrino-mode. The fit includes 32 nuisance parameters for the fluxes which are constrained by the fit, reducing the flux uncertainties by around 60\%. In previous analyses the flux was not constrained by the fit since this procedure could introduce a model dependencies~\cite{Abe:2016tmq}. For this reason dedicated mock data studies have been performed as described in \cref{sec:anaStrategy}.

\textbf{Detector Systematic Uncertainties.} Detector uncertainties can be grouped into three categories depending on the way they are propagated: efficiency-like, observable variation and normalization systematics. The systematics belonging to the first group have been propagated by applying a weight that depends on one or more observables, the second by adjusting the observables and reapplying the selection, the last by applying a single weight applied to all events. Dedicated data and MC samples have been used to quantify the detector uncertainties in the modeling of FGD and TPC responses, of neutrino interactions outside of the FGD1 FV, pion and proton secondary interactions. The differences between data and MC observed in control samples have been applied as correction factors to the nominal MC to take into account the observed discrepancies, while the error on these factors has been taken as detector systematic uncertainty. The \numu and \barnumu selections are affected by the same detector uncertainties since they employ similar features of the sub-detectors. The dominant systematics are due to the uncertainties on the amount of background from interactions occurring outside of the fiducial volume, the modeling of the pion secondary interactions and the TPC PID. The detector systematics have been stored in a covariance matrix corresponding to the uncertainties on the total number of reconstructed events in each bin and in each signal and control regions. The systematics on the cross sections have been propagated by repeating the fit over many toy MC data sets where the detector parameters have been varied according with their covariance matrix but have been kept fixed in the fit. This choice has been driven by the necessity to ensure the stability and convergence of the fit by reducing the number of nuisance parameters.
The uncertainty associated with the number of targets has been computed separately. A covariance matrix has been formed from the uncertainties on the areal densities of all the elements present in FGD1 and, the uncertainty calculated from toy experiments sampling such covariance matrix.

\begin{table}[ht!]
	\caption{Prior values and errors of the cross section model parameters used in this analysis.}
	\centering
	\renewcommand{\arraystretch}{1.3} 
	\setlength{\tabcolsep}{4pt}
	\begin{ruledtabular}
		\begin{tabular}{c|c|c}
			Parameter                                            & Prior & Error \\ 
			\hline   
			$M_A^{QE}$ (GeV/c$^2$)                                   & 1.2 & 0.3\\
			$p_F^C$ (MeV/c)                                                & 217  & 30 \\
			$E_B^C$  (MeV)                                            & 25  & 9 \\
			2p2h $\nu$                                          & 1     & 1 \\
			2p2h \barnu                                         & 1 & 1 \\
			$C_A^5$  (GeV/c$^2$)                        & 1.01 & 0.12 \\
			$M_A^{Res}$ (GeV/c$^2$)                & 0.95 & 0.15 \\
			I$_{1/2}$                                            & 1.3 & 0.2 \\
			DIS Multiple pion                                 & 0.0 & 0.4 \\
			CC Coherent on C                                    & 1.0 & 1.0 \\		
			CC-1$\pi$ $E_\nu < $ 2.5 GeV                 & 1.0 & 0.5 \\		
			CC-1$\pi$ $E_{\bar\nu} < $ 2.5 GeV      & 1.0 & 1.0 \\		
			CC-1$\pi$ $E_\nu >$ 2.5 GeV                 & 1.0 & 0.5 \\
			CC-1$\pi$ $E_{\bar\nu}  >$ 2.5 GeV      &   1.0 & 1.0 \\
			CC Multile $\pi$                                       & 1.0 & 0.5 \\
			CC-DIS $\nu$                                           &  1.0 & 0.035 \\
			CC-DIS \barnu                                        &   1.0 & 0.065 \\
			NC Coherent                                         & 1.0 & 0.3 \\
			NC Other                                                & 1.0 & 0.3 \\
			Pion production                                  & 0.0     & 0.5  \\
			Pion absorption                                    & 0.0    & 0.41  \\
			Pion quasi-elastic int. for $p_{\pi}\:<$ 500 MeV/c &  0.0     & 0.41\\
			Pion quasi-elastic int. for $p_{\pi}\:>$ 400 MeV/c &  0.0     & 0.34\\
			Pion charge exchange for $p_{\pi}\:<$ 500 MeV/c  & 0.0    & 0.57 \\
			Pion charge exchange for $p_{\pi}\:>$ 400 MeV/c & 0.0    & 0.28 \\
		\end{tabular}
	\end{ruledtabular}
	\label{tbl:xsecparam}
\end{table}

\begin{figure*}[!htbp]
	\centering
	\includegraphics[width=1.\linewidth]{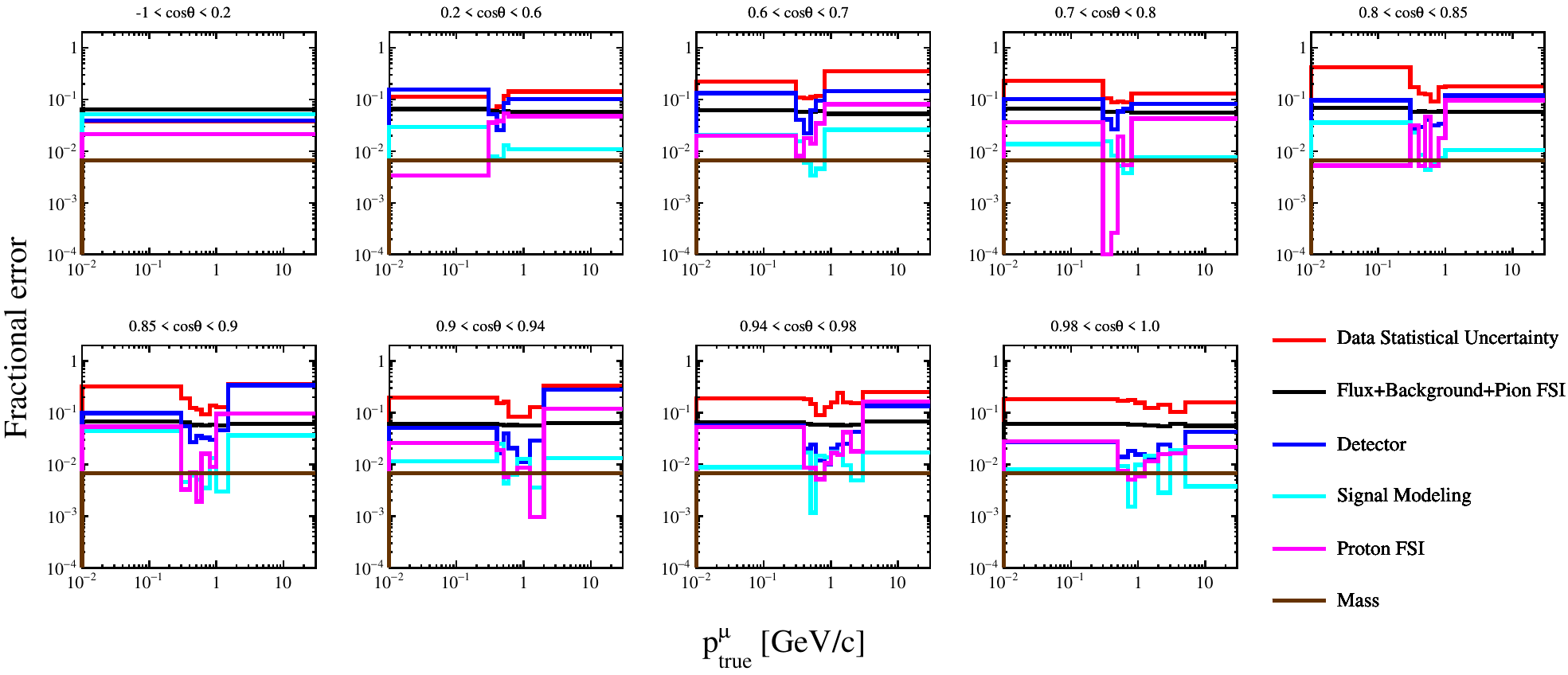}
	\caption{Summary of all of the systematic uncertainties for the $\nu_{\mu}$ \cczeropi cross section, in bins of true muon kinematics.}
	\label{fig:nuerror}
\end{figure*}

\begin{figure*}[!htbp]
	\centering
	\includegraphics[width=1.\linewidth]{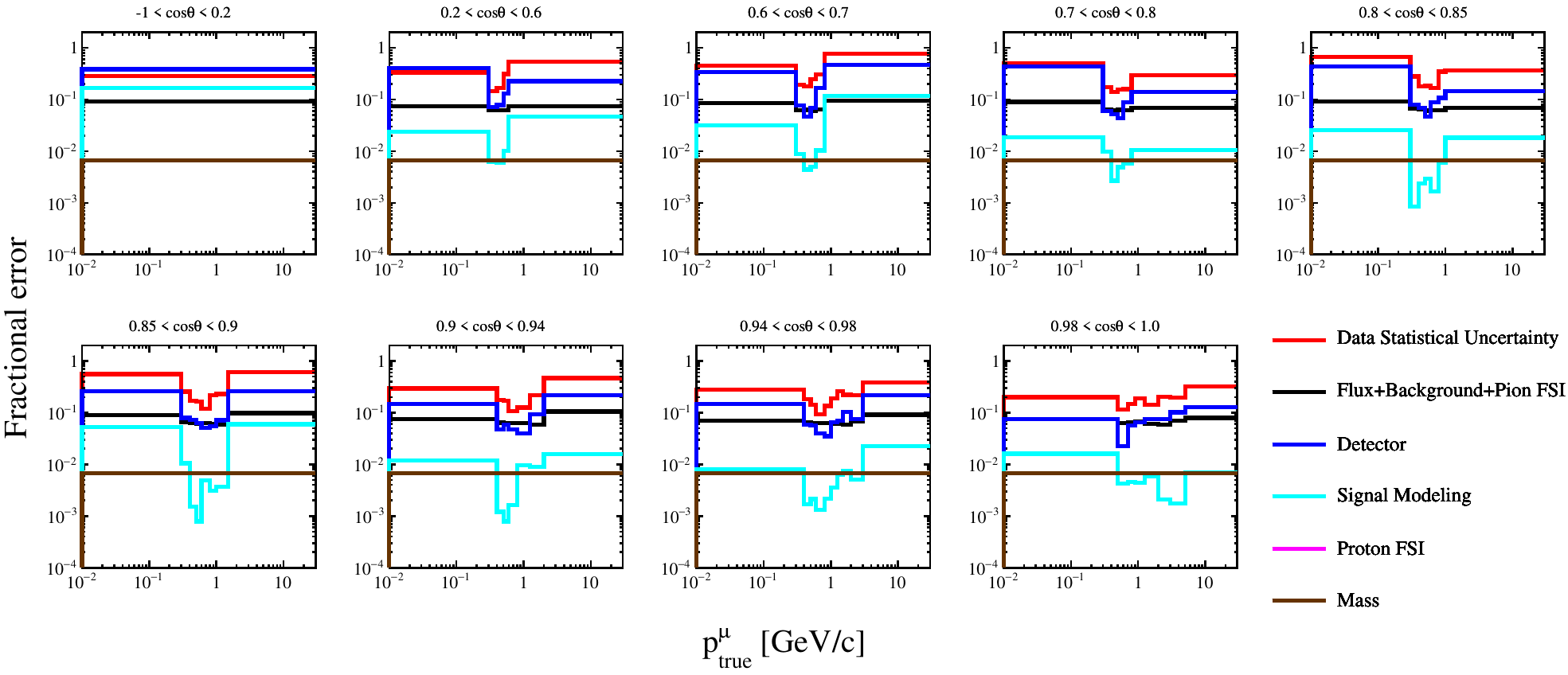}
	\caption{Summary of all of the systematic uncertainties for the $\bar{\nu}_{\mu}$ \cczeropi cross section, in bins of true muon kinematics.}
	\label{fig:antinuerror}
\end{figure*}

\textbf{Modeling of Signal and Background.} The signal efficiency and number of background events in each bin are affected by uncertainties in our cross-section model. \cref{tbl:xsecparam} summarizes the cross-section parameters used for this analysis along with their prior value and error. The parameters include shape variations of the CCQE cross section ($M_{A}^{QE}$, Fermi momentum $p_{F}^{C}$, binding energy $E_{B}^{C}$) and the normalization of the amount of 2p2h interactions in neutrino and antineutrino. Fermi momentum and binding energy variations are modeled using RFG. The signal modeling parameters only affect the efficiency and are not constrained by the fit to avoid model dependencies. Other parameters control the shape and normalization of the background processes:  the axial mass $M_{A}^{Res}$ and the form factor $C_{A}^{5}$ control the shape of the RES cross section; I$_{1/2}$ the normalization of non-resonant pion production; \textit{CC-1$\pi$} the normalization of such background in different (anti)neutrino energy ranges; \textit{DIS Multiple pion}, \textit{CC-DIS $\nu$} and \textit{CC-DIS \barnu} the normalization of the DIS; \textit{CC Coherent on C} the normalization of such process; \textit{NC Coherent} and \textit{NC Other} the normalization of NC interactions. Pions that are produced in neutrino interactions can be affected by FSI as they leave the nuclear medium, changing their kinematics, charge and multiplicity. Dedicated systematic parameters have been included in the fit to describe the pion production, absorption, charge exchange and quasi-elastic scattering of the exiting pions. Again these modify not only the selected number of events, but the selection efficiency as well. Similarly protons are also subjected to FSI: the uncertainty is evaluated by comparing two different \textsc{NuWro}~\cite{Golan:2012wx} MC simulations\footnote{The version used is \texttt{11q}. The models implemented in this version are the same as present in version \texttt{18.02.1} described in \cref{sec:resultscomp}.}, with and without FSI. The difference in the efficiency as a function of the muon kinematics between the two simulations has been taken as the uncertainty due to the proton FSI. To be conservative, it has been added in quadrature to the other efficiency uncertainties. Since in \barnumu interactions proton FSI has a negligible impact, this uncertainty has been added to \numu cross section only. The cross section parameters have been propagated by throwing from Gaussian distributions that have as mean and sigma the prior and error values reported in \cref{tbl:xsecparam}. 

All of the systematic errors in each bin are summarized in \cref{fig:nuerror,fig:antinuerror}. In most bins, the dominant uncertainty is due to the statistical error on data. The systematic uncertainties are typically dominated by the detector systematics. The modeling errors are generally subdominant and smaller than 10\% and closer to 1\% in regions of high purity. Uncertainties related to the flux, modeling of the background and pion FSI have been propagated together since they are anti-correlated. The errors on the sum, difference and asymmetry has been computed numerically from toy experiments sampling the covariance matrix that includes the uncertainty and correlations between the two cross sections. 

\section{Results and comparisons with models}
\label{sec:resultscomp}
\begin{figure*}[h!]
	\centering
	\includegraphics[width=0.36\linewidth]{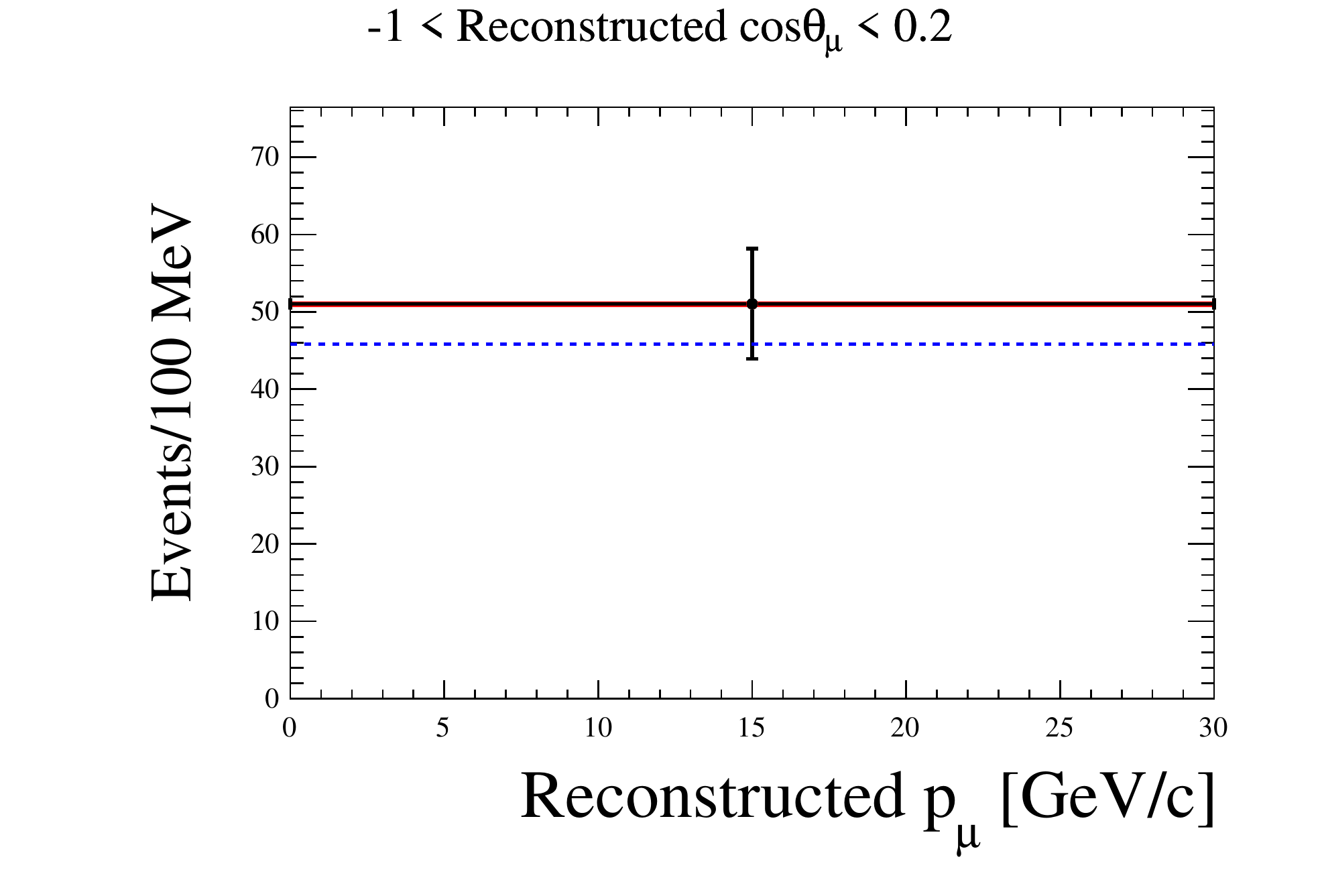}	\includegraphics[width=0.36\linewidth]{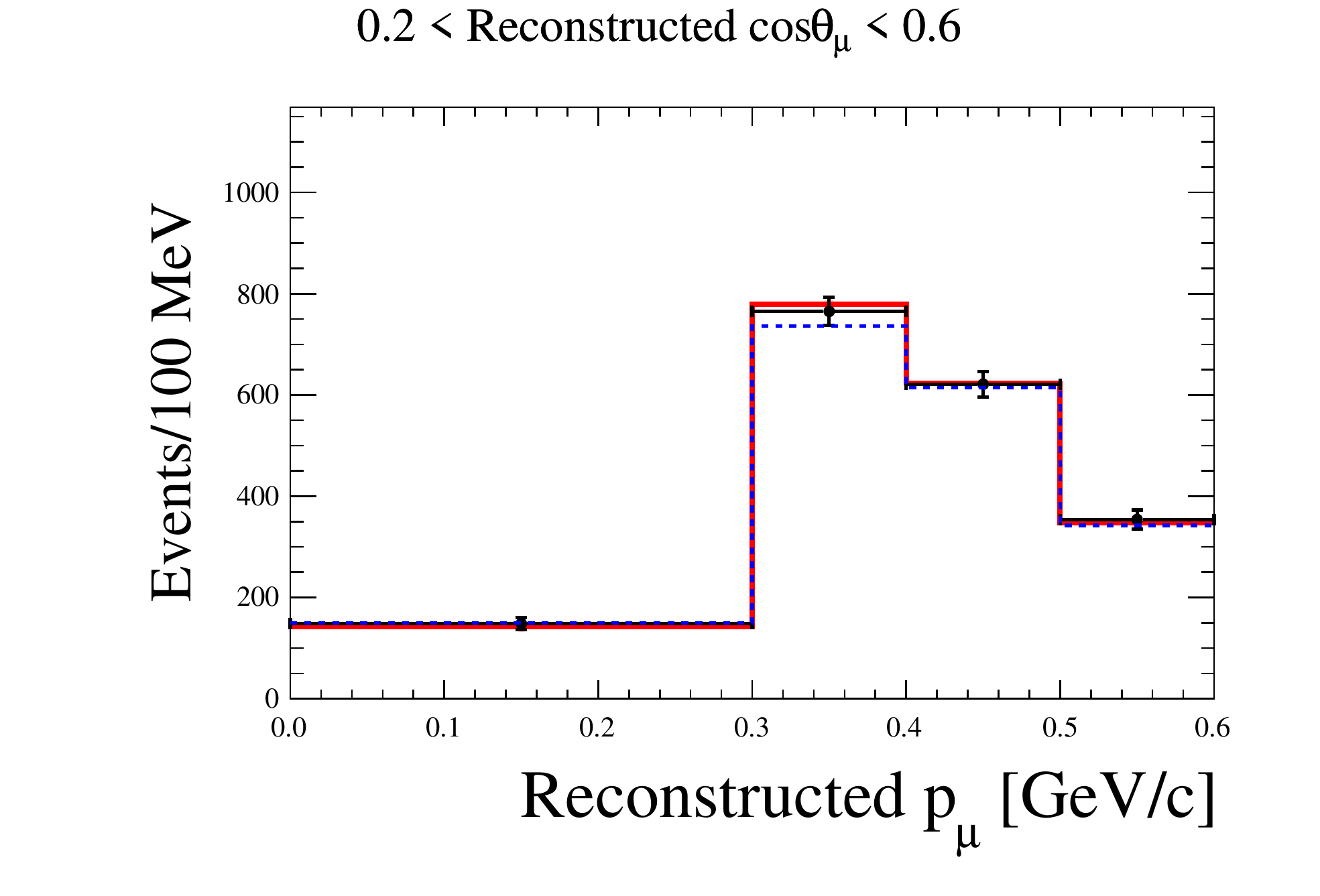}	\includegraphics[width=0.36\linewidth]{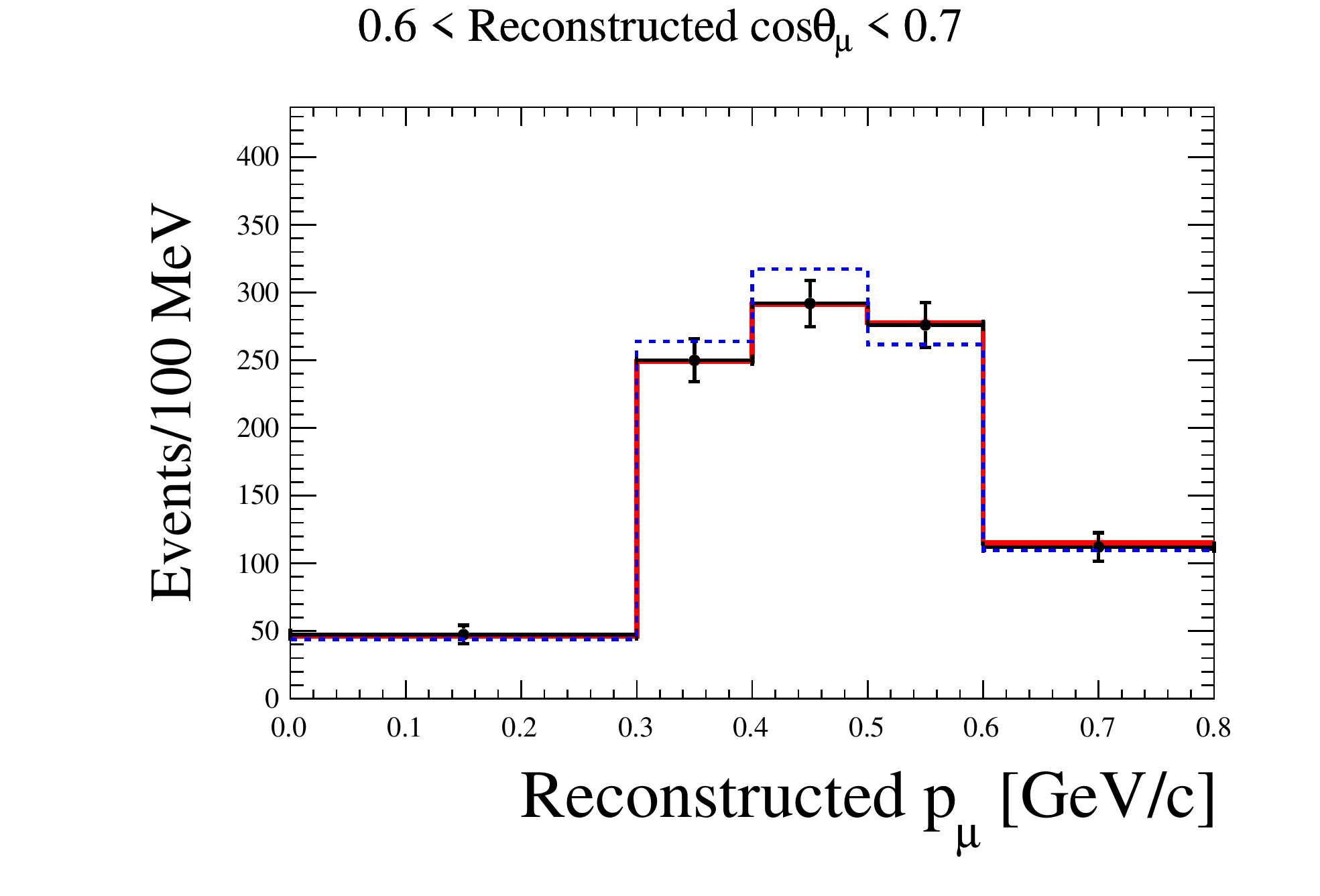}	\includegraphics[width=0.36\linewidth]{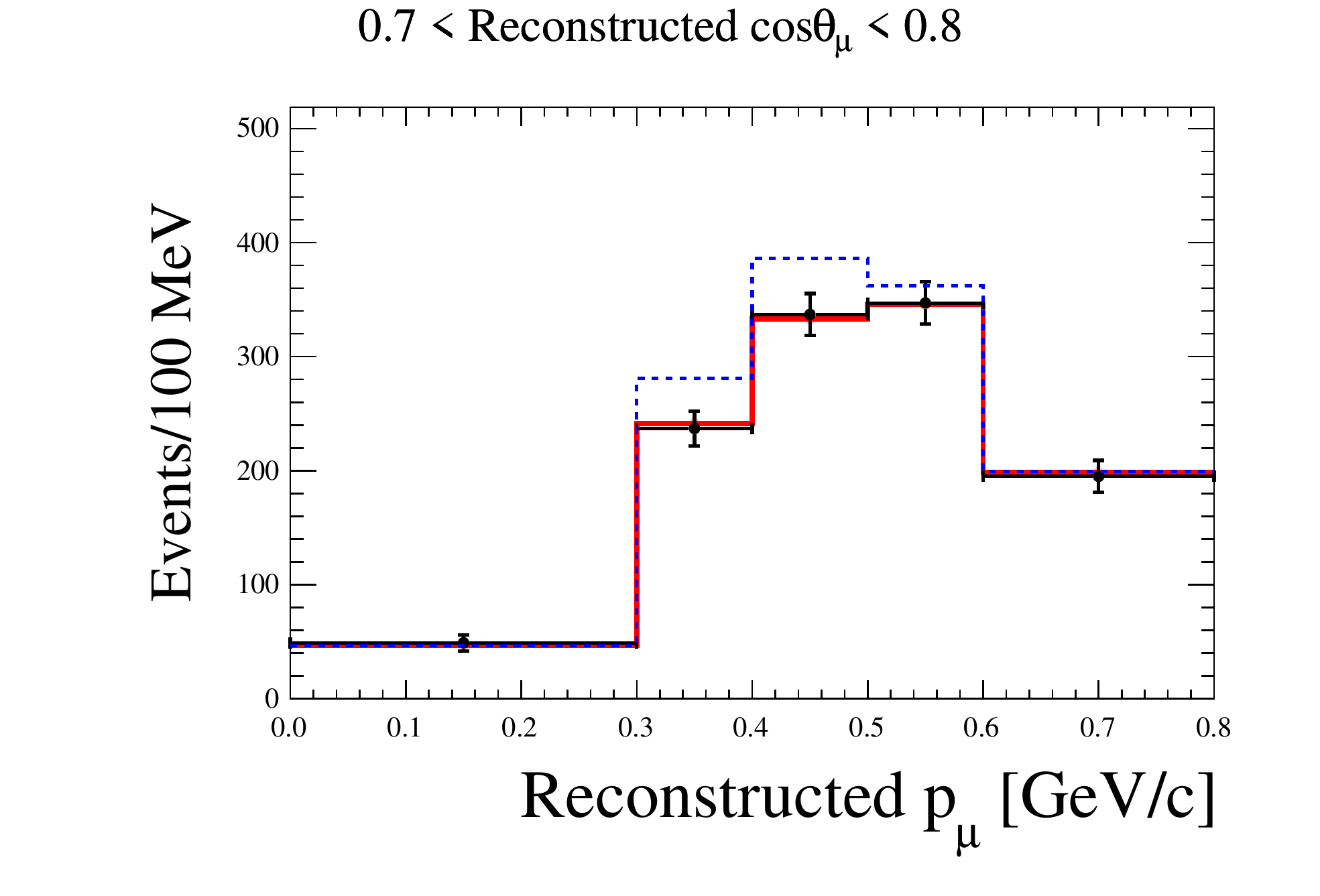}	\includegraphics[width=0.36\linewidth]{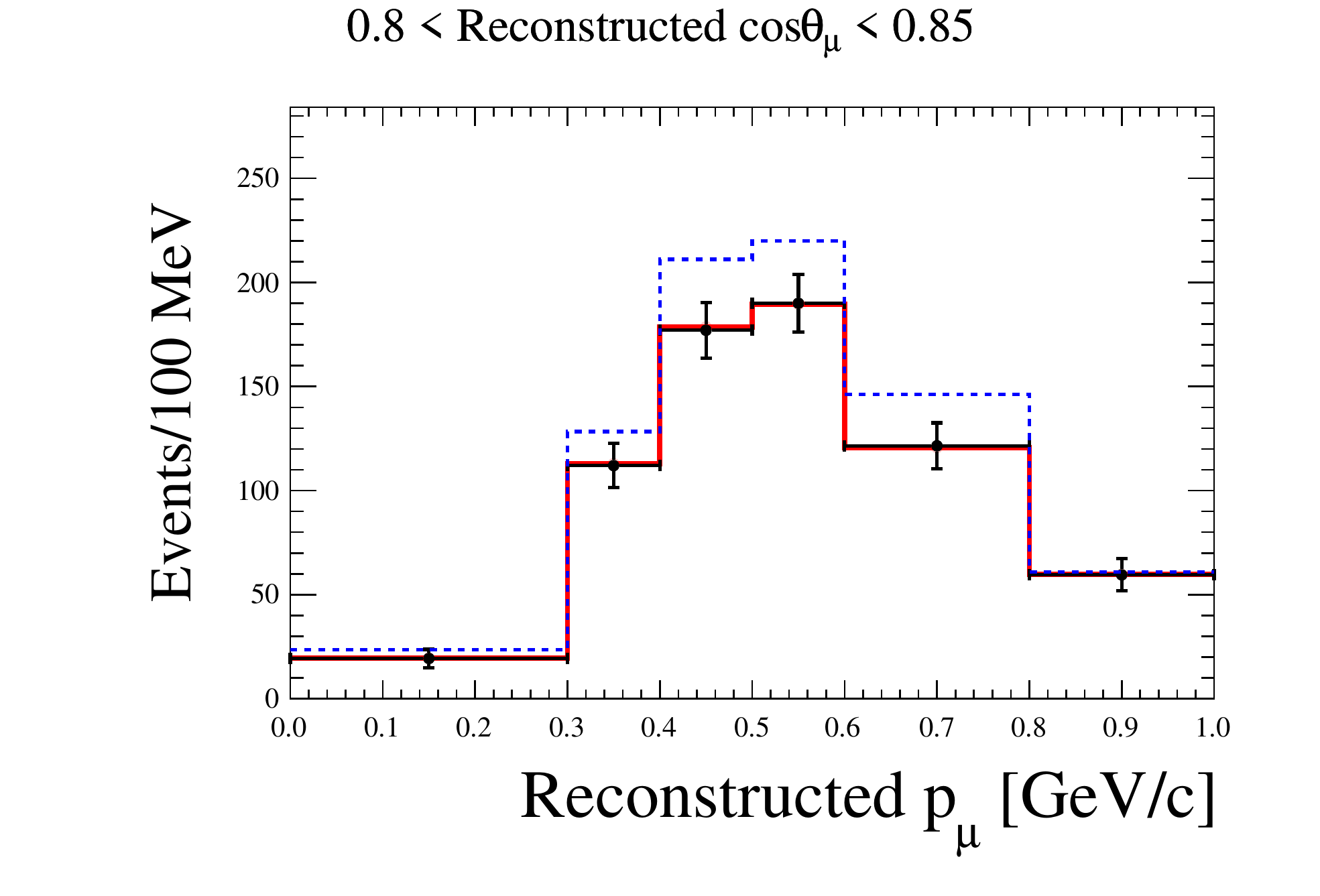}	\includegraphics[width=0.36\linewidth]{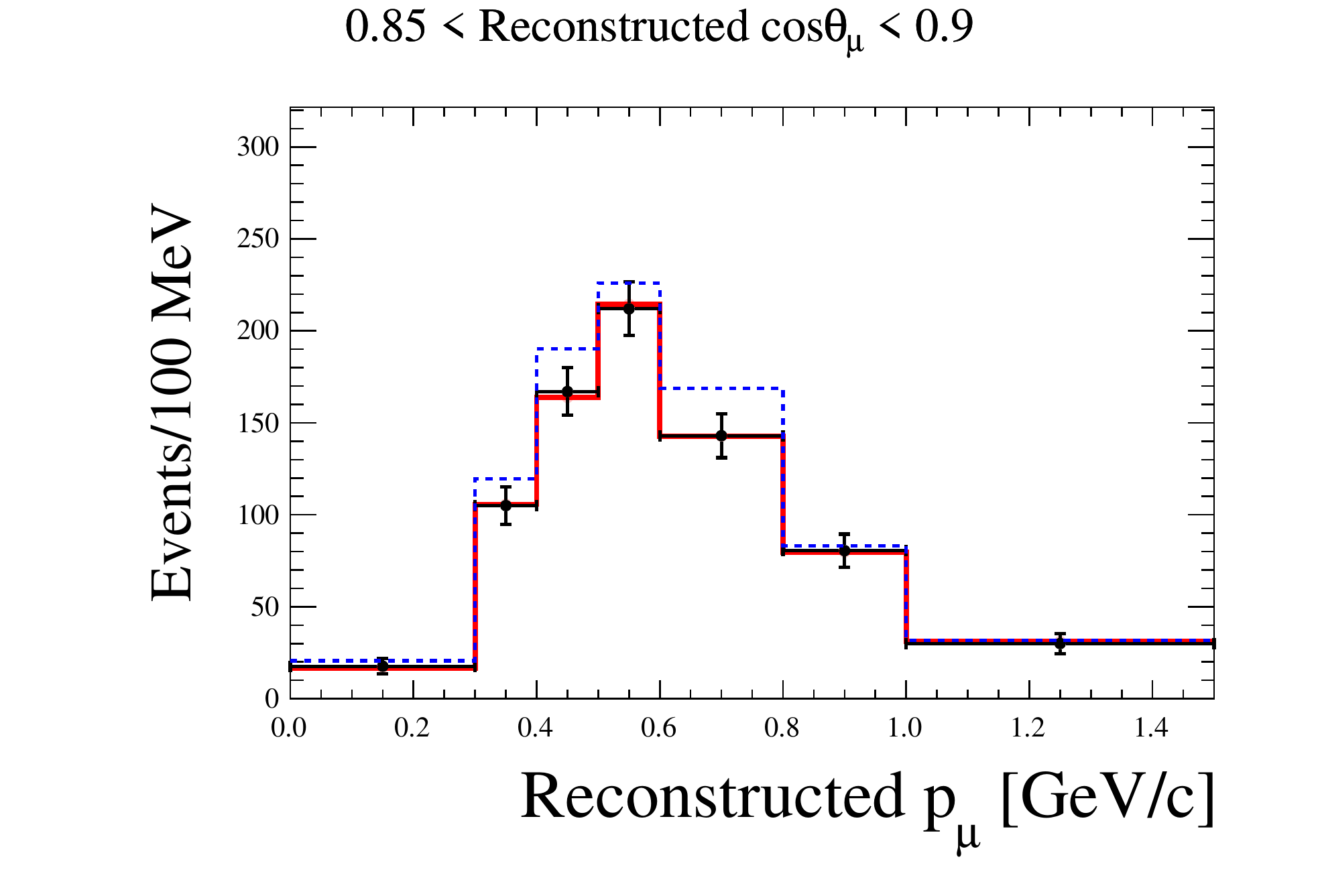}	\includegraphics[width=0.36\linewidth]{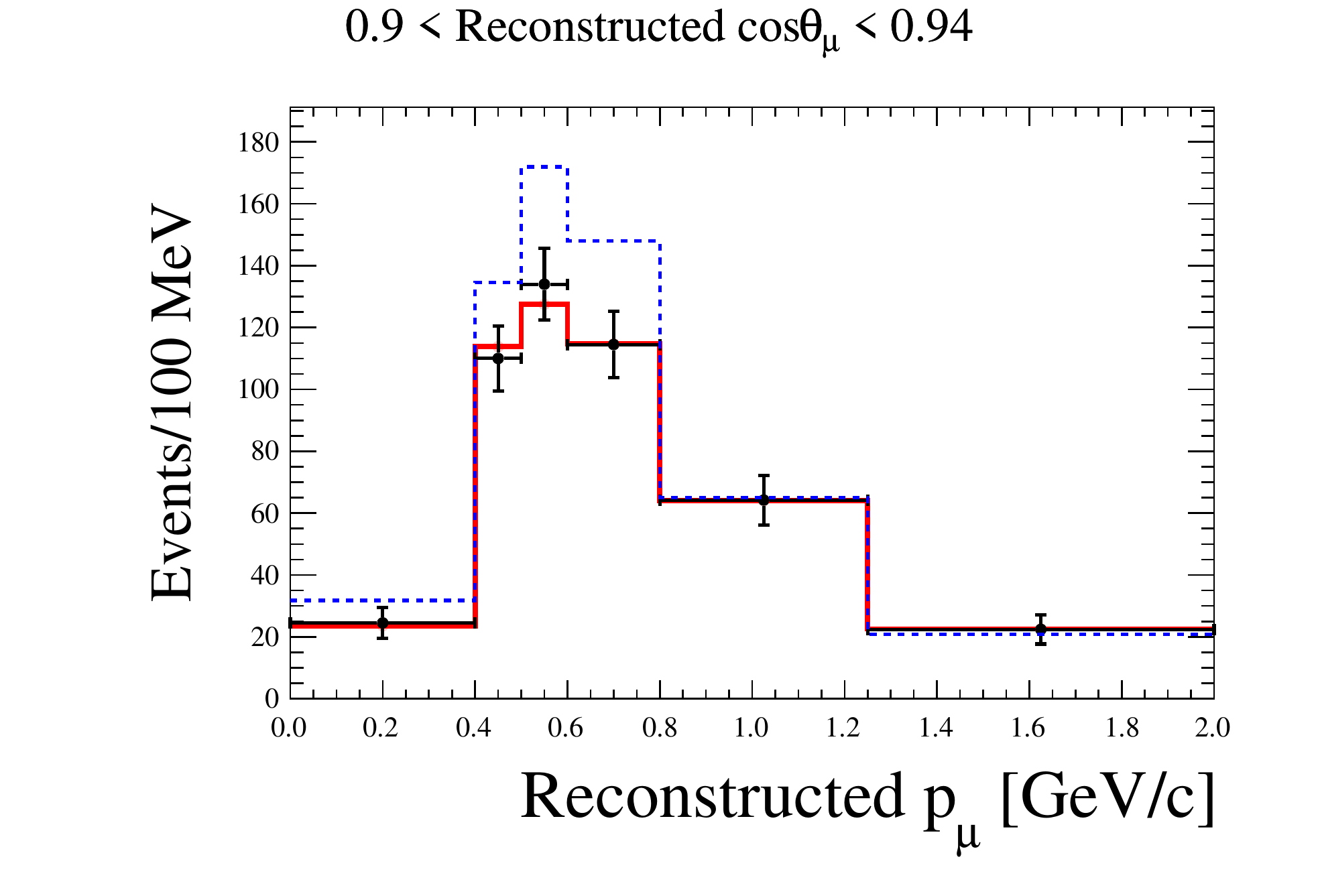}	\includegraphics[width=0.36\linewidth]{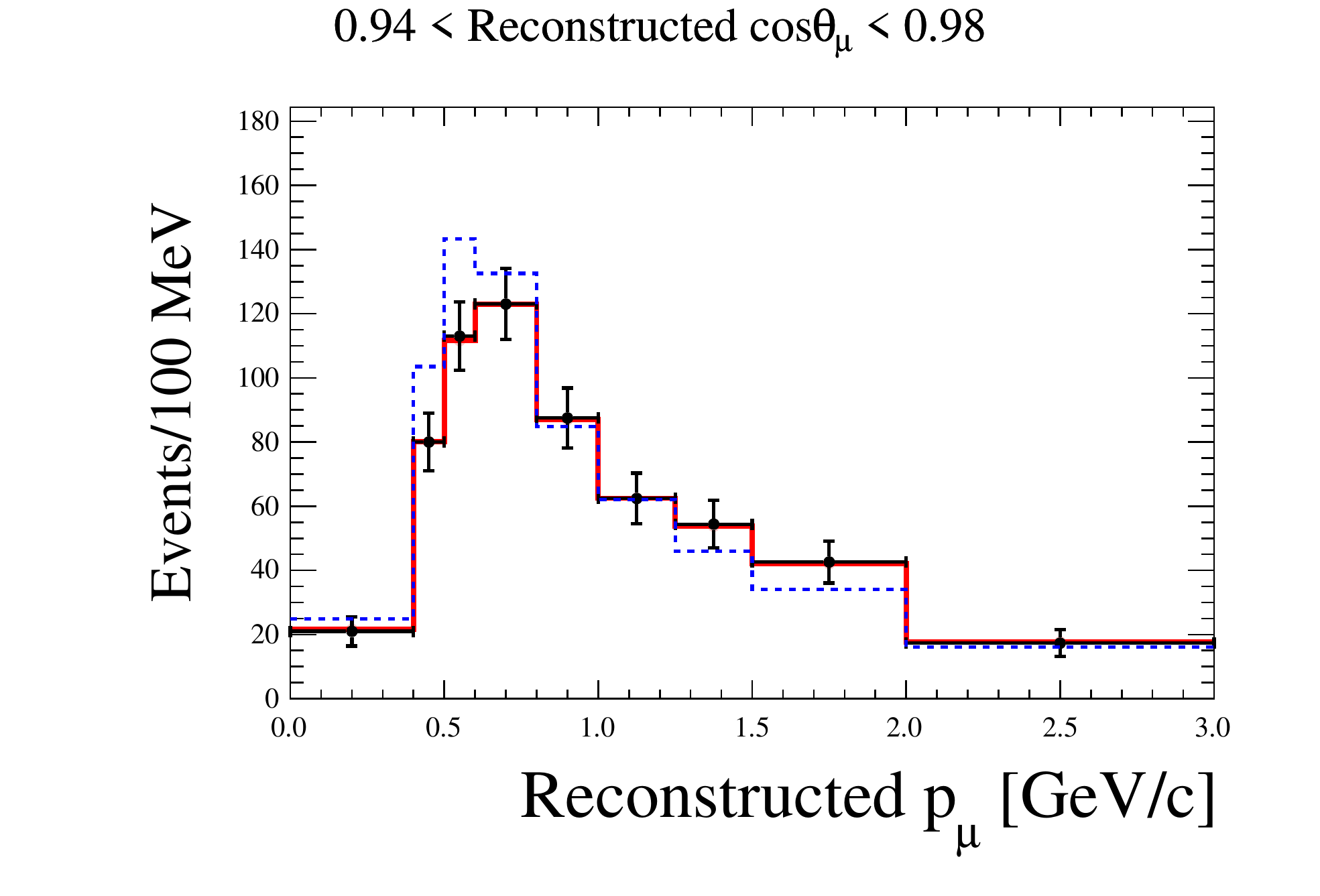}
	\includegraphics[width=0.36\linewidth]{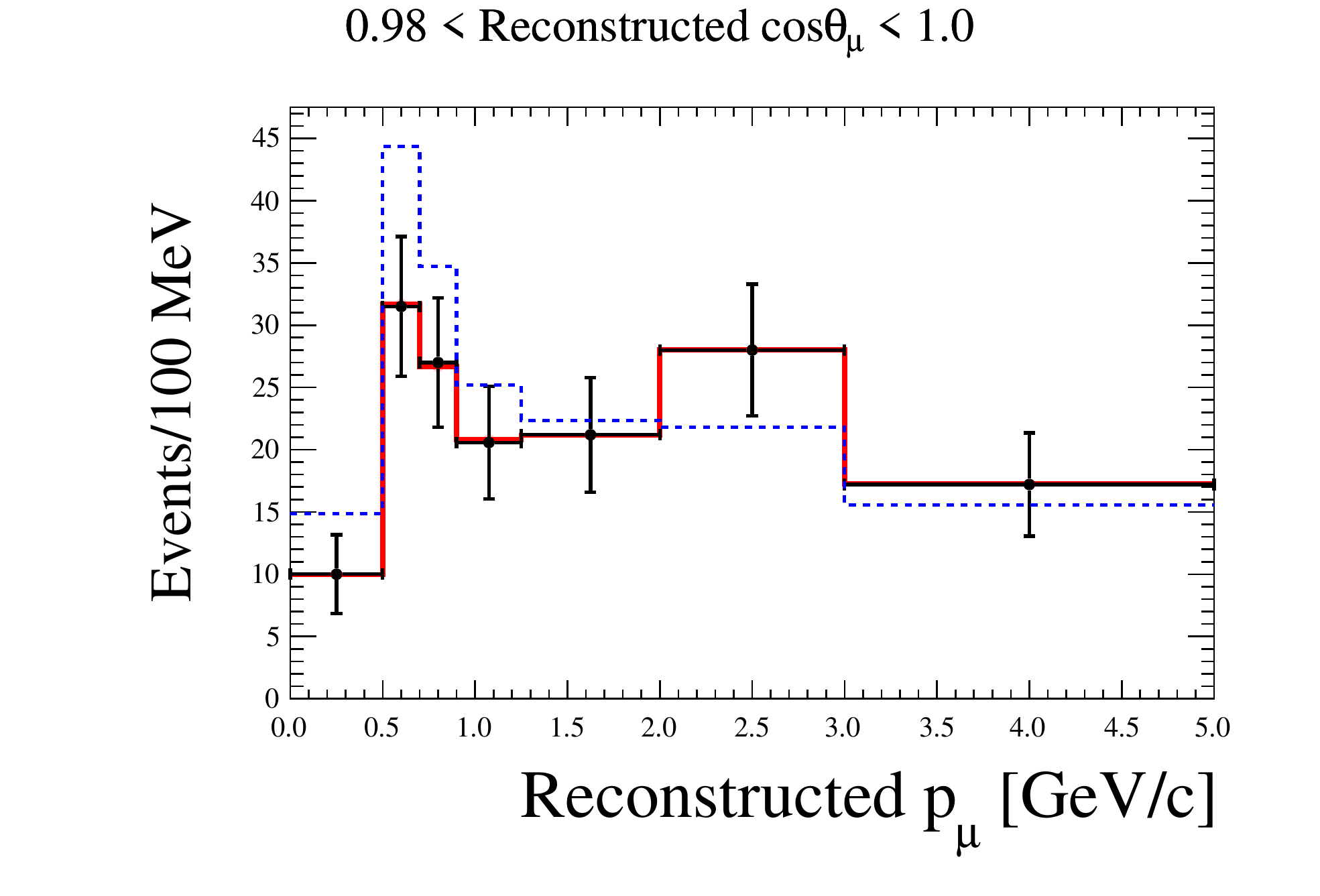}
	\includegraphics[width=0.36\linewidth]{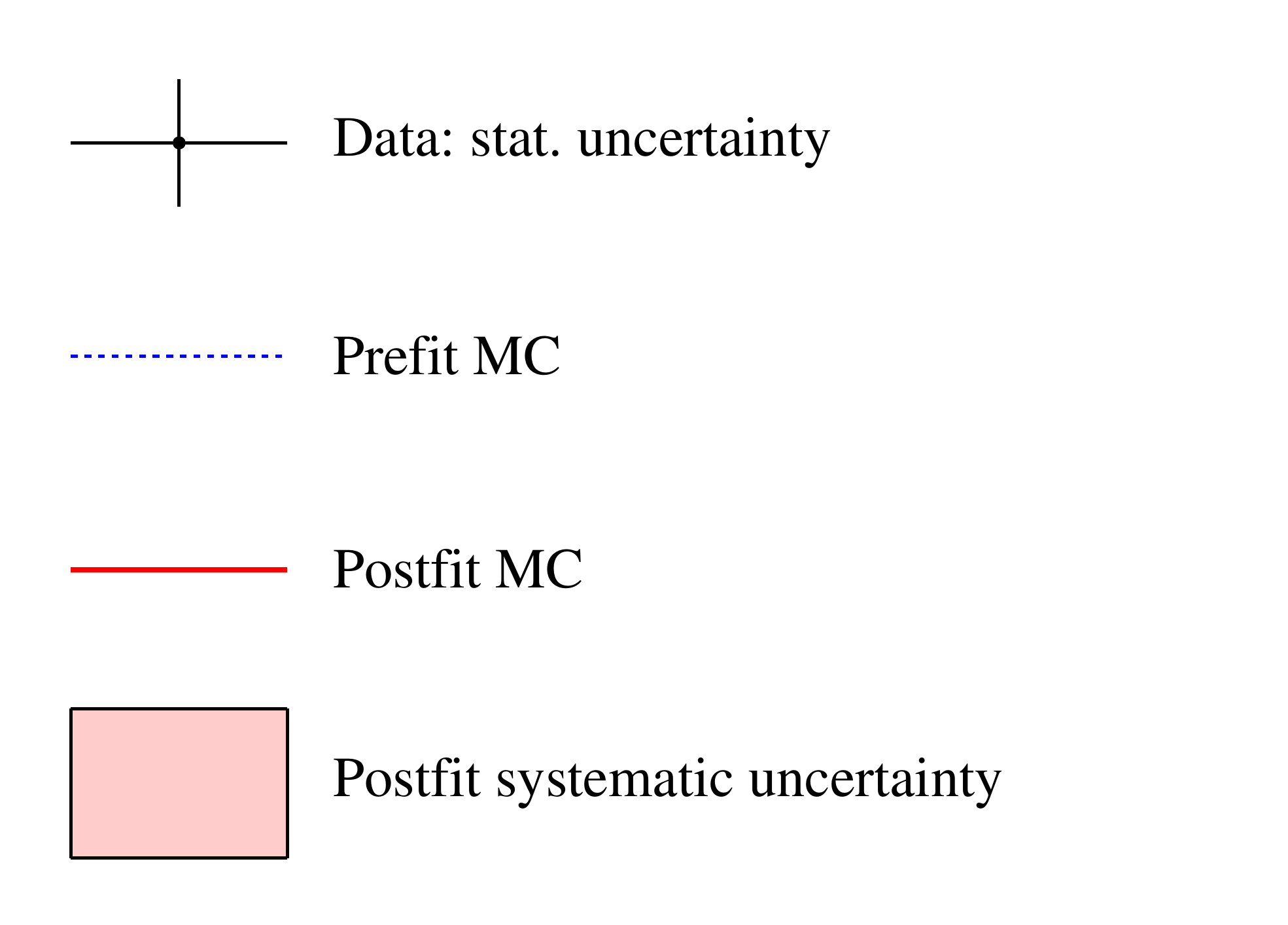}
	\caption{Distribution of events in \numu signal samples added together as a function of reconstructed muon kinematics compared with the MC prediction before the fit (dotted blue line), and after the fit (solid red line) including systematics errors indicated by the pink band. The data are shown in black with statistical errors.}
	\label{fig:numucc0pirecopostfitvsdata}
\end{figure*}

\begin{figure*}[h!]
	\centering
	\includegraphics[width=0.36\linewidth]{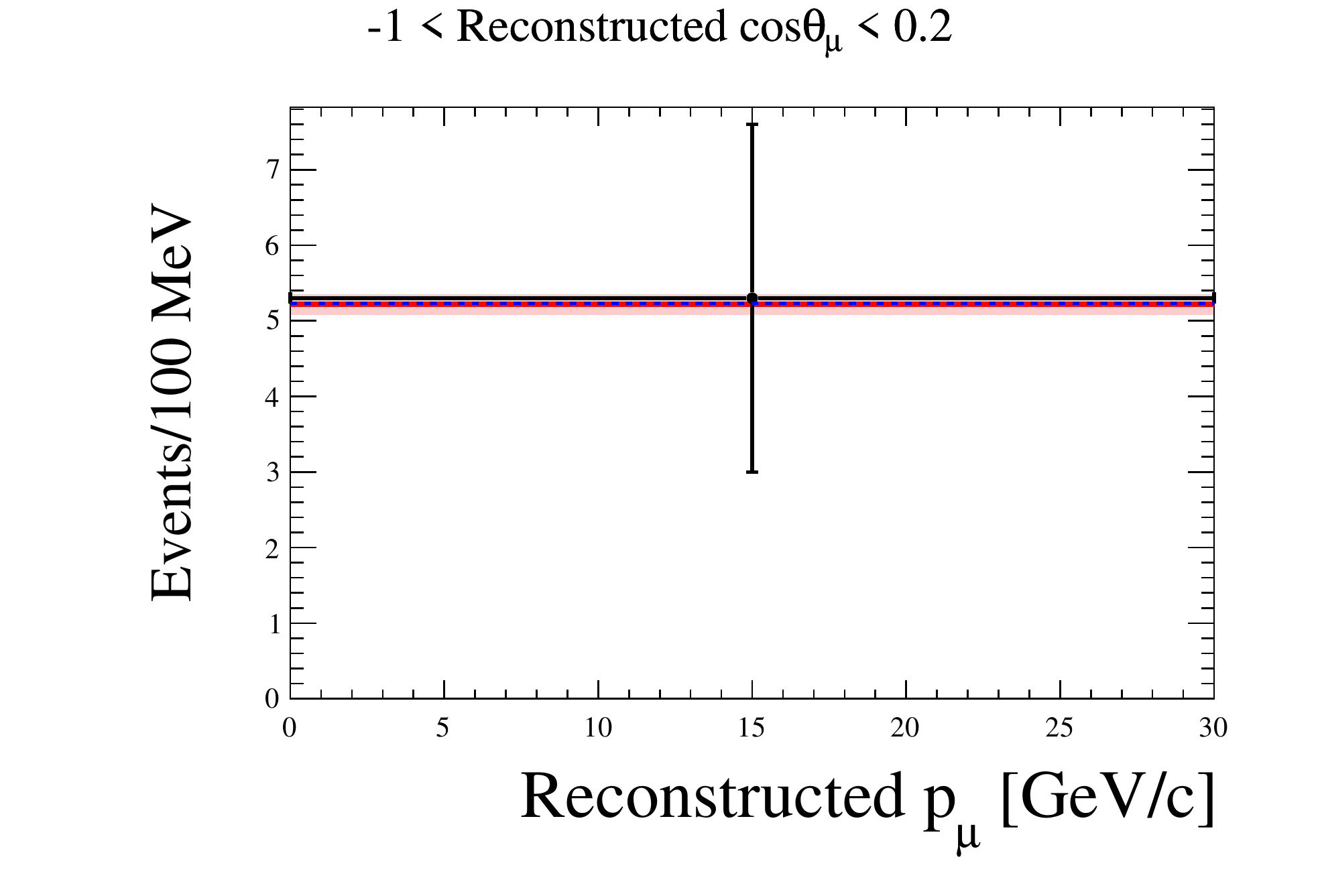}	\includegraphics[width=0.36\linewidth]{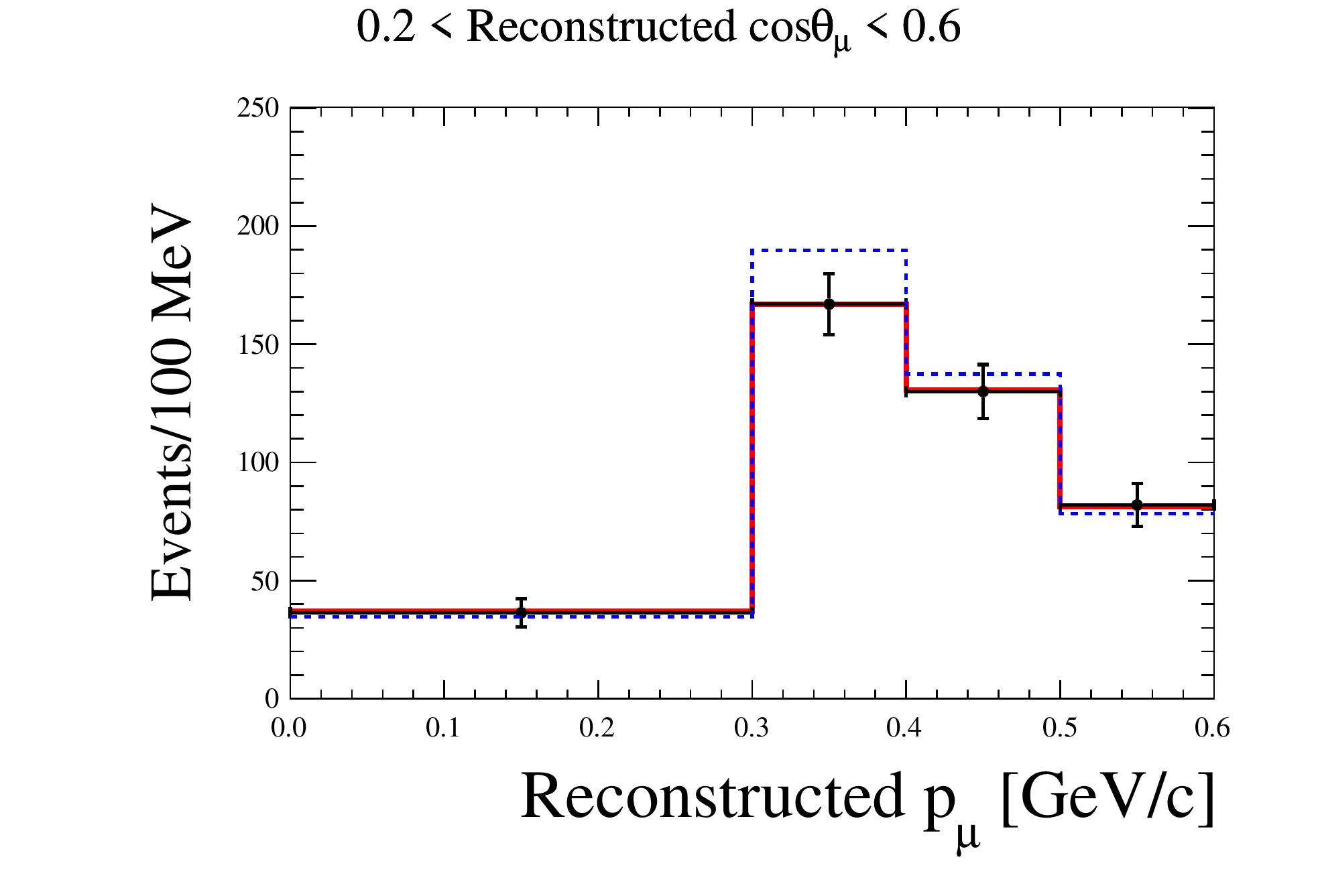}	\includegraphics[width=0.36\linewidth]{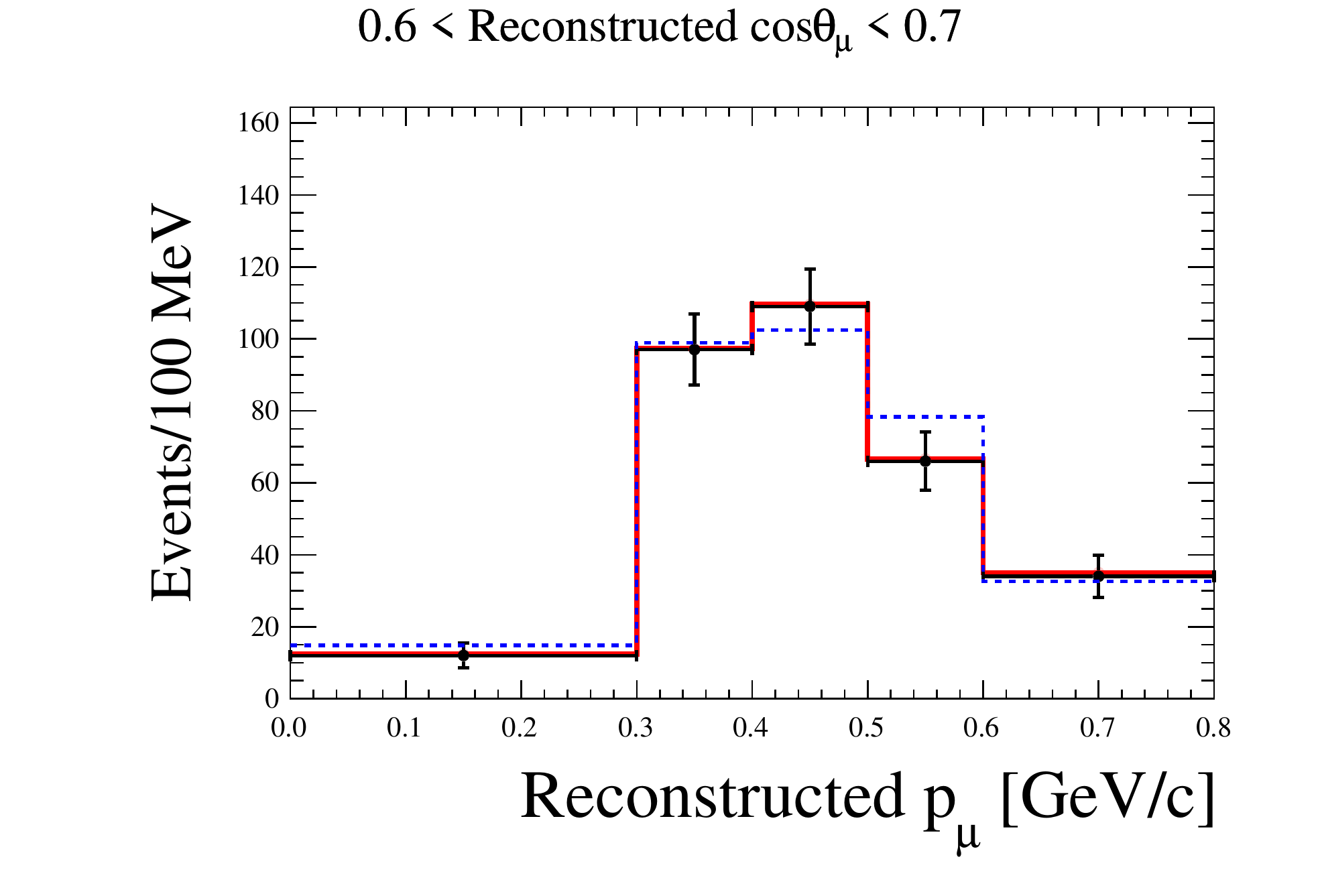}	\includegraphics[width=0.36\linewidth]{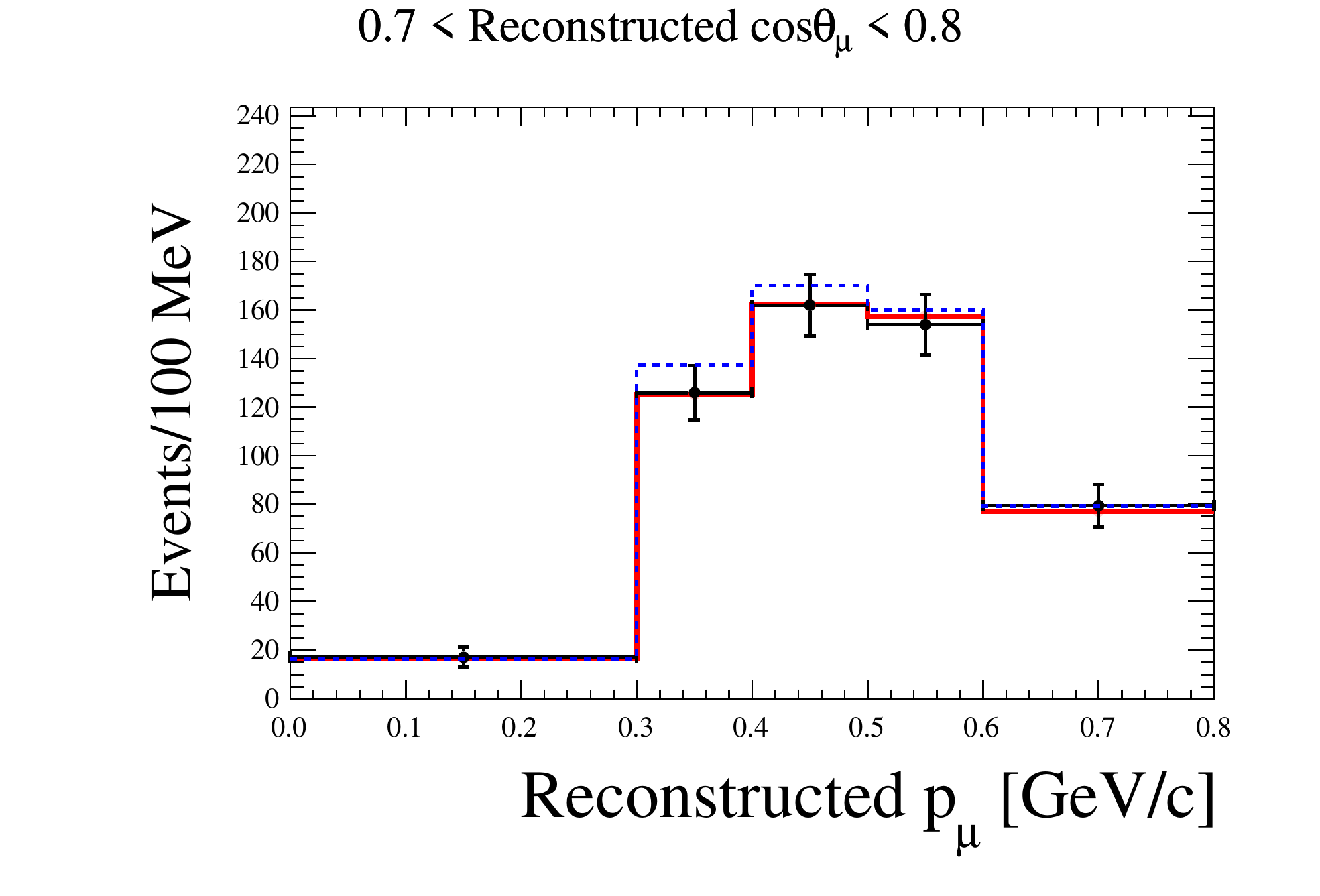}	\includegraphics[width=0.36\linewidth]{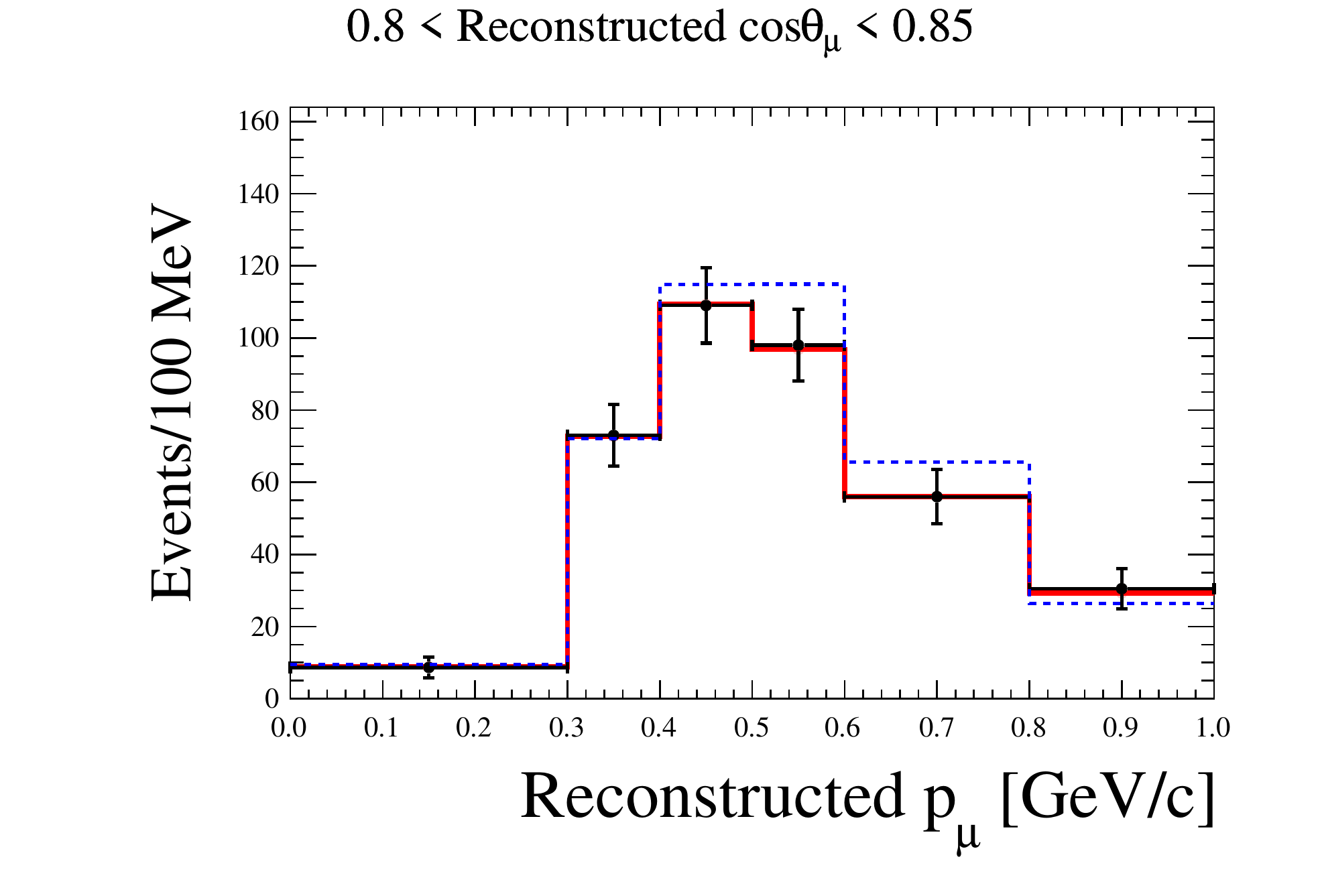}	\includegraphics[width=0.36\linewidth]{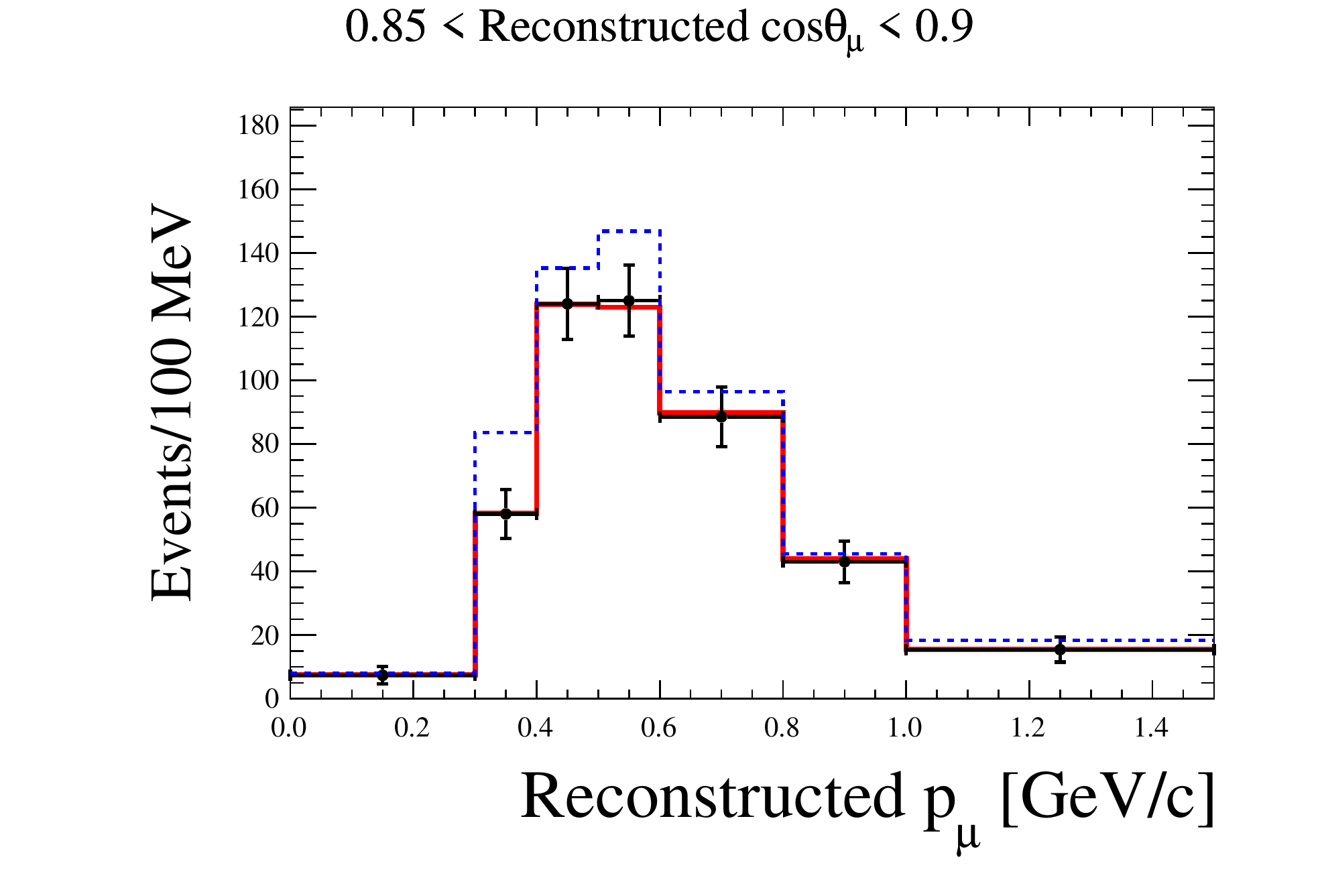}	\includegraphics[width=0.36\linewidth]{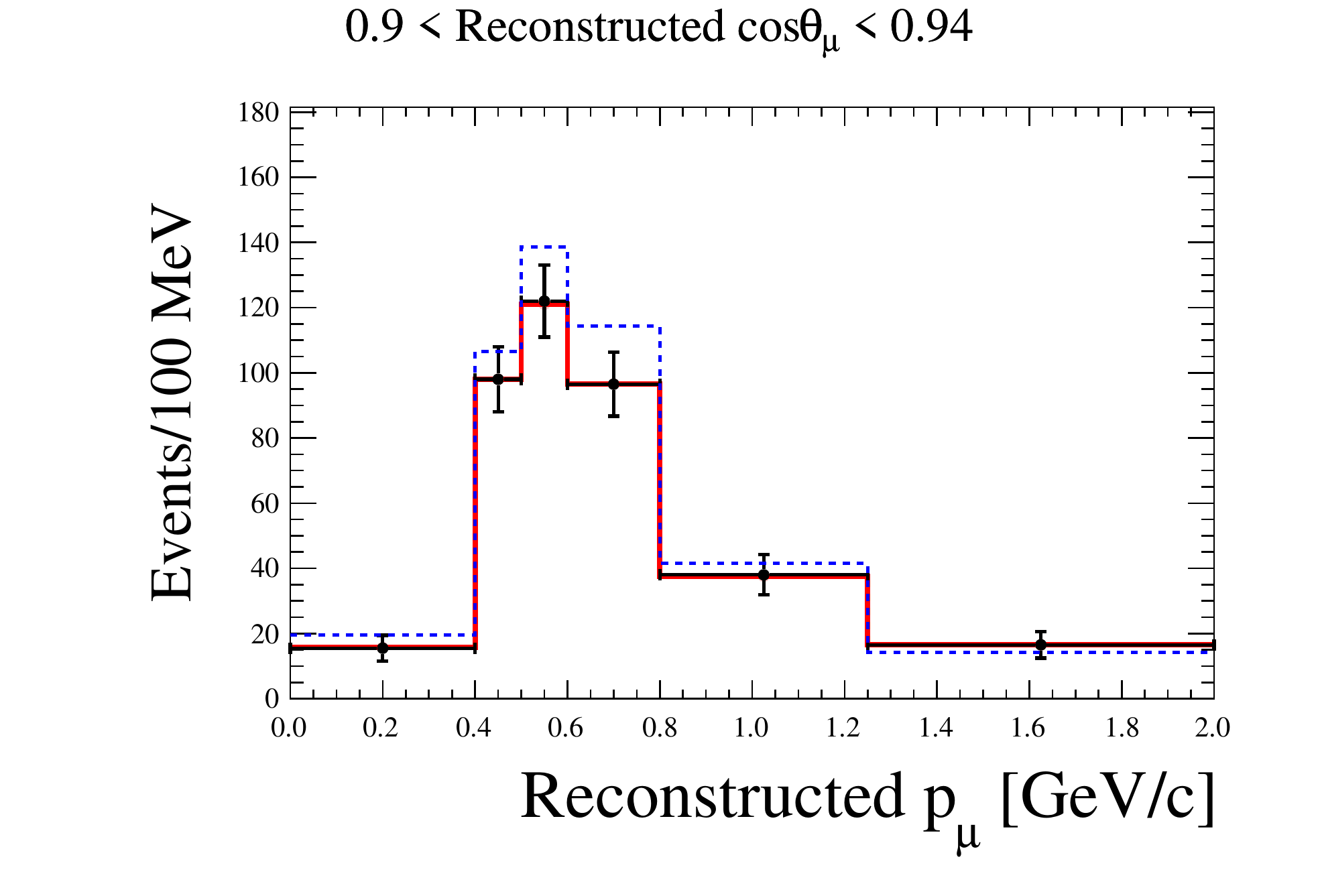}	\includegraphics[width=0.36\linewidth]{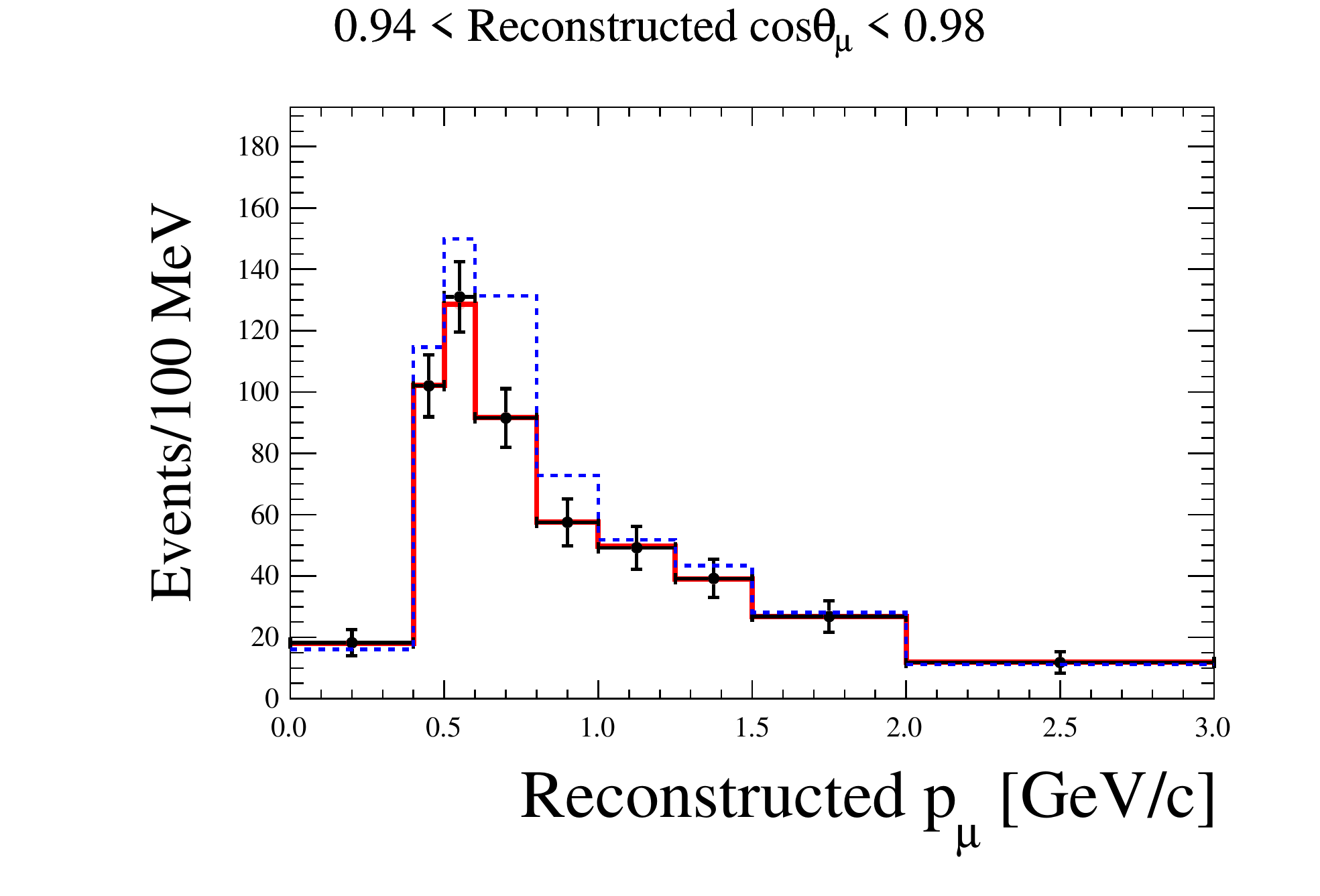}
	\includegraphics[width=0.36\linewidth]{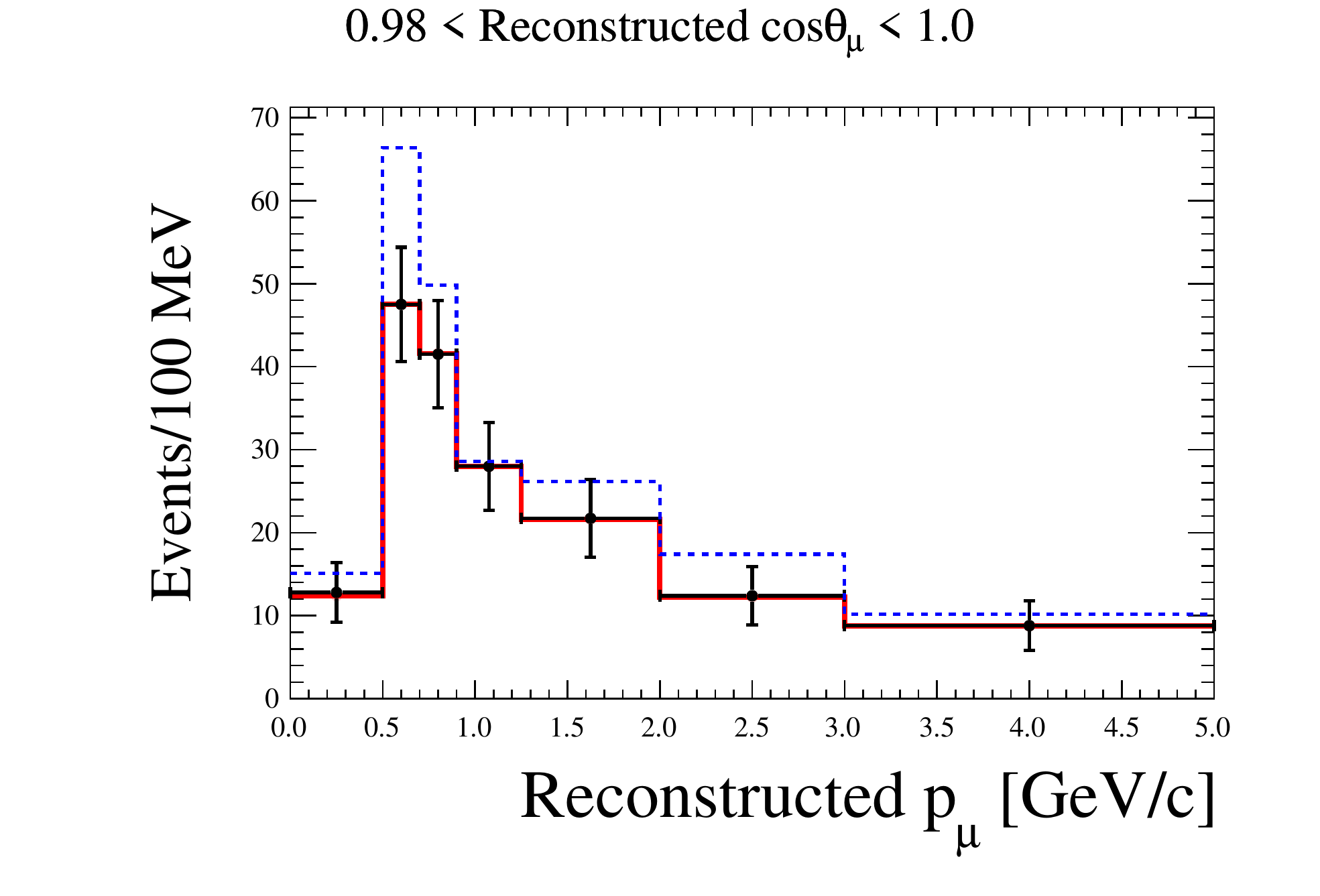}
	\includegraphics[width=0.36\linewidth]{LegendFitVsData}
	\caption{Distribution of events in \barnumu signal samples added together as a function of reconstructed muon kinematics compared with the MC prediction before the fit (dotted blue line), and after the fit (solid red line) including systematics errors indicated by the pink band. The data are shown in black with statistical errors.}
	\label{fig:antinumucc0pirecopostfitvsdata}
\end{figure*}

\begin{figure*}[th!]
	\centering
	\begin{overpic}[width=0.40\linewidth]{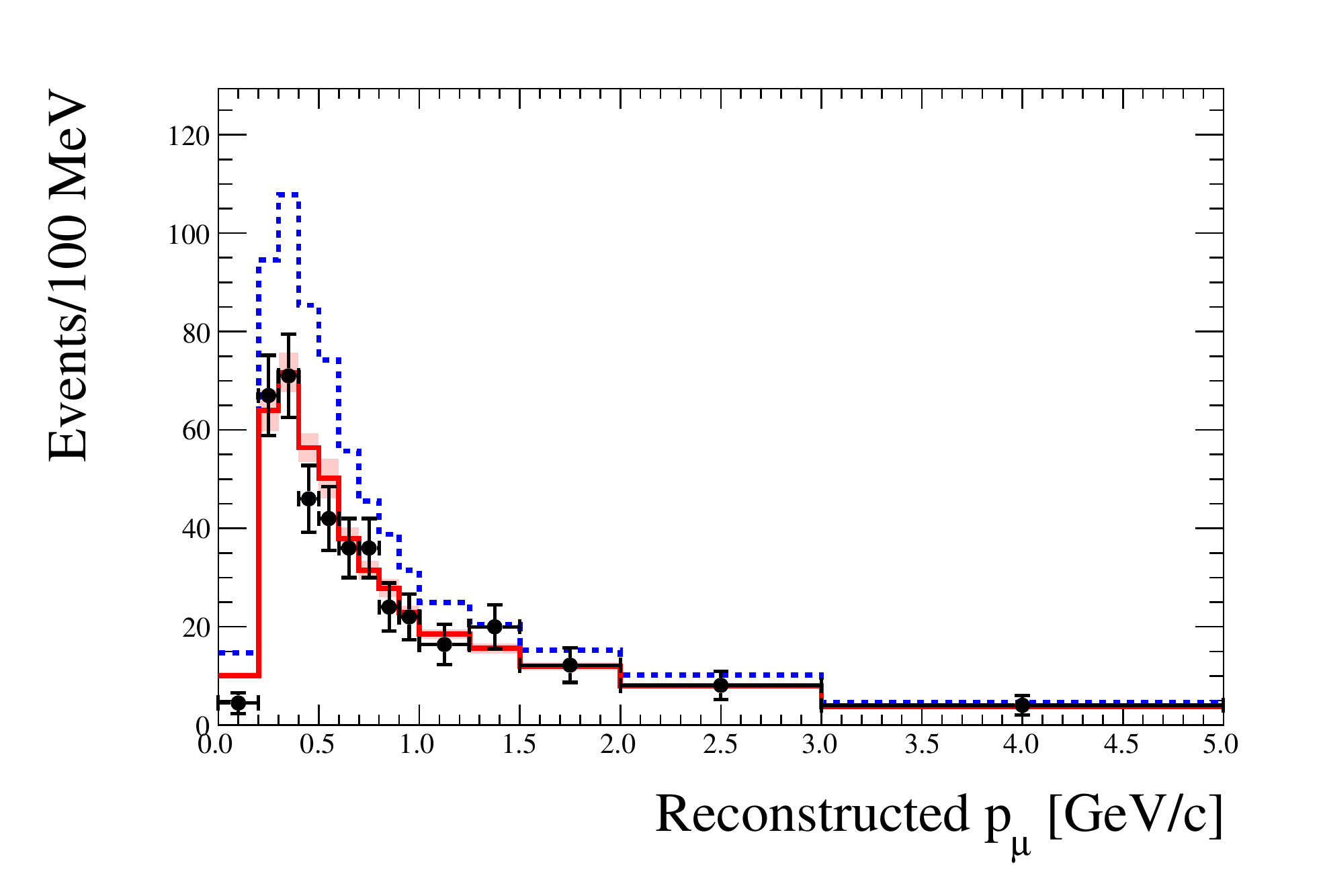}
		\put (42,22) {\includegraphics[width=0.20\linewidth]{LegendFitVsData}}
	\end{overpic}
	\includegraphics[width=0.40\linewidth]{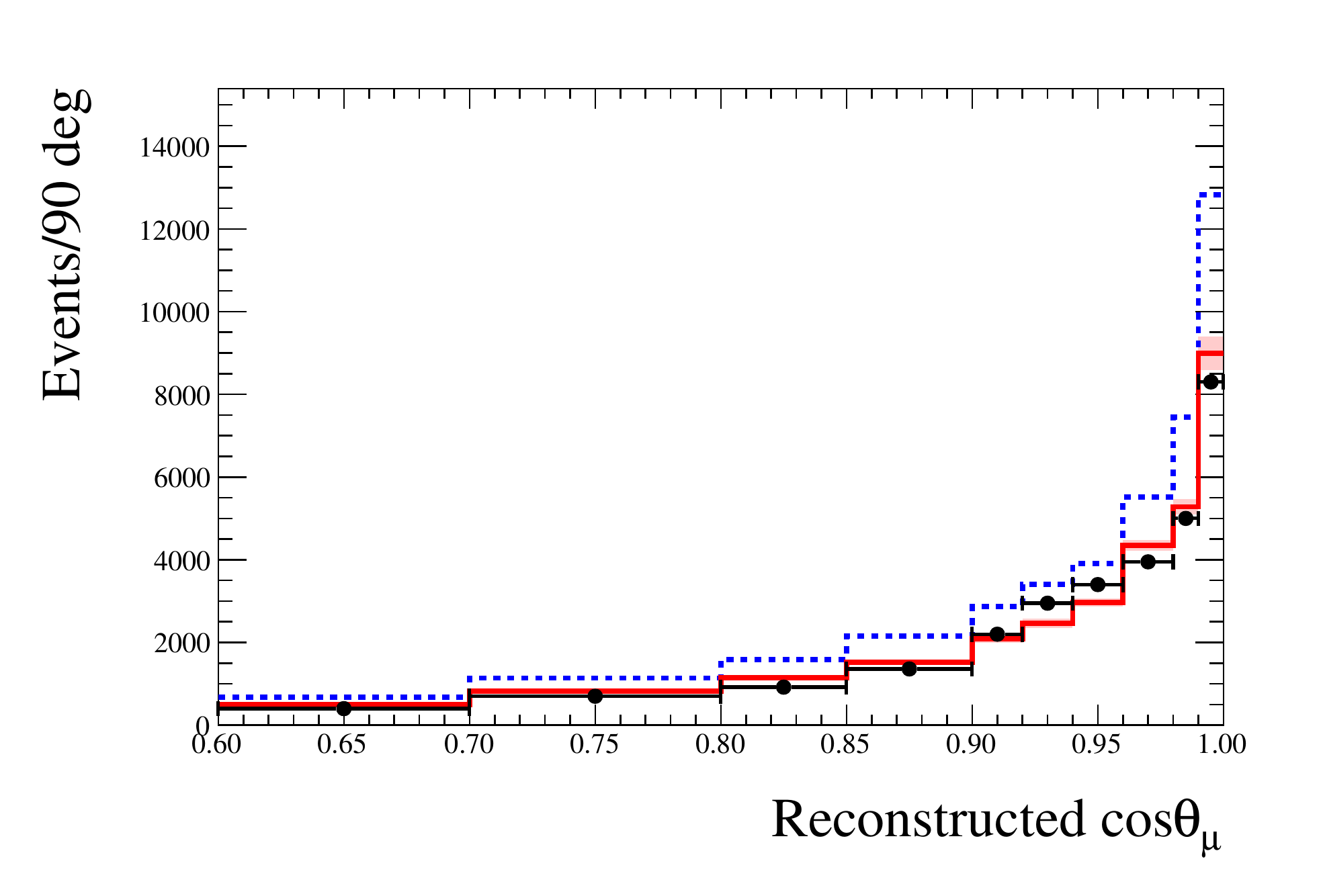}
	
	\includegraphics[width=0.40\linewidth]{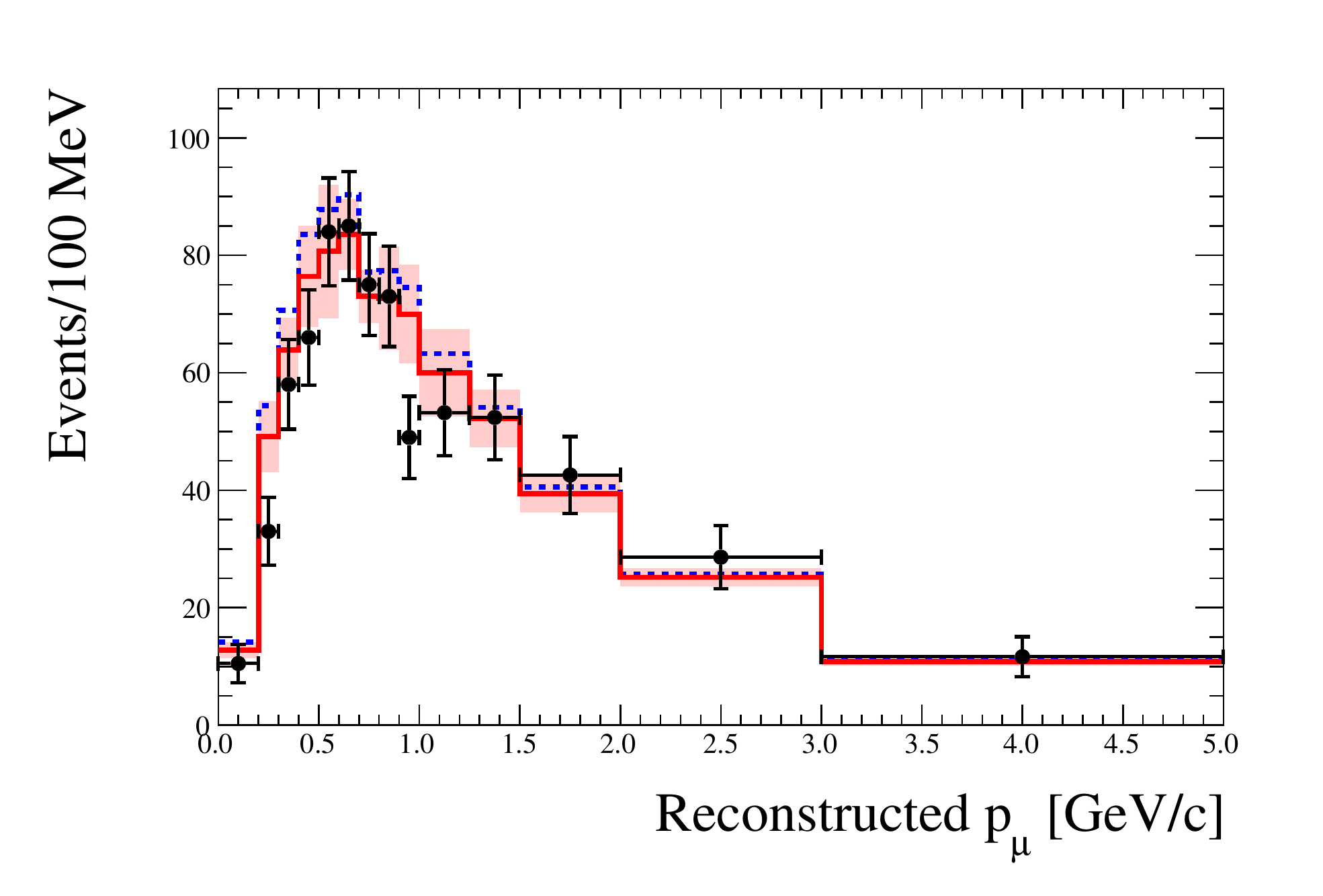}
	\includegraphics[width=0.40\linewidth]{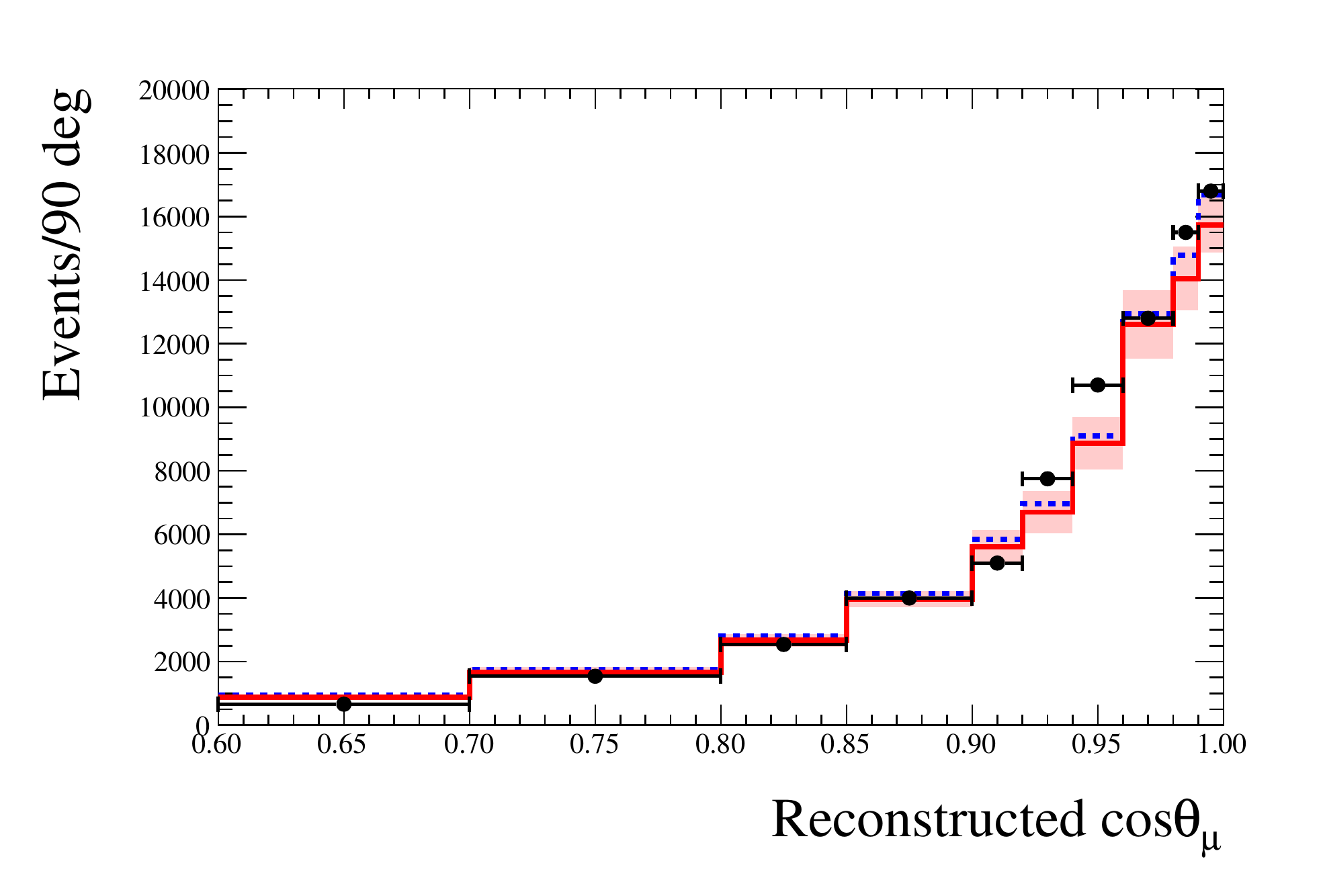}
	
	\caption{Distribution of events in the \numu \cconepiplus (top) and \ccother control samples (bottom), as a function of muon momentum (left) and muon $\cos\theta$ (right) compared with the MC predictions before the fit (dotted blue line), and those after the fit (solid red line) including systematics errors indicated by the pink band. The data are shown in black with statistical errors.}
	\label{fig:numucontrolsamples}
\end{figure*}

\begin{figure*}[th!]
	\centering
	\begin{overpic}[width=0.40\linewidth]{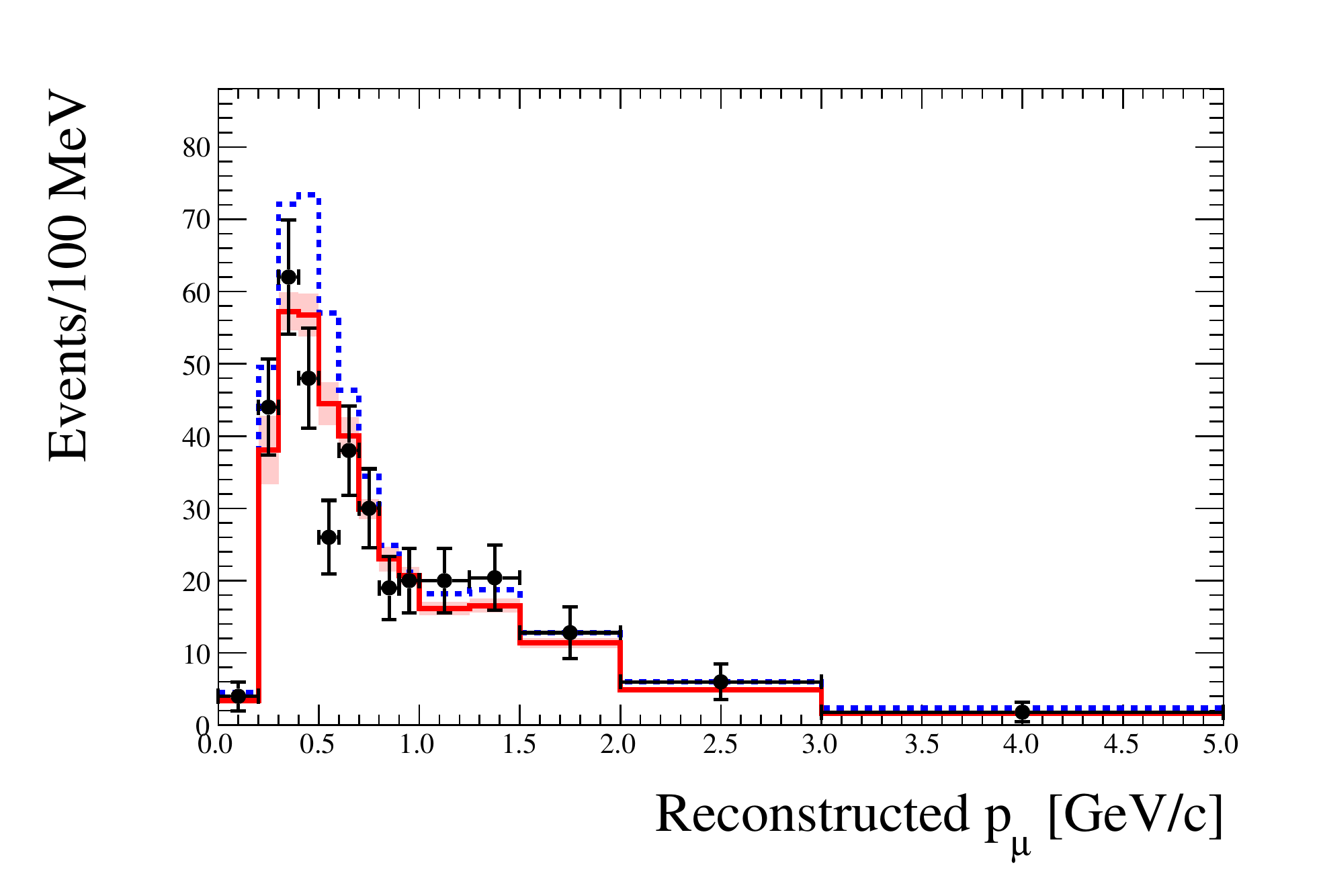}
		\put (42,22) {\includegraphics[width=0.20\linewidth]{LegendFitVsData}}
	\end{overpic}
	\includegraphics[width=0.40\linewidth]{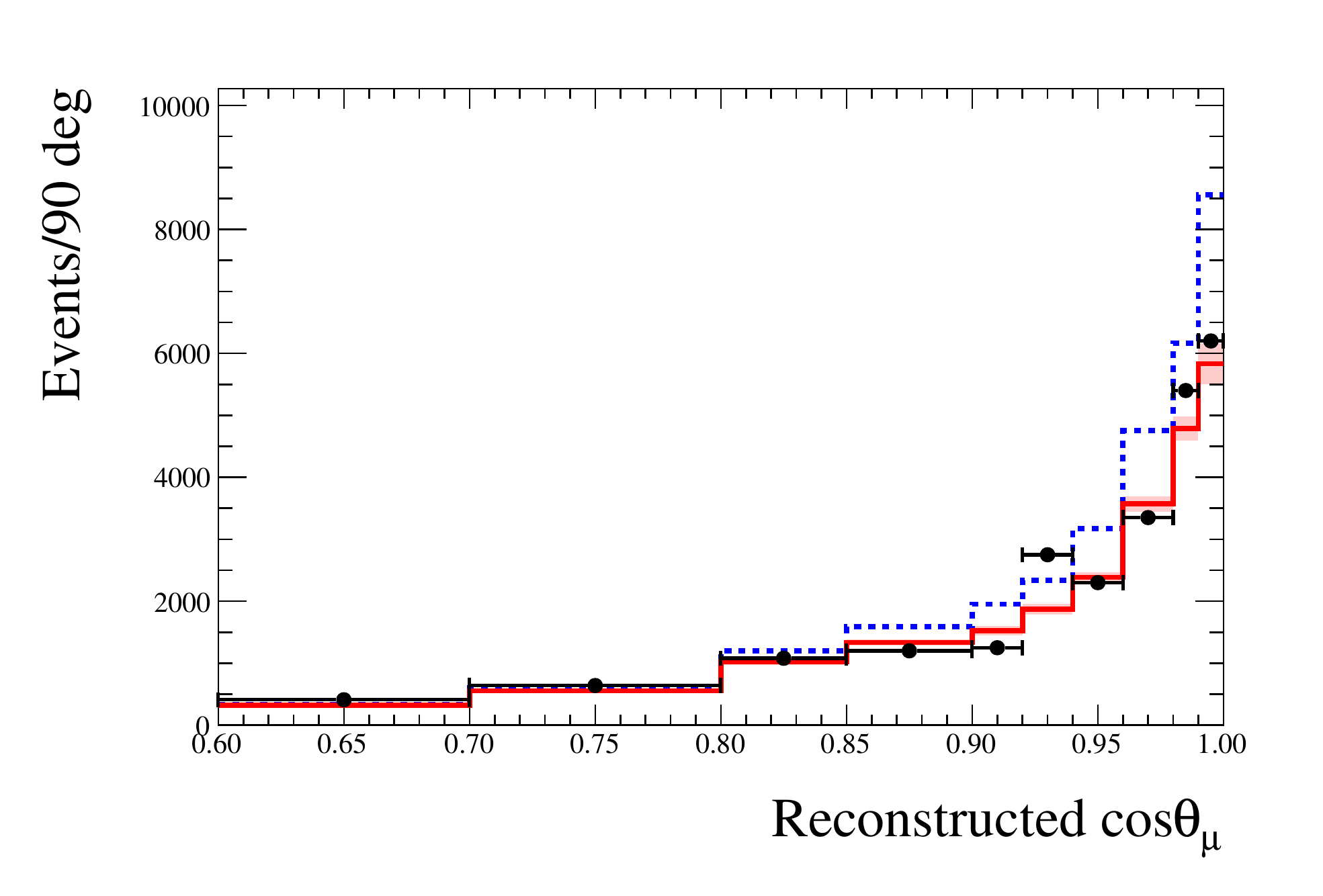}
	
	\includegraphics[width=0.40\linewidth]{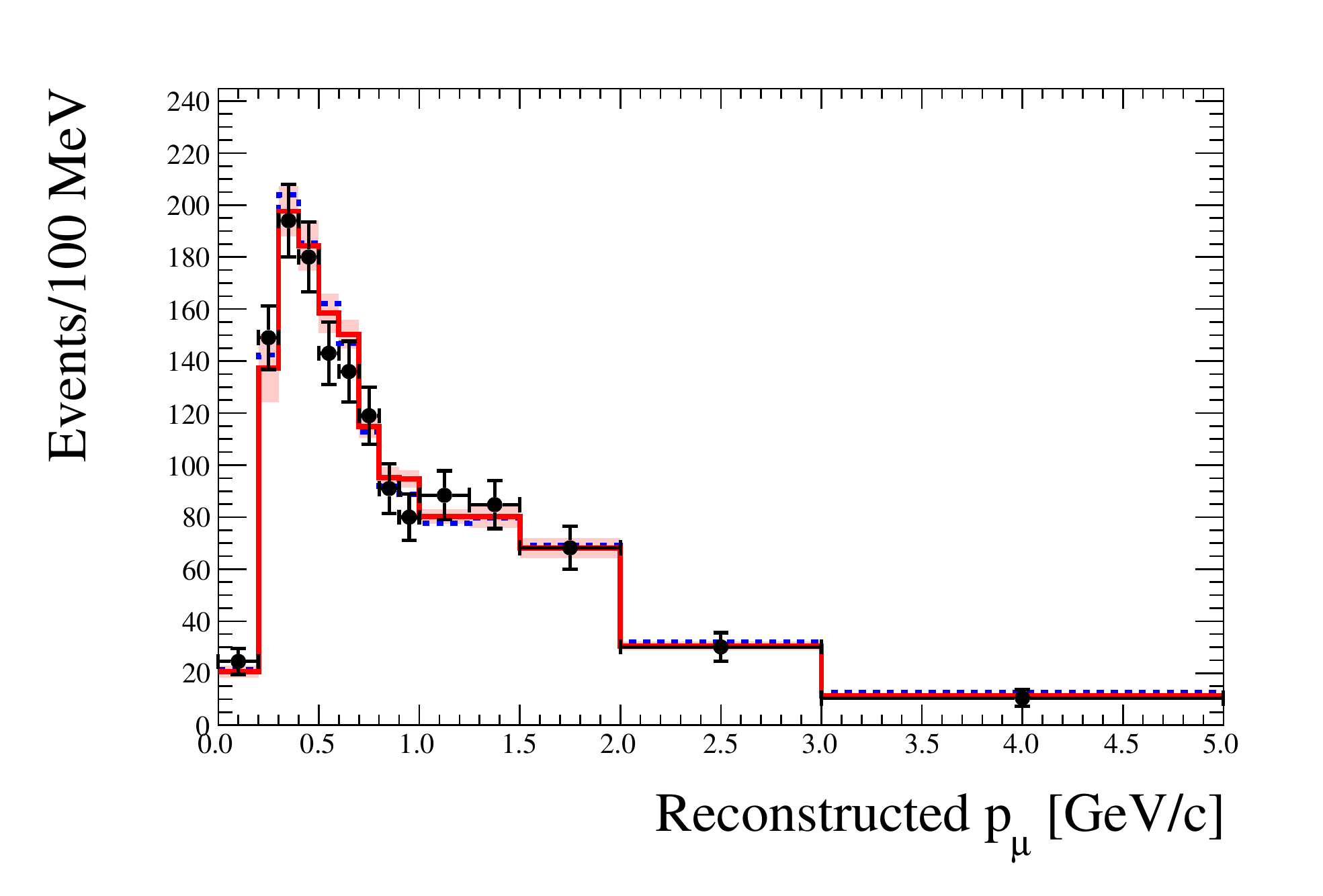}
	\includegraphics[width=0.40\linewidth]{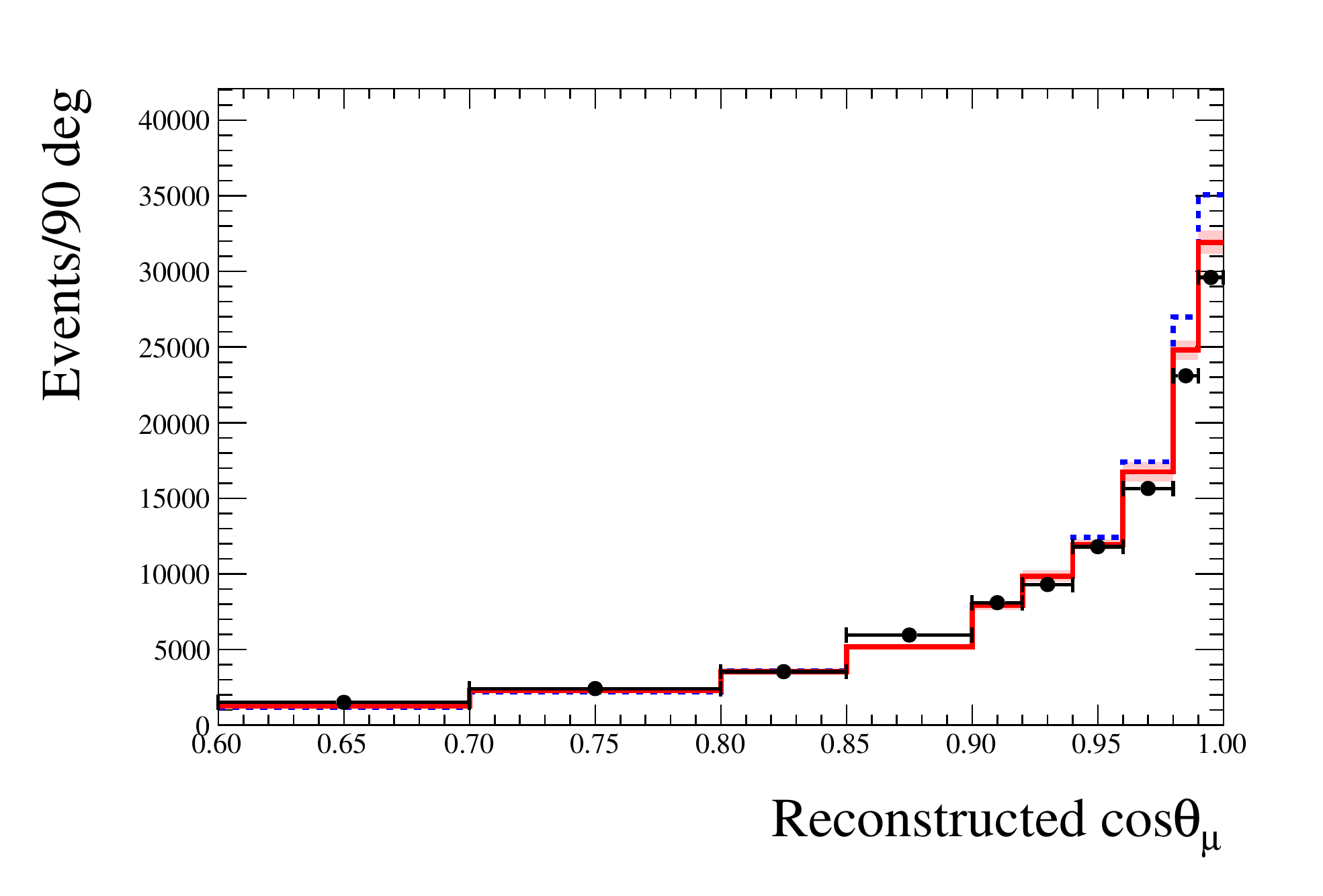}
	
	\caption{Distribution of events in the \barnumu \cconepiminus (top) and \ccother control samples (bottom), as a function of muon momentum (left) and muon $\cos\theta$ (right) compared with the MC predictions before the fit (dotted blue line), and those after the fit (solid red line) including systematics errors indicated by the pink band. The data are shown in black with statistical errors.}
	\label{fig:antinumucontrolsamples}
\end{figure*}

The distribution of reconstructed events in bins used to evaluate the cross sections and in the background control samples is shown in \cref{fig:numucc0pirecopostfitvsdata,fig:antinumucc0pirecopostfitvsdata}. The data are compared to MC predictions before and after the fit. The fit is able to reproduce the observed distributions in data by varying the parameters of interest to describe the signal cross section and the nuisance parameters describing the systematic uncertainties, as explained in \cref{sec:anaStrategy}. As \cref{fig:numucontrolsamples,fig:antinumucontrolsamples} show, the large discrepancy in the pre-fit MC prediction in the CC1$\pi^\pm$ control region is well corrected by the fit by varying the nuisance parameters describing the pion production cross-section listed in \cref{tbl:xsecparam}.

In the following, the measured cross sections, and their combinations, are compared to different (anti)neutrino-interaction models using the framework \textsc{Nuisance}~\cite{Stowell:2016jfr} and the agreement is quantified by the $\chi^2$ statistic. Since the \numu and the \barnumu cross sections are extracted simultaneously, a global $\chi^2$ computed using the full covariance matrix, i.e. the one containing the correlation between the two cross sections, is reported. It should be noted that, especially for the neutrino and antineutrino cross sections, as well as their sum, the overall normalization uncertainty (fully correlated between bins) constitutes a relatively large fraction of the uncertainty. In particular it contributes to 48\% and 35\% of the total systematic uncertainty for neutrino and antineutrino cross sections respectively and to 49\% for the sum, while for the difference it decreases to 19\% and for the asymmetry to 5\%. Therefore the $\chi^2$ statistics may suffer from Peelle's Pertinent Puzzle~\cite{ppp} and may not be a reliable estimation of the data-MC agreement. This issue does not affect the shape-only $\chi^2$ which is reported as well. 

The models considered for the data-MC comparisons are as follows:
\begin{itemize}
	\item \textsc{Neut} (version~\texttt{5.4.1}) Local Fermi Gas (LFG) assuming an axial mass $M_A^{QE}=1.03$~GeV/c$^2$, and corrections from the Random Phase Approximation (RPA) approach with and without 2p2h. A 1p1h and 2p2h model is used in this case from Ref.~\cite{Nieves:2011pp};
	\item \textsc{NuWro} (version~\texttt{18.02.1}) LFG~\cite{Golan:2012wx} assuming an axial mass $M_A^{QE}=1.03$~GeV/c$^2$ with 2p2h and RPA corrections also from Ref.~\cite{Nieves:2011pp};
	\item \textsc{Genie} (version~\texttt{3.00.04}, configuration \texttt{G18\_10b\_000\_00}) LFG assuming an axial mass $M_A^{QE}=0.99$~GeV/c$^2$ with 2p2h and RPA corrections from Ref.~\cite{Nieves:2011pp};
	\item \textsc{NuWro} Spectral Function (SF), as developed in Ref.~\cite{Benhar:1994hw}, using the same 2p2h model as \textsc{Neut};
	\item \textsc{GiBUU 2019} LFG in a coordinate- and momentum-dependent nuclear potential, as described in Ref.~\cite{Gallmeister:2016dnq}, using a 2p2h model based on Ref.~\cite{OConnell:1972edu} and further tuned in Ref.~\cite{Dolan:2018sbb}, which uses the T2K measurements of final-state muon and proton kinematics and correlations in charged-current pionless interactions discussed in Ref.~\cite{Abe:2018pwo};
	\item \textsc{SuSav2} is a complete implementation of the SuSAv2 model~\cite{Gonzalez-Jimenez:2014eqa,Megias:2014qva,Megias:2016lke,Megias:2016fjk} in GENIE, as described in~\cite{Dolan:2019bxf}, where 1p1h is based on the Relativistic Mean Field approach~\cite{Caballero:2005sj} and 2p2h is based on the calculation from Ref.~\cite{Simo:2016ikv}. The pion production and FSI models are the same as in \textsc{Genie} version~\texttt{3.00.04}, configuration \texttt{G18\_10b\_000\_00}; 
	\item Martini \textit{et al.} is the model described in Ref.~\cite{Martini:2009uj}. It employs a LFG 1p1h model and RPA corrections including contribution from 2p2h.
\end{itemize}
The contribution of pion production, subsequently reabsorbed by FSI, is included in all the generators but not in the model by Martini \textit{et al.}. It accounts for about 10\%(5\%) of the neutrino (antineutrino) measured cross section and, in order to properly compare this model with others, a prediction of this component obtained using \textsc{Neut} version~\texttt{5.4.1} has been added on top of the Martini \textit{et al.} prediction. This model is also missing antineutrino interactions on hydrogen which have been added to the antineutrino \cczeropi cross section using the same strategy described above. 

\begin{figure*}[h!]
	\centering
	\includegraphics[width=0.36\linewidth]{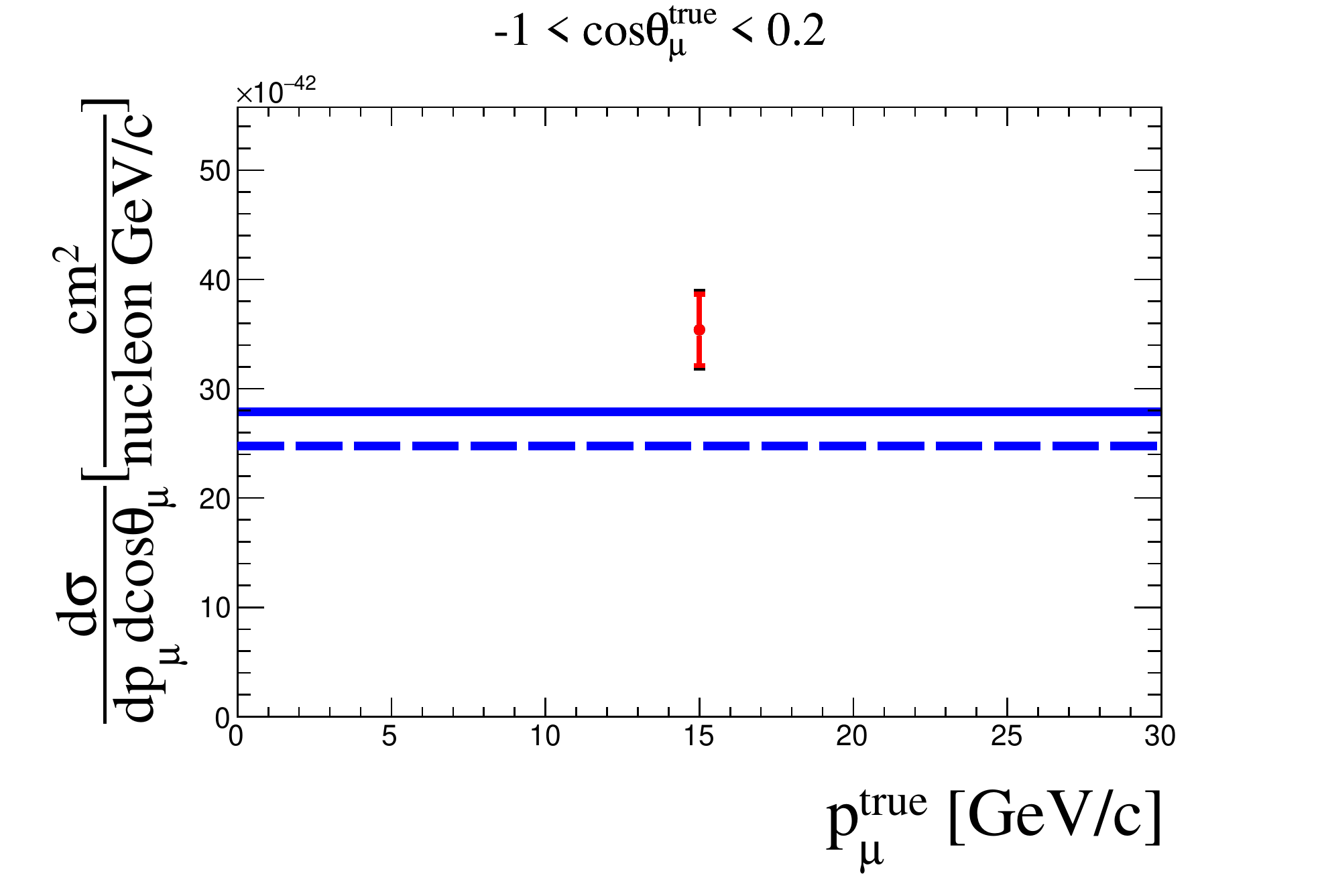}
	\includegraphics[width=0.36\linewidth]{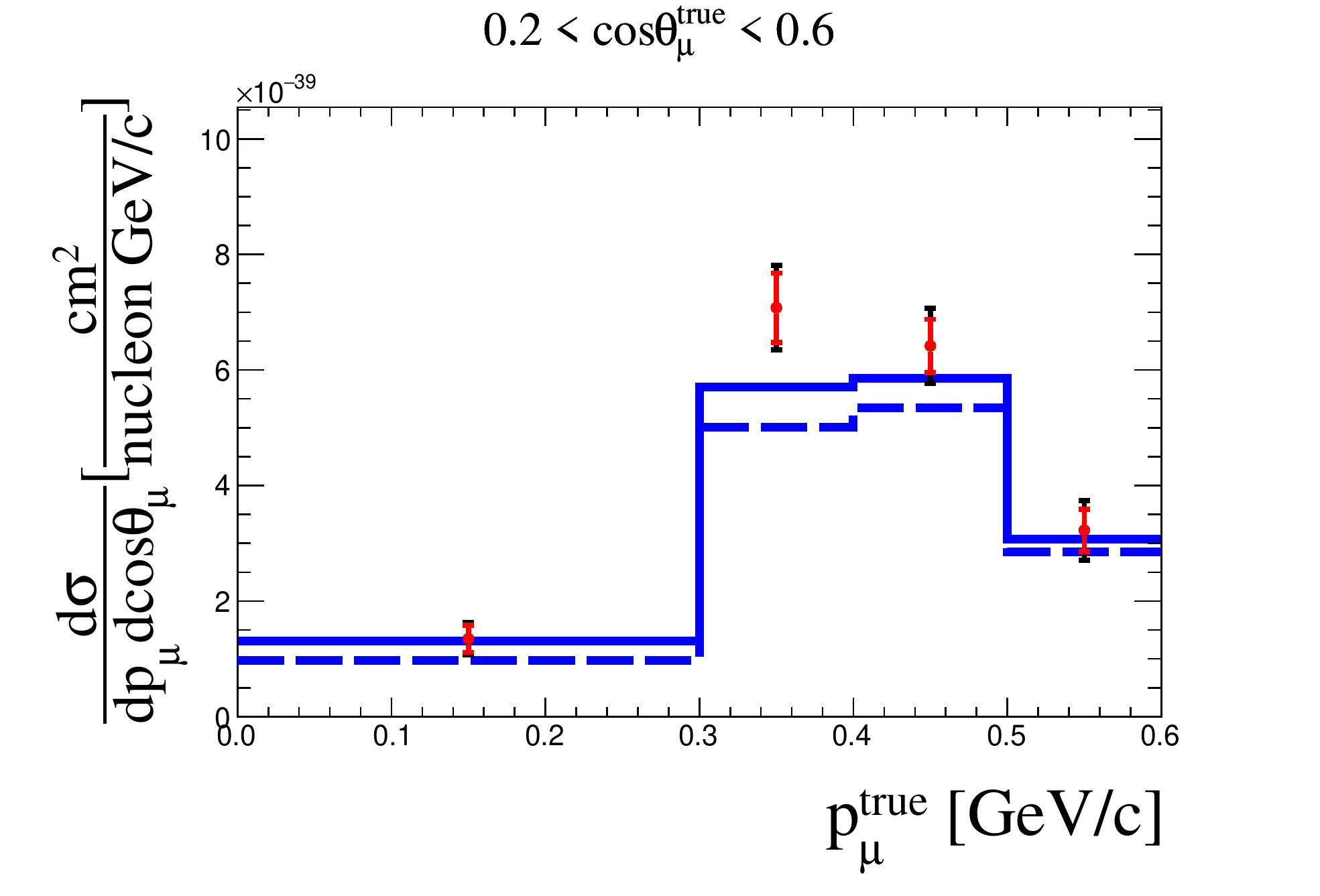}
	\includegraphics[width=0.36\linewidth]{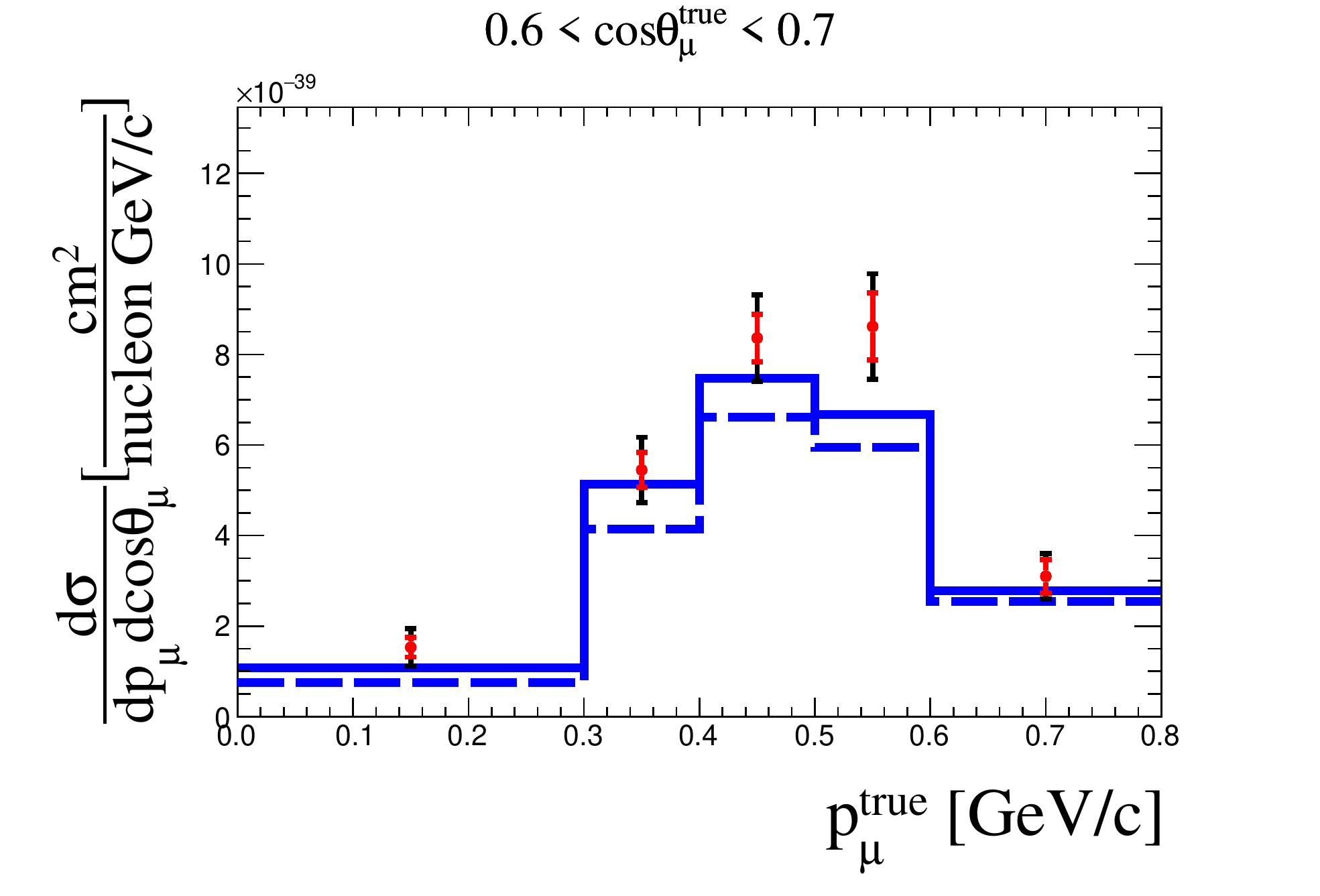}
	\includegraphics[width=0.36\linewidth]{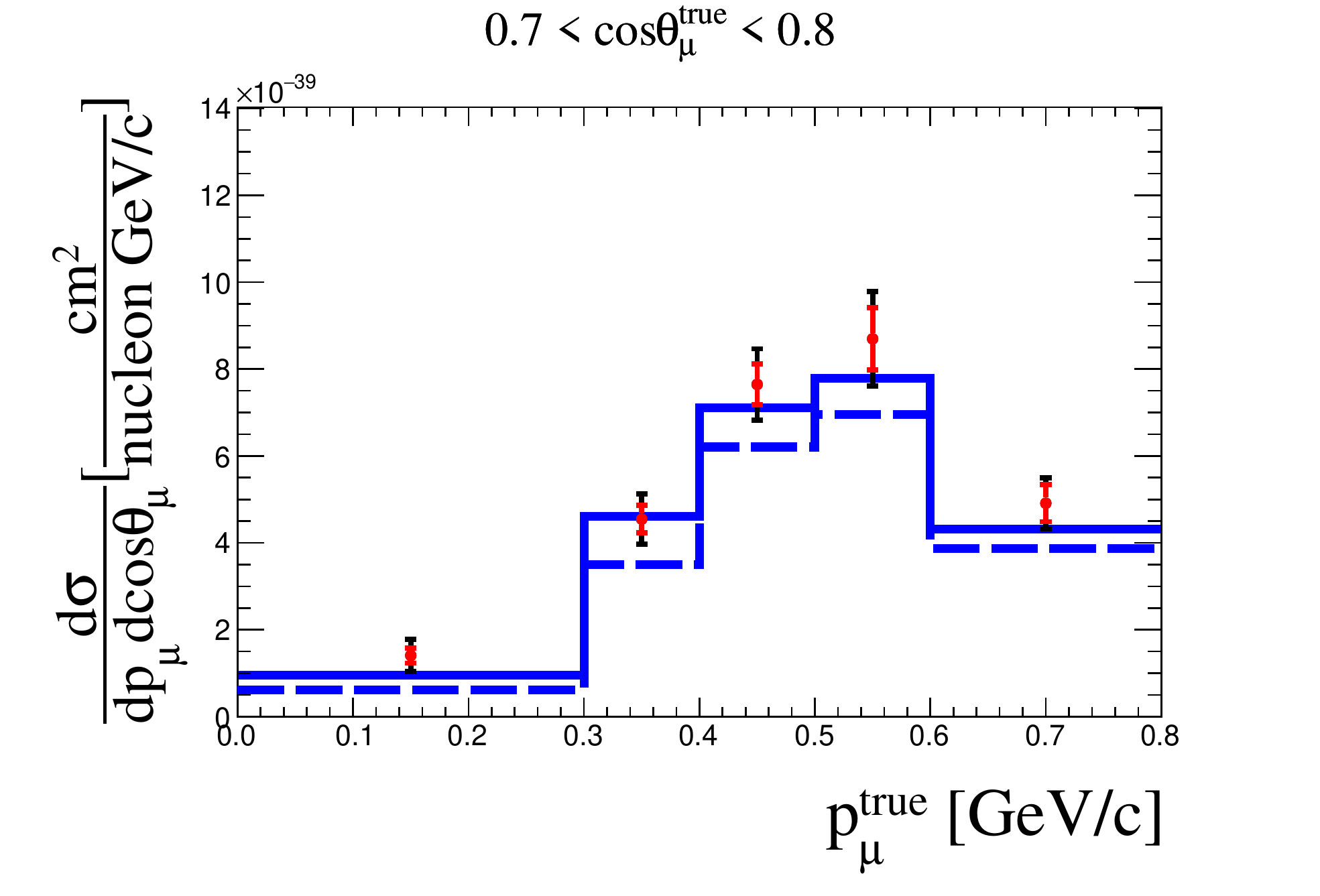}
	\includegraphics[width=0.36\linewidth]{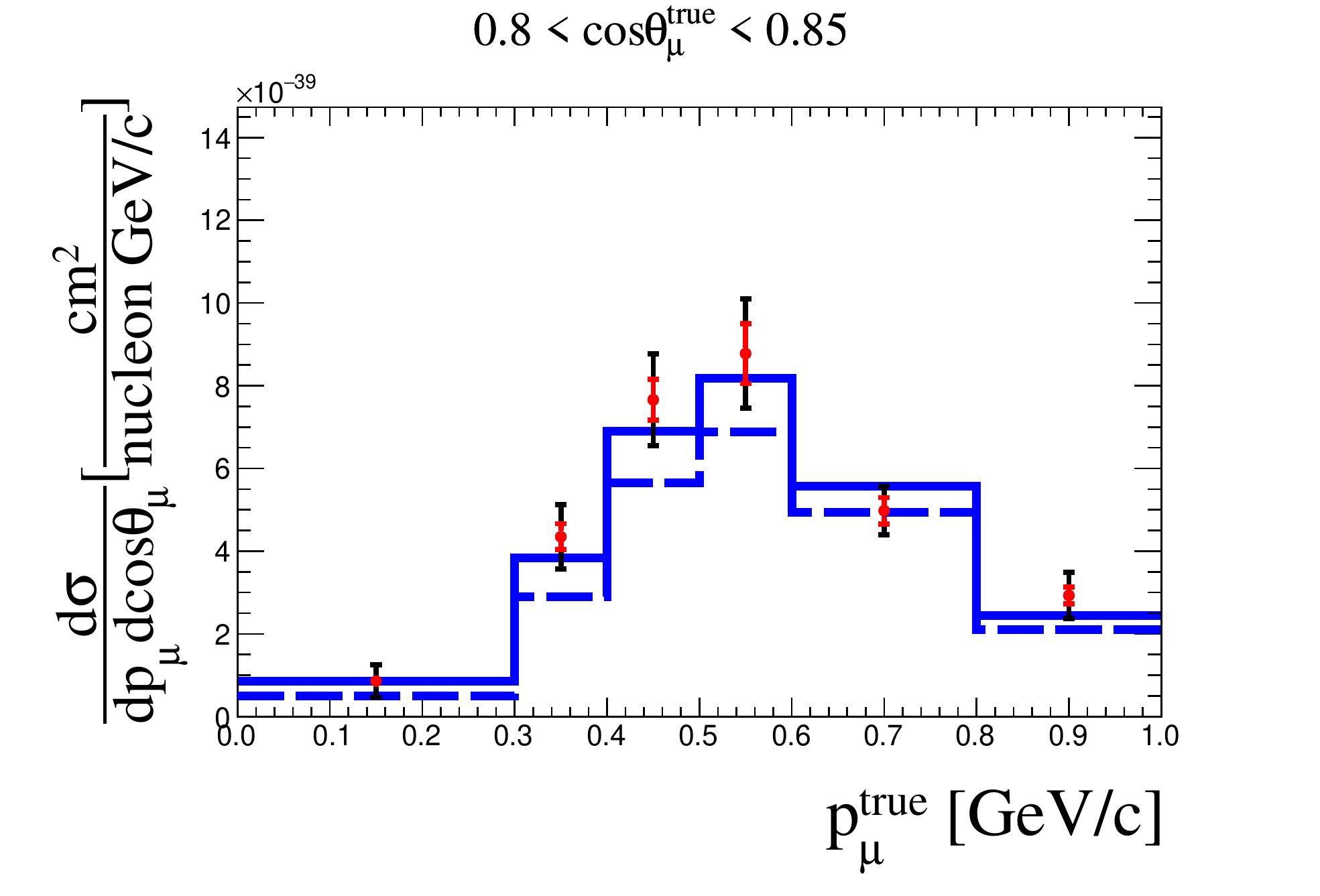}
	\includegraphics[width=0.36\linewidth]{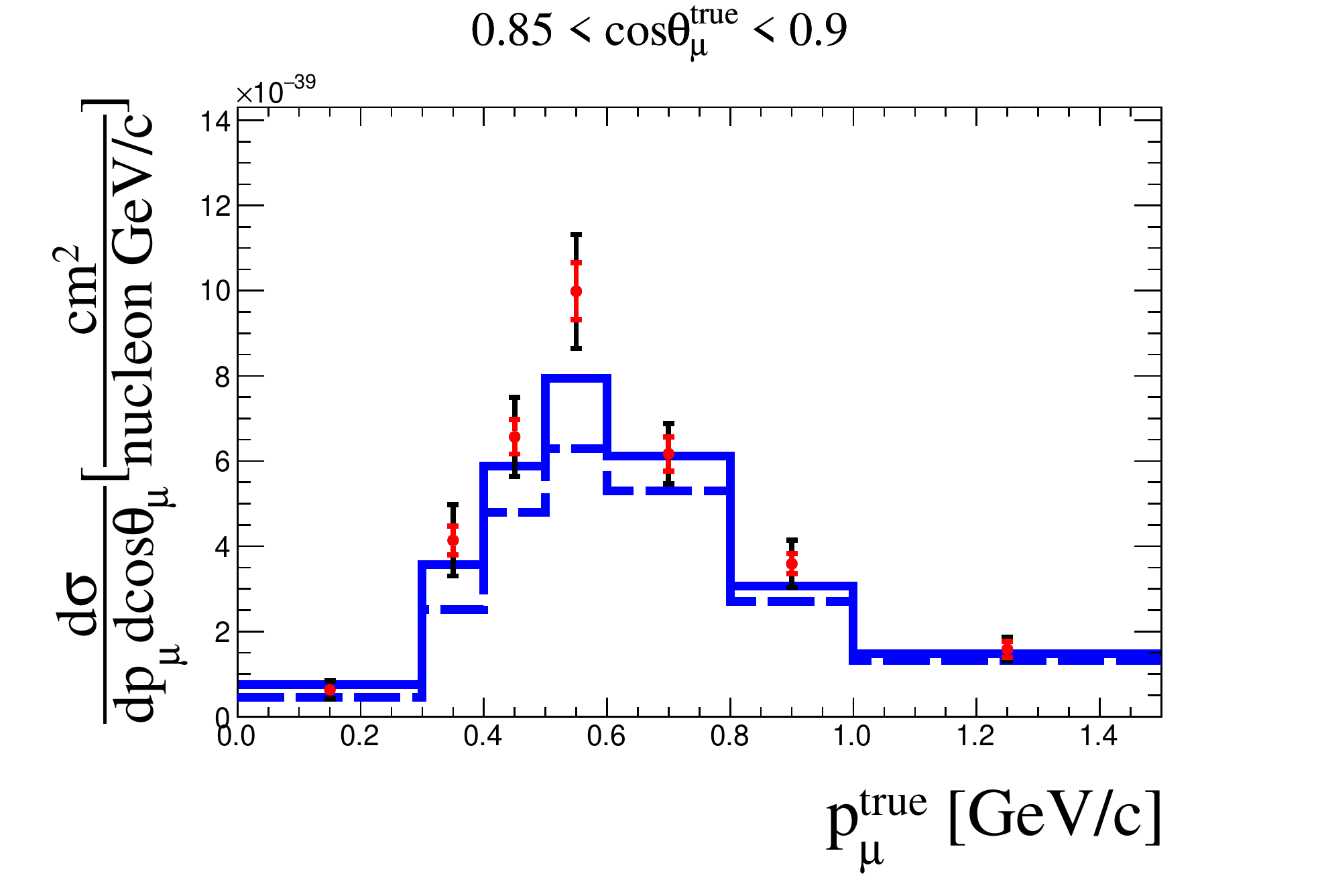}
	\includegraphics[width=0.36\linewidth]{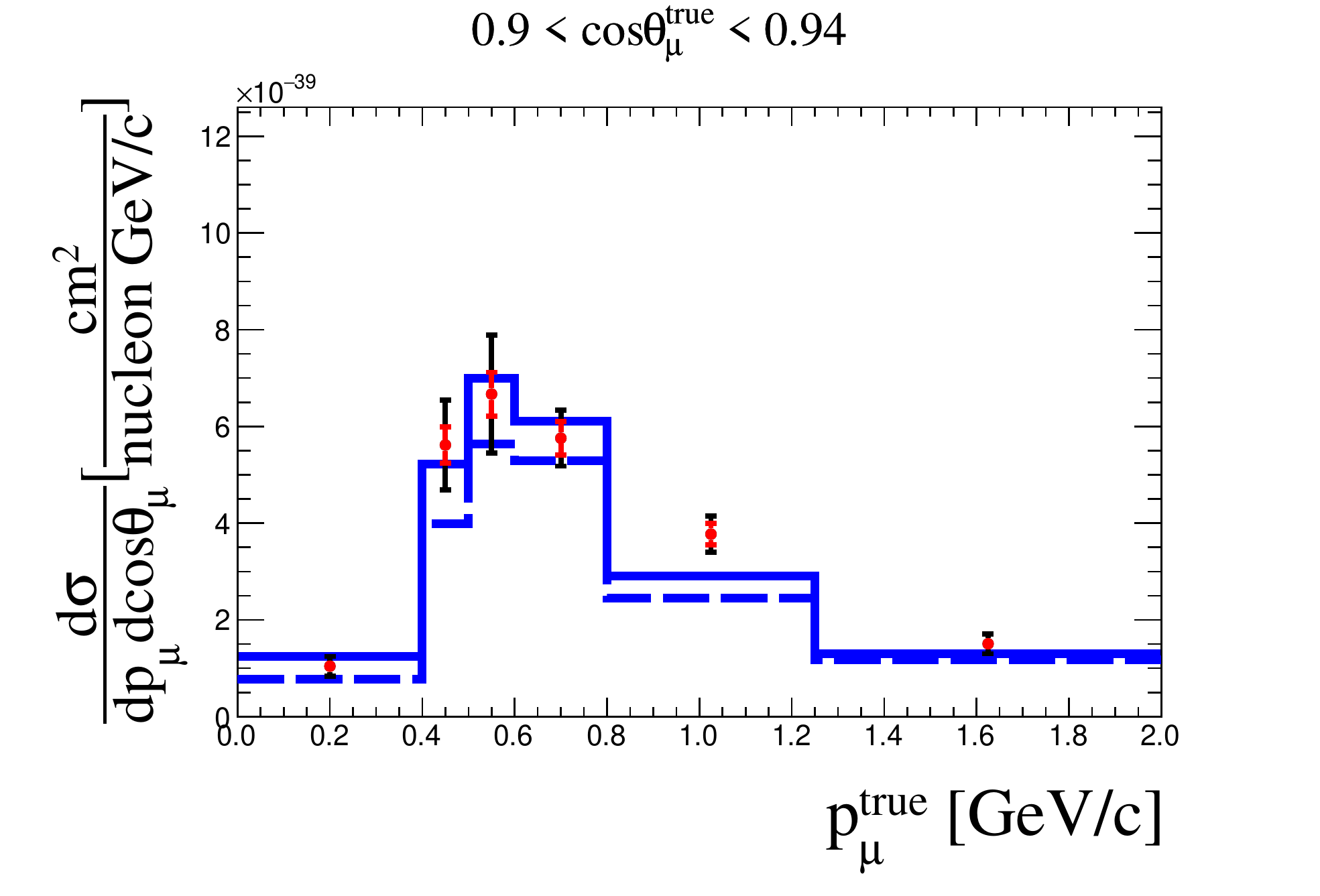}			
	\includegraphics[width=0.36\linewidth]{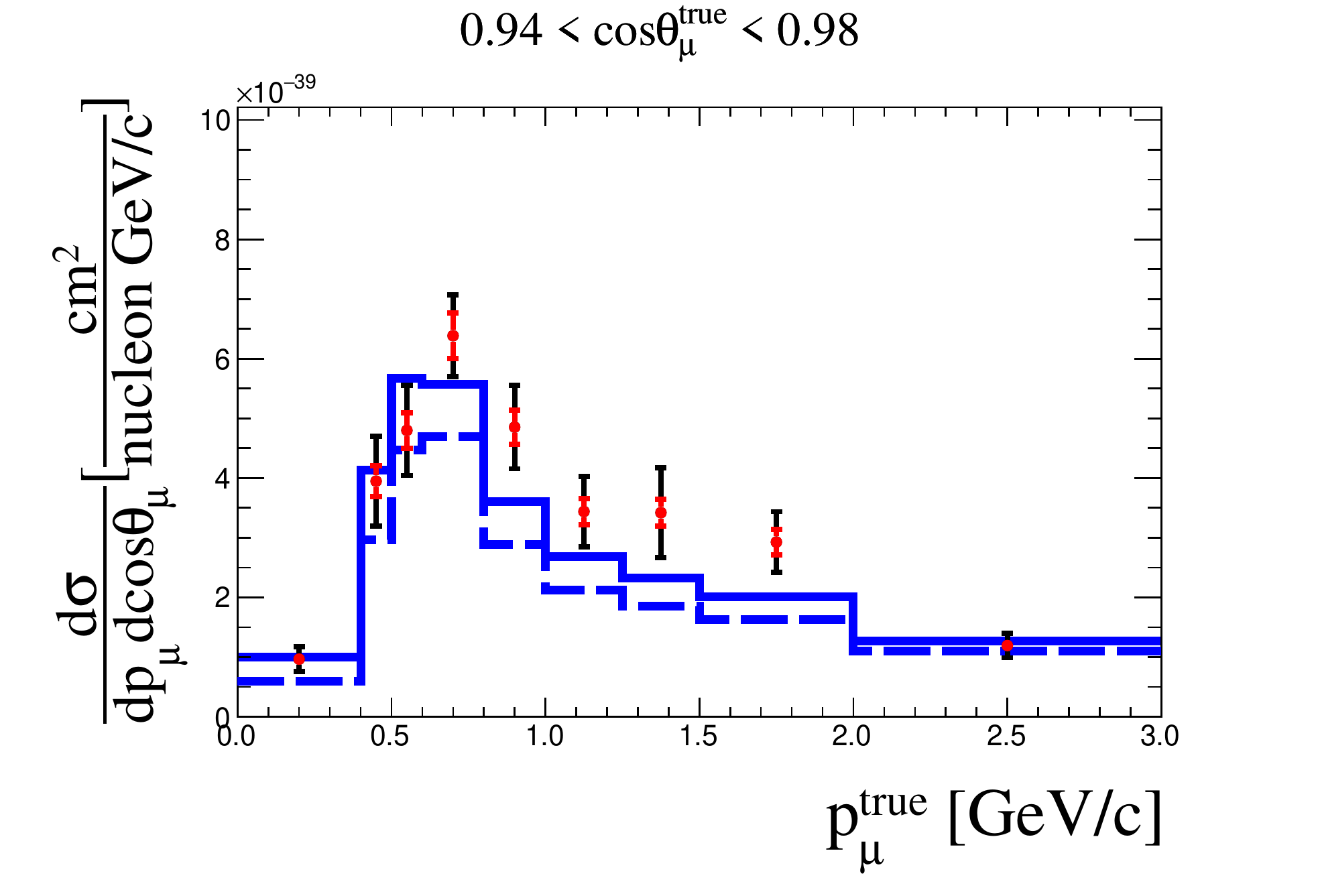}
	\includegraphics[width=0.36\linewidth]{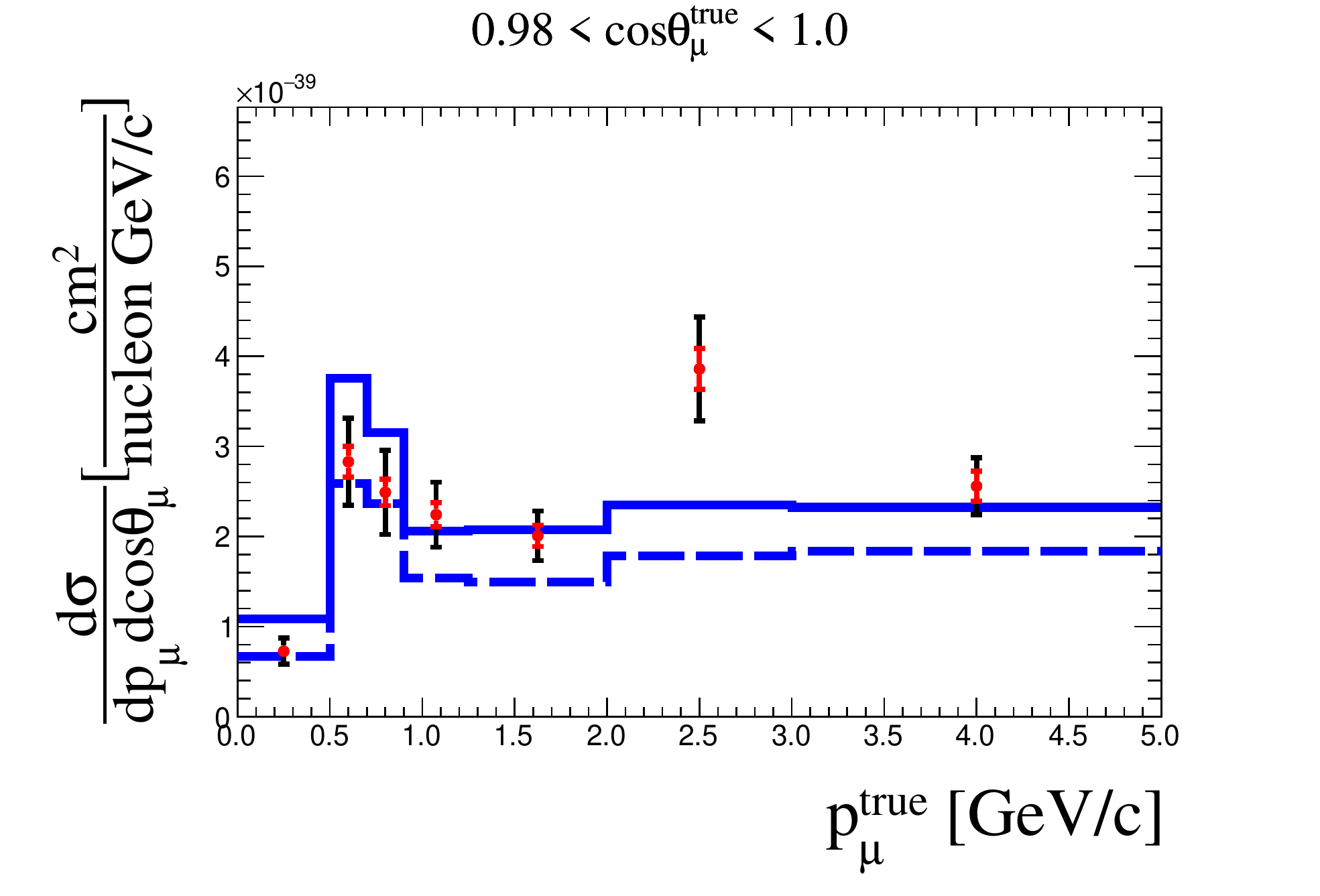}	
	\includegraphics[width=0.36\linewidth]{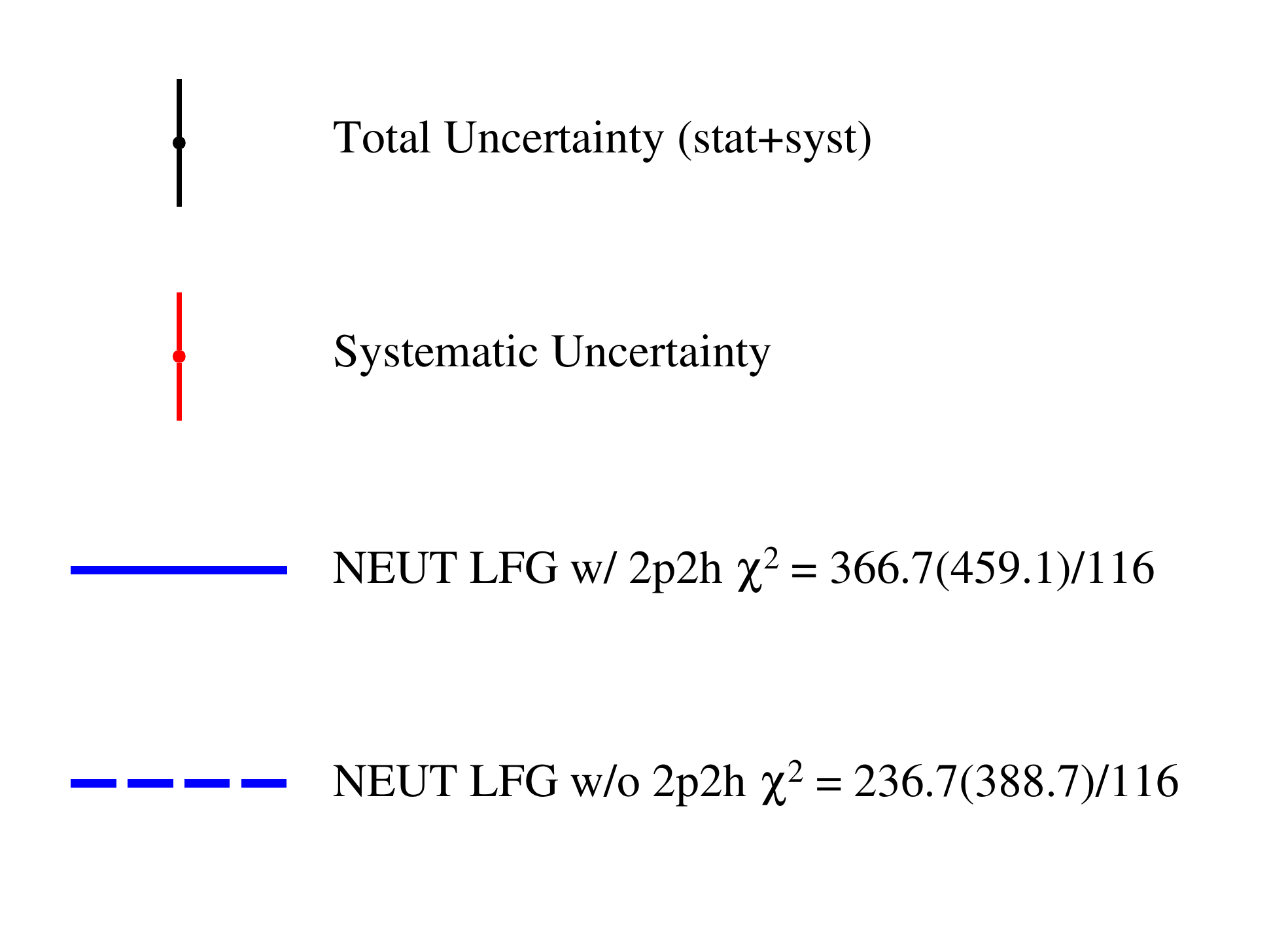}				
	\caption{Measured \numu \cczeropi double-differential cross-section per nucleon in bins of true muon kinematics with systematic uncertainty (red bars) and total (stat.+syst.) uncertainty (black bars). The results are compared to \textsc{Neut} version~\texttt{5.4.1}, which uses an LFG+RPA model, with (solid line) and without (dashed line) 2p2h. The full and shape-only (in parenthesis) $\chi2$ are reported. The last bin in momentum is not displayed for readability.}
	\label{fig:numucc0pixsecneut2p2h}
\end{figure*}

\begin{figure*}[h!]
	\centering
	\includegraphics[width=0.36\linewidth]{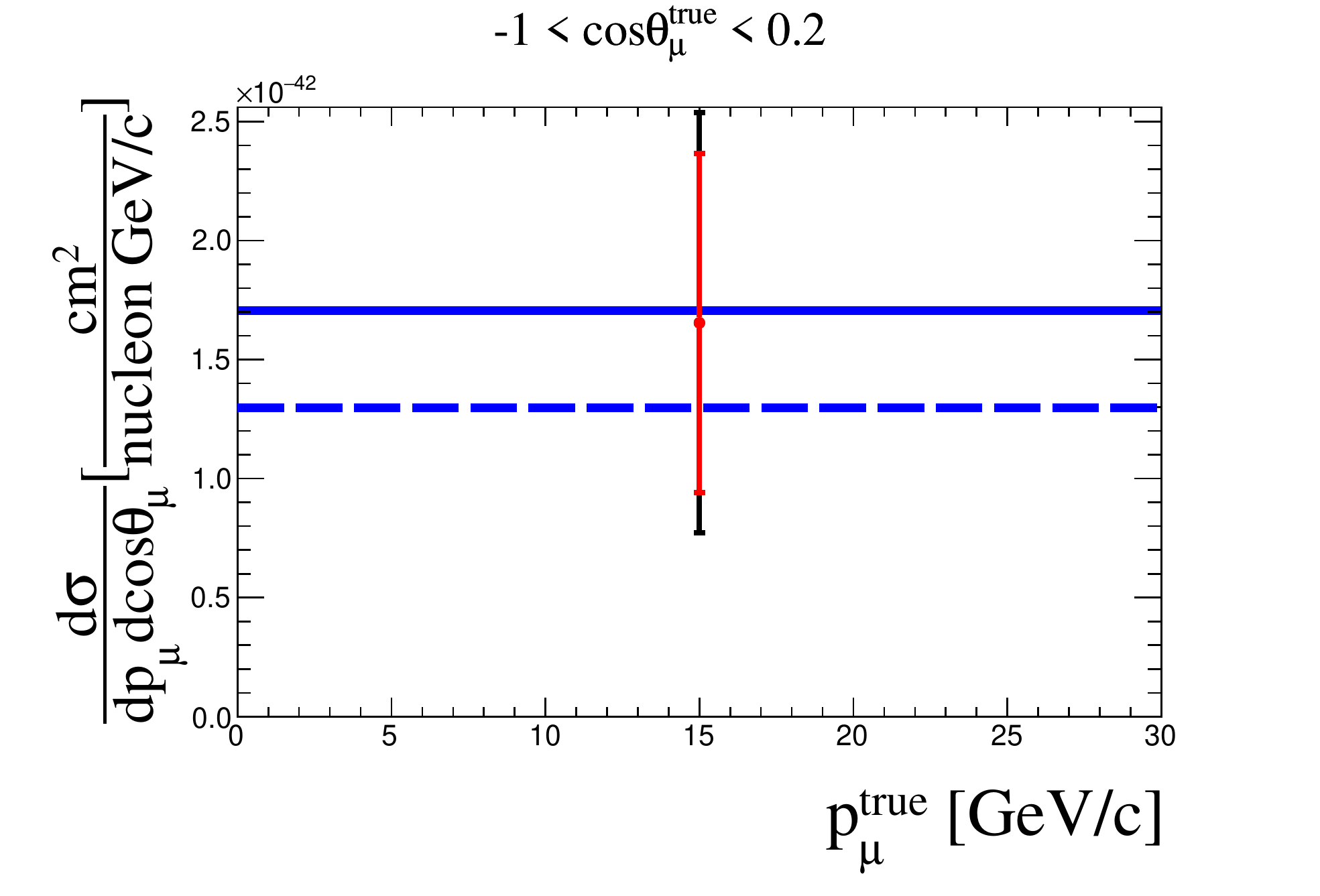}
	\includegraphics[width=0.36\linewidth]{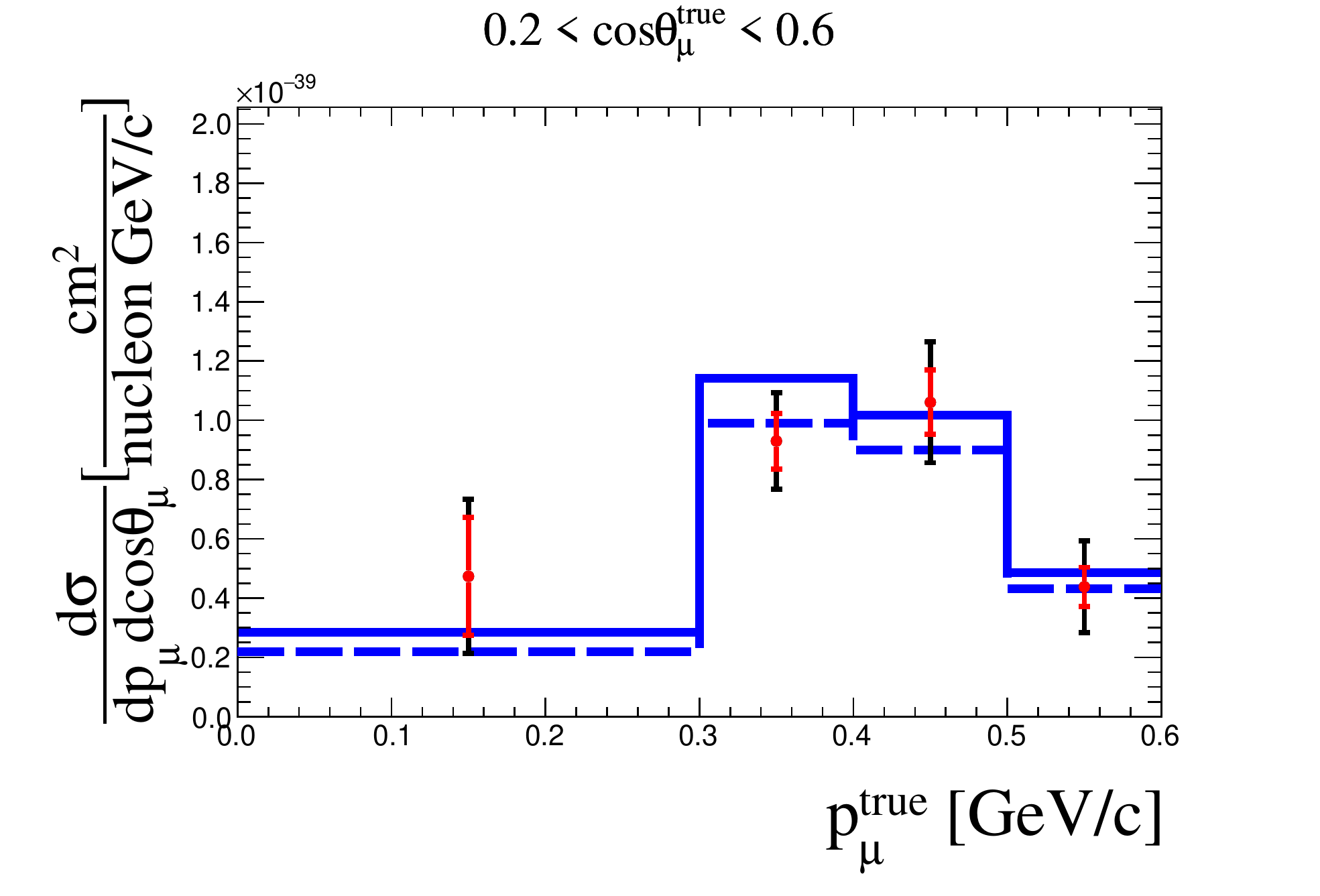}
	\includegraphics[width=0.36\linewidth]{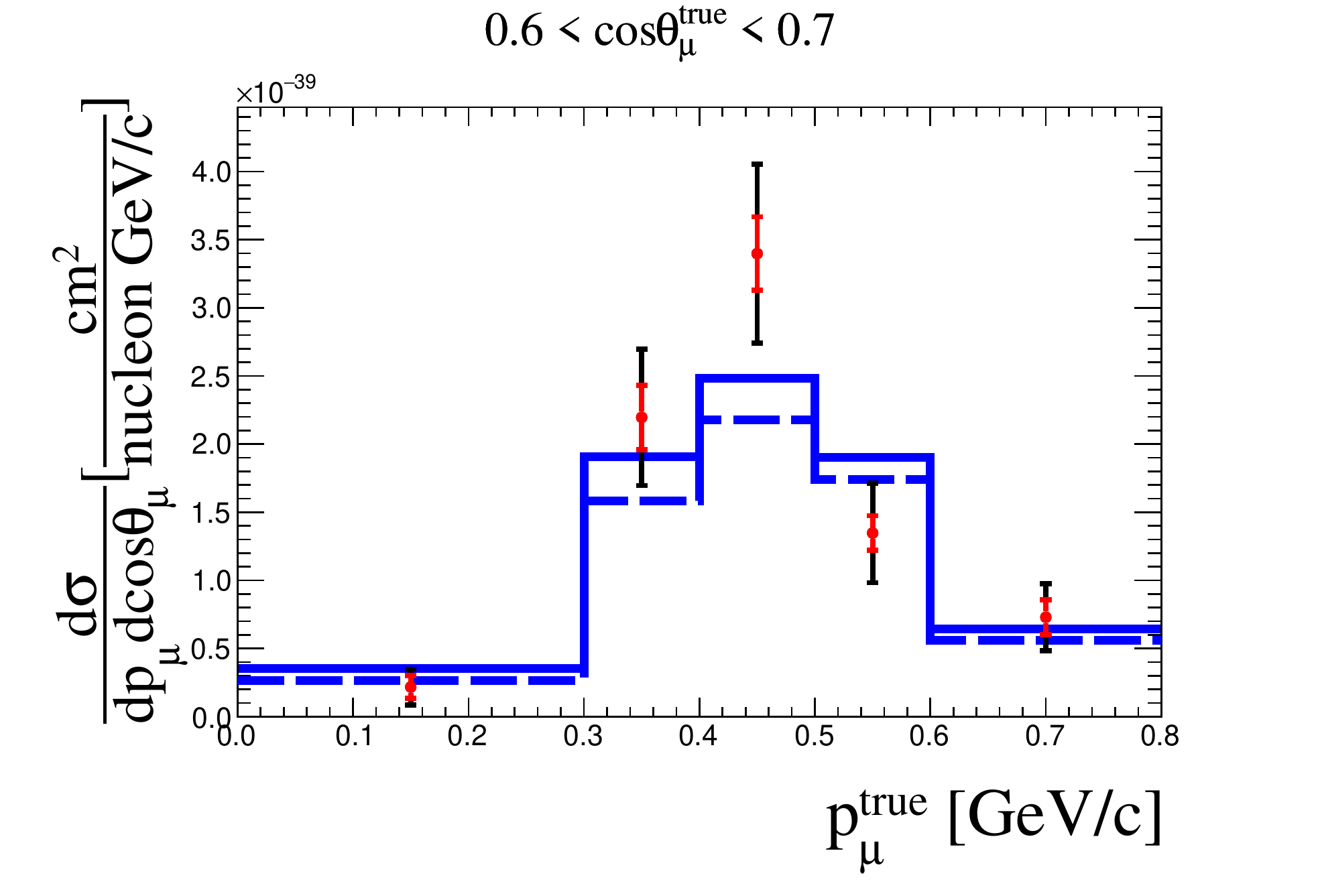}
	\includegraphics[width=0.36\linewidth]{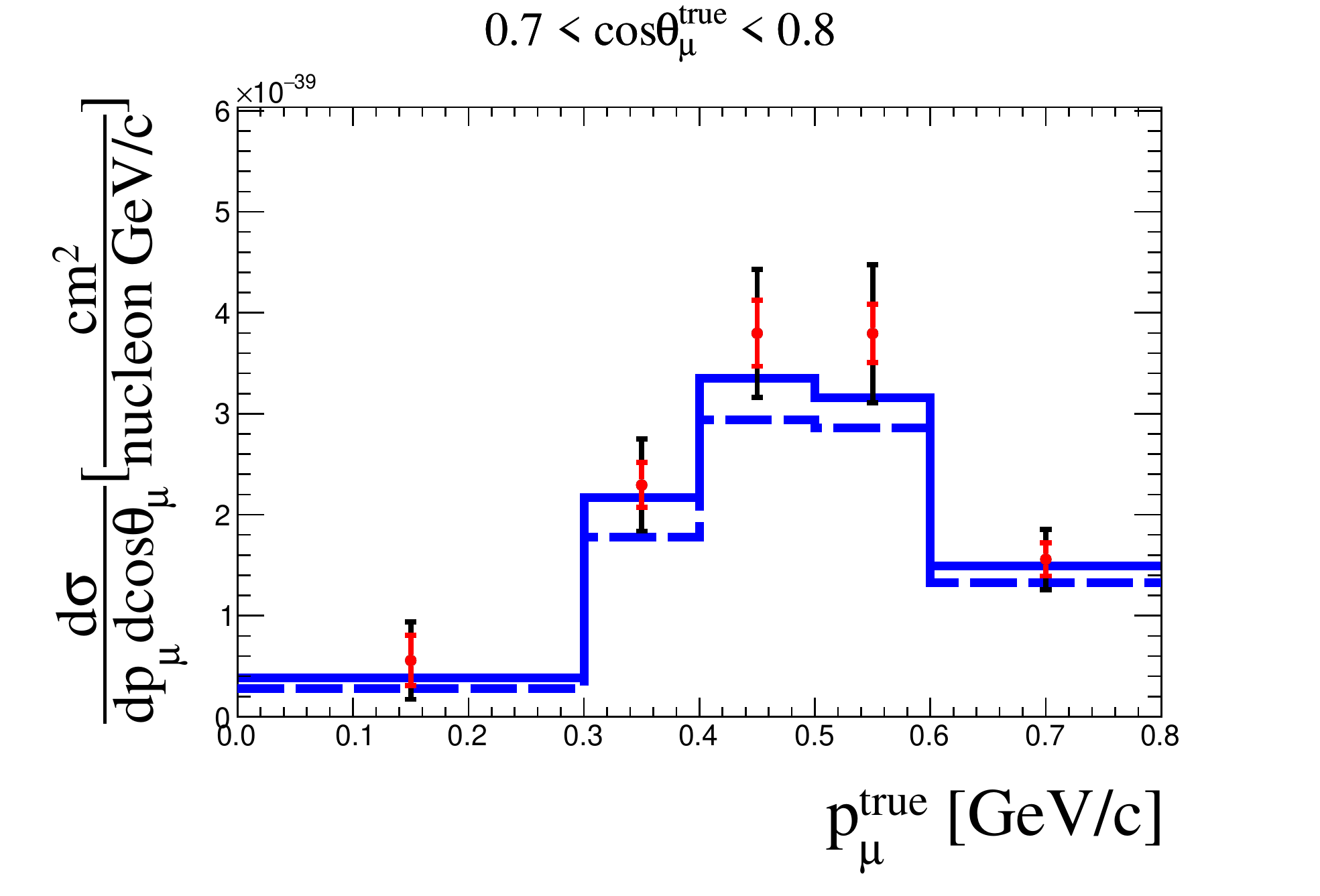}
	\includegraphics[width=0.36\linewidth]{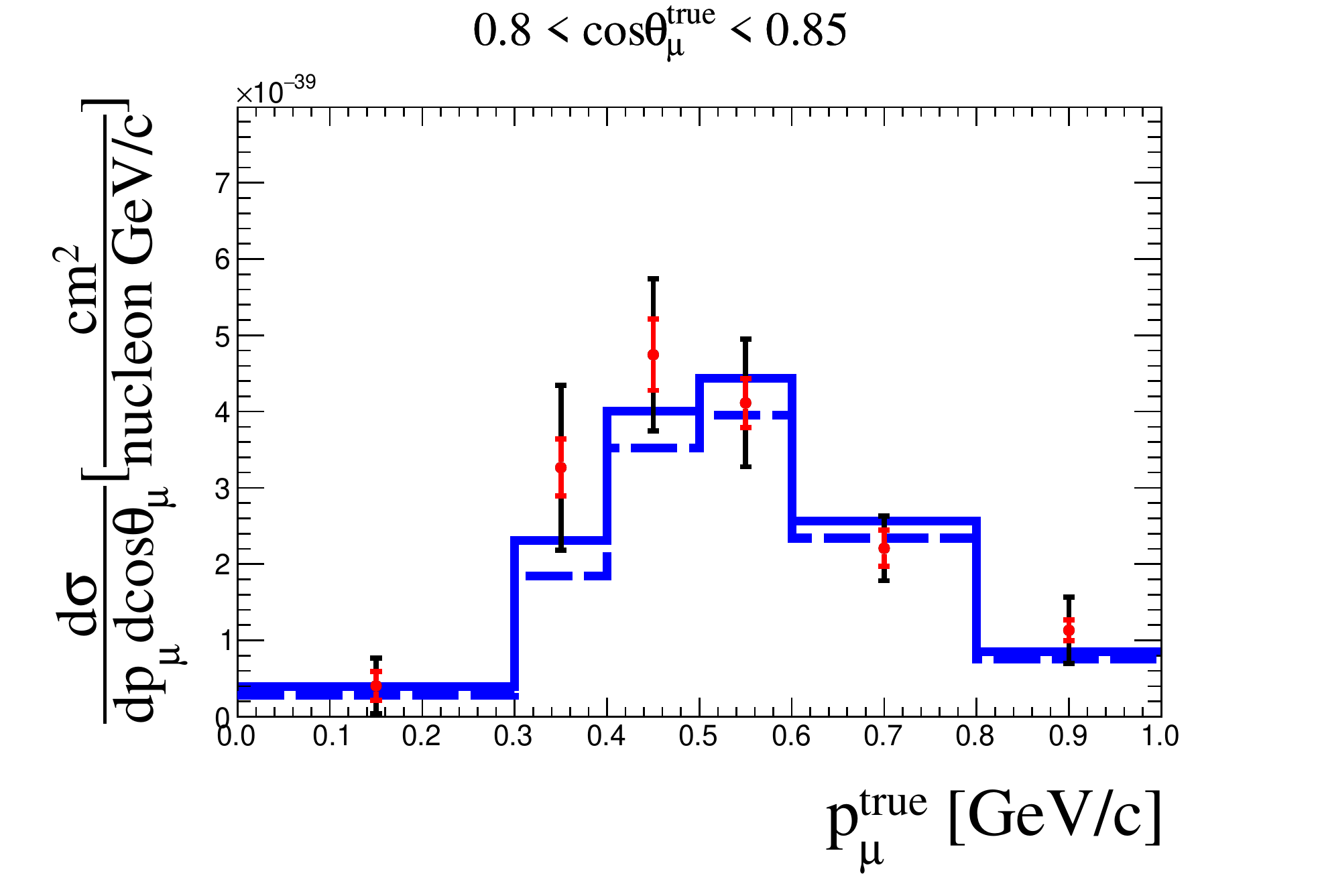}
	\includegraphics[width=0.36\linewidth]{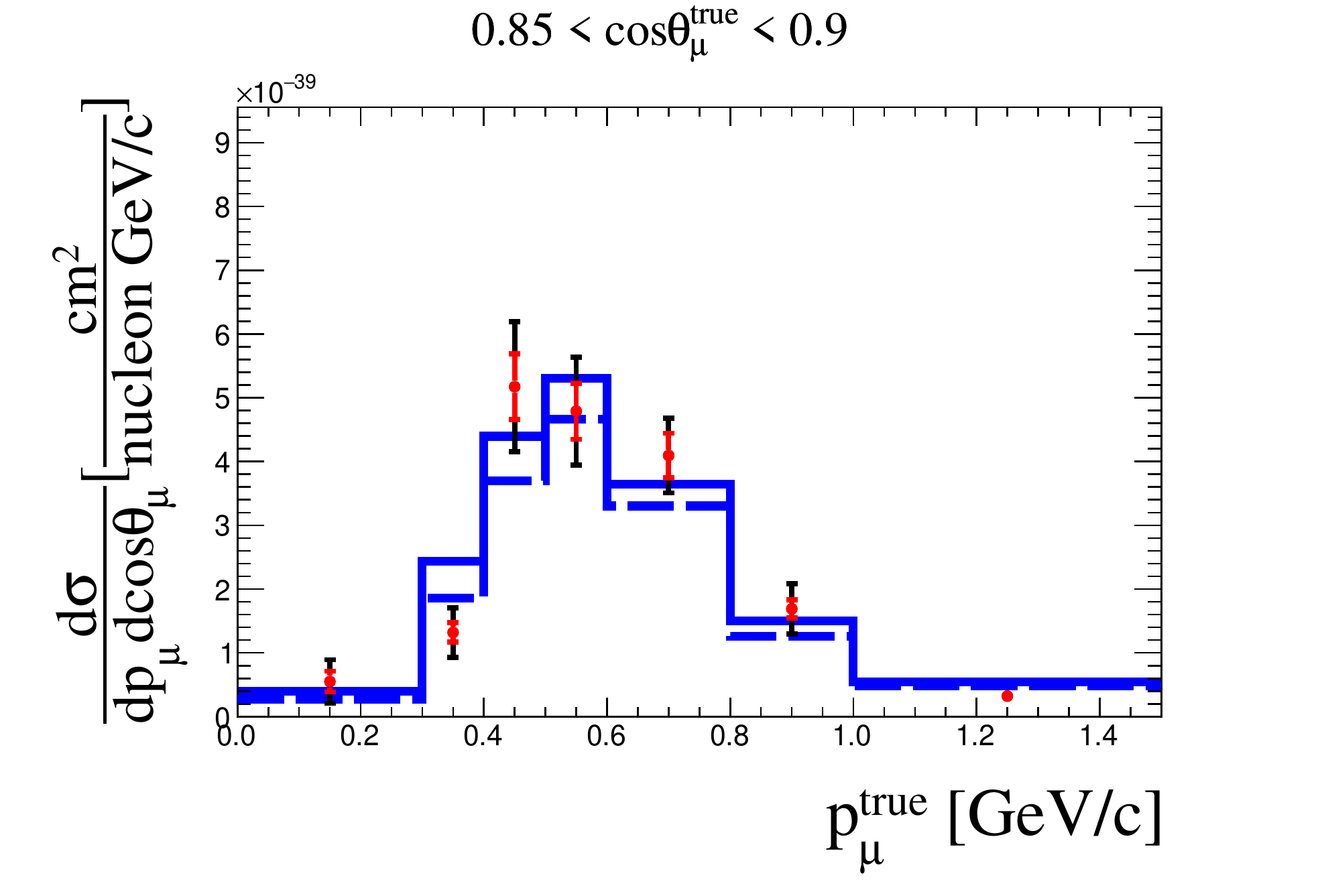}
	\includegraphics[width=0.36\linewidth]{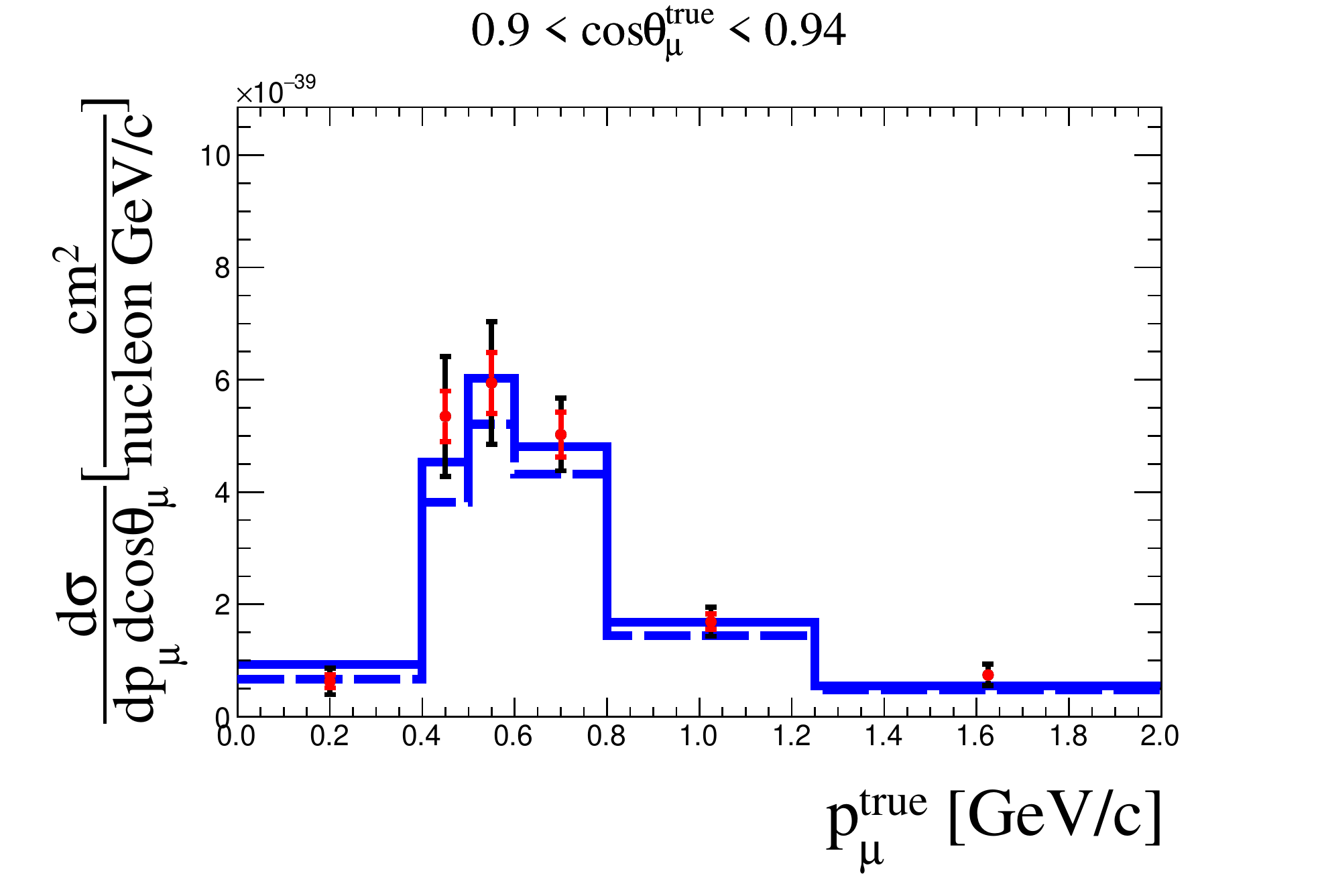}			
	\includegraphics[width=0.36\linewidth]{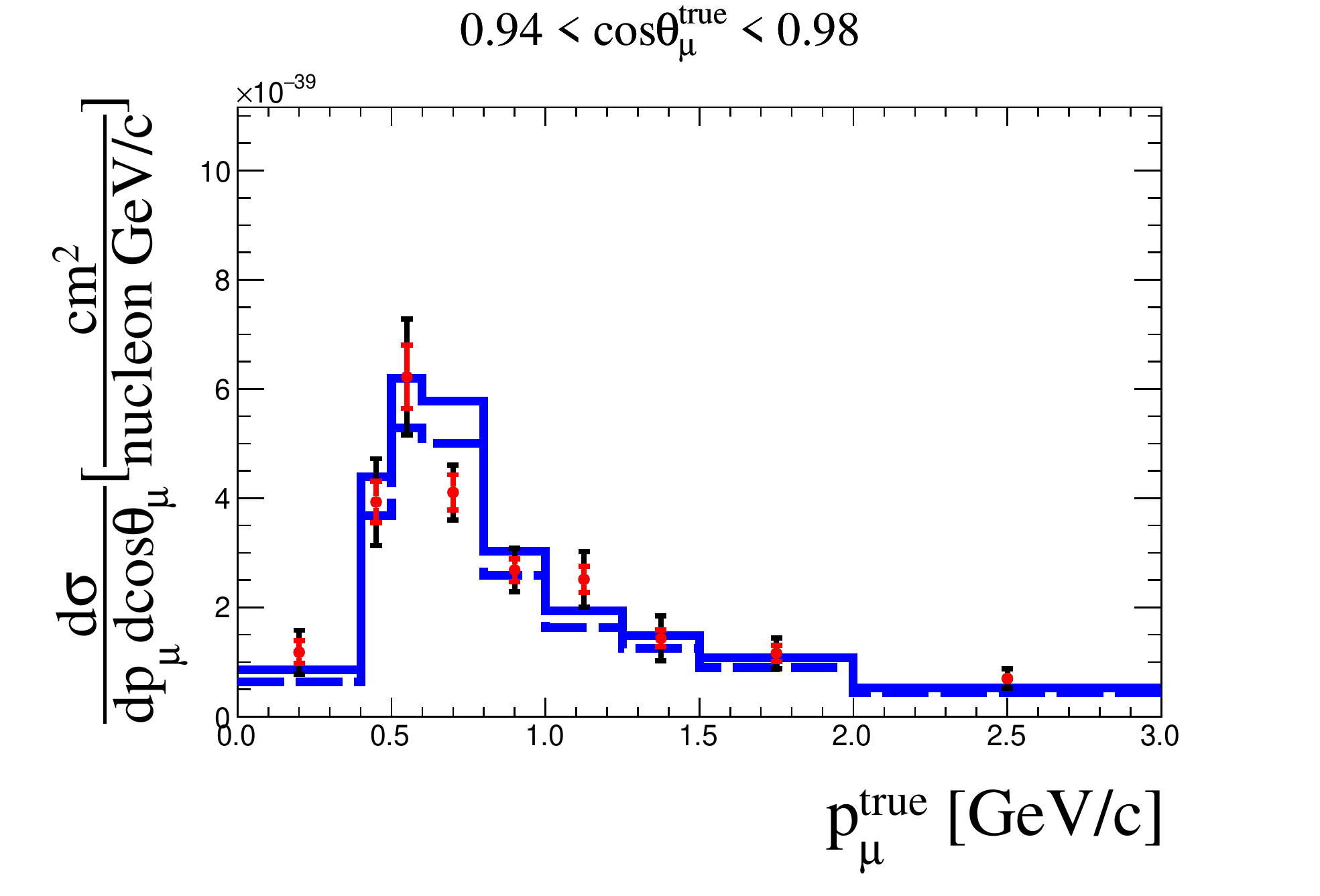}
	\includegraphics[width=0.36\linewidth]{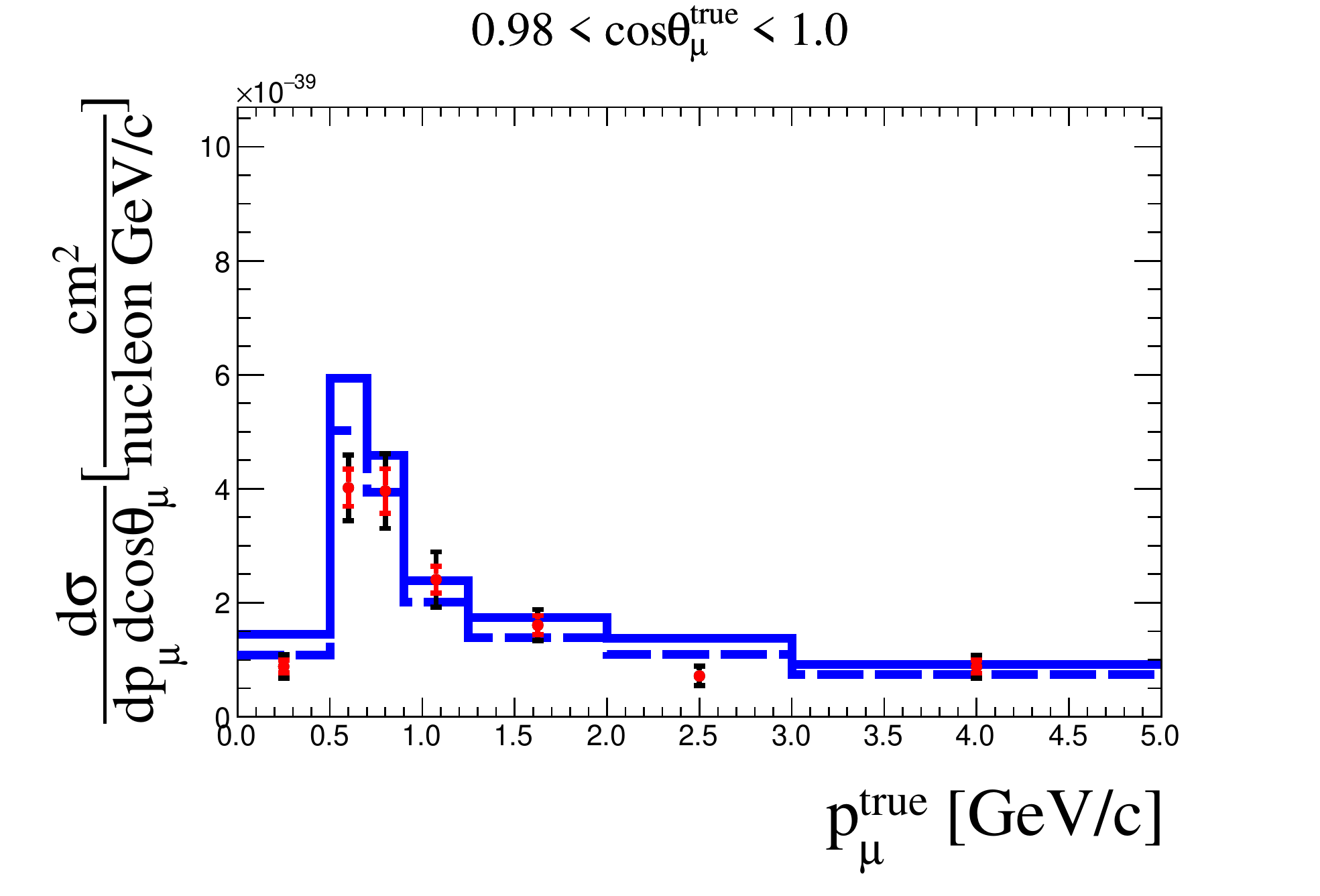}	
	\includegraphics[width=0.36\linewidth]{XsecLegendNEUT2p2h}								
	\caption{Measured \barnumu \cczeropi double-differential cross-section per nucleon in bins of true muon kinematics with systematic uncertainty (red bars) and total (stat.+syst.) uncertainty (black bars). The results are compared to \textsc{Neut} version~\texttt{5.4.1}, which uses an LFG+RPA model, with (solid line) and without (dashed line) 2p2h. The full and shape-only (in parenthesis) $\chi2$ are reported. The last bin in momentum is not displayed for readability.}
	\label{fig:antinumucc0pixsecneut2p2h}
\end{figure*}

\begin{figure*}[h!]
	\centering
	\includegraphics[width=0.36\linewidth]{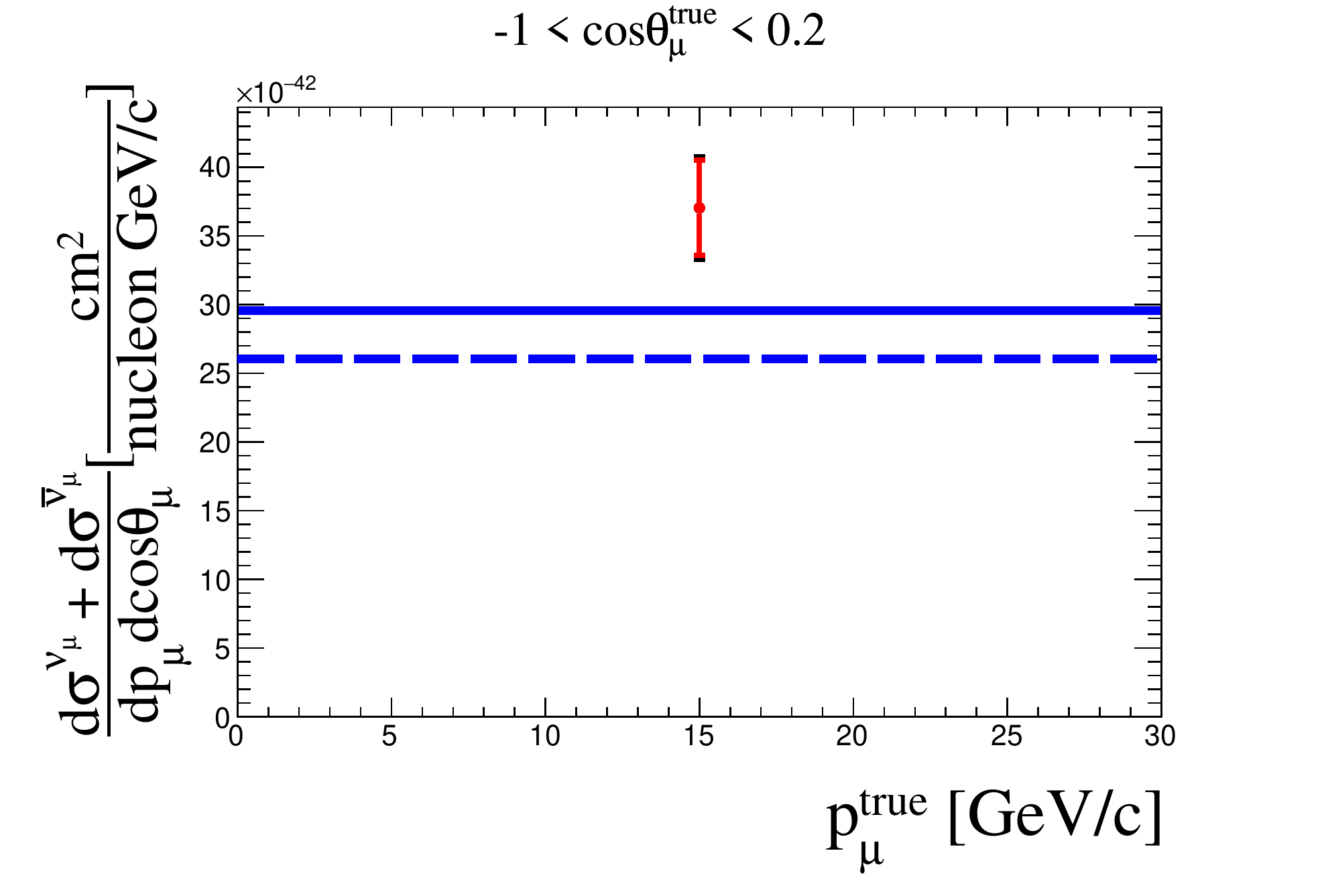}
	\includegraphics[width=0.36\linewidth]{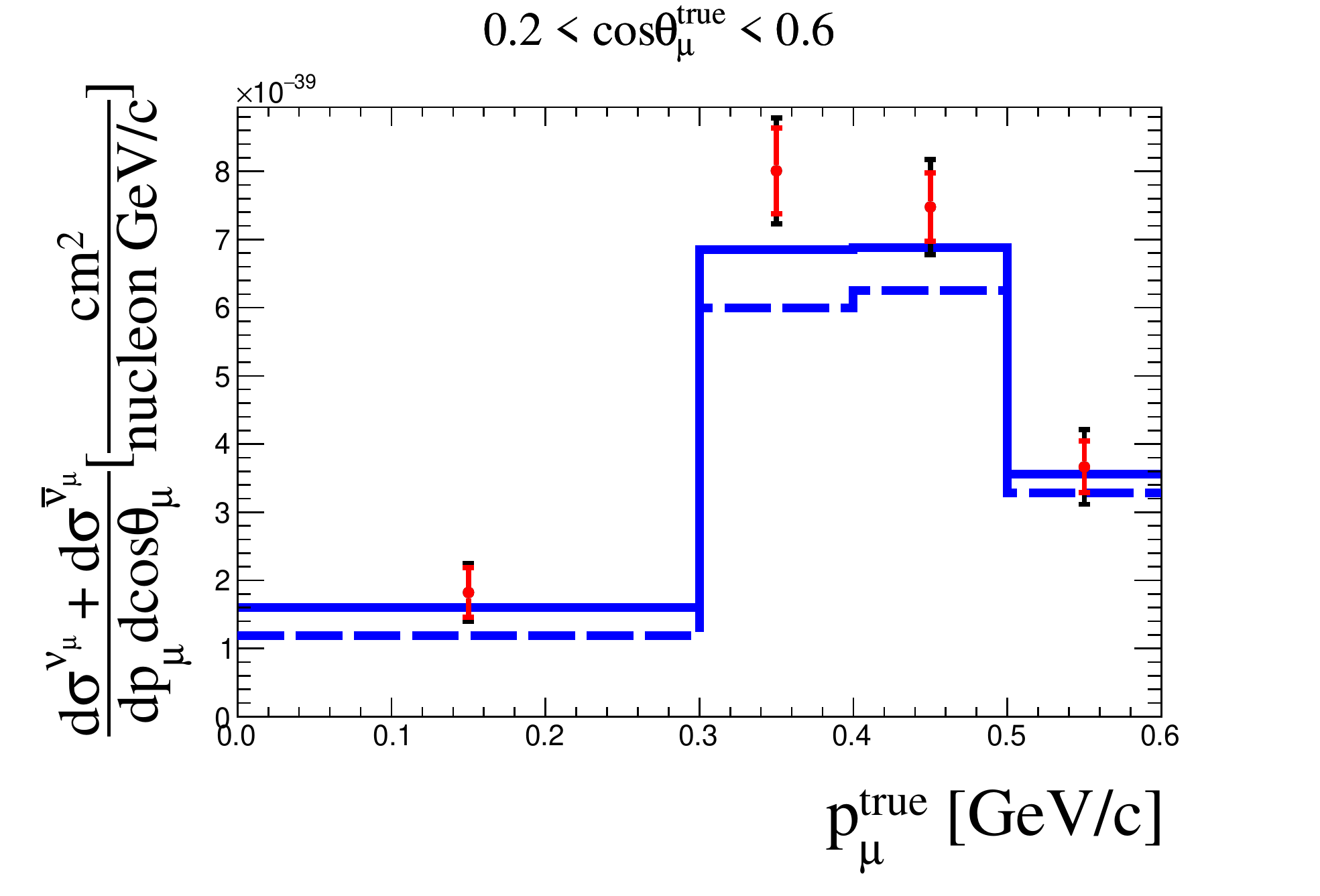}
	\includegraphics[width=0.36\linewidth]{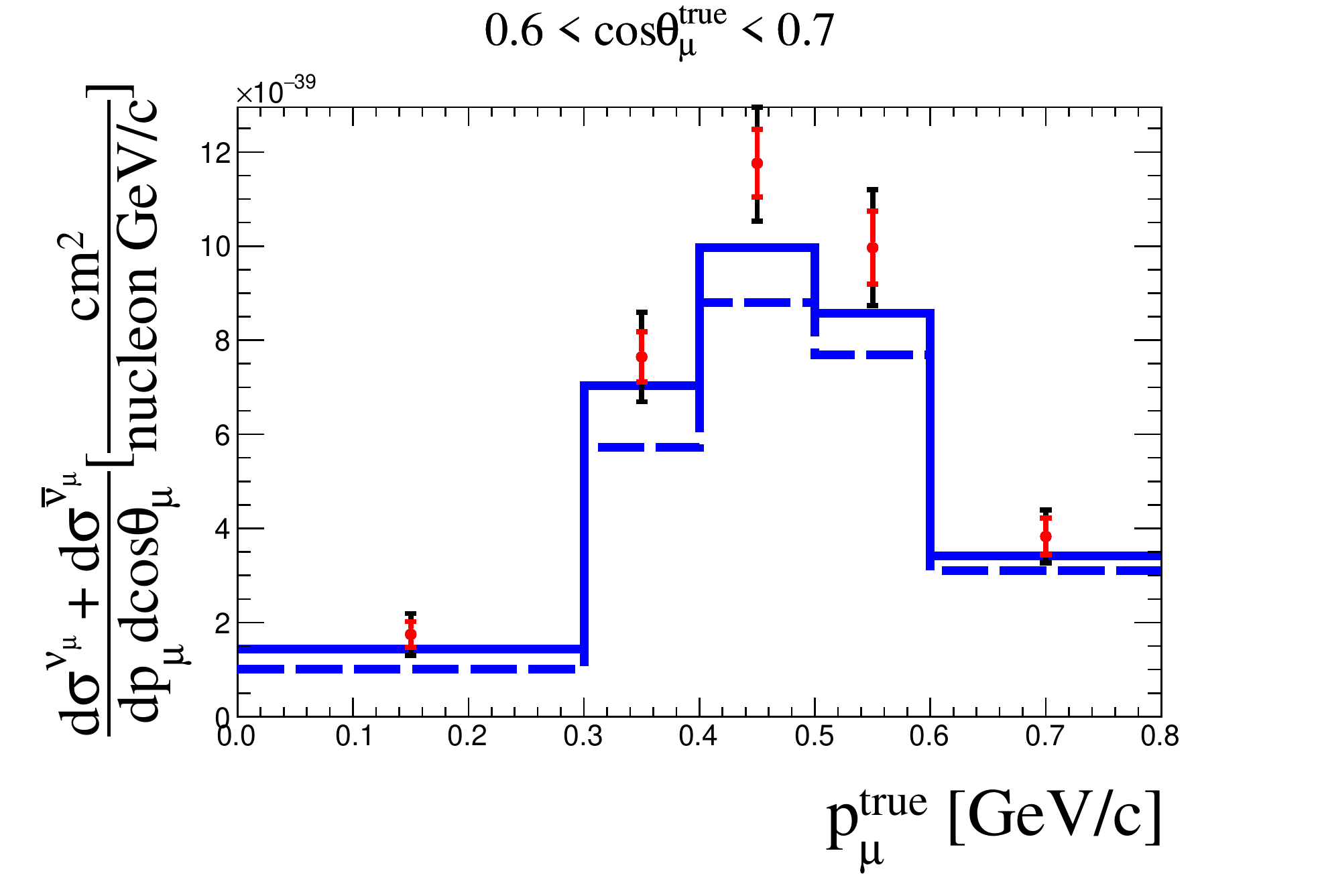}
	\includegraphics[width=0.36\linewidth]{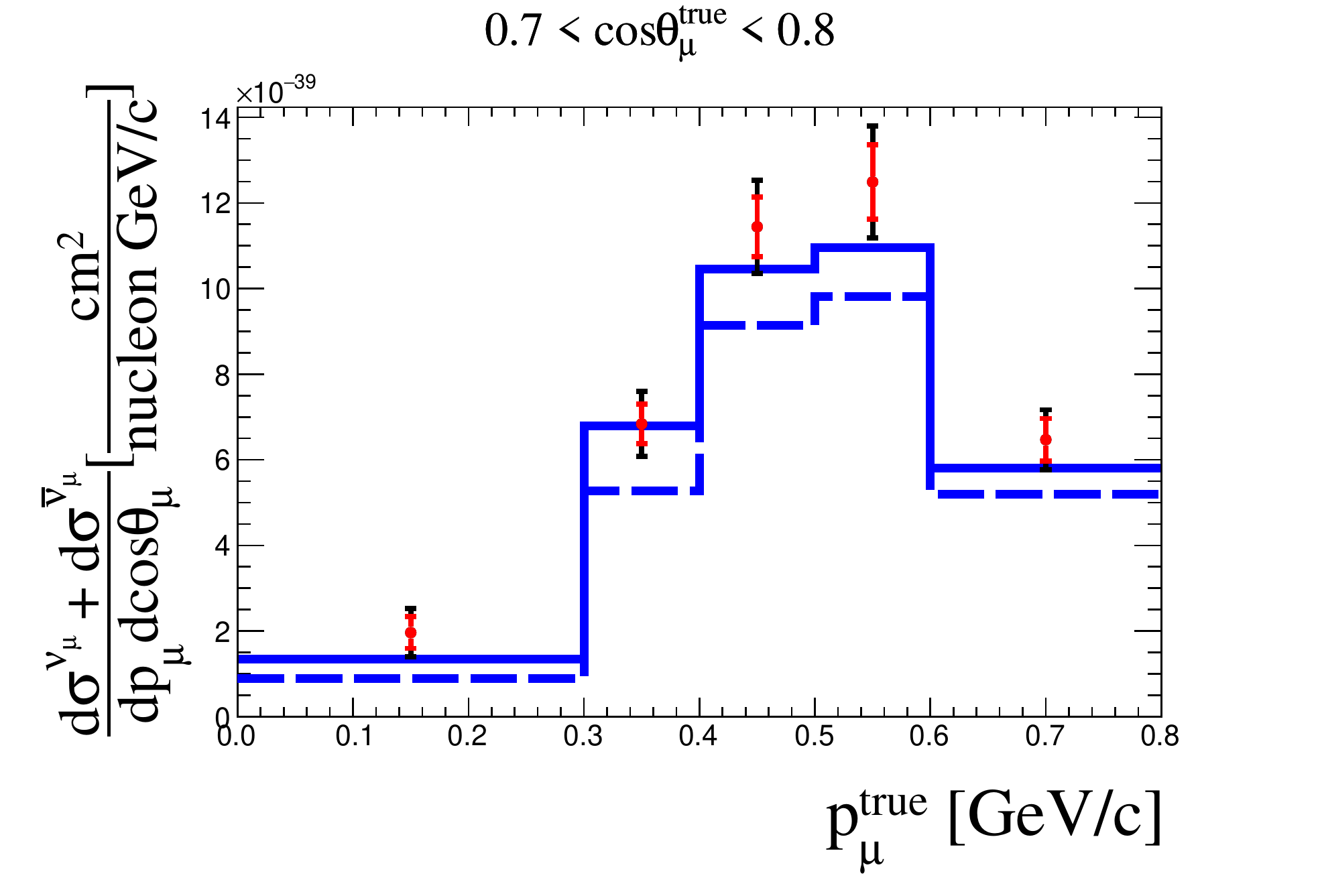}
	\includegraphics[width=0.36\linewidth]{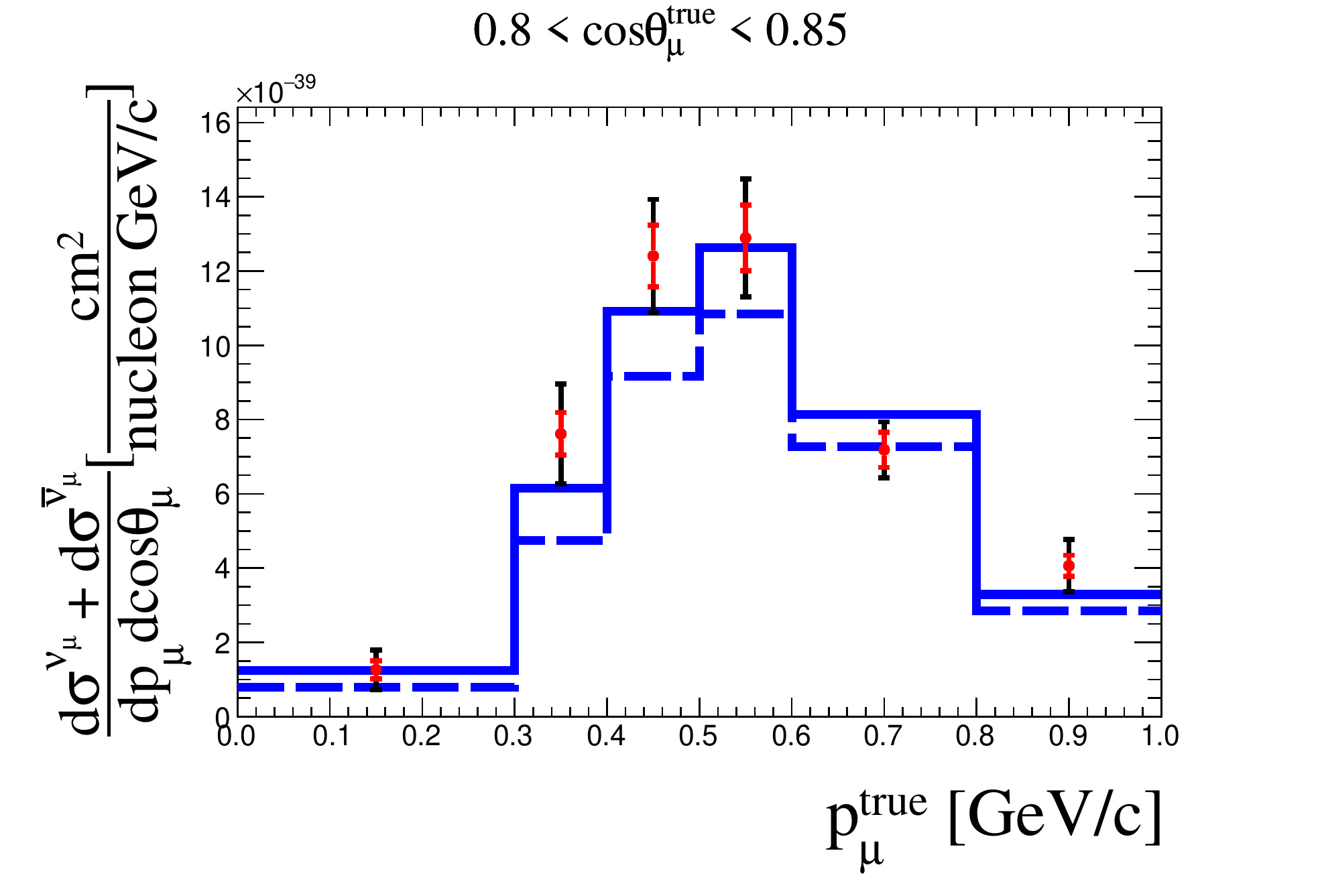}
	\includegraphics[width=0.36\linewidth]{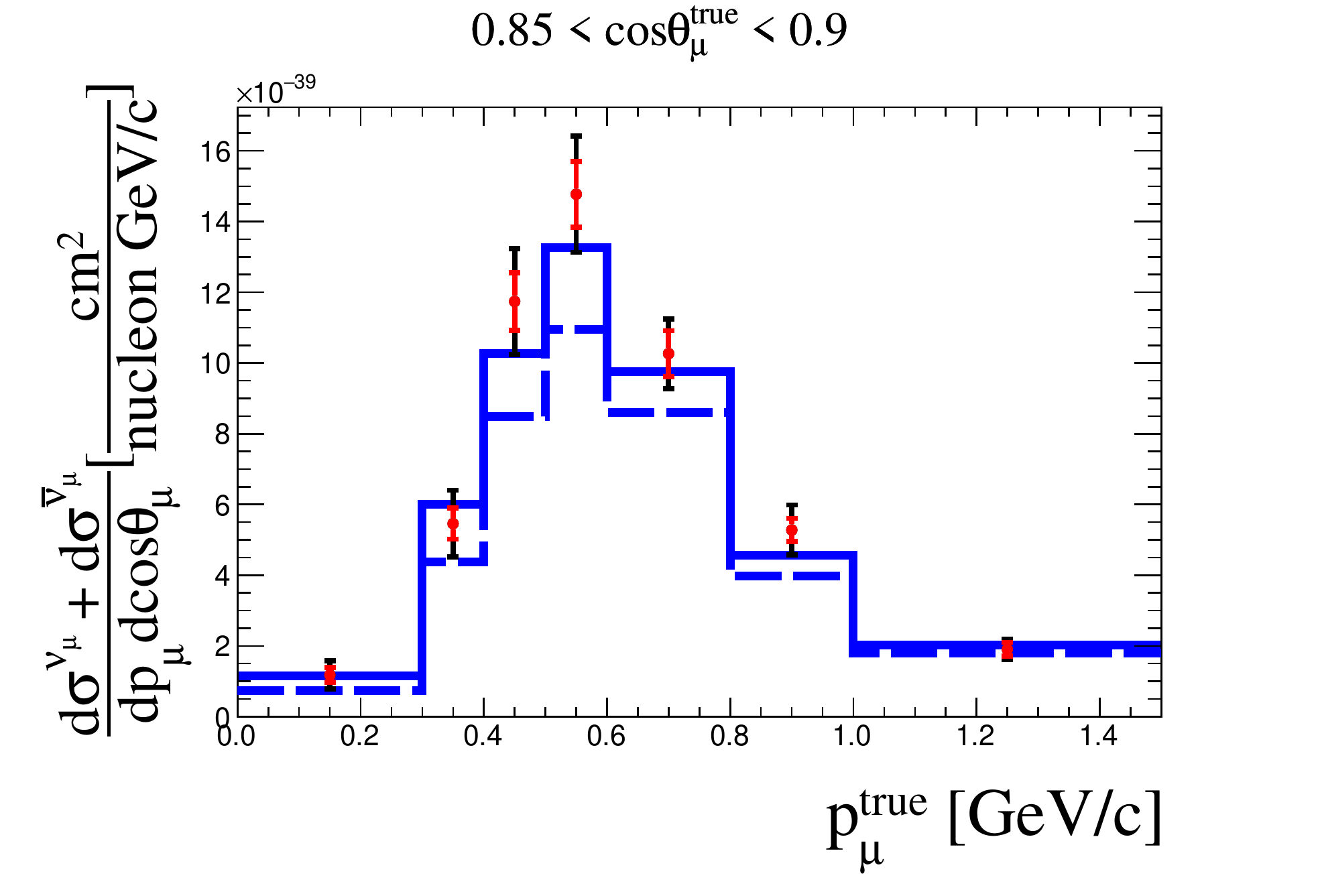}
	\includegraphics[width=0.36\linewidth]{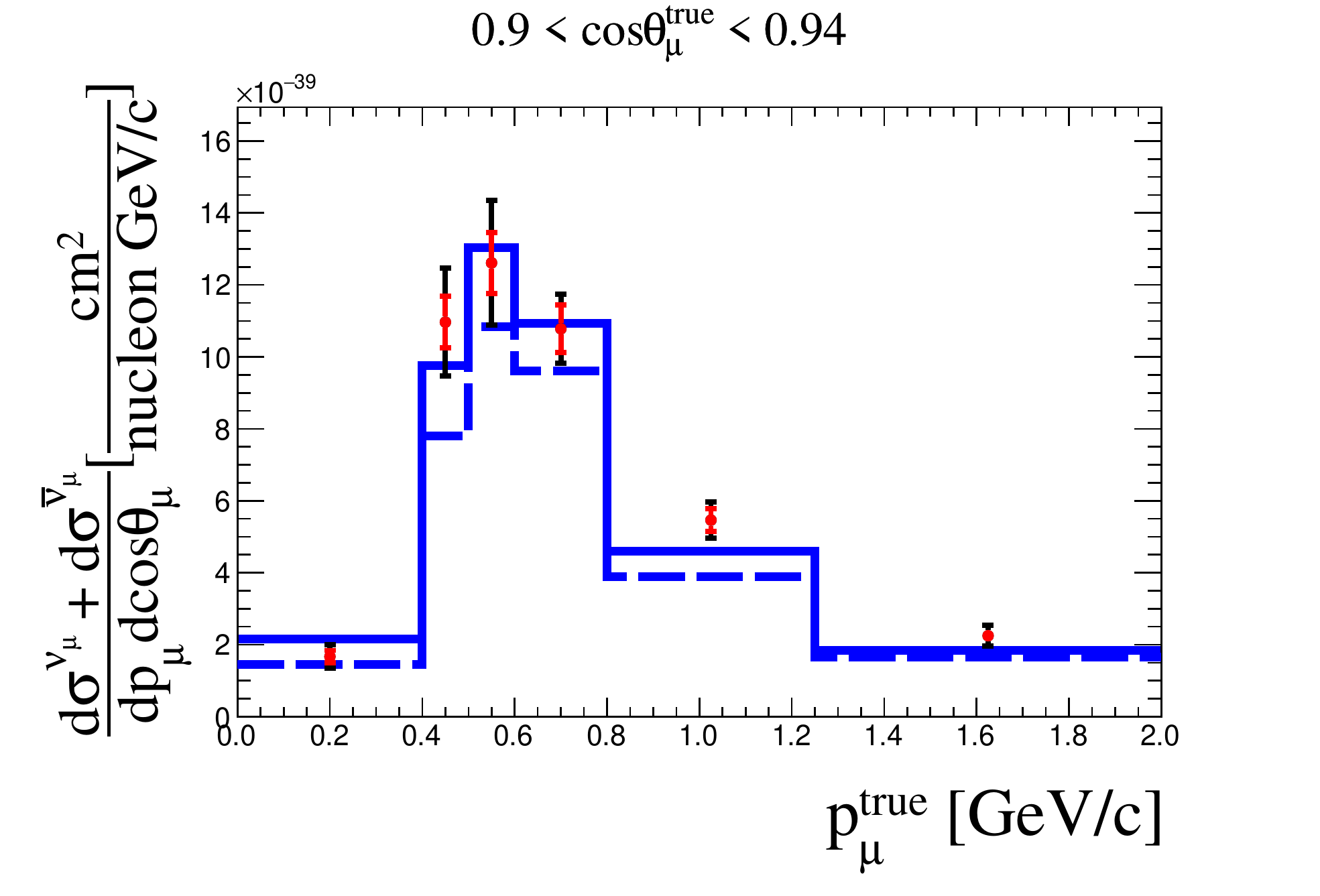}
	\includegraphics[width=0.36\linewidth]{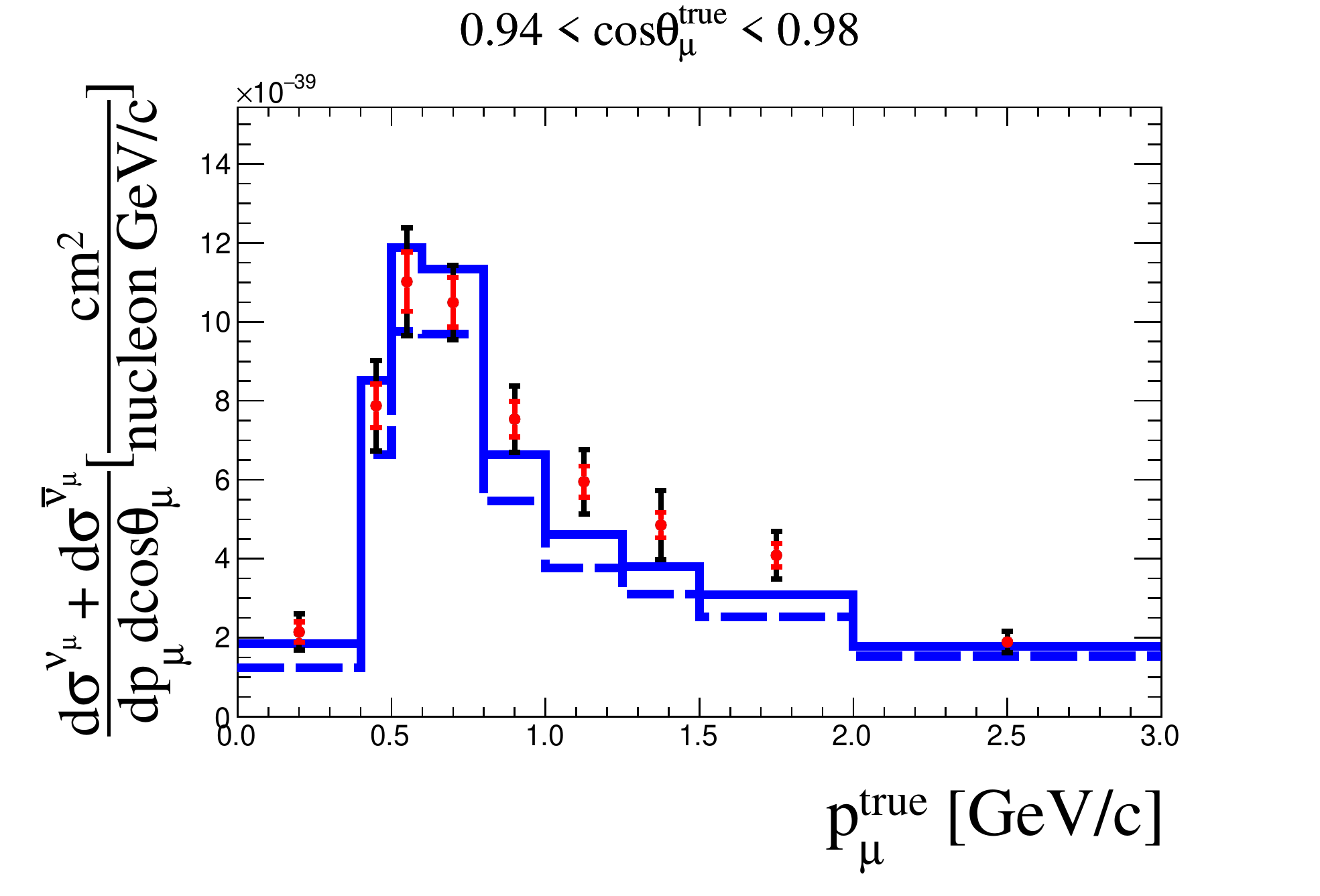}
	\includegraphics[width=0.36\linewidth]{XsecSumNEUT2p2h_cosbin_7}	
	\includegraphics[width=0.36\linewidth]{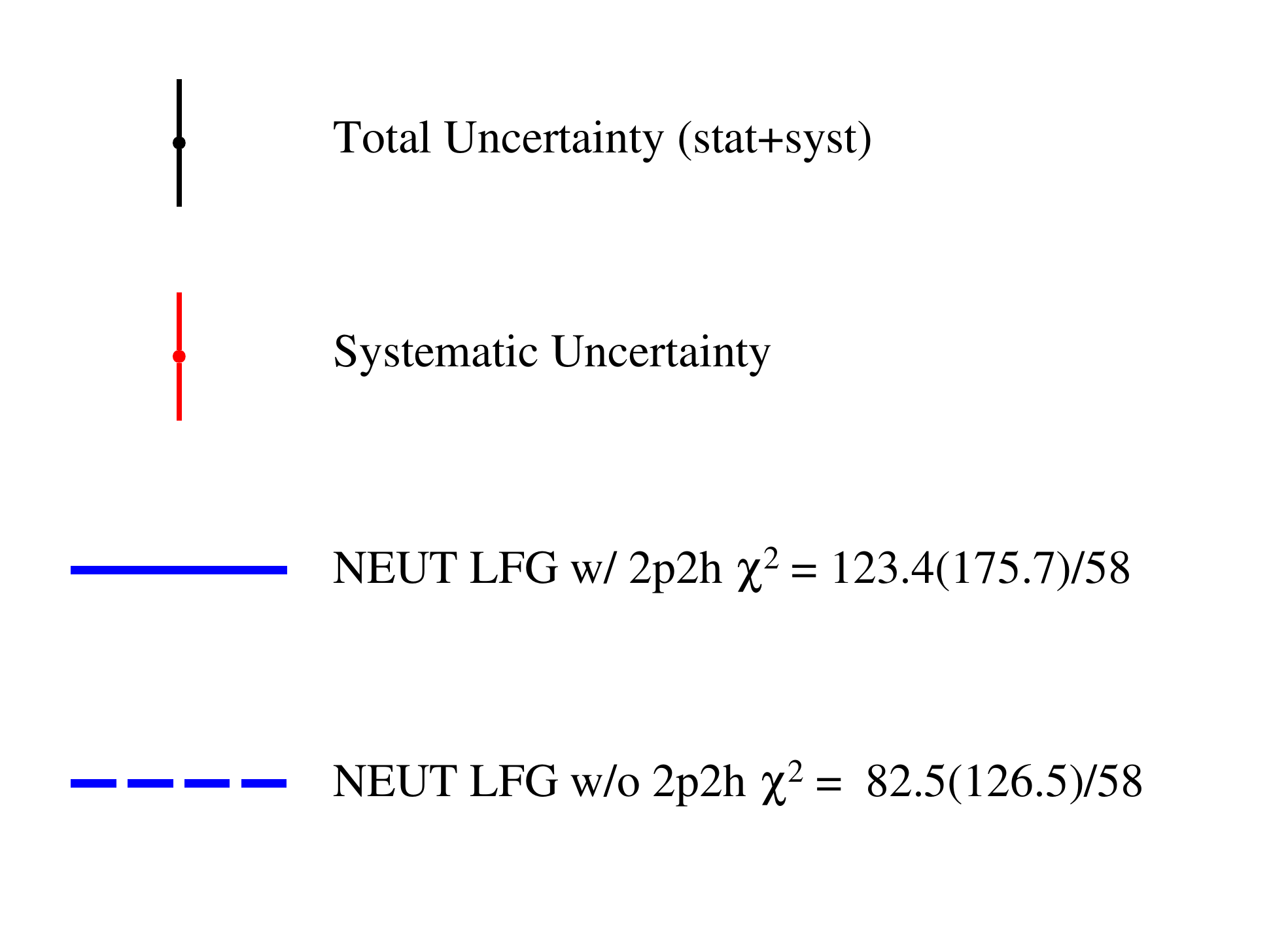}																
	\caption{Measured double-differential \numu + \barnumu \cczeropi cross-section sum in bins of true muon kinematics with systematic uncertainty (red bars) and total (stat.+syst.) uncertainty (black bars). The results are compared to \textsc{Neut} version~\texttt{5.4.1}, which uses an LFG+RPA model, with (solid line) and without (dashed line) 2p2h. The full and shape-only (in parenthesis) $\chi2$ are reported. The last bin in momentum is not displayed for readability.}
	\label{fig:xsecsumneut2p2h}
\end{figure*}

\begin{figure*}[h!]
	\centering
	\includegraphics[width=0.36\linewidth]{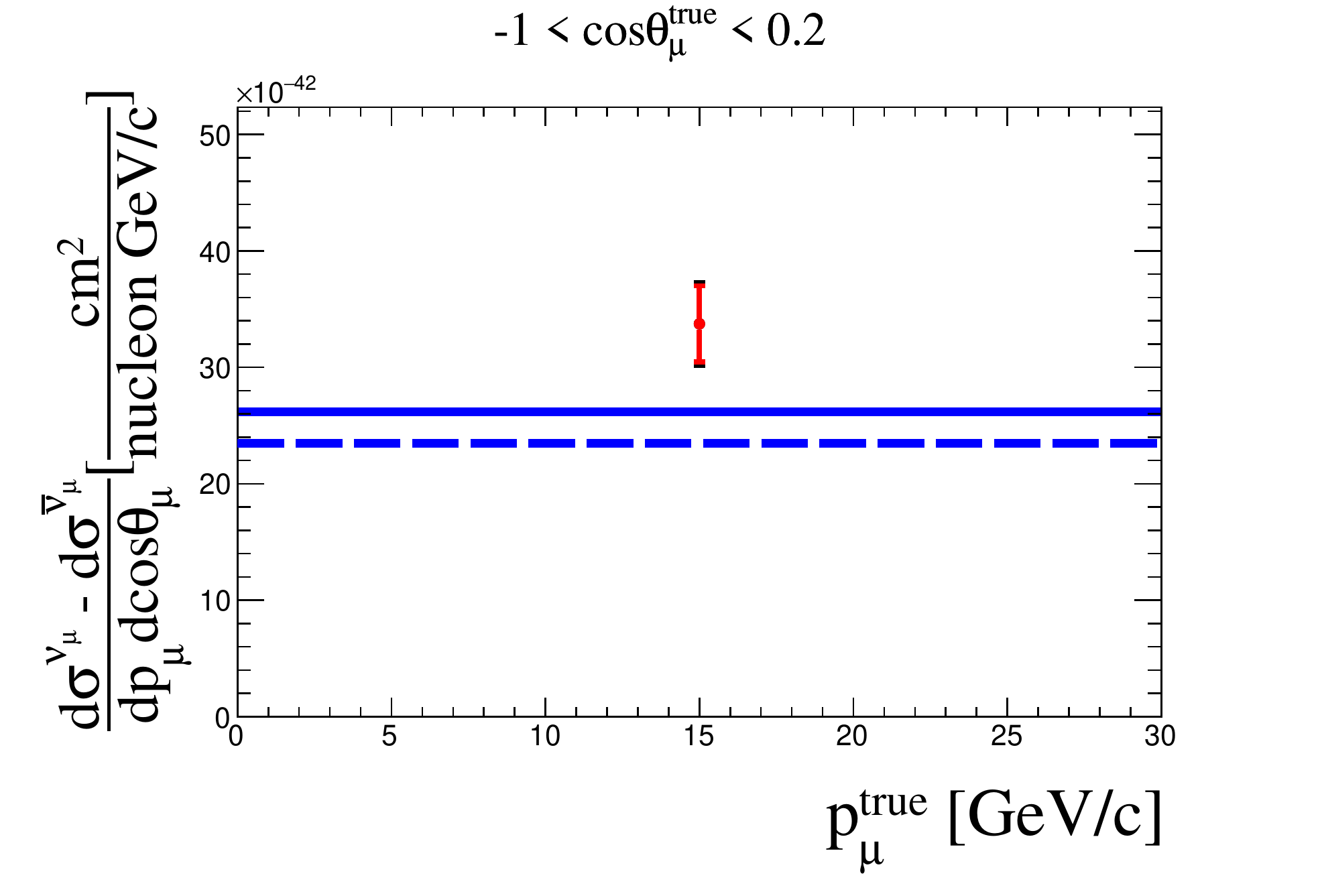}
	\includegraphics[width=0.36\linewidth]{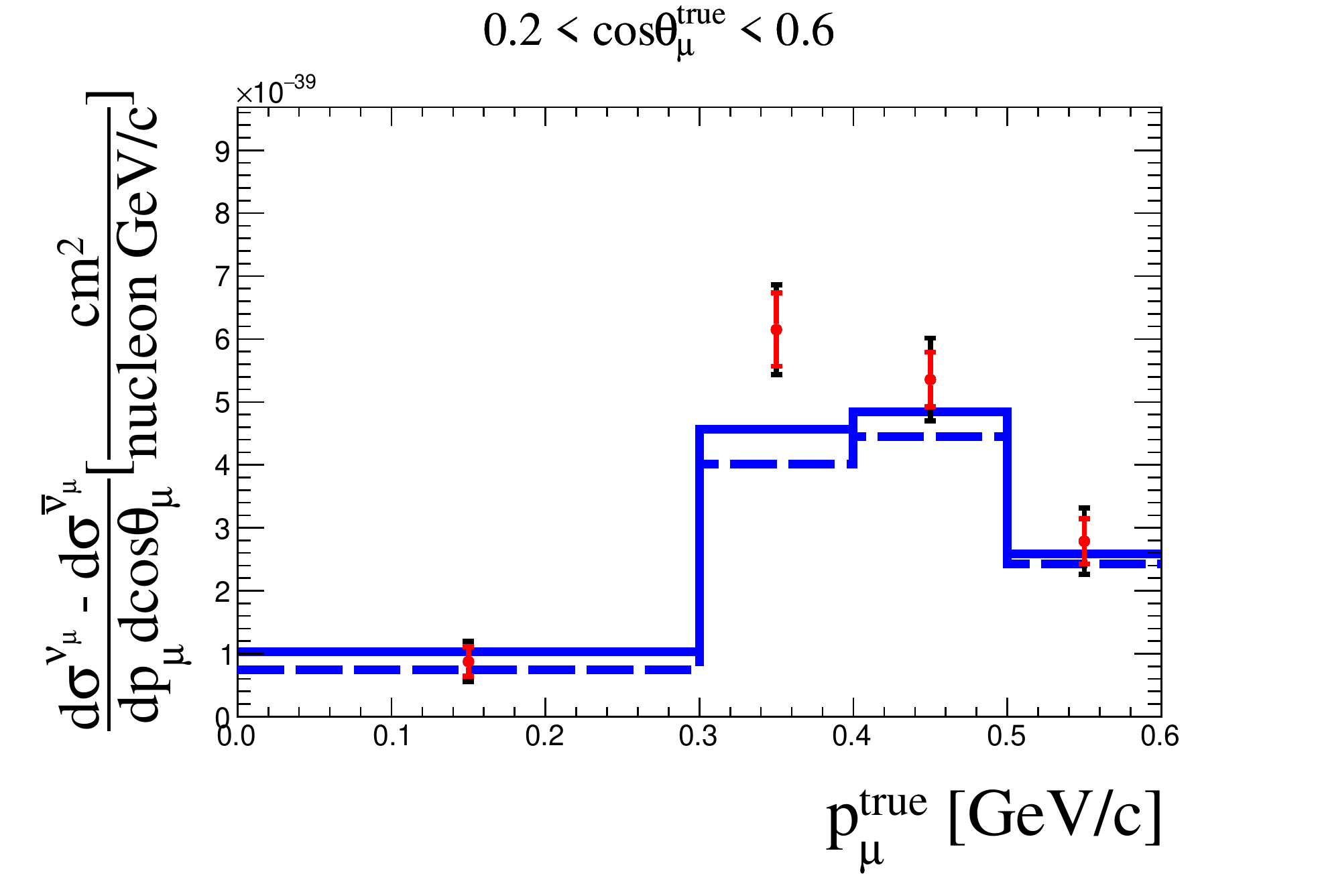}
	\includegraphics[width=0.36\linewidth]{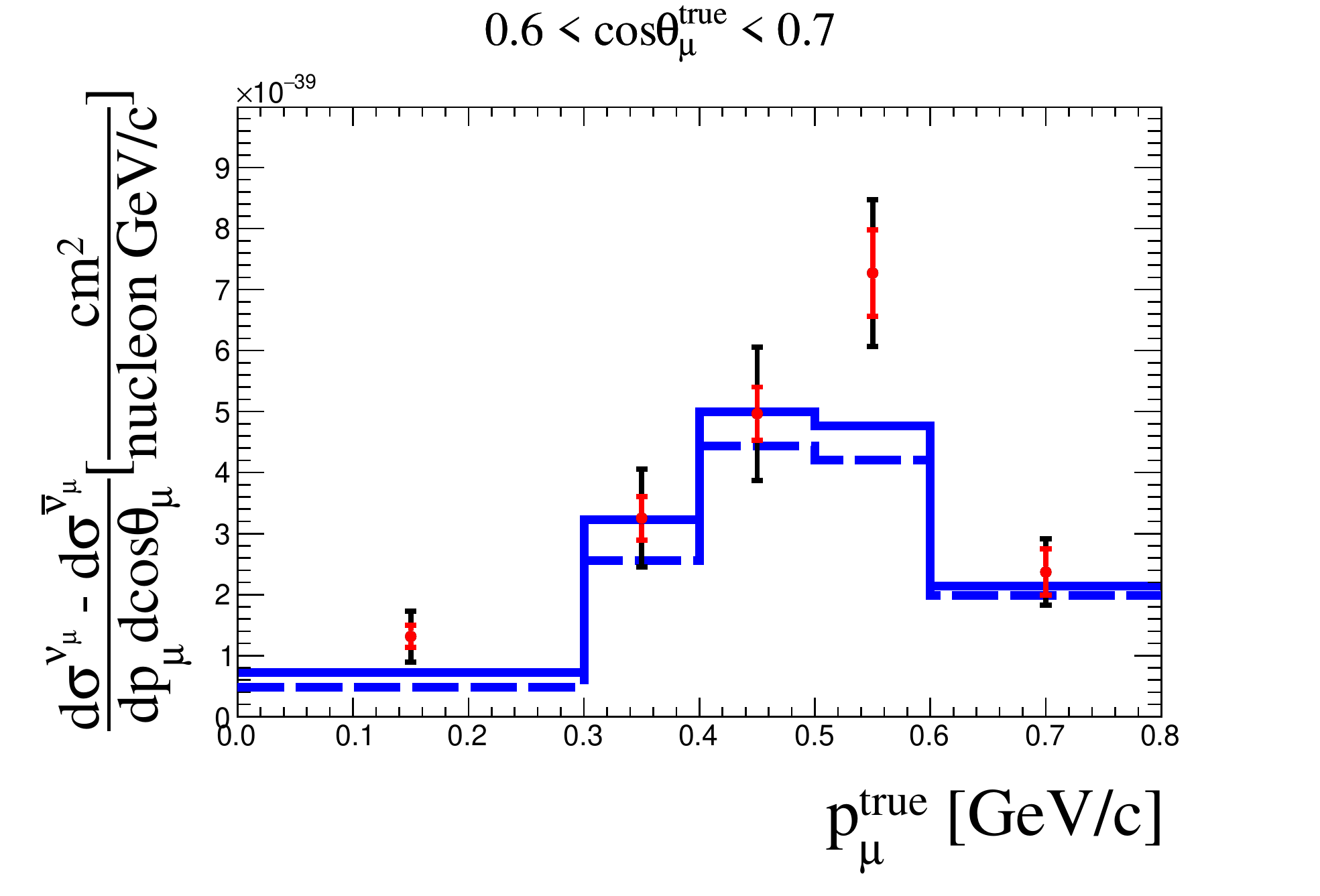}
	\includegraphics[width=0.36\linewidth]{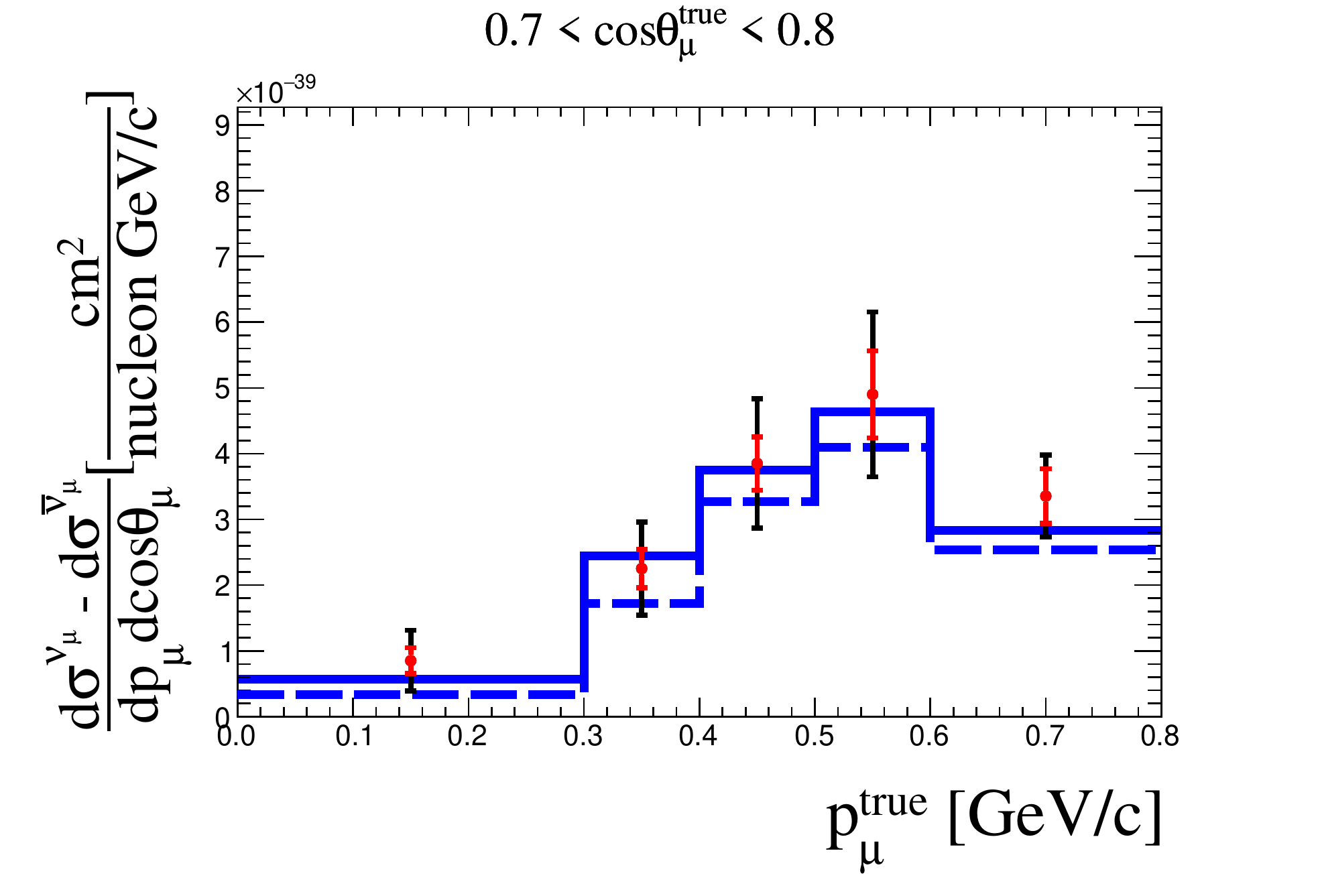}
	\includegraphics[width=0.36\linewidth]{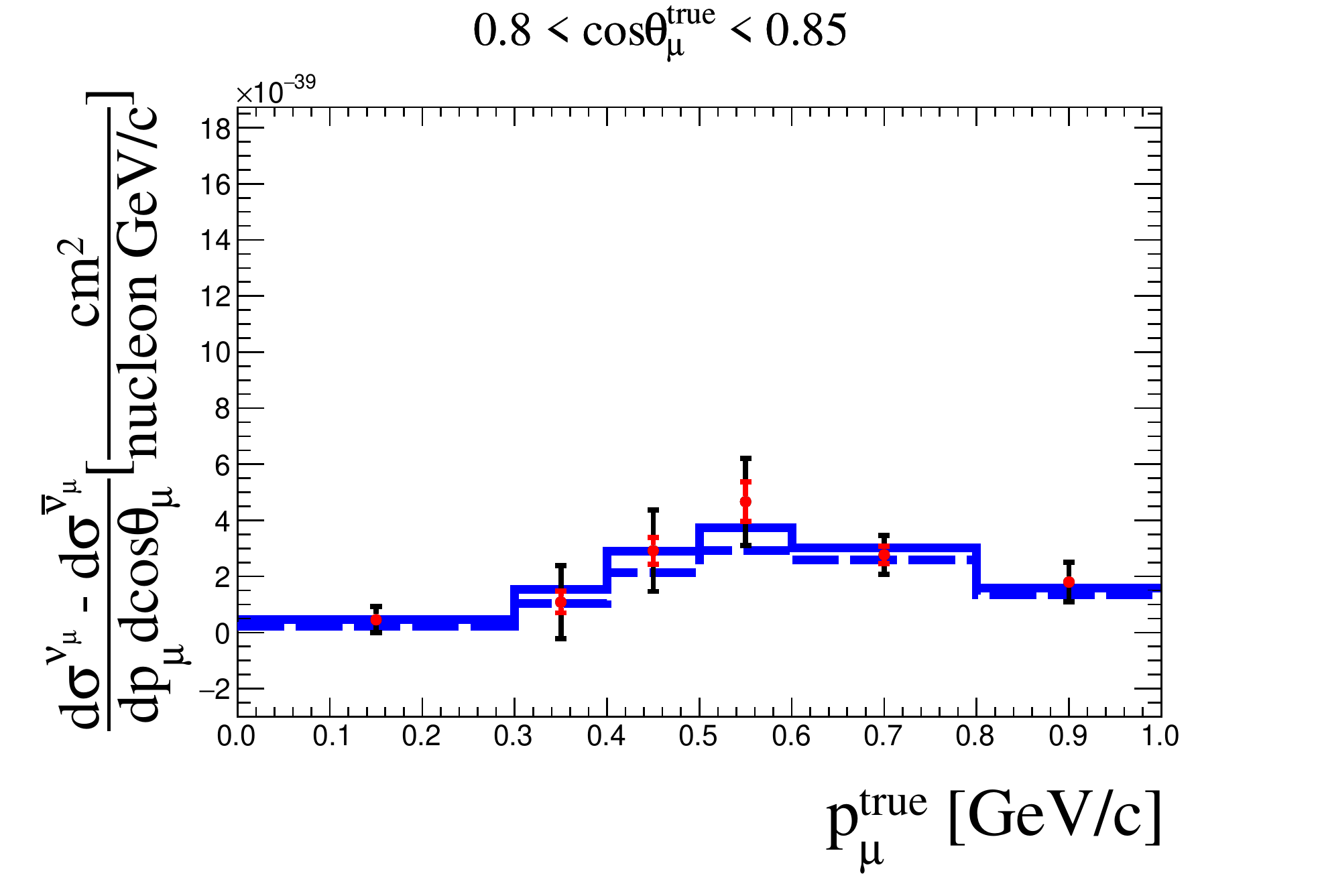}
	\includegraphics[width=0.36\linewidth]{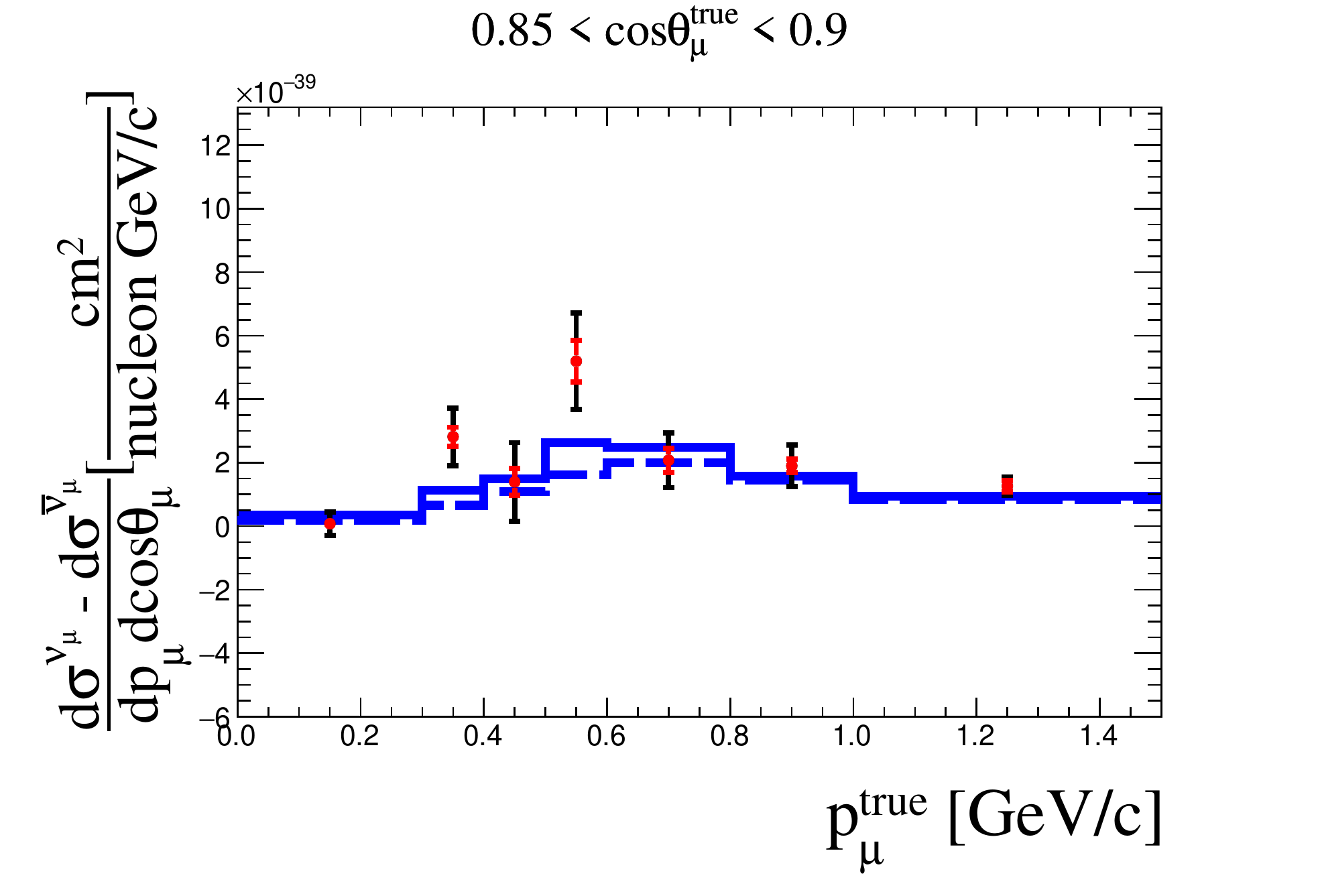}
	\includegraphics[width=0.36\linewidth]{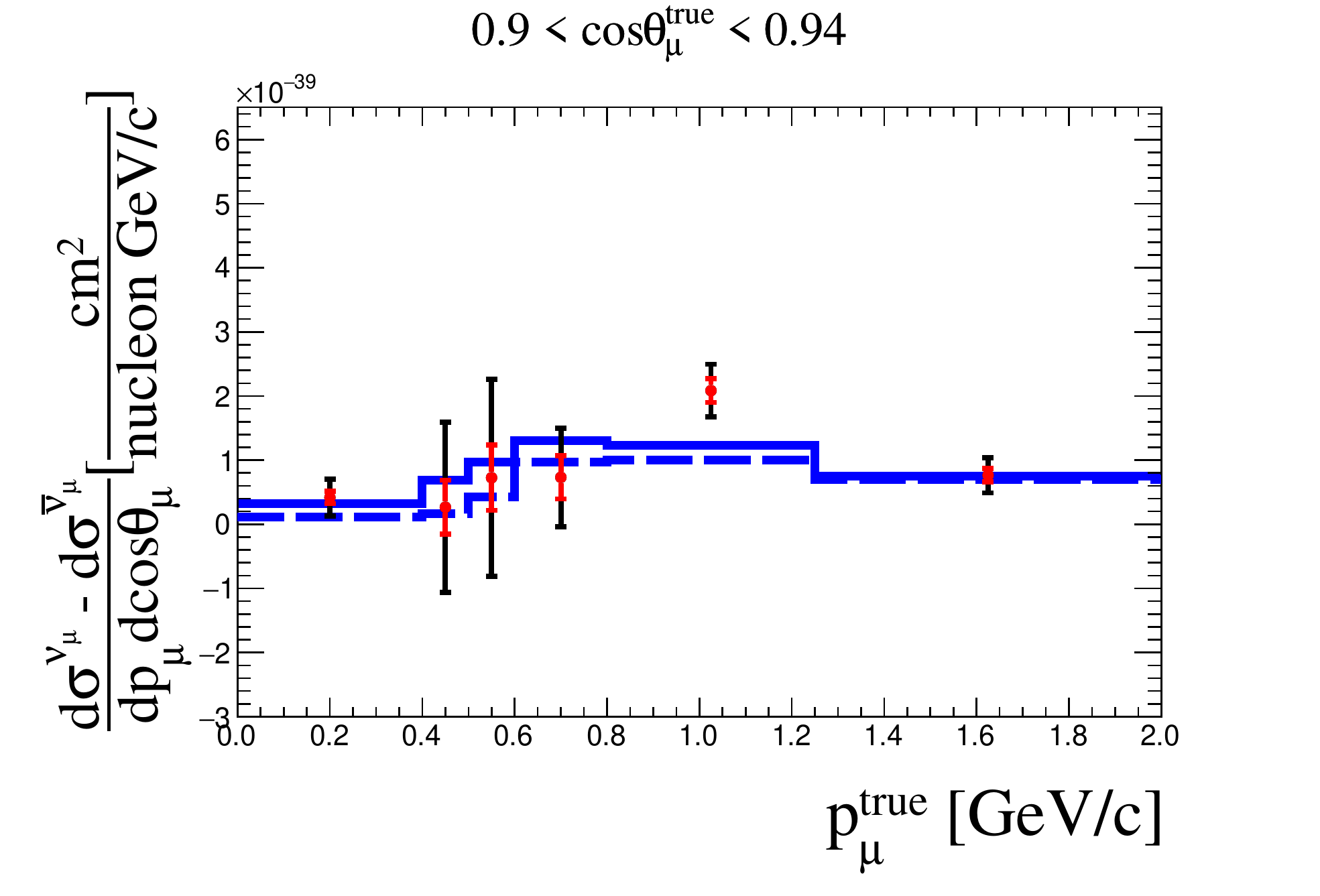}
	\includegraphics[width=0.36\linewidth]{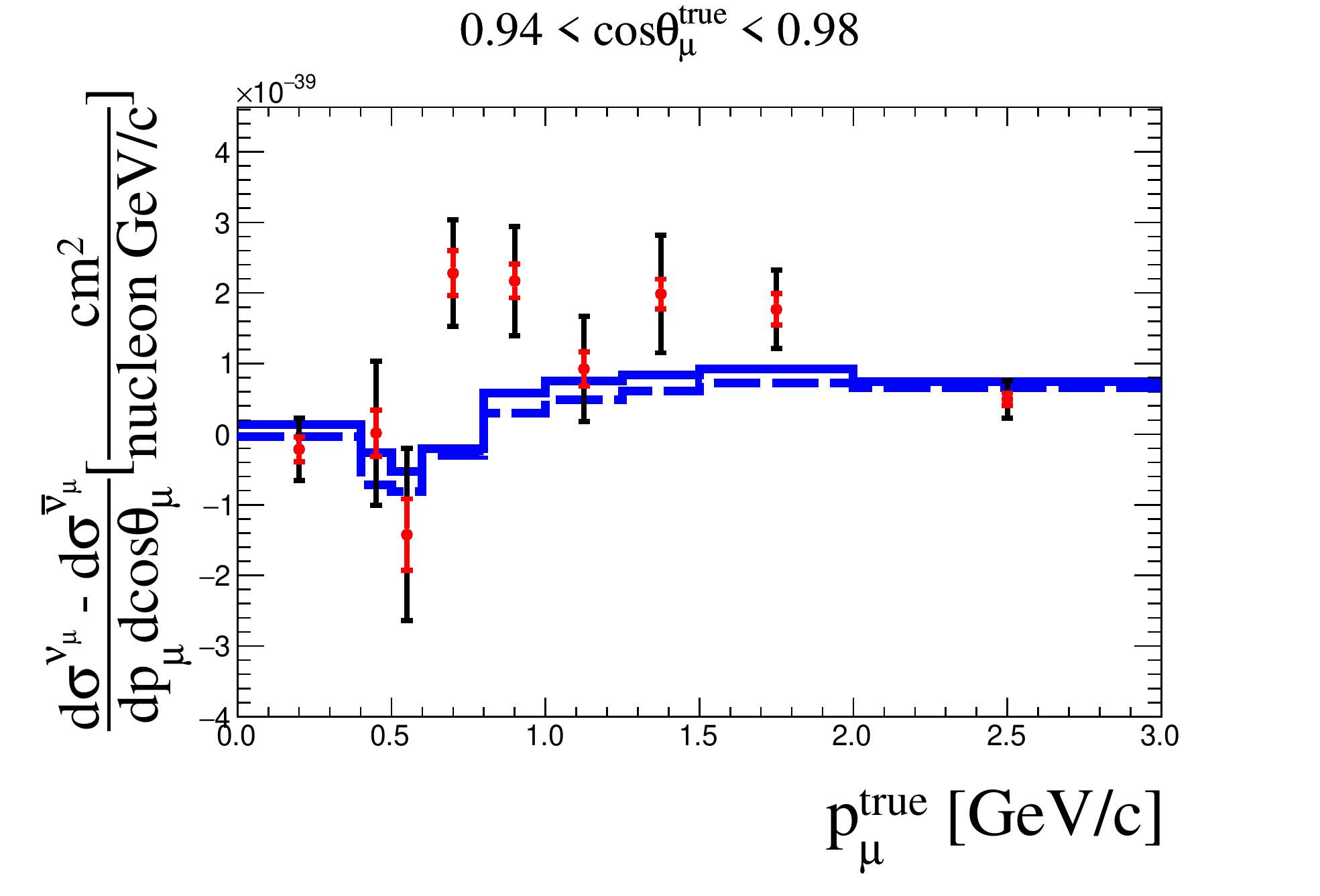}	
	\includegraphics[width=0.36\linewidth]{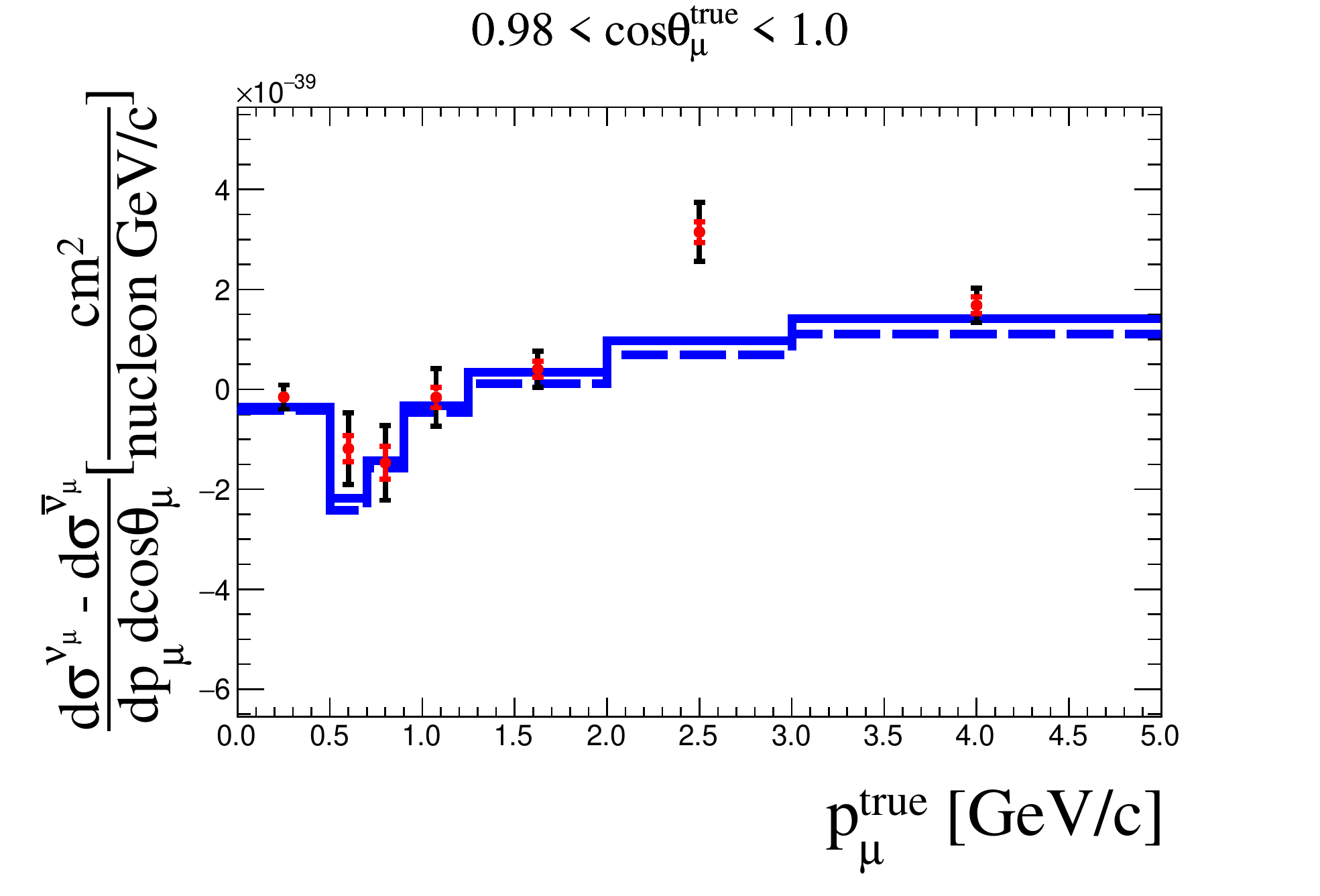}
	\includegraphics[width=0.36\linewidth]{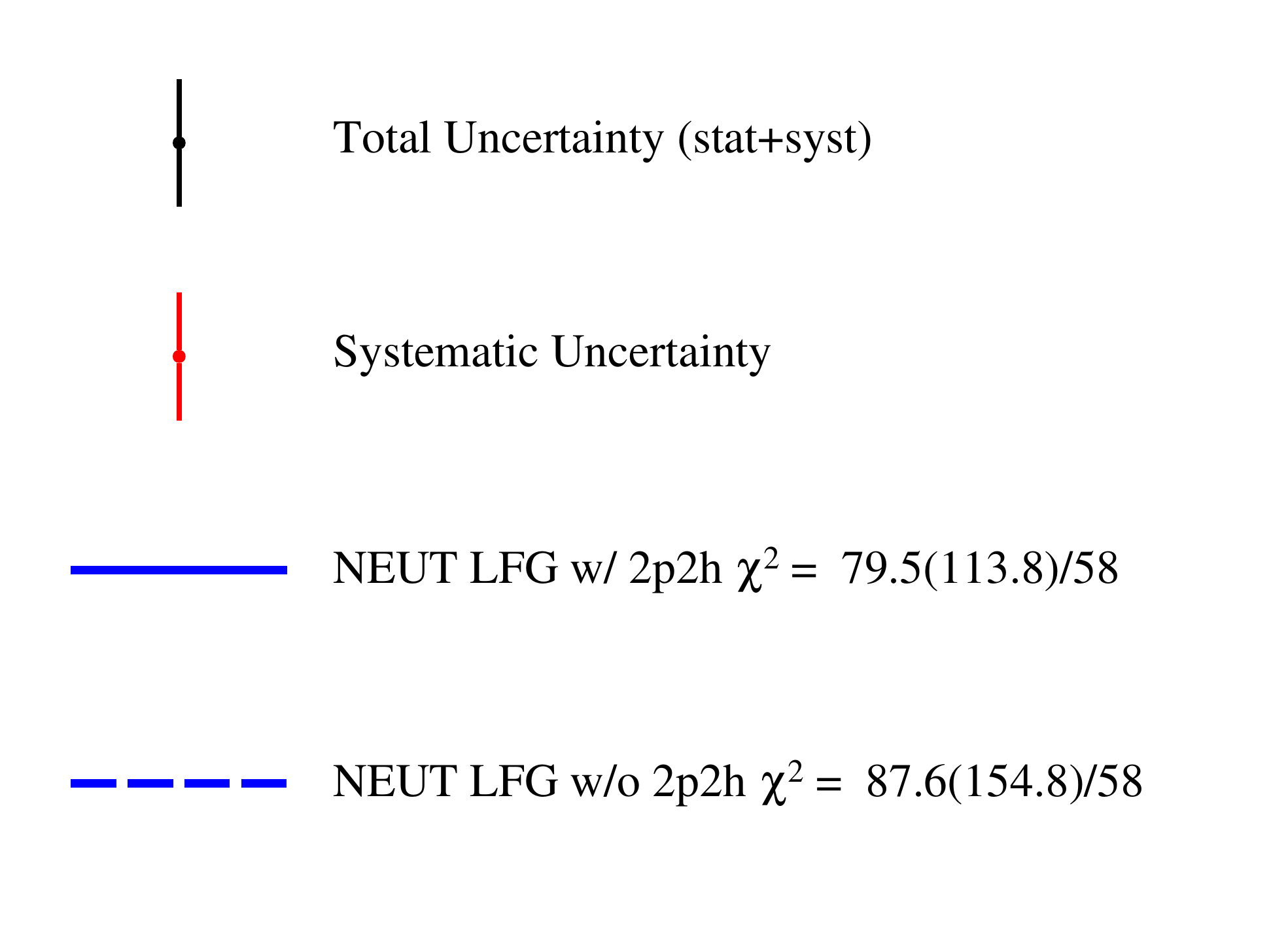}	
	\caption{Measured double-differential \numu - \barnumu \cczeropi cross-section difference in bins of true muon kinematics with systematic uncertainty (red bars) and total (stat.+syst.) uncertainty (black bars). The results are compared to \textsc{Neut} version~\texttt{5.4.1}, which uses an LFG+RPA model, with (solid line) and without (dashed line) 2p2h. The full and shape-only (in parenthesis) $\chi2$ are reported. The last bin in momentum is not displayed for readability.}
	\label{fig:xsecdifneut2p2h}
\end{figure*}

\begin{figure*}[h!]
	\centering
	\includegraphics[width=0.36\linewidth]{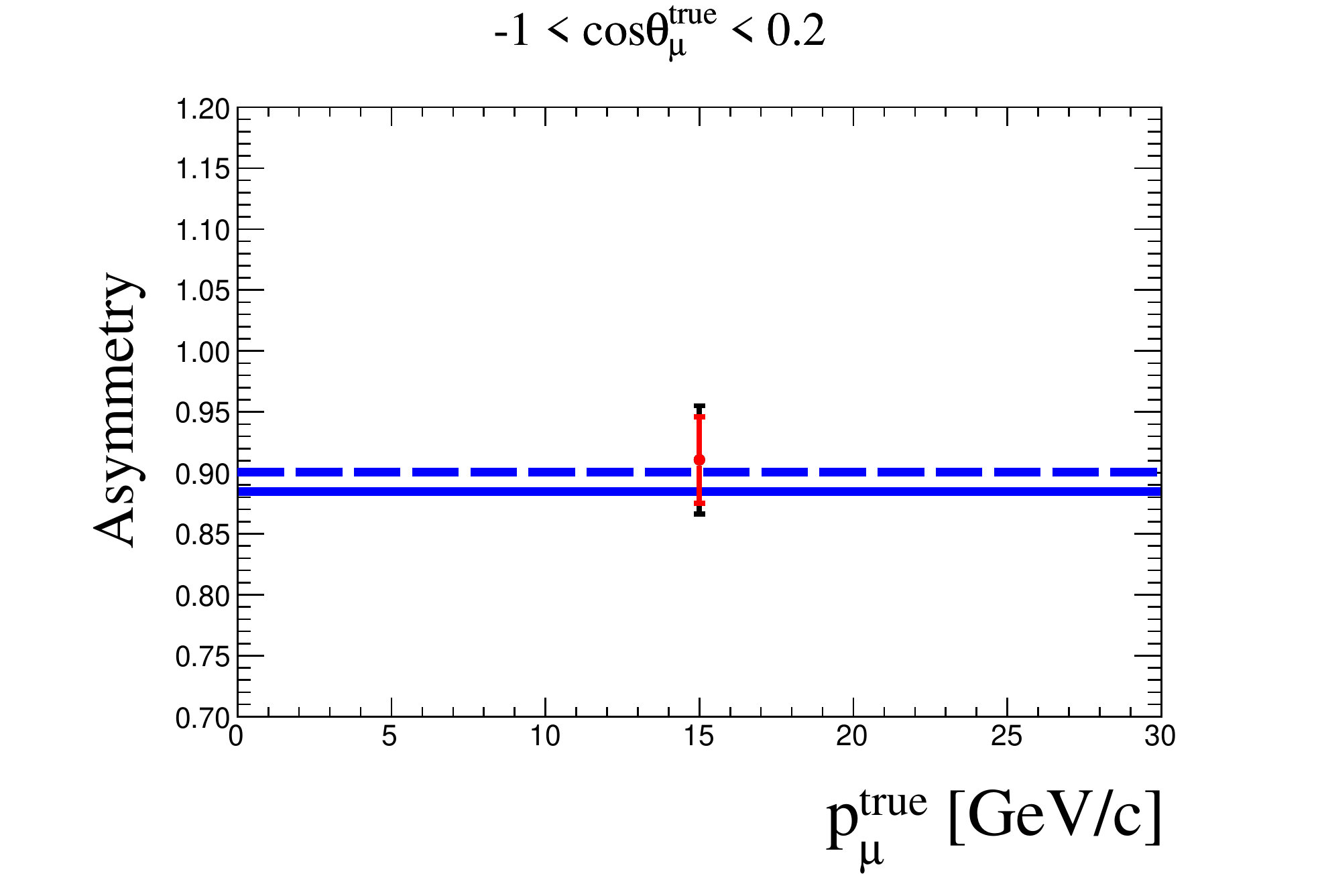}
	\includegraphics[width=0.36\linewidth]{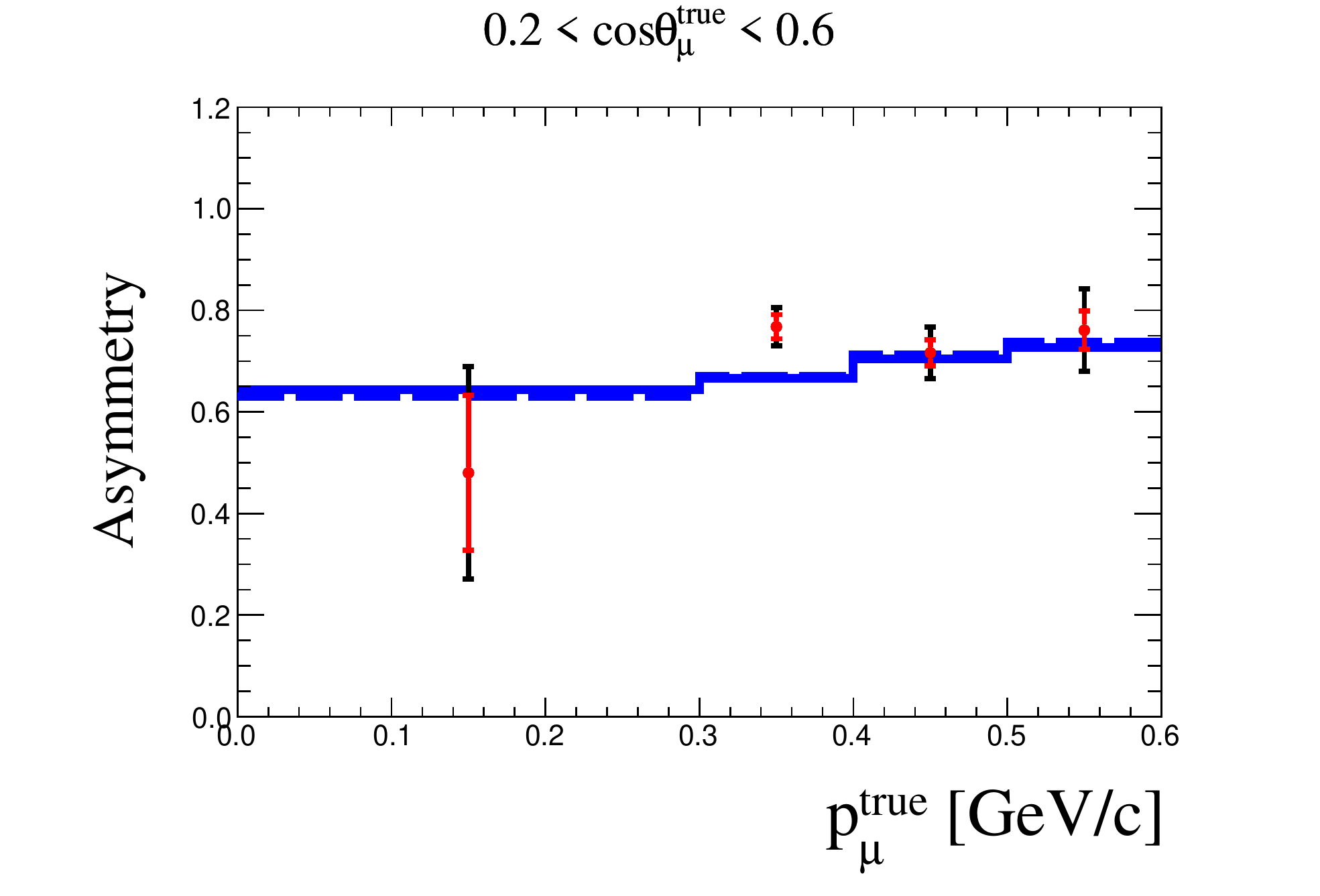}
	\includegraphics[width=0.36\linewidth]{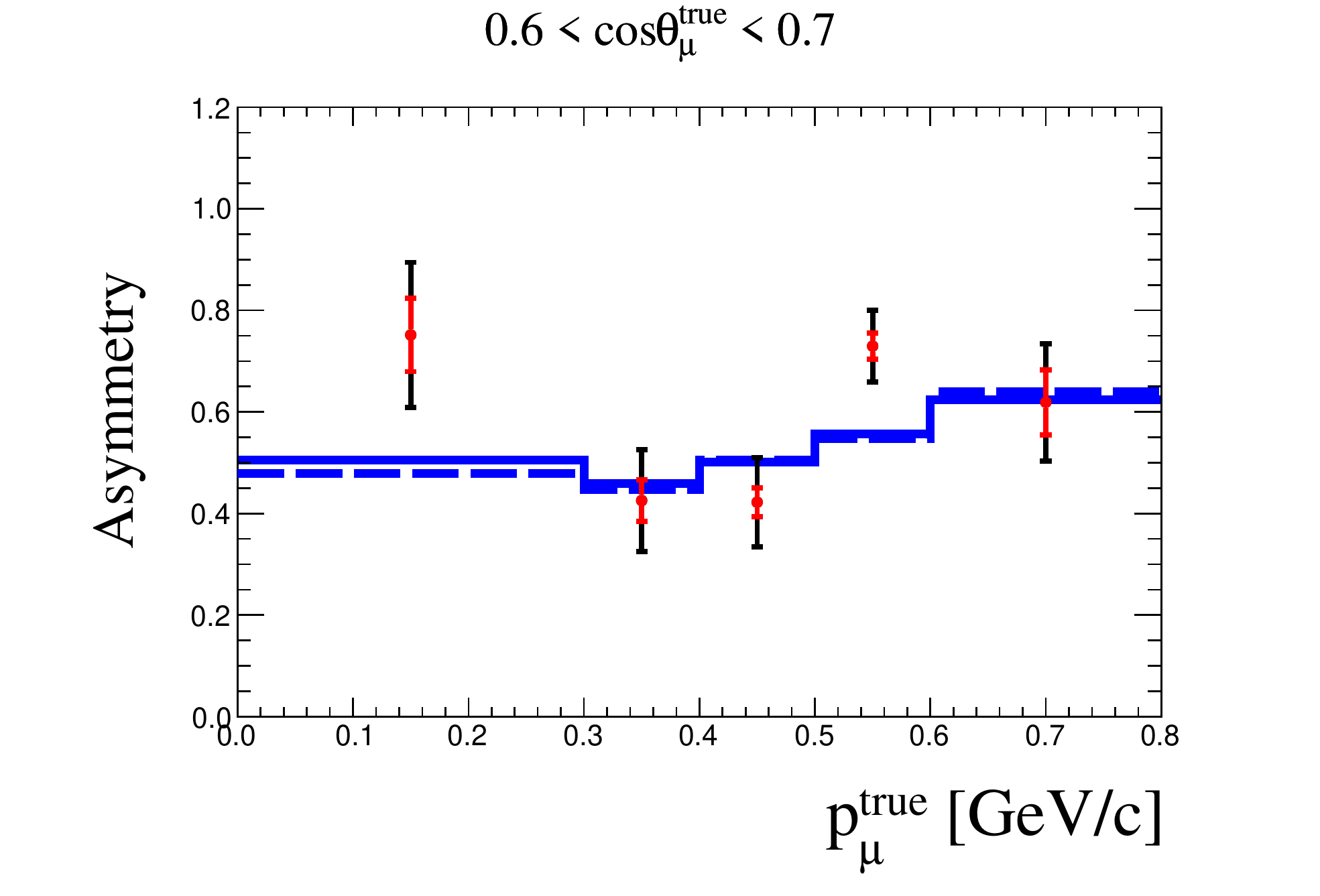}
	\includegraphics[width=0.36\linewidth]{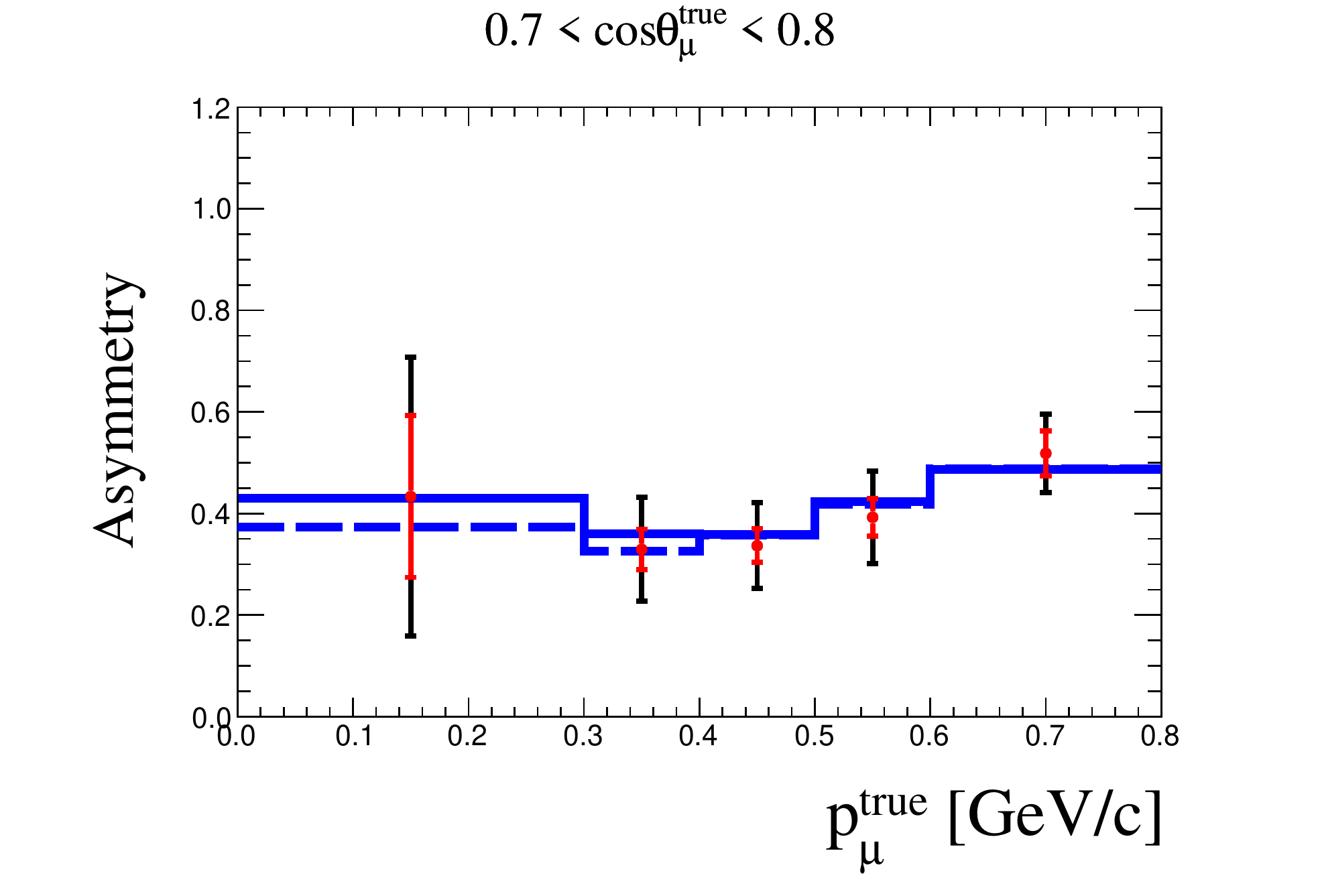}
	\includegraphics[width=0.36\linewidth]{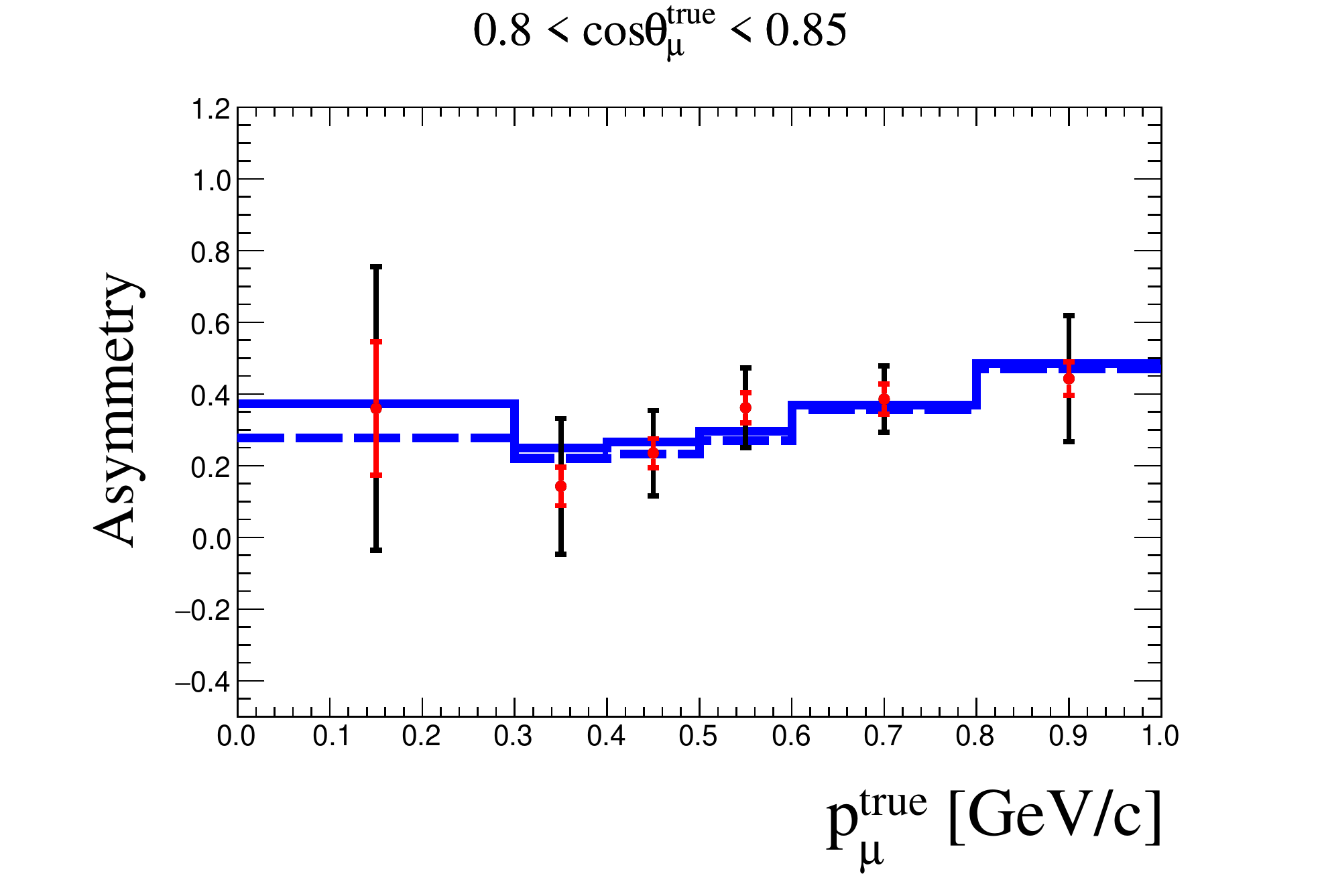}
	\includegraphics[width=0.36\linewidth]{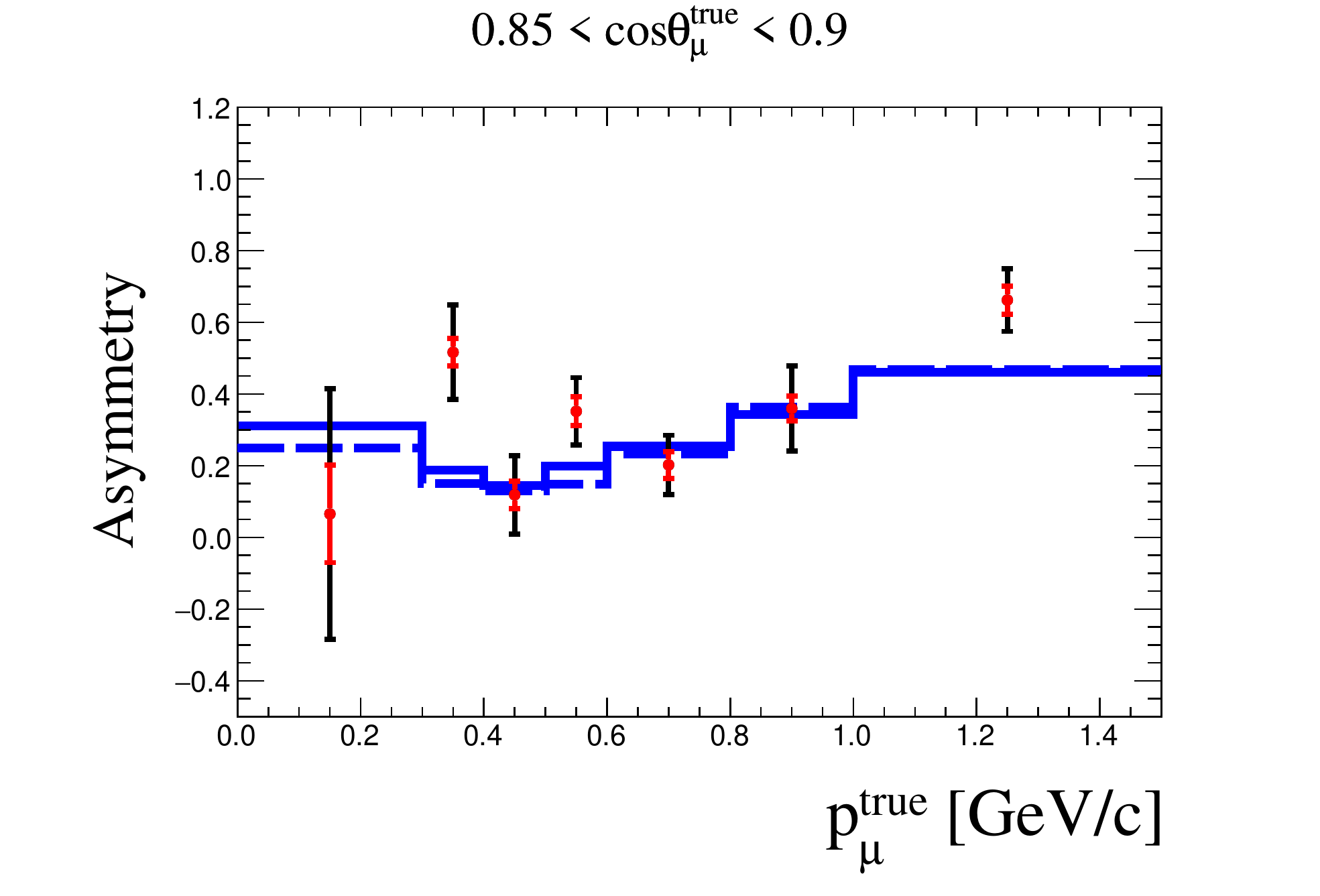}
	\includegraphics[width=0.36\linewidth]{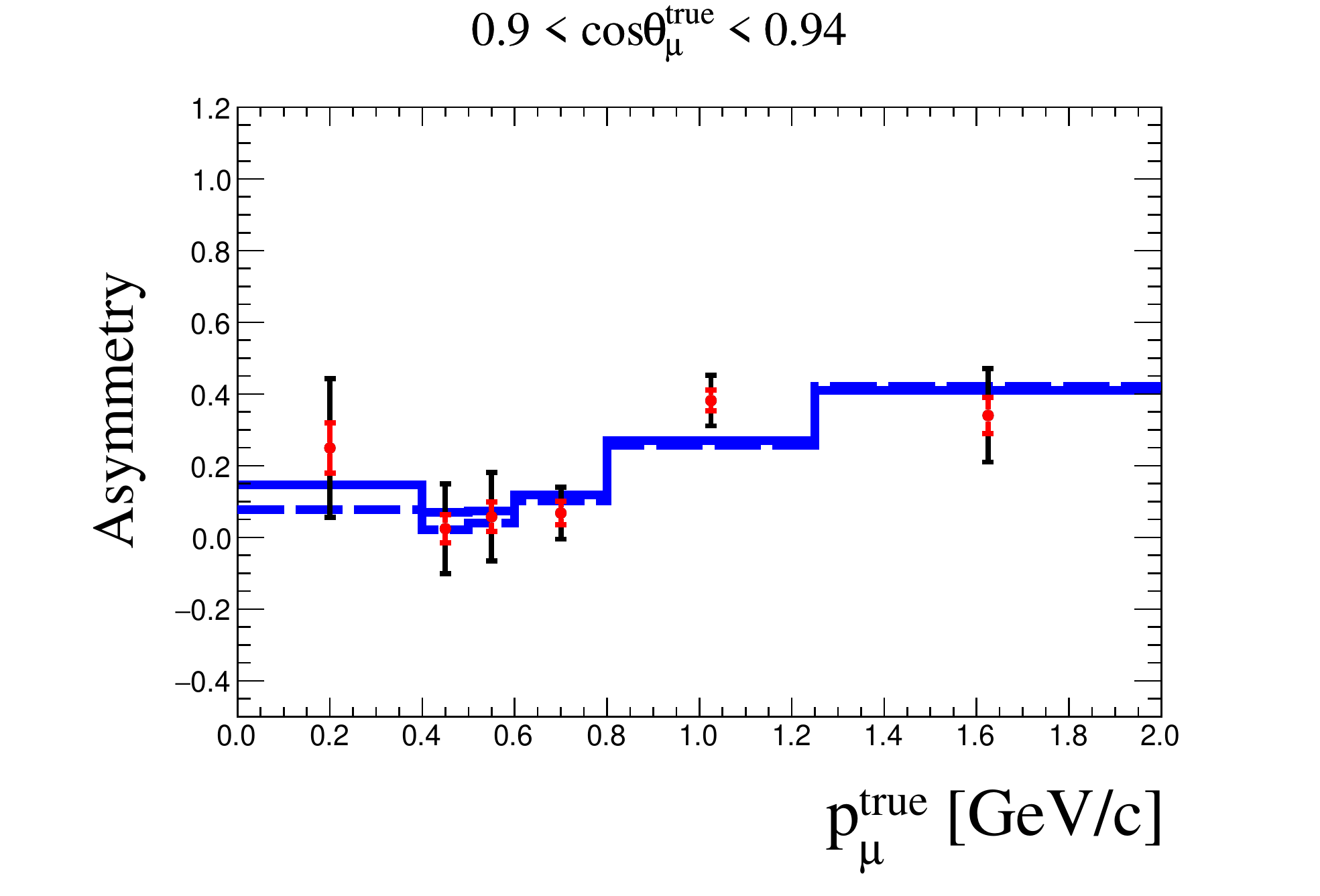}
	\includegraphics[width=0.36\linewidth]{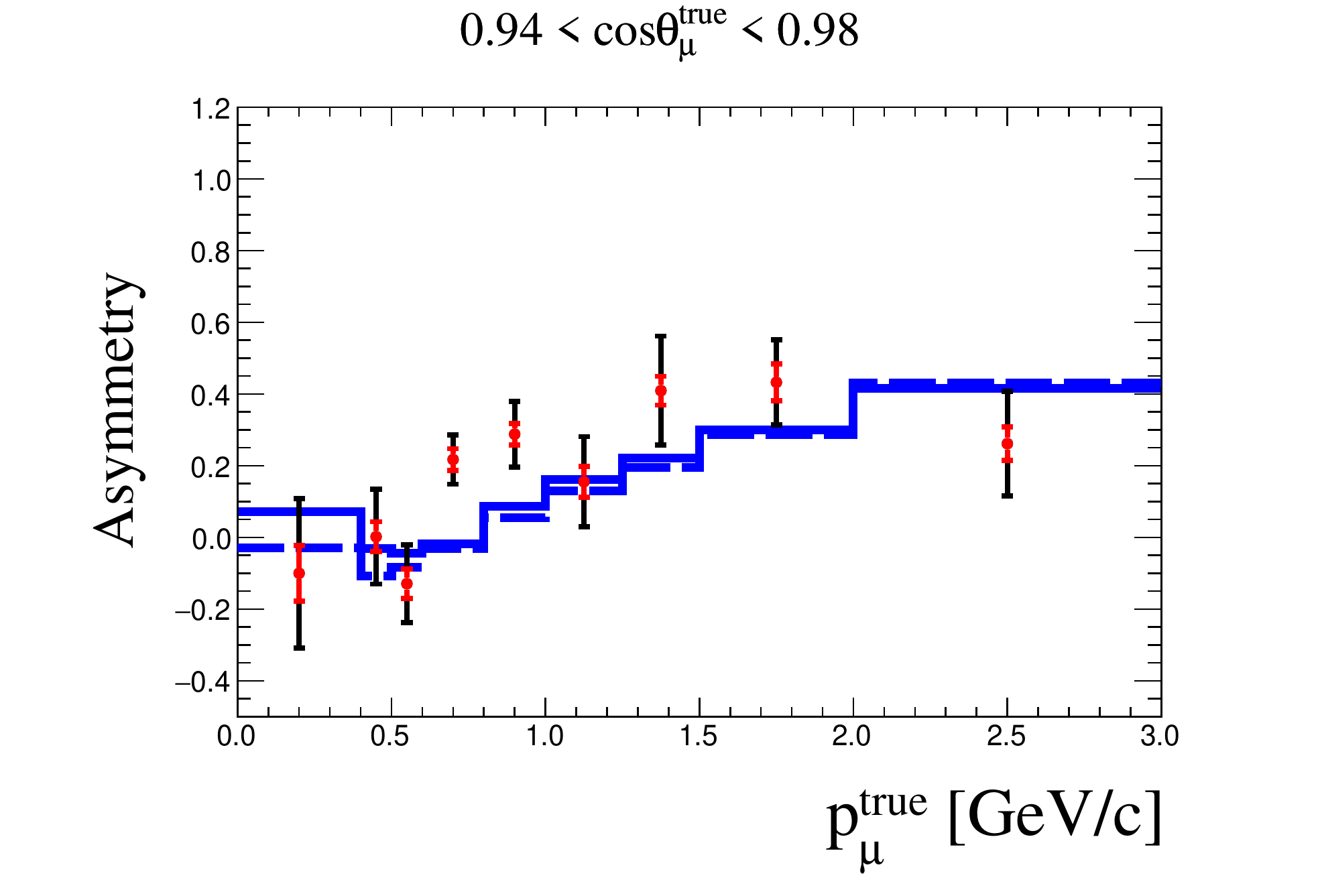}	
	\includegraphics[width=0.36\linewidth]{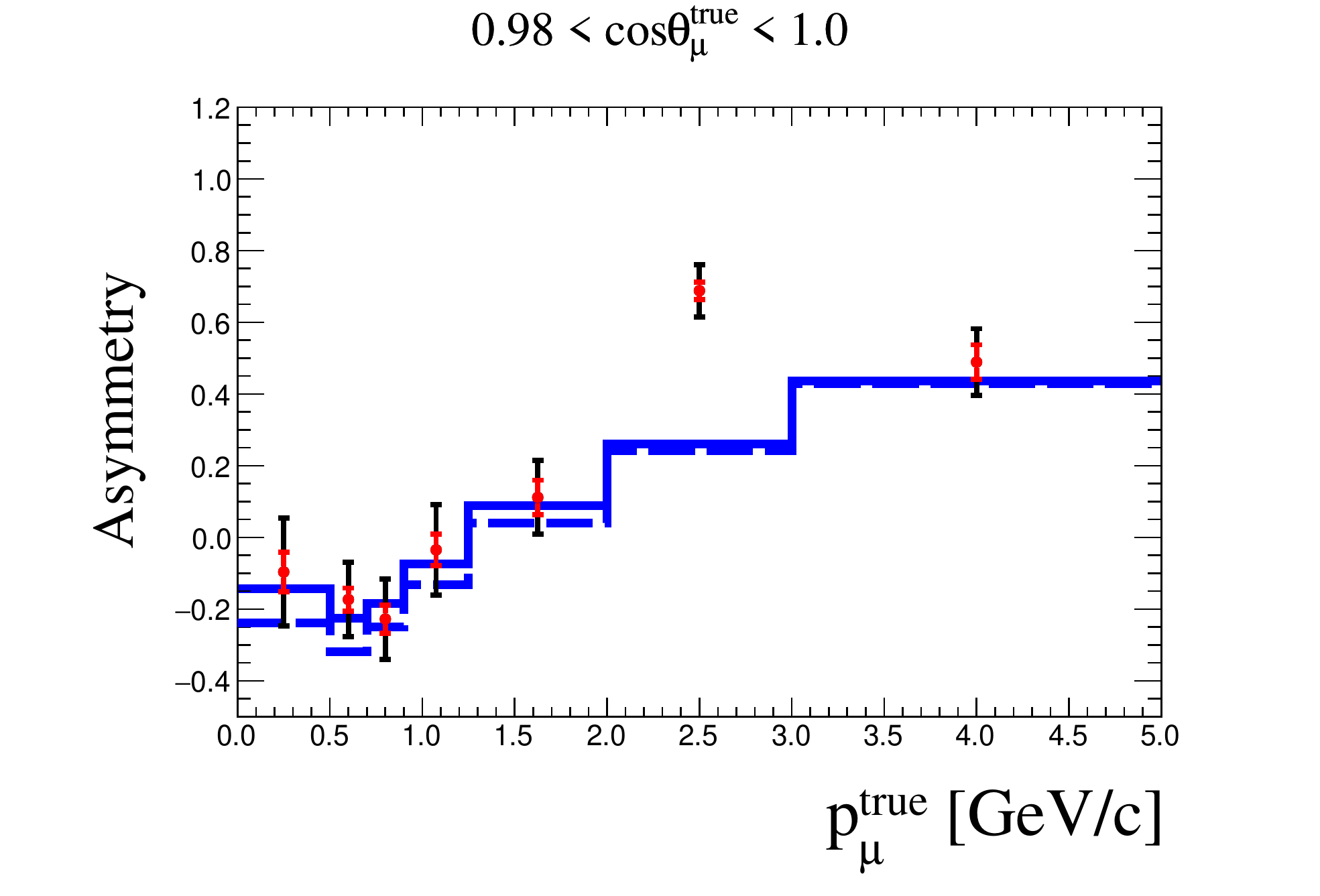}
	\includegraphics[width=0.36\linewidth]{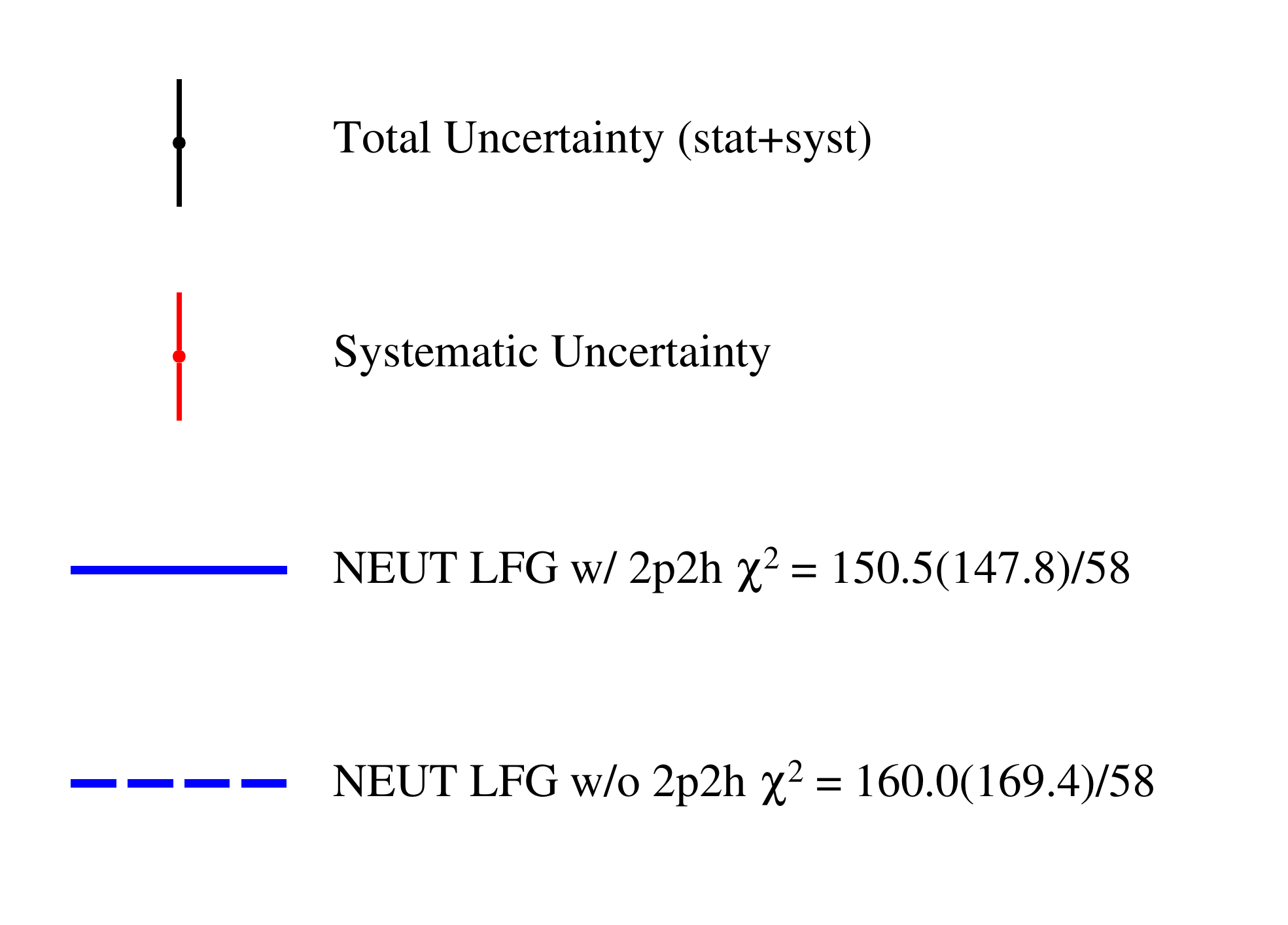}	
	\caption{Measured double-differential \cczeropi cross-section asymmetry in bins of true muon kinematics with systematic uncertainty (red bars) and total (stat.+syst.) uncertainty (black bars). The results are compared to \textsc{Neut} version~\texttt{5.4.1}, which uses an LFG+RPA model, with (solid line) and without (dashed line) 2p2h. The full and shape-only (in parenthesis) $\chi2$ are reported. The last bin in momentum is not displayed for readability.}
	\label{fig:xsecasyneut2p2h}
\end{figure*}

\begin{figure*}[h!]
	\centering
	\includegraphics[width=0.36\linewidth]{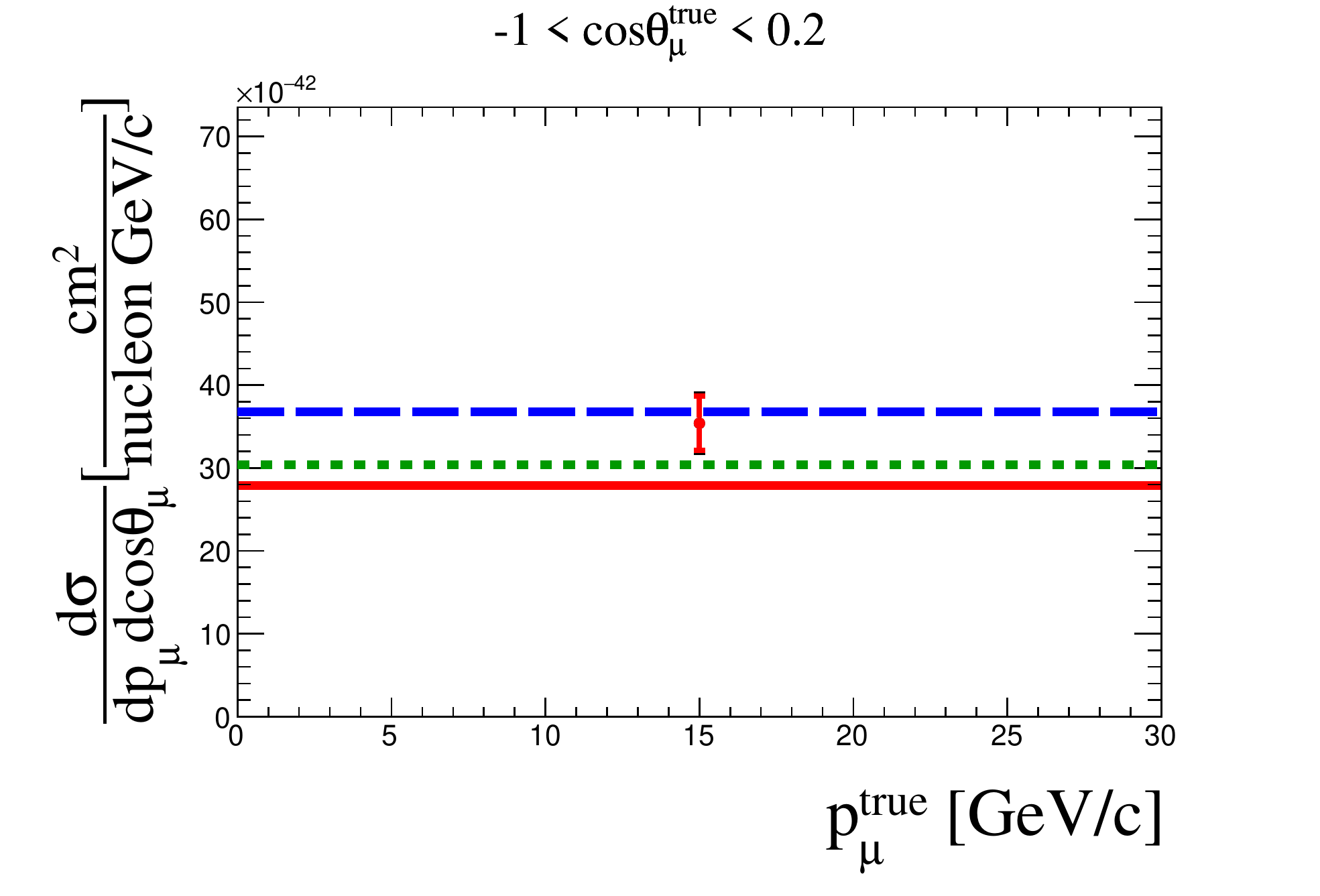}
	\includegraphics[width=0.36\linewidth]{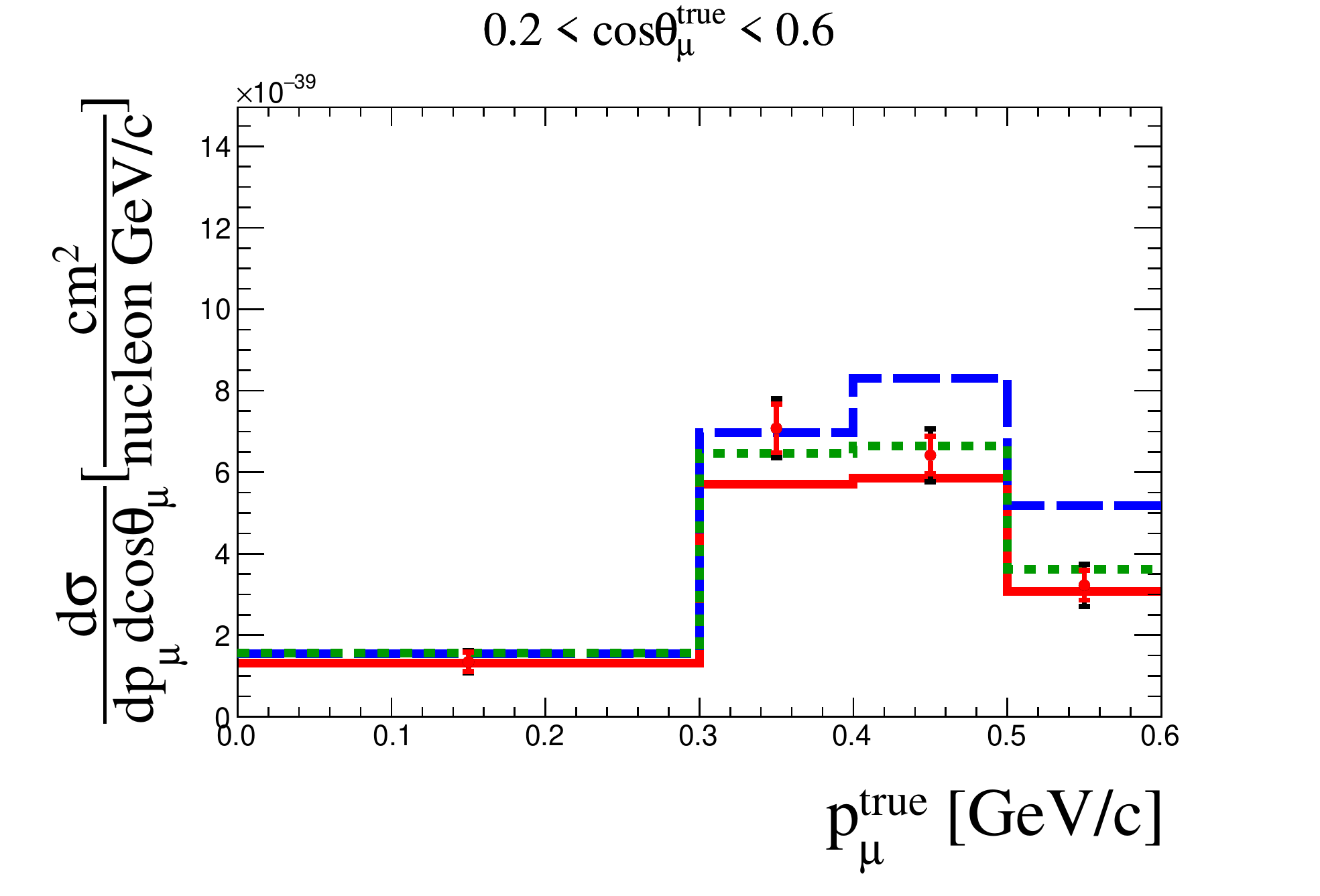}	
	\includegraphics[width=0.36\linewidth]{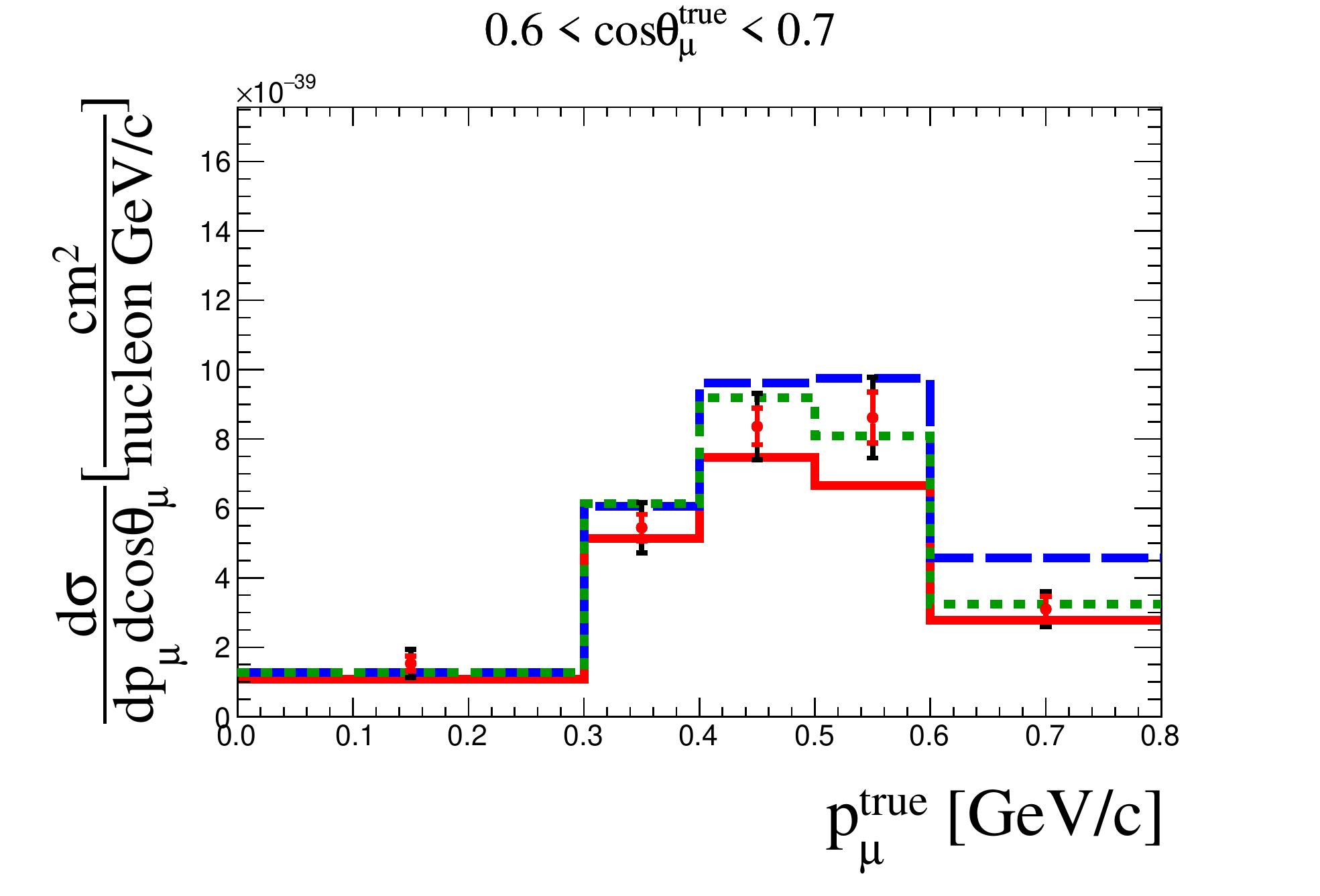}	
	\includegraphics[width=0.36\linewidth]{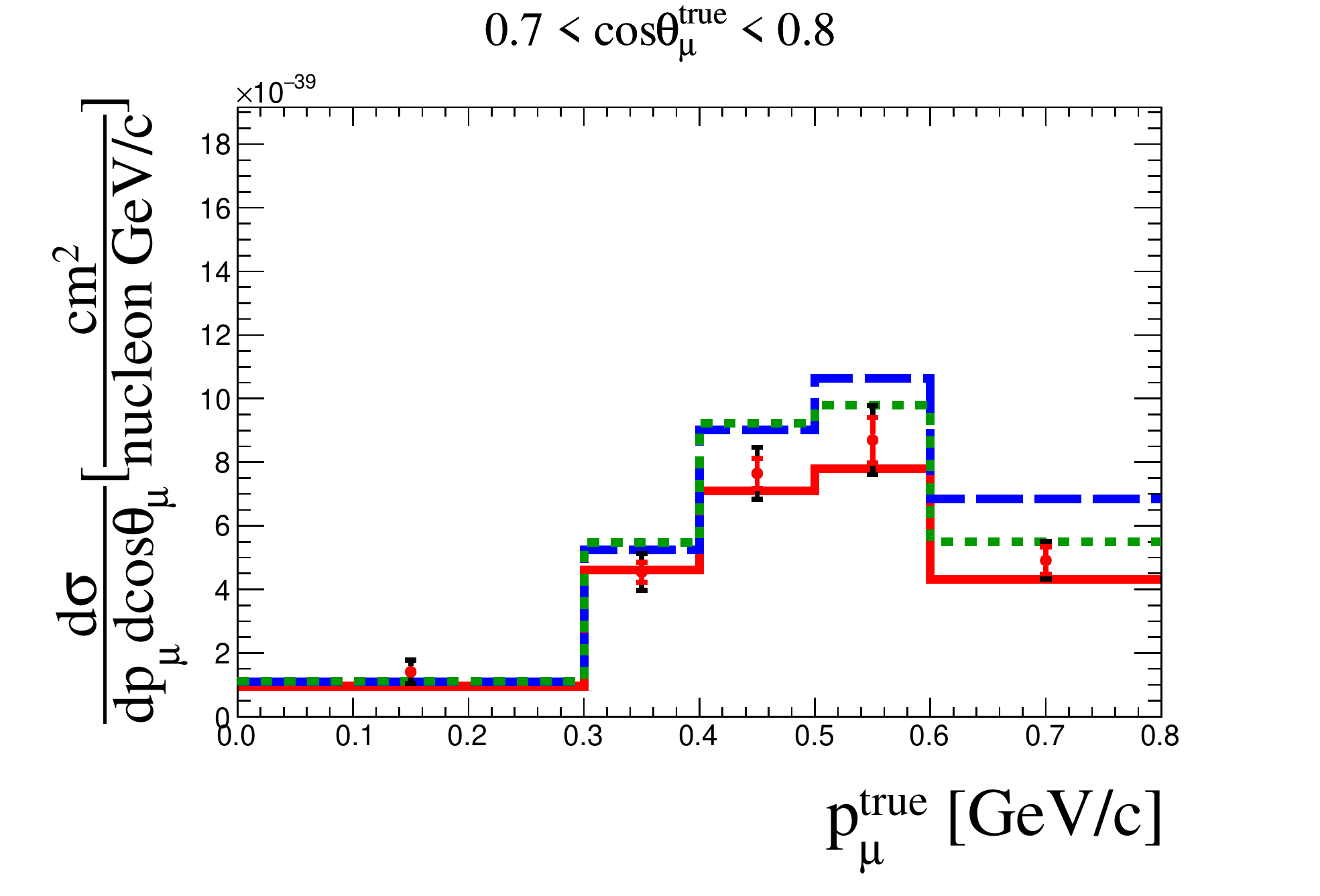}	
	\includegraphics[width=0.36\linewidth]{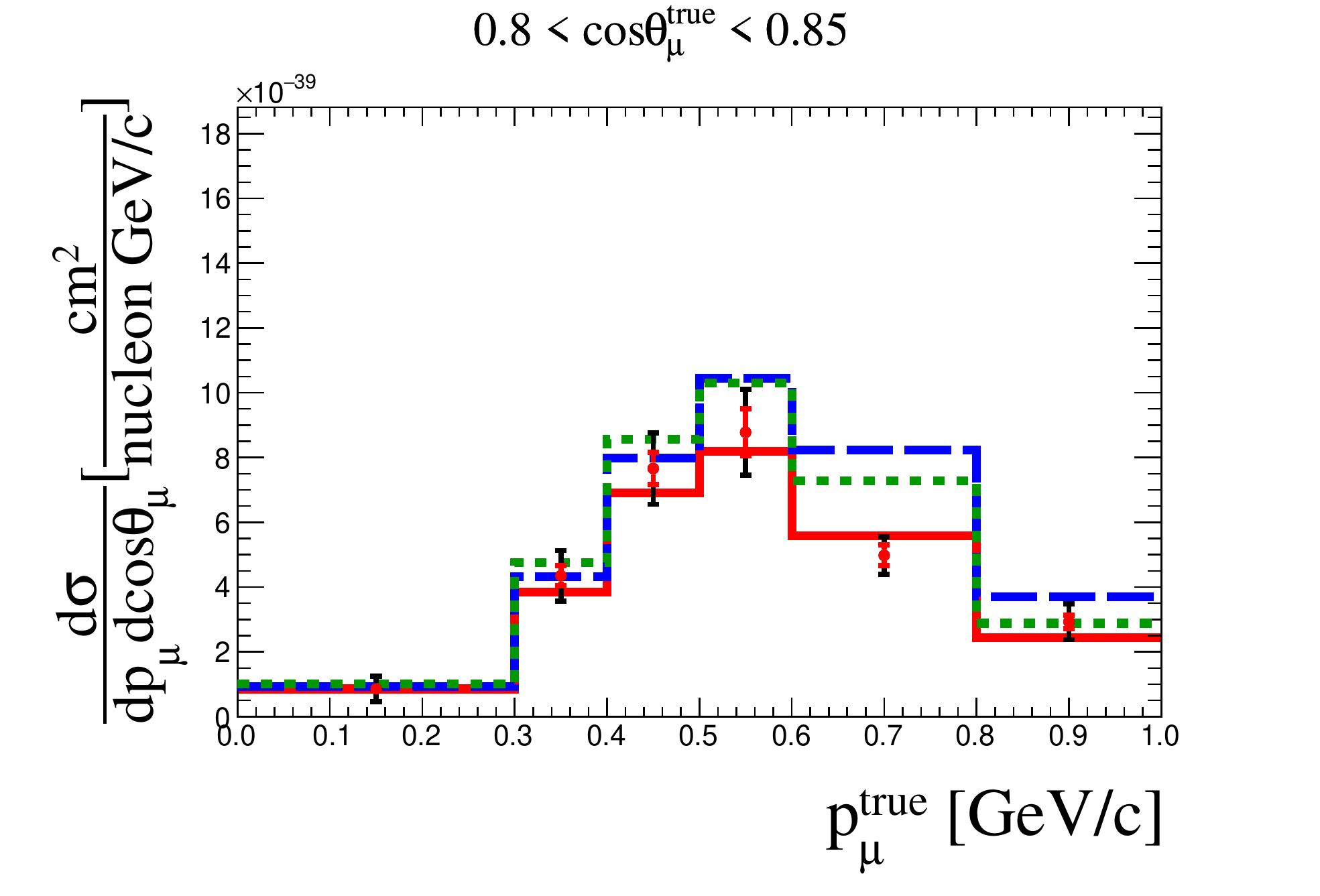}	
	\includegraphics[width=0.36\linewidth]{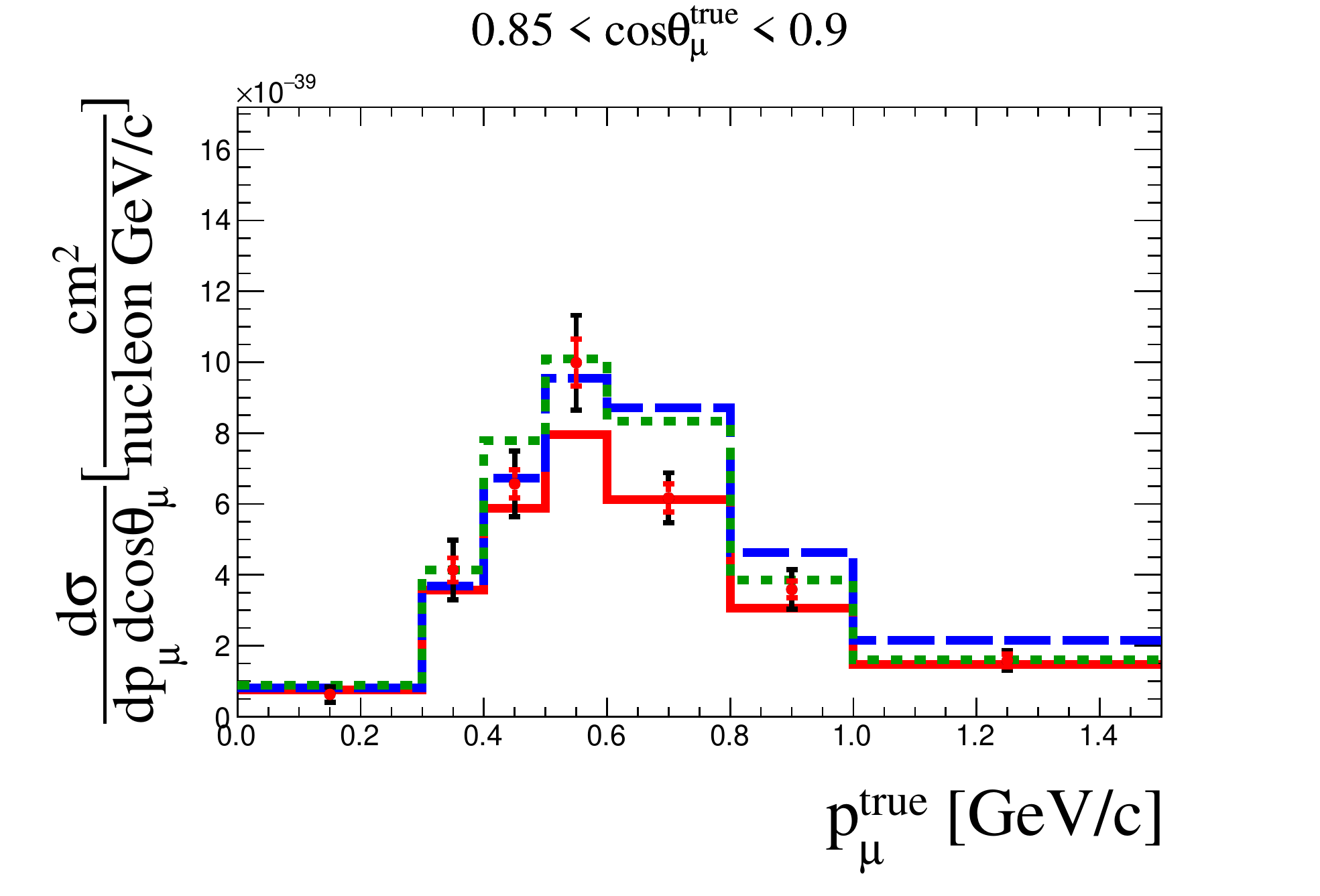}	
	\includegraphics[width=0.36\linewidth]{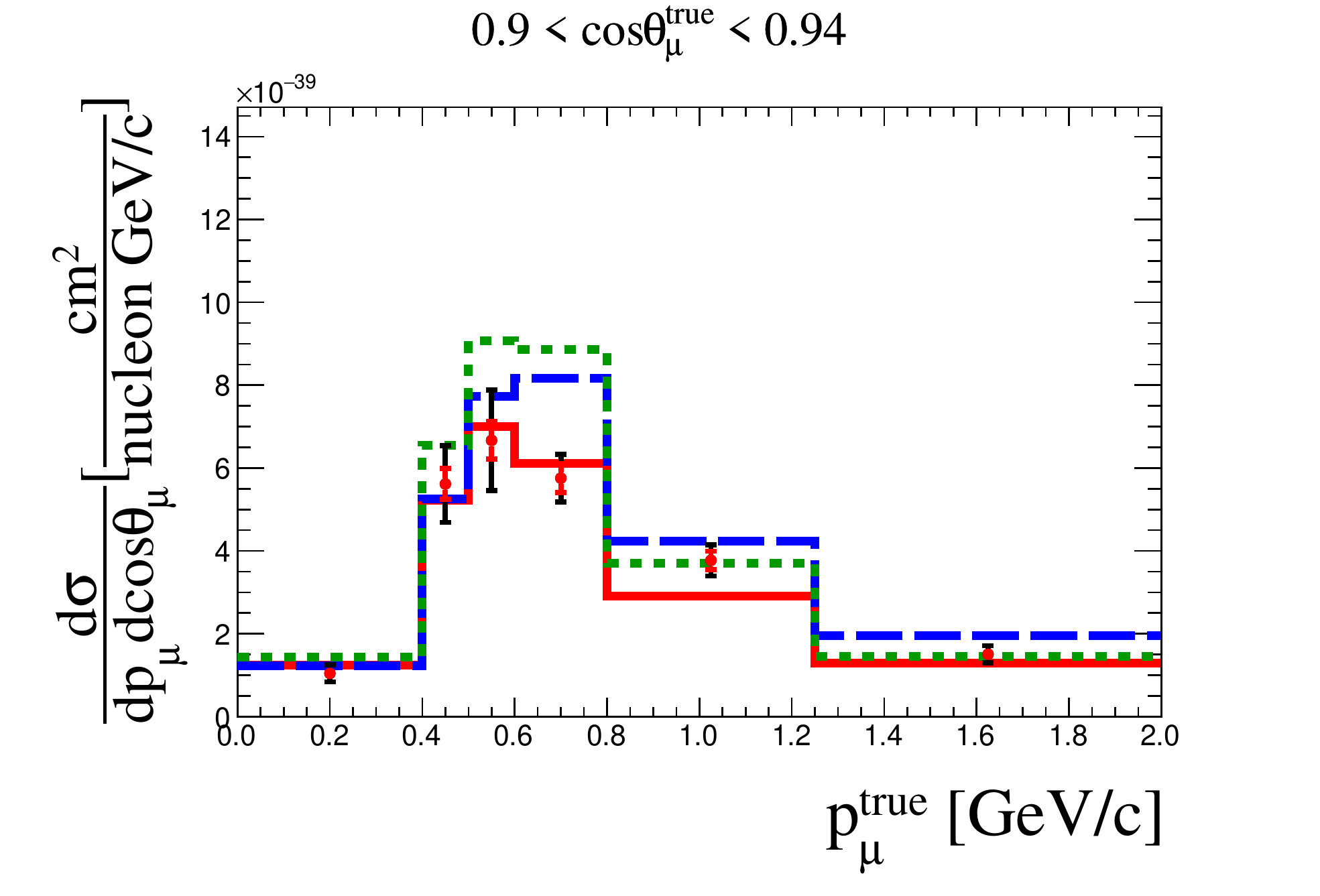}	
	\includegraphics[width=0.36\linewidth]{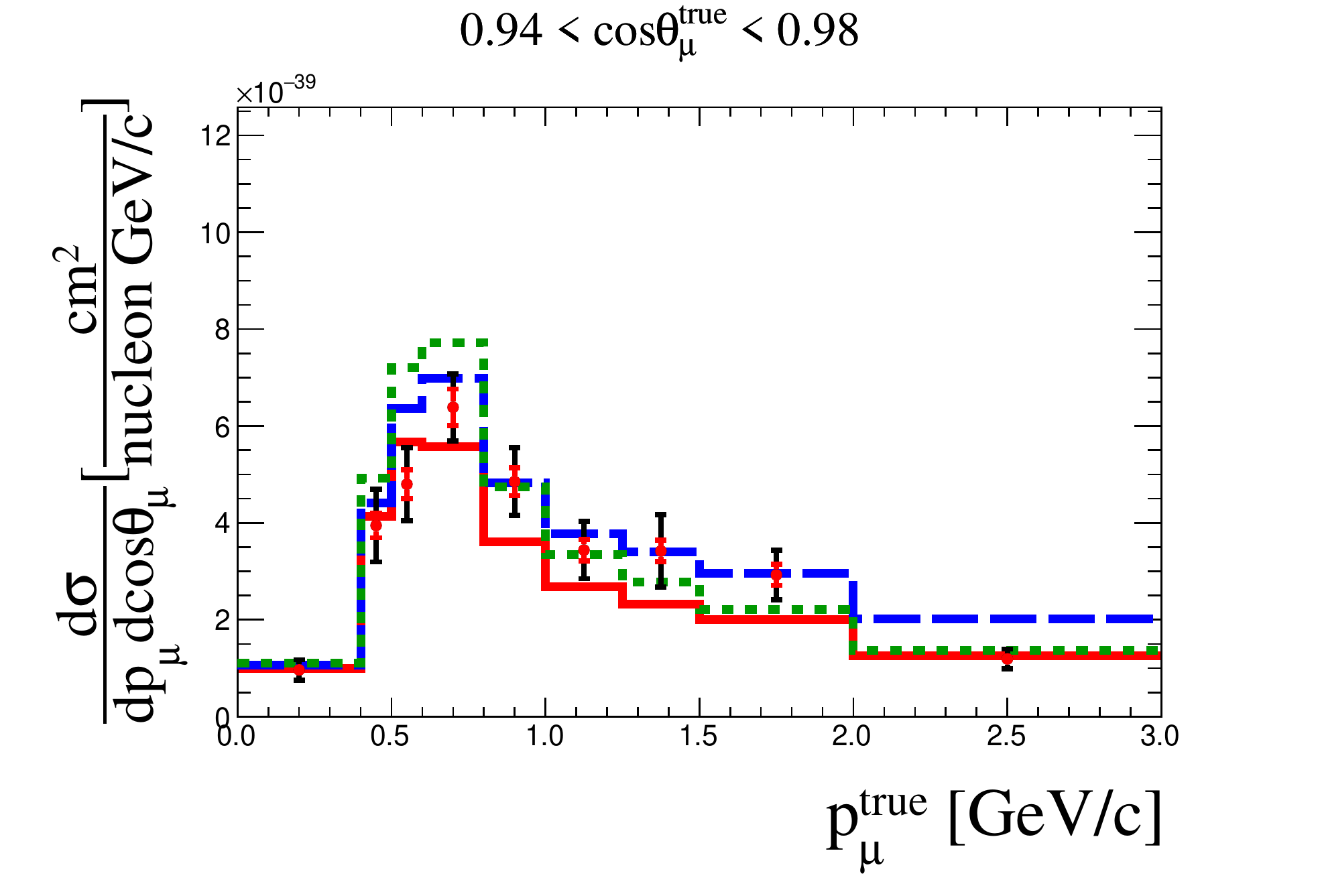}
	\includegraphics[width=0.36\linewidth]{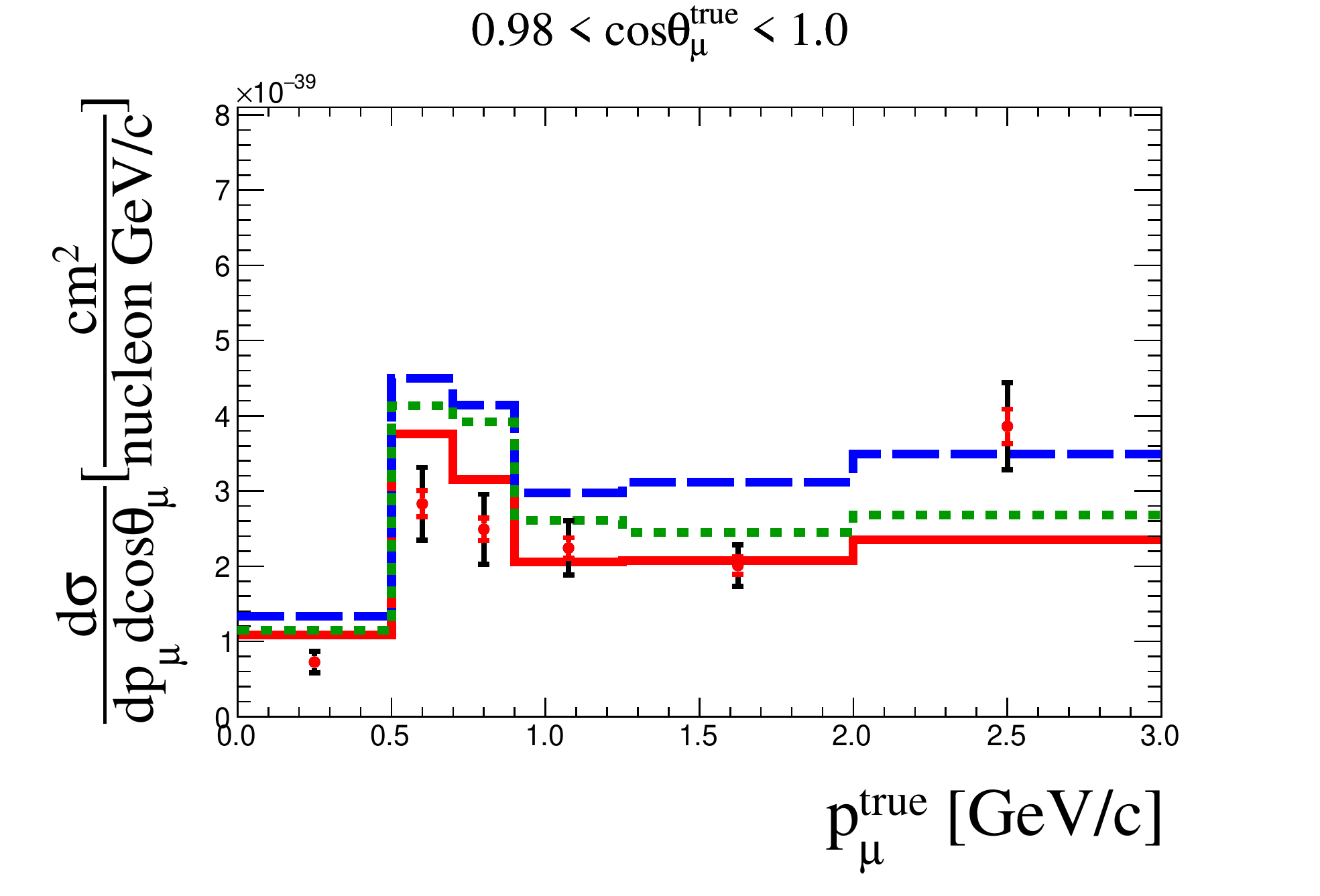}	
	\includegraphics[width=0.36\linewidth]{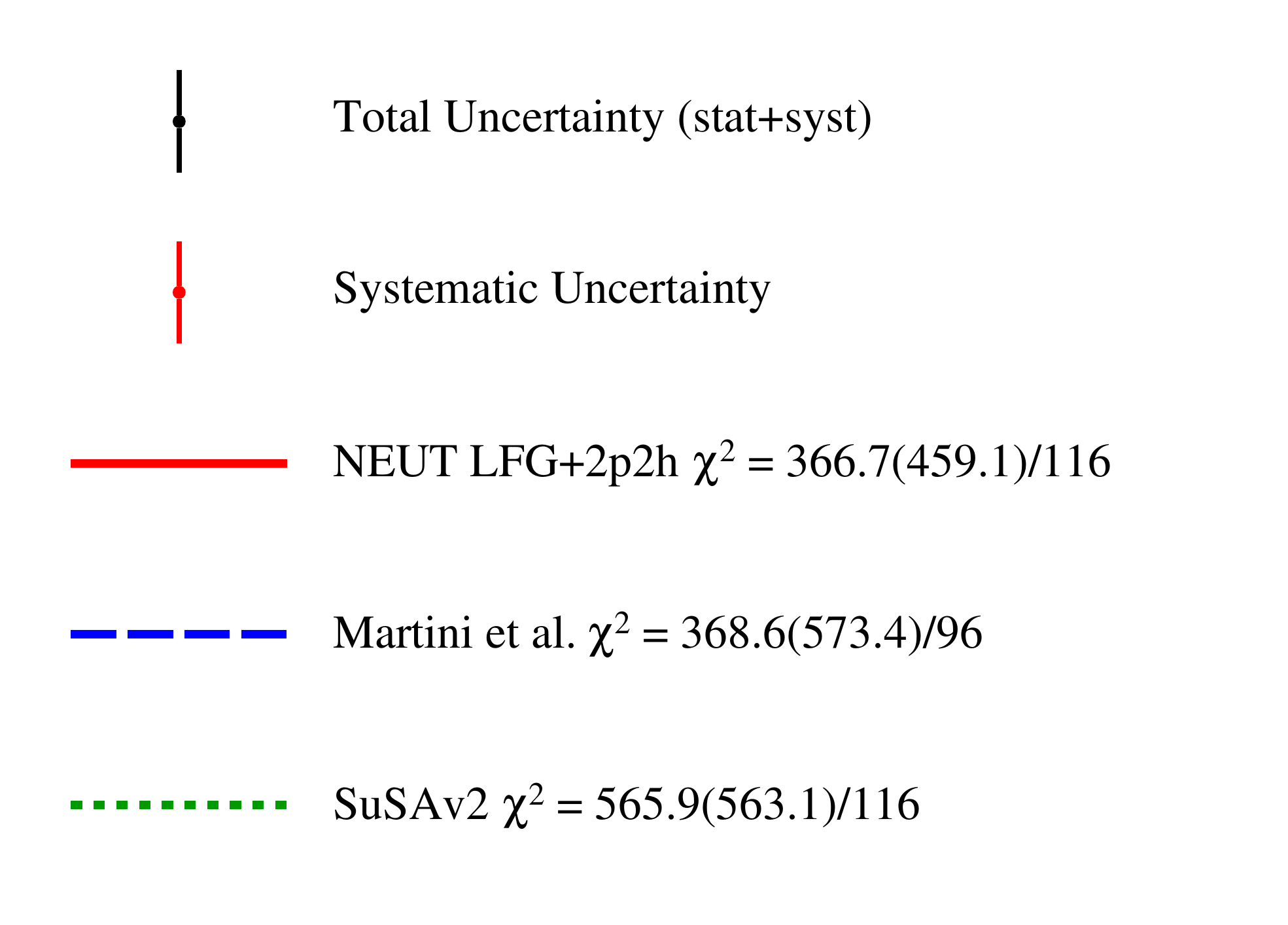}		
	\caption{Measured \numu \cczeropi double-differential cross-section per nucleon in bins of true muon kinematics with systematic uncertainty (red bars) and total (stat.+syst.) uncertainty (black bars). The results are compared to \textsc{Neut} version~\texttt{5.4.1}, which uses an LFG+RPA model with 2p2h (solid red line), Martini \textit{et al.} (dashed blue line) and \textsc{SuSAv2} (green dashed line) models. The full and shape-only (in parenthesis) $\chi2$ are reported. The last bin in momentum is not displayed for readability.}
	\label{fig:numucc0piNMS}
\end{figure*}

\begin{figure*}[h!]
	\centering
	\includegraphics[width=0.36\linewidth]{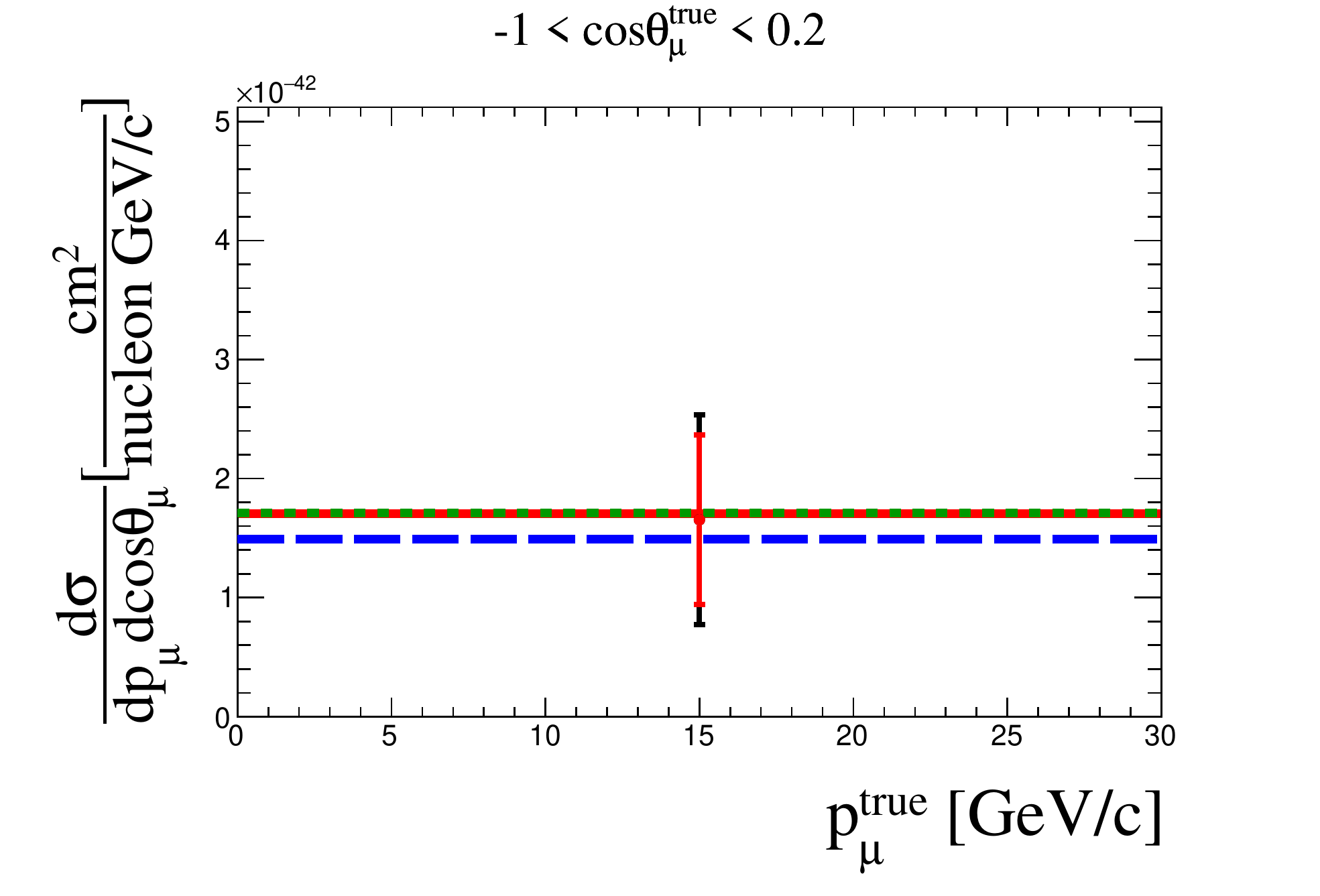}
	\includegraphics[width=0.36\linewidth]{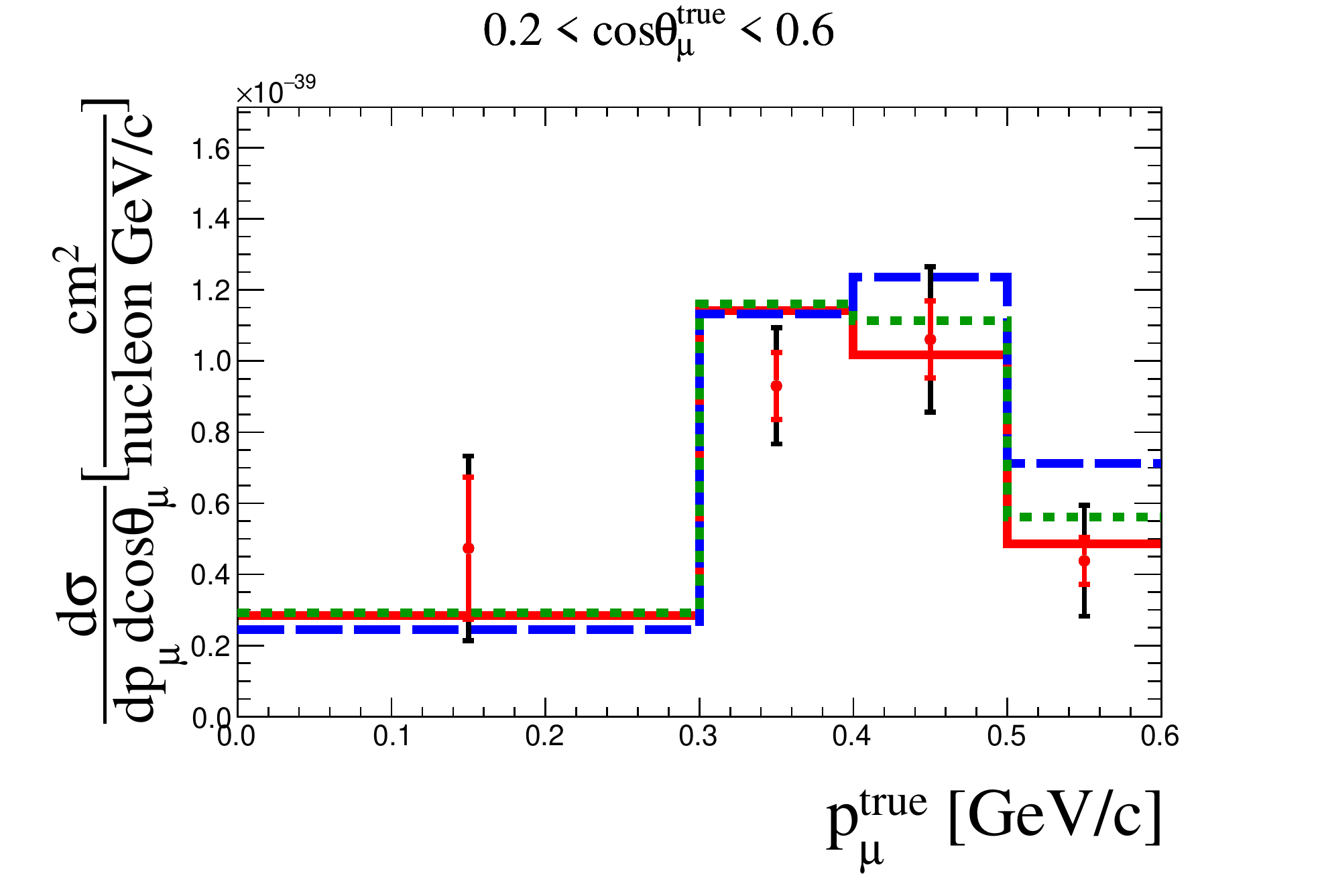}	
	\includegraphics[width=0.36\linewidth]{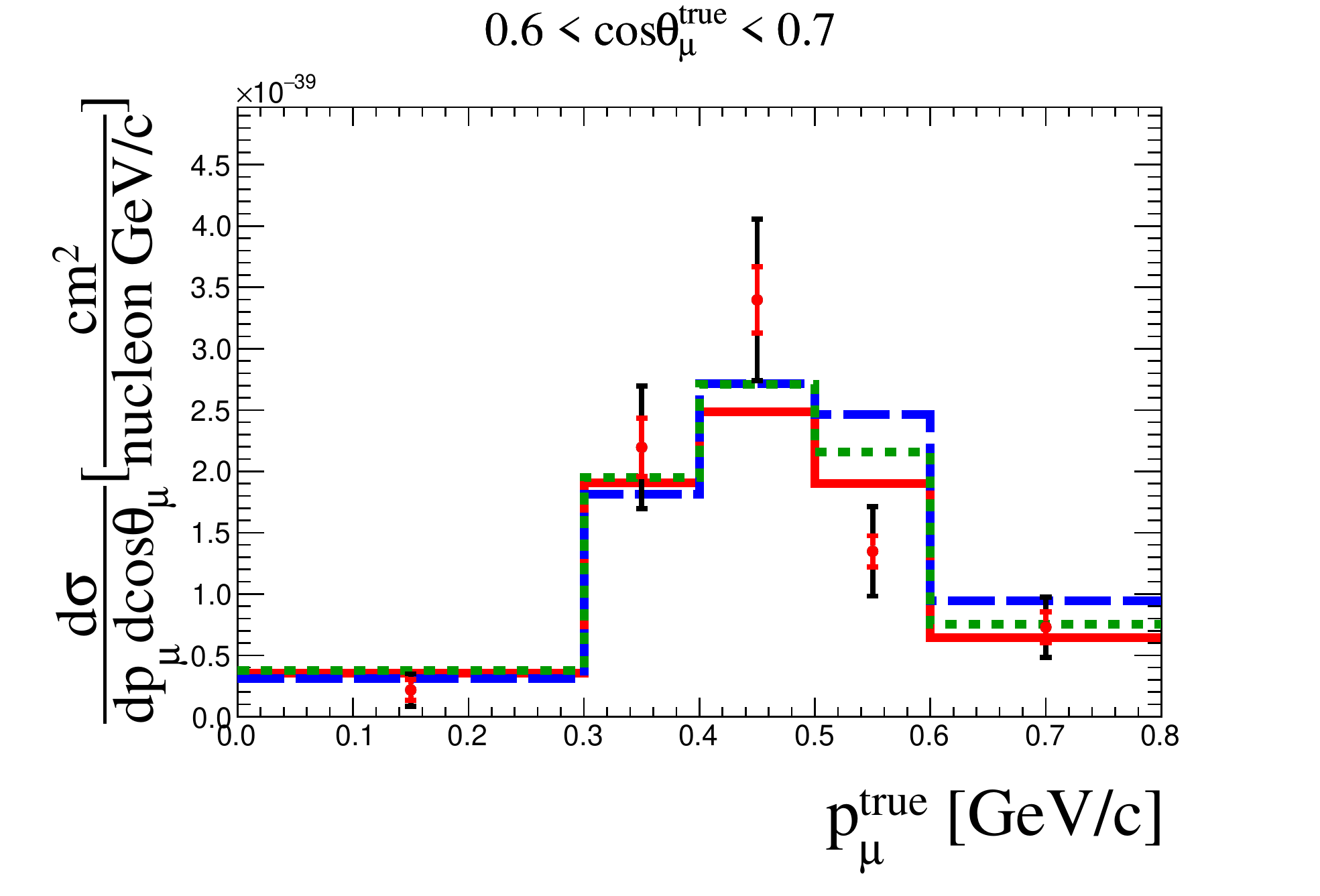}	
	\includegraphics[width=0.36\linewidth]{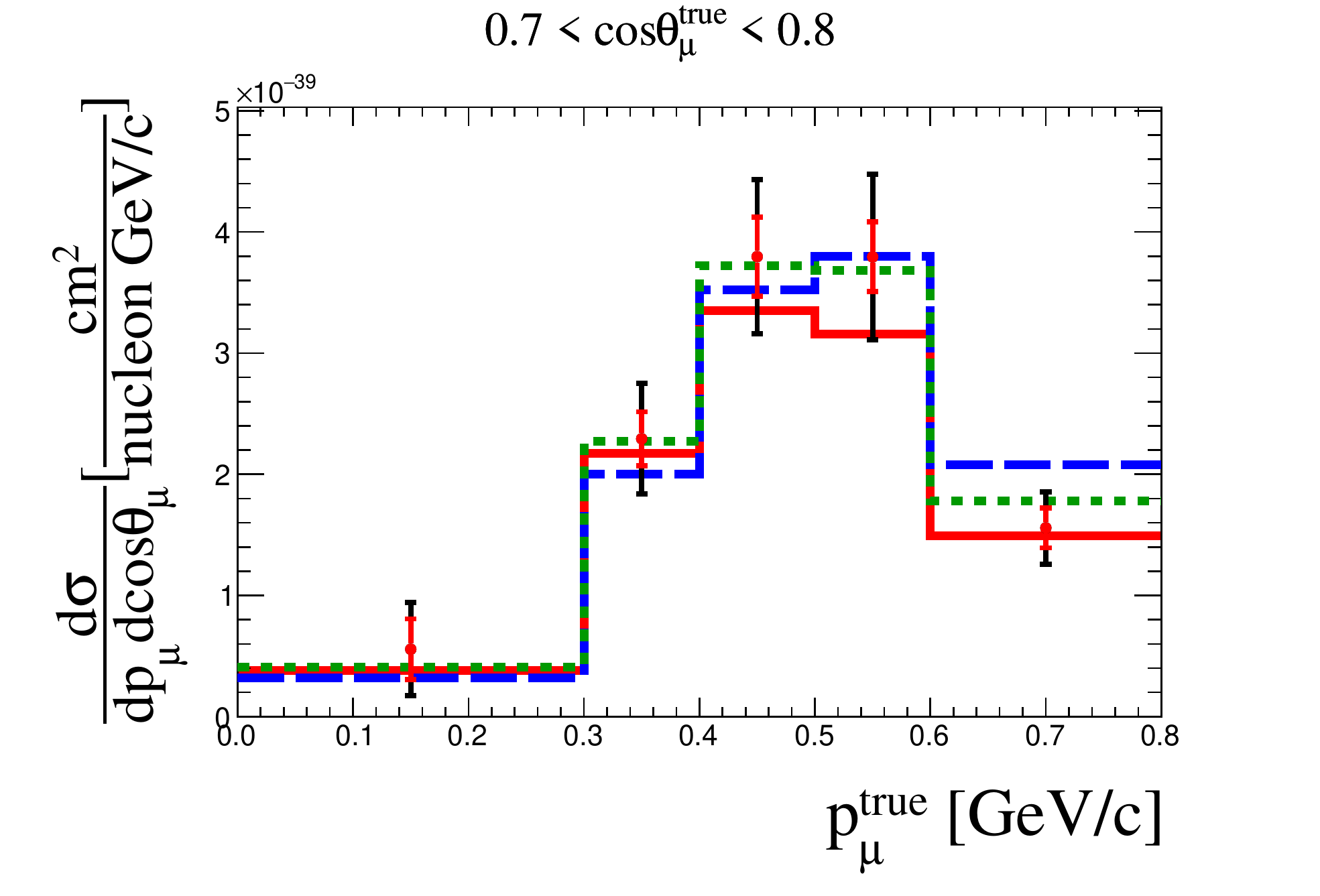}	
	\includegraphics[width=0.36\linewidth]{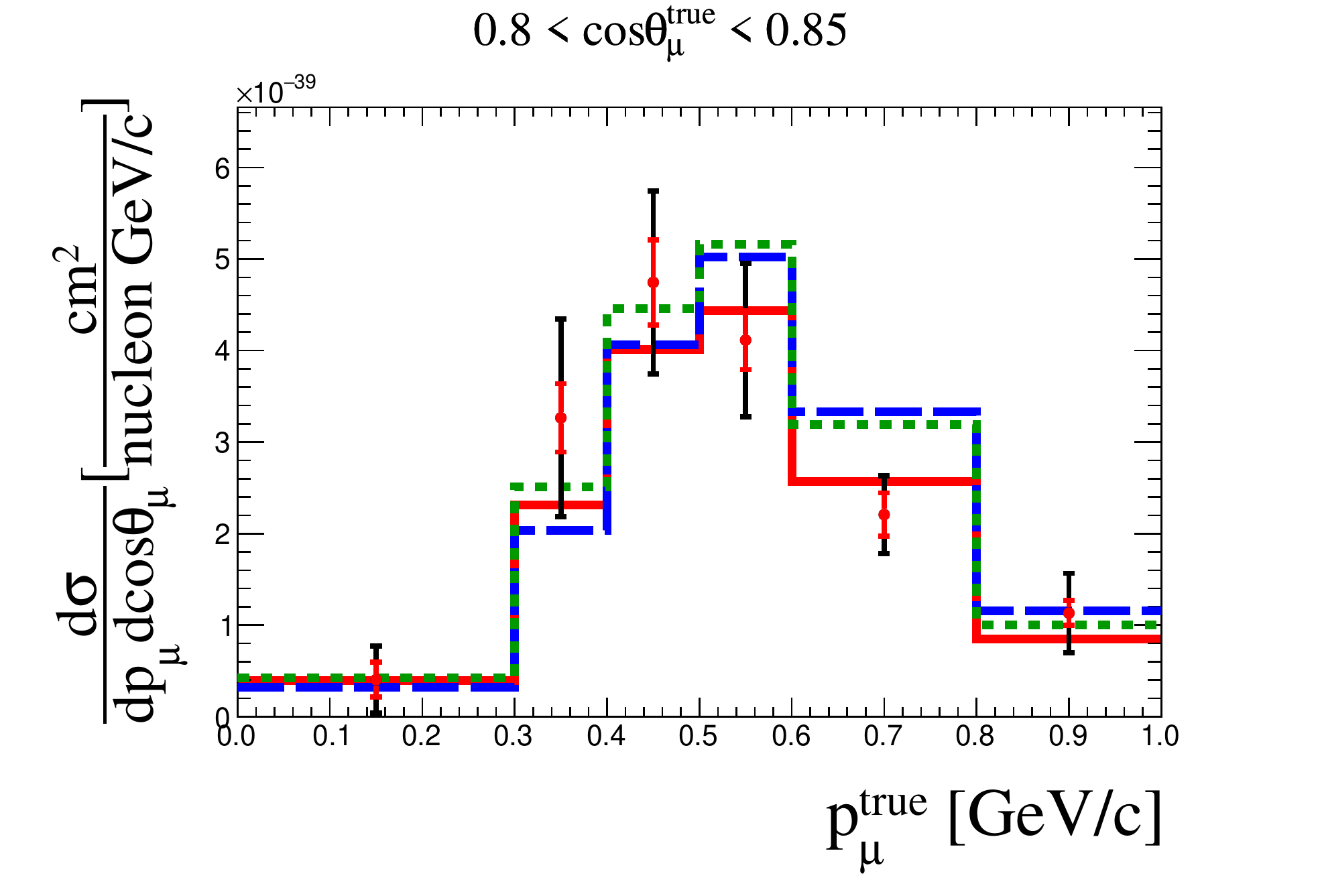}	
	\includegraphics[width=0.36\linewidth]{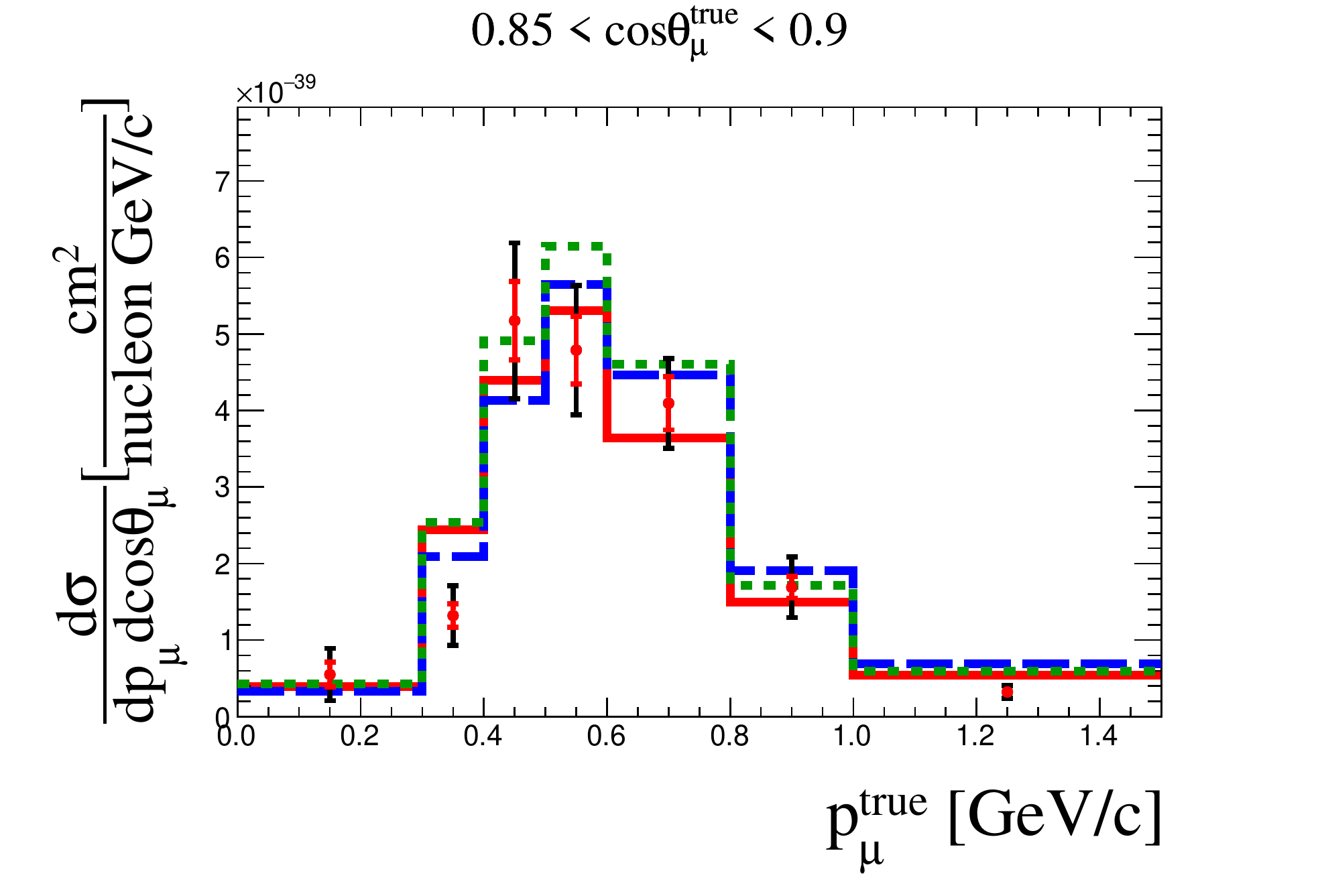}	
	\includegraphics[width=0.36\linewidth]{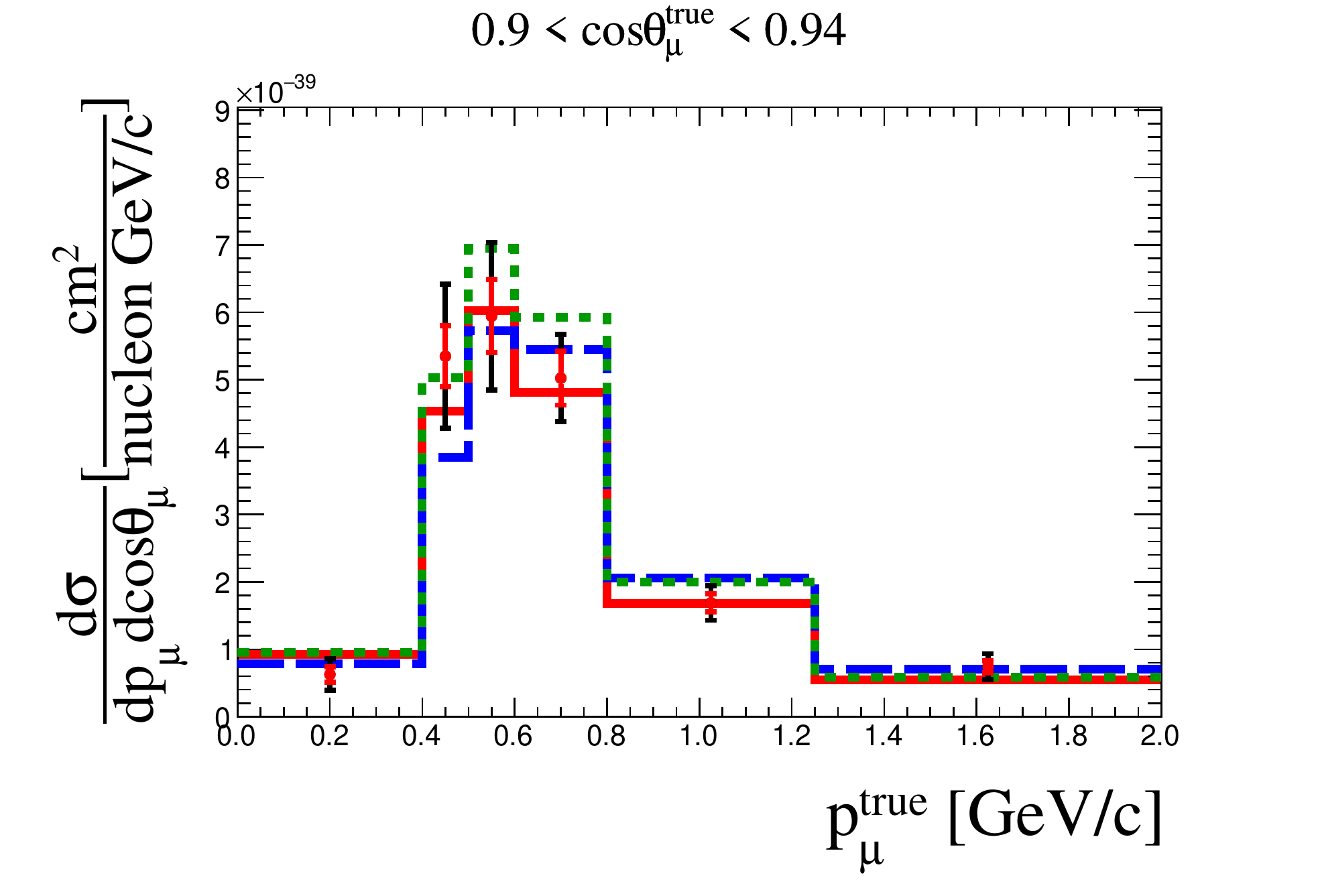}	
	\includegraphics[width=0.36\linewidth]{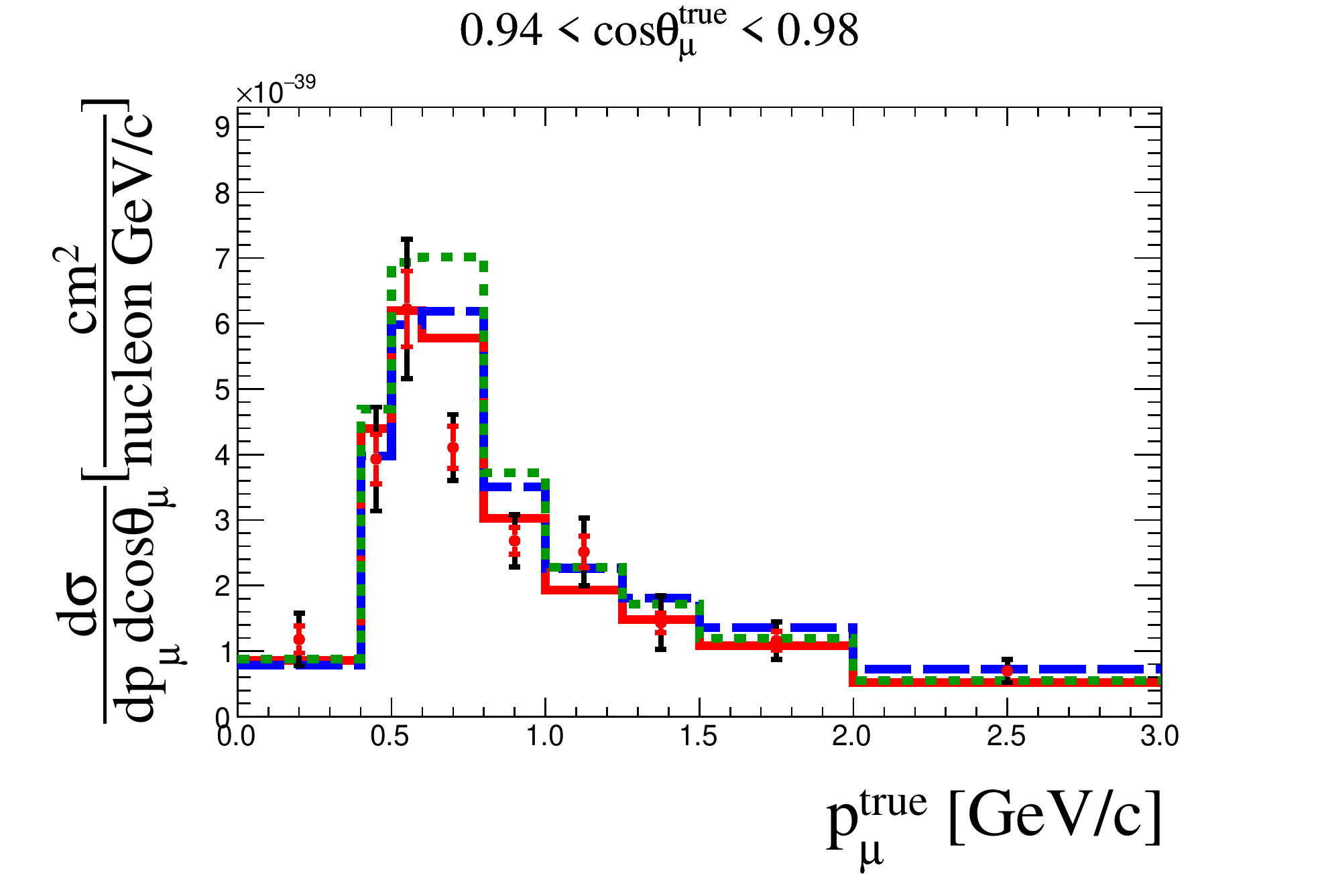}
	\includegraphics[width=0.36\linewidth]{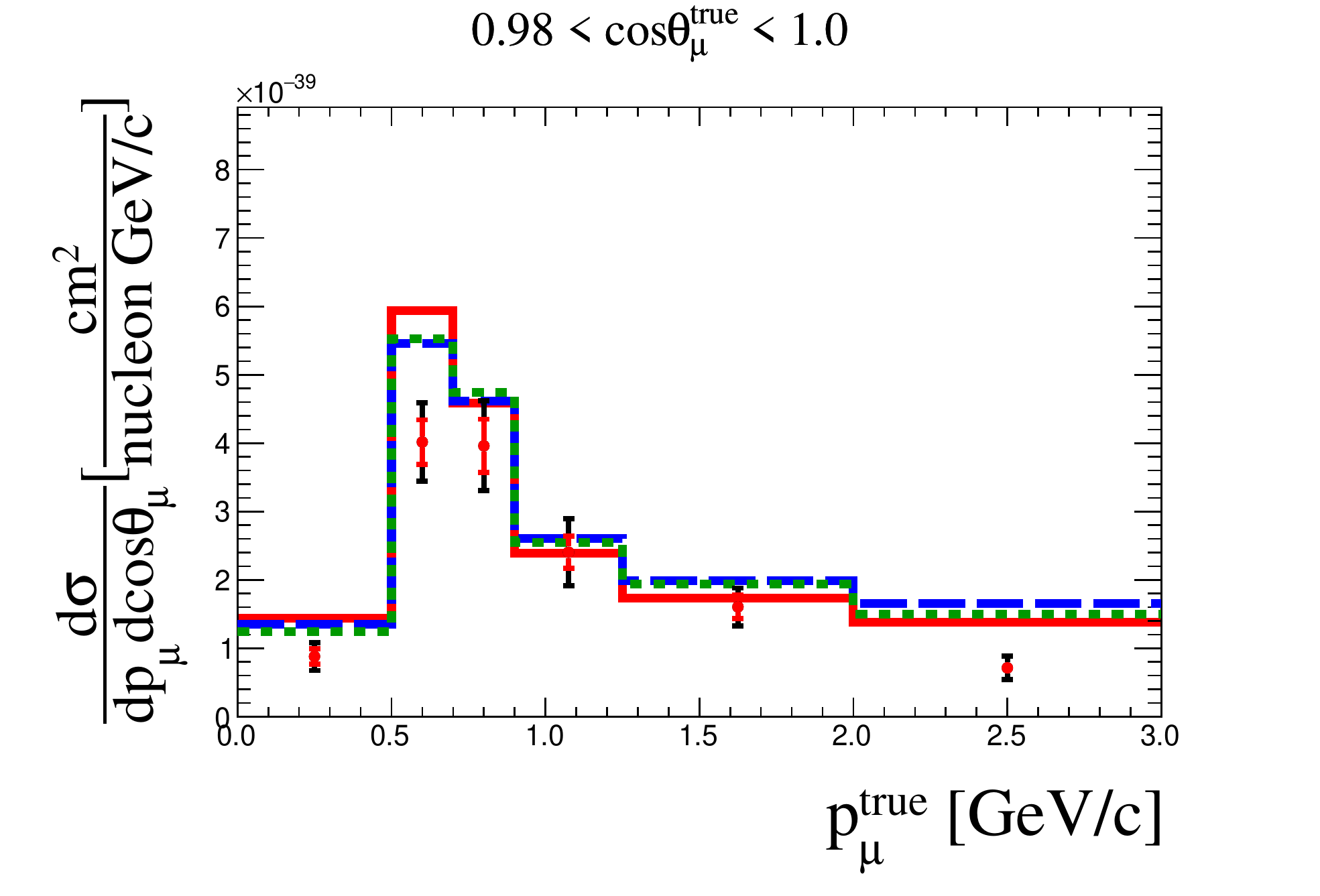}	
	\includegraphics[width=0.36\linewidth]{XsecLegendMartiniNievesSuSAv2}		
	\caption{Measured \barnumu \cczeropi double-differential cross-section per nucleon in bins of true muon kinematics with systematic uncertainty (red bars) and total (stat.+syst.) uncertainty (black bars). The results are compared to \textsc{Neut} version~\texttt{5.4.1}, which uses an LFG+RPA model with 2p2h (solid red line), Martini \textit{et al.} (dashed blue line) and \textsc{SuSAv2} (green dashed line) models. The full and shape-only (in parenthesis) $\chi2$ are reported. The last bin in momentum is not displayed for readability.}
	\label{fig:antinumucc0piNMS}
\end{figure*}

\begin{figure*}[h!]
	\centering
	\includegraphics[width=0.36\linewidth]{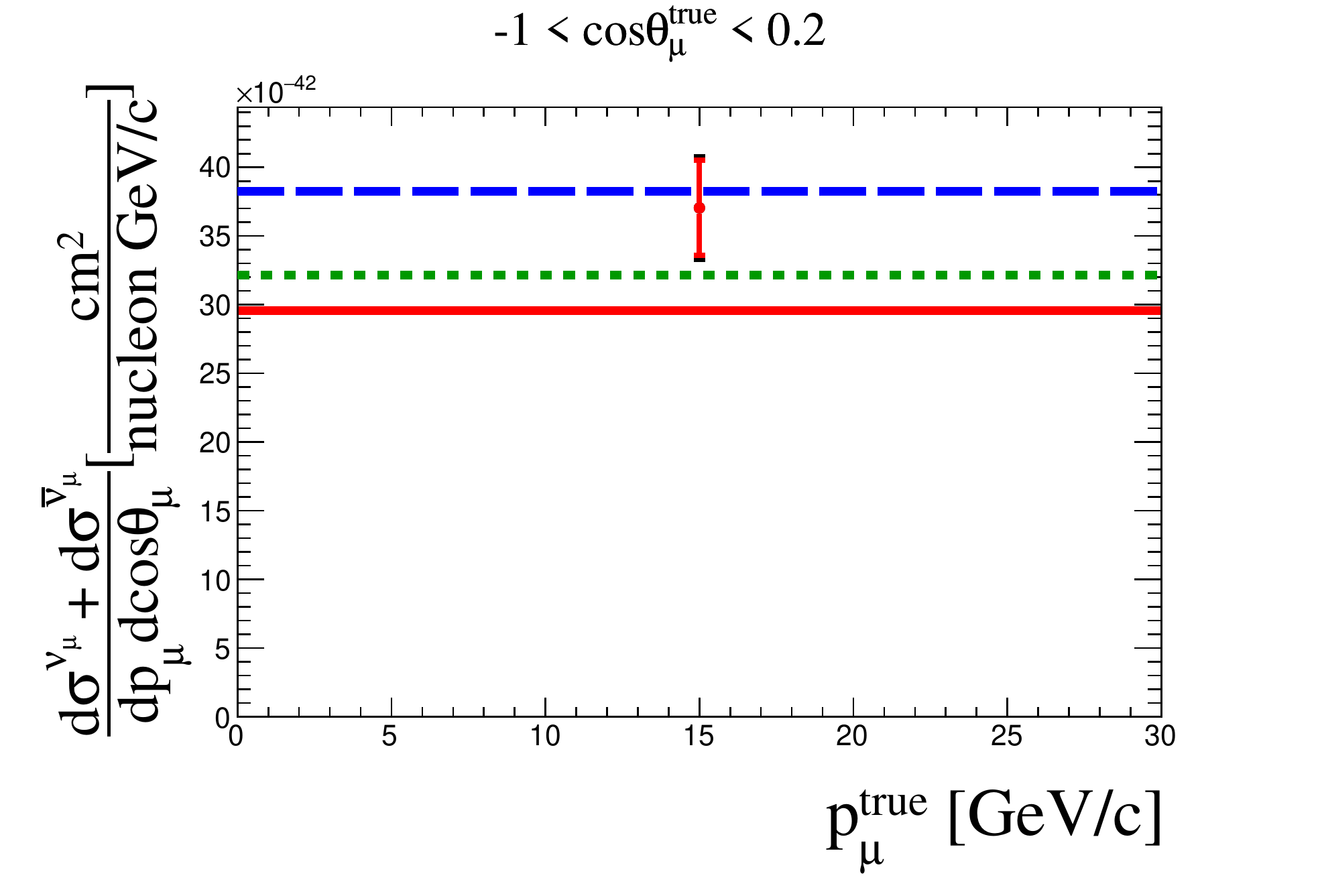}
	\includegraphics[width=0.36\linewidth]{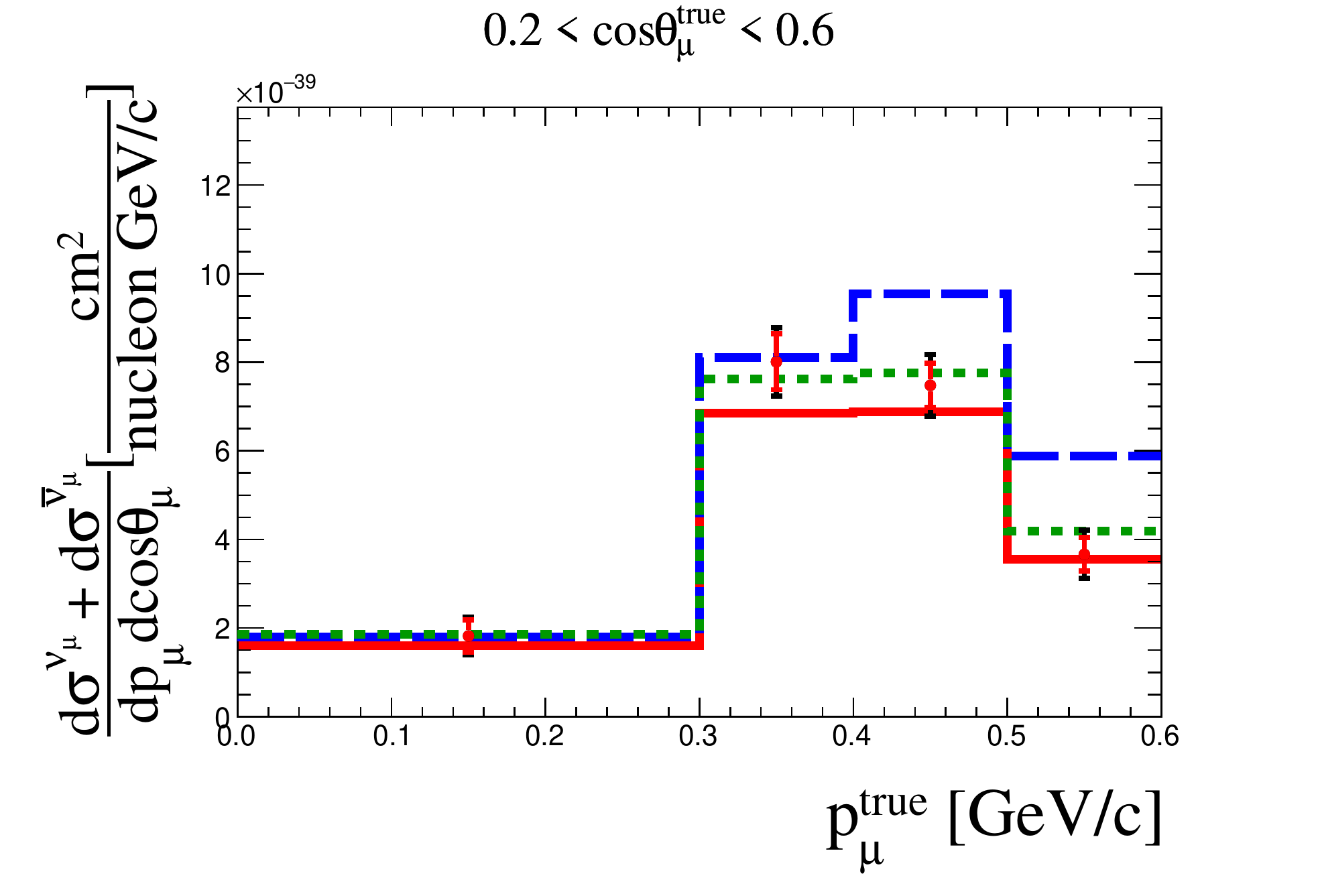}
	\includegraphics[width=0.36\linewidth]{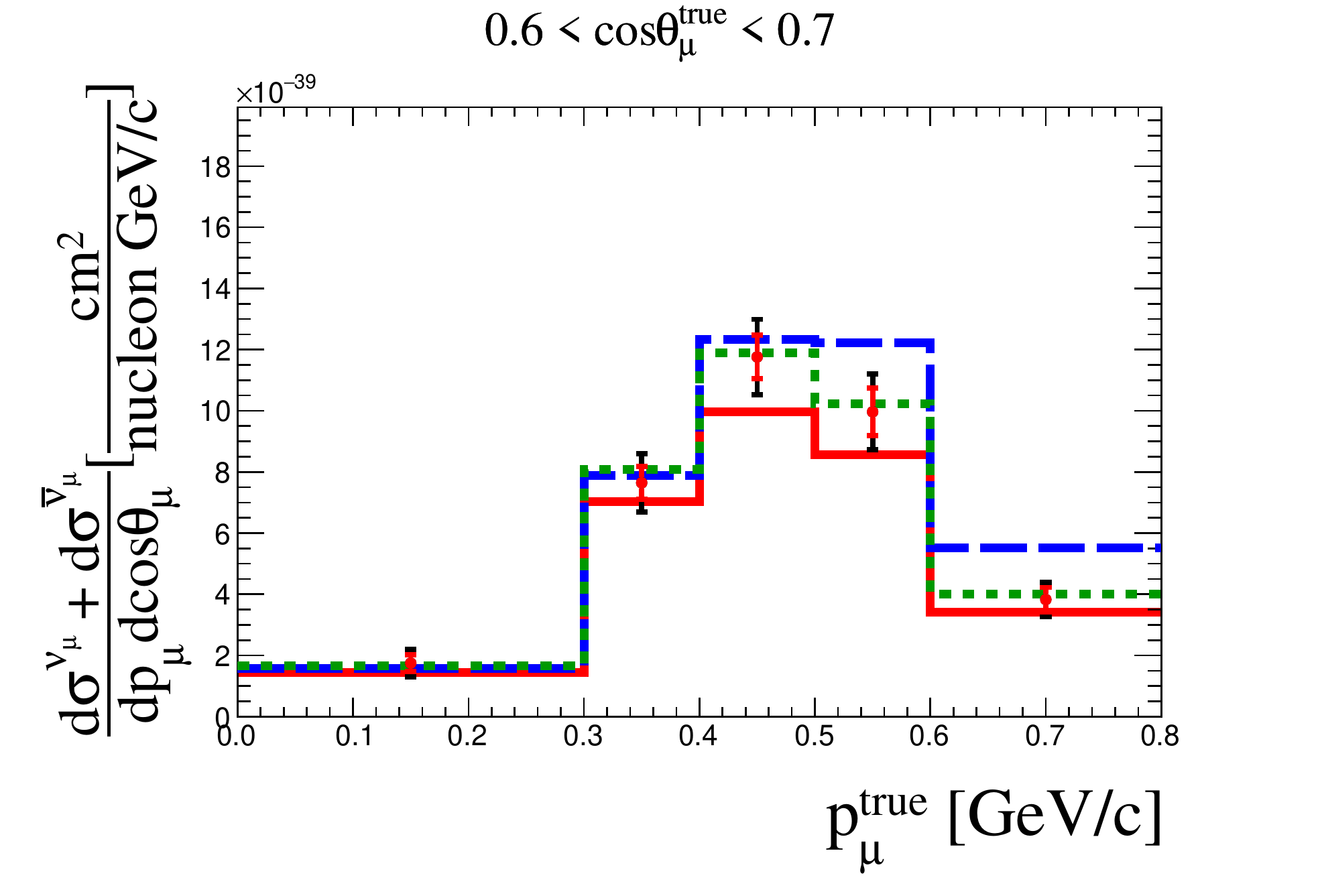}
	\includegraphics[width=0.36\linewidth]{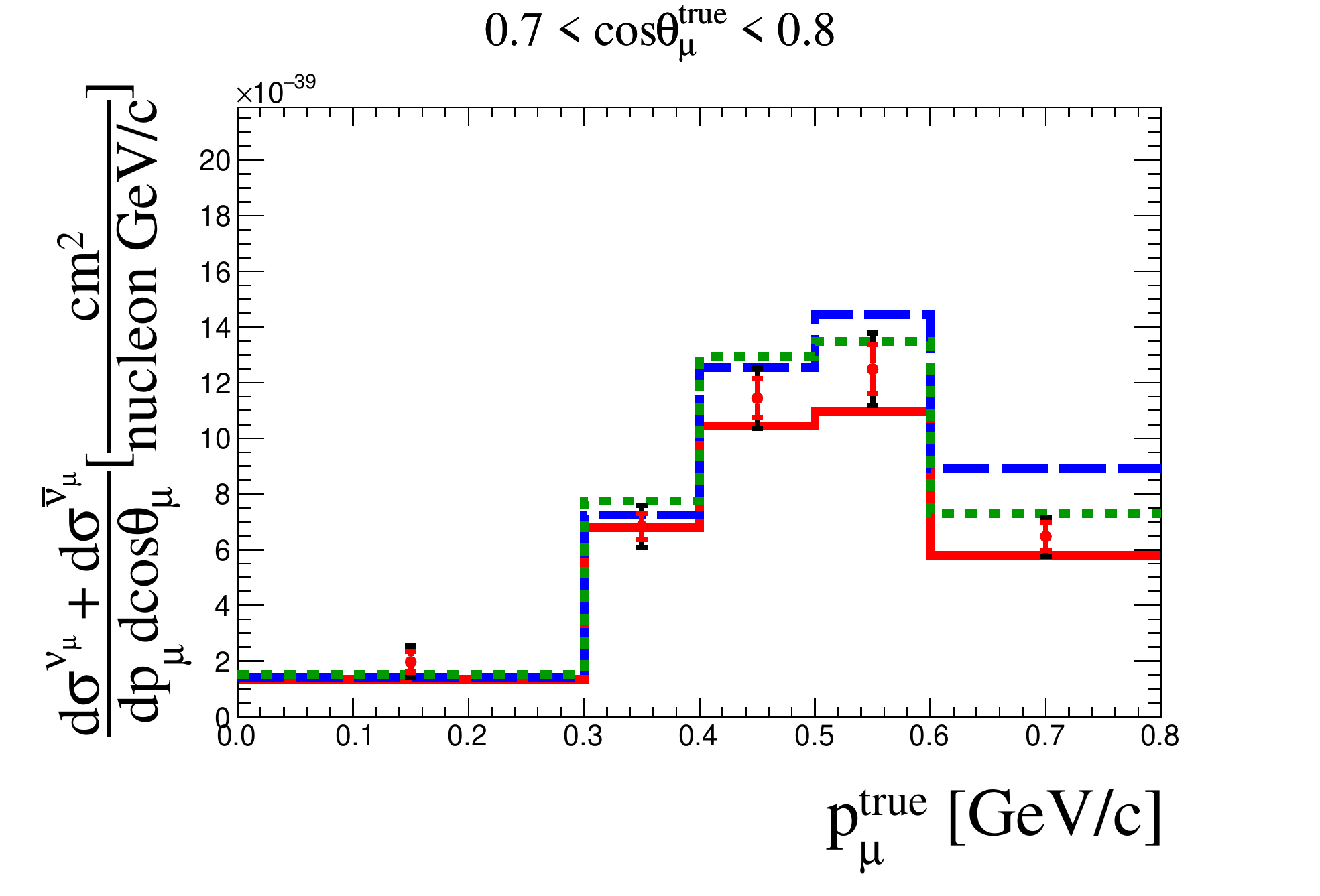}
	\includegraphics[width=0.36\linewidth]{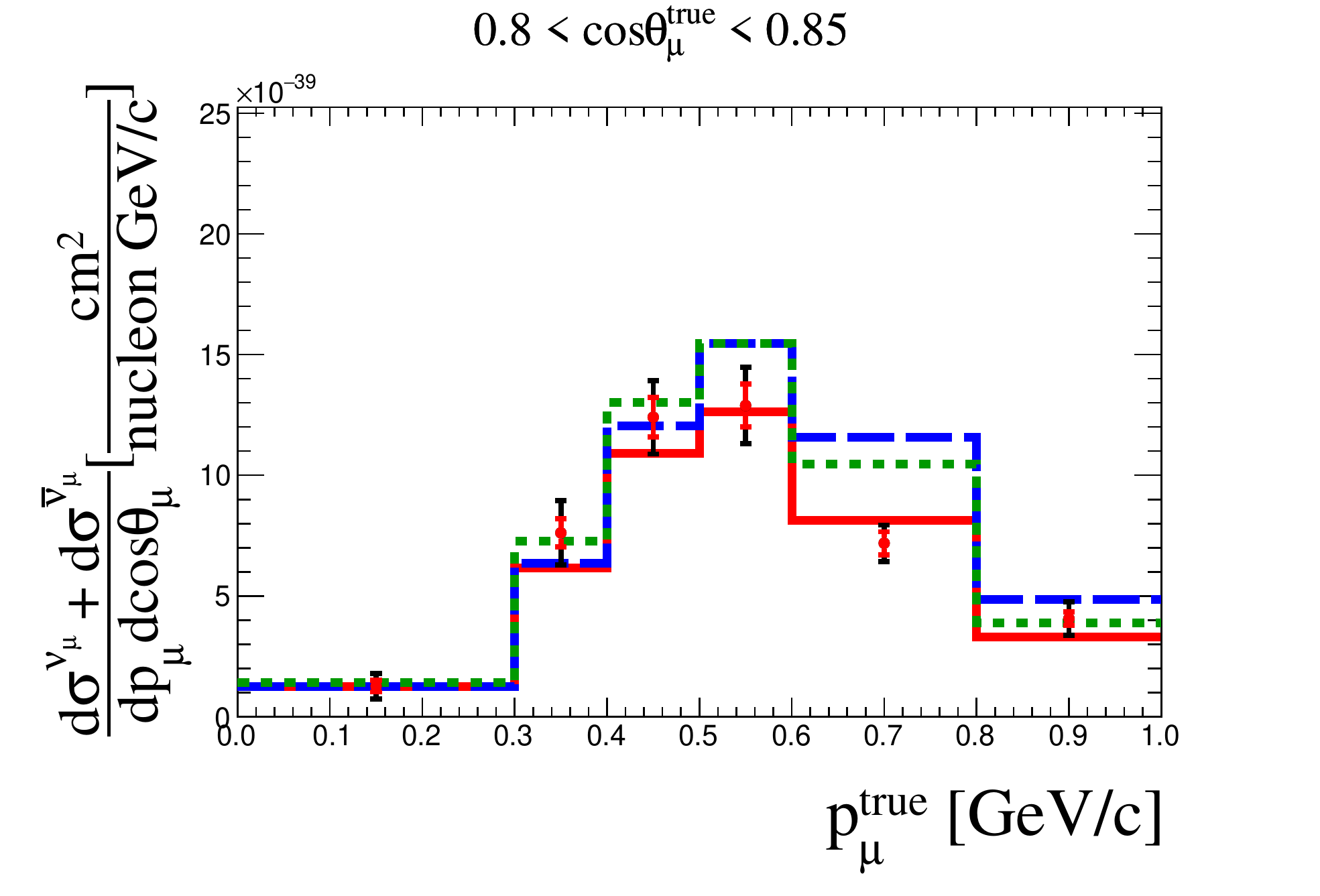}	
	\includegraphics[width=0.36\linewidth]{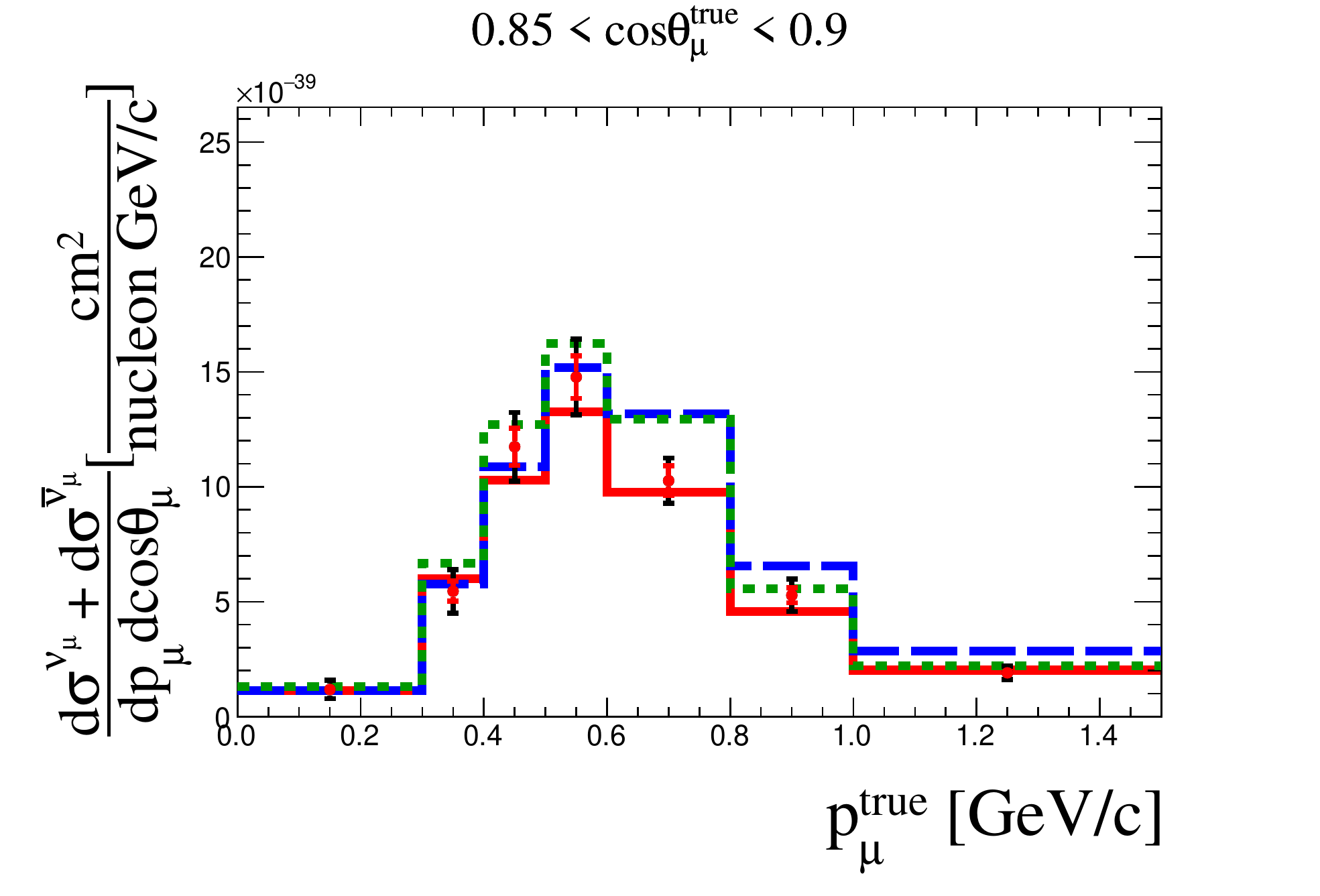}	
	\includegraphics[width=0.36\linewidth]{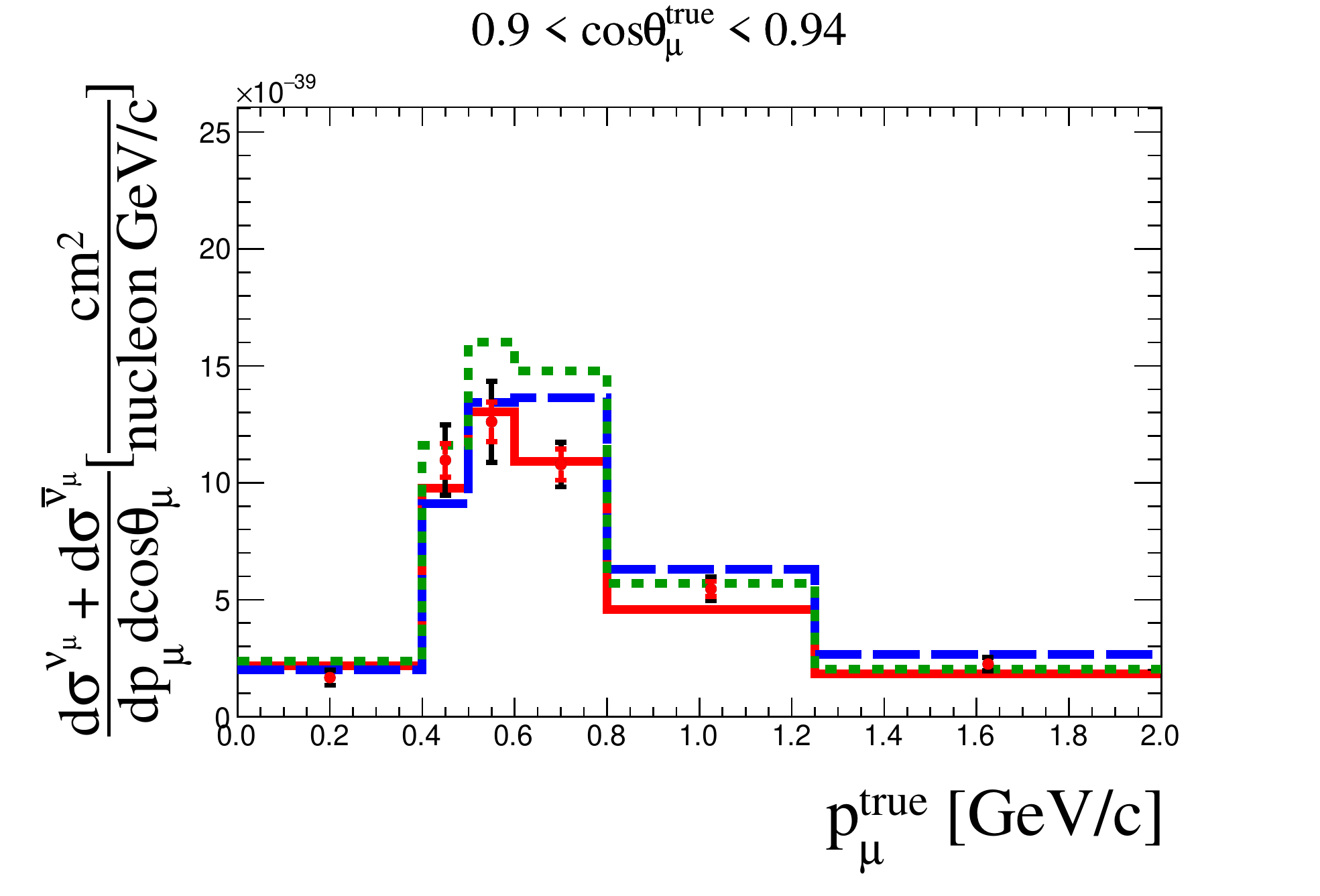}	
	\includegraphics[width=0.36\linewidth]{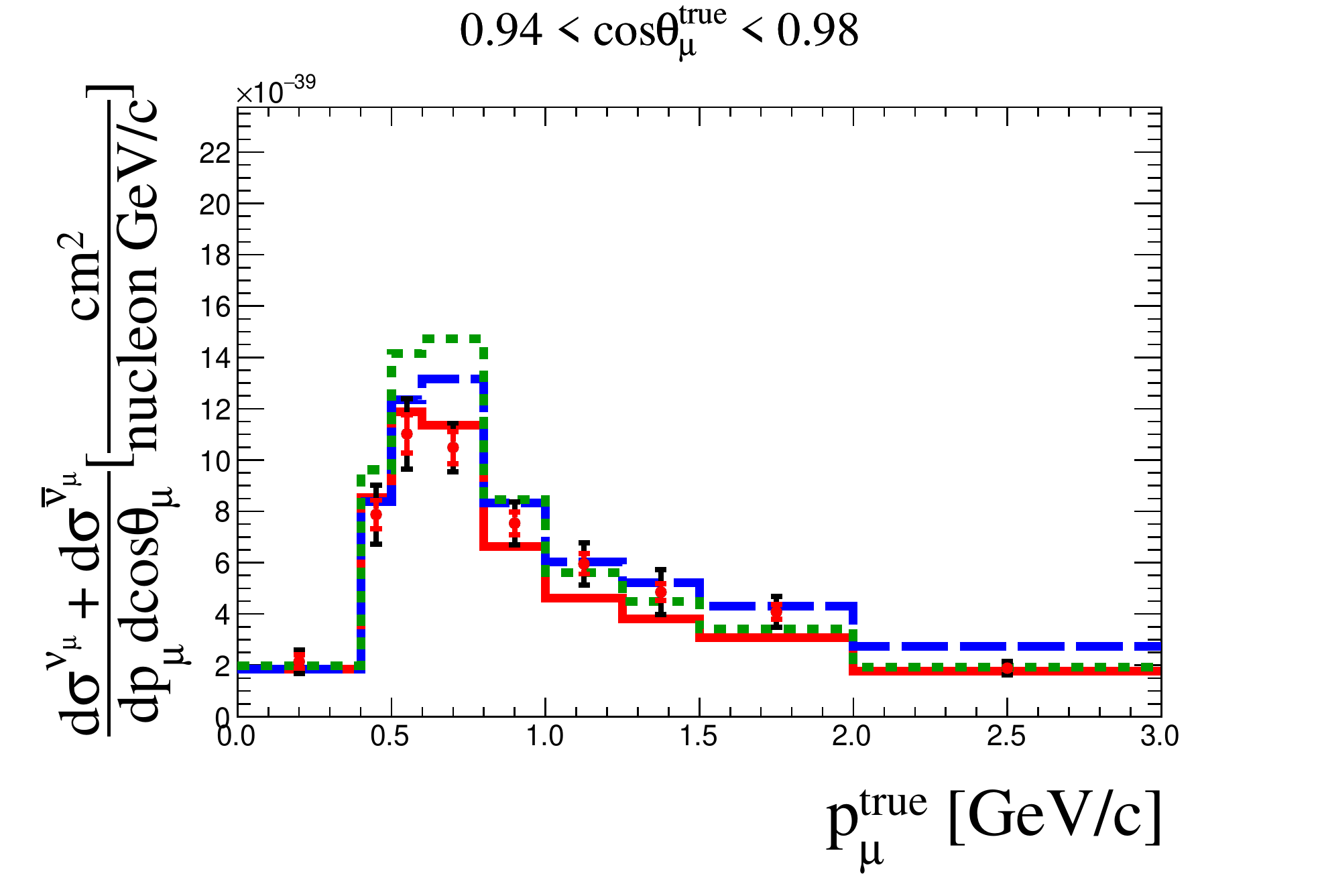}	
	\includegraphics[width=0.36\linewidth]{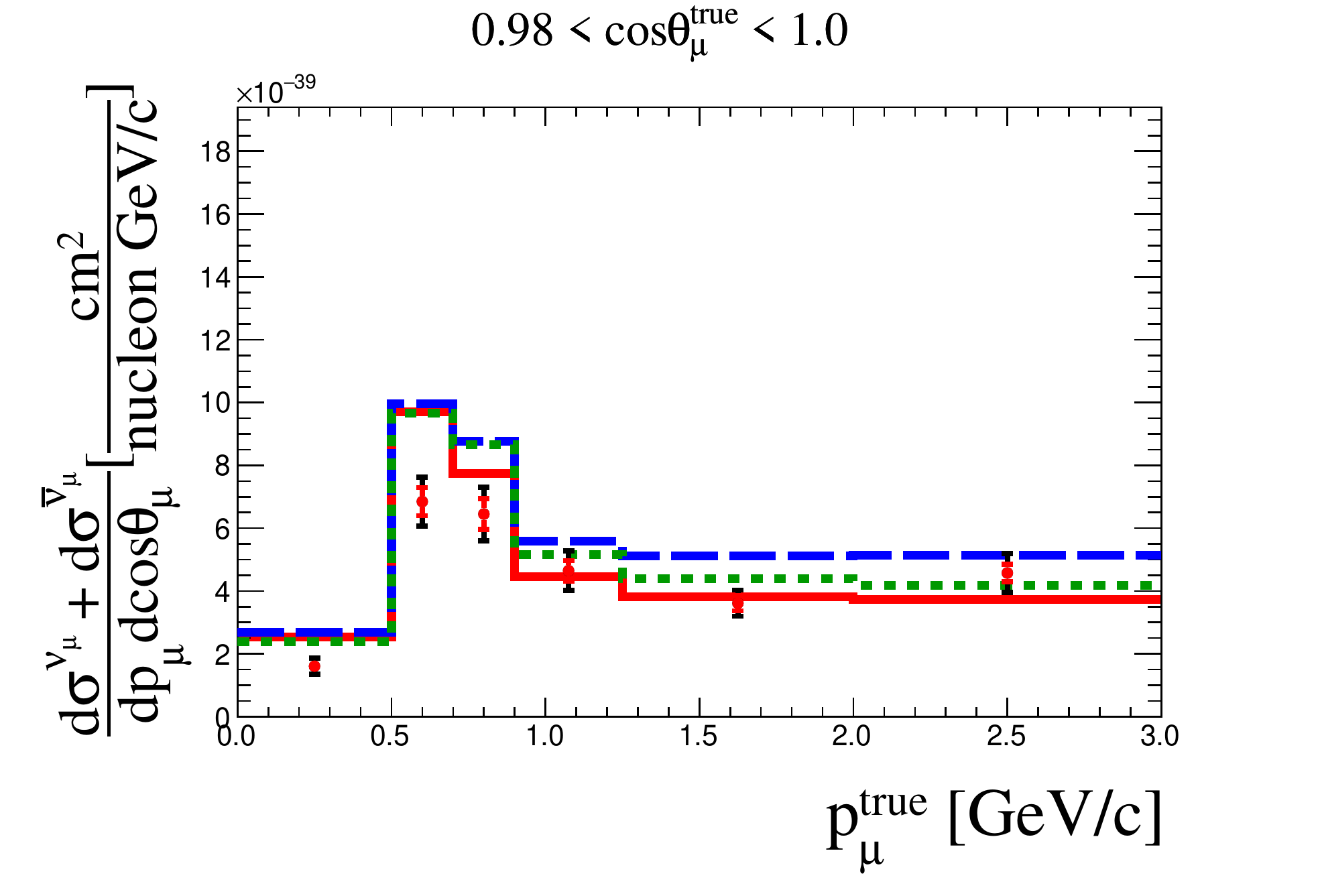}	
	\includegraphics[width=0.36\linewidth]{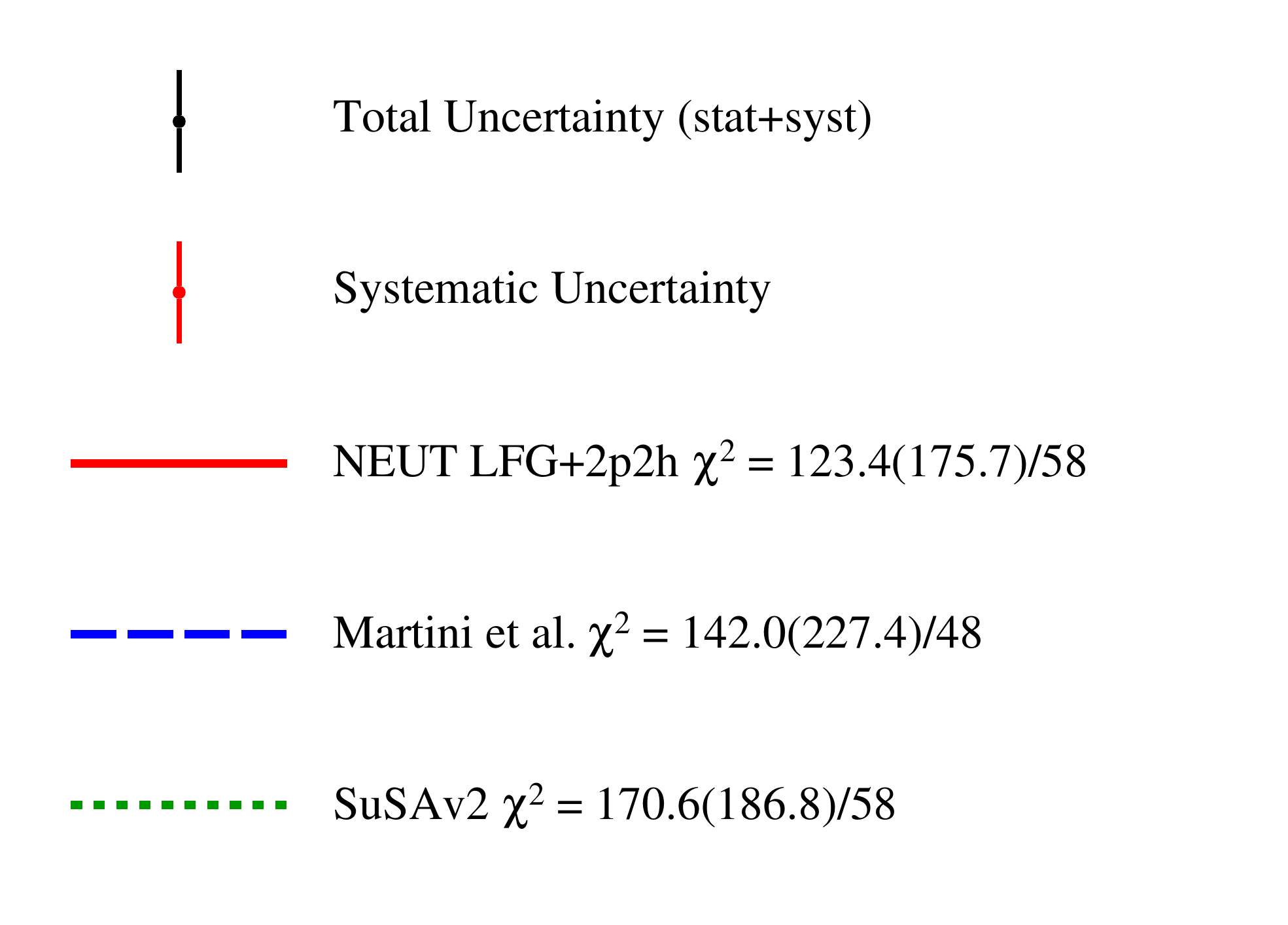}		
	\caption{Measured double-differential \numu + \barnumu \cczeropi cross-section sum in bins of true muon kinematics with systematic uncertainty (red bars) and total (stat.+syst.) uncertainty (black bars). The results are compared to \textsc{Neut} version~\texttt{5.4.1}, which uses an LFG+RPA model with 2p2h (solid red line), Martini \textit{et al.} (dashed blue line) and \textsc{SuSAv2} (green dashed line) models. The full and shape-only (in parenthesis) $\chi2$ are reported. The last bin in momentum is not displayed for readability.}
	\label{fig:sumnumucc0piNMS}
\end{figure*}

\begin{figure*}[h!]
	\centering
	\includegraphics[width=0.36\linewidth]{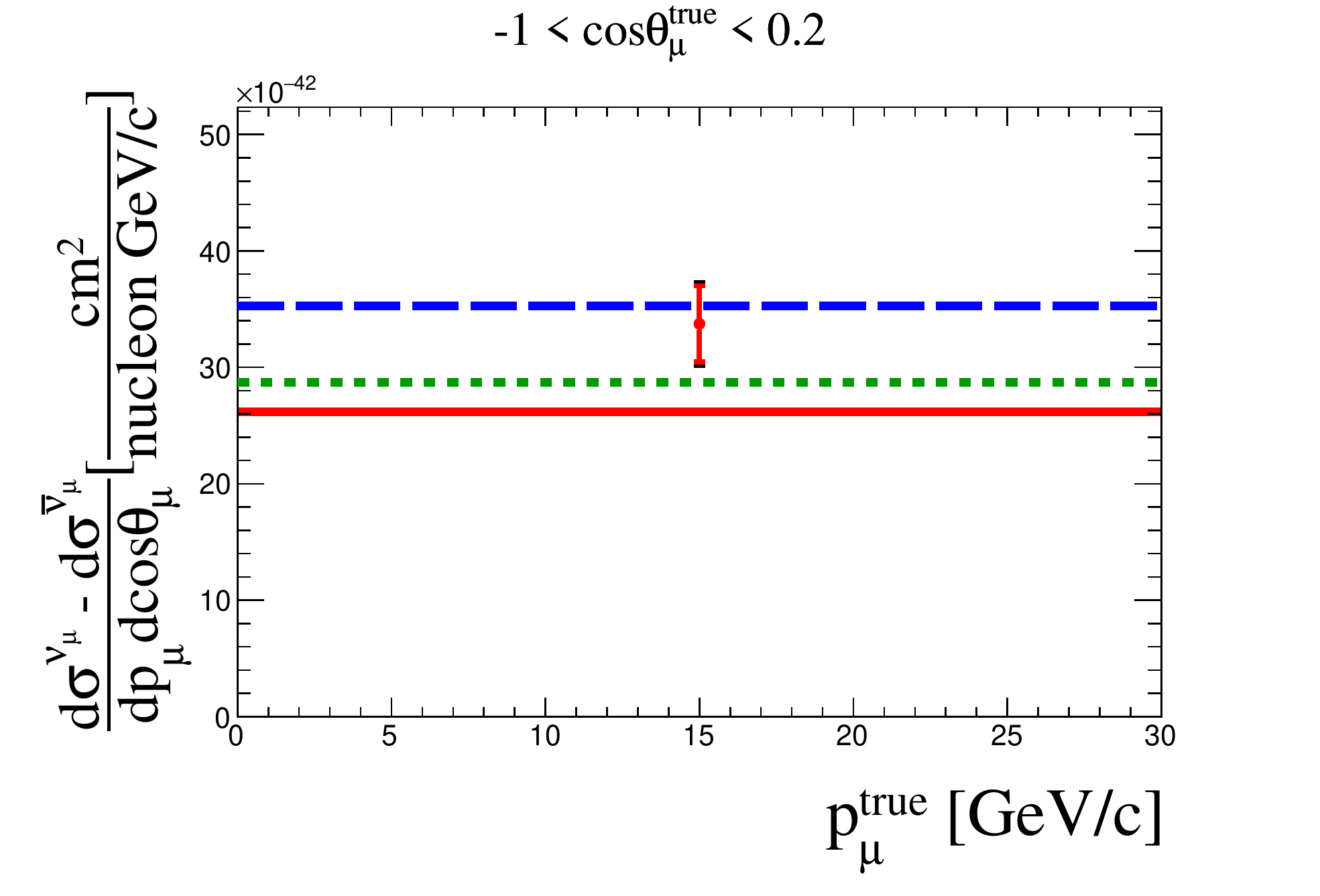}
	\includegraphics[width=0.36\linewidth]{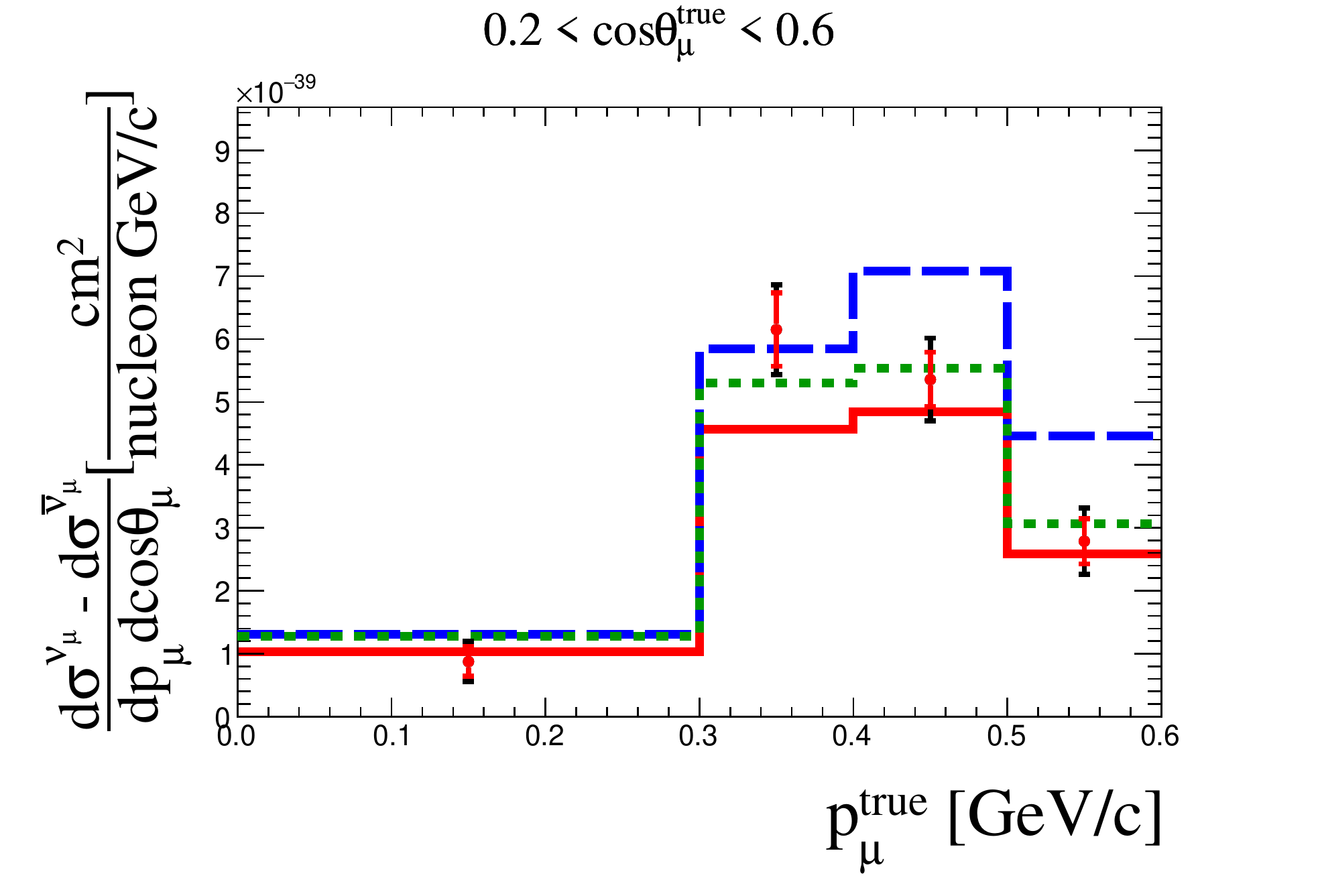}
	\includegraphics[width=0.36\linewidth]{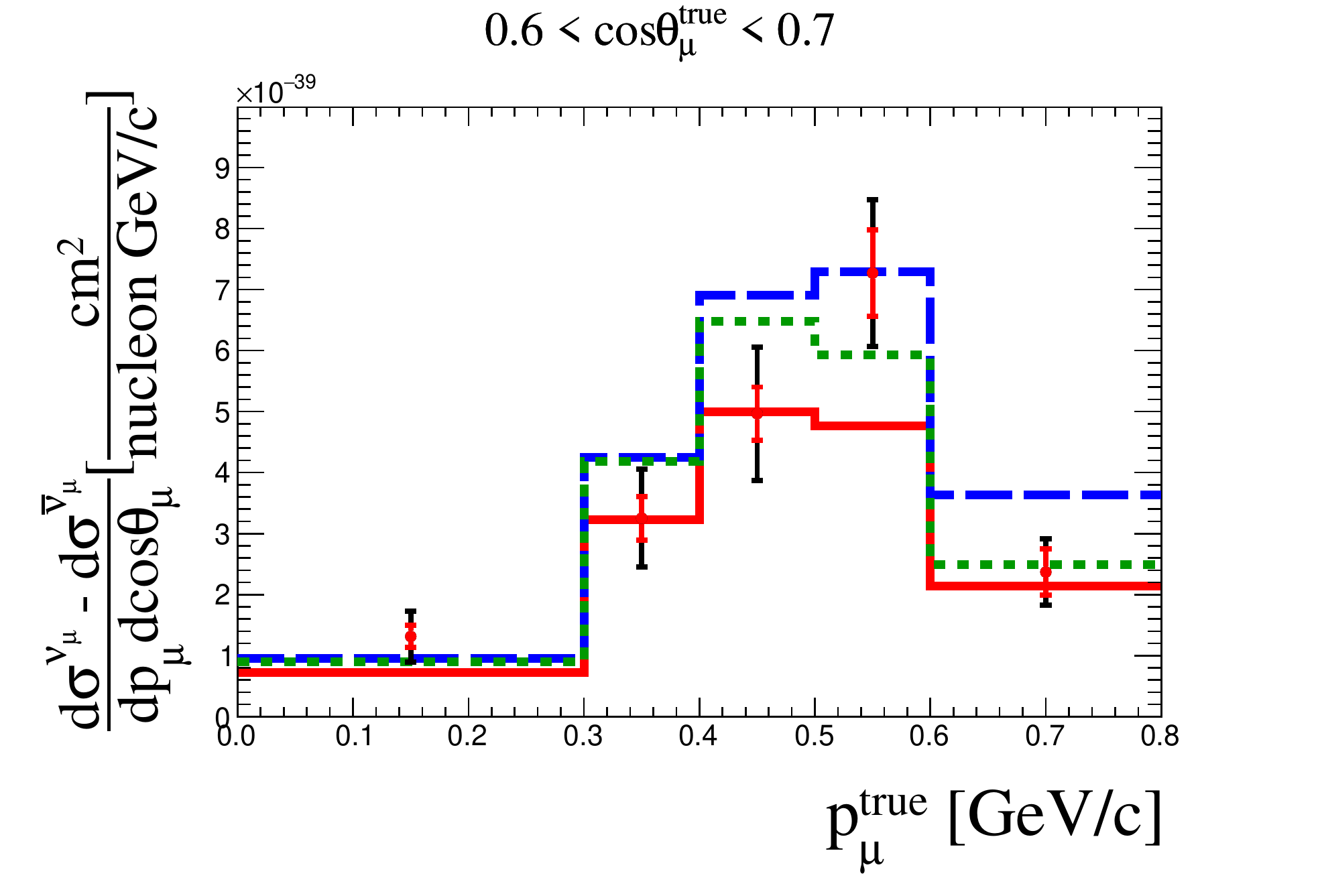}
	\includegraphics[width=0.36\linewidth]{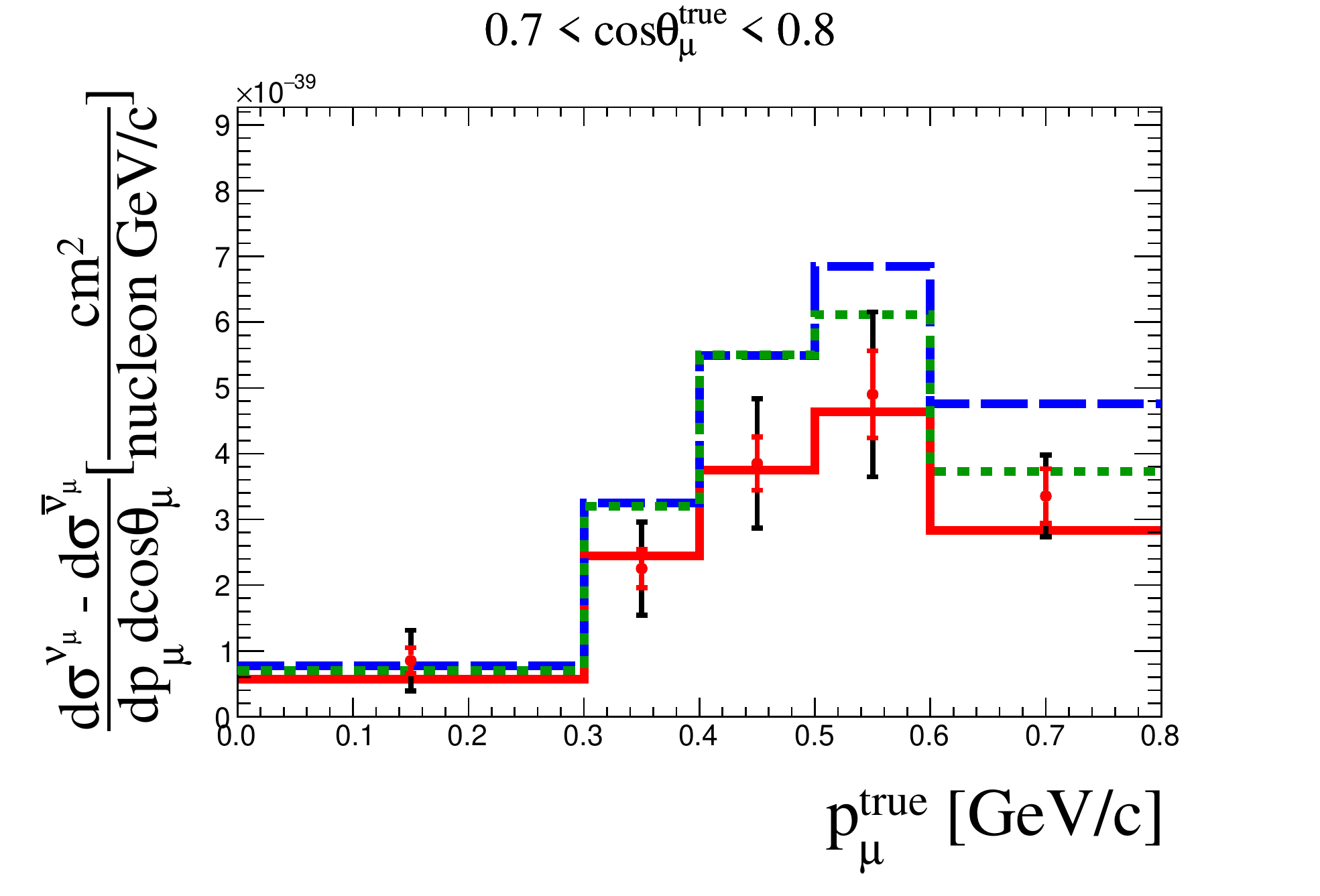}
	\includegraphics[width=0.36\linewidth]{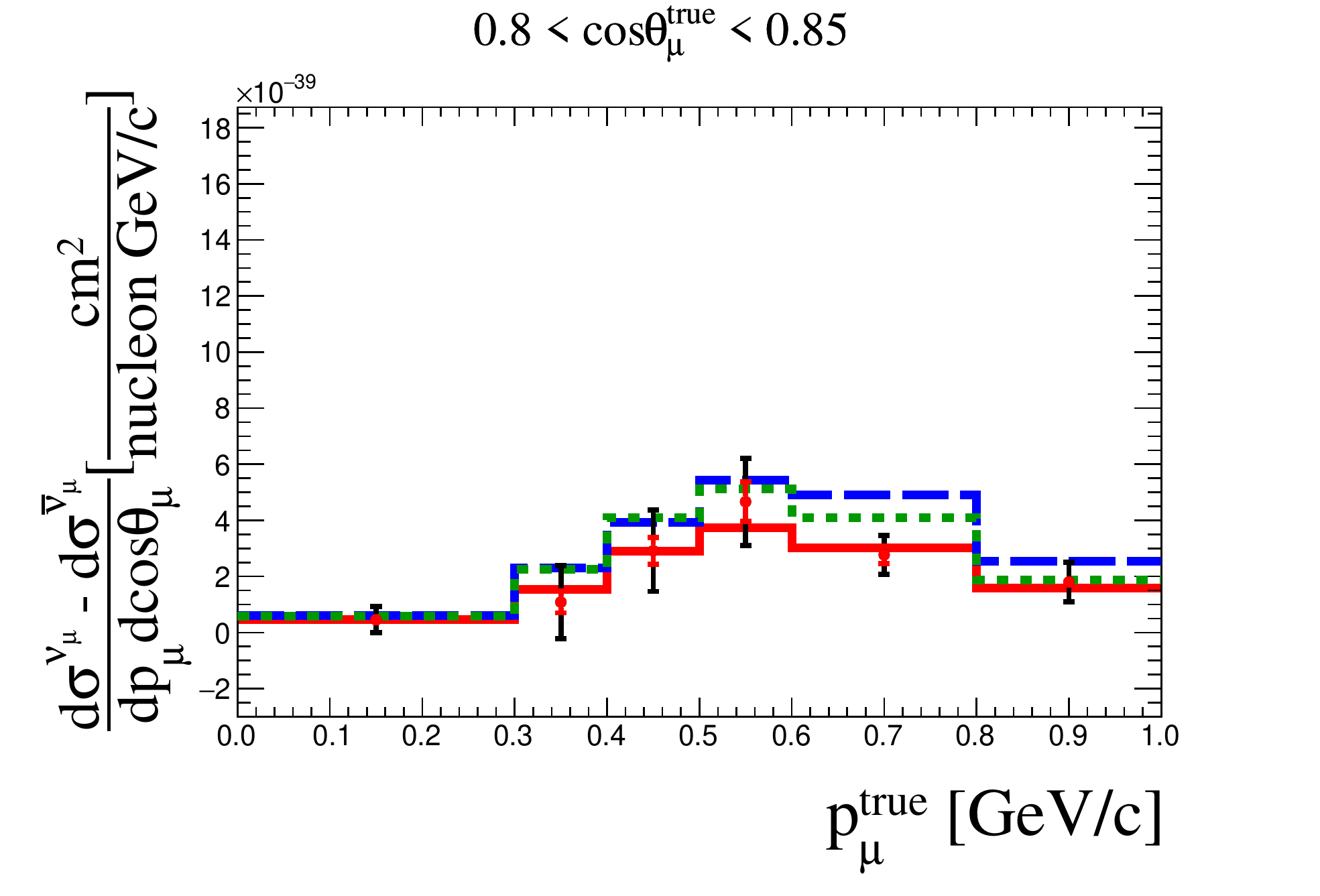}	
	\includegraphics[width=0.36\linewidth]{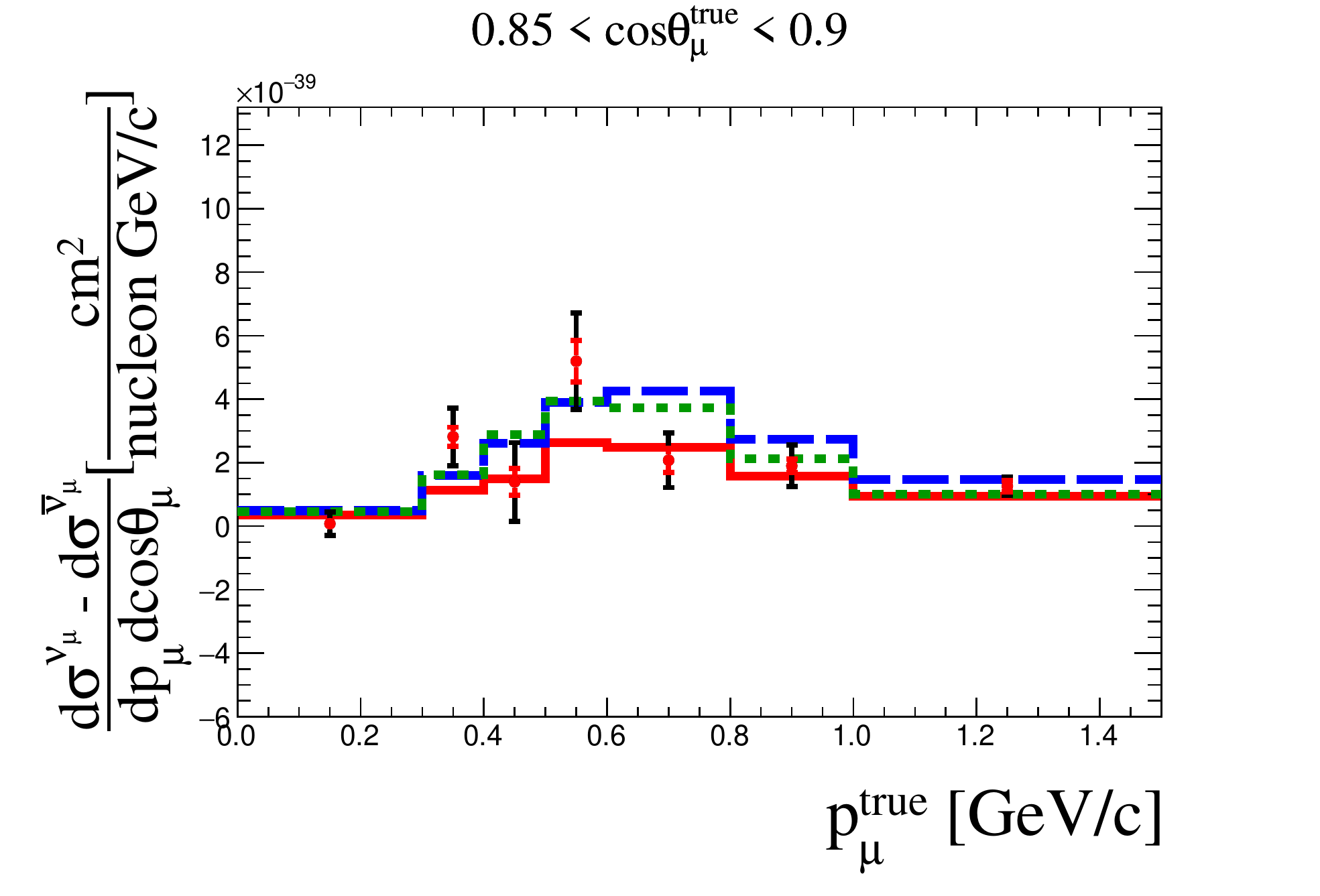}	
	\includegraphics[width=0.36\linewidth]{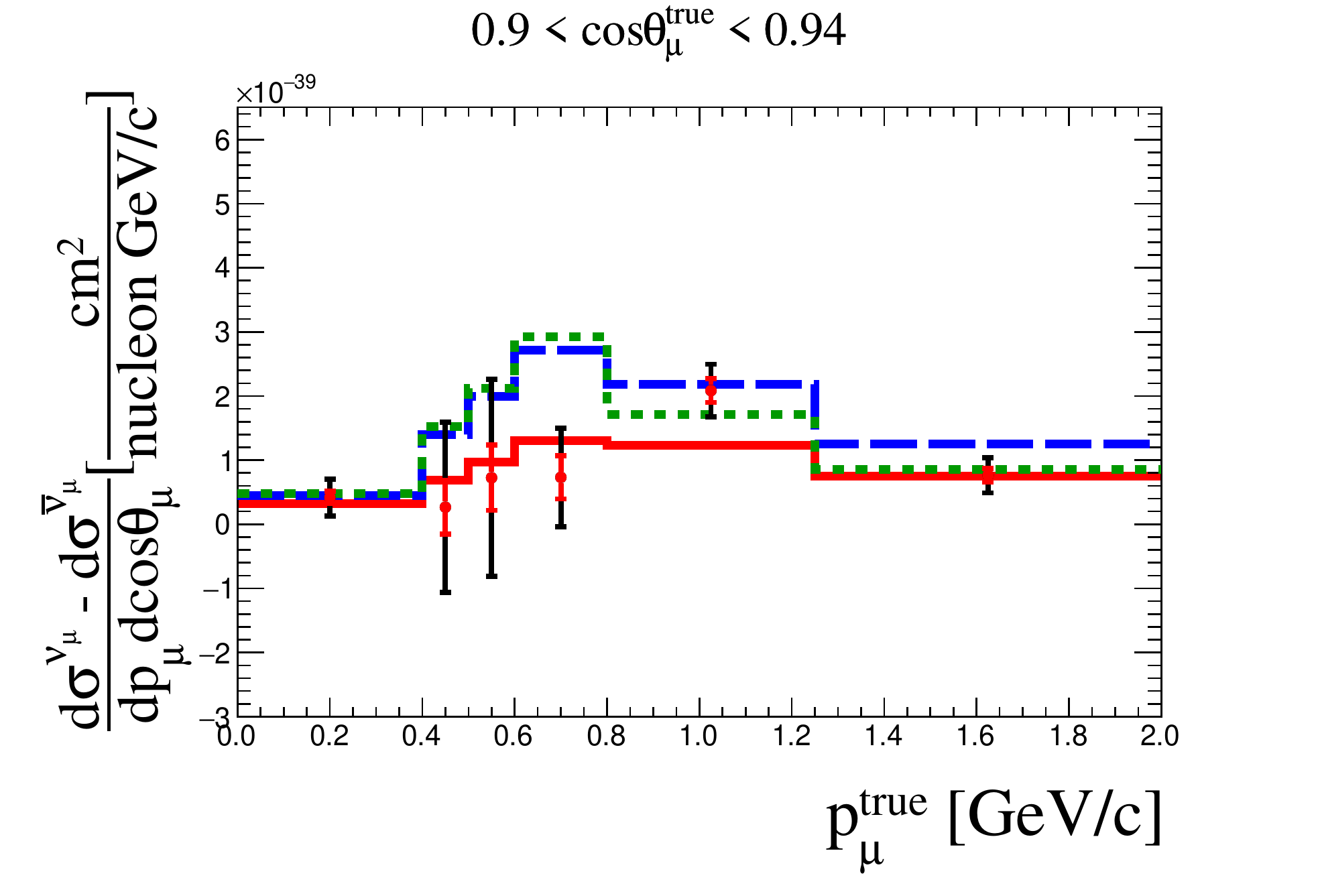}	
	\includegraphics[width=0.36\linewidth]{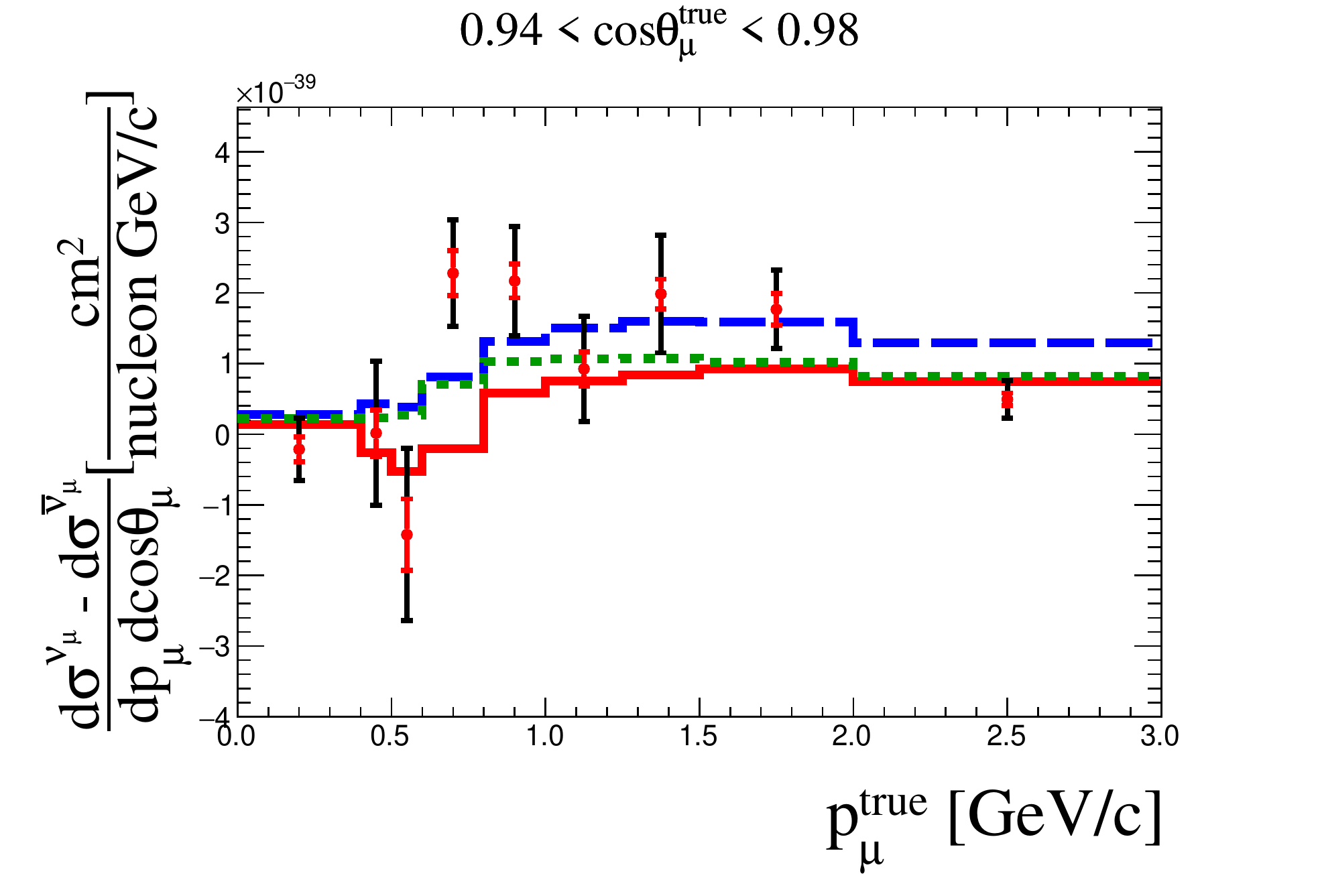}	
	\includegraphics[width=0.36\linewidth]{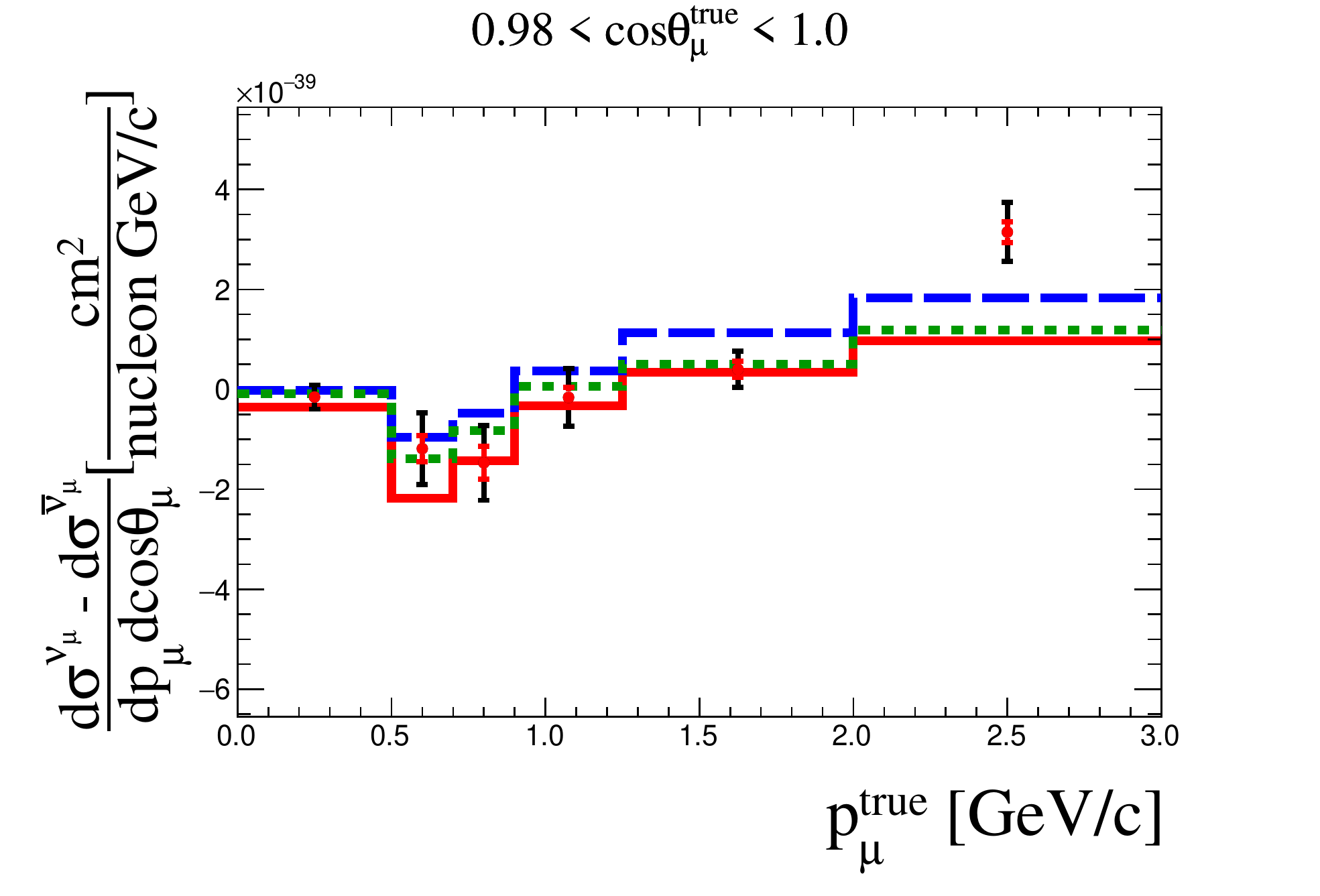}
	\includegraphics[width=0.36\linewidth]{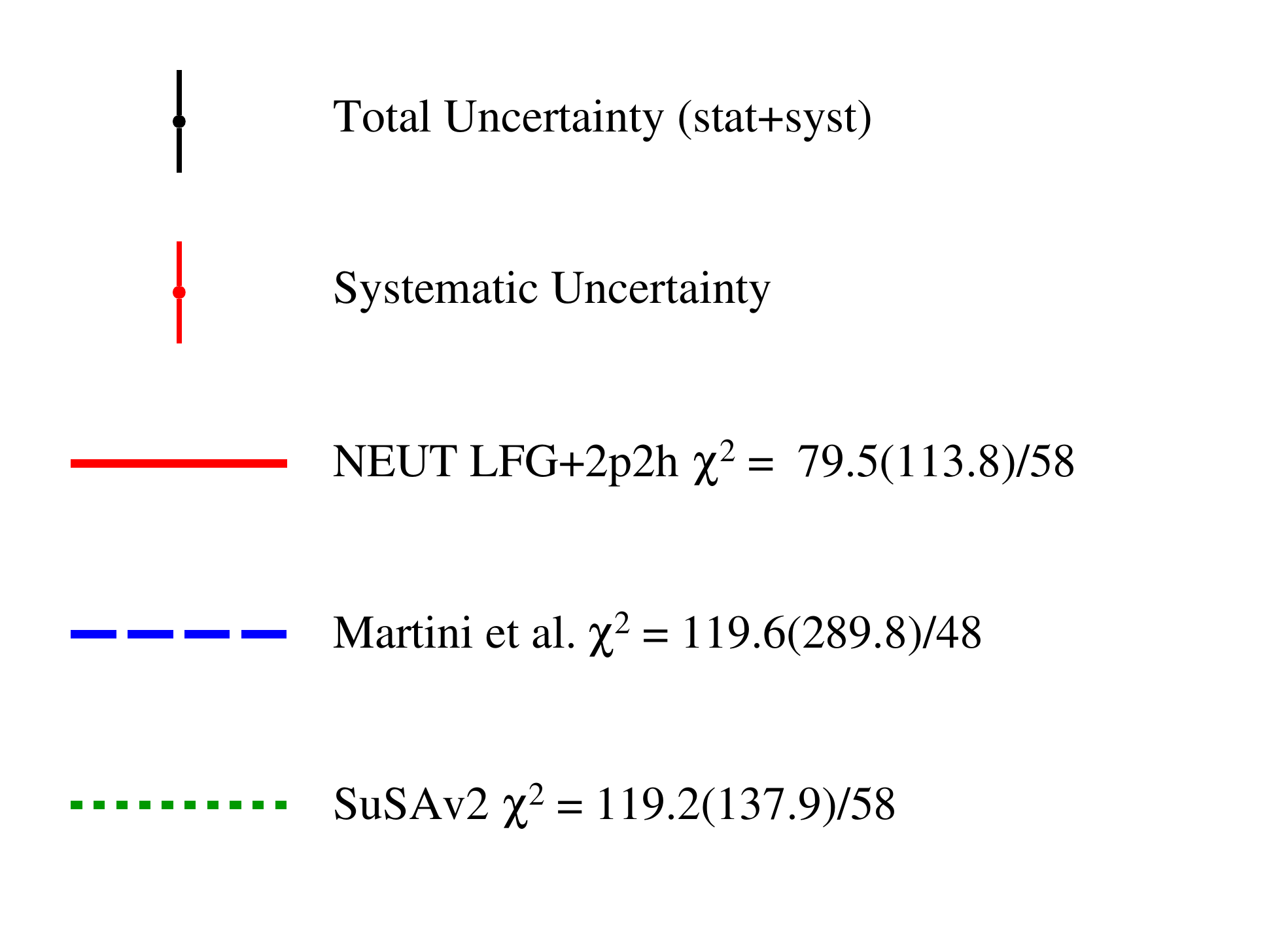}		
	\caption{Measured double-differential \numu - \barnumu \cczeropi cross-section difference in bins of true muon kinematics with systematic uncertainty (red bars) and total (stat.+syst.) uncertainty (black bars). The results are compared to \textsc{Neut} version~\texttt{5.4.1}, which uses an LFG+RPA model with 2p2h (solid red line), Martini \textit{et al.} (dashed blue line) and \textsc{SuSAv2} (green dashed line) models. The full and shape-only (in parenthesis) $\chi2$ are reported. The last bin in momentum is not displayed for readability.}
	\label{fig:difnumucc0piNMS}
\end{figure*}

\begin{figure*}[h!]
	\centering
	\includegraphics[width=0.36\linewidth]{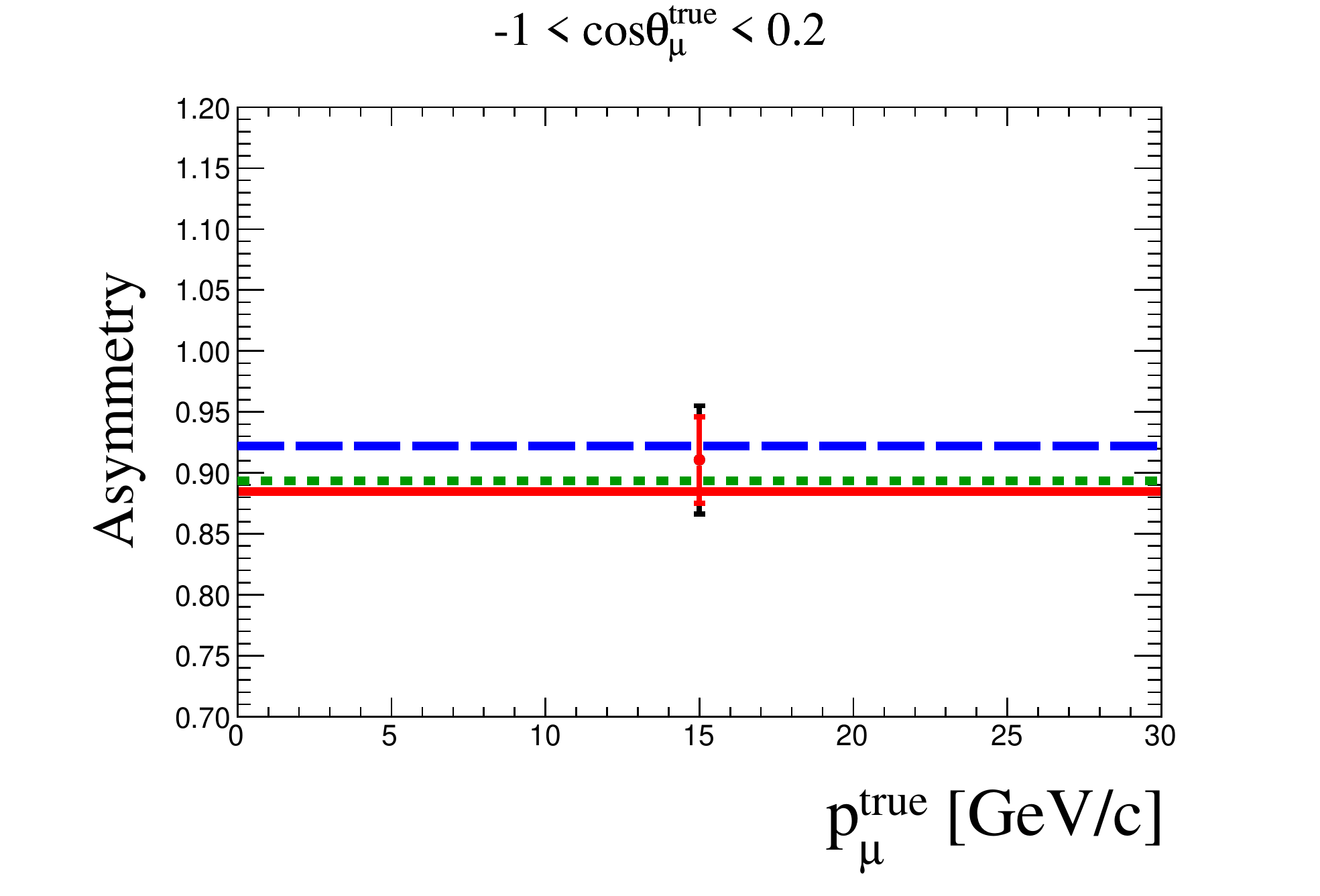}
	\includegraphics[width=0.36\linewidth]{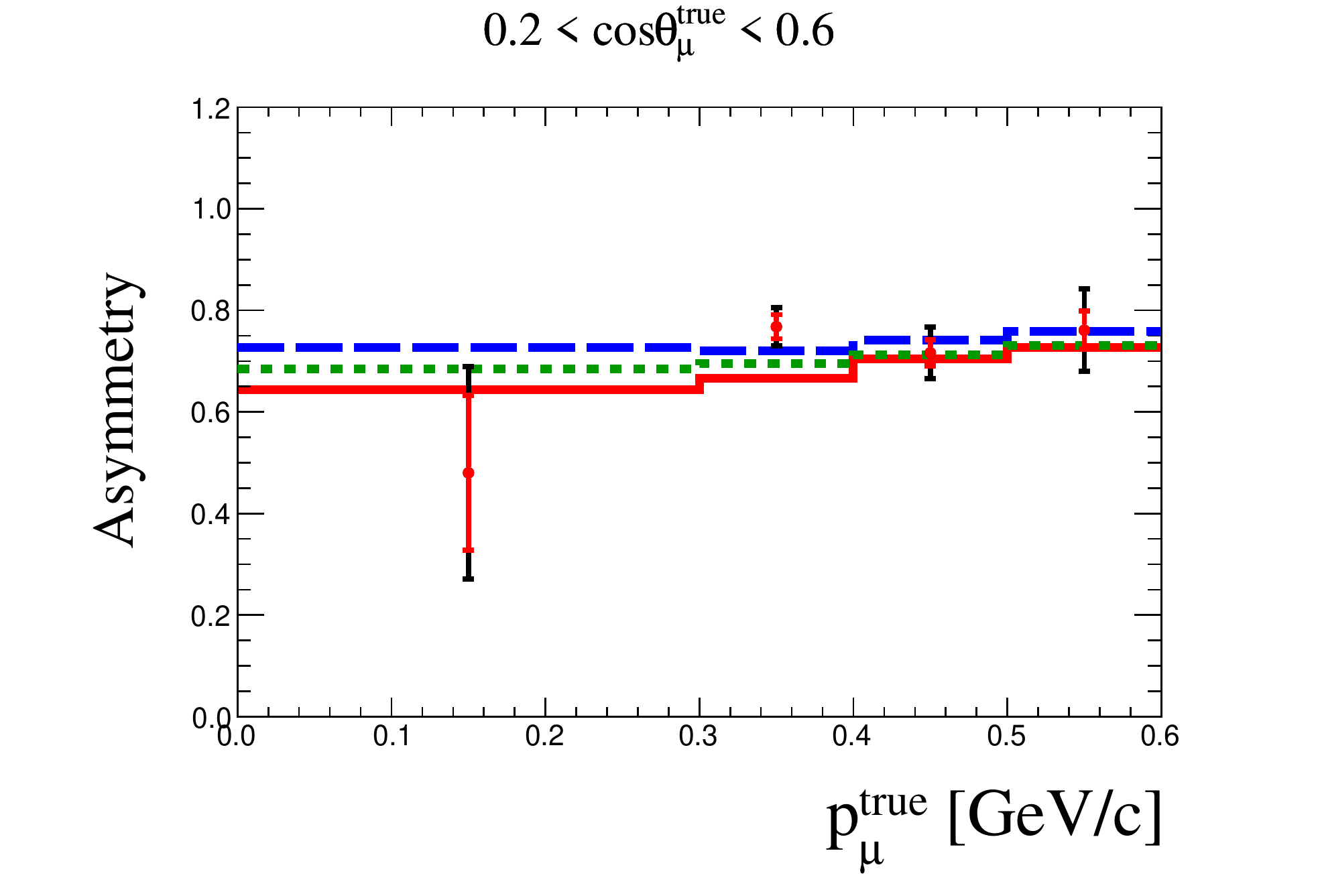}
	\includegraphics[width=0.36\linewidth]{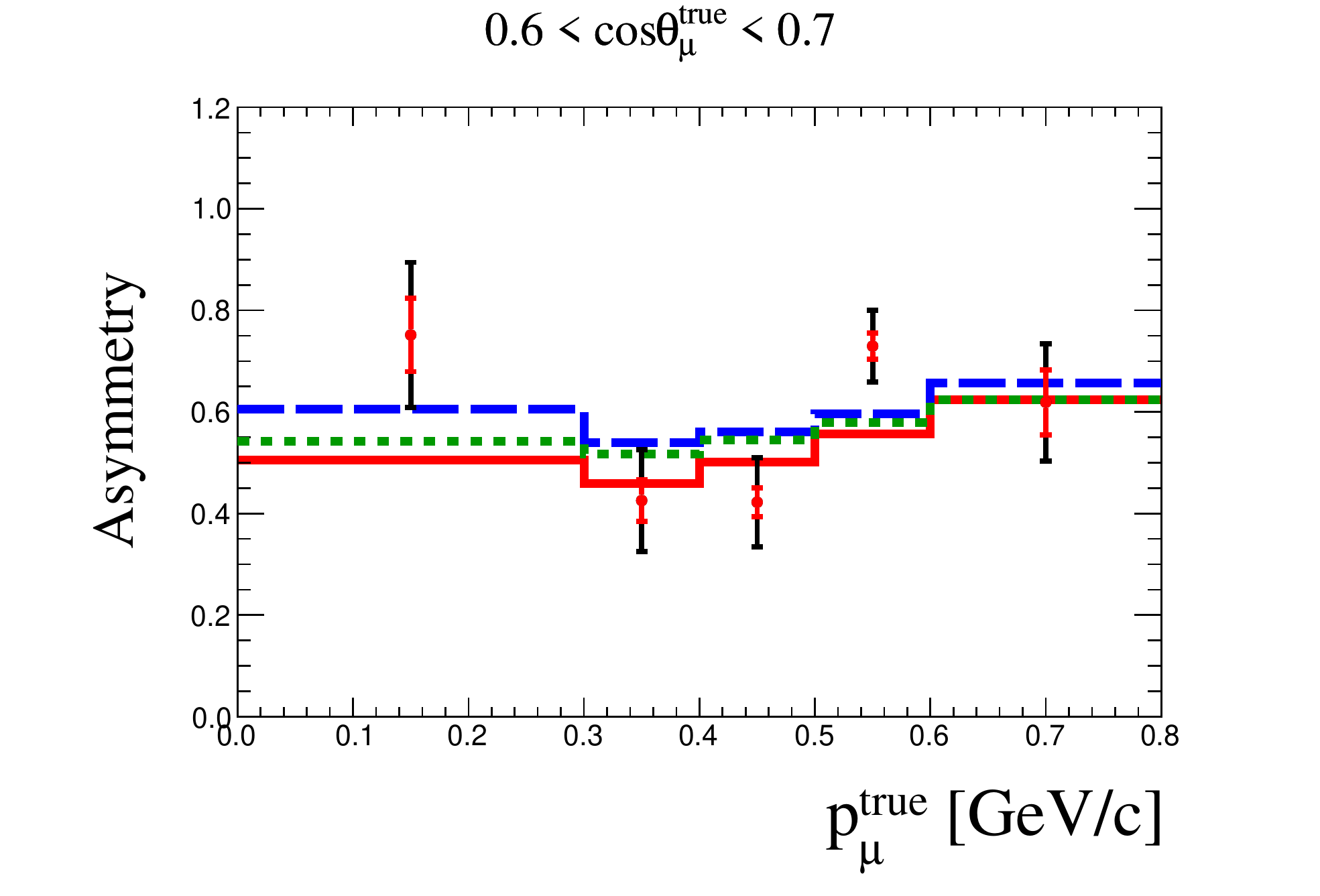}
	\includegraphics[width=0.36\linewidth]{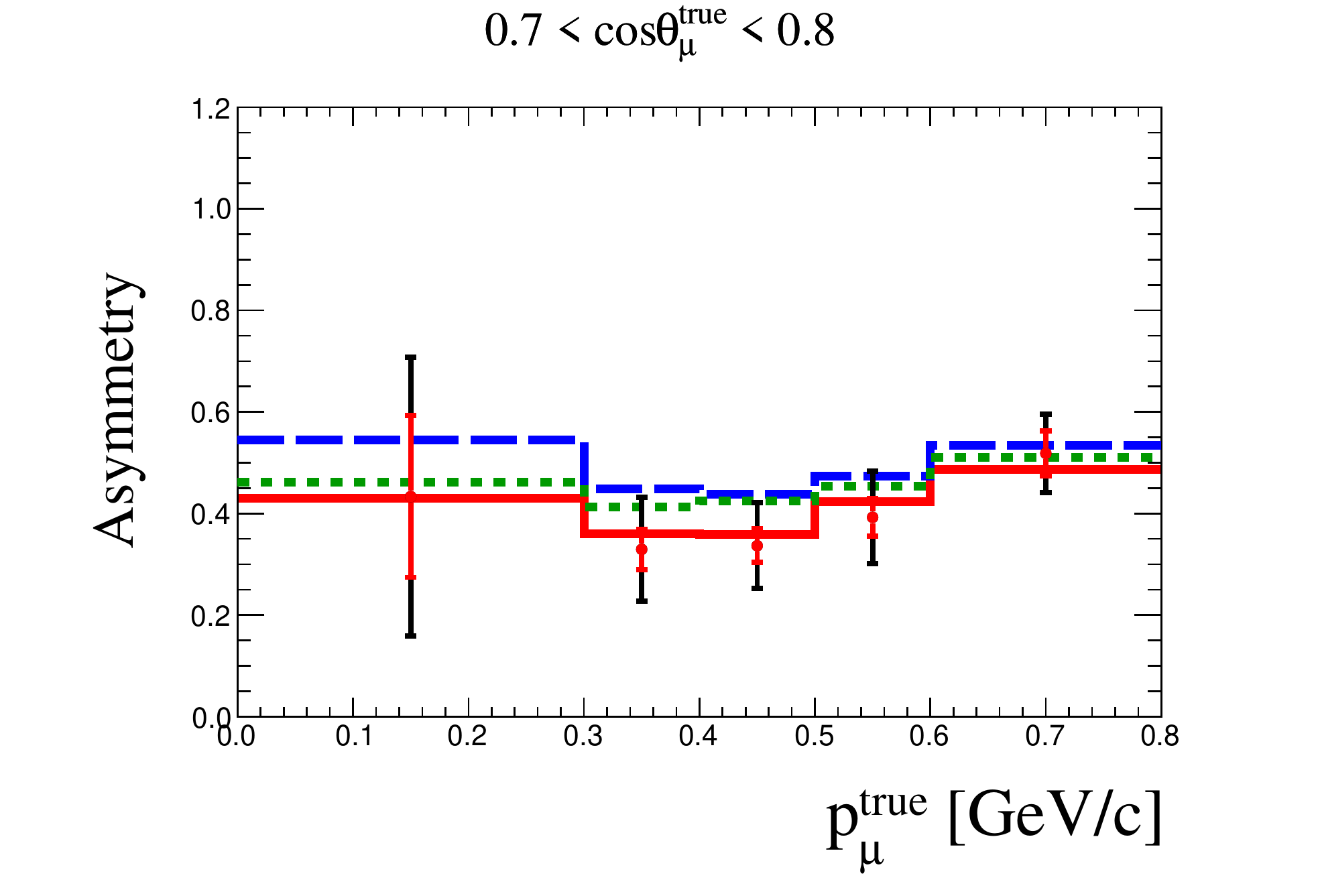}
	\includegraphics[width=0.36\linewidth]{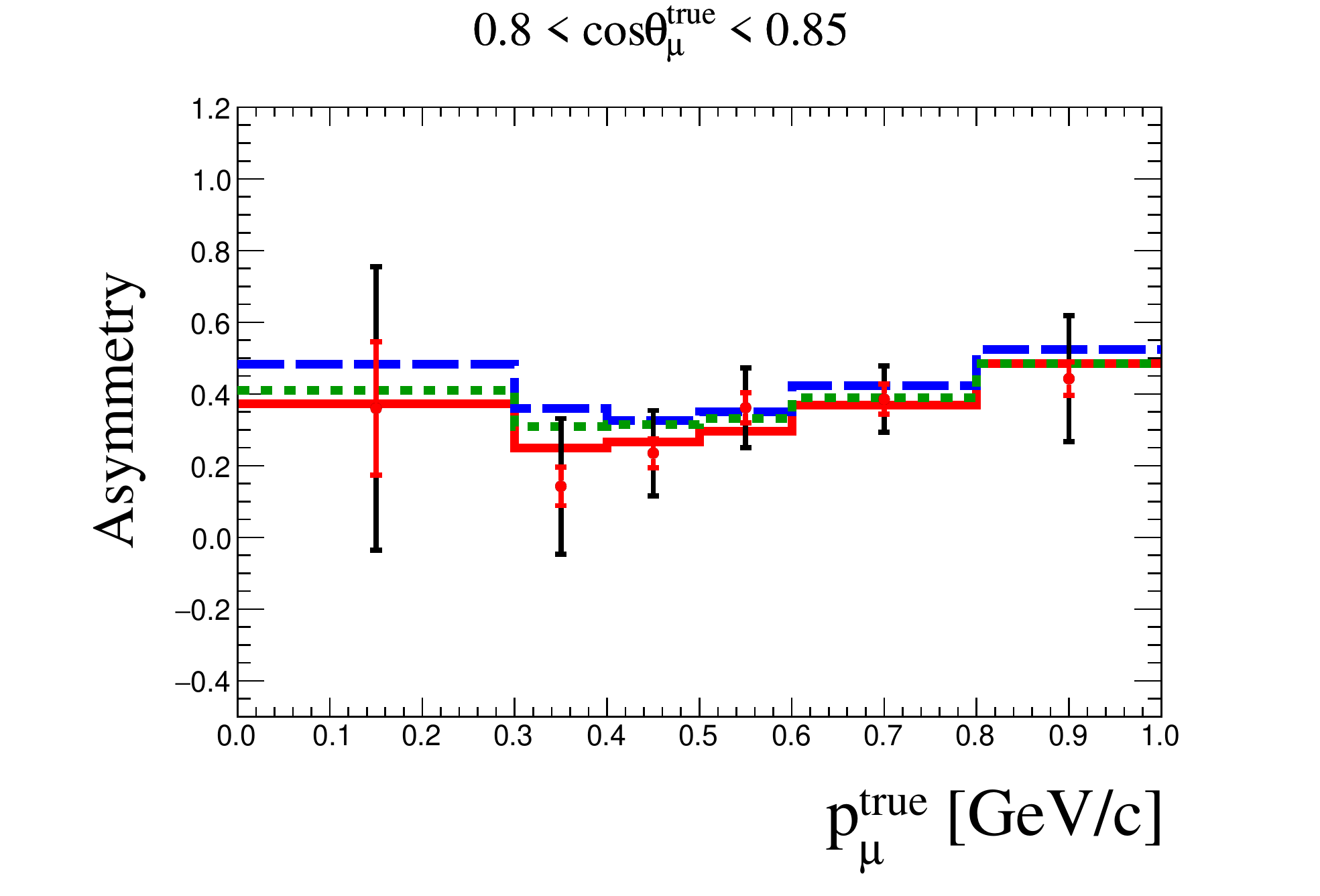}	
	\includegraphics[width=0.36\linewidth]{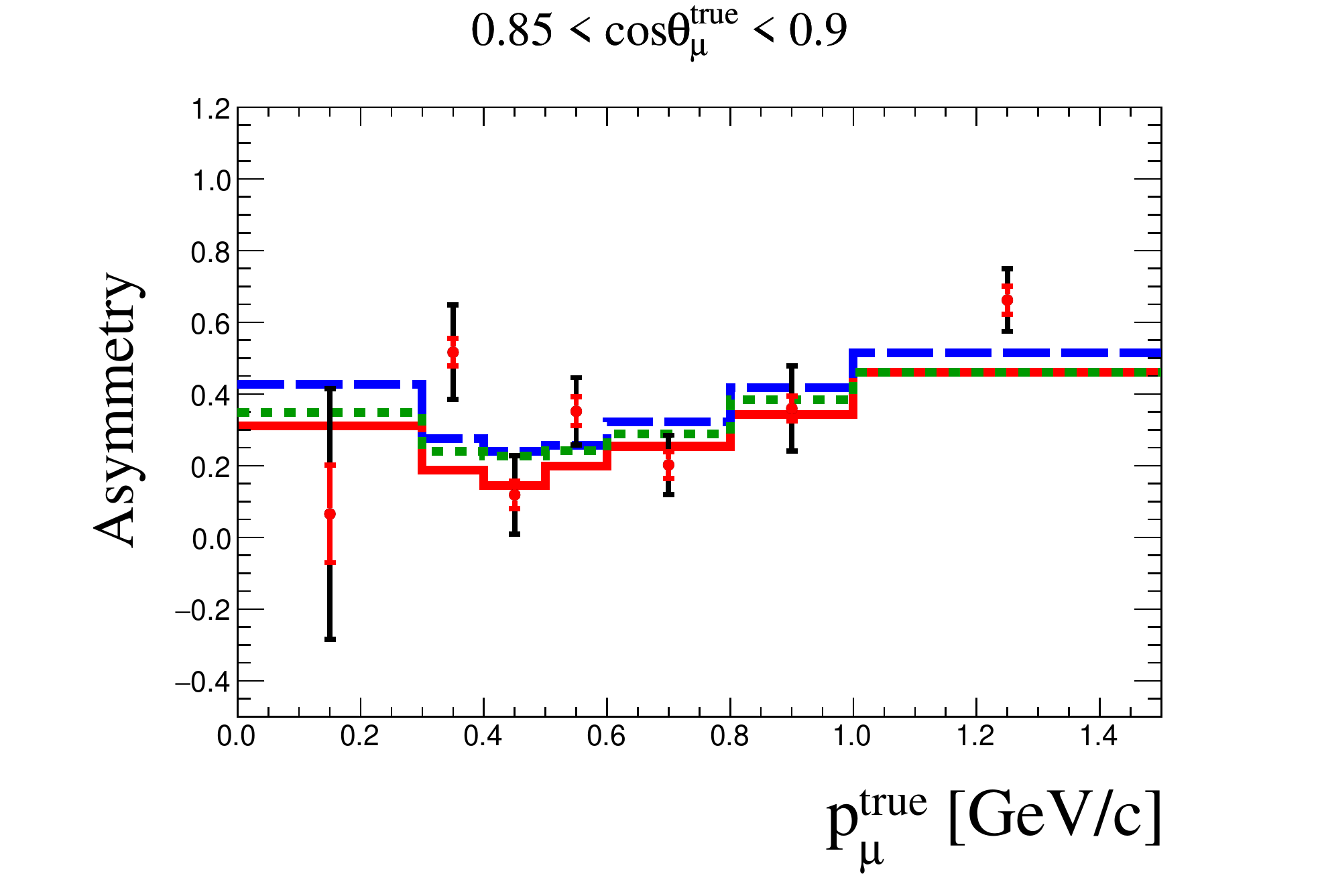}
	\includegraphics[width=0.36\linewidth]{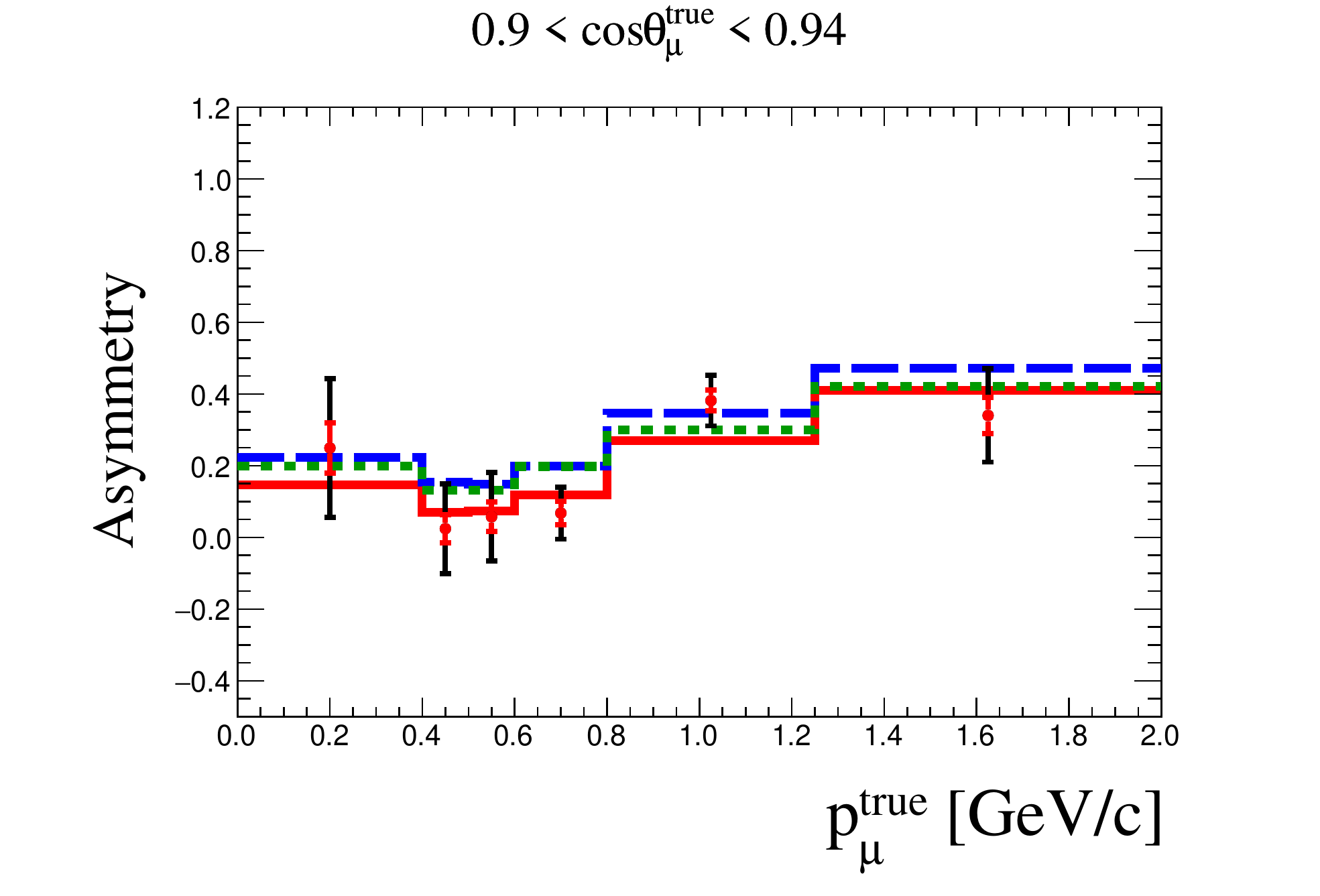}	
	\includegraphics[width=0.36\linewidth]{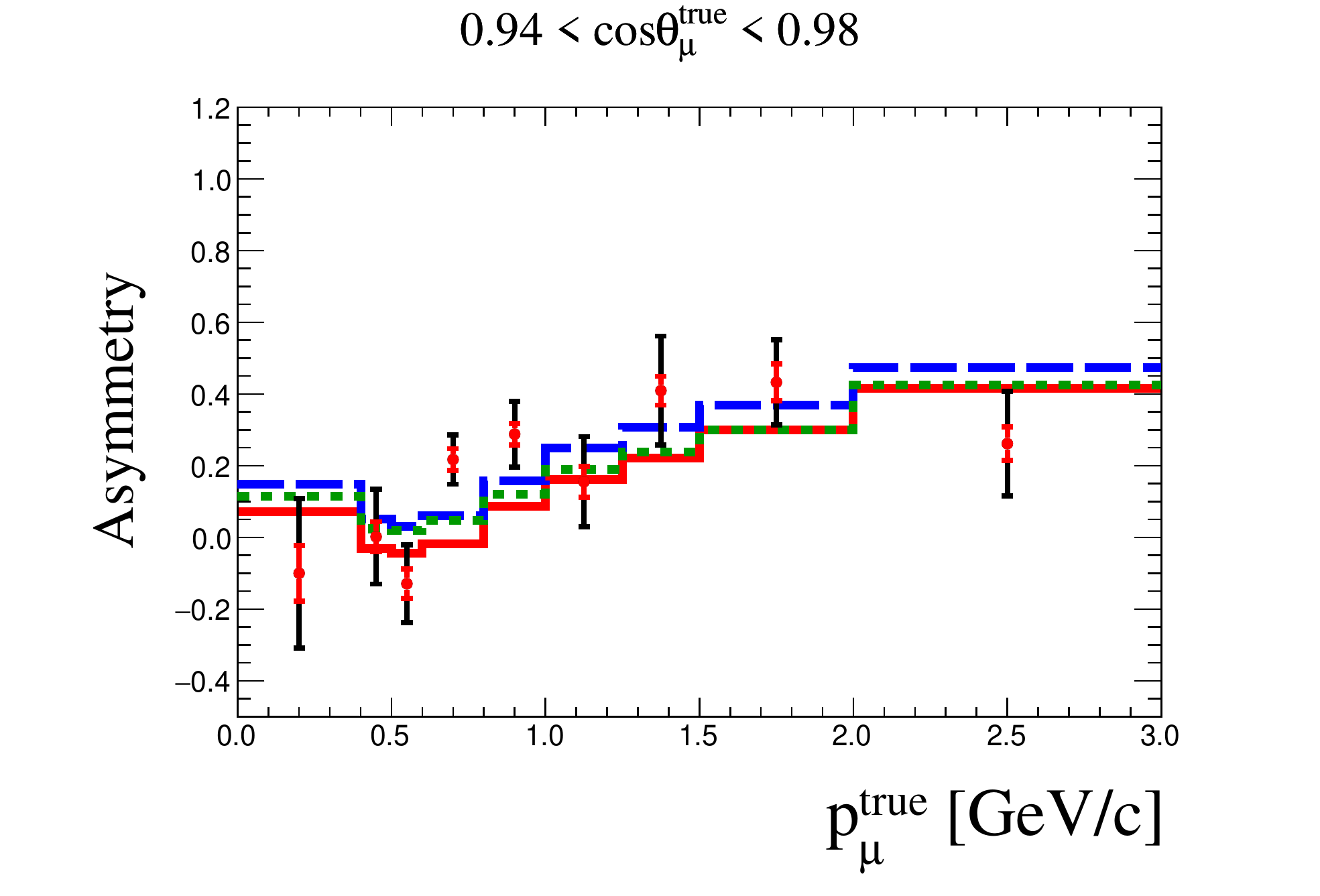}
	\includegraphics[width=0.36\linewidth]{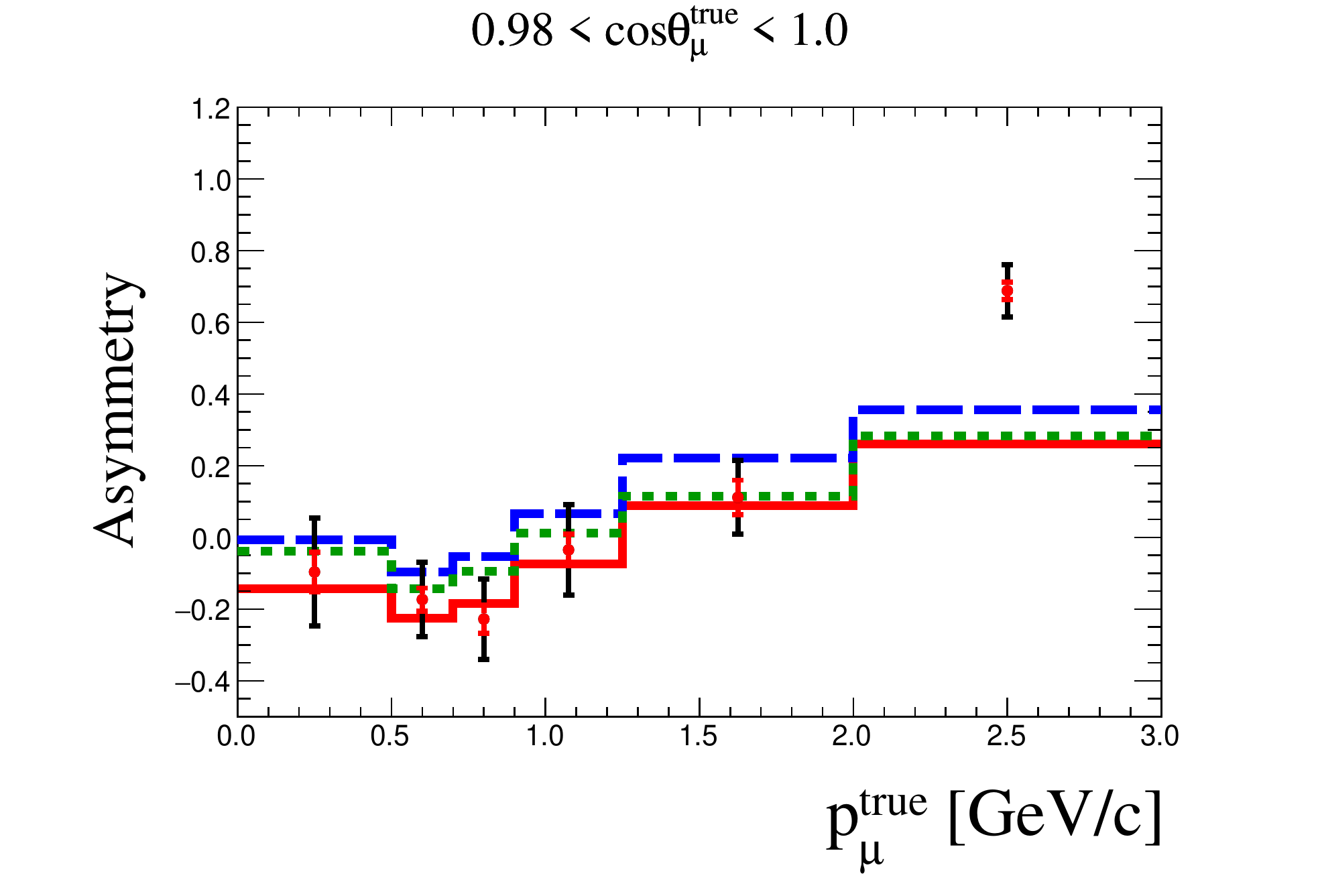}	
	\includegraphics[width=0.36\linewidth]{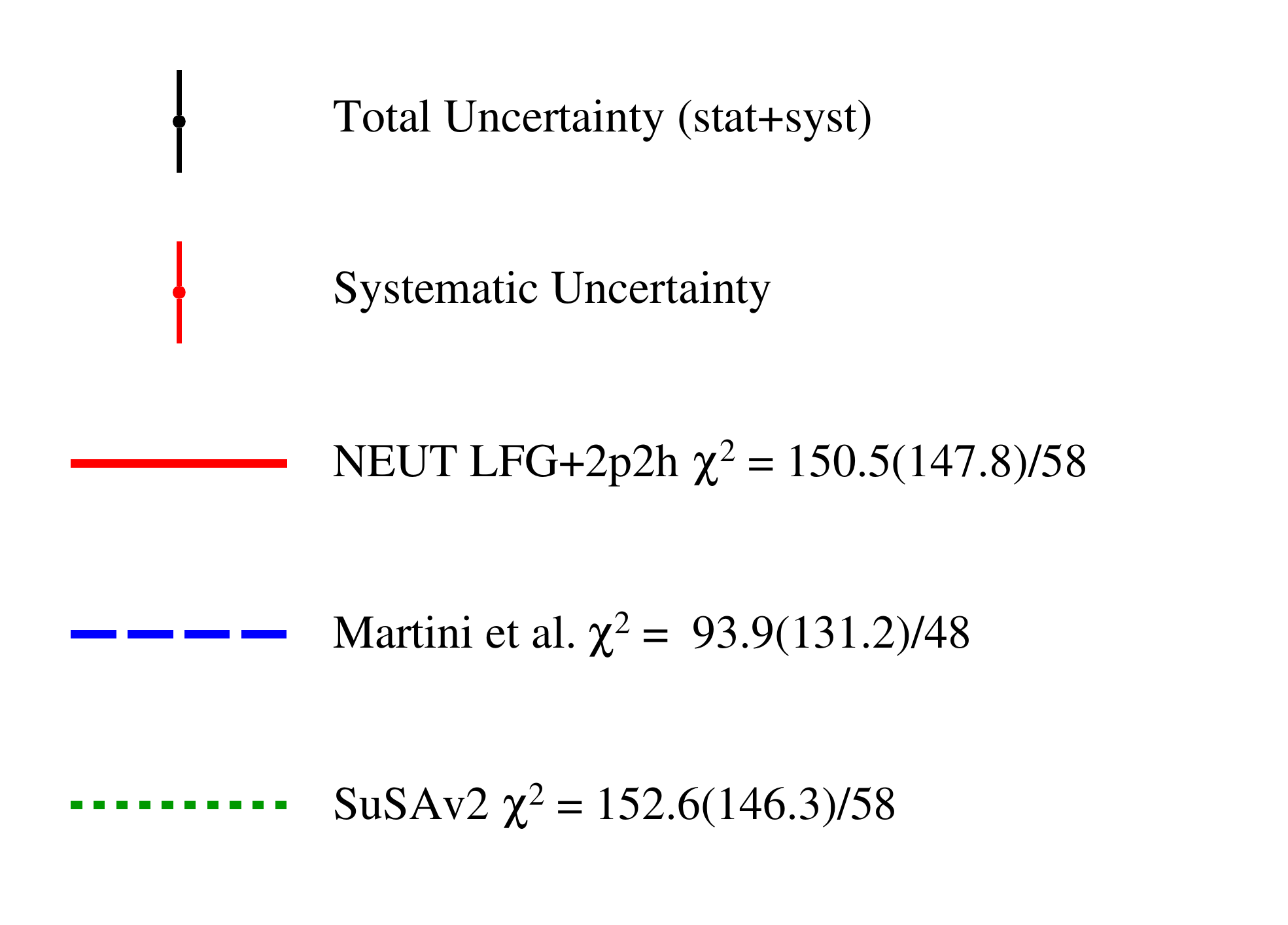}		
	\caption{Measured double-differential \cczeropi cross-section asymmetry in bins of true muon kinematics with systematic uncertainty (red bars) and total (stat.+syst.) uncertainty (black bars). The results are compared to \textsc{Neut} version~\texttt{5.4.1}, which uses an LFG+RPA model with 2p2h (solid red line), Martini \textit{et al.} (dashed blue line) and \textsc{SuSAv2} (green dashed line) models. The full and shape-only (in parenthesis) $\chi2$ are reported. The last bin in momentum is not displayed for readability.}
	\label{fig:asynumucc0piNMS}
\end{figure*}

\begin{figure*}[h!]
	\centering
	\includegraphics[width=0.36\linewidth]{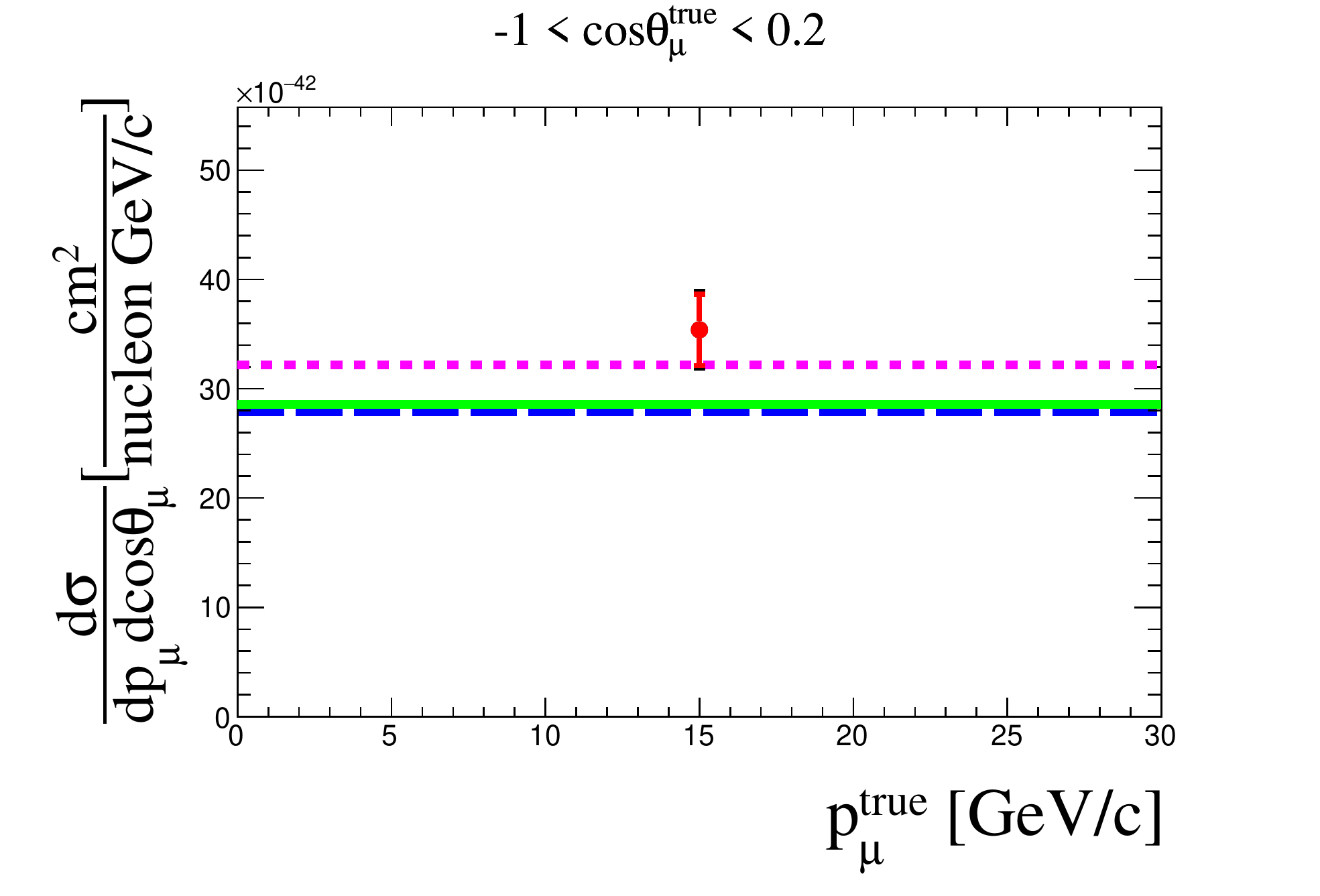}
	\includegraphics[width=0.36\linewidth]{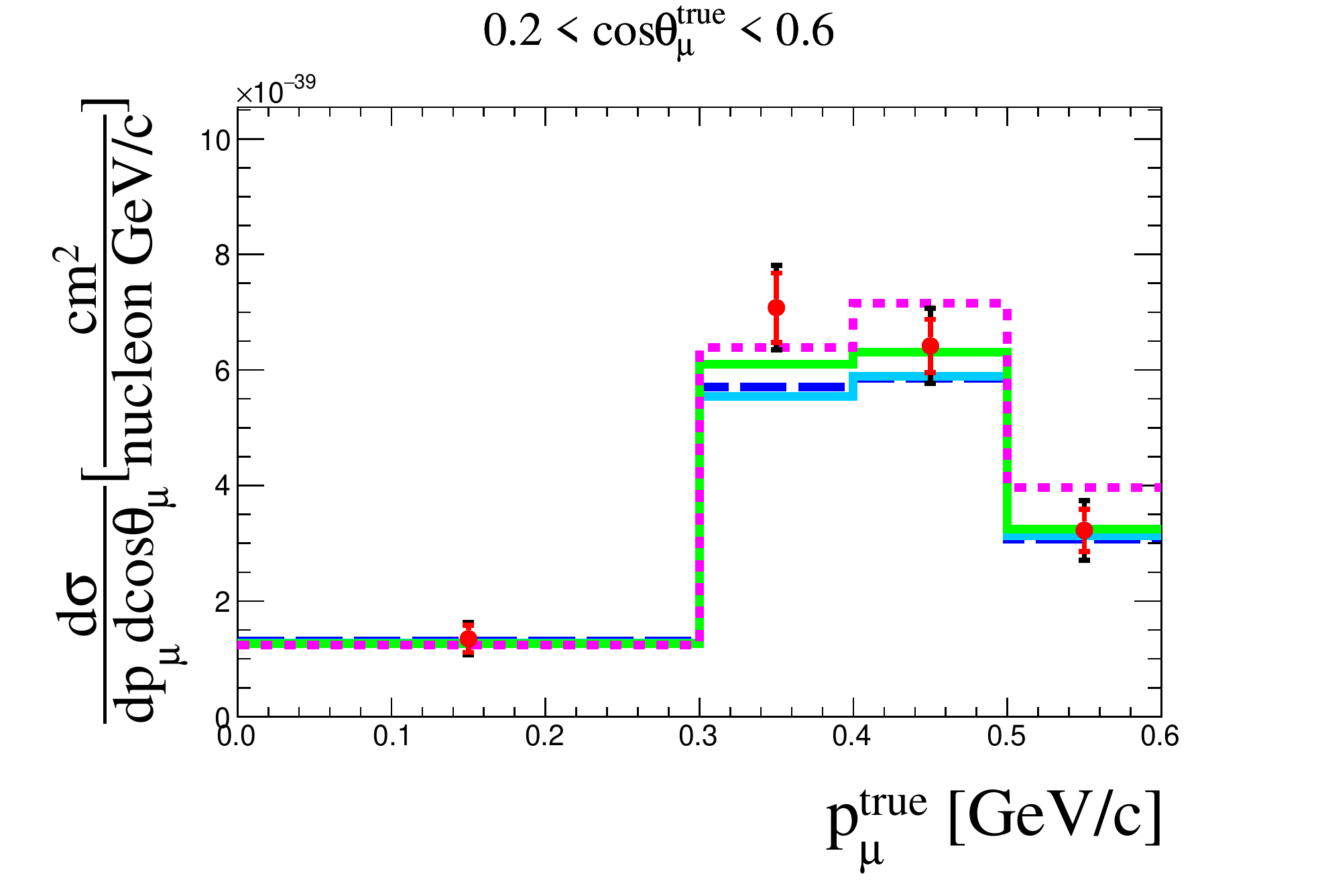}
	\includegraphics[width=0.36\linewidth]{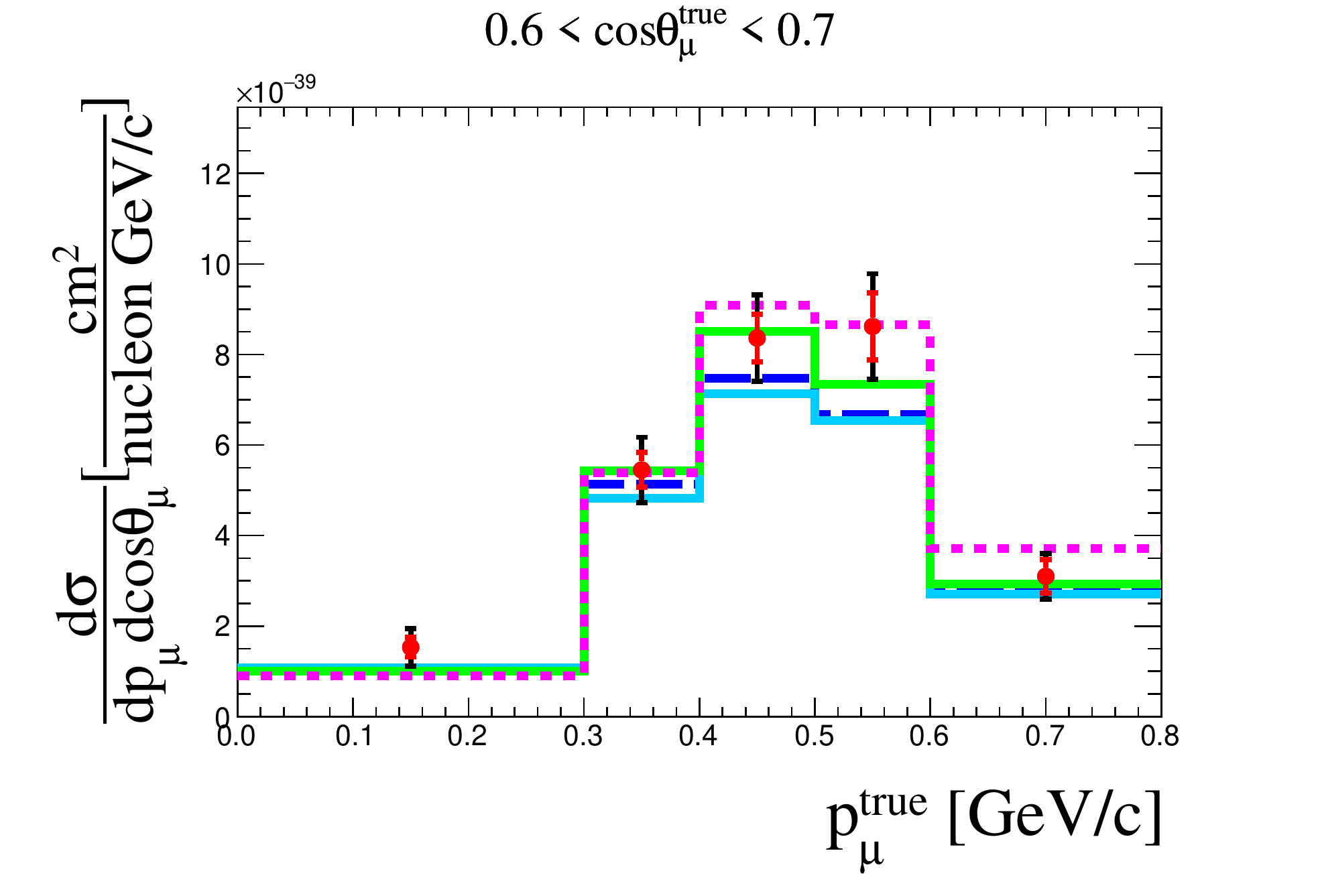}
	\includegraphics[width=0.36\linewidth]{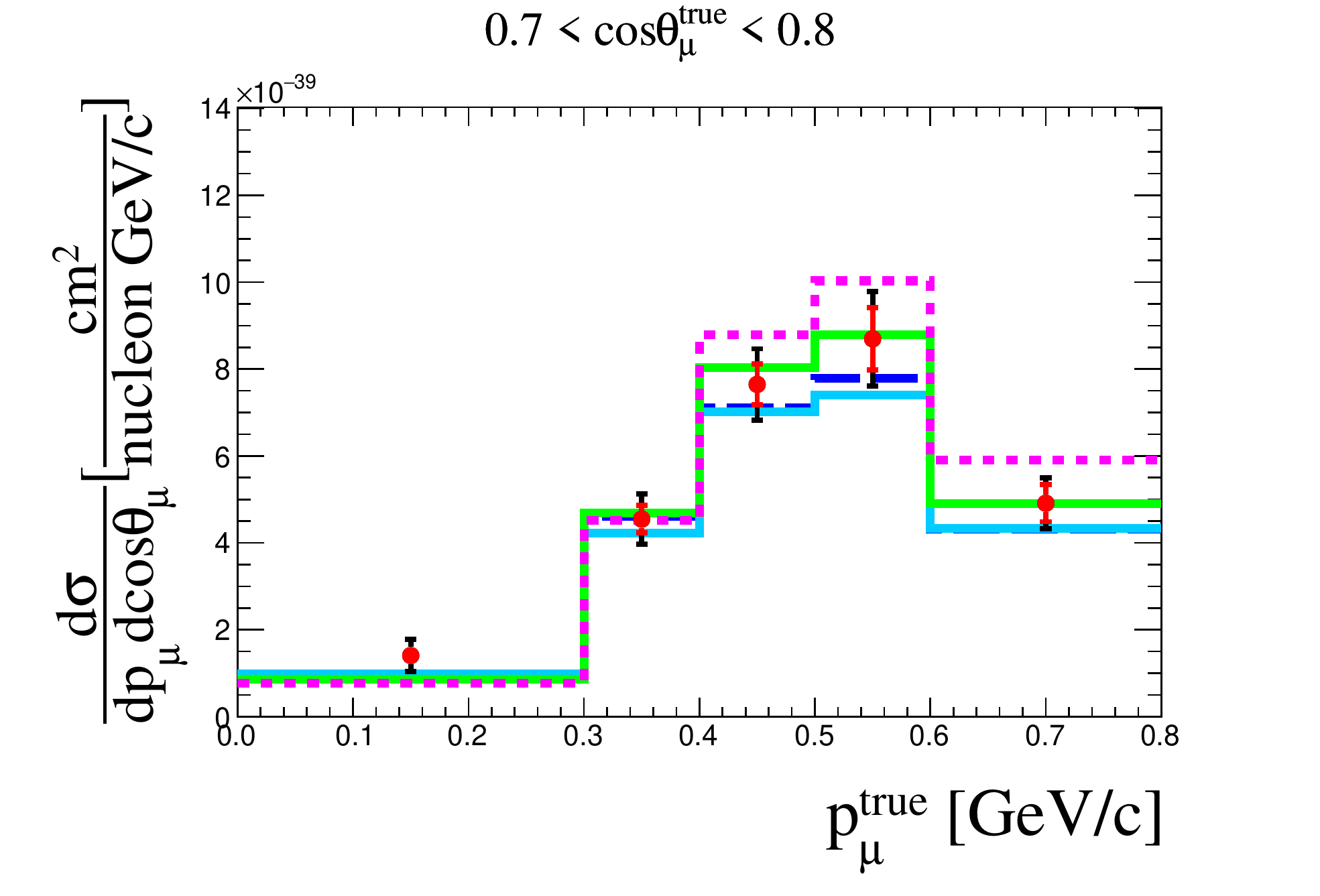}	
	\includegraphics[width=0.36\linewidth]{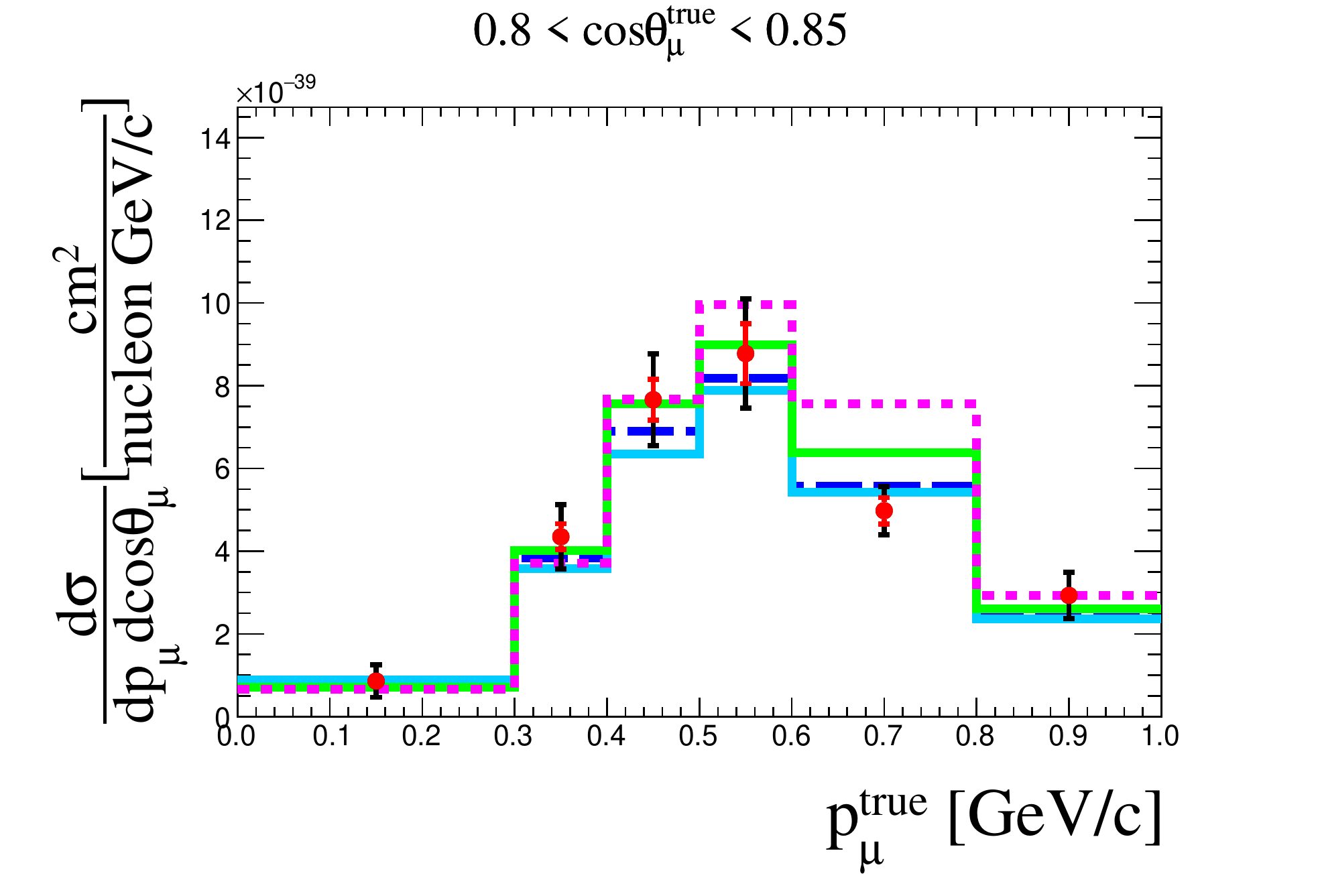}
	\includegraphics[width=0.36\linewidth]{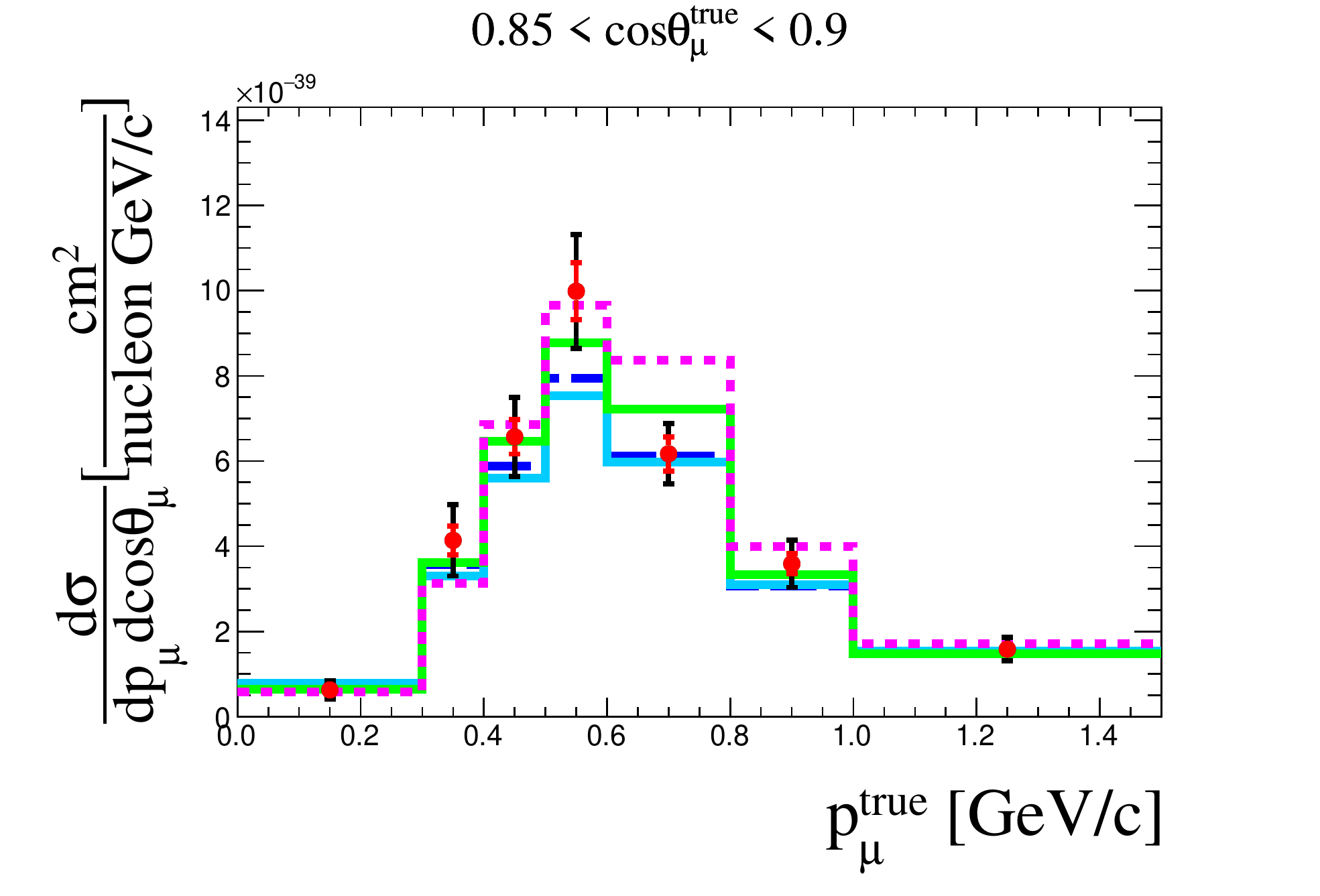}
	\includegraphics[width=0.36\linewidth]{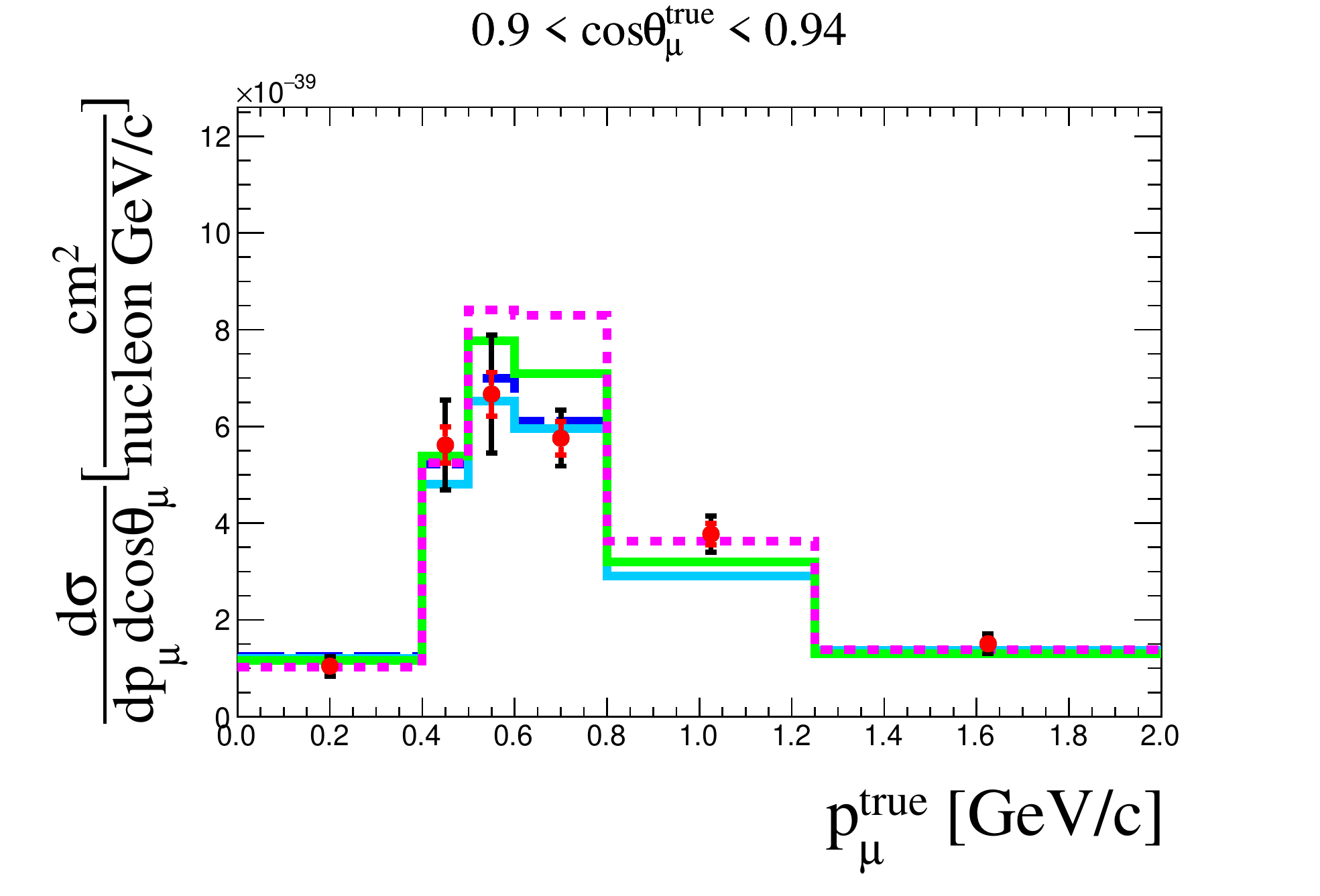}
	\includegraphics[width=0.36\linewidth]{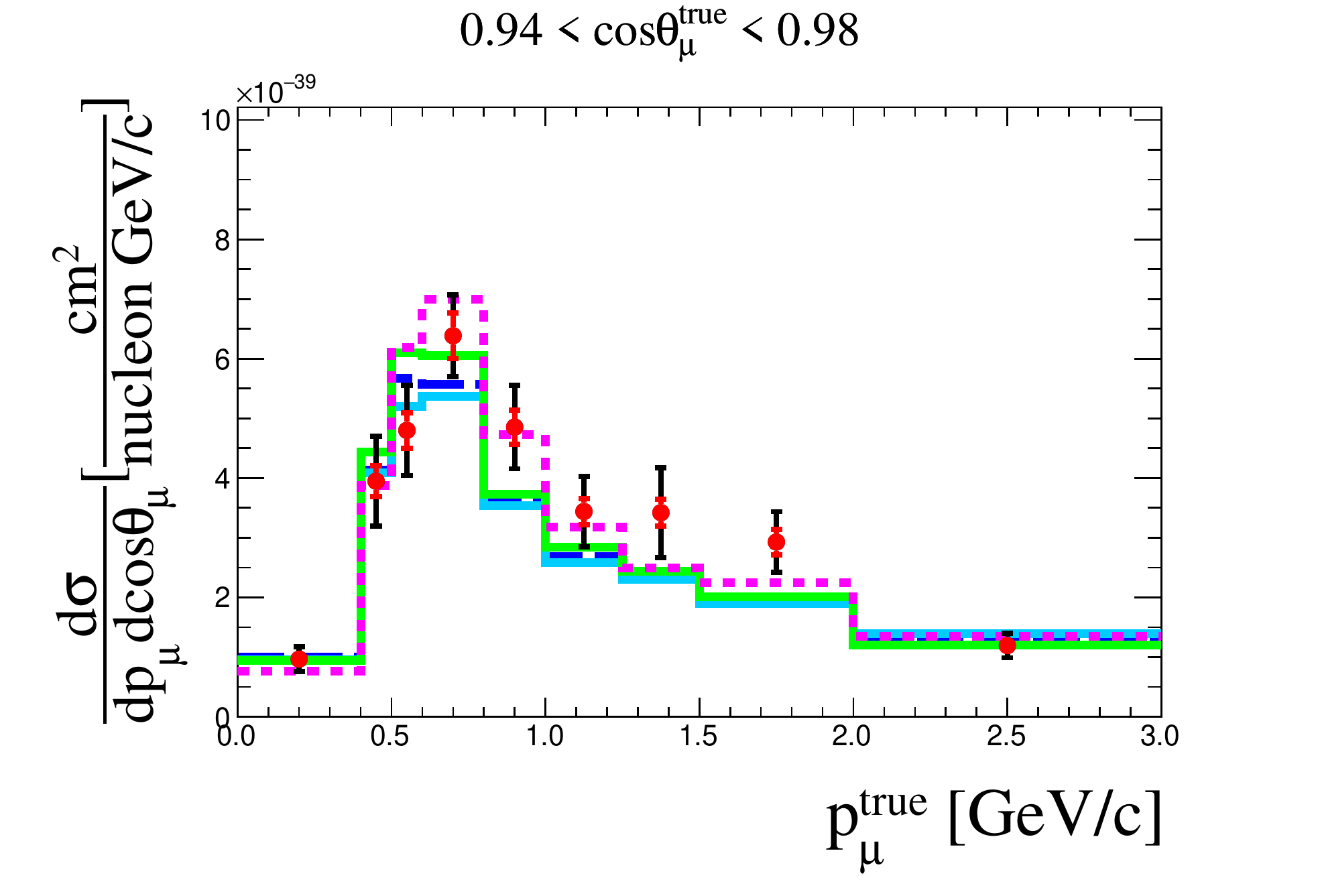}
	\includegraphics[width=0.36\linewidth]{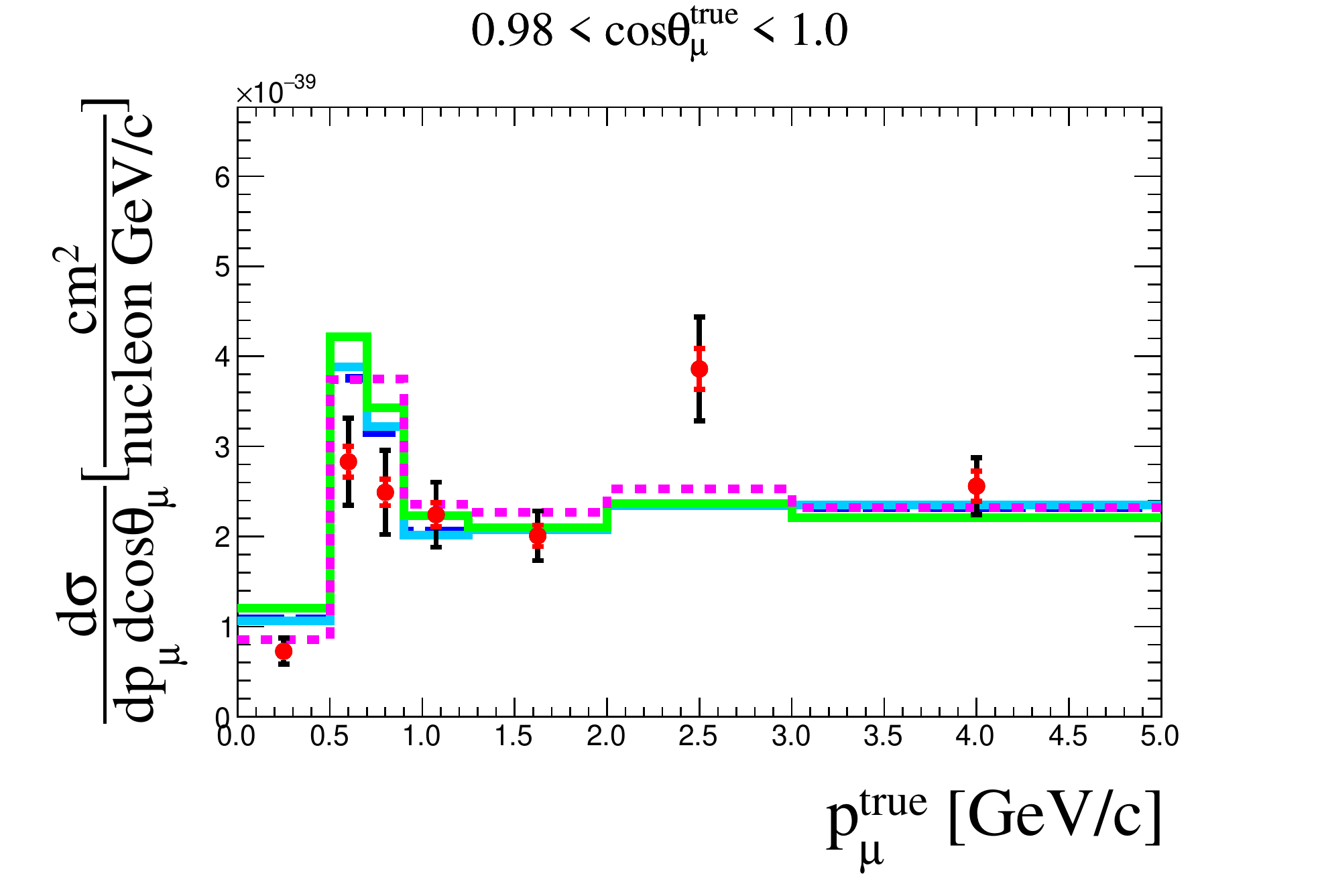}		
	\includegraphics[width=0.36\linewidth]{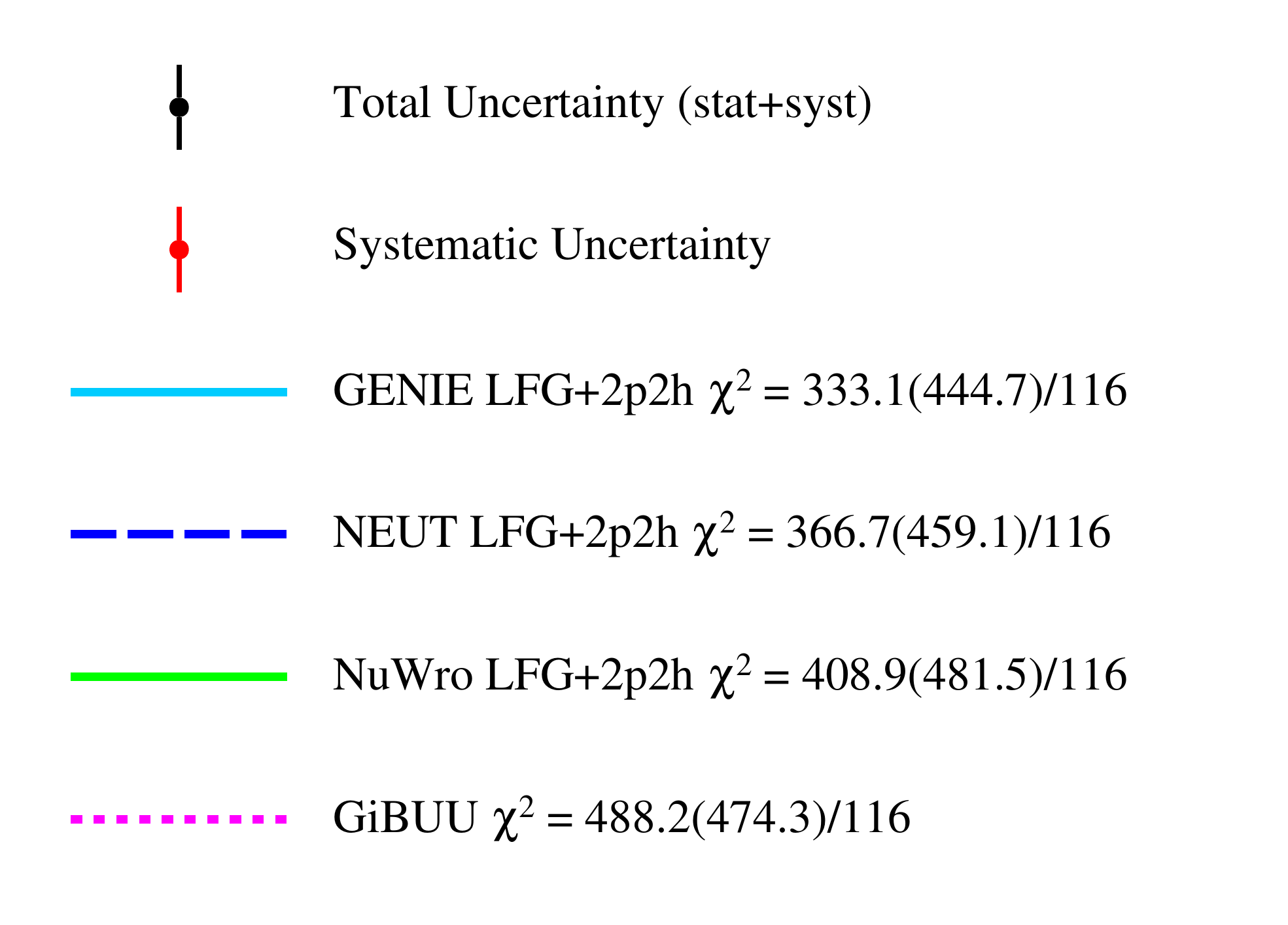}				
	\caption{Measured \numu \cczeropi double-differential cross-section per nucleon in bins of true muon kinematics with systematic uncertainty (red bars) and total (stat.+syst.) uncertainty (black bars). The result is compared with \textsc{Neut} (dashed blue line), \textsc{NuWro} version~\texttt{18.02.1}  (green solid line) and \textsc{GiBUU}~\texttt{2019} (pink dotted line) prediction. All generators use an LFG+RPA model that includes 2p2h. The full and shape-only (in parenthesis) $\chi2$ are reported. The last bin in momentum is not displayed for readability.}
	\label{fig:numucc0pixsecgibuuneutnuwro}
\end{figure*}

\begin{figure*}[h!]
	\centering
	\includegraphics[width=0.36\linewidth]{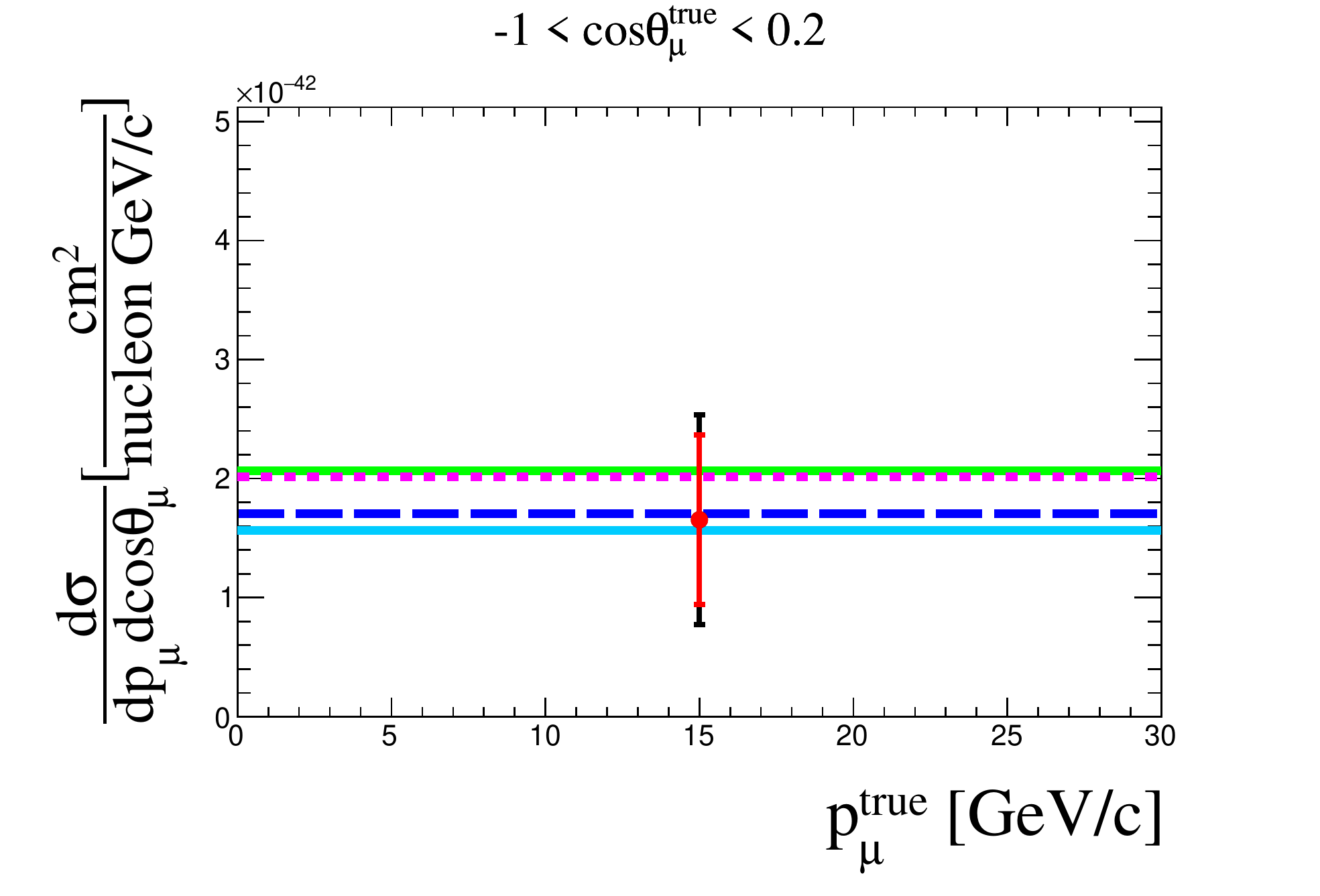}
	\includegraphics[width=0.36\linewidth]{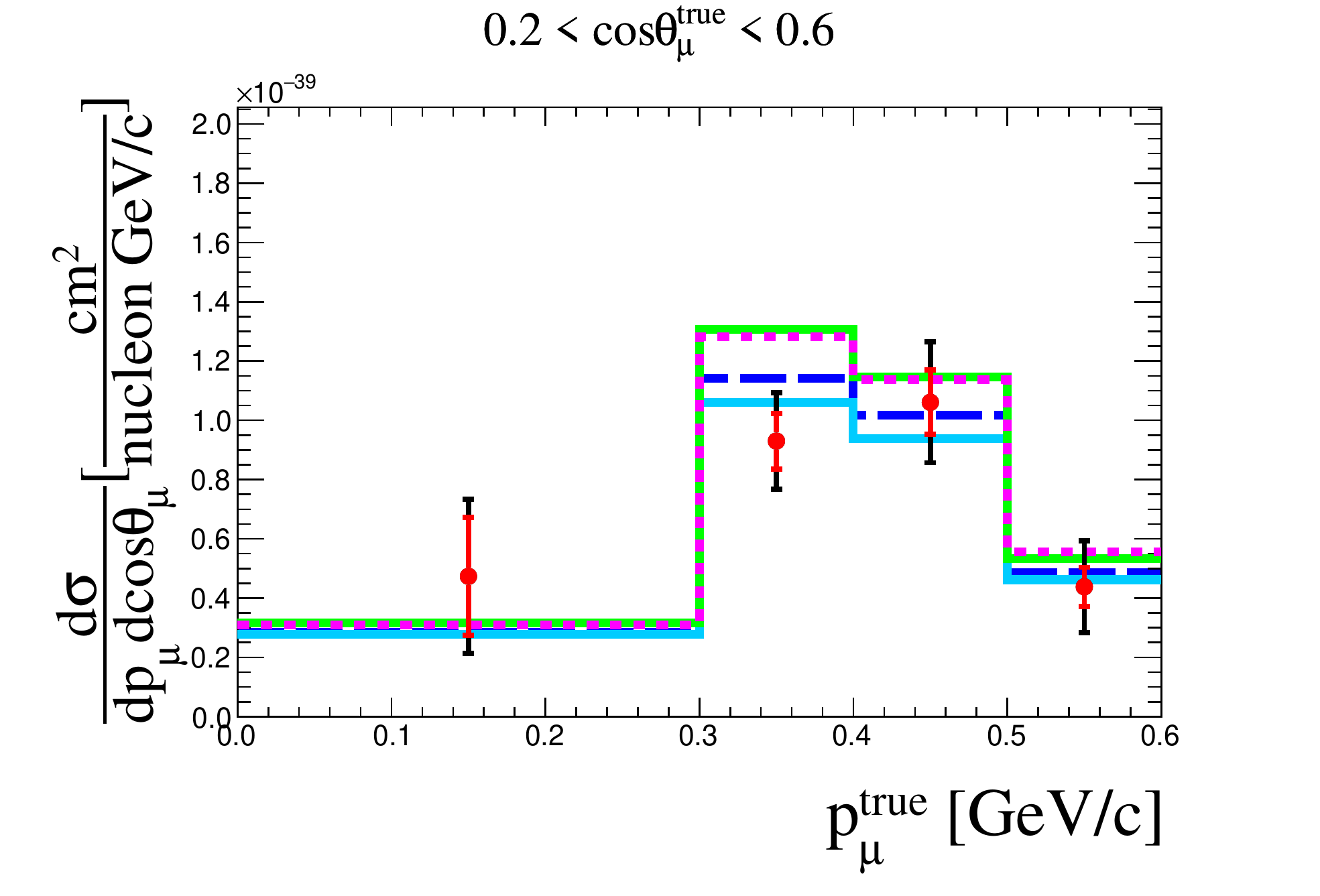}
	\includegraphics[width=0.36\linewidth]{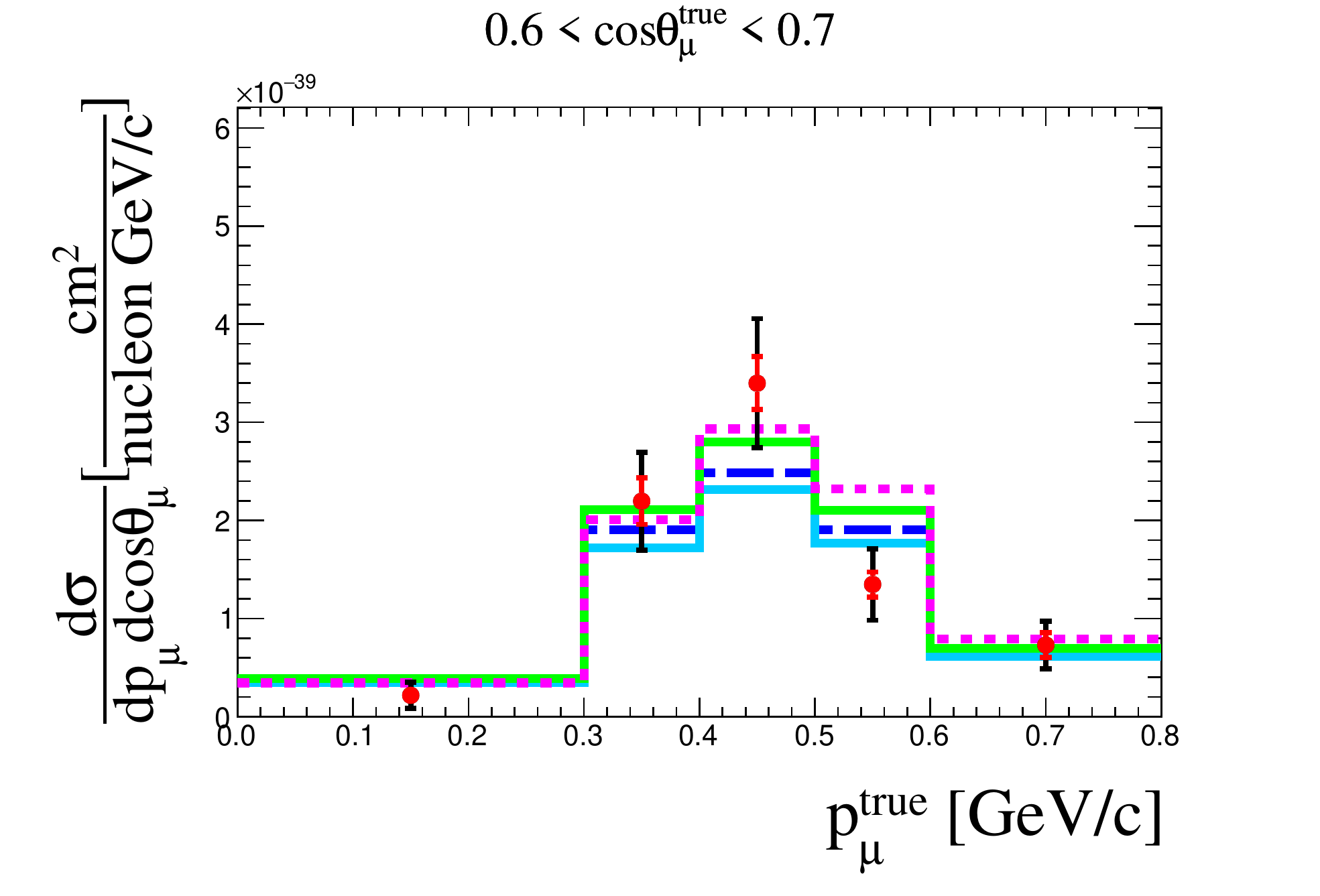}
	\includegraphics[width=0.36\linewidth]{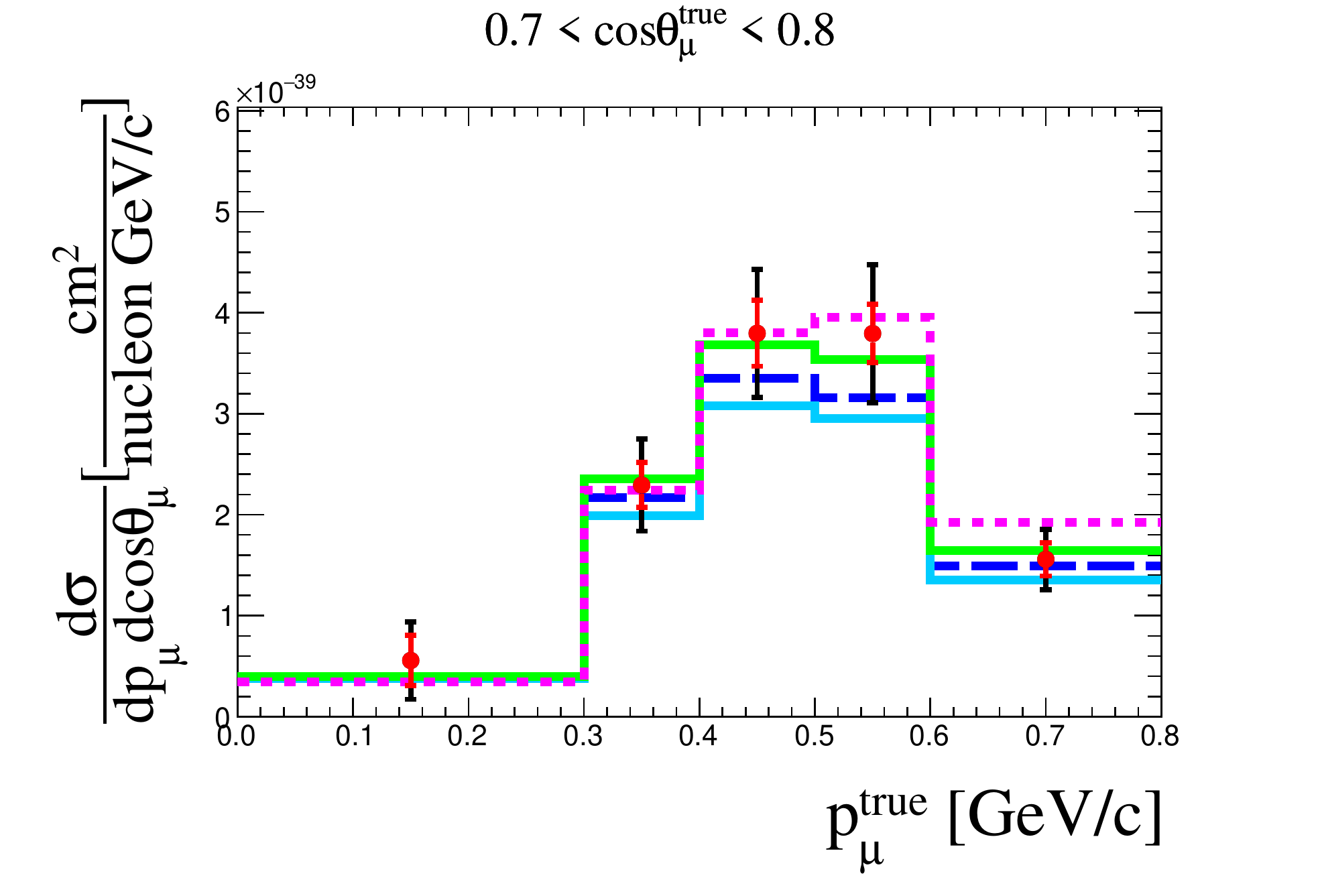}	
	\includegraphics[width=0.36\linewidth]{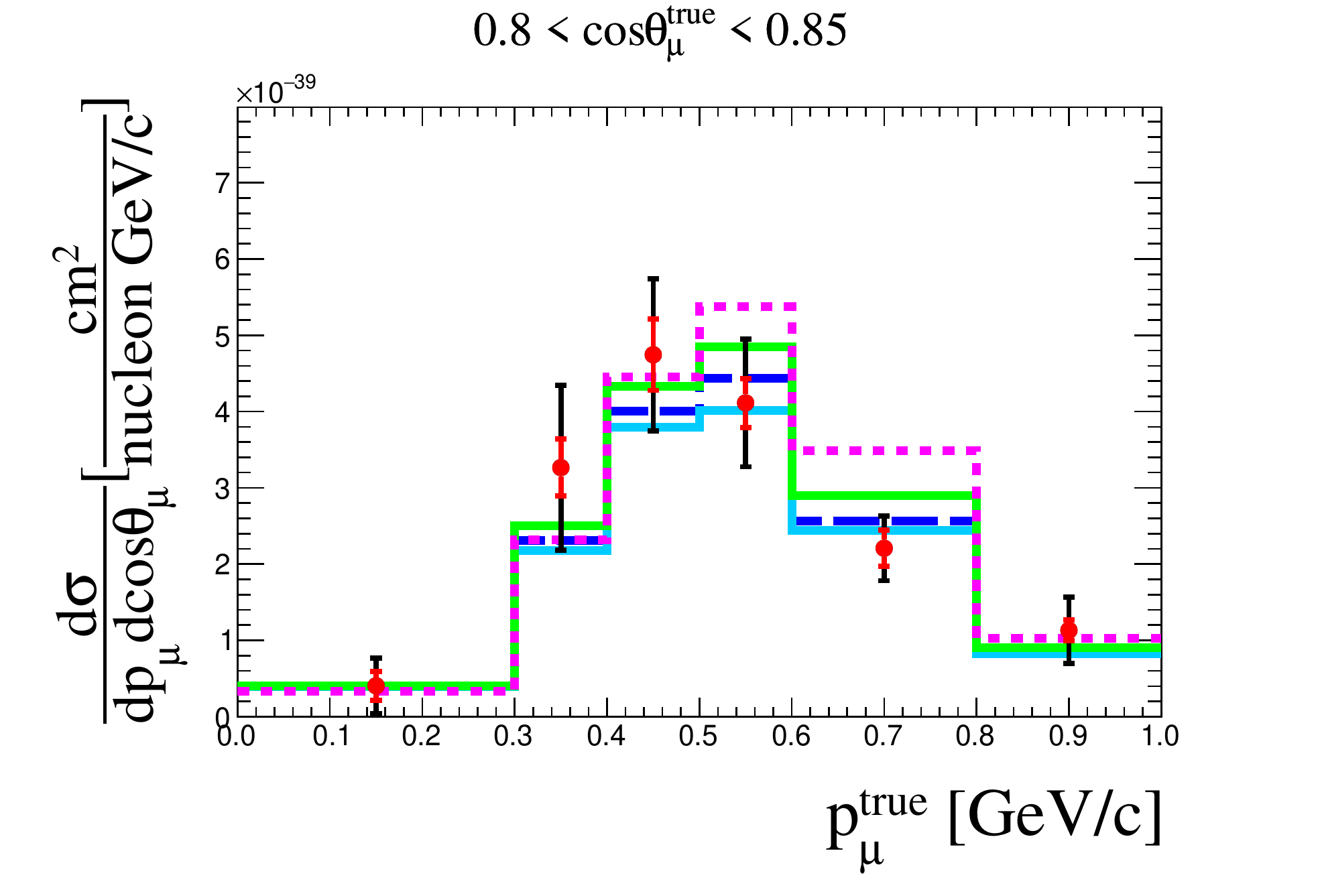}
	\includegraphics[width=0.36\linewidth]{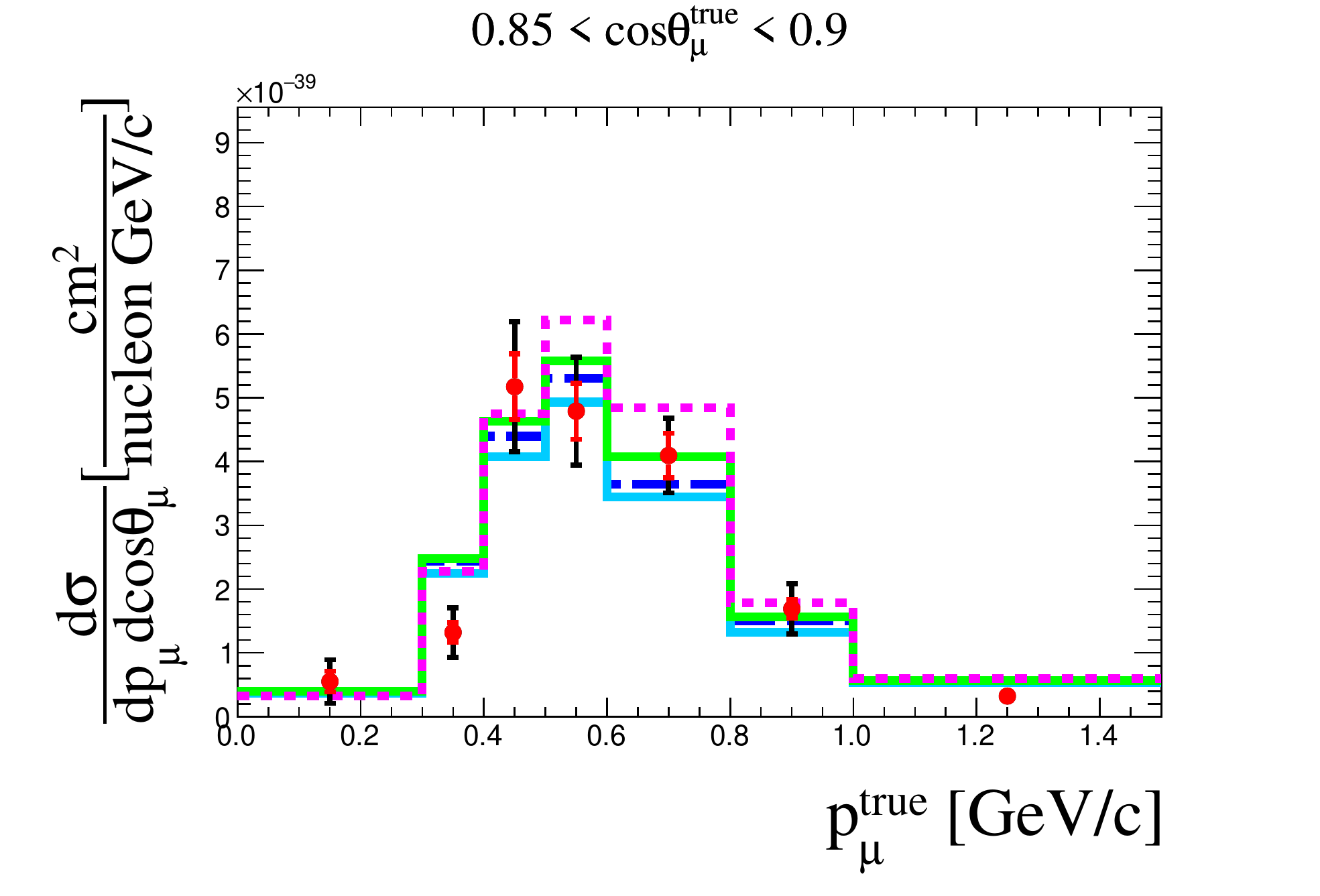}
	\includegraphics[width=0.36\linewidth]{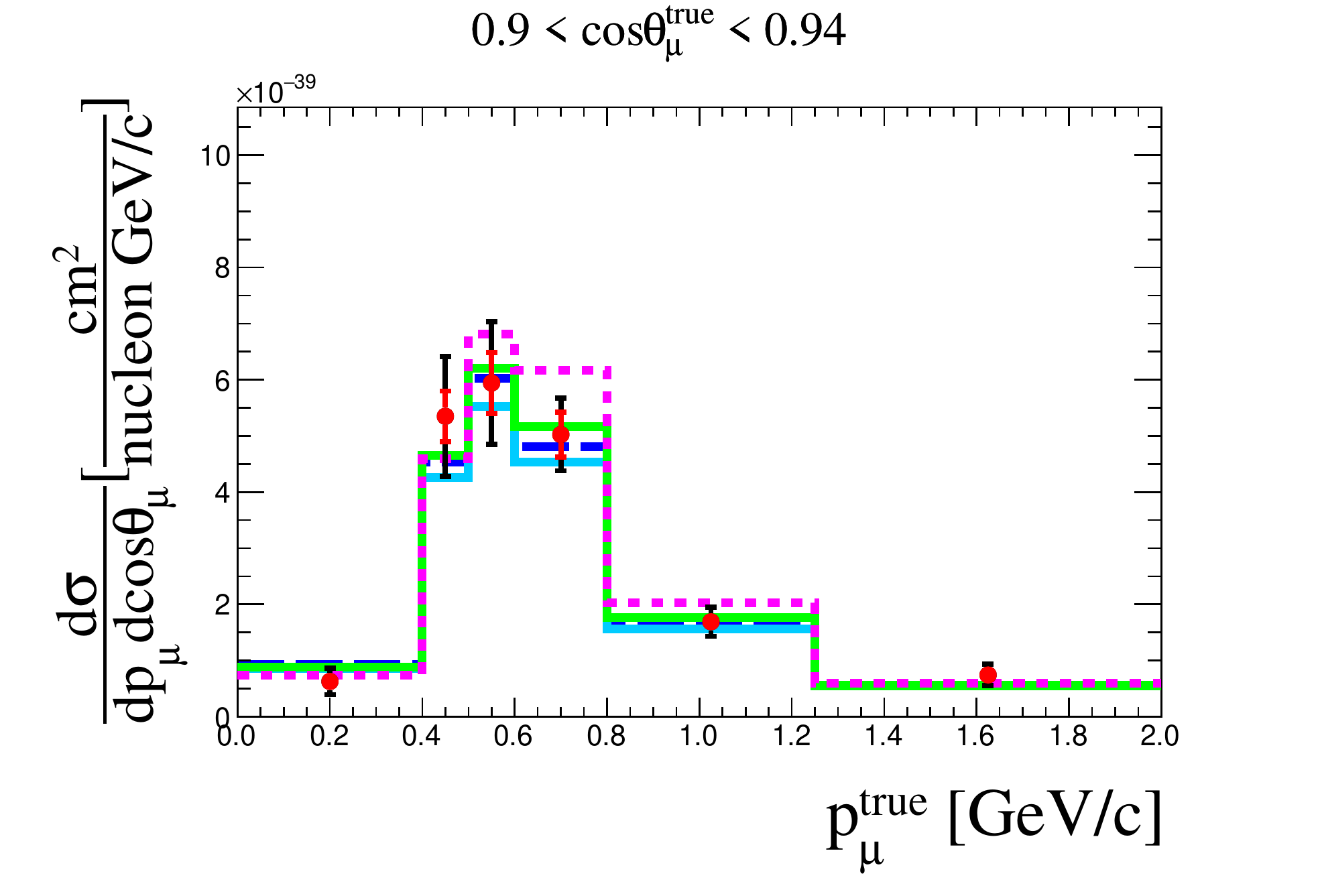}
	\includegraphics[width=0.36\linewidth]{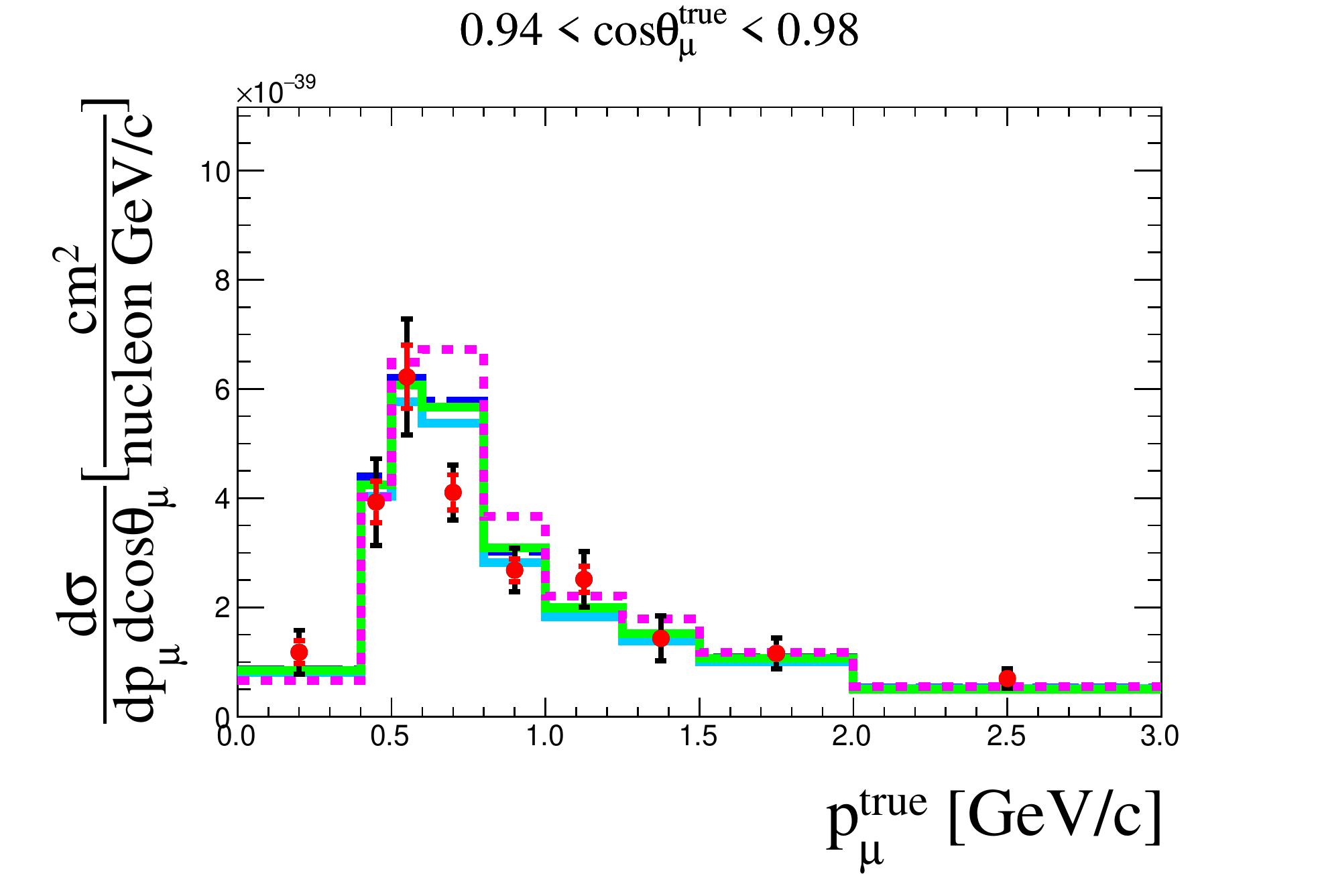}
	\includegraphics[width=0.36\linewidth]{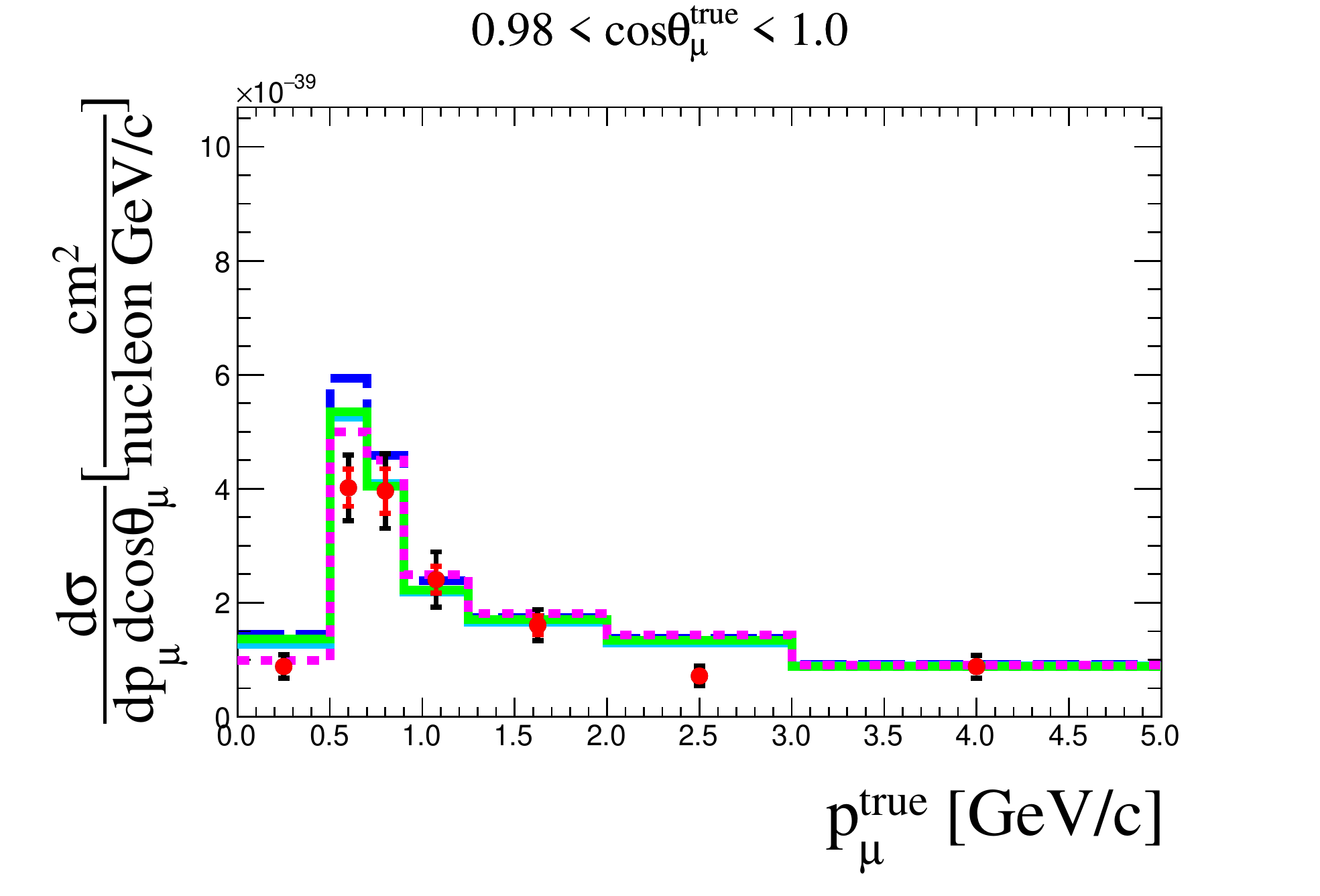}		
	\includegraphics[width=0.36\linewidth]{XsecLegendGNNG}									
	\caption{Measured \barnumu \cczeropi double-differential cross-section per nucleon in bins of true muon kinematics with systematic uncertainty (red bars) and total (stat.+syst.) uncertainty (black bars). The result is compared with \textsc{Neut} (dashed blue line), \textsc{NuWro} version~\texttt{18.02.1}  (green solid line) and \textsc{GiBUU}~\texttt{2019} (pink dotted line) prediction. All generators use an LFG+RPA model that includes 2p2h. The full and shape-only (in parenthesis) $\chi2$ are reported. The last bin in momentum is not displayed for readability.}
	\label{fig:antinumucc0pixsecgibuuneutnuwro}
\end{figure*}

\begin{figure*}[h!]
	\centering
	\includegraphics[width=0.36\linewidth]{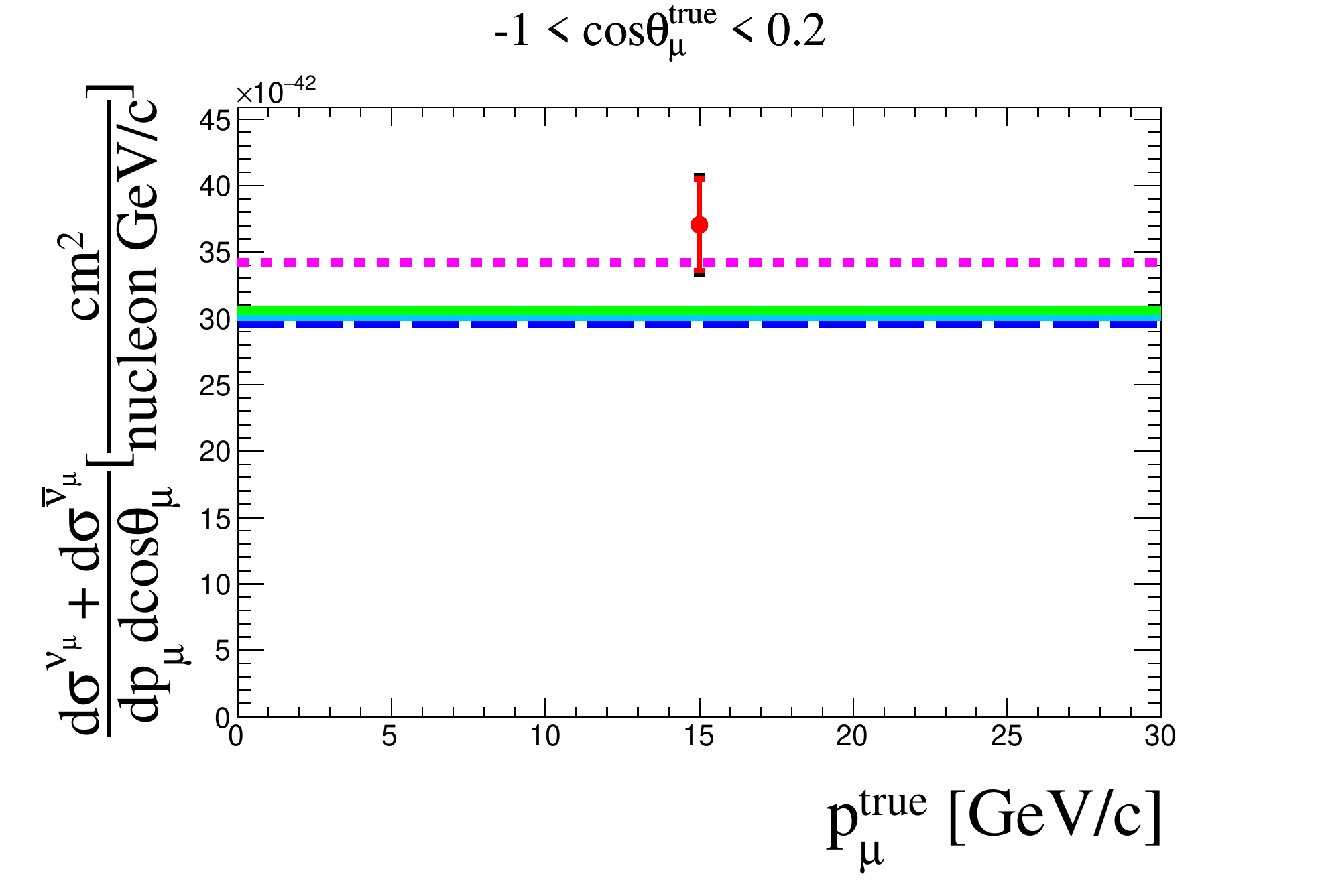}
	\includegraphics[width=0.36\linewidth]{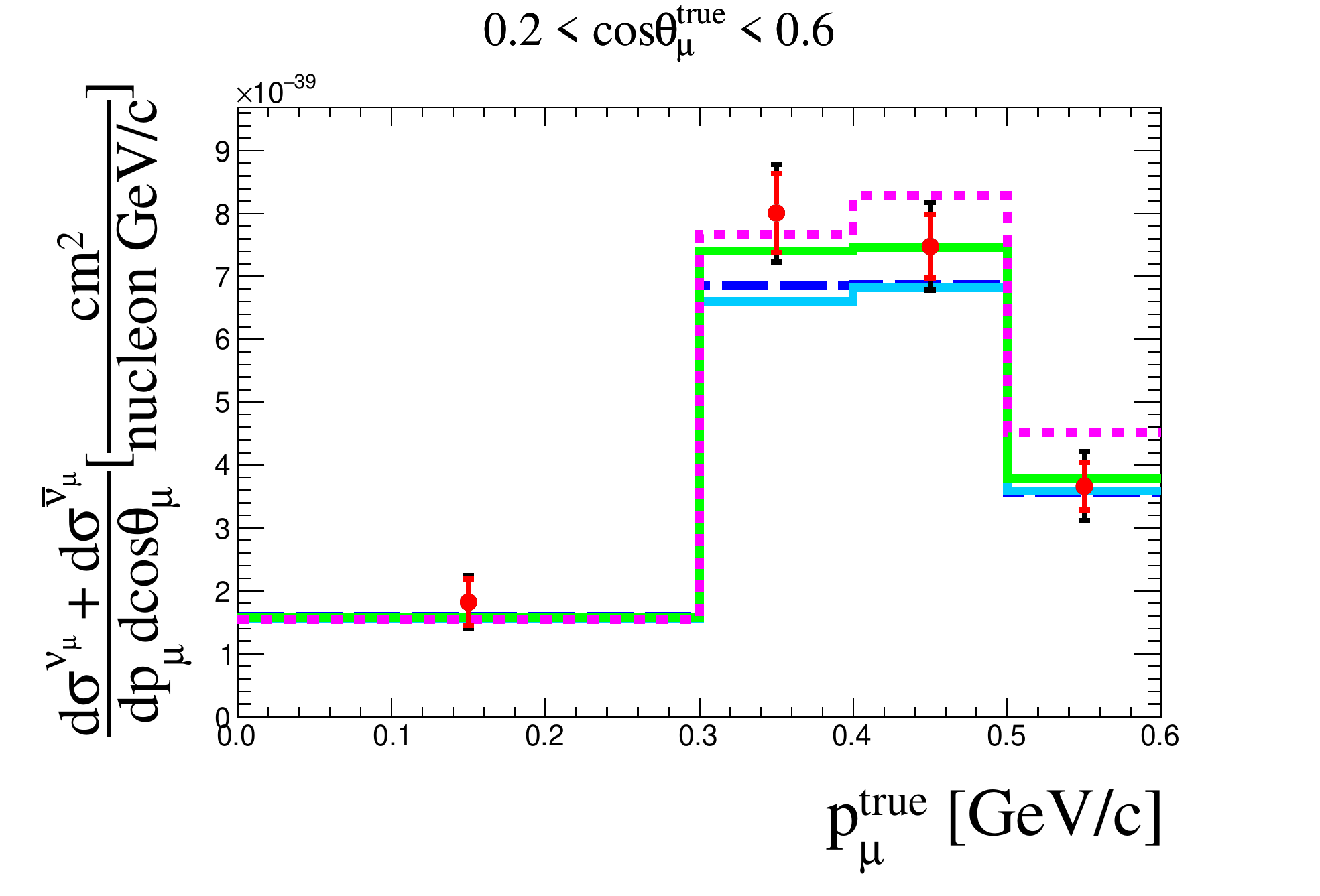}	
	\includegraphics[width=0.36\linewidth]{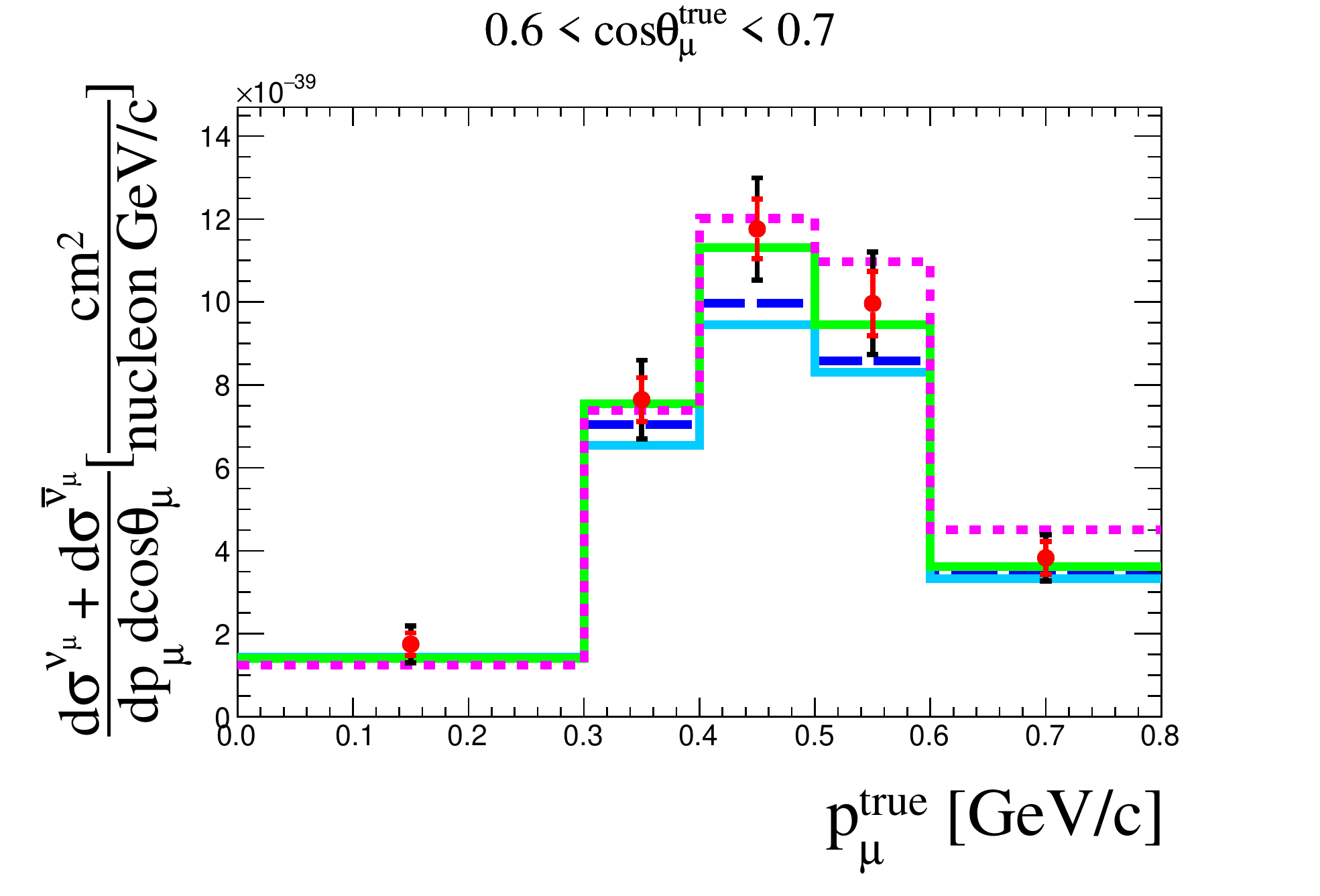}
	\includegraphics[width=0.36\linewidth]{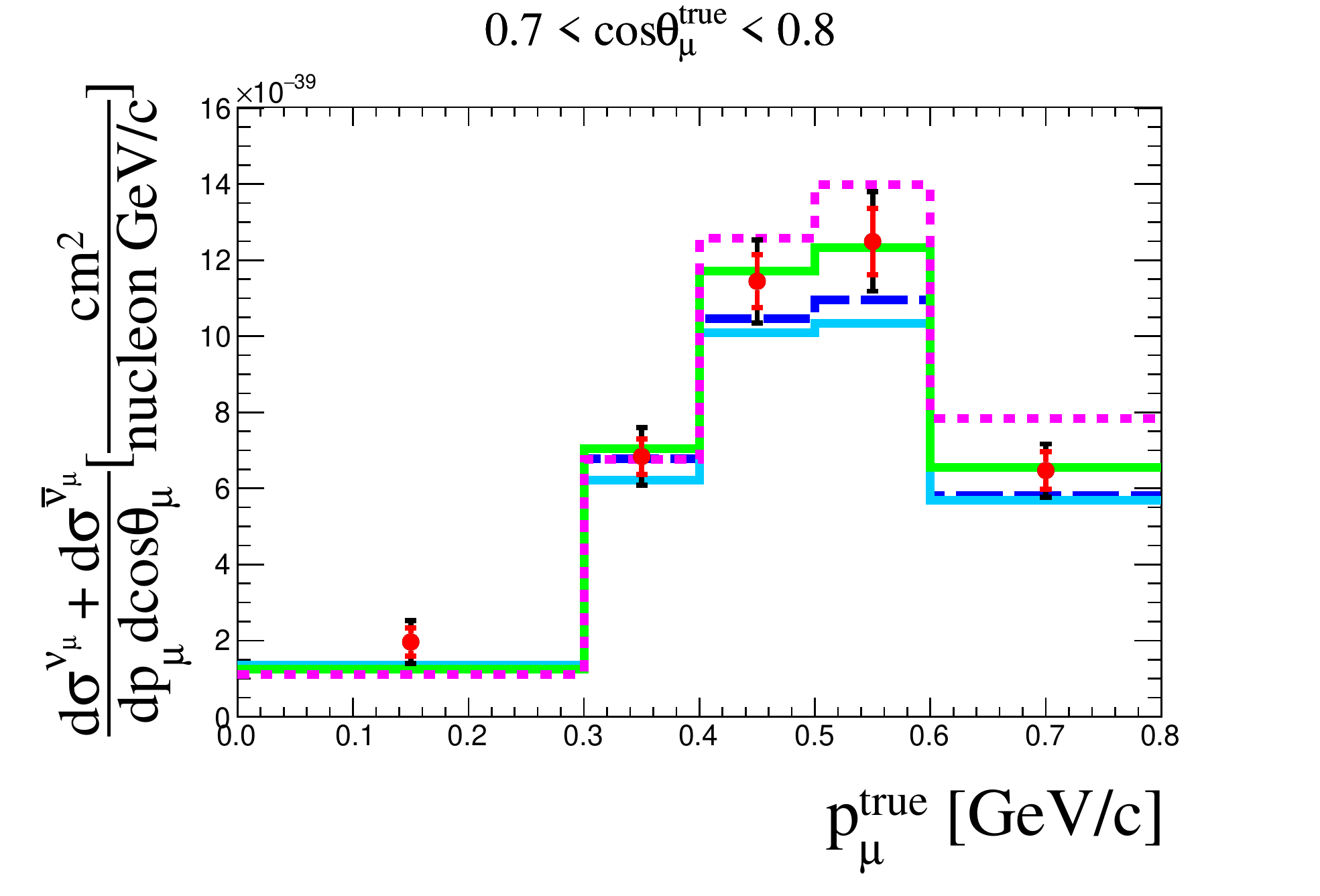}
	\includegraphics[width=0.36\linewidth]{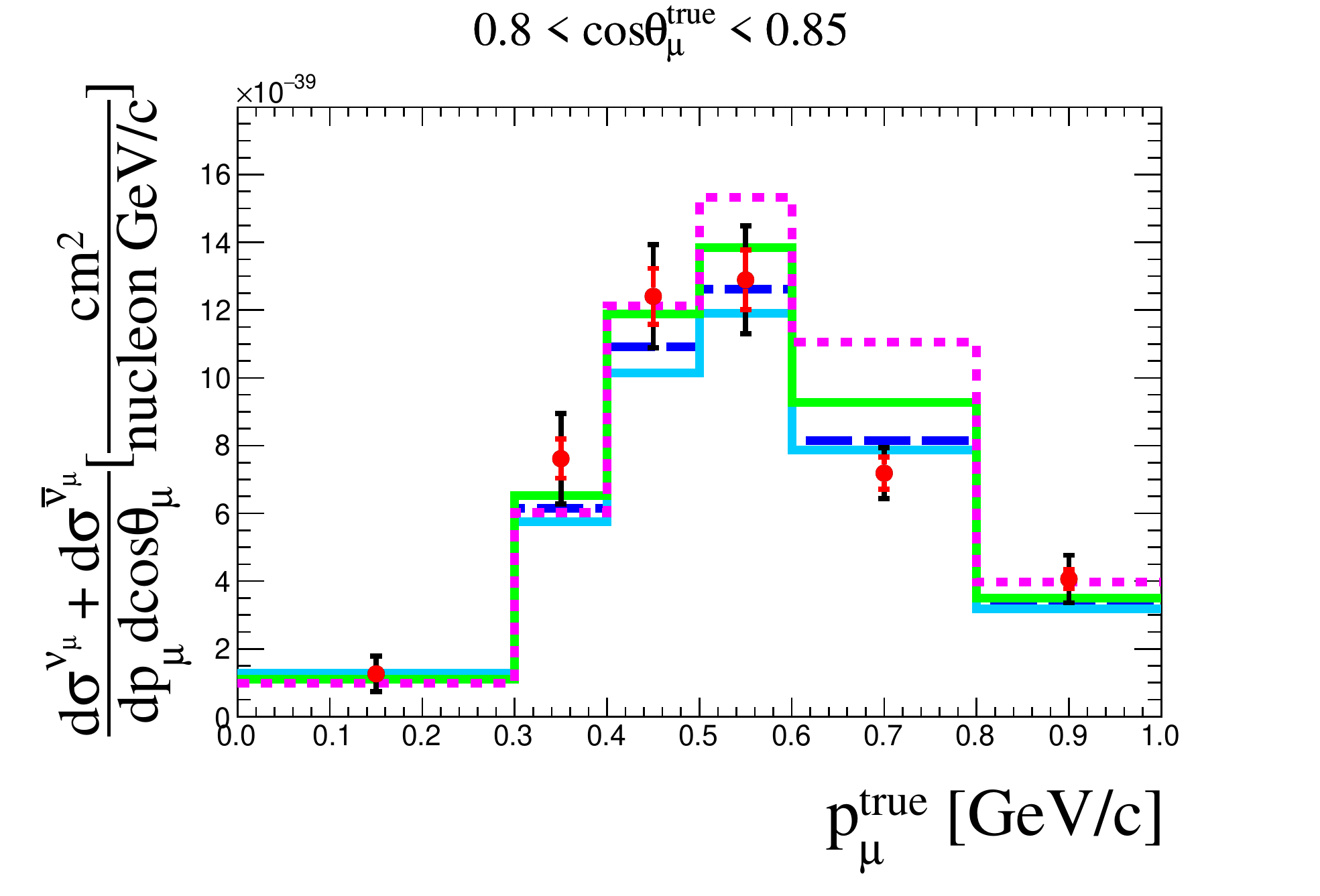}
	\includegraphics[width=0.36\linewidth]{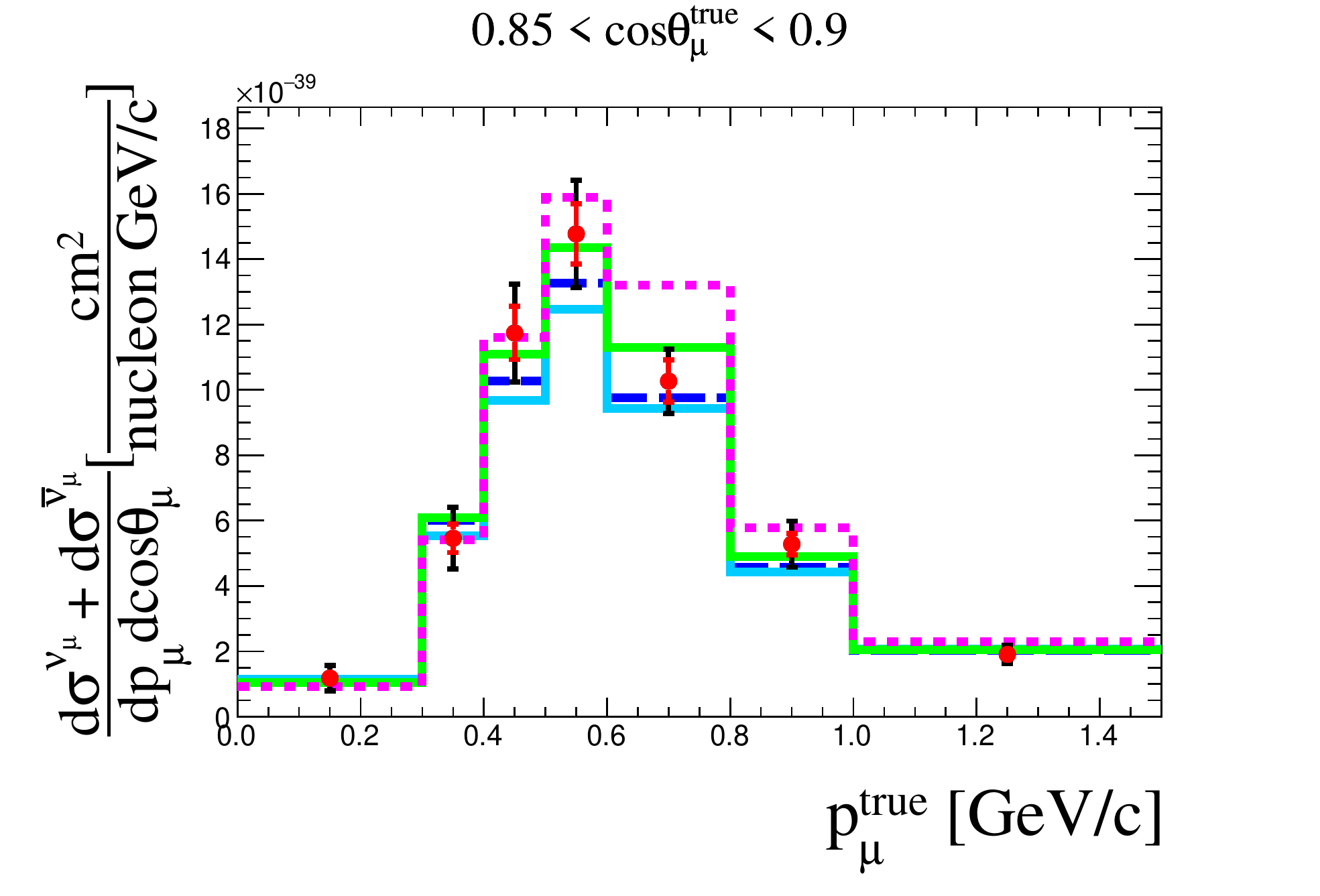}
	\includegraphics[width=0.36\linewidth]{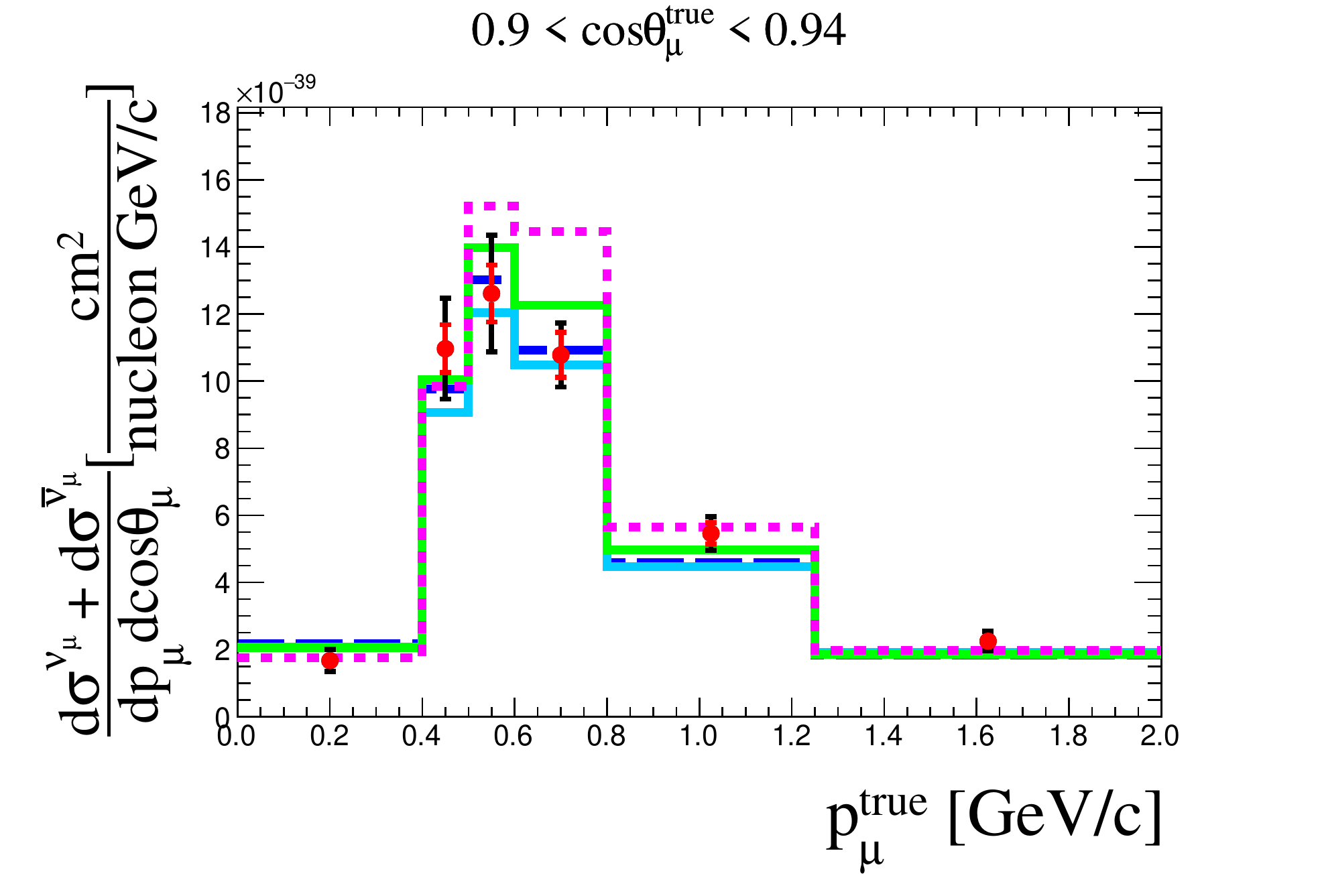}
	\includegraphics[width=0.36\linewidth]{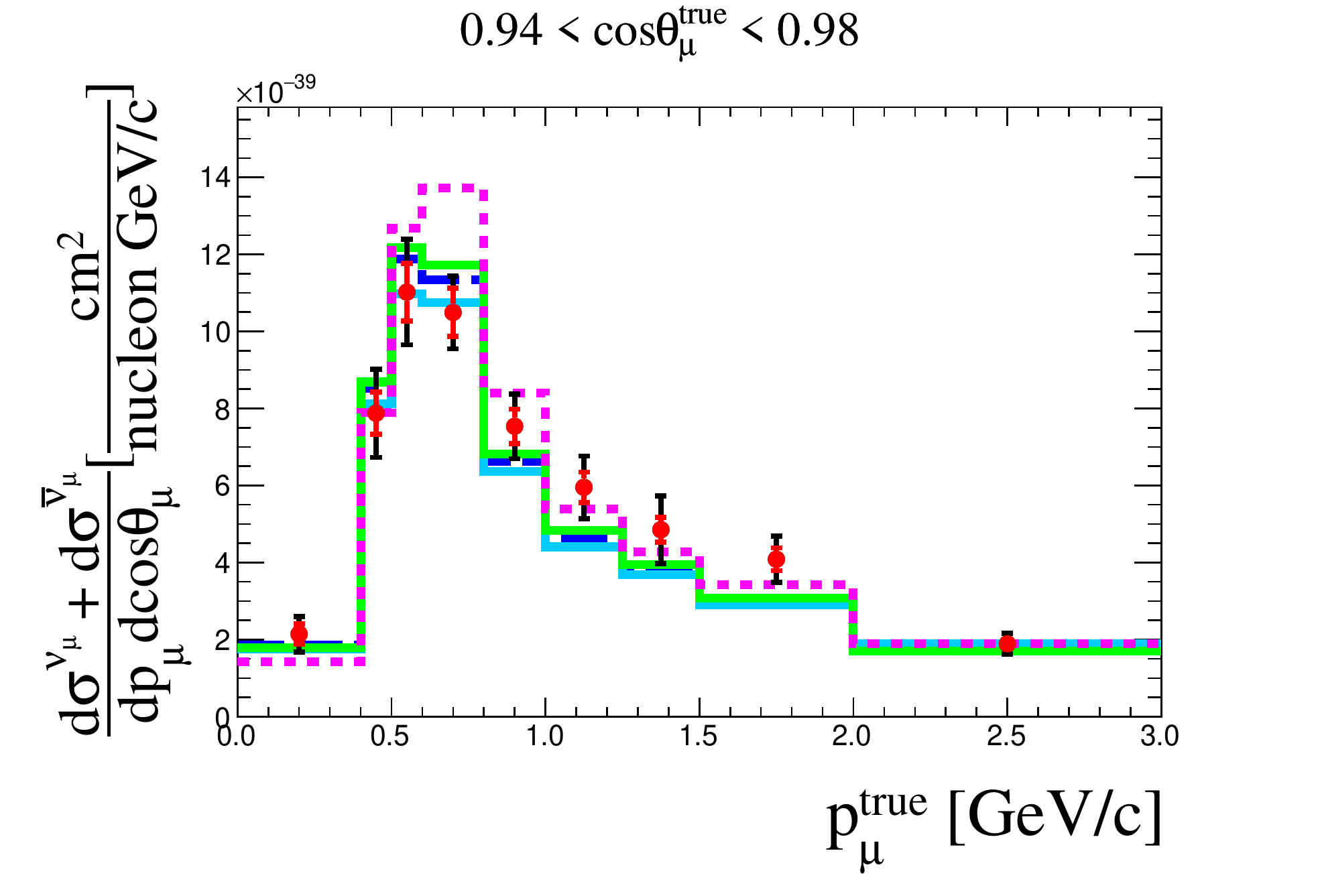}
	\includegraphics[width=0.36\linewidth]{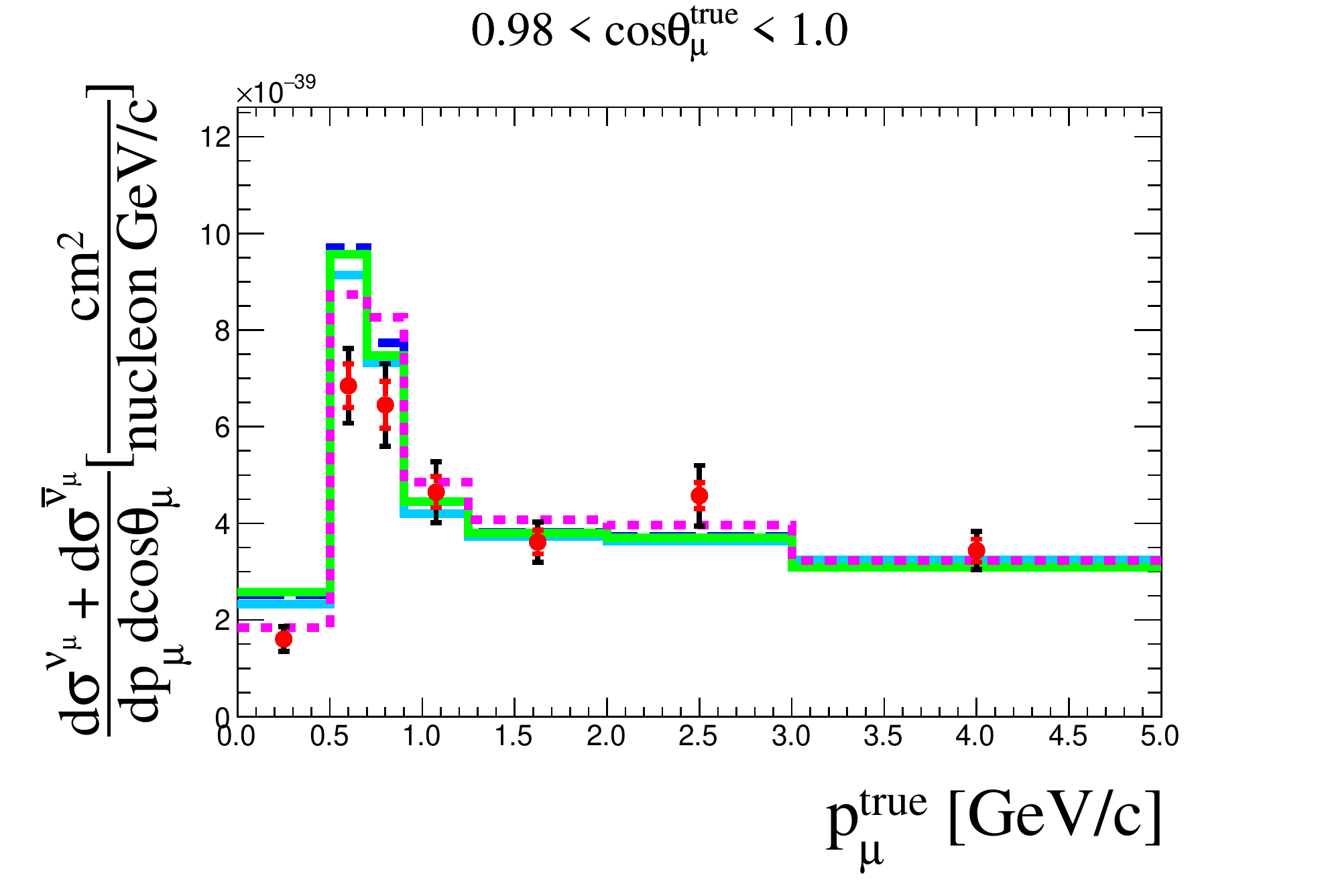}	
	\includegraphics[width=0.36\linewidth]{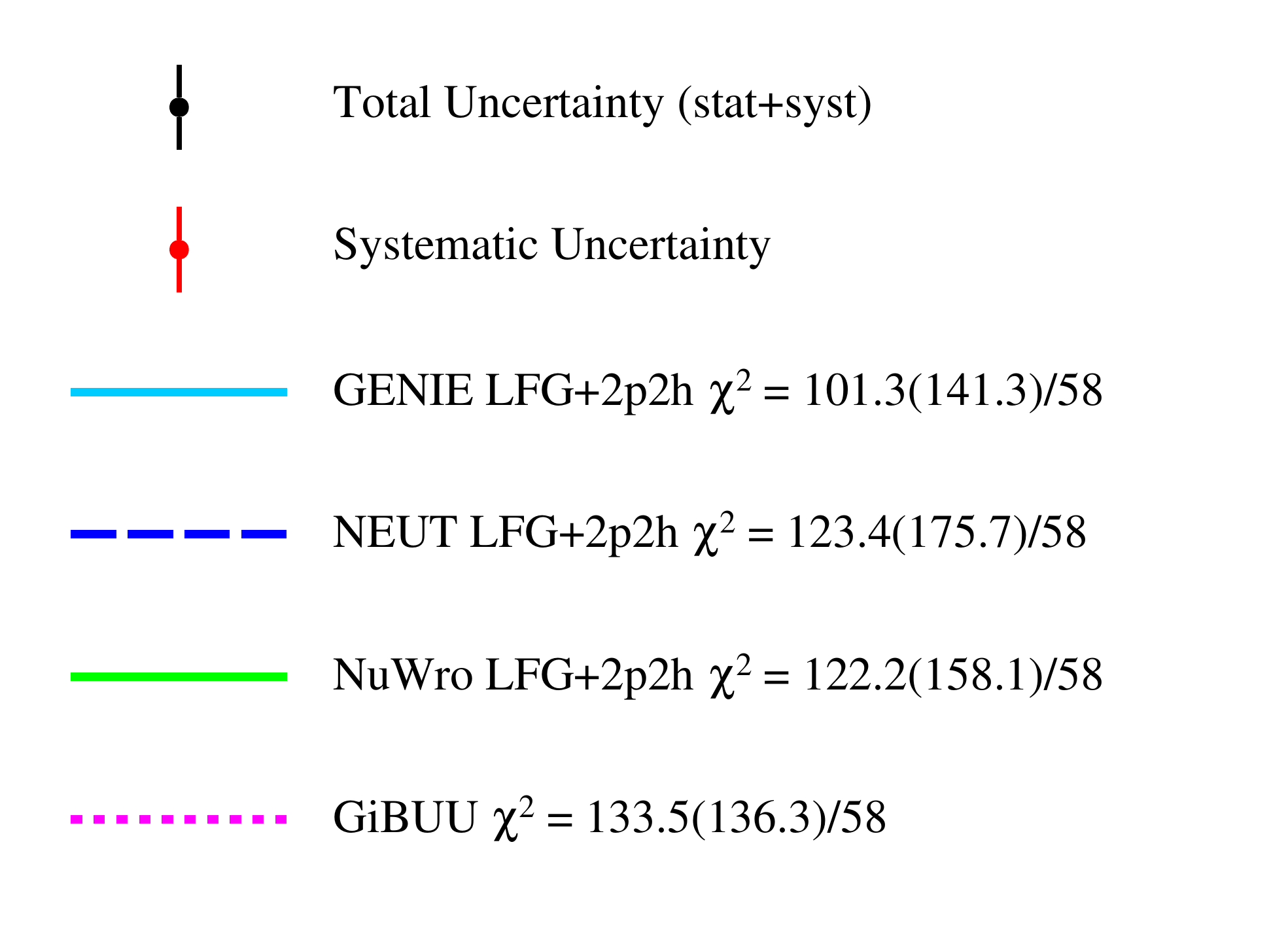}							
	\caption{Measured double-differential \numu + \barnumu \cczeropi cross-section sum in bins of true muon kinematics with systematic uncertainty (red bars) and total (stat.+syst.) uncertainty (black bars). The result is compared with \textsc{Neut} (dashed blue line), \textsc{NuWro} version~\texttt{18.02.1}  (green solid line) and \textsc{GiBUU}~\texttt{2019} (pink dotted line) prediction. All generators use an LFG+RPA model that includes 2p2h. The full and shape-only (in parenthesis) $\chi2$ are reported. The last bin in momentum is not displayed for readability.}
	\label{fig:xsecsumgibuuneutnuwro}
\end{figure*}

\begin{figure*}[h!]
	\centering
	\includegraphics[width=0.36\linewidth]{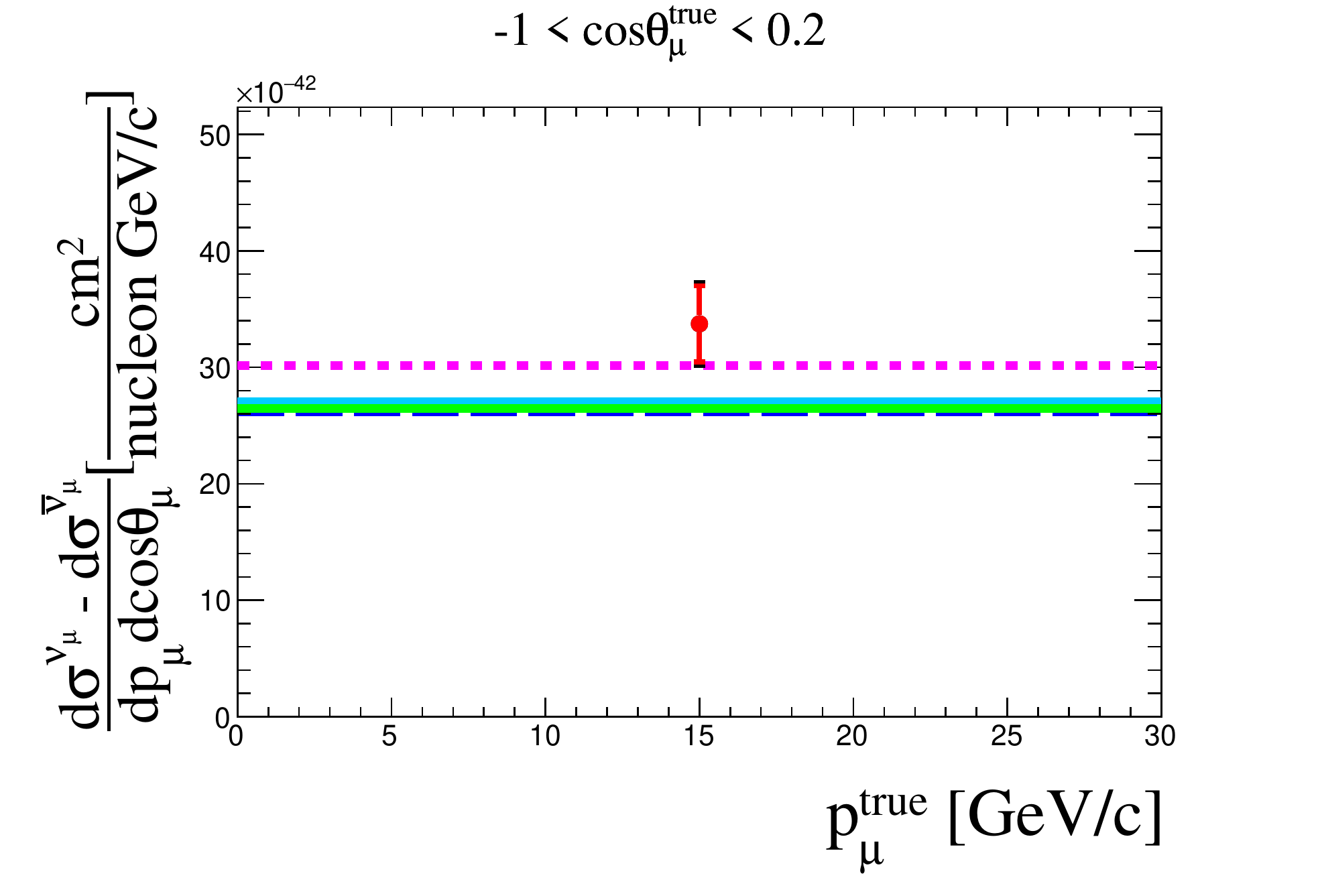}
	\includegraphics[width=0.36\linewidth]{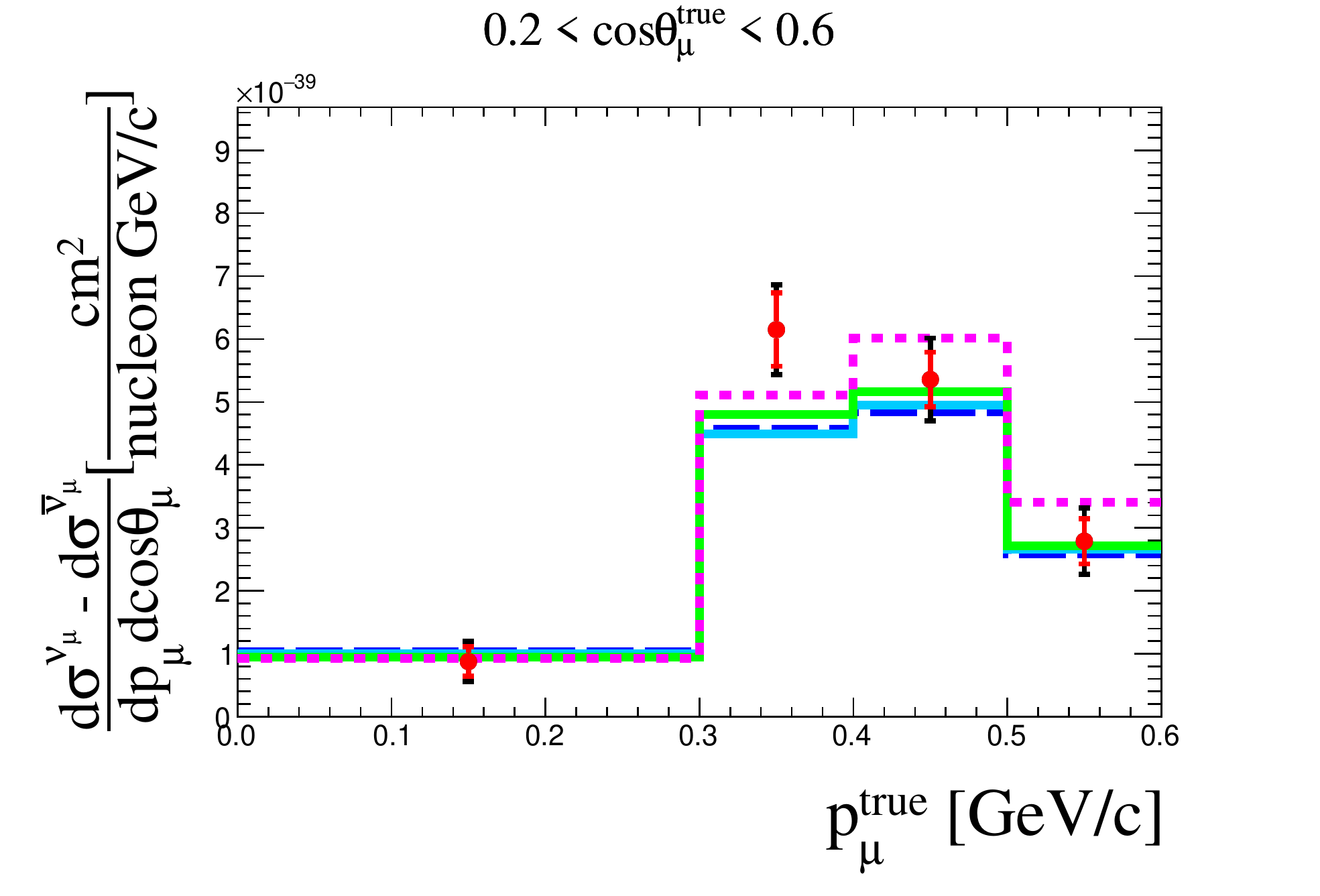}
	\includegraphics[width=0.36\linewidth]{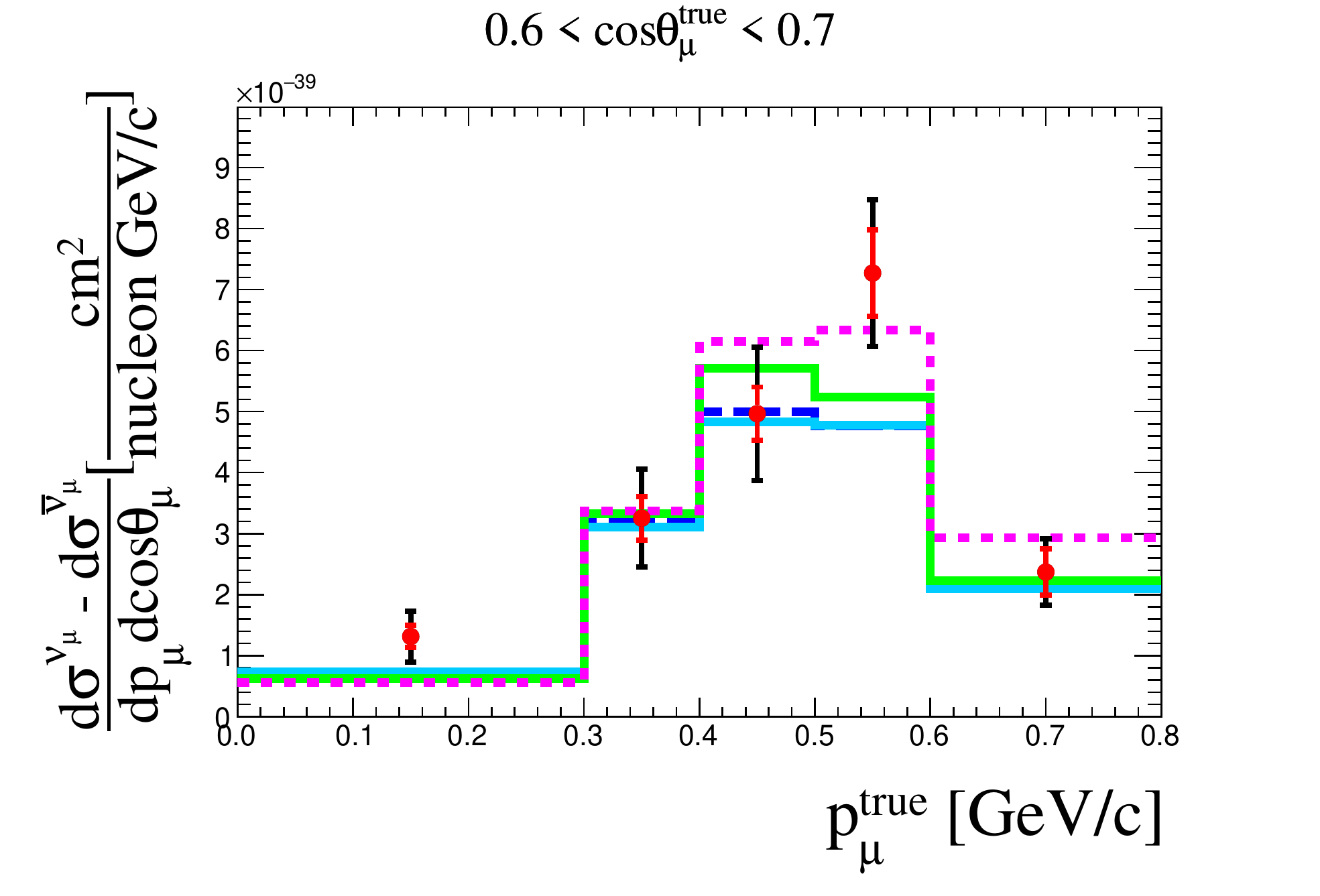}
	\includegraphics[width=0.36\linewidth]{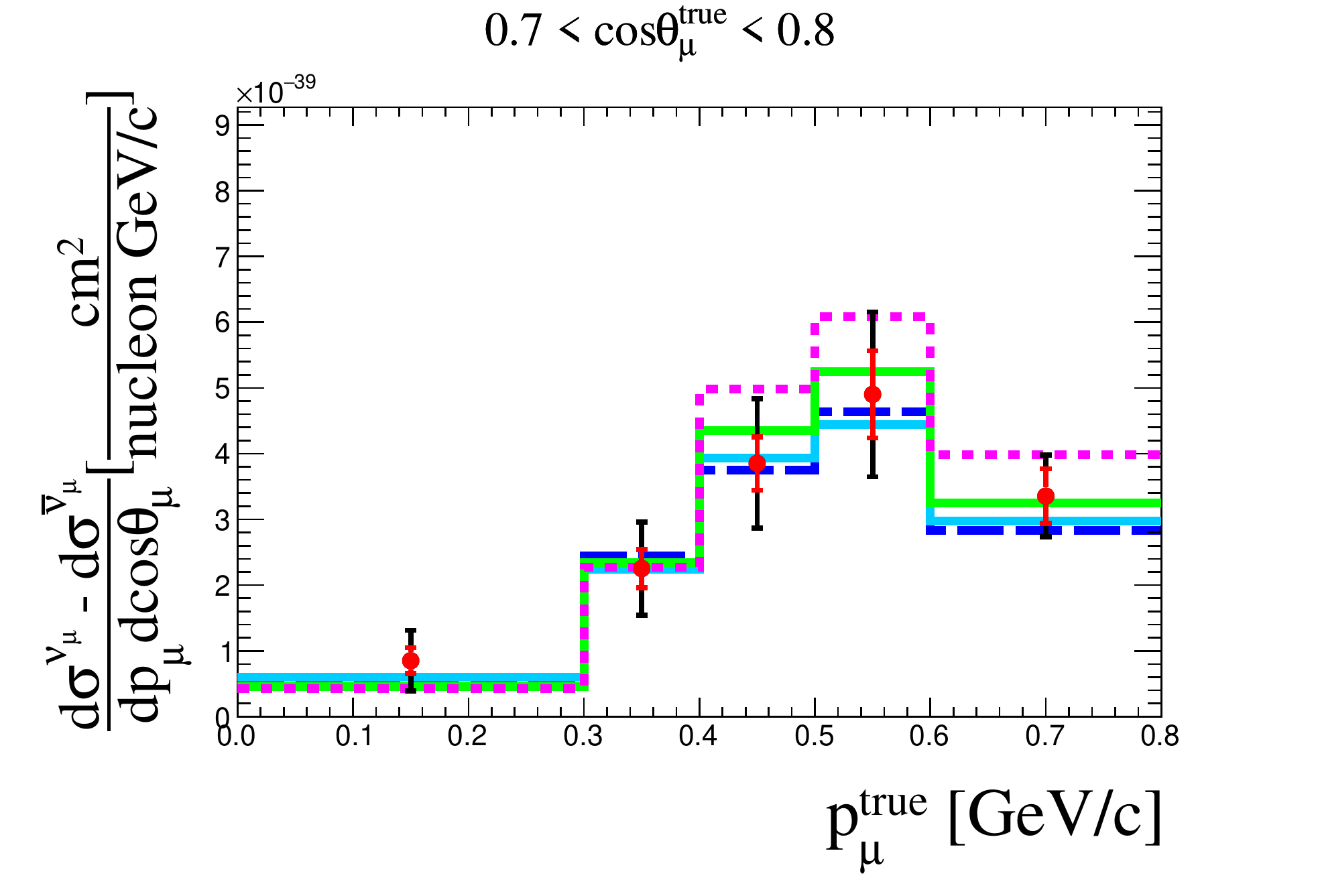}
	\includegraphics[width=0.36\linewidth]{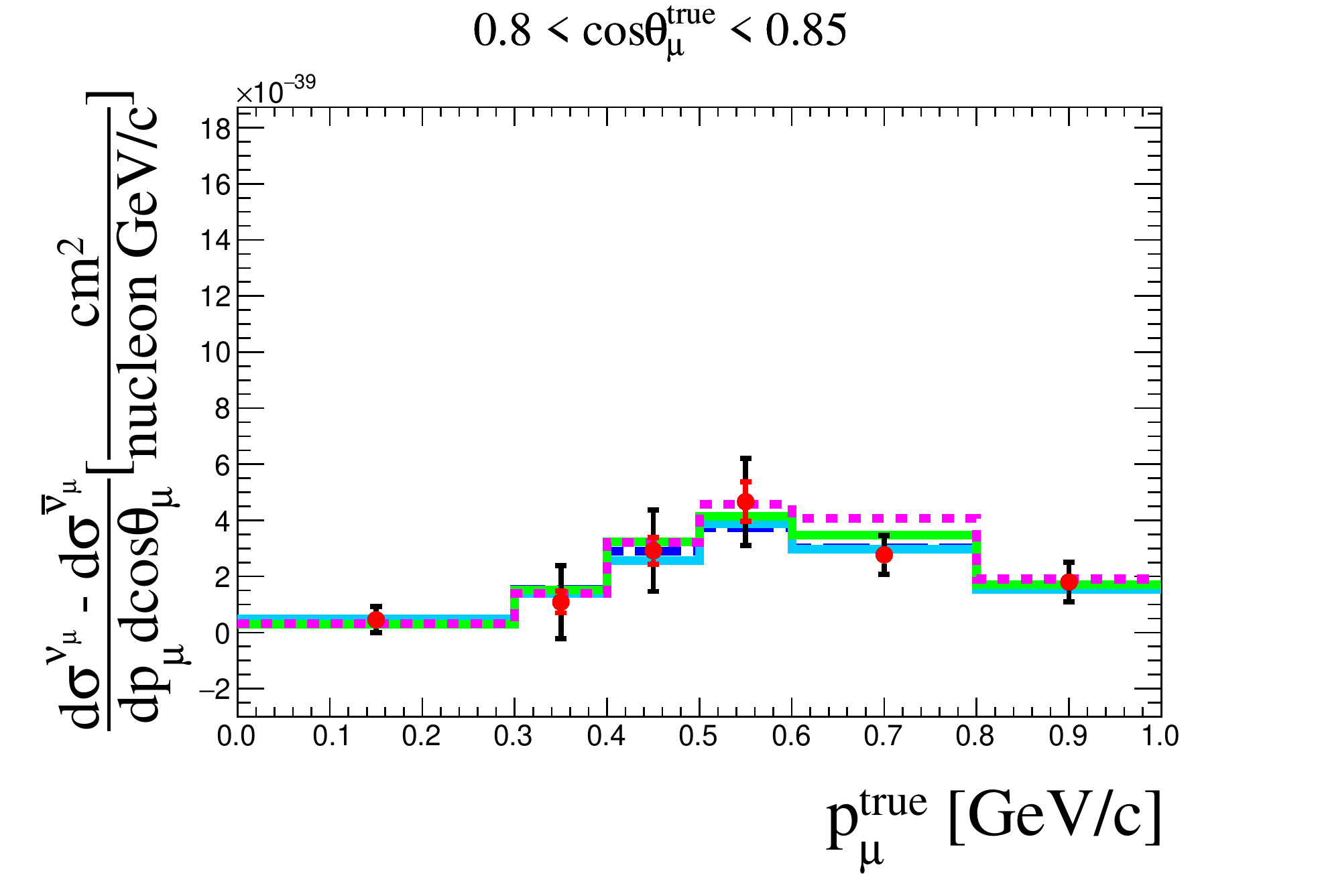}
	\includegraphics[width=0.36\linewidth]{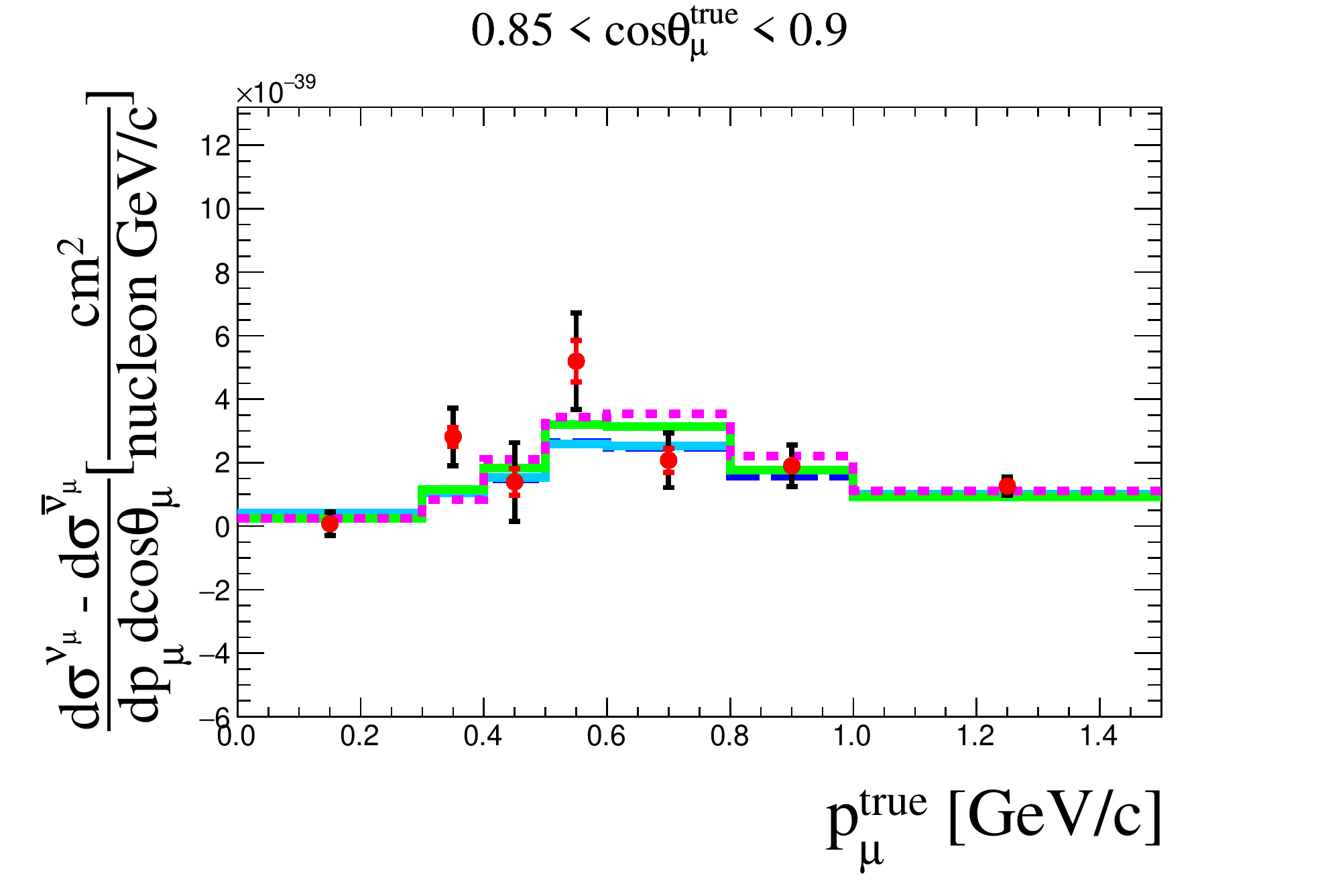}
	\includegraphics[width=0.36\linewidth]{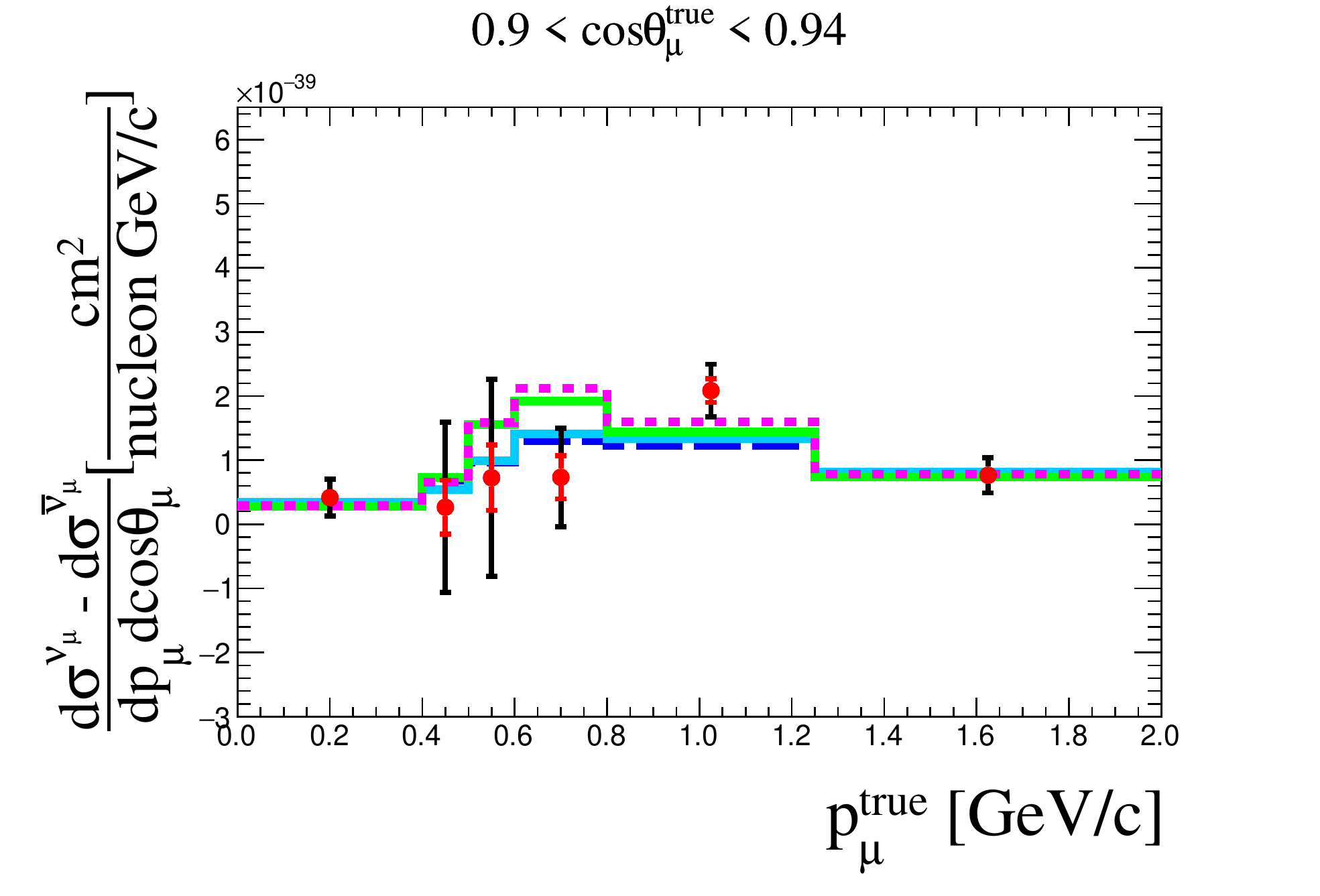}
	\includegraphics[width=0.36\linewidth]{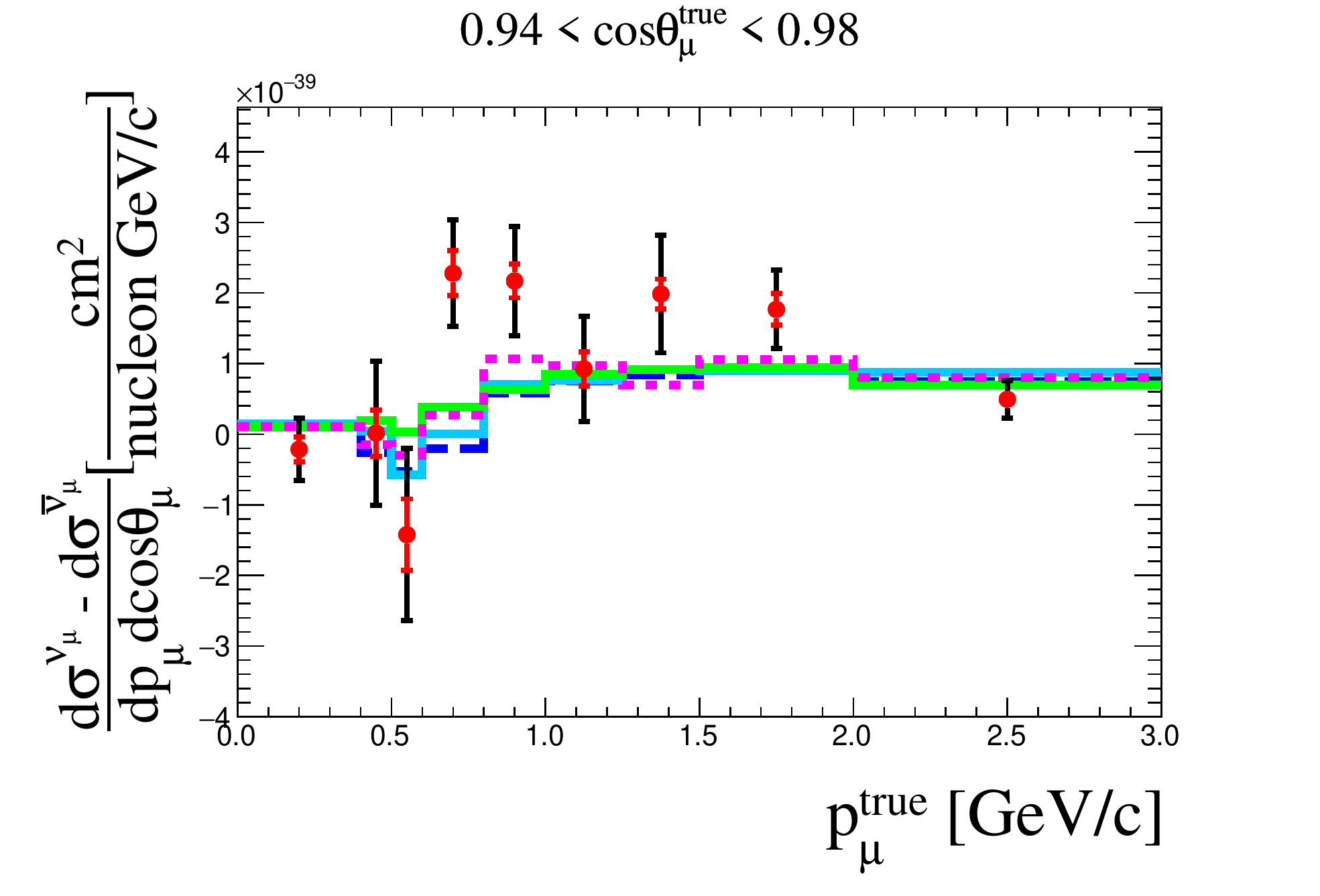}
	\includegraphics[width=0.36\linewidth]{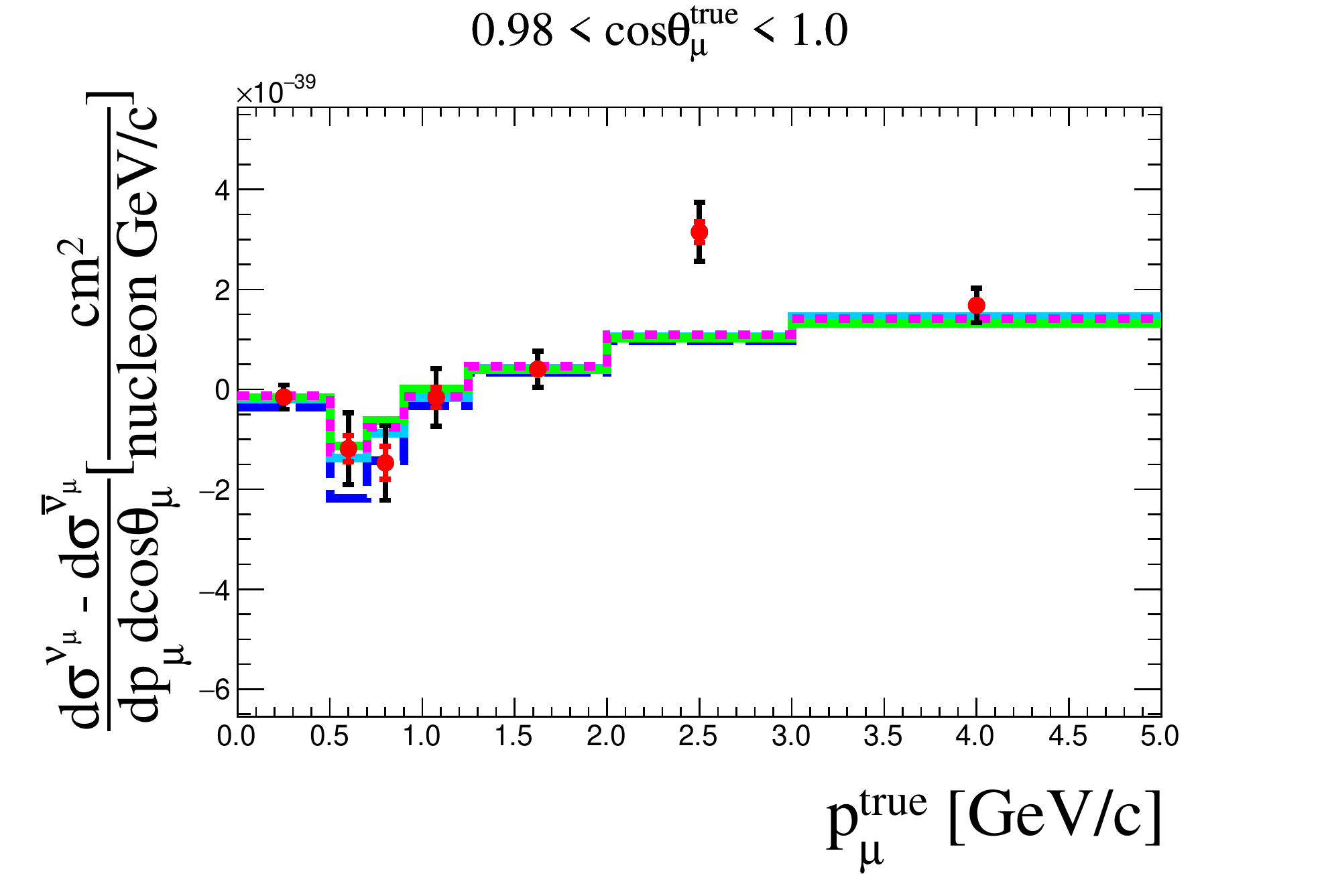}	
	\includegraphics[width=0.36\linewidth]{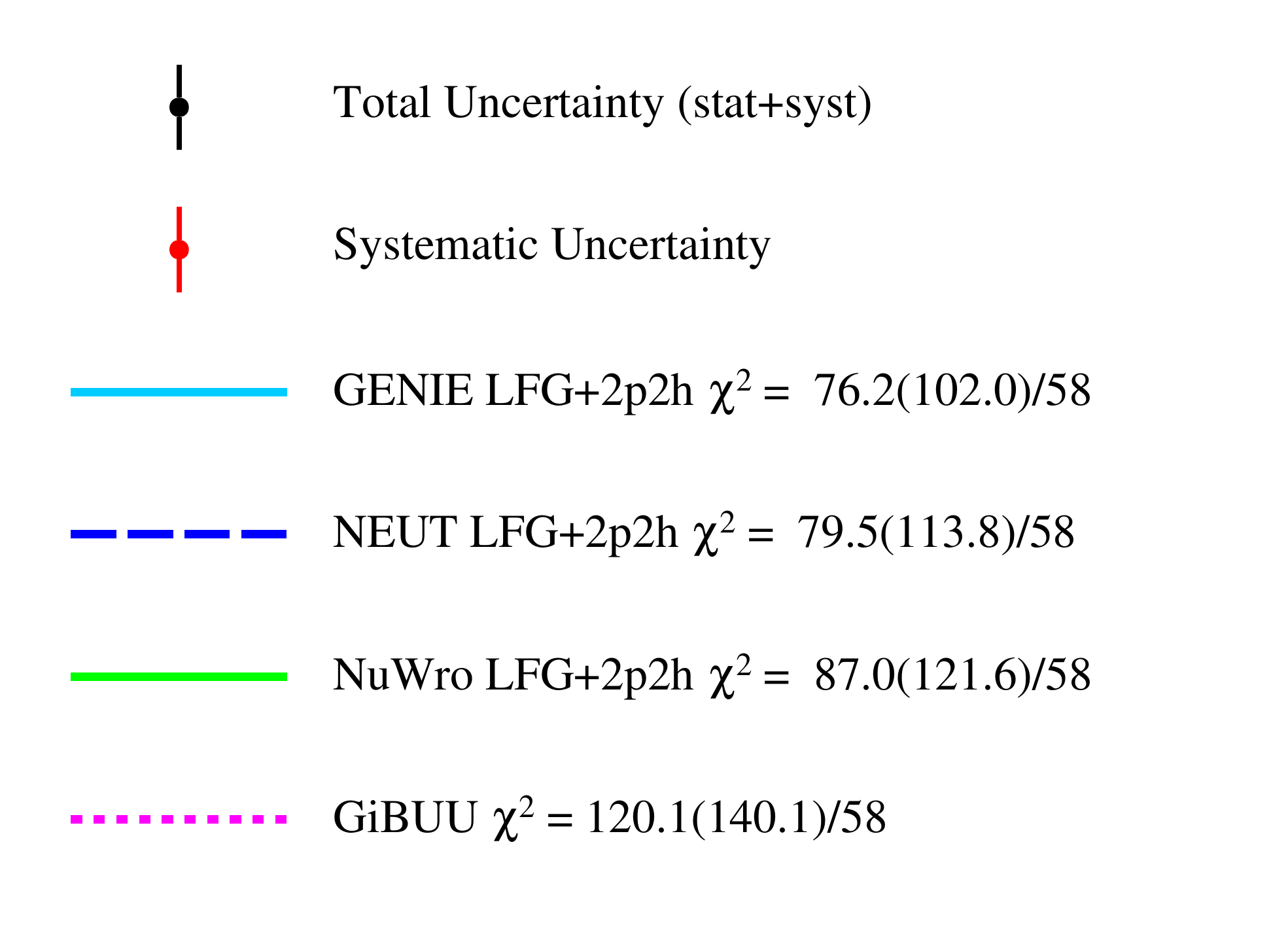}				
	\caption{Measured double-differential \numu - \barnumu \cczeropi cross-section difference in bins of true muon kinematics with systematic uncertainty (red bars) and total (stat.+syst.) uncertainty (black bars). The result is compared with \textsc{Neut} (dashed blue line), \textsc{NuWro} version~\texttt{18.02.1}  (green solid line) and \textsc{GiBUU}~\texttt{2019} (pink dotted line) prediction. All generators use an LFG+RPA model that includes 2p2h. The full and shape-only (in parenthesis) $\chi2$ are reported. The last bin in momentum is not displayed for readability.}
	\label{fig:xsecdifgibuuneutnuwro}
\end{figure*}

\begin{figure*}[h!]
	\centering
	\includegraphics[width=0.36\linewidth]{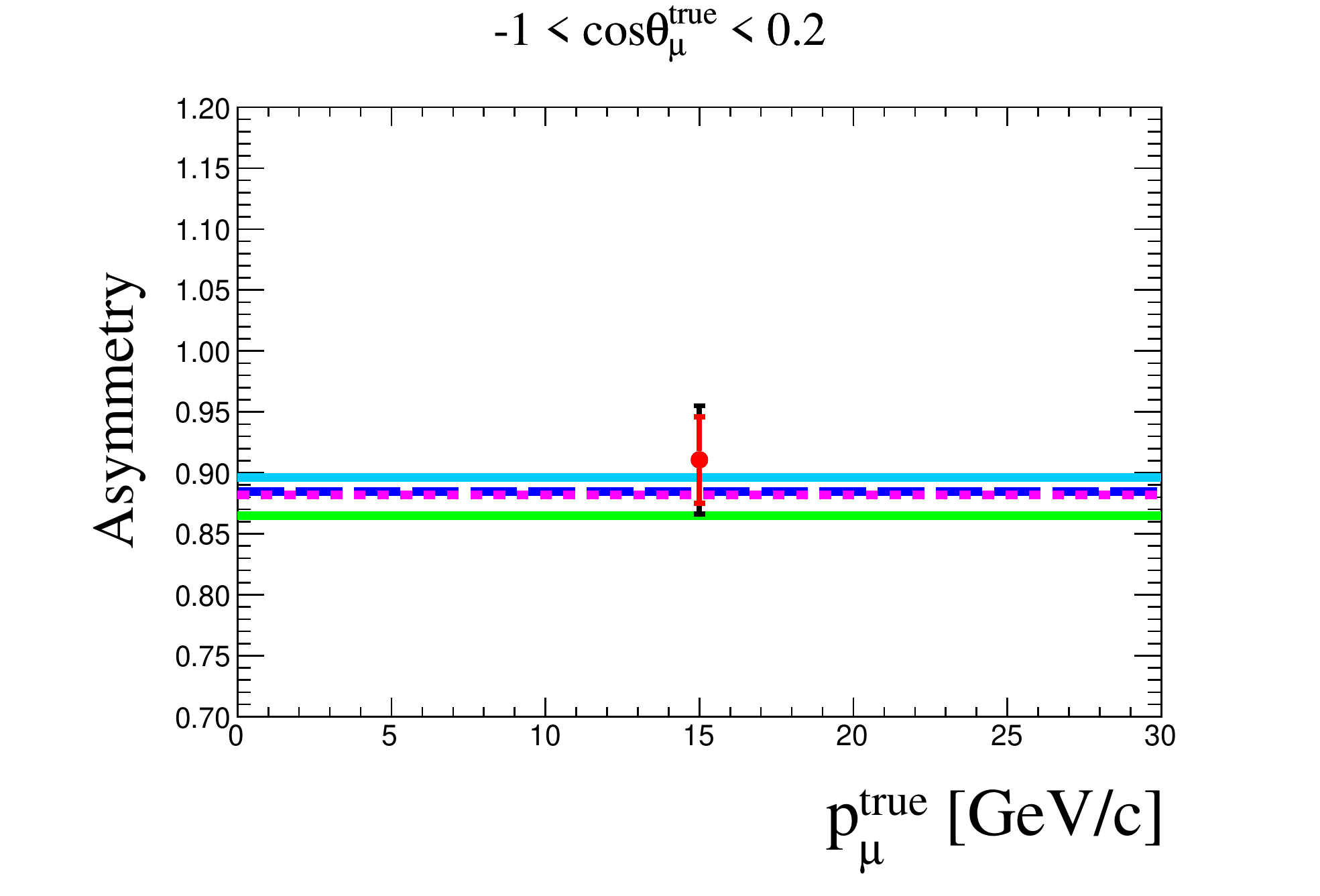}
	\includegraphics[width=0.36\linewidth]{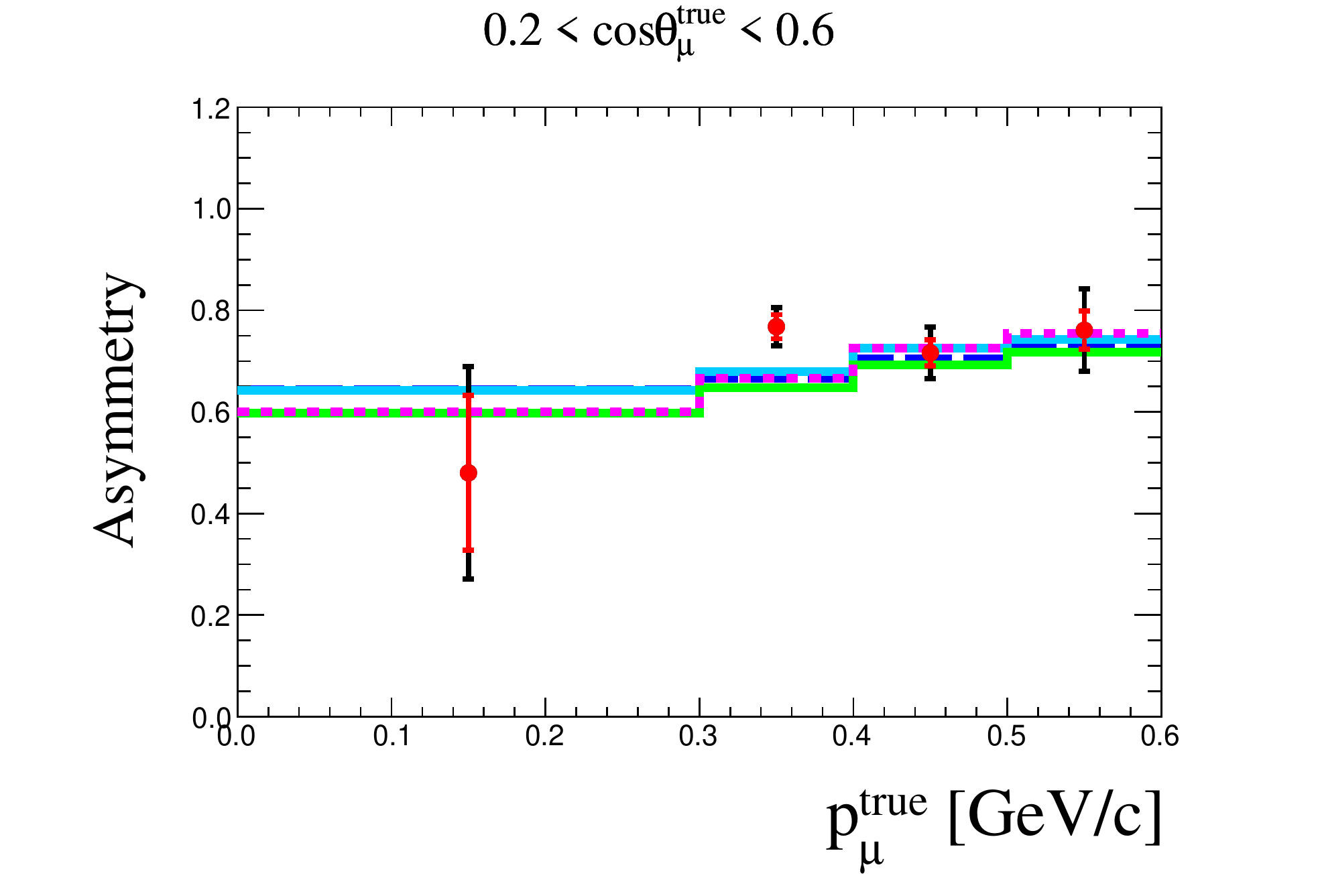}
	\includegraphics[width=0.36\linewidth]{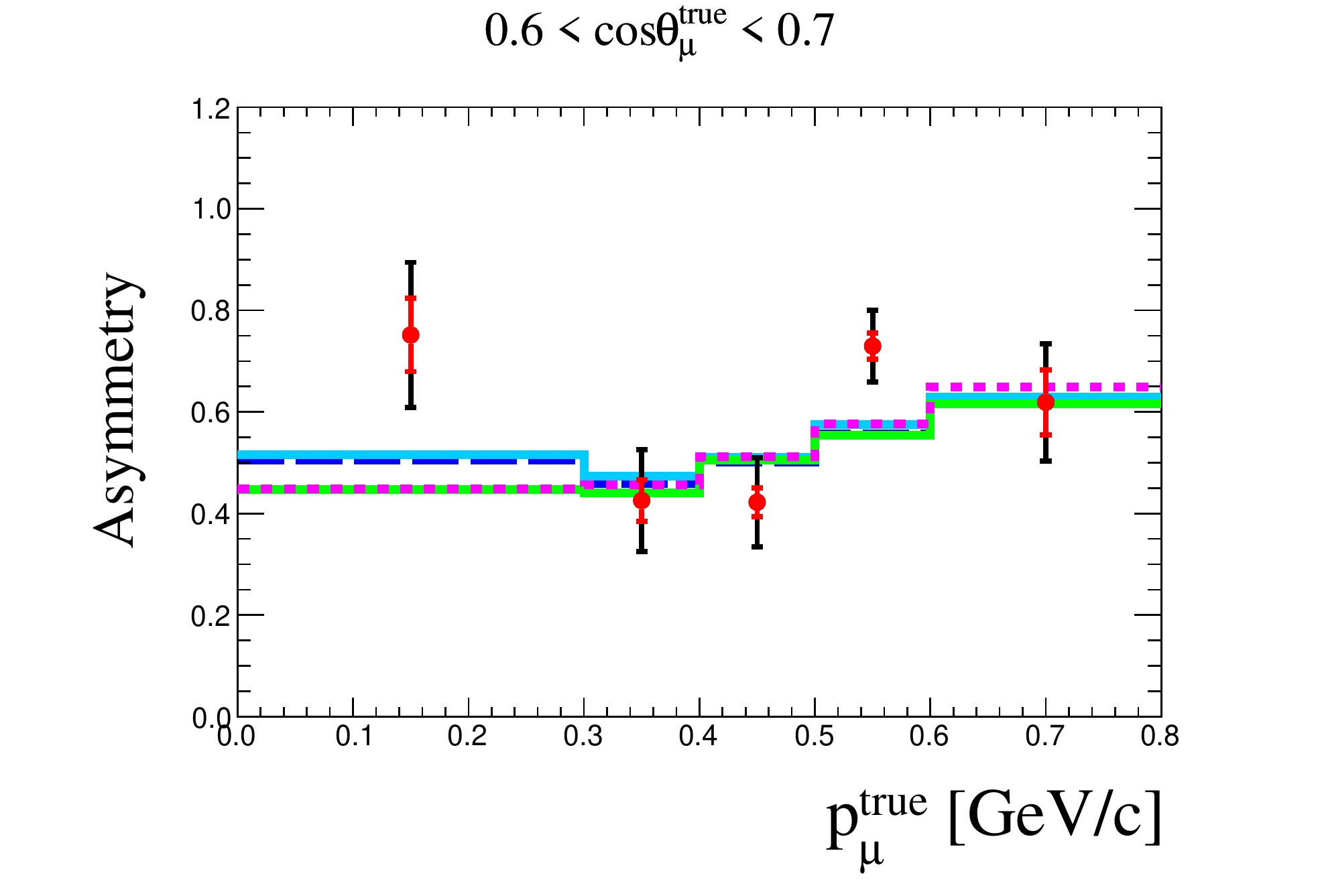}
	\includegraphics[width=0.36\linewidth]{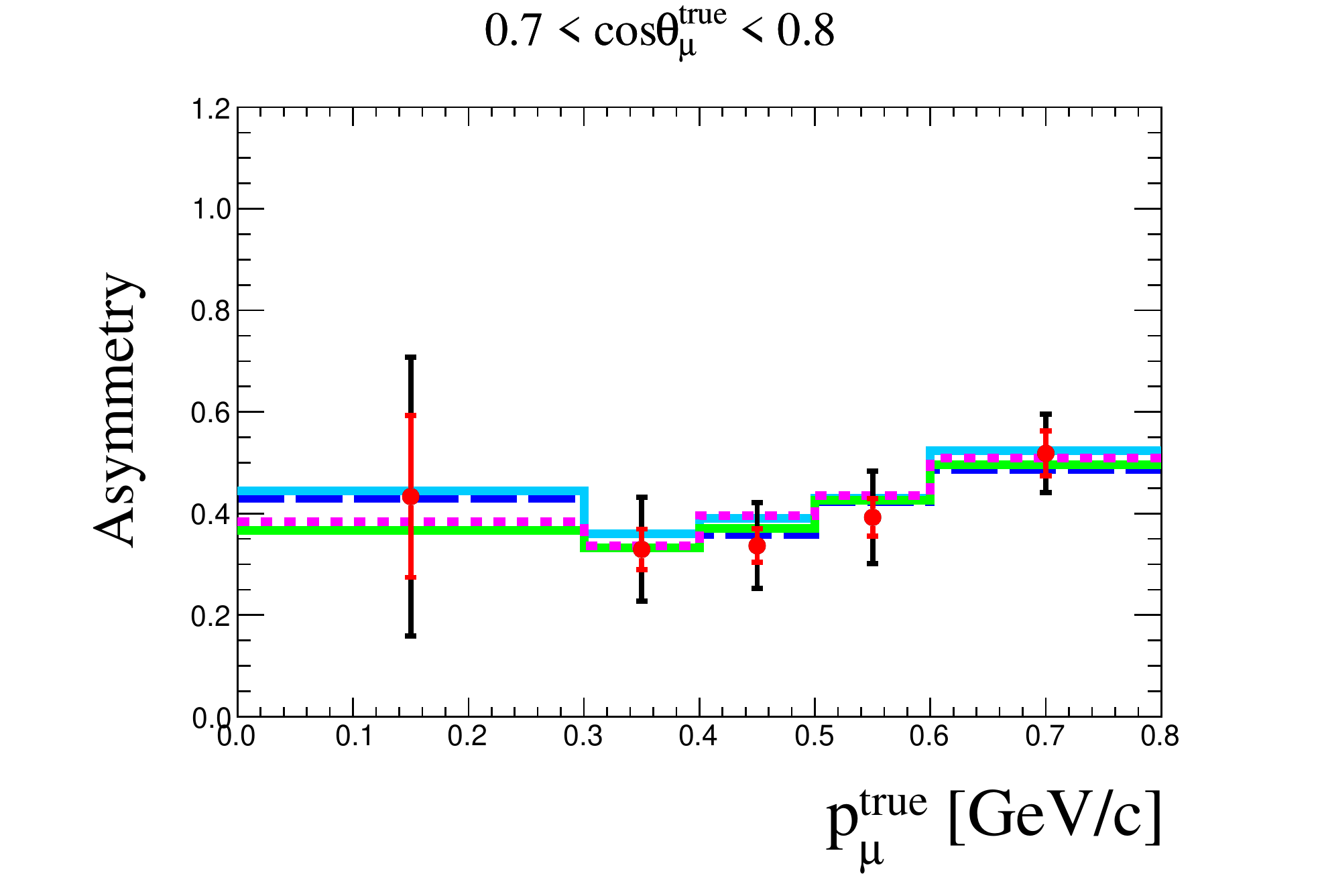}
	\includegraphics[width=0.36\linewidth]{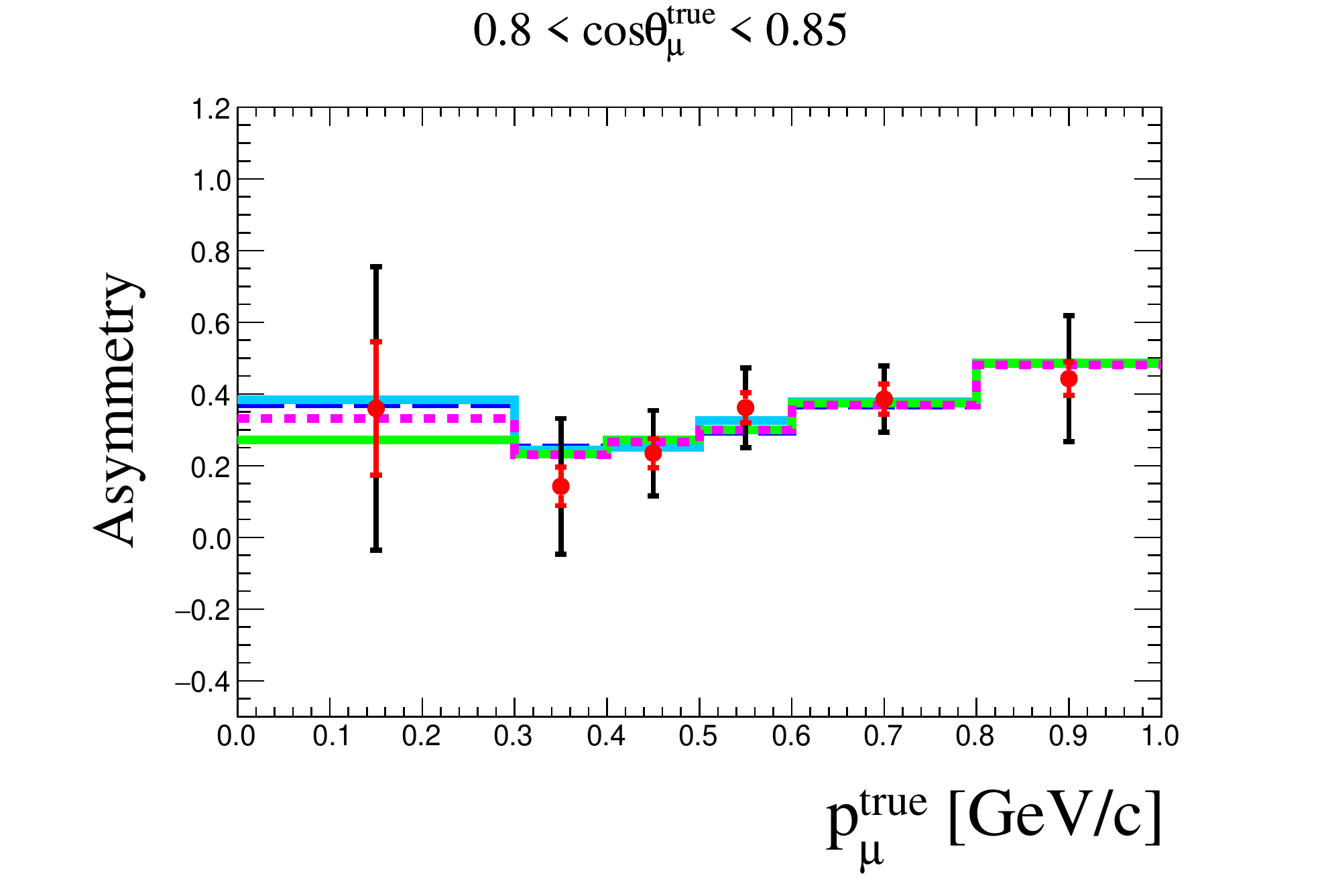}
	\includegraphics[width=0.36\linewidth]{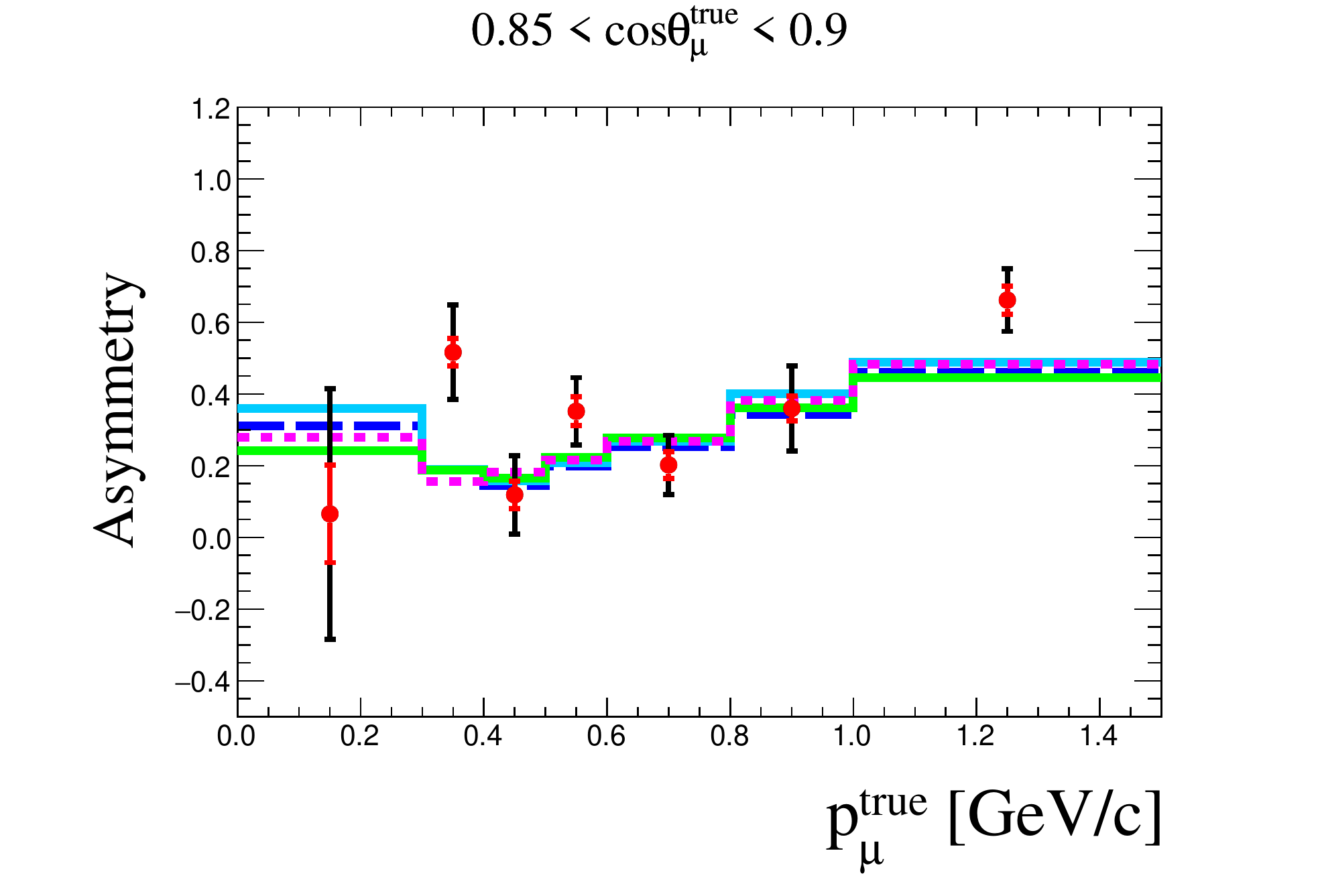}
	\includegraphics[width=0.36\linewidth]{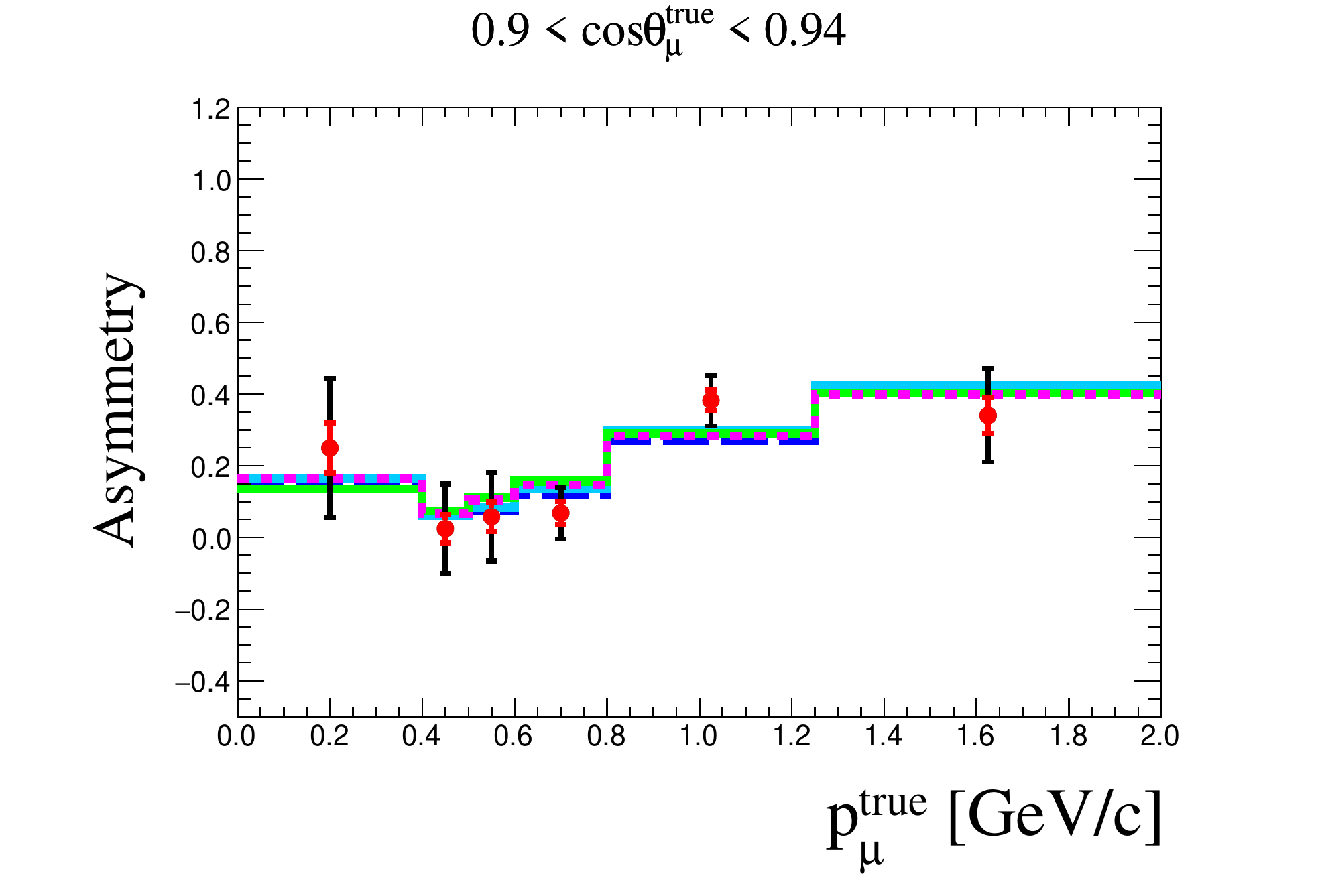}
	\includegraphics[width=0.36\linewidth]{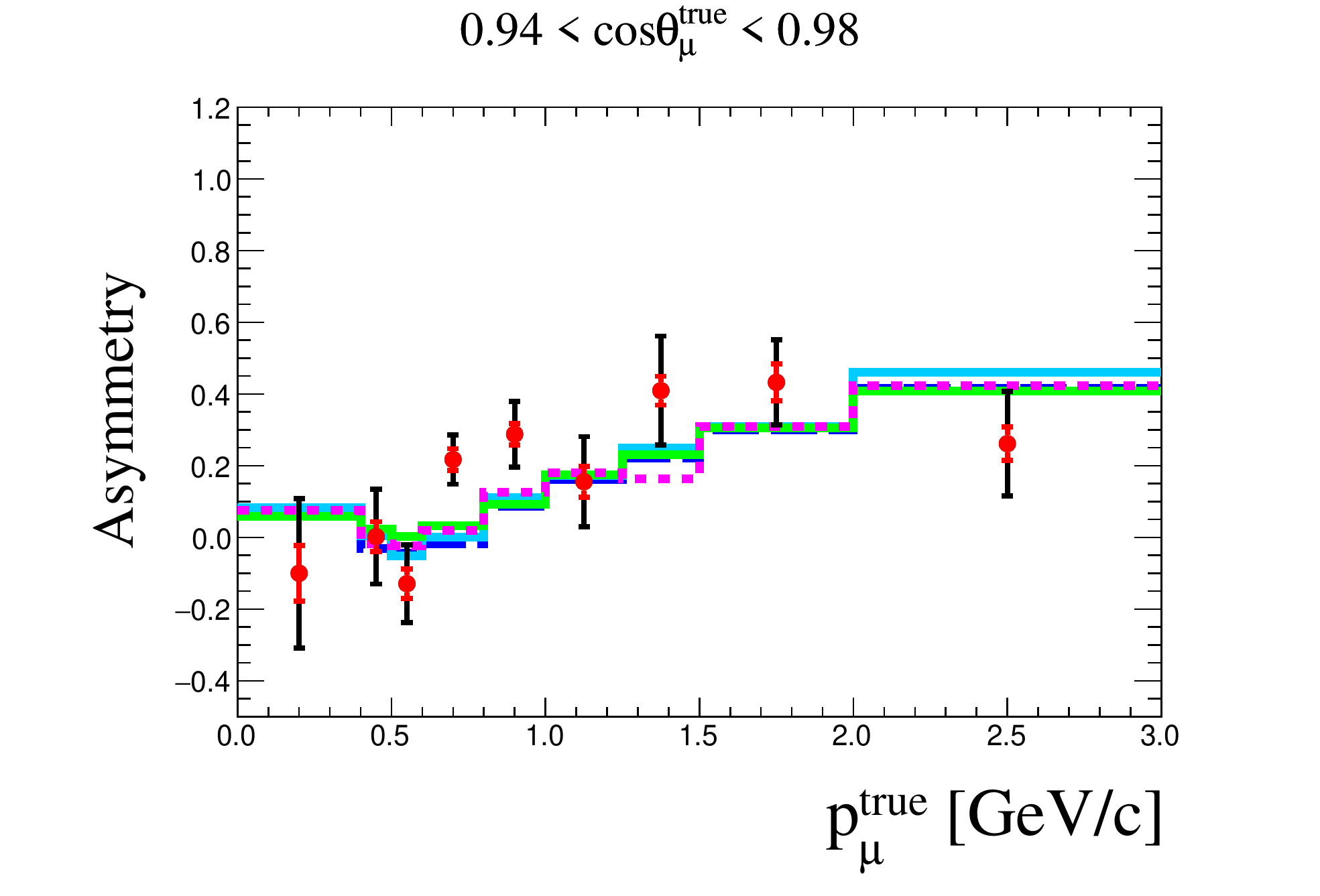}			
	\includegraphics[width=0.36\linewidth]{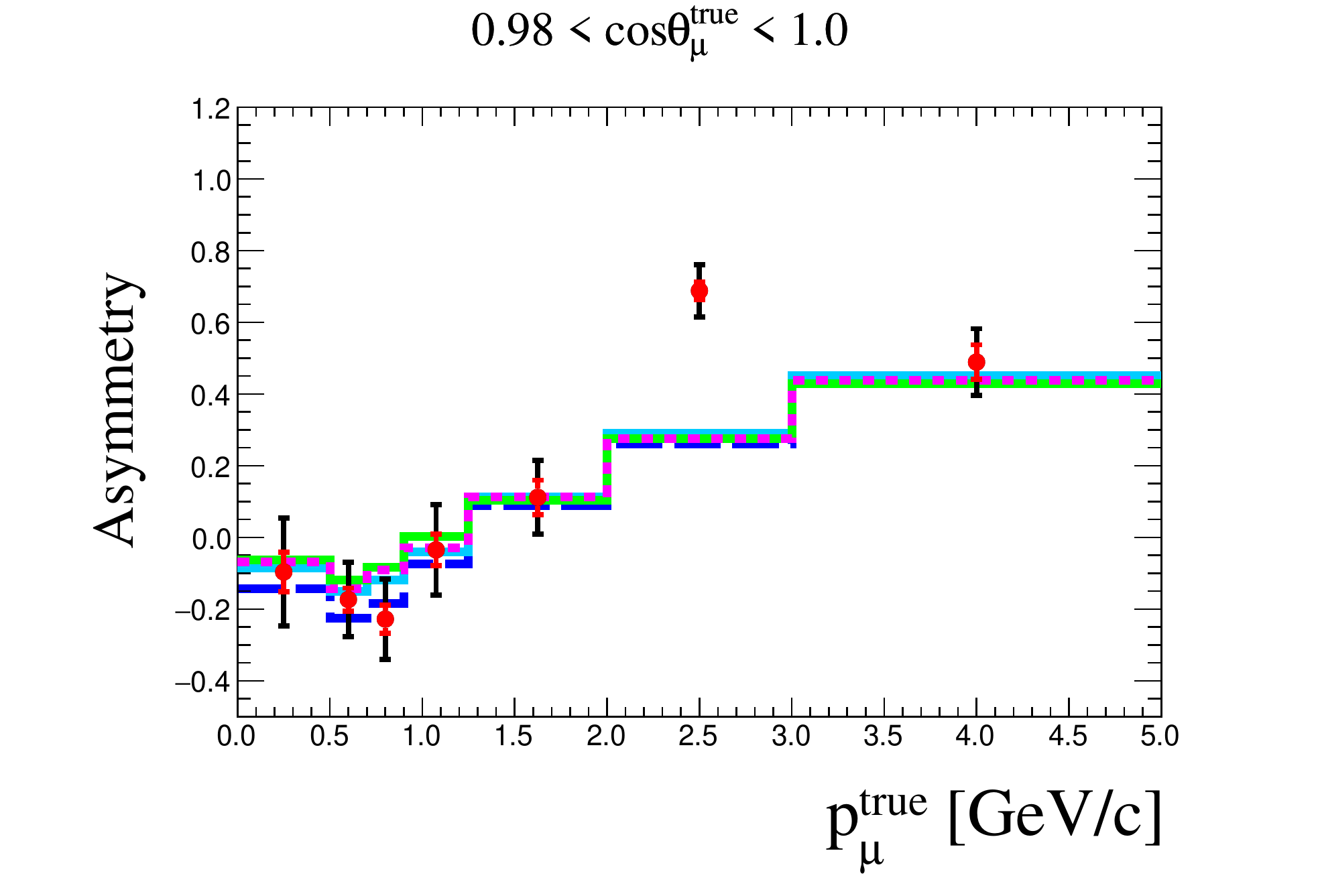}	
	\includegraphics[width=0.36\linewidth]{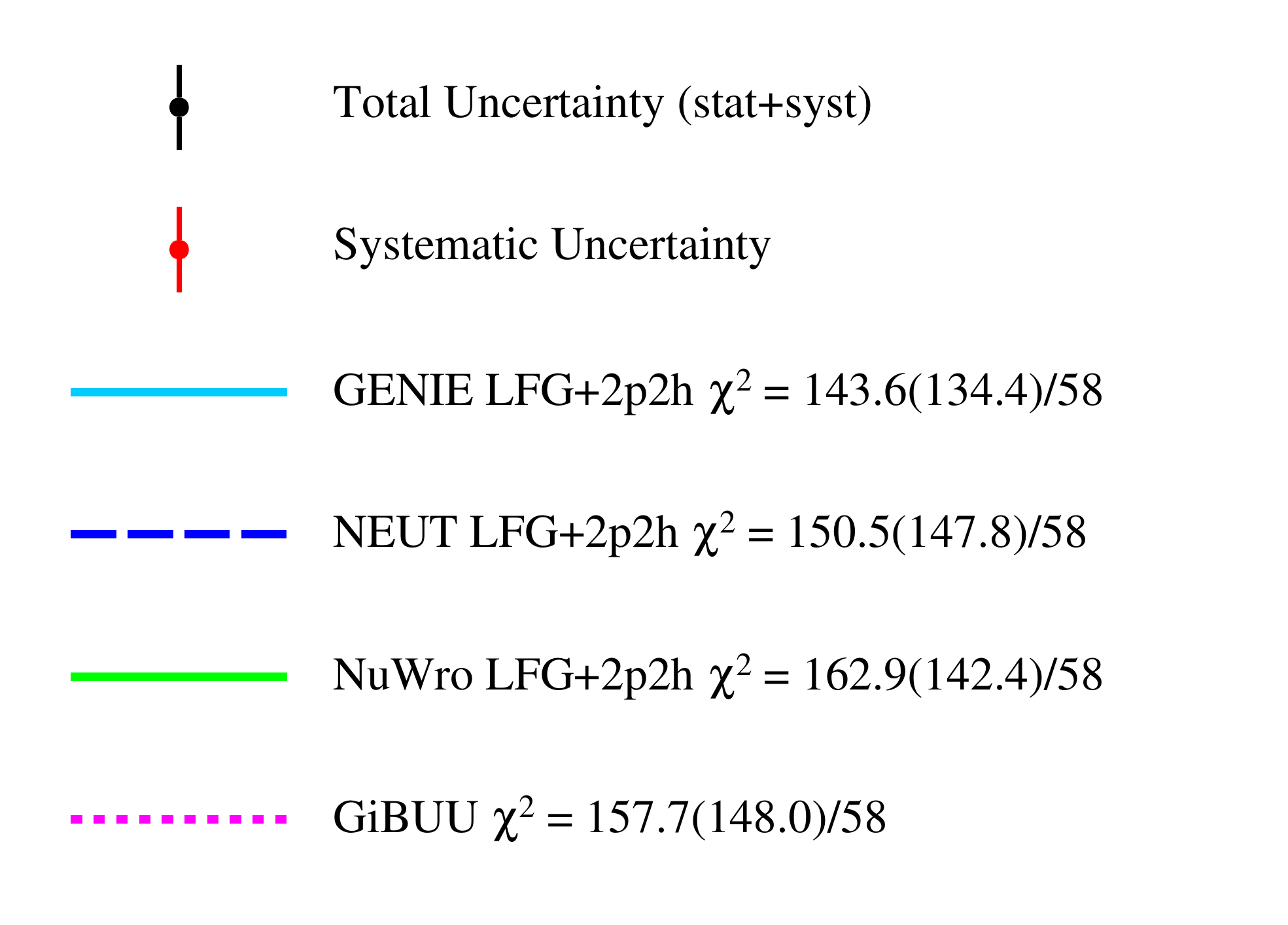}	
	\caption{Measured double-differential \cczeropi cross-section asymmetry in bins of true muon kinematics with systematic uncertainty (red bars) and total (stat.+syst.) uncertainty (black bars). The result is compared with \textsc{Neut} (dashed blue line), \textsc{NuWro} version~\texttt{18.02.1}  (green solid line) and \textsc{GiBUU}~\texttt{2019} (pink dotted line) prediction. All generators use an LFG+RPA model that includes 2p2h. The full and shape-only (in parenthesis) $\chi2$ are reported. The last bin in momentum is not displayed for readability.}
	\label{fig:xsecasygibuuneutnuwro}
\end{figure*}

\begin{figure*}[h!]
	\centering
	\includegraphics[width=0.36\linewidth]{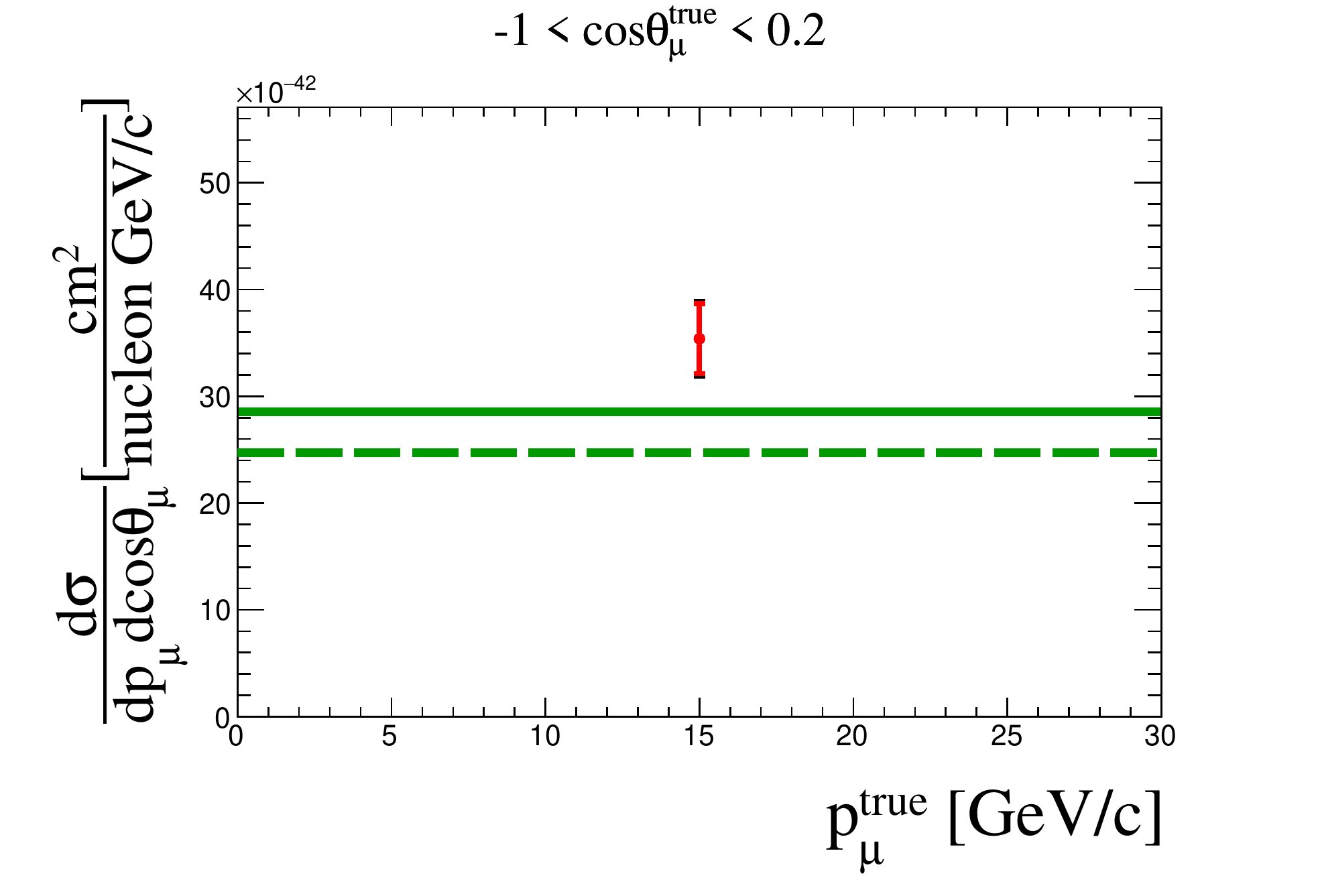}
	\includegraphics[width=0.36\linewidth]{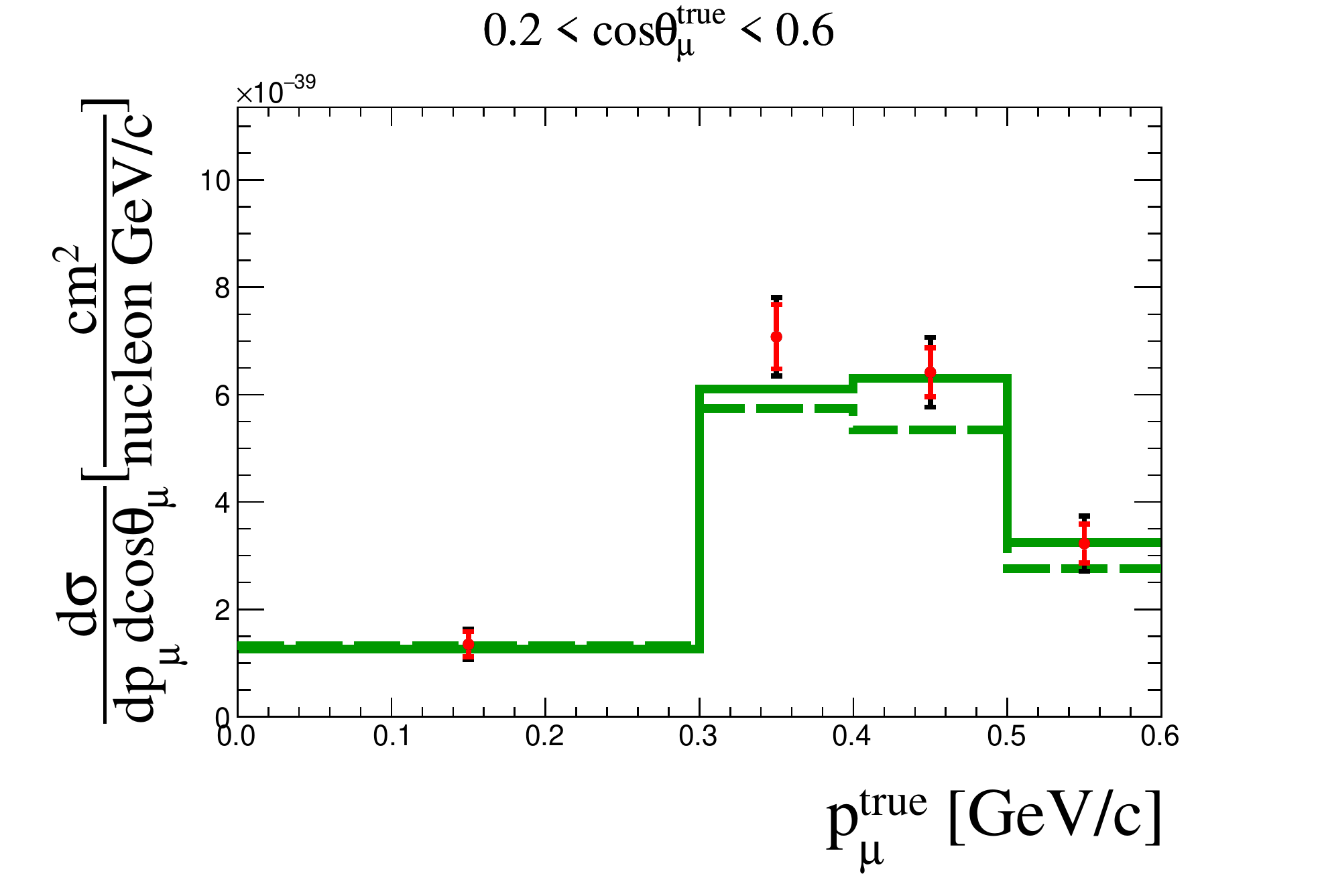}
	\includegraphics[width=0.36\linewidth]{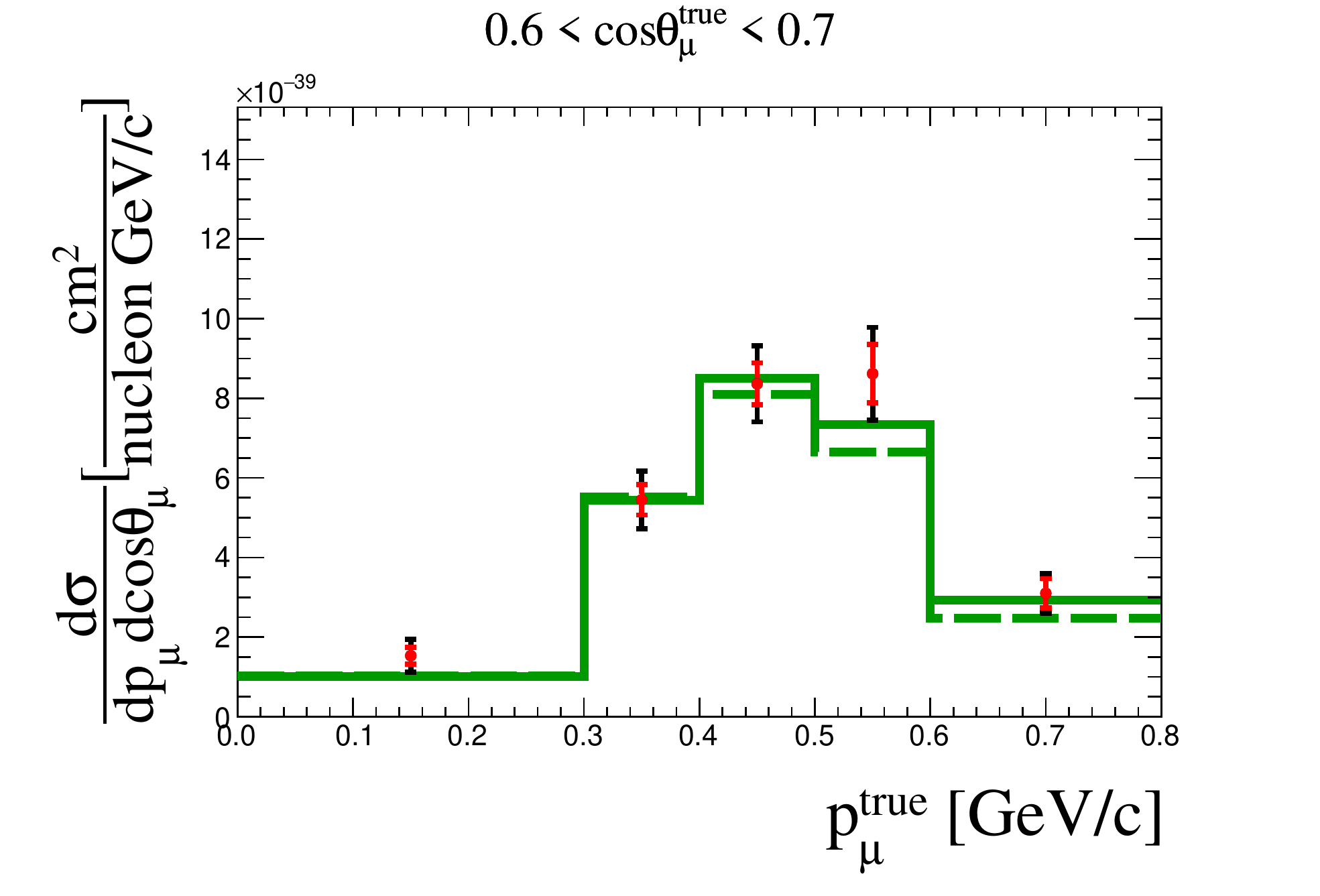}
	\includegraphics[width=0.36\linewidth]{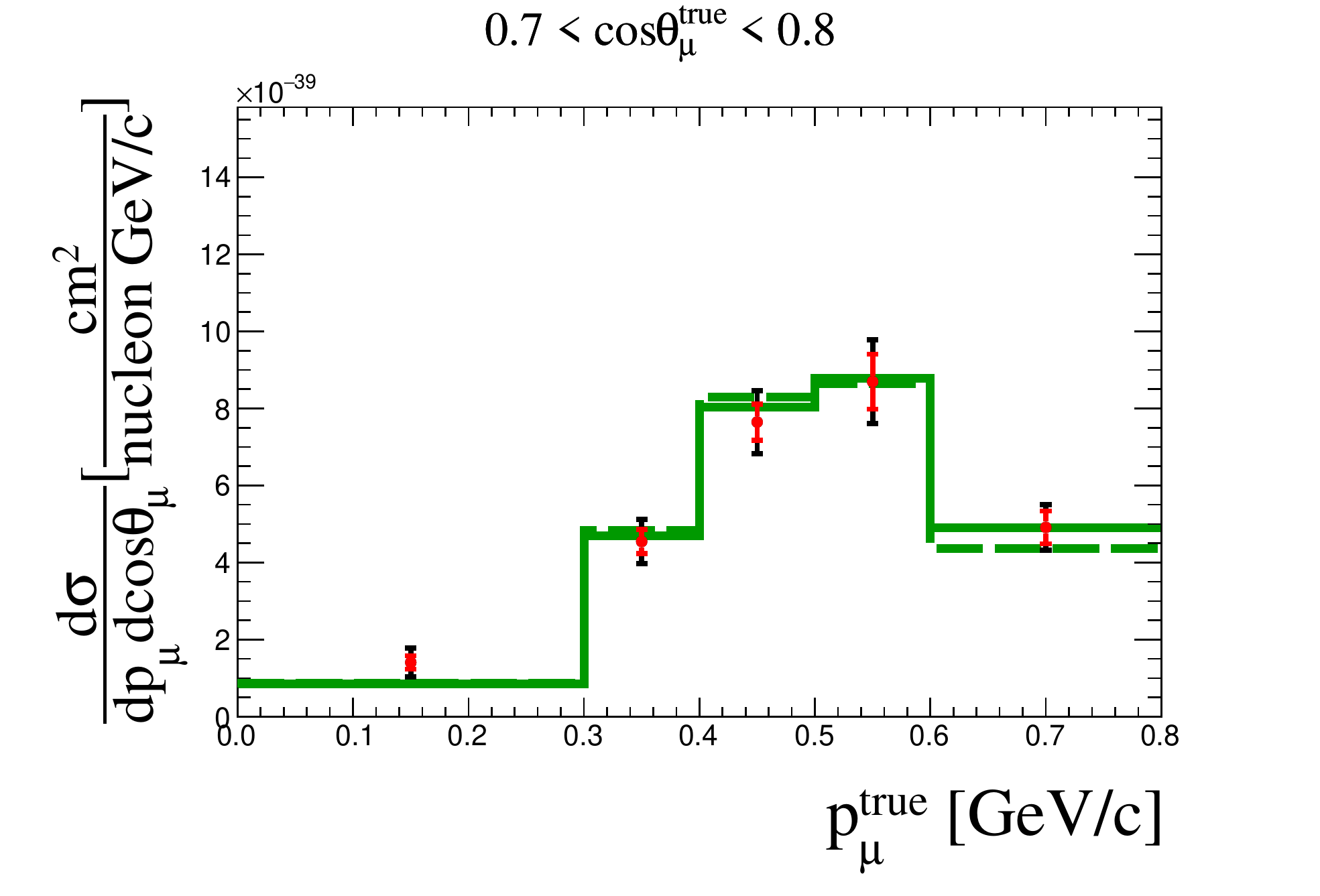}
	\includegraphics[width=0.36\linewidth]{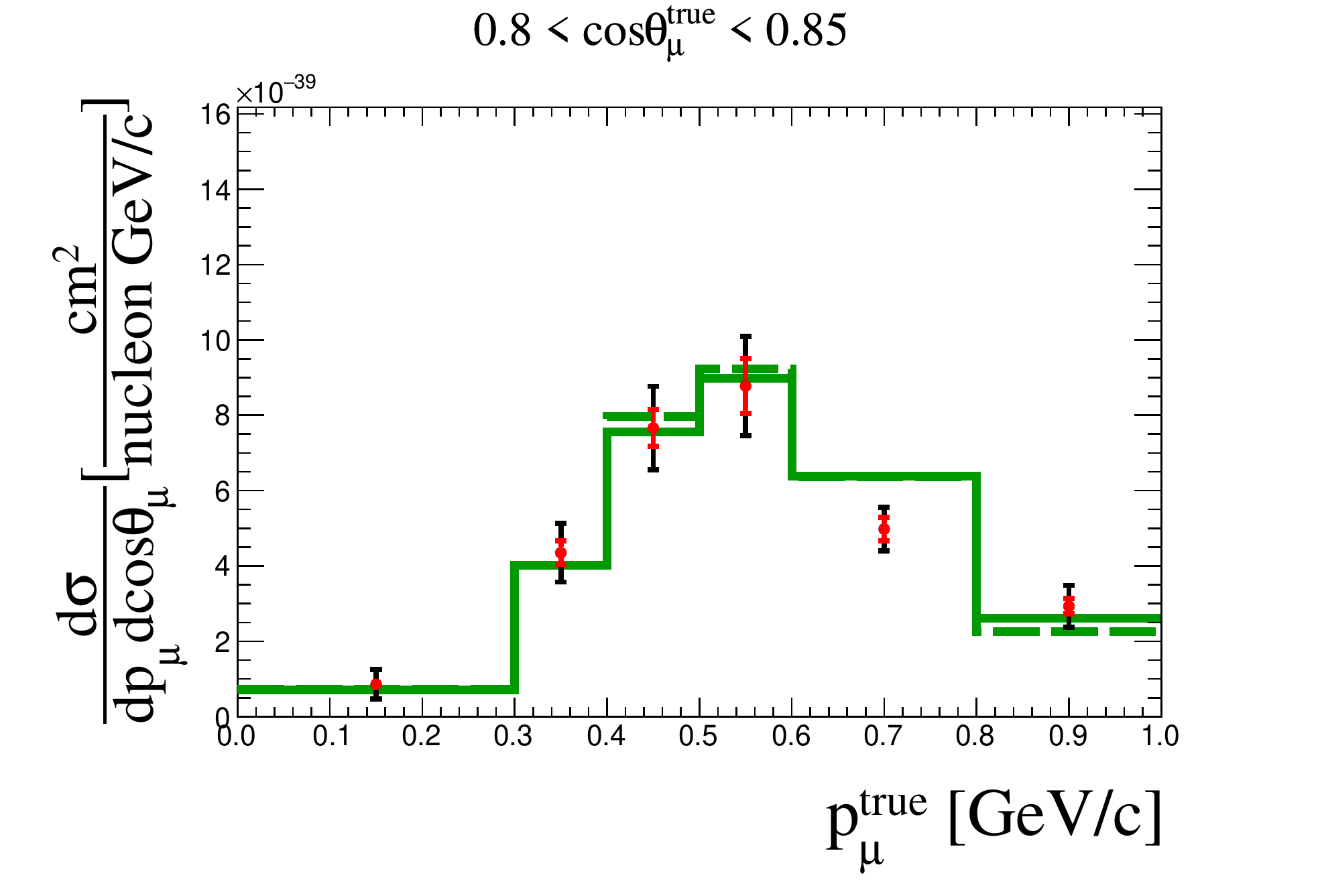}
	\includegraphics[width=0.36\linewidth]{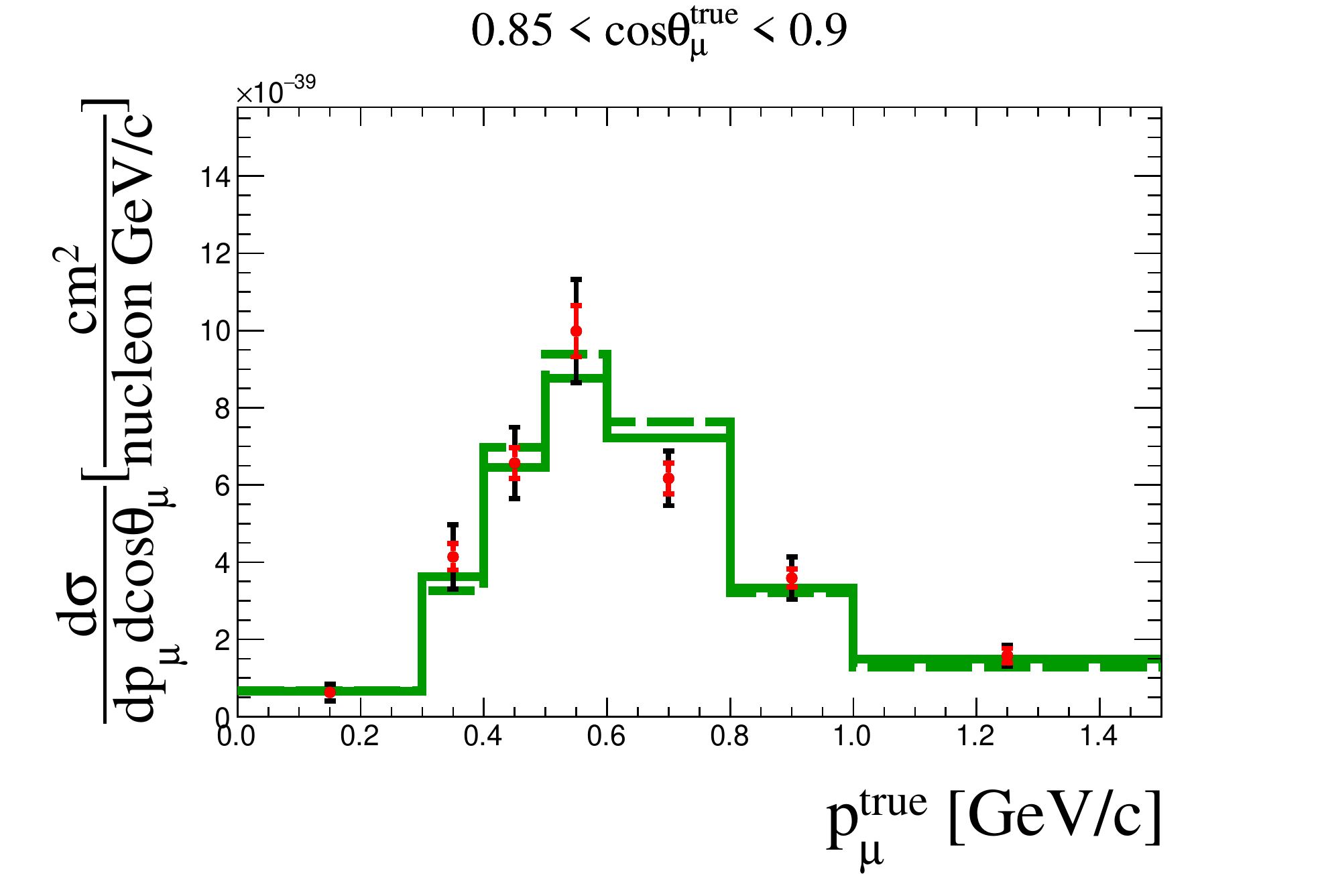}
	\includegraphics[width=0.36\linewidth]{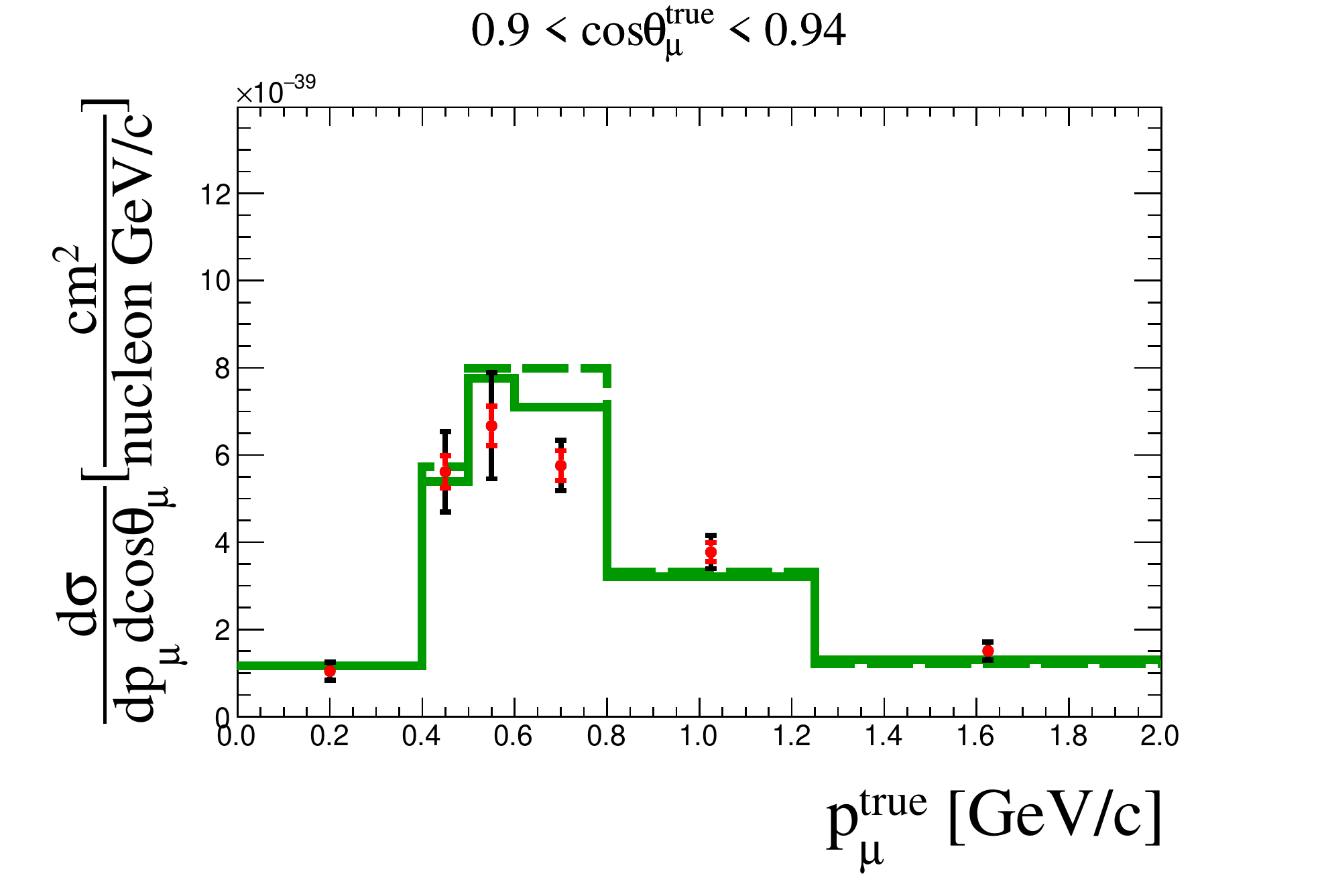}
	\includegraphics[width=0.36\linewidth]{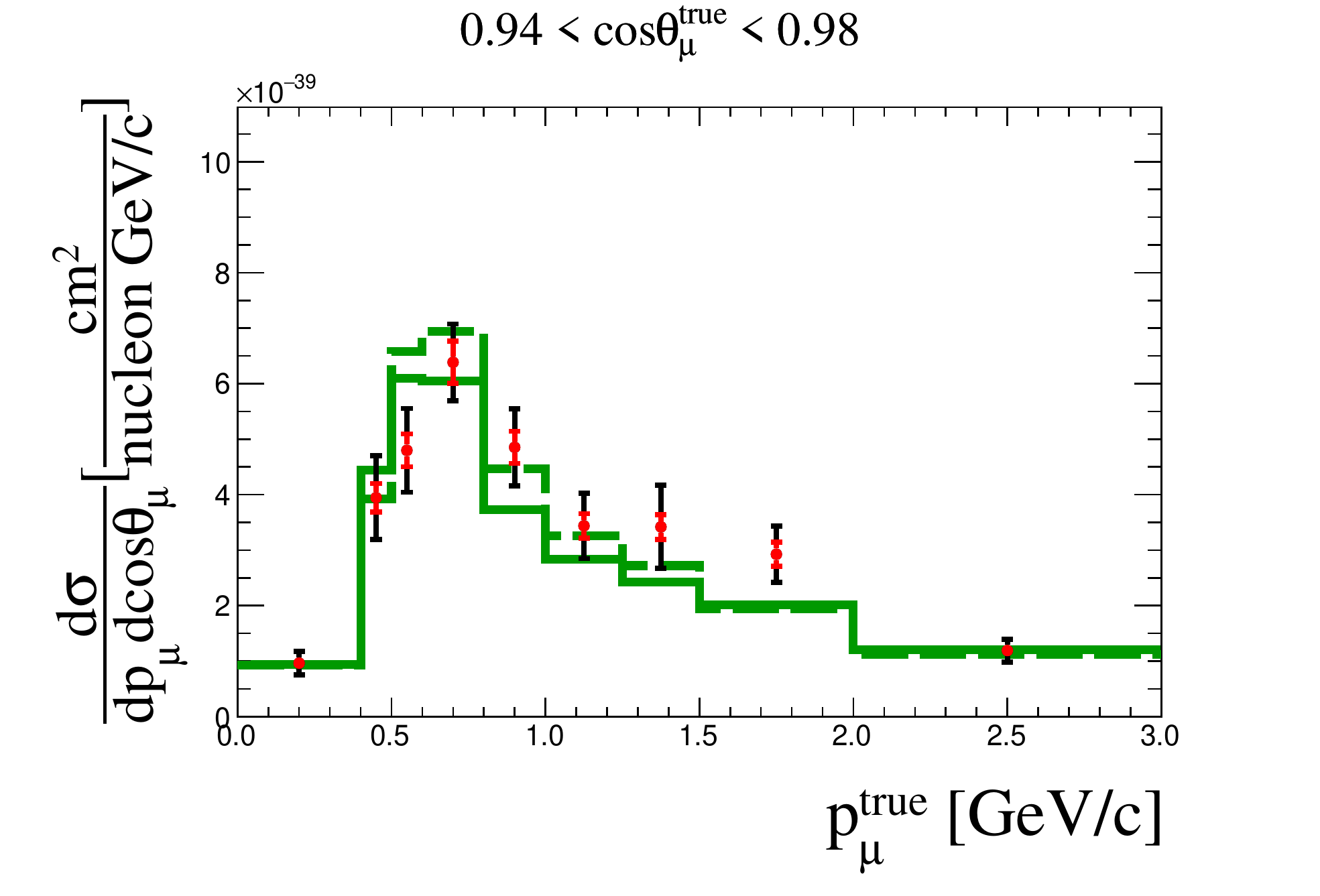}
	\includegraphics[width=0.36\linewidth]{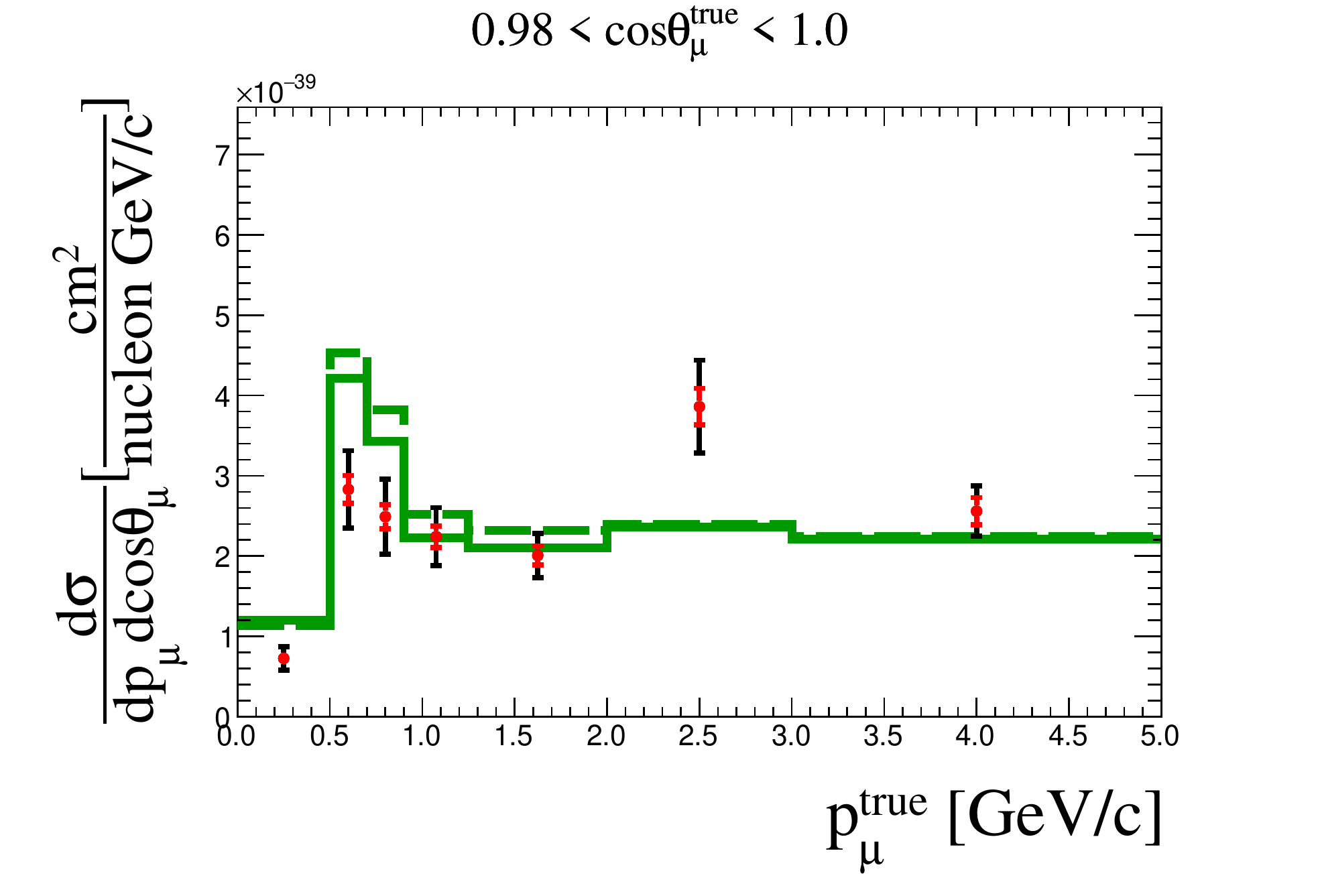}
	\includegraphics[width=0.36\linewidth]{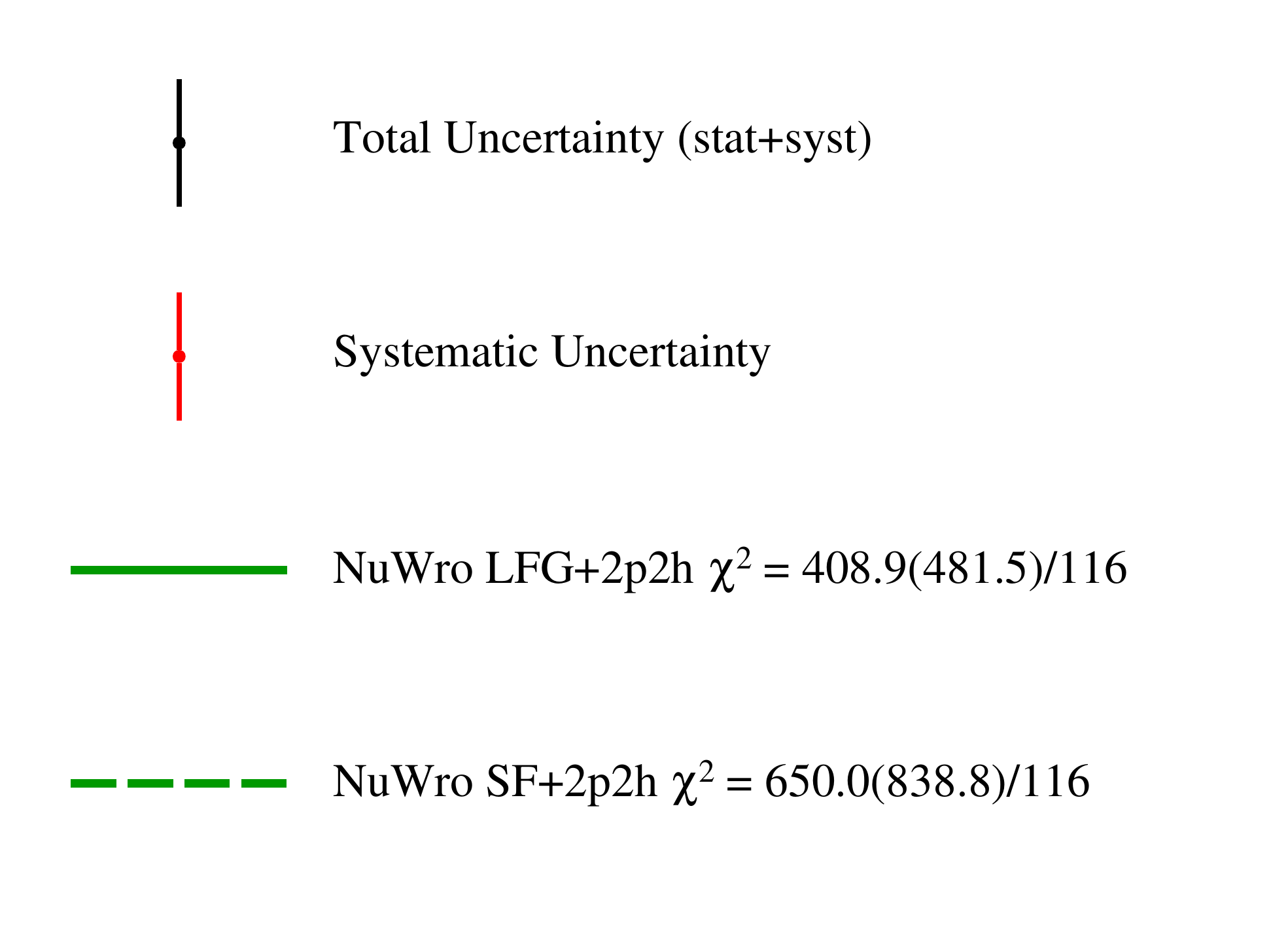}
	\caption{Measured \numu \cczeropi double-differential cross-section per nucleon in bins of true muon kinematics with systematic uncertainty (red bars) and total (stat.+syst.) uncertainty (black bars). The result is compared with \textsc{NuWro} version~\texttt{18.02.1} with LFG+RPA (green solid line) and with the SF nuclear model (green dashed line), both including 2p2h predictions. The full and shape-only (in parenthesis) $\chi2$ are reported. The last bin in momentum is not displayed for readability.}
	\label{fig:numucc0pixsecnuwro}
\end{figure*}

\begin{figure*}[h!]
	\centering
	\includegraphics[width=0.36\linewidth]{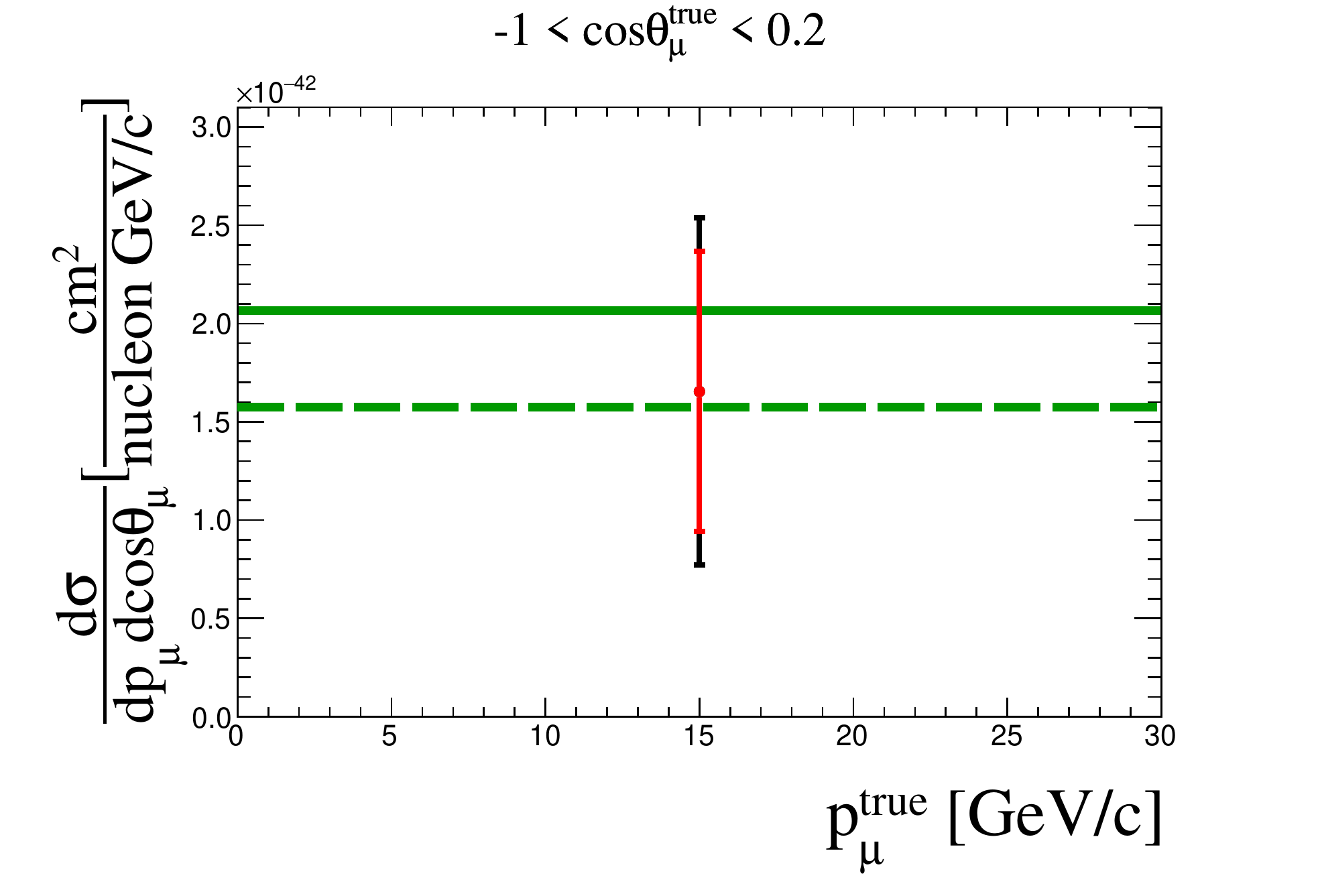}
	\includegraphics[width=0.36\linewidth]{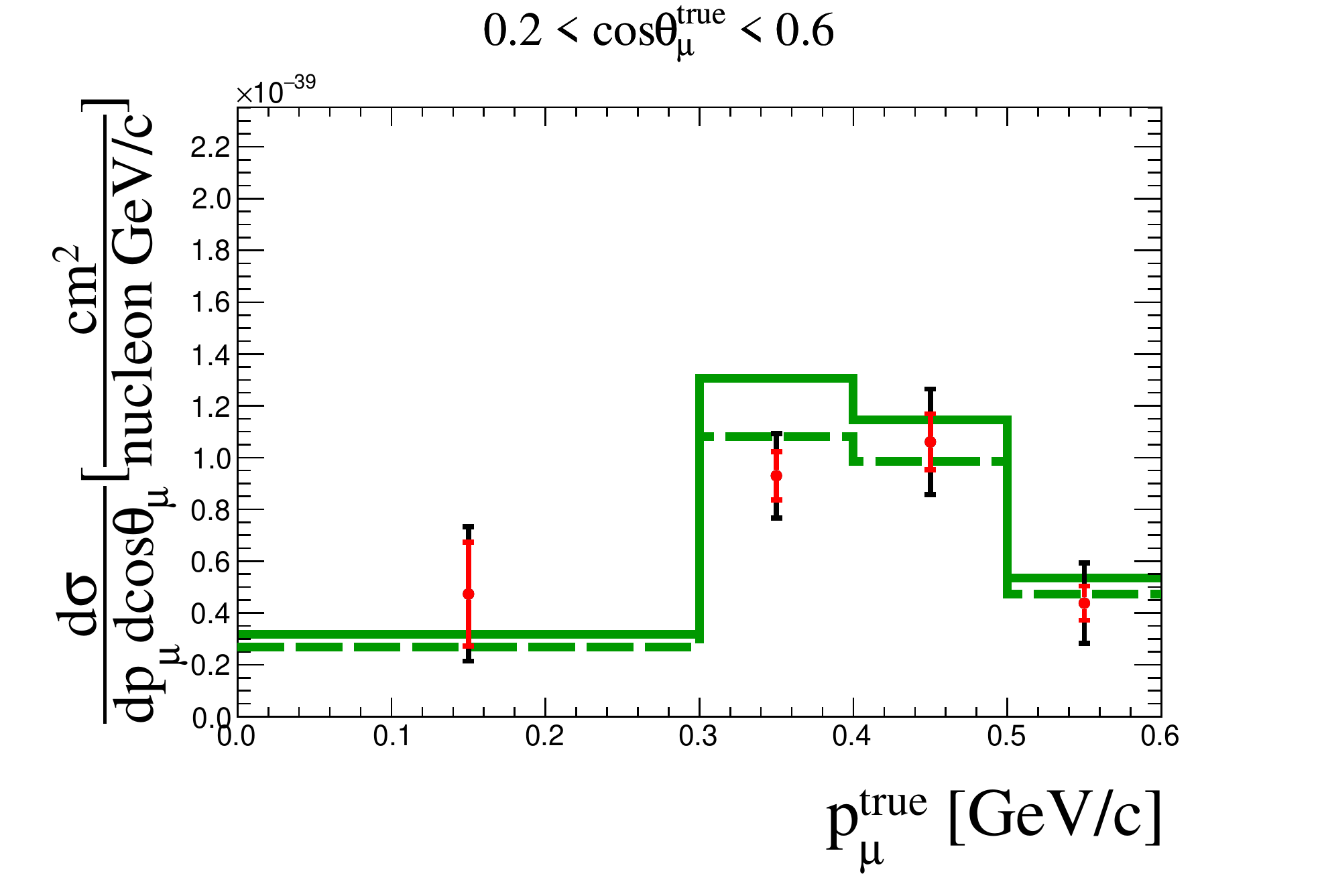}
	\includegraphics[width=0.36\linewidth]{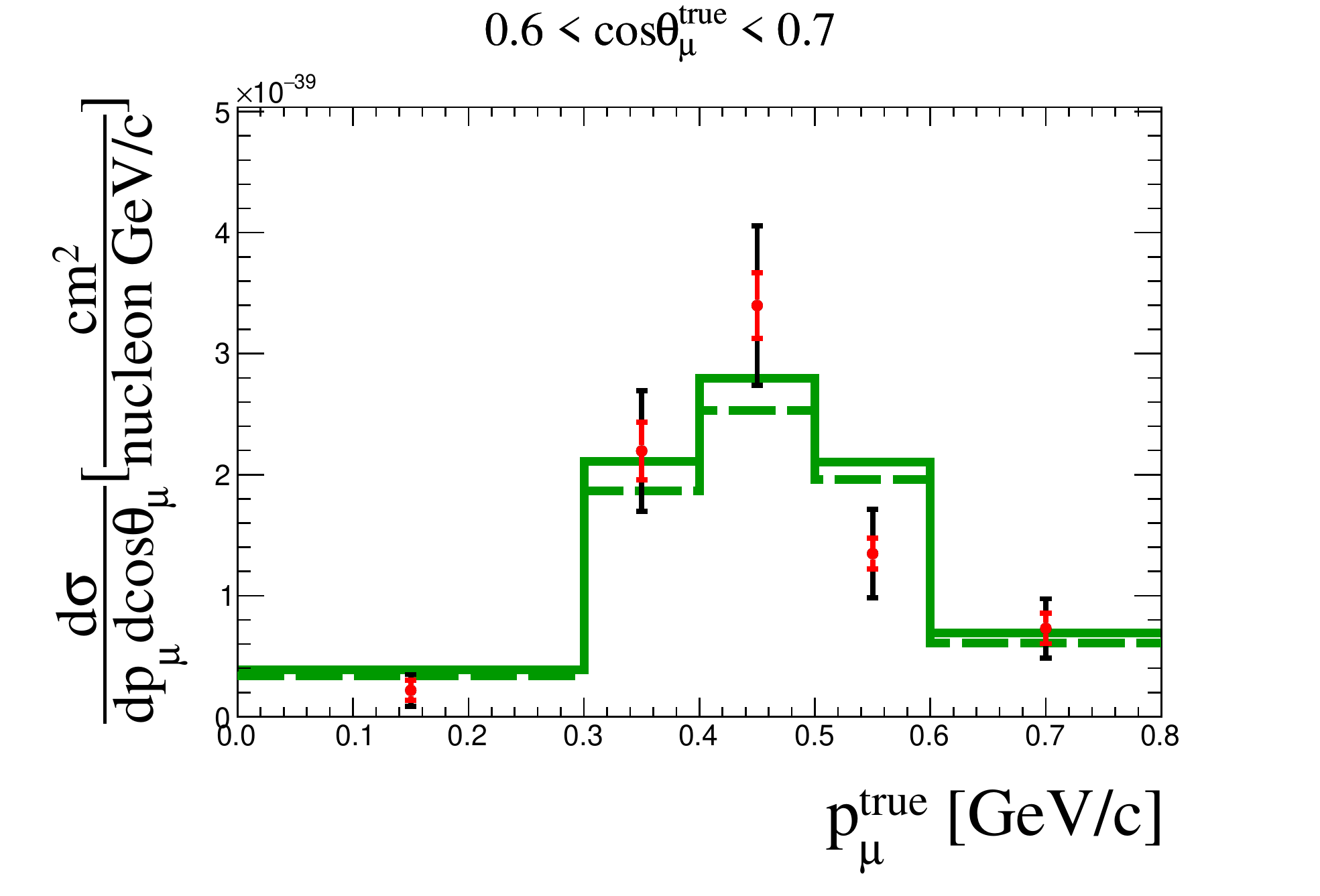}
	\includegraphics[width=0.36\linewidth]{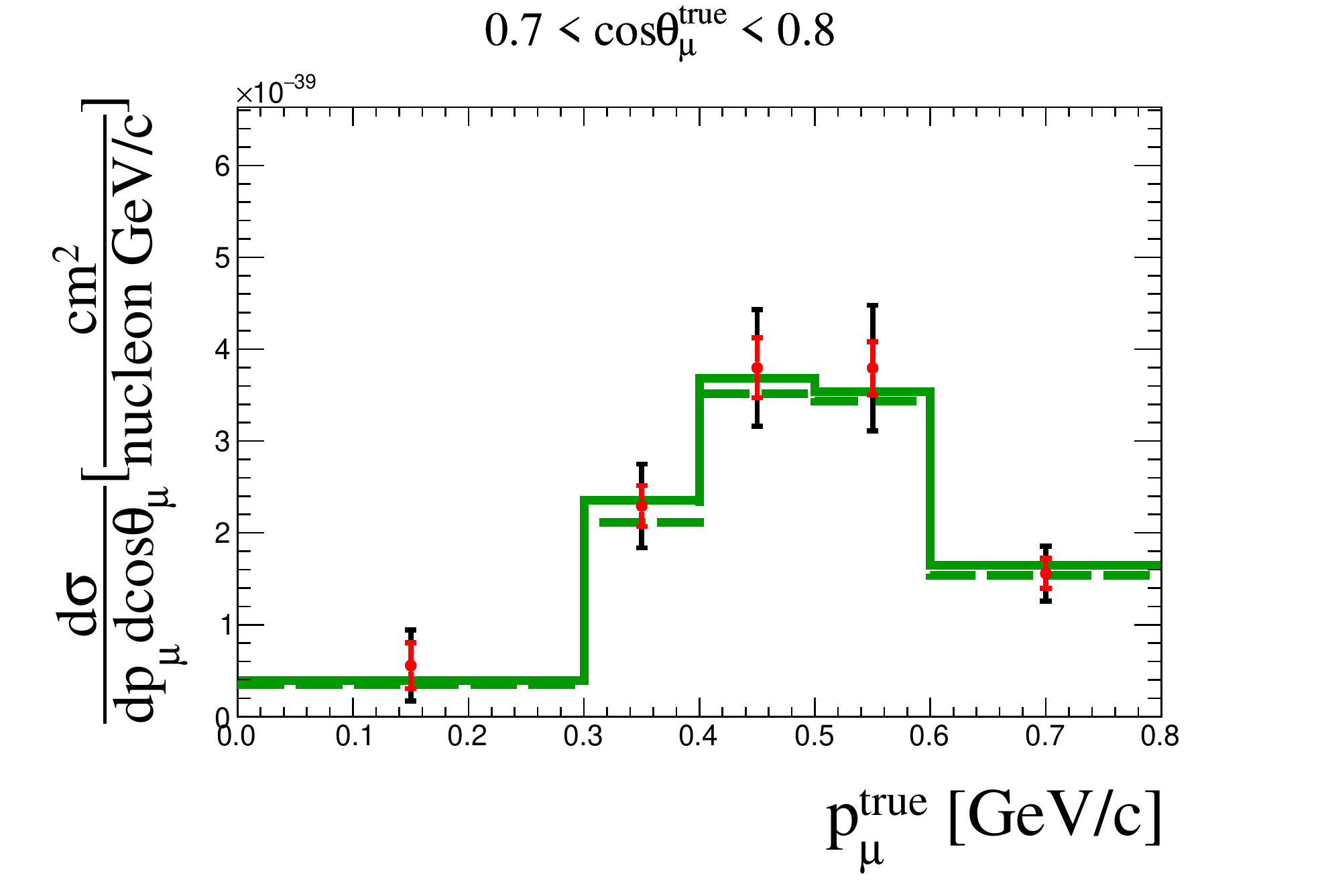}
	\includegraphics[width=0.36\linewidth]{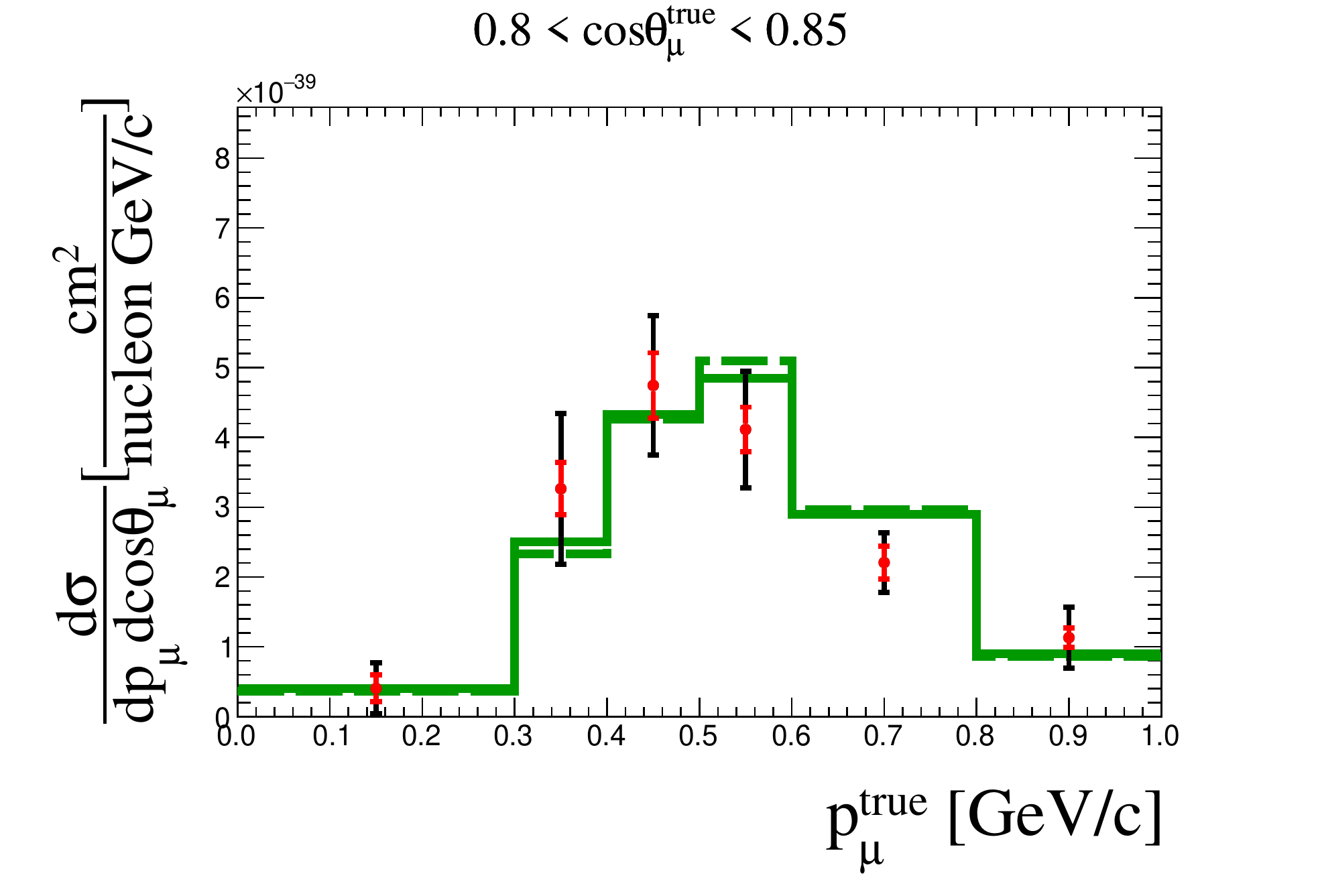}
	\includegraphics[width=0.36\linewidth]{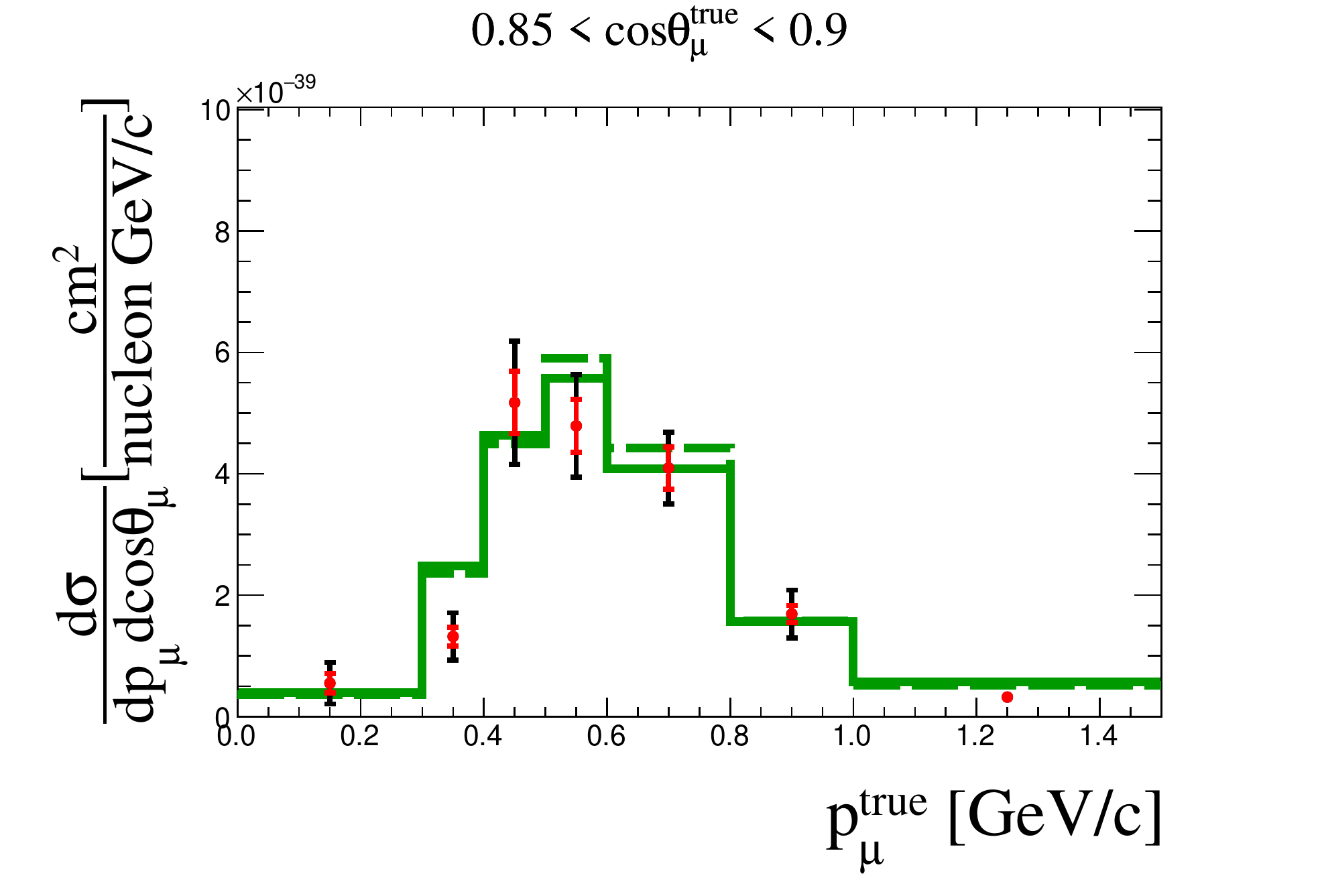}
	\includegraphics[width=0.36\linewidth]{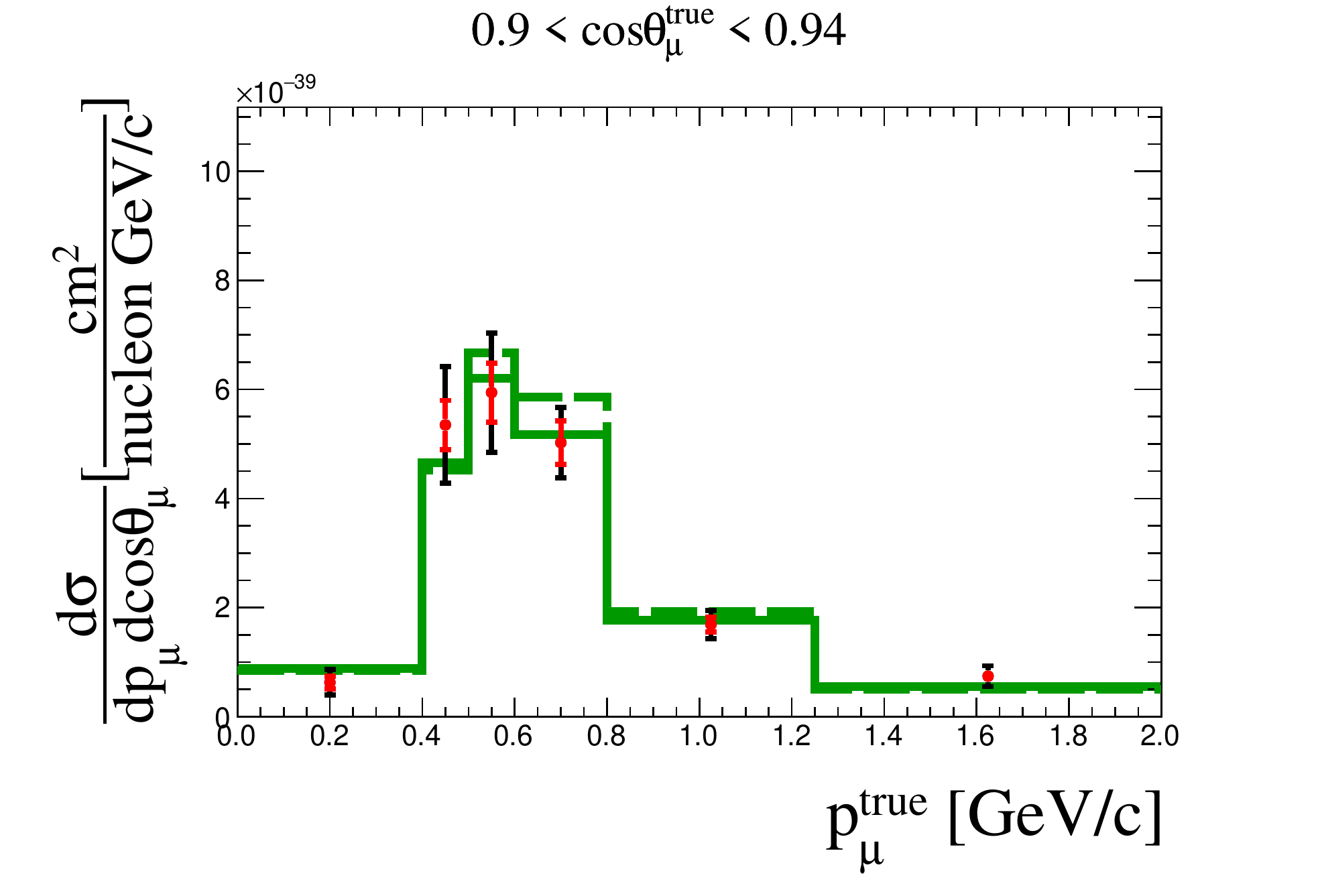}
	\includegraphics[width=0.36\linewidth]{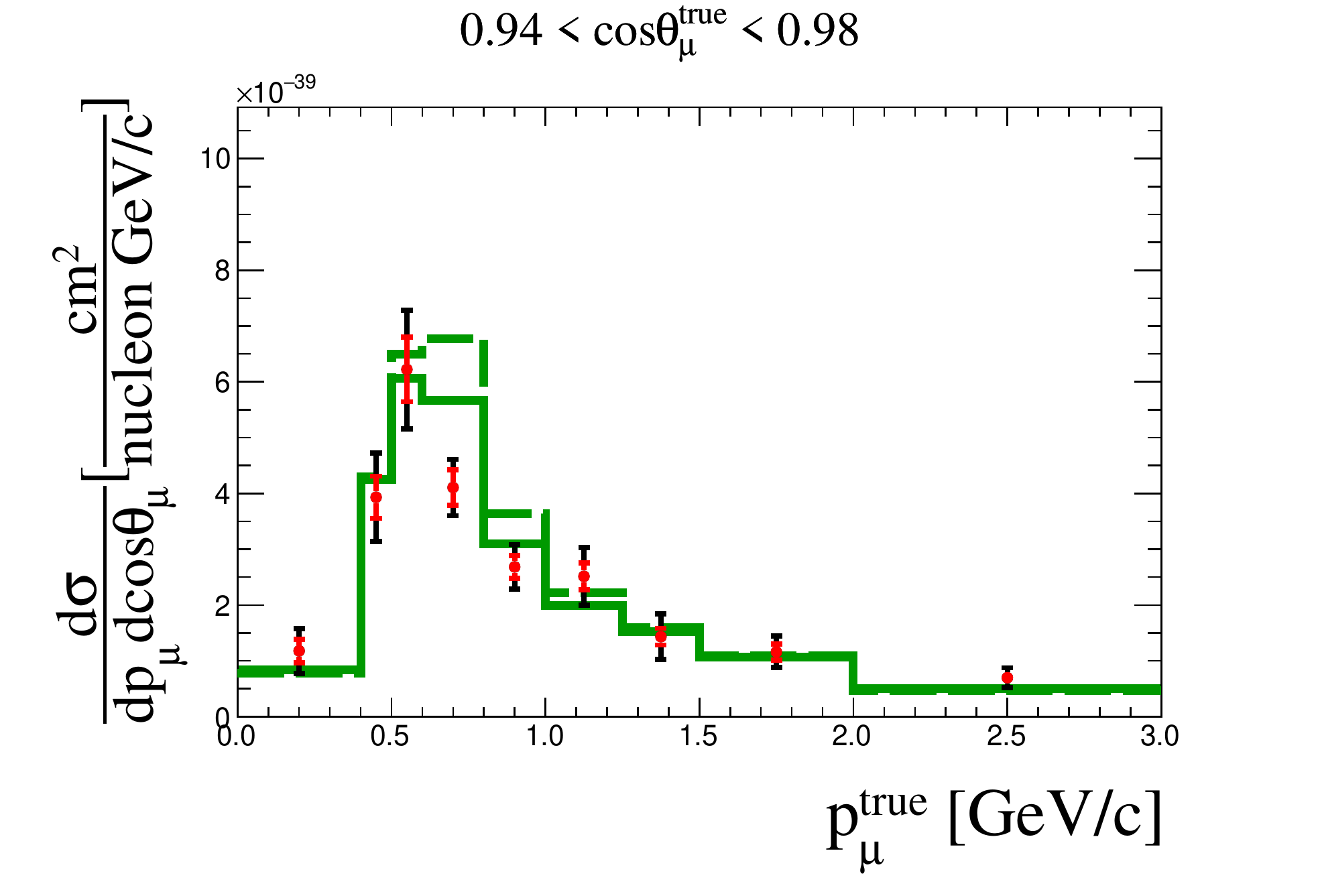}
	\includegraphics[width=0.36\linewidth]{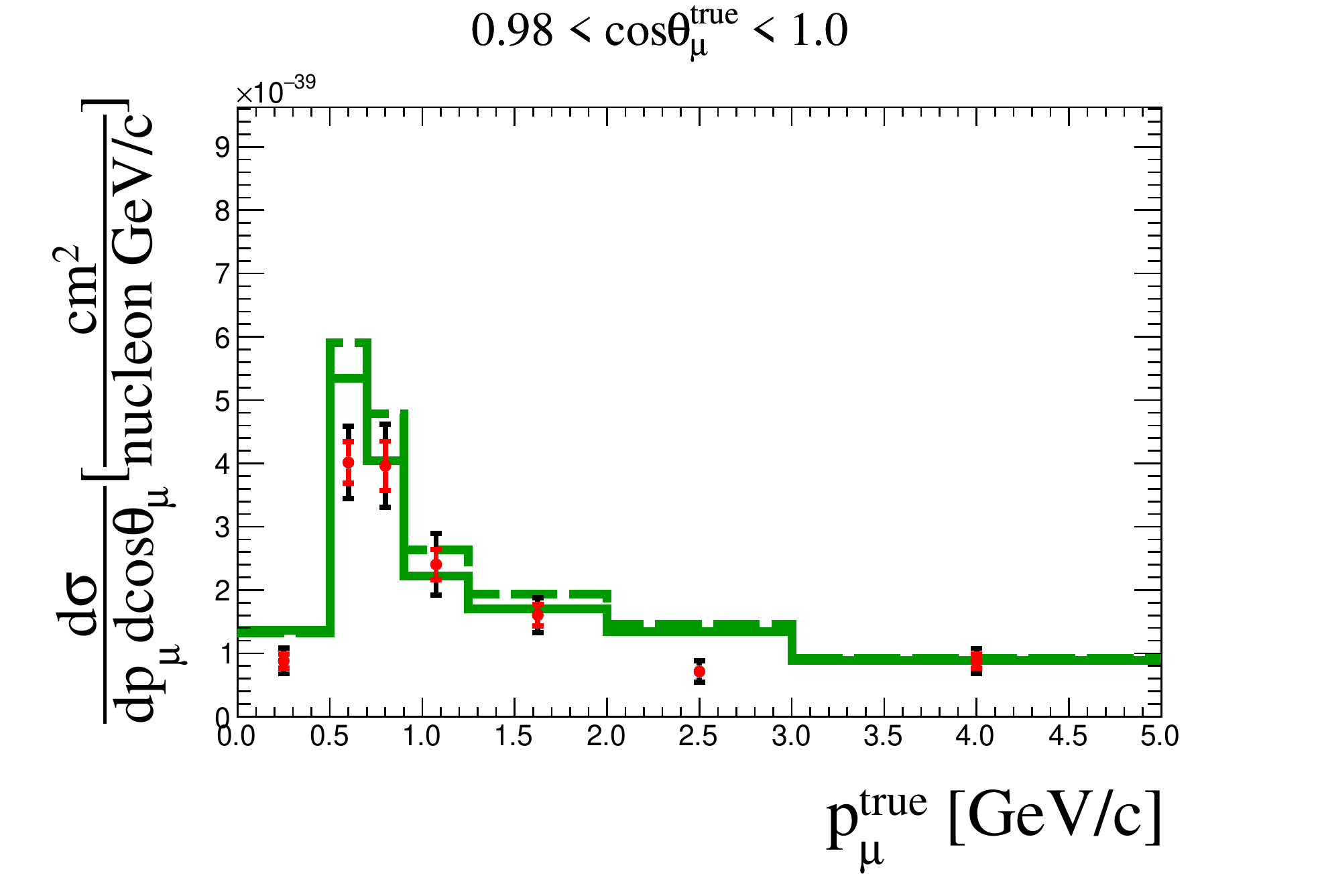}
	\includegraphics[width=0.36\linewidth]{XsecLegendNuWro}
	\caption{Measured \barnumu \cczeropi double-differential cross-section per nucleon in bins of true muon kinematics with systematic uncertainty (red bars) and total (stat.+syst.) uncertainty (black bars). The result is compared with \textsc{NuWro} version~\texttt{18.02.1} with LFG+RPA (green solid line) and with the SF nuclear model (green dashed line), both including 2p2h predictions. The full and shape-only (in parenthesis) $\chi2$ are reported. The last bin in momentum is not displayed for readability.}
	\label{fig:antinumucc0pixsecnuwro}
\end{figure*}

\begin{figure*}[h!]
	\centering
	\includegraphics[width=0.36\linewidth]{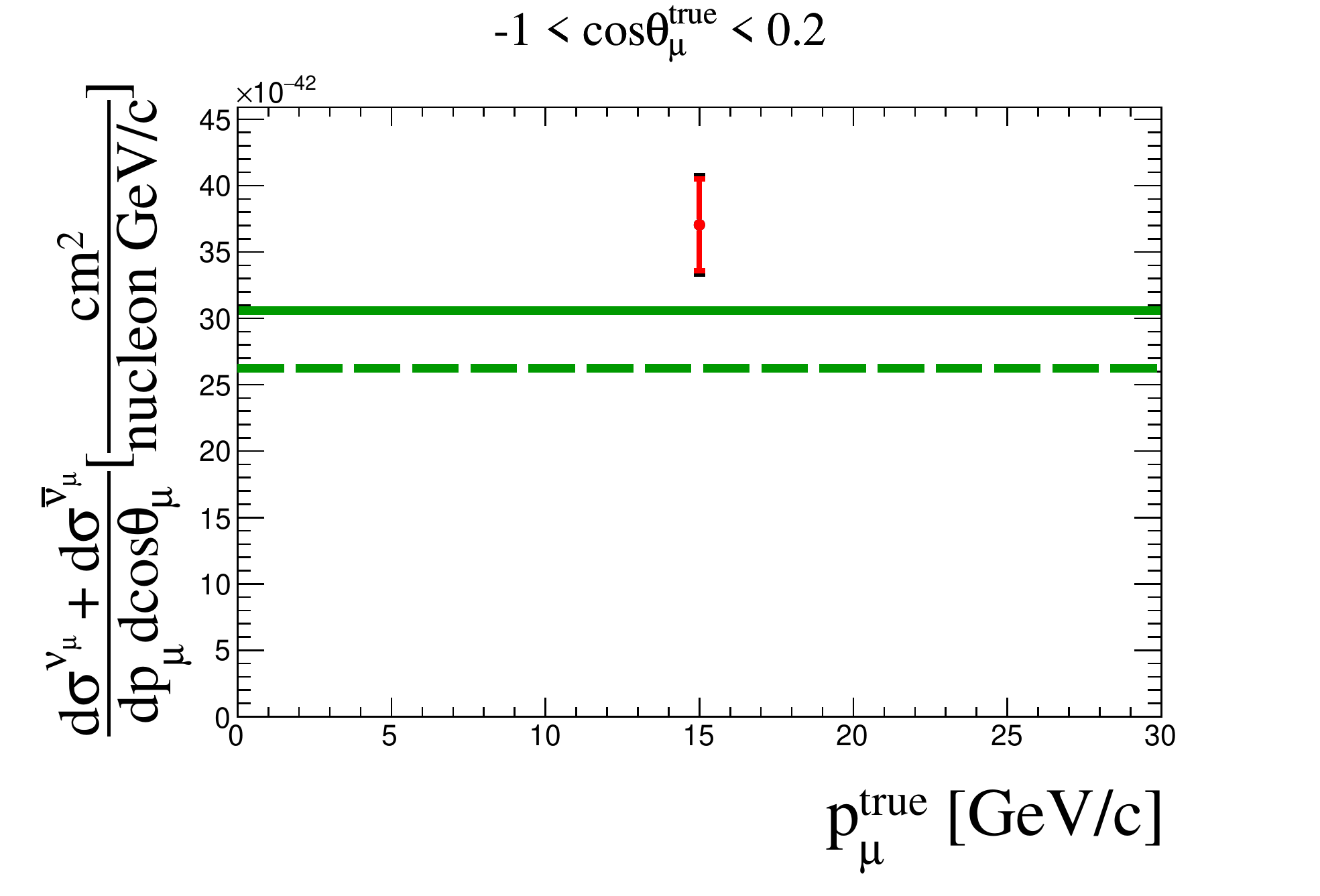}
	\includegraphics[width=0.36\linewidth]{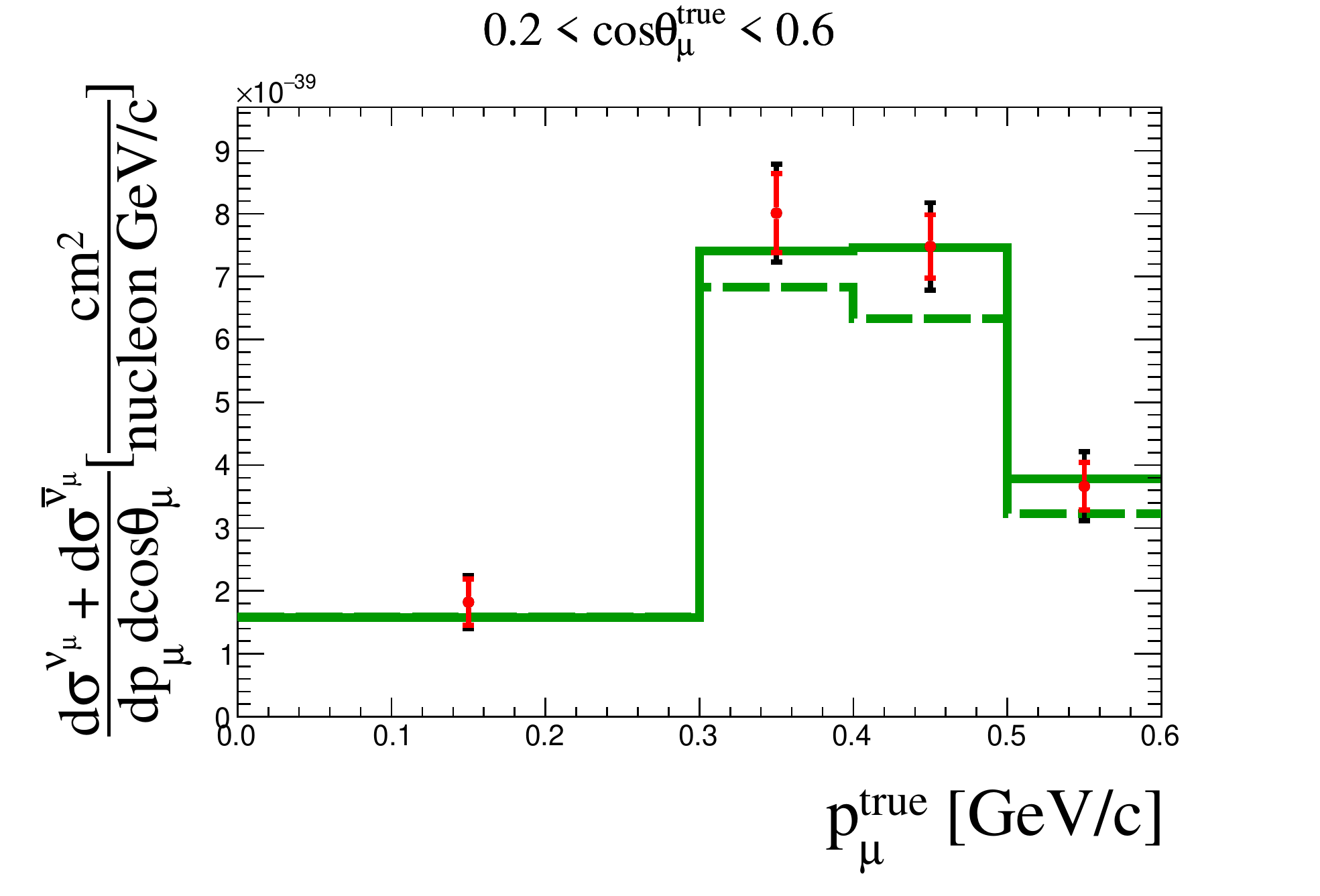}
	\includegraphics[width=0.36\linewidth]{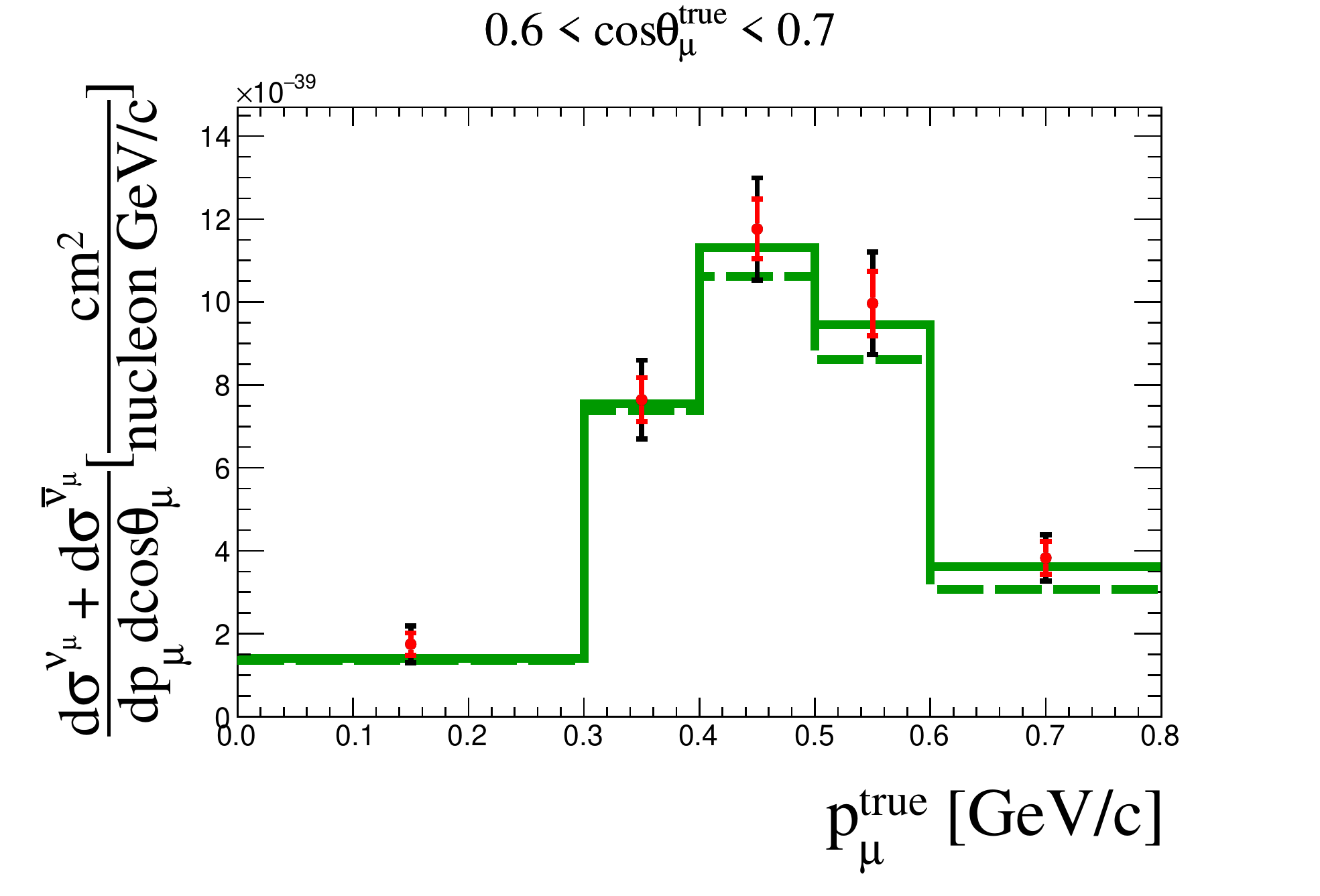}
	\includegraphics[width=0.36\linewidth]{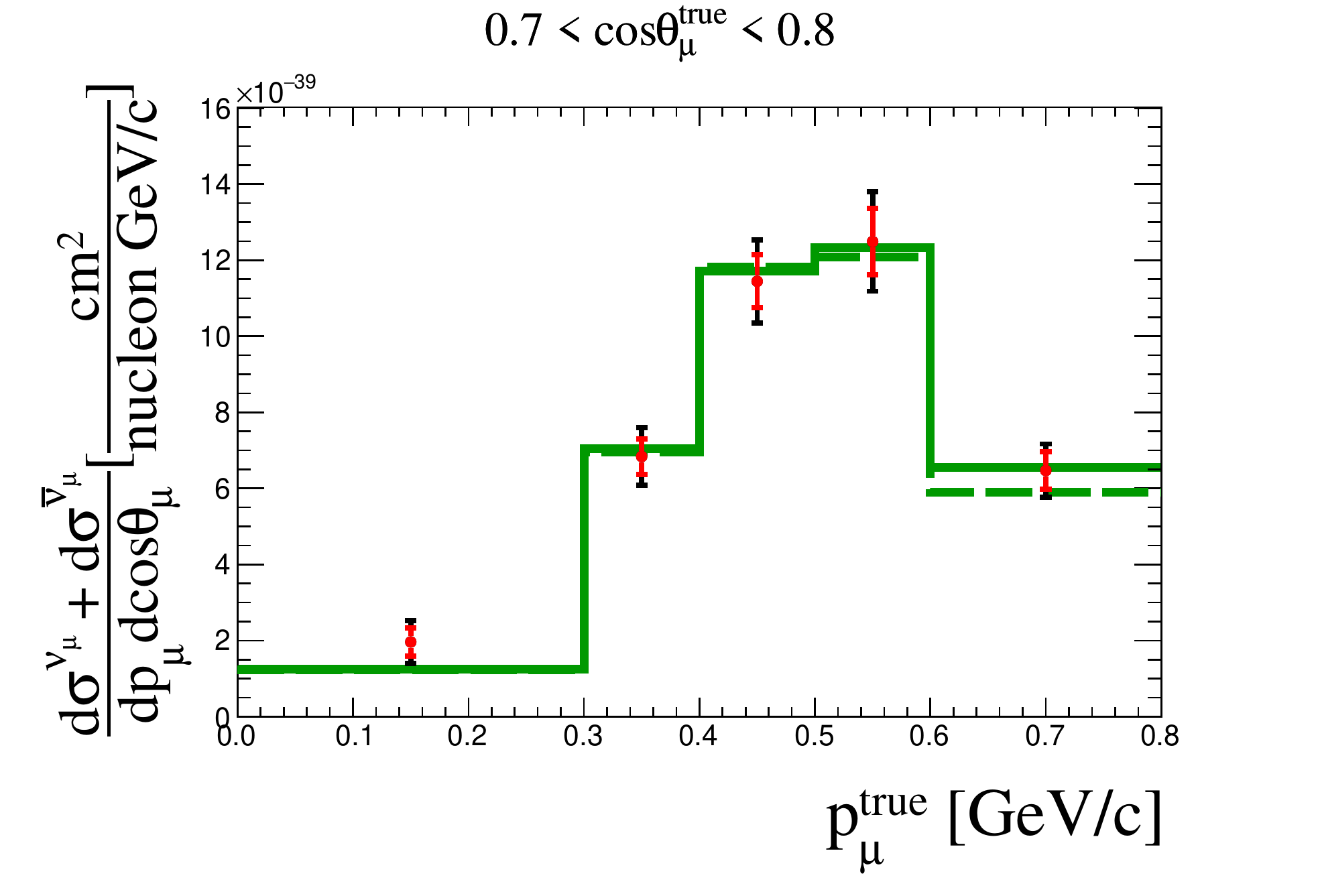}
	\includegraphics[width=0.36\linewidth]{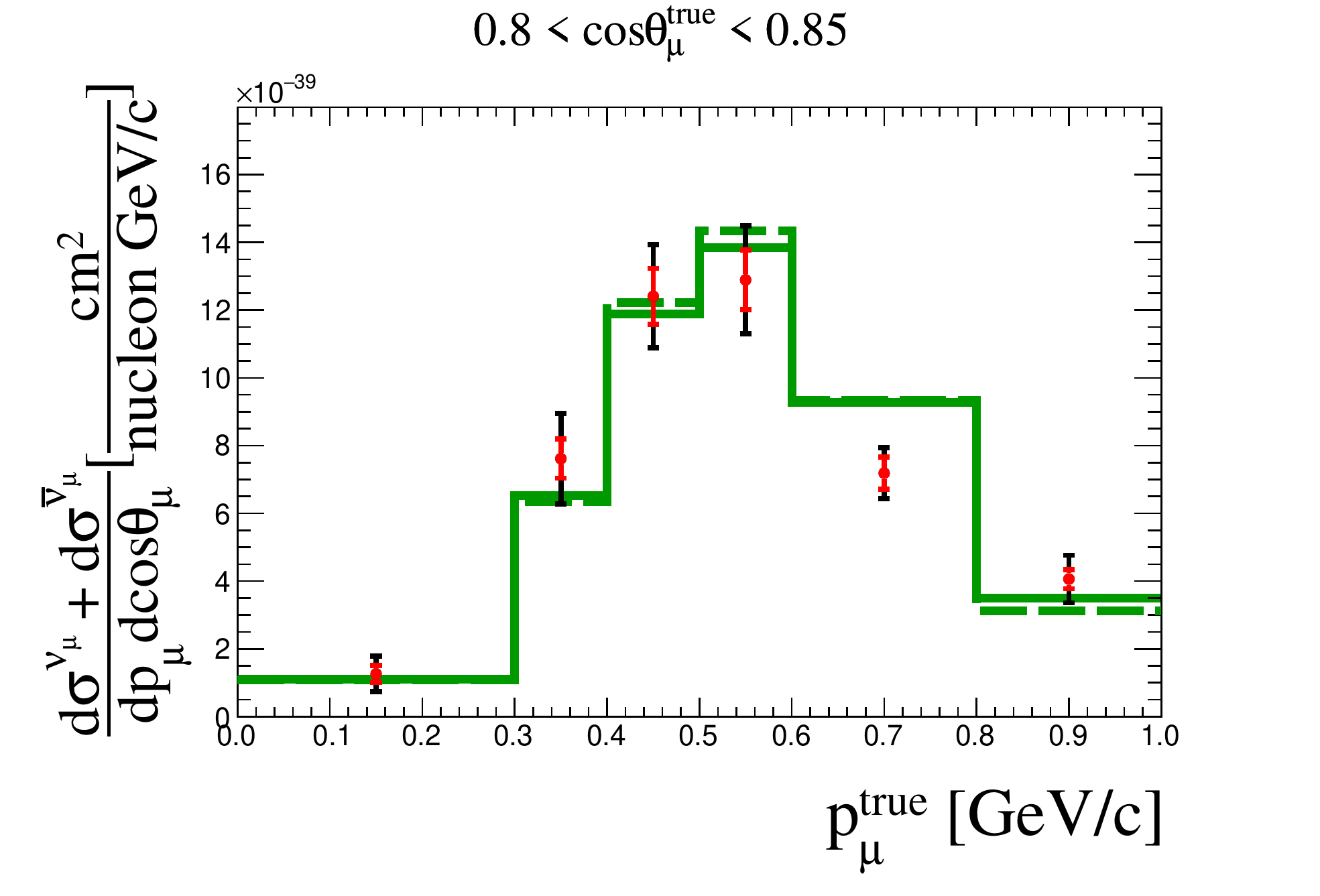}
	\includegraphics[width=0.36\linewidth]{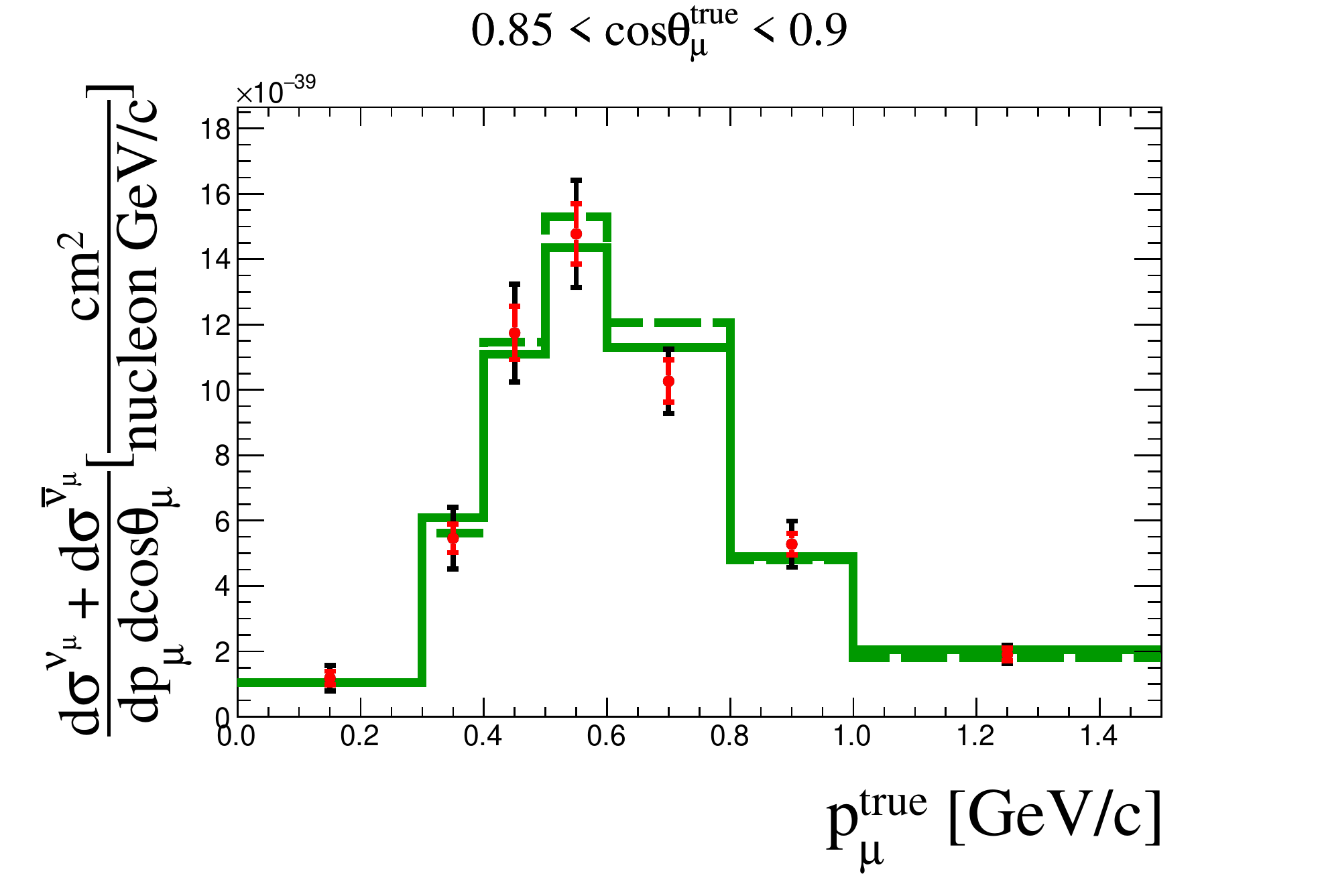}
	\includegraphics[width=0.36\linewidth]{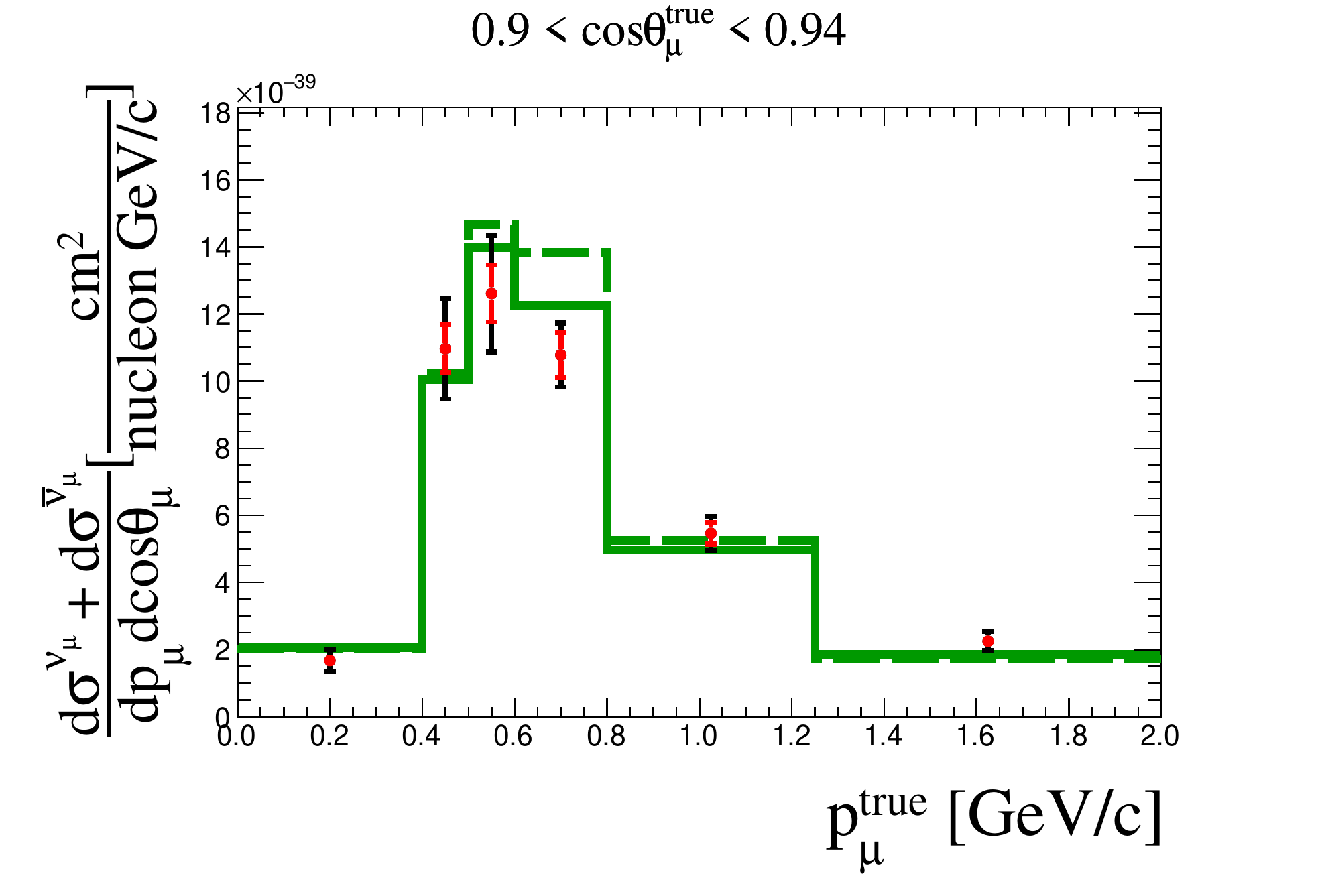}
	\includegraphics[width=0.36\linewidth]{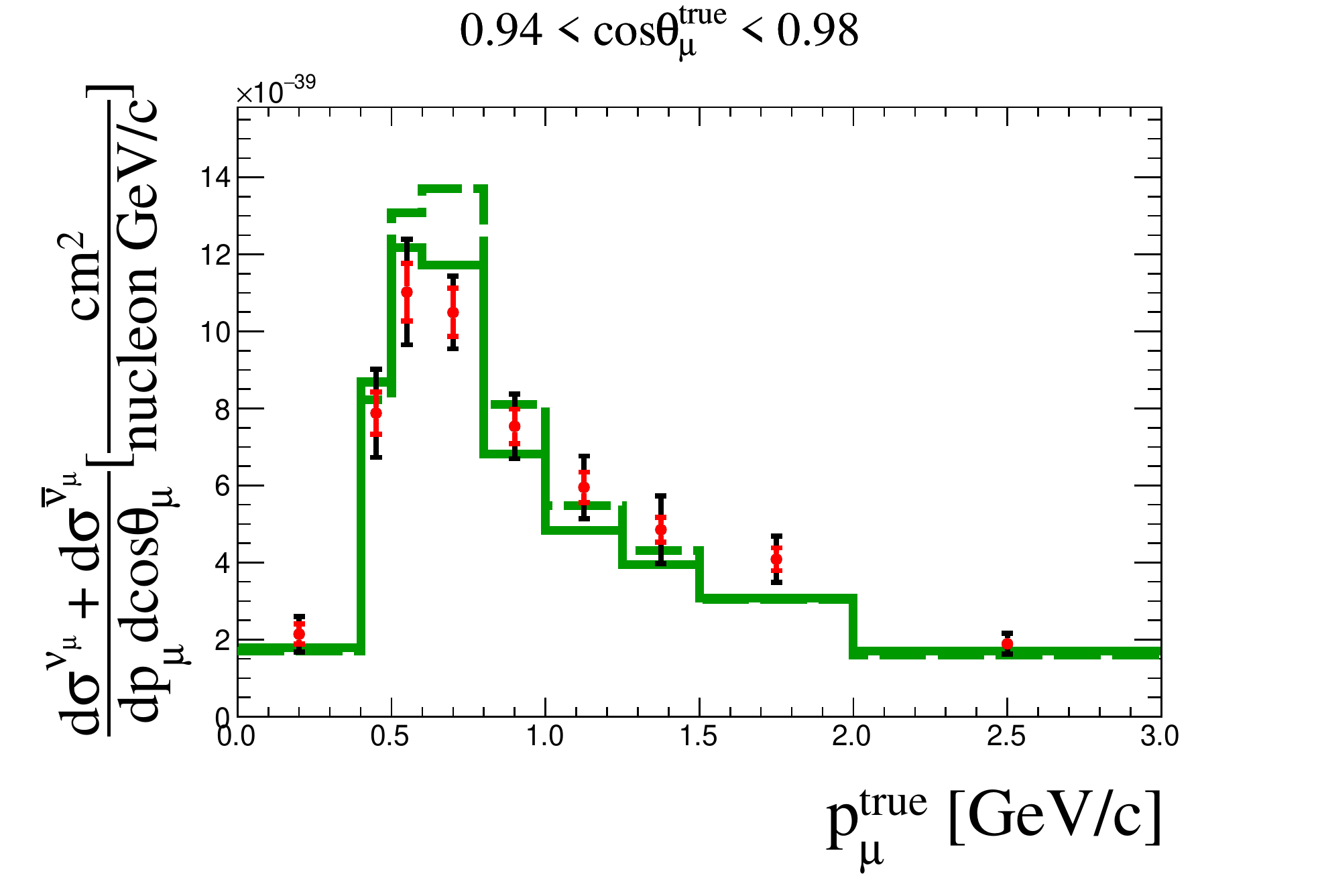}
	\includegraphics[width=0.36\linewidth]{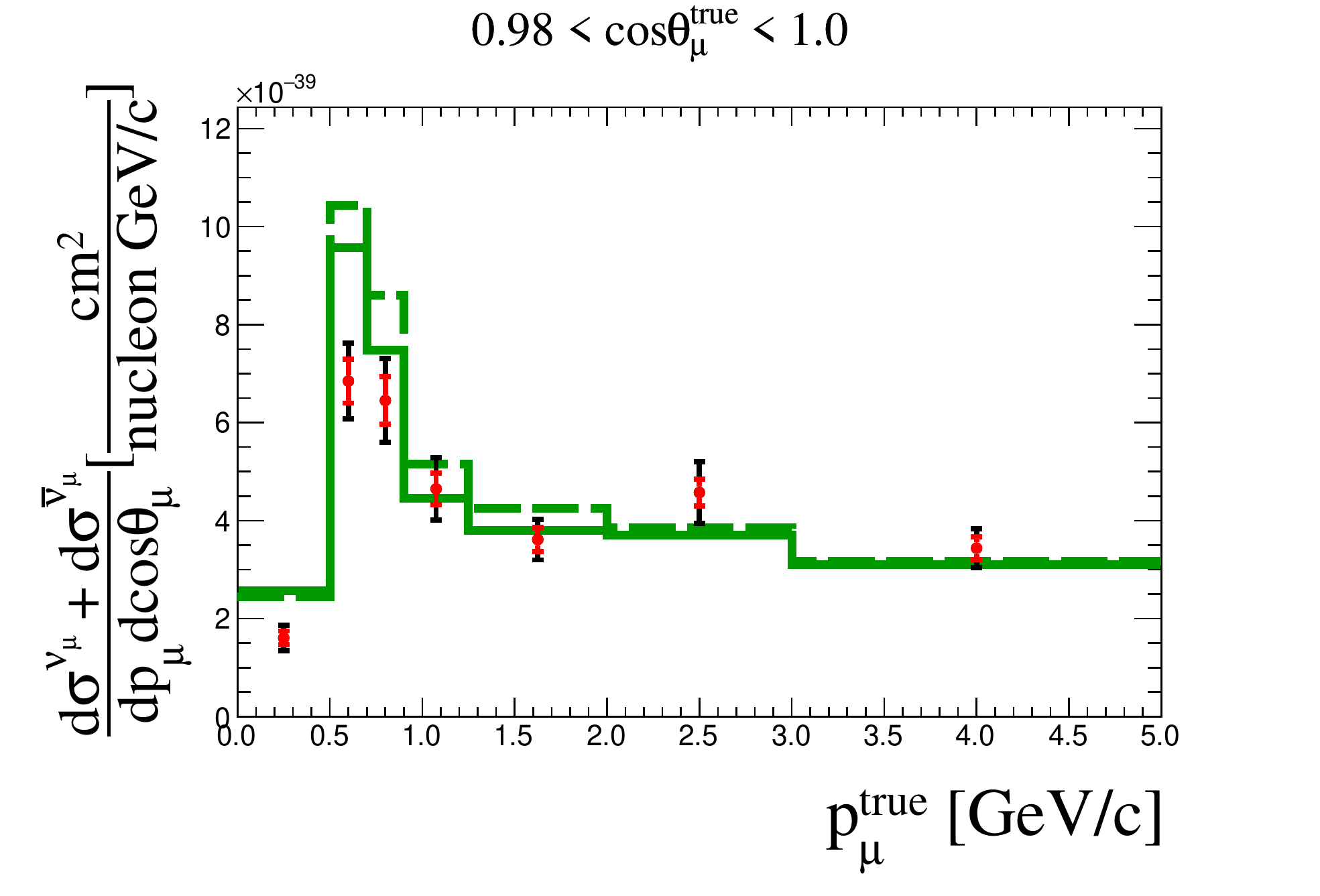}
	\includegraphics[width=0.36\linewidth]{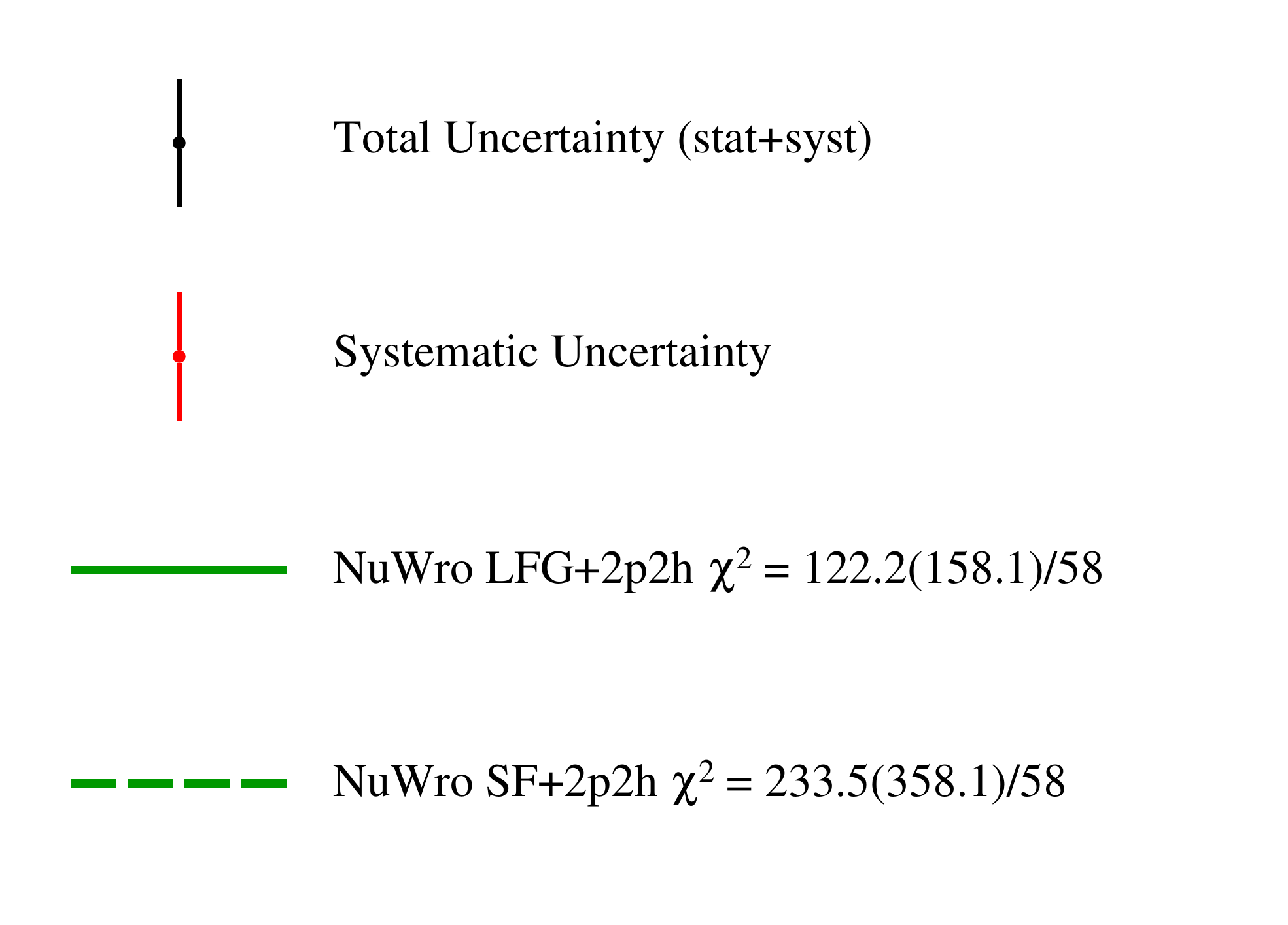}
	\caption{Measured double-differential \numu + \barnumu \cczeropi cross-section sum in bins of true muon kinematics with systematic uncertainty (red bars) and total (stat.+syst.) uncertainty (black bars). The result is compared with \textsc{NuWro} version~\texttt{18.02.1} with LFG+RPA (green solid line) and with the SF nuclear model (green dashed line), both including 2p2h predictions. The full and shape-only (in parenthesis) $\chi2$ are reported. The last bin in momentum is not displayed for readability.}
	\label{fig:xsecsumnuwro}
\end{figure*}

\begin{figure*}[h!]
	\centering
	\includegraphics[width=0.36\linewidth]{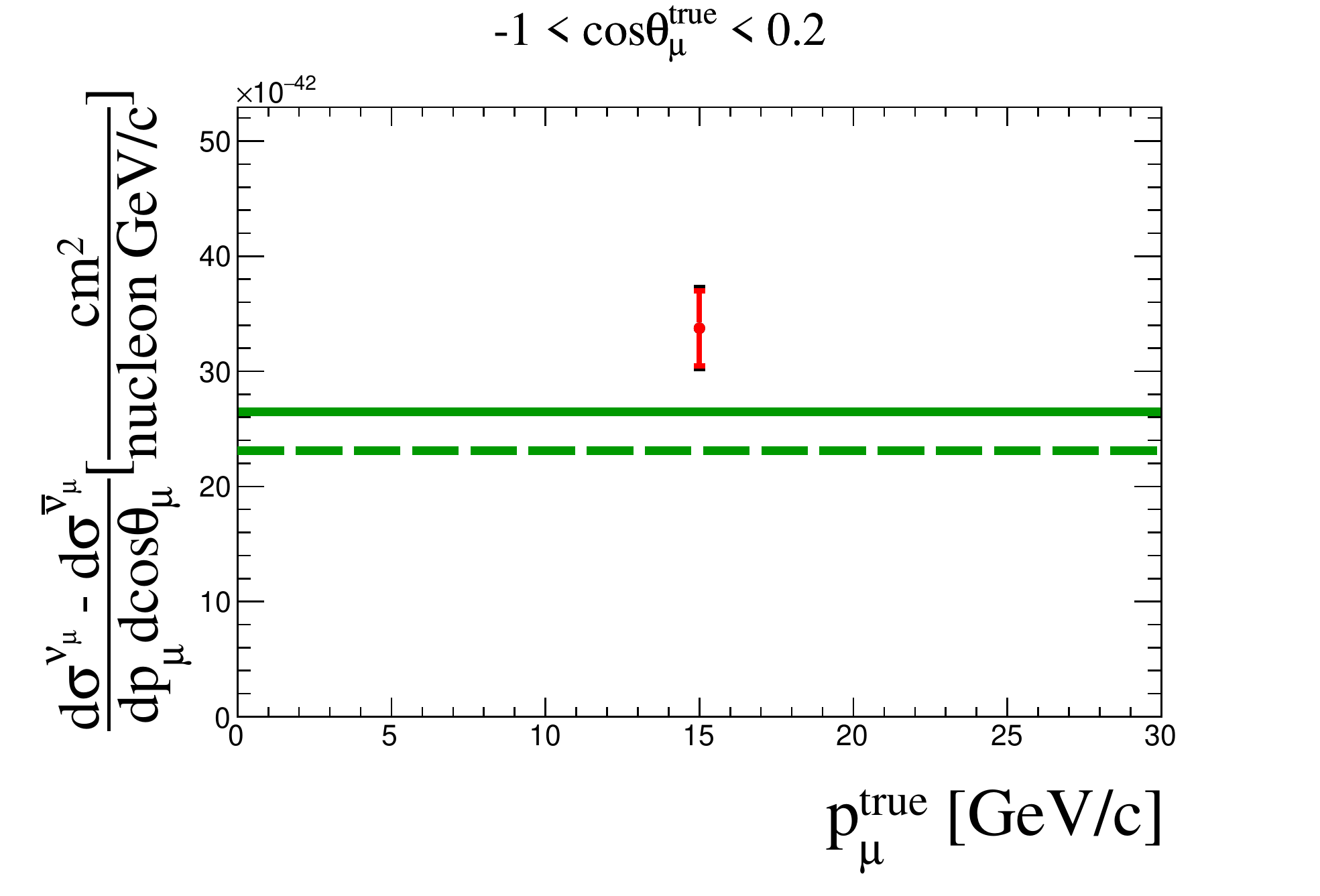}
	\includegraphics[width=0.36\linewidth]{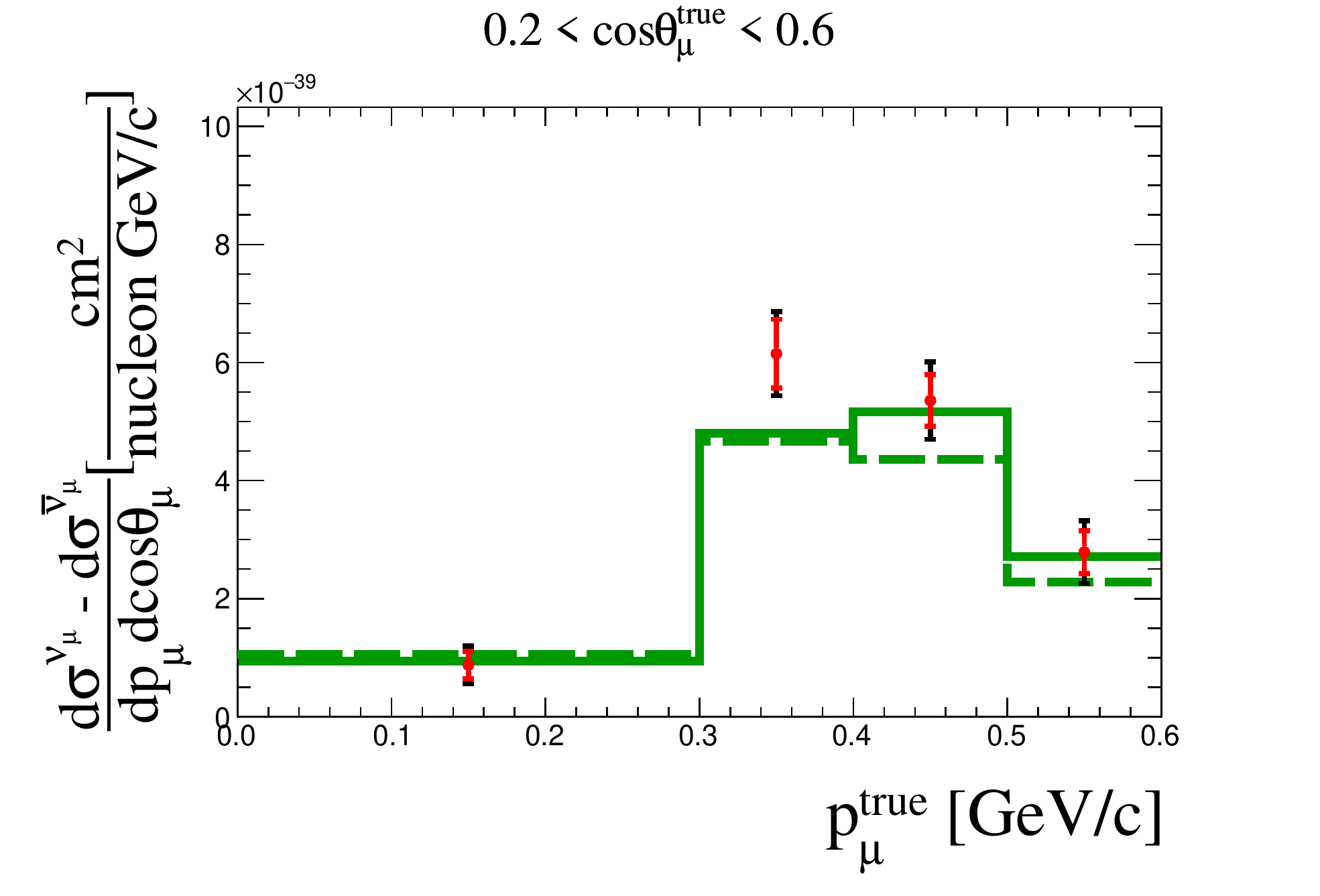}
	\includegraphics[width=0.36\linewidth]{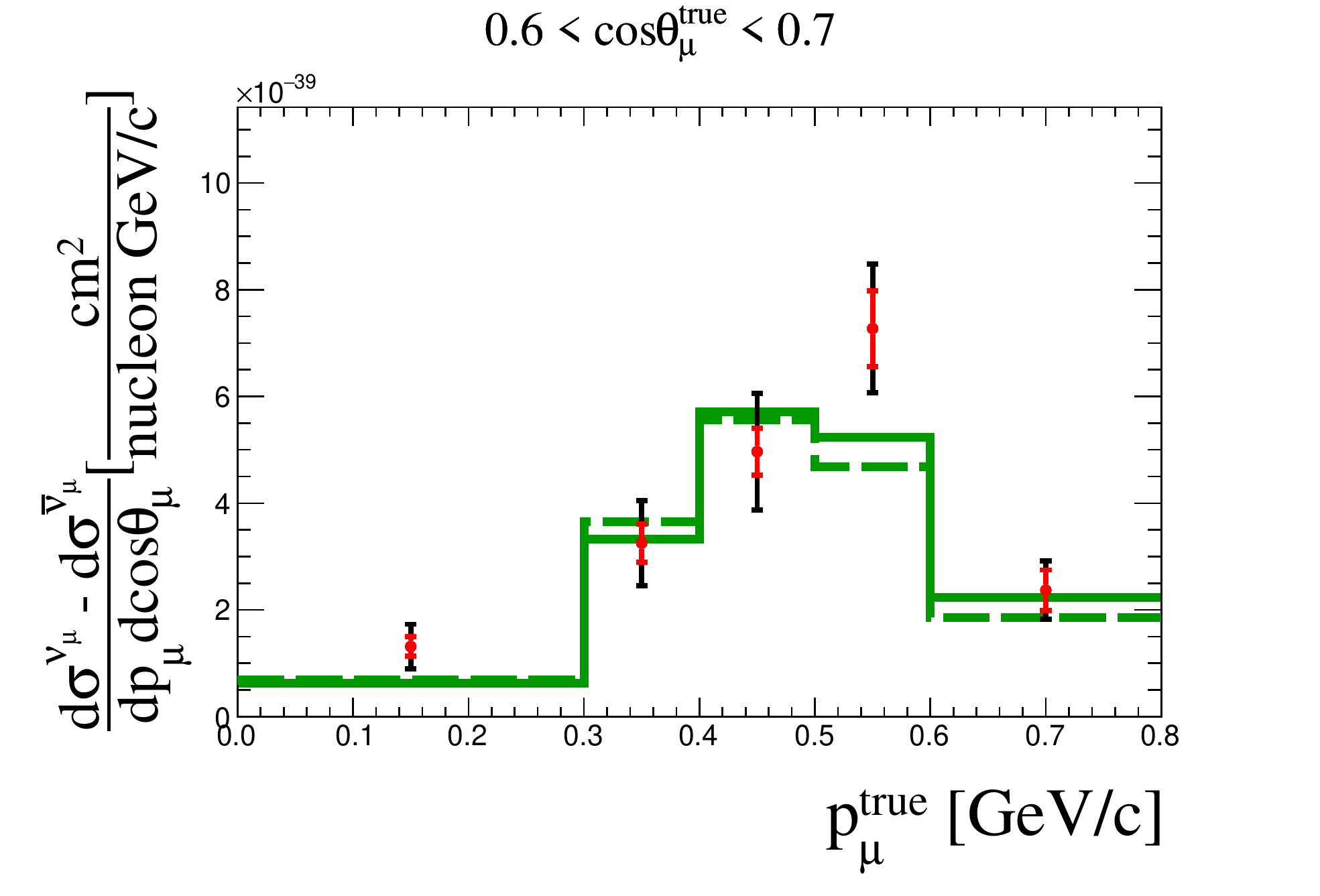}
	\includegraphics[width=0.36\linewidth]{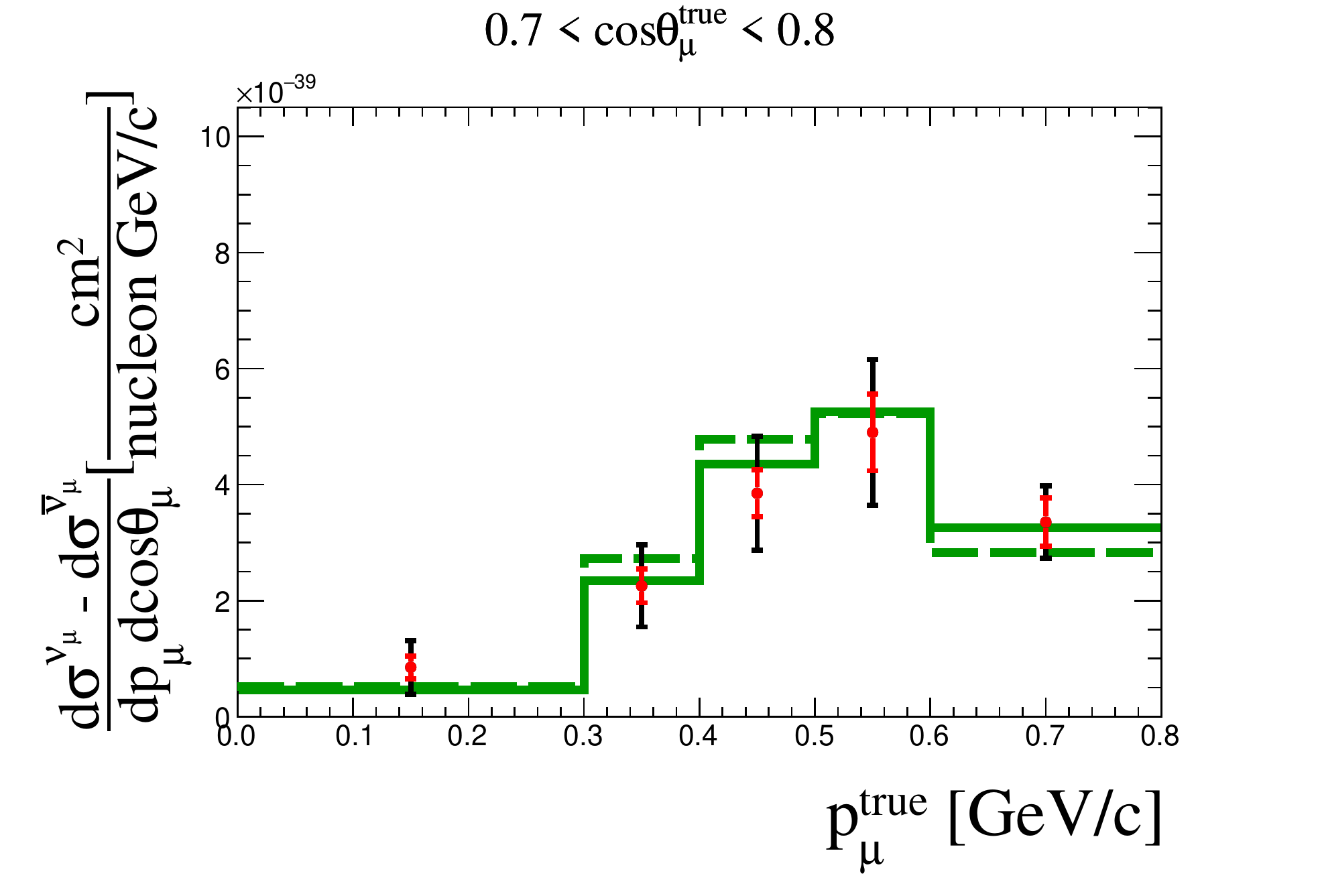}
	\includegraphics[width=0.36\linewidth]{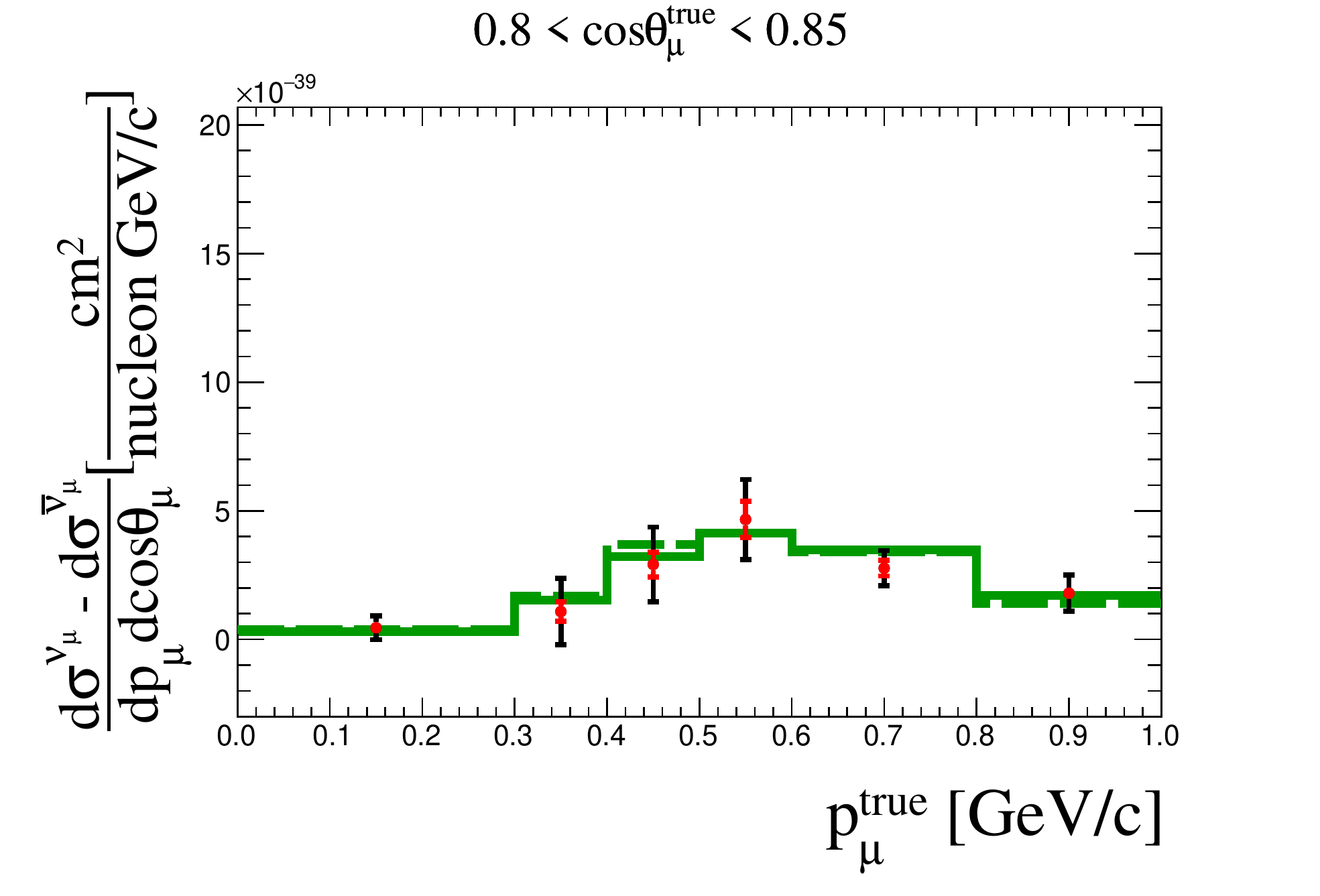}
	\includegraphics[width=0.36\linewidth]{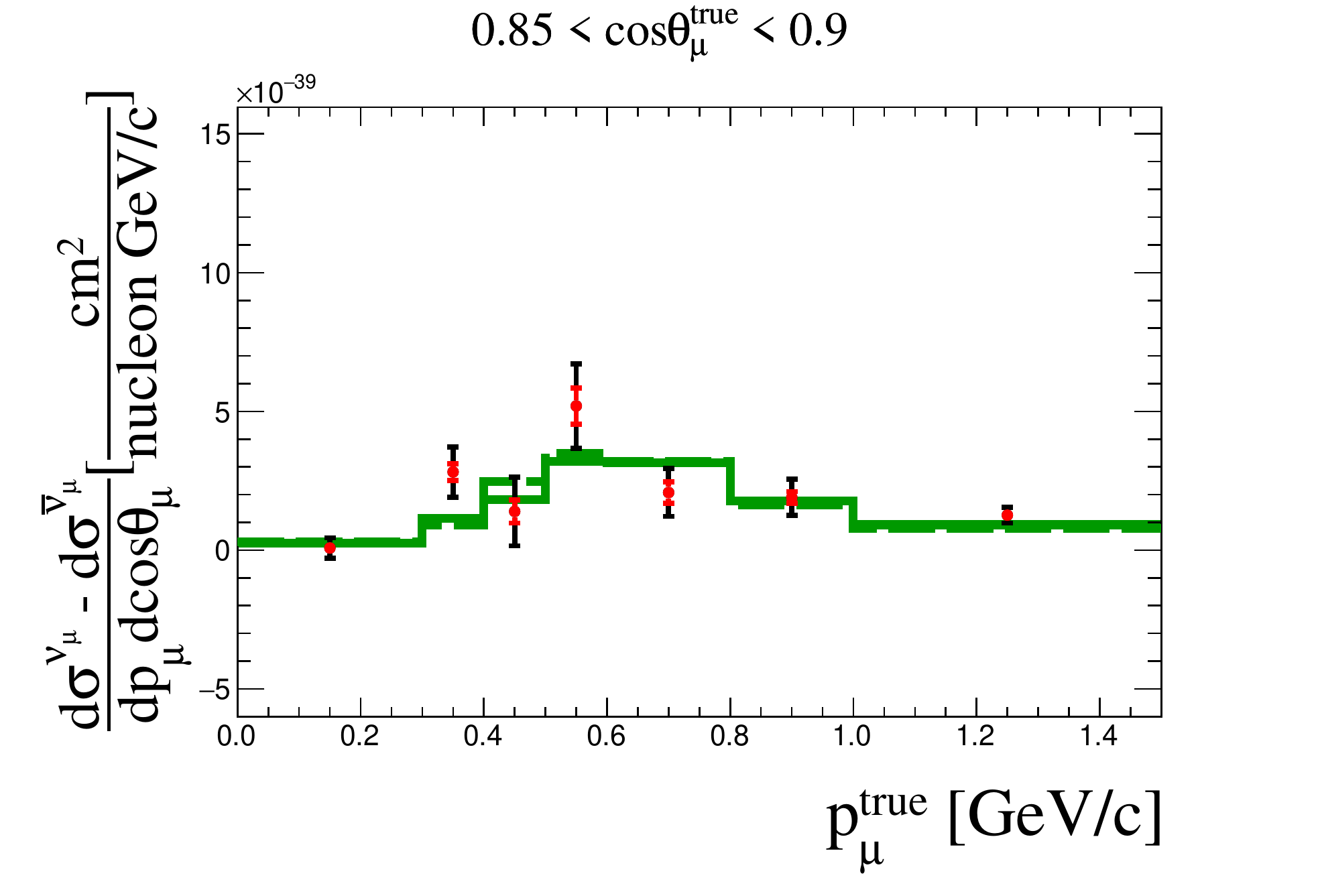}
	\includegraphics[width=0.36\linewidth]{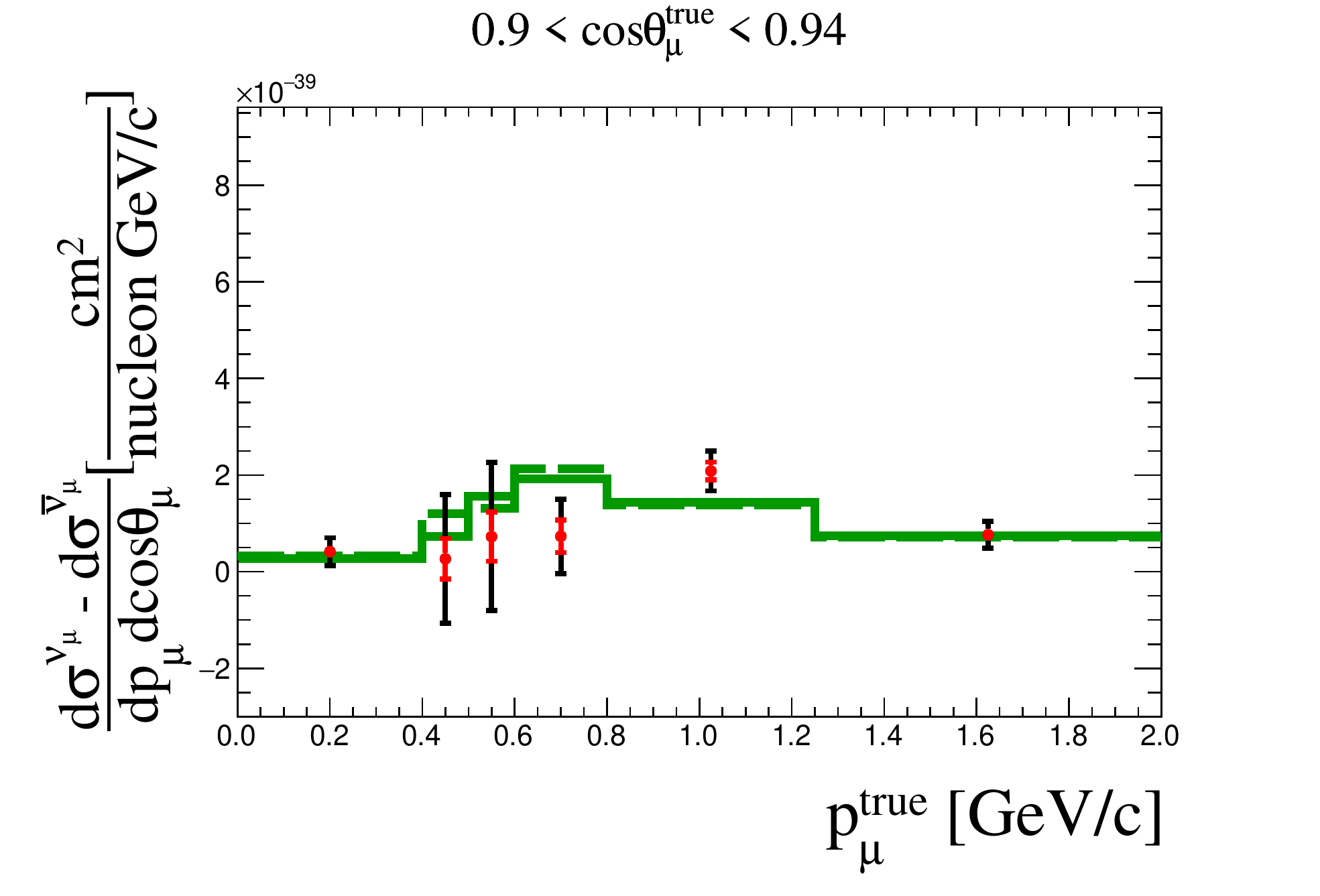}
	\includegraphics[width=0.36\linewidth]{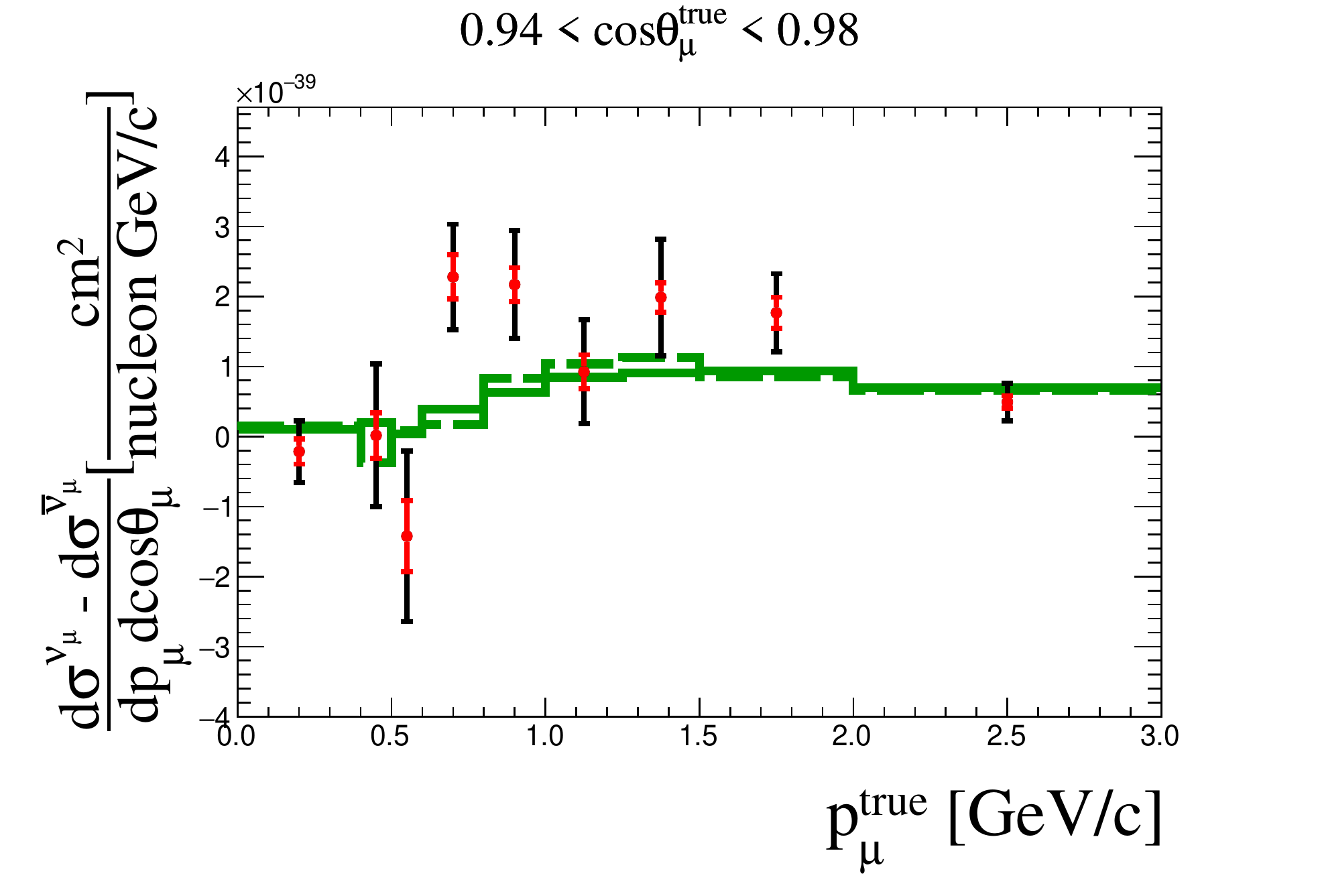}
	\includegraphics[width=0.36\linewidth]{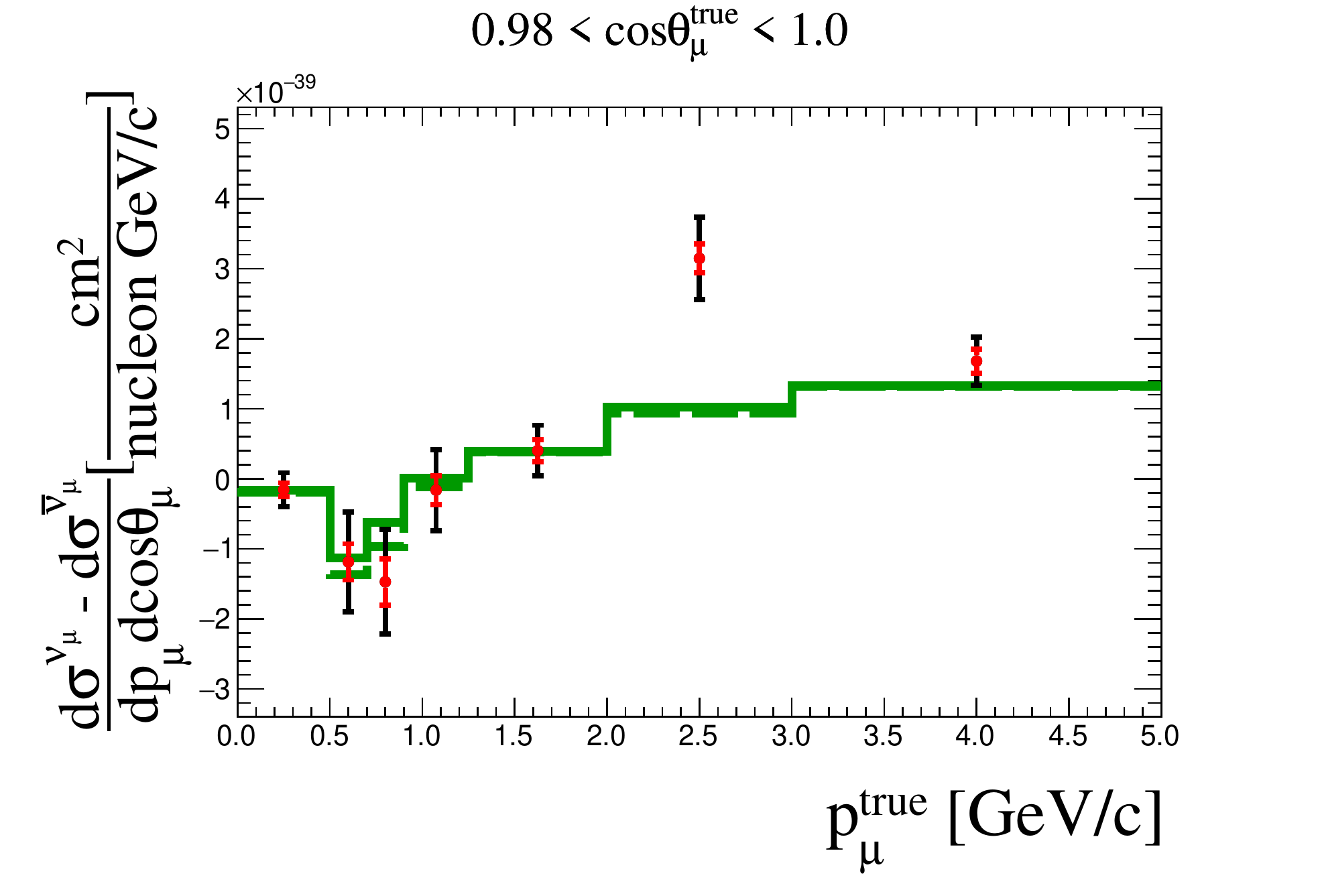}
	\includegraphics[width=0.36\linewidth]{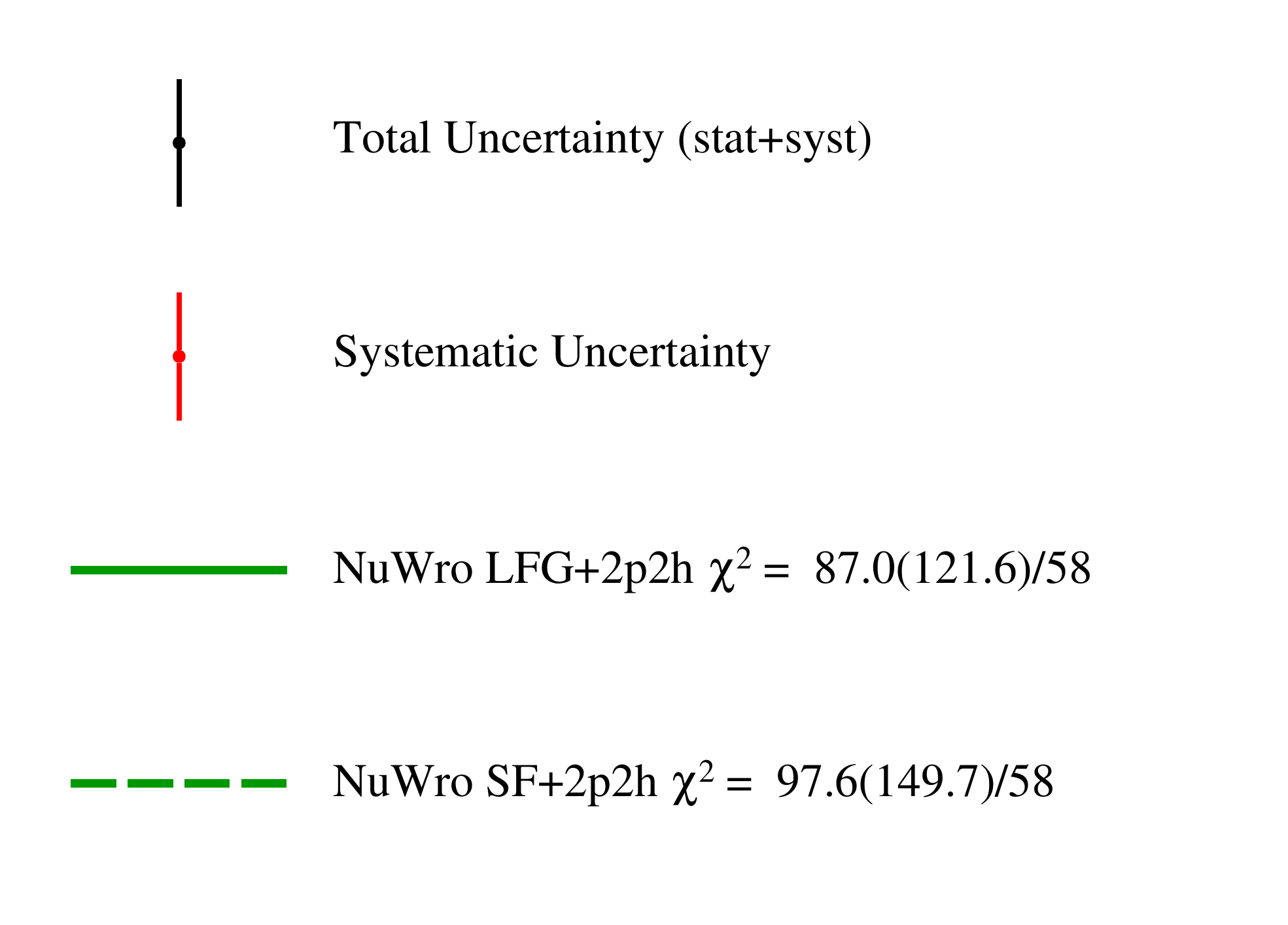}
	\caption{Measured double-differential \numu - \barnumu \cczeropi cross-section difference in bins of true muon kinematics with systematic uncertainty (red bars) and total (stat.+syst.) uncertainty (black bars). The result is compared with \textsc{NuWro} version~\texttt{18.02.1} with LFG+RPA (green solid line) and with the SF nuclear model (green dashed line), both including 2p2h predictions. The full and shape-only (in parenthesis) $\chi2$ are reported. The last bin in momentum is not displayed for readability.}
	\label{fig:xsedifnuwro}
\end{figure*}

\begin{figure*}[h!]
	\centering
	\includegraphics[width=0.36\linewidth]{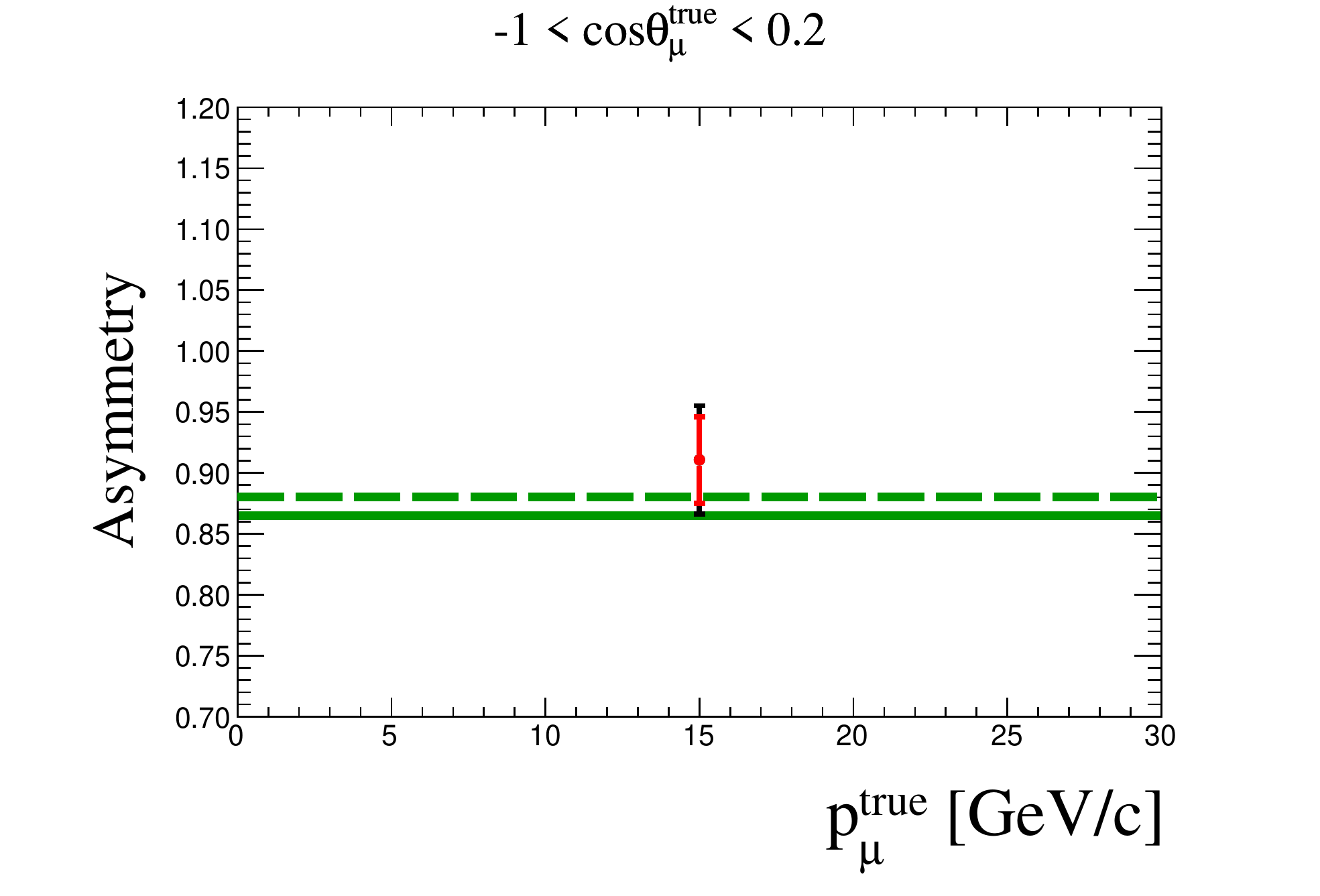}
	\includegraphics[width=0.36\linewidth]{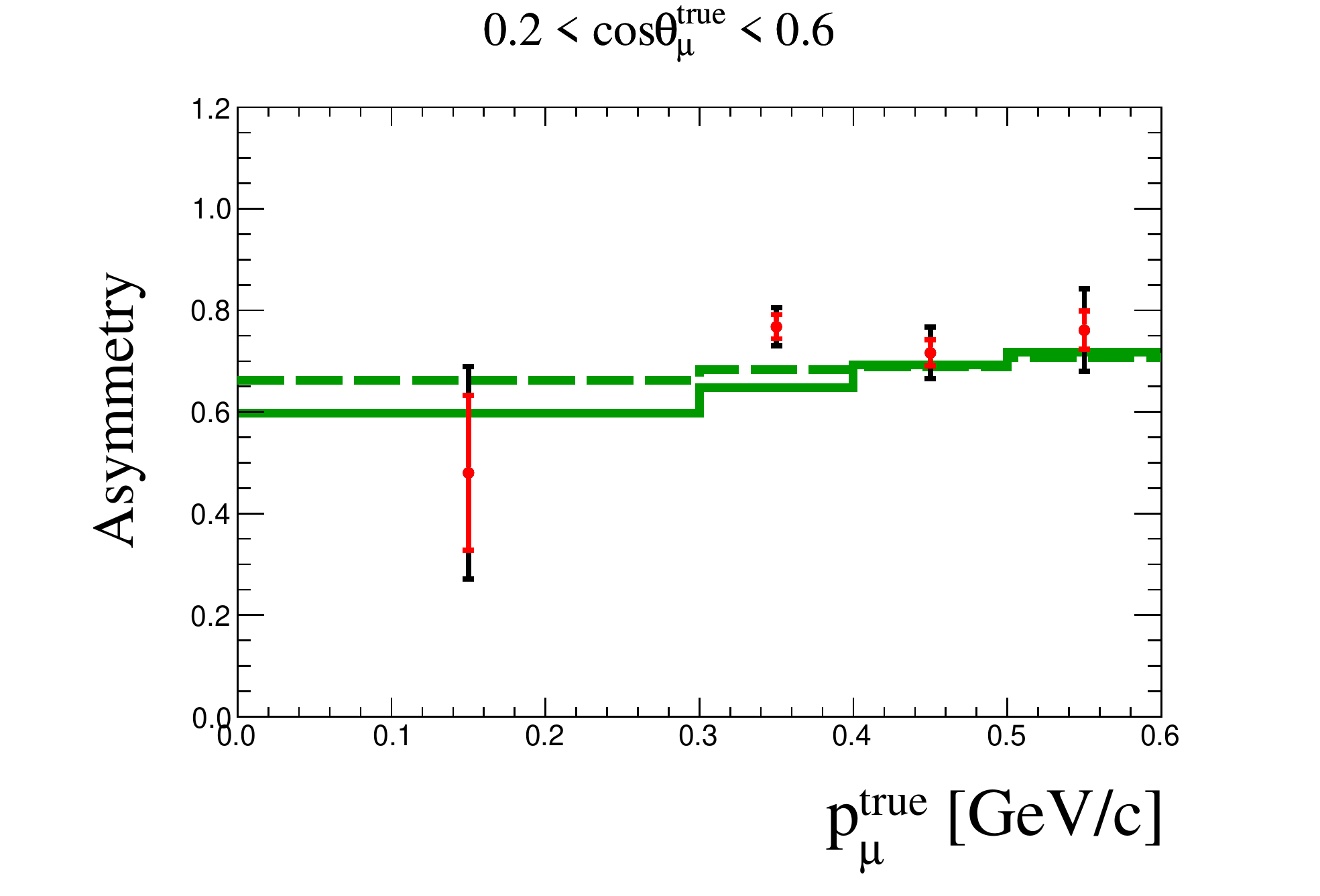}
	\includegraphics[width=0.36\linewidth]{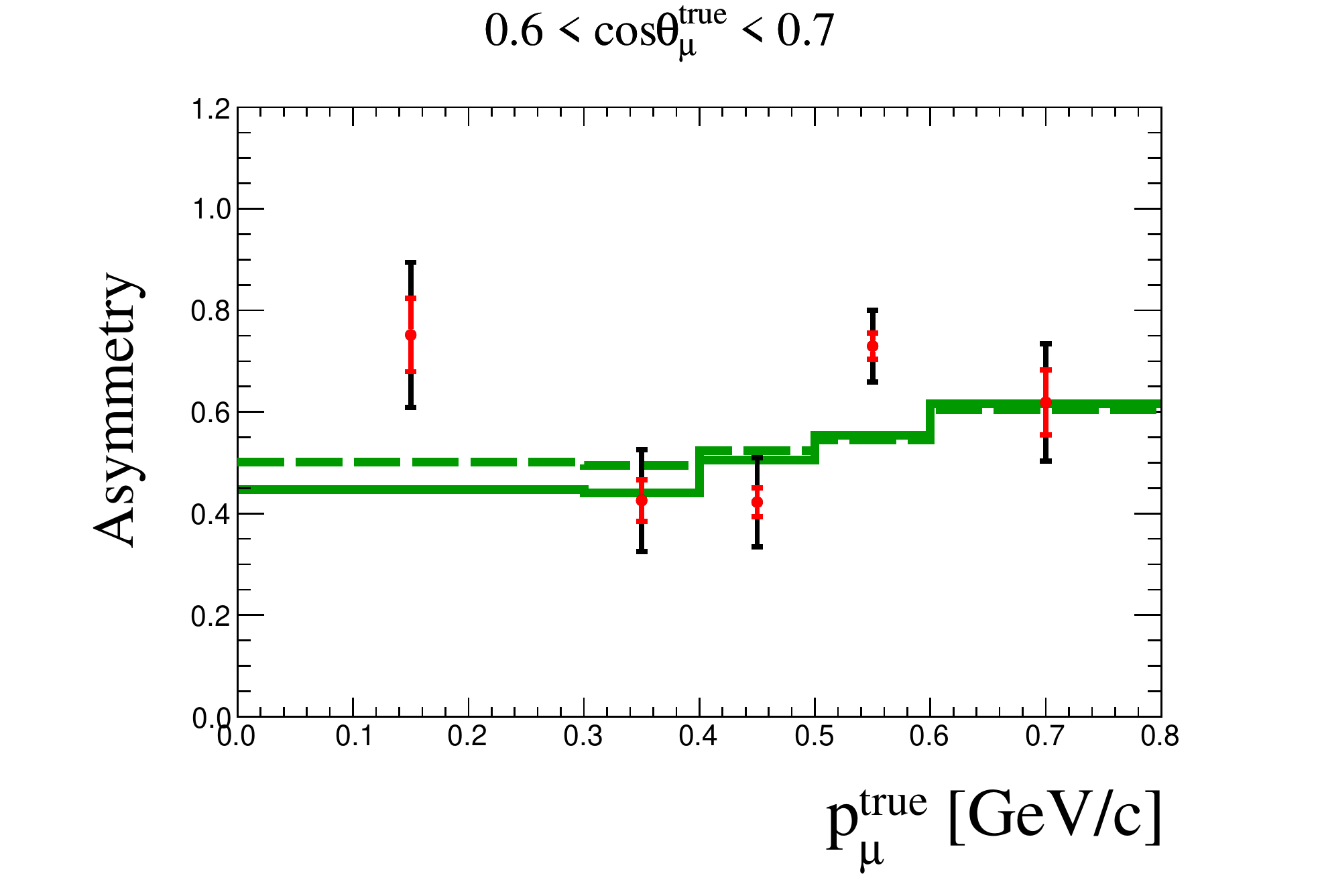}
	\includegraphics[width=0.36\linewidth]{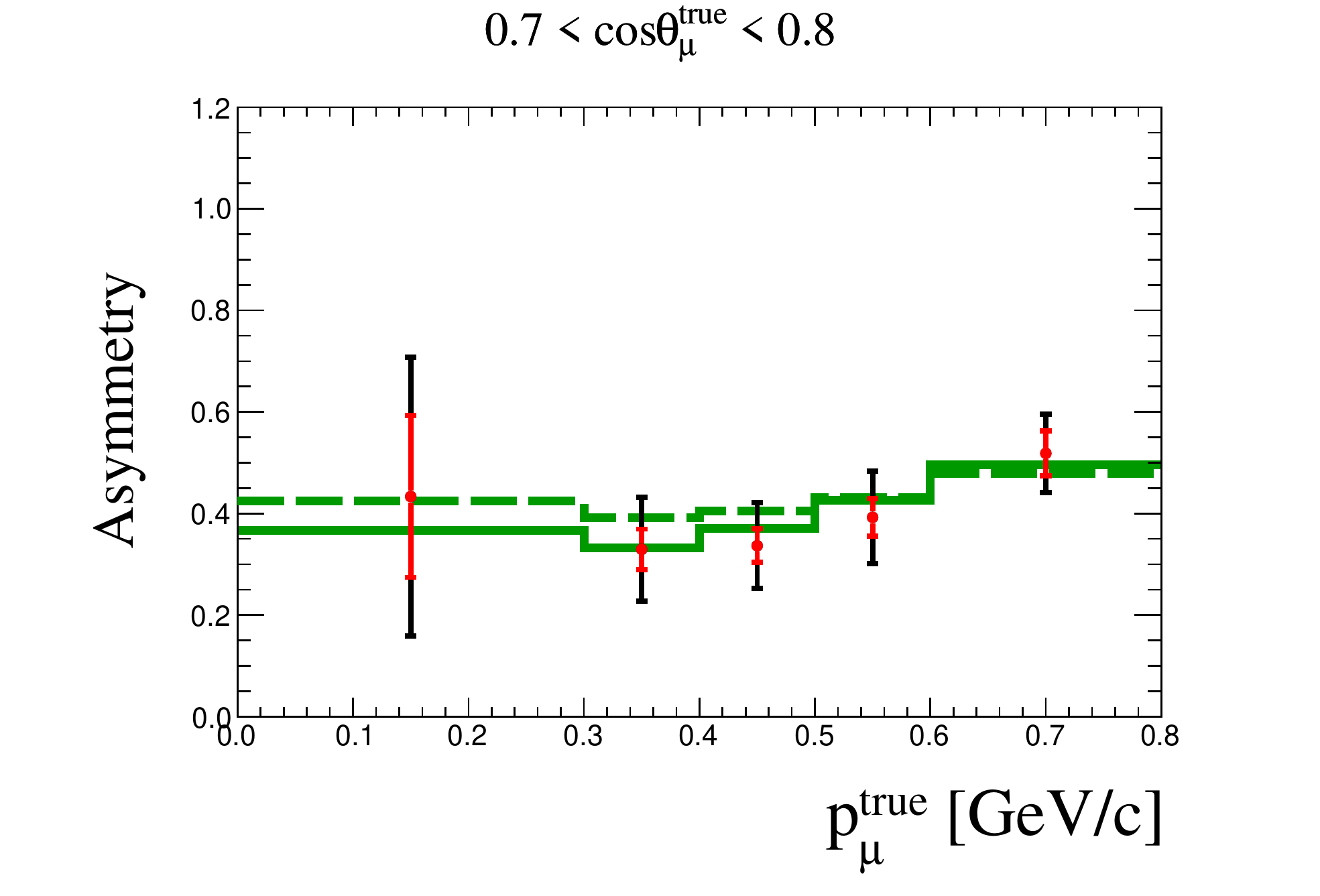}
	\includegraphics[width=0.36\linewidth]{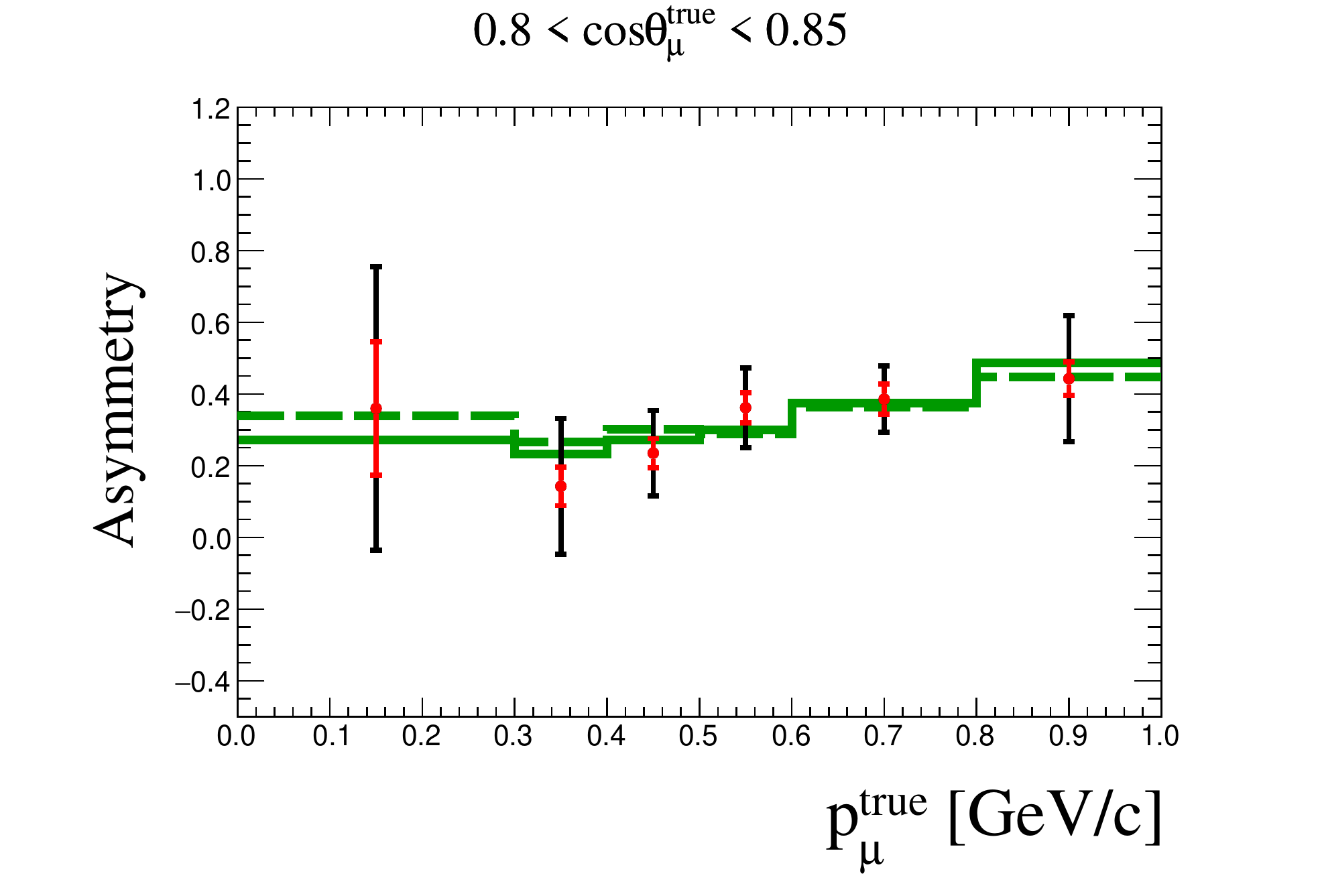}
	\includegraphics[width=0.36\linewidth]{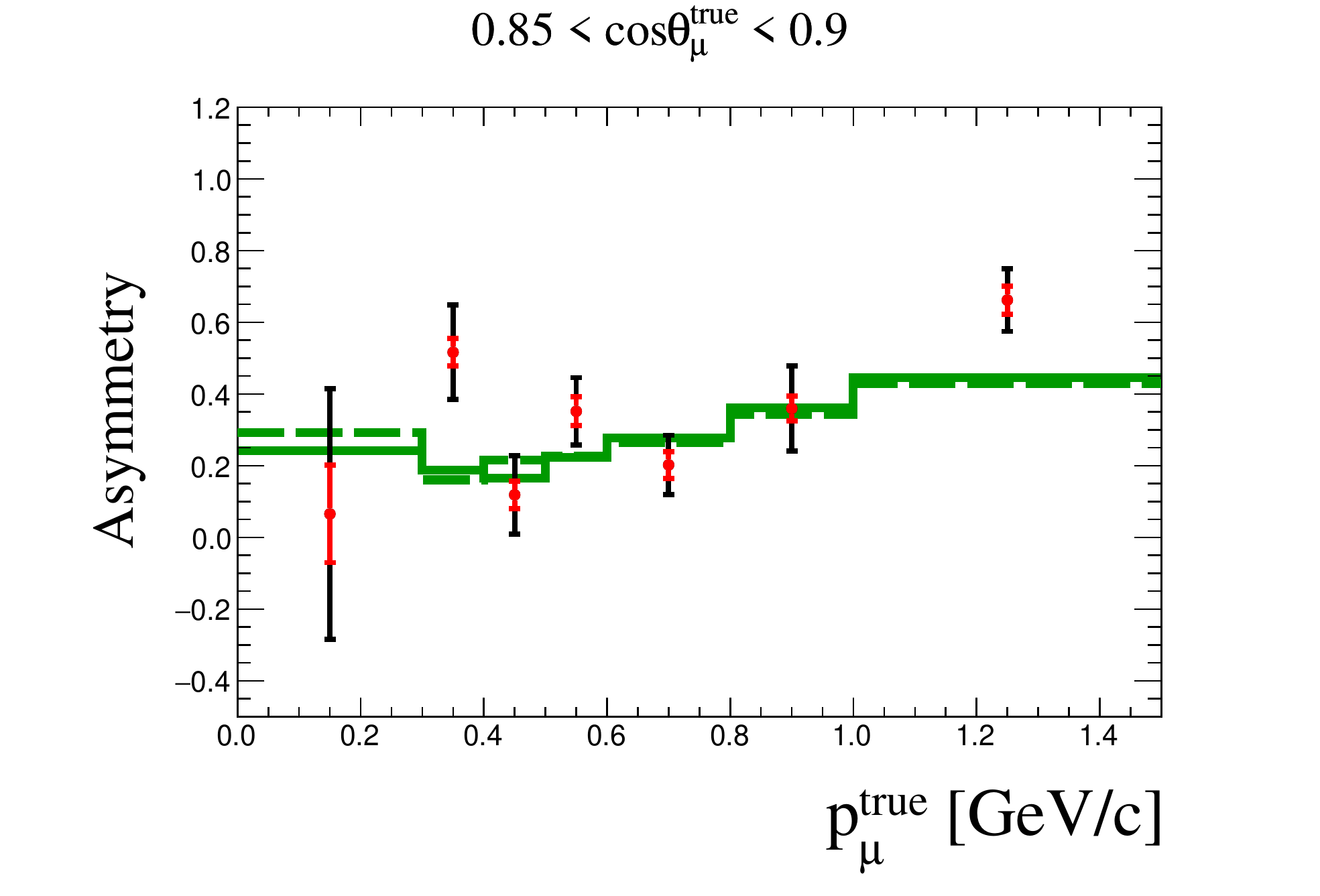}
	\includegraphics[width=0.36\linewidth]{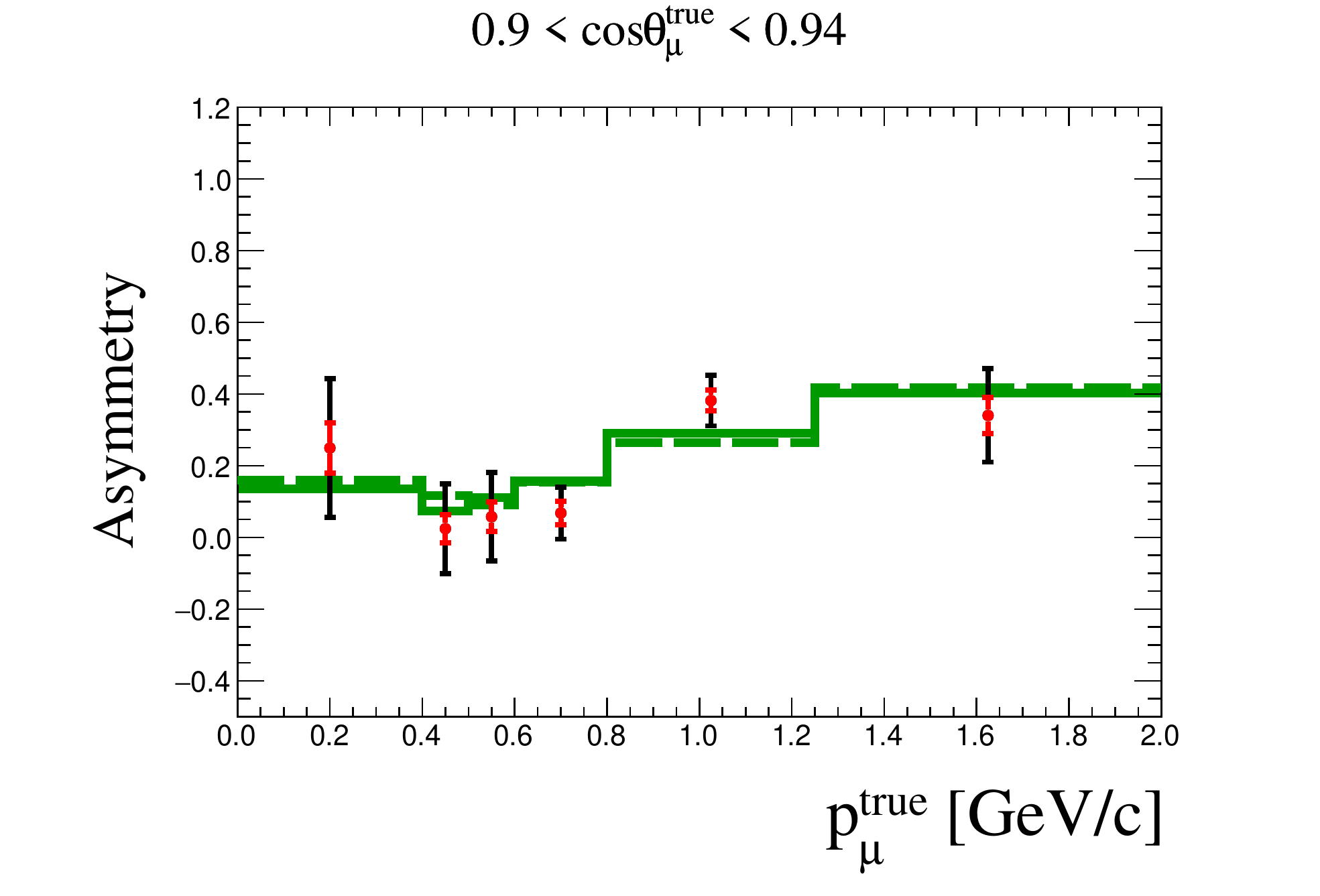}
	\includegraphics[width=0.36\linewidth]{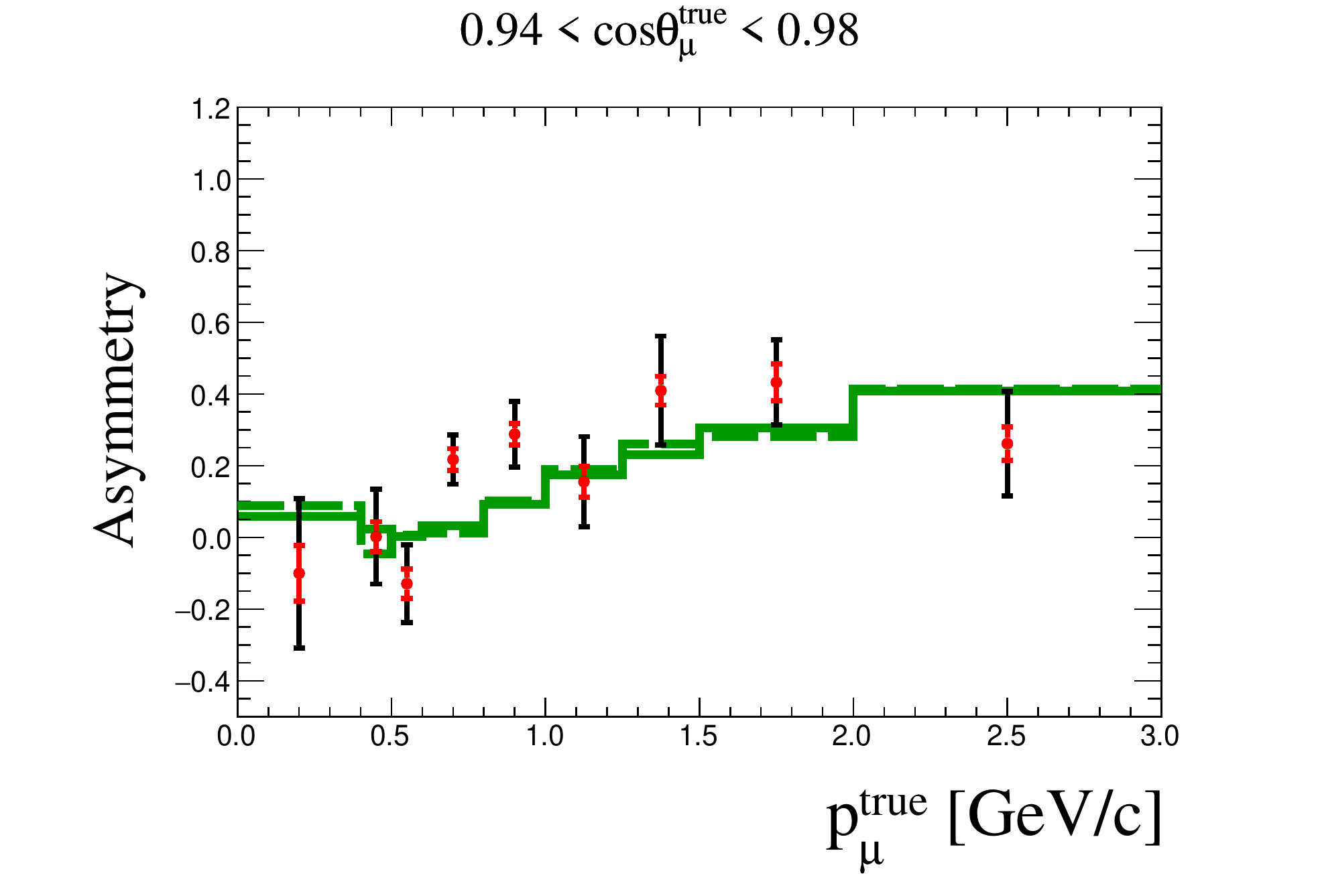}
	\includegraphics[width=0.36\linewidth]{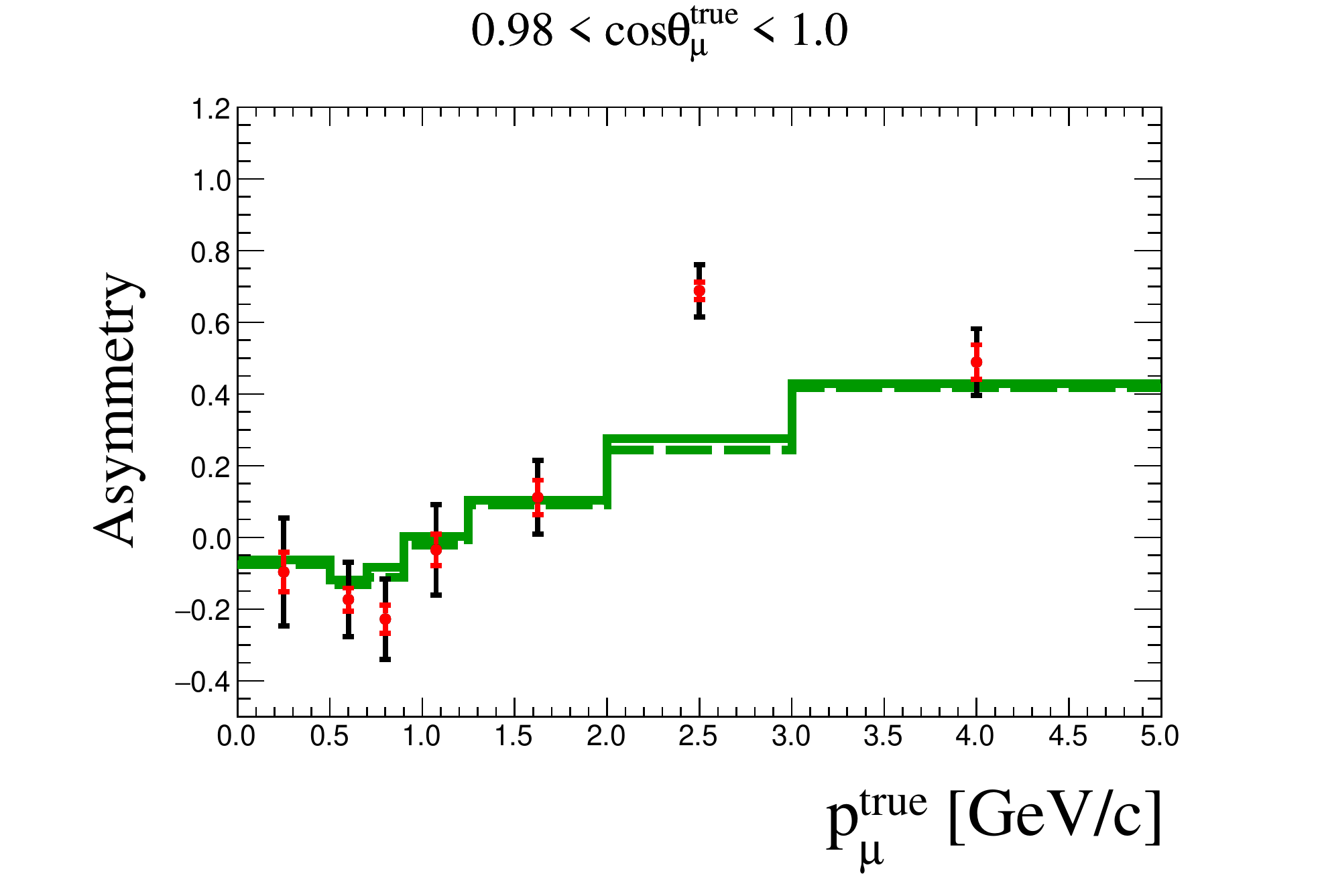}
	\includegraphics[width=0.36\linewidth]{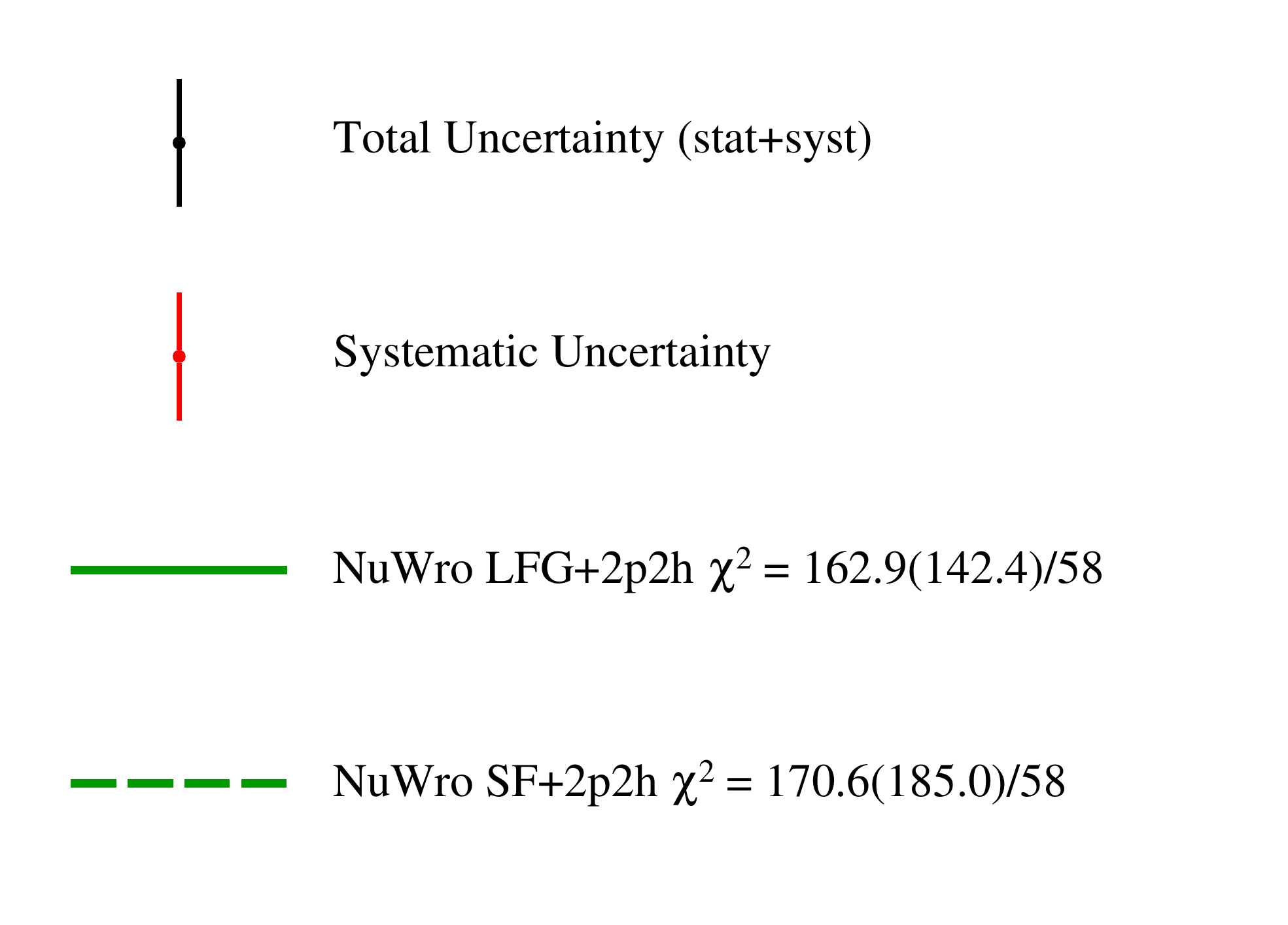}
	\caption{Measured double-differential \cczeropi cross-section asymmetry in bins of true muon kinematics with systematic uncertainty (red bars) and total (stat.+syst.) uncertainty (black bars). The result is compared with \textsc{NuWro} version~\texttt{18.02.1} with LFG+RPA (green solid line) and with the SF nuclear model (green dashed line), both including 2p2h predictions. The full and shape-only (in parenthesis) $\chi2$ are reported. The last bin in momentum is not displayed for readability.}
	\label{fig:xseasynuwro}
\end{figure*}

The comparisons with the models described above are shown in \crefrange{fig:numucc0pixsecneut2p2h}{fig:xseasynuwro}. The full and shape-only $\chi^2$ are reported in the legends (shape-only $\chi^2$ is reported in parenthesis) and are summarized in \cref{tab:chi2}. In \cref{tab:redchi2} the reduced $\chi^2$ is reported as well.

In order to evaluate the sensitivity to the 2p2h process, the measured cross sections, and their combinations, are compared in \crefrange{fig:numucc0pixsecneut2p2h}{fig:xsecasyneut2p2h} to \textsc{Neut} LFG with and without 2p2h. The full and shape-only $\chi^2$ show that the sensitivity is limited. Some conclusions can be drawn looking at each angular bin. In the intermediate and high-angle region both neutrino and antineutrino data tend to prefer the presence of 2p2h, as already shown in the previous T2K neutrino analysis~\cite{Abe:2016tmq}. The $\chi^2$ in each angular bin has been computed, further confirming the preference for the presence of 2p2h in the intermediate and high-angle region. The effect is particularly evident in the sum of the neutrino and antineutrino cross sections, where the statistical uncertainty is smaller. For instance, in the angular bin 0.6 $<\cos\theta_\mu<$ 0.7 the reduced $\chi^2$ is 0.8 and 2.4 with and without 2p2h respectively. On the other hand, a clear overestimation of the cross section is visible in the forward region below 1~GeV, both for neutrinos and antineutrinos. This may point to incorrect 1p1h predictions, notably in the region of small energy transfer to the nucleus, where the treatment of various nuclear effects, like binding energy, is crucial. This issue is further discussed below, in the comparison to different 1p1h models. As expected the neutrino-antineutrino cross section difference emphasizes the 2p2h cross section, due to the change of sign of the axial-vector component. The statistical and systematic uncertainties, which are dominated by the flux, are still too large for a measurement of this component. Future foreseen reduction of such uncertainties, with more ND280 data and relying on NA61/SHINE T2K replica target data for flux tuning~\cite{Berns:2018tap}, will improve the sensitivity to the axial-vector 2p2h component. In some bins the difference is negative since antineutrinos can interact with the hydrogen of the hydrocarbon molecule, leading to a cross-section for antineutrino higher than for neutrino. The neutrino-antineutrino cross-section asymmetry shows a very small 2p2h dependence. The fractional change of the asymmetry with and without 2p2h is very small, except in the low momentum region where, at forward angle, it may reach 50\%. The sensitivity to such observable is drastically limited by the statistical uncertainty. Despite most of the systematic uncertainties cancel out due to the correlation between neutrino and antineutrino, a residual not correlated detector systematic dominate the systematic error, driven by the differences the \numu and \barnumu event selections.

A more sophisticated assessment of the 2p2h sensitivity is shown in \crefrange{fig:numucc0piNMS}{fig:asynumucc0piNMS}, where the results are compared to different 2p2h models. The 2p2h model in \textsc{Neut} and the 2p2h model by Martini \textit{et al.}~\cite{Martini:2009uj} are both implemented on top of a similar 1p1h LFG model while the \textsc{SuSav2} model includes different 1p1h~\cite{Caballero:2005sj} and 2p2h~\cite{Simo:2016ikv} predictions. For the comparison with the model from Martini et al. the number of degrees of freedom (ndof) have been reduced to 96 for the cross sections and to 48 for their combinations in terms of sum, difference and asymmetry (w.r.t. 116 and 58 respectively) because the model predicts the cross section only for muon momentum lower than 3 GeV/c. Thus 10 high-momentum bins have been removed from the covariance matrix to compute the full $\chi^2$. Similarly, a complete shape-only covariance matrix has been obtained and those 10 bins have been removed afterwards to compute the shape-only $\chi^2$. An extended implementation of this model would be crucial for a better comparison with other models. None of the model is able to well describe the measured neutrino and antineutrino cross-sections in the entire phase space. As previously mentioned, the disagreement with cross-section measurements can be interpreted both in terms of 1p1h or 2p2h mismodeling. On the other hand, the various 2p2h models have quite different predictions for the axial-vector component, making the measurement of the neutrino-antineutrino cross-section difference a powerful probe to test the physics implemented in the different 2p2h predictions. 

To further investigate the dependence of the results on the 1p1h model, the measured cross-sections, and their combinations, are compared to different LFG implementations in \crefrange{fig:numucc0pixsecgibuuneutnuwro}{fig:xsecasygibuuneutnuwro}. 
The \textsc{Neut}, \textsc{NuWro} and \textsc{Genie} LFG implementations differ mainly in the treatment of the nucleon binding energy. 
None of the generators is able to describe the measured neutrino and antineutrino cross-sections in the entire phase space. Among the different combinations the cross-sections difference show the lowest full $\chi^2$ in the comparison with \textsc{Genie}.

The measured cross-sections, and their combinations, are also compared to a SF model in \crefrange{fig:numucc0pixsecnuwro}{fig:xseasynuwro}. The SF cross-section shows a different angular dependence than the LFG one: smaller for the backward and high-angle region and larger in the forward region. Interestingly, while SF is a more sophisticated model, the full $\chi^2$ is the largest (see \cref{tab:chi2}). This may be due to an incomplete implementation of SF or to the merging with a 2p2h simulation modeled using RFG as nuclear model. The difference between LFG and SF tends to cancel in the neutrino-antineutrino cross-section difference and asymmetry. A more complete implementation of an SF model (including a 2p2h contribution) is likely required to investigate this further.

The integrated cross sections per nucleon and their combinations are reported in \cref{tab:integrated} and compared with the model described above. The \numu \cczeropi integrated cross section is compatible with the one reported in previous T2K published analyses~\cite{Abe:2016tmq,Abe:2018pwo}. It is striking that the models which exhibit best agreement in shape and in normalization are different, calling for further measurements with smaller systematic uncertainties and further model development.

In summary, even if some conclusion can be drawn looking at the comparisons in some angular bins, none of the models is able to simultaneously describe \numu and \barnumu \cczeropi cross sections in all the phase space. Among the different combination, the difference between neutrino and antineutrino cross sections shows interesting sensitivity to different 2p2h models, which is limited by large uncertainties.

The poor (anti)neutrino-nucleus interaction modeling highlighted by this analysis is a limiting factor for the future neutrino oscillation experiments that have as primary goal the measurement of the CP violation, calling for a deeper understanding of the underlying processes involved in (anti)neutrino-nucleus interactions and for new cross-section analyses with larger statistics and improved systematic uncertainties.

\begin{table*}[ht!]
	\centering
	\setlength{\extrarowheight}{.1cm}
	\begin{ruledtabular}
		\caption{$\chi^2$ values for different generators and models. The number of degrees of freedom is 116 for the combined $\chi^2$ (96 for Martini \textit{et al.}) and 58 for the sum, difference and asymmetry (48 for Martini \textit{et al.}).}
		\label{tab:chi2}
		\begin{tabular}{lcc|cc|cc|cc}
			Generator/model        &  \multicolumn{2}{c}{Cross section}   &  \multicolumn{2}{c}{Sum}  &  \multicolumn{2}{c}{Difference}  &   \multicolumn{2}{c}{Asymmetry}  \\ 
			&  Full $\chi^2$ & Shape-only $\chi^2$  &  Full $\chi^2$ & Shape-only $\chi^2$  &  Full $\chi^2$ & Shape-only $\chi^2$ &  Full $\chi^2$ & Shape-only $\chi^2$ \\ 
			\hline
			\textsc{Genie} LFG w/ 2p2h       &  333.1 & 444.7   & 101.3 & 141.3  & 76.2 & 102.0 &  143.6  & 134.4 \\    		                                               		
			\textsc{Neut} LFG w/ 2p2h       &  366.7 & 459.1   & 123.4 & 175.7  & 79.5 & 113.8 &  150.5  &  147.8 \\ 
			\textsc{Neut} LFG w/o 2p2h     &  236.7 & 388.7   &  82.5 & 126.5  & 87.6  & 154.8 & 160.0    & 169.4 \\ 
			\textsc{NuWro} LFG w/ 2p2h   &  408.9 &  481.5    &  122.2 & 158.1 &  87.0  & 121.6 &  162.9  & 142.4 \\ 
			\textsc{NuWro} SF w/ 2p2h     &  650.0 &  838.8   &  233.5 & 358.1 & 97.6   & 149.7 &  170.6  & 185.0 \\ 
			\textsc{GiBUU}                        &  488.2 &  474.3    &  133.5 & 136.3 & 120.1 & 140.1 &  157.7  & 148.0 \\ 	
			Martini \textit{et al.}                 &  368.6 & 573.4    &  142.0 & 227.4 & 119.6 & 289.8 & 93.9 & 131.2\\ 
			\textsc{SuSAv2}                     &  565.9  &  563.1    & 170.6 & 186.8   & 119.2  & 137.9 & 152.6 & 146.3 \\ 
		\end{tabular}
	\end{ruledtabular}
\end{table*}

\begin{table*}[ht!]
	\centering
	\setlength{\extrarowheight}{.1cm}
	\begin{ruledtabular}	
		\caption{Reduced $\chi^2$ values for different generators and models.}
		\label{tab:redchi2}
		\begin{tabular}{lcc|cc|cc|cc}
			Generator/model        &  \multicolumn{2}{c}{Cross section}   &  \multicolumn{2}{c}{Sum}  &  \multicolumn{2}{c}{Difference}  &  \multicolumn{2}{c}{Asymmetry}  \\ 
			&  Full $\chi^2$ & Shape-only $\chi^2$  &  Full $\chi^2$ & Shape-only $\chi^2$  &  Full $\chi^2$ & Shape-only $\chi^2$ &  Full $\chi^2$ & Shape-only $\chi^2$ \\ 
			\hline   		                                               		
			\textsc{Genie} LFG w/ 2p2h       & 2.9 & 3.8 & 1.7 & 2.4  & 1.3 & 1.8 &  2.5 &   2.3\\ 
			\textsc{Neut} LFG w/ 2p2h       &  3.2 & 4.0 & 2.1 & 3.0  & 1.4 & 2.0 &  2.6  &  2.5\\ 
			\textsc{Neut} LFG w/o 2p2h     &  2.0 & 3.3 &  1.4 & 2.2  & 1.5  & 2.7 & 2.7 &  2.9\\ 
			\textsc{NuWro} LFG w/ 2p2h   &  3.5 &  4.1 &  2.1 & 2.7 &  1.5  & 2.1 &  2.8 &  2.4\\ 
			\textsc{NuWro} SF w/ 2p2h     &  5.6 &  7.2 &  4.0 & 6.2 & 1.7   & 2.6 &  2.9 &  3.2\\ 
			\textsc{GiBUU}                         &  4.2 &  4.1 &  2.3 & 2.3 & 2.1 & 2.4 &  2.7  & 2.5\\ 			
			Martini \textit{et al.}                 &  3.8 & 6.0 &  3.0 & 4.7  & 2.5  & 6.0 & 2.0 & 2.7 \\ 
			\textsc{SuSAv2}                       &  4.9  &  4.8 & 2.9 & 3.2   & 2.0  & 2.4 & 2.6 & 2.5\\ 
		\end{tabular}
	\end{ruledtabular}
\end{table*}

\begin{table*}[ht!]
	\centering
	\begin{ruledtabular}
		\caption{Integrated \numu, \barnumu cross sections and their combinations. On the first row are reported the values computed on data, while on the other rows for different generators and models.}
		\label{tab:integrated}
		\setlength{\extrarowheight}{.1cm}
		\begin{tabular}{lccccc}
			\multirow{2}{*}{} 	&   \numu $\times$10$^{-39}$ & \barnumu $\times$10$^{-39}$ & Sum $\times$10$^{-39}$ & Difference $\times$10$^{-39}$ & Asymmetry \\ 
			& cm$^2$/nucleon & cm$^2$/nucleon & cm$^2$/nucleons & cm$^2$/nucleon & \\
			\hline   
			\multirow{2}{*}{Data}  &  4.35  $\pm$ 0.06(\textit{stat.})& 1.30 $\pm$ 0.04(\textit{stat.}) & 5.65 $\pm$ 0.07(\textit{stat.}) & 3.05 $\pm$ 0.07(\textit{stat.})& 0.54 $\pm$ 0.01(\textit{stat.})\\
			&  $\pm$ 0.30(\textit{syst.}) & $\pm$ 0.10(\textit{syst.}) & $\pm$ 0.30(\textit{syst.}) & $\pm$ 0.20(\textit{syst.}) & $\pm$ 0.02(\textit{syst.}) \\
			\textsc{Genie} LFG w/ 2p2h        &   3.76   & 1.14  & 4.90 &  2.62  & 0.53 \\
			\textsc{Neut} LFG w/ 2p2h        &   3.74   & 1.21  & 4.95  &  2.53  & 0.51 \\
			\textsc{Neut} LFG w/o 2p2h   &  3.20   & 1.03  & 4.23 & 2.17 & 0.51 \\
			\textsc{NuWro} LFG w/ 2p2h    &  3.91    & 1.28 & 5.19  & 2.63   & 0.51 \\
			\textsc{NuWro} SF w/ 2p2h      & 3.68    &  1.25 & 4.93 &  2.43  & 0.49 \\
			\textsc{GiBUU}                     &   4.33   & 1.34 & 5.67 &  2.99 & 0.53 \\
			Martini                                  &  4.50   & 1.16 & 5.67  & 3.34  & 0.59 \\
			\textsc{SuSAv2}                   &  4.35  & 1.35 & 5.70 & 3.00 & 0.53 \\
		\end{tabular}
	\end{ruledtabular}	
\end{table*}

\section{Conclusions}
\label{sec:conclusions}

The T2K experiment has measured the first combined double-differential $\nu_{\mu}$ and $\bar{\nu}_{\mu}$ cross sections with no pions in the final state in the full phase space using $5.8\times 10^{20}$ POT of neutrino data and $6.2\times10^{20}$ POT of antineutrino data. The inclusion of ToF, in the selection of backward-going and high-angle tracks, enabled the exploration of the full phase space with better efficiency over previously reported T2K measurements of neutrino cross sections~\cite{Abe:2016tmq}. The sum, difference and asymmetry of neutrino and antineutrino cross sections were measured, including full treatment of the correlations between the neutrino and antineutrino samples. Such observables have been compared with different models to shed light on the nuclear effects involved in the (anti)neutrino-nucleus interactions. Although none of the models considered in this work are able to describe the full phase space of the neutrino and antineutrino \cczeropi cross section, it is difficult to determine the source of the problem. A precise understanding of this mis-modeling may be of critical importance for the next generation of neutrino oscillation experiments. Further investigation would benefit from smaller uncertainties and a mitigation of some of the approximations built into generator implementations of the models. 

This analysis opens the road to joint cross-section measurements putting together different samples to minimize systematic uncertainties and to account properly for correlated systematics, enabling more complete and precise tuning of neutrino-nucleus interactions. A promising observable, measured here for the first time, is the difference between neutrino and antineutrino cross sections which shows interesting sensitivity to different 2p2h models, that can be further explored with more statistics and improved systematics uncertainties.

The data release for the results presented in this analysis is posted at the link in Ref.~\cite{datarelease}. It contains the \numu and \barnumu double-differential cross sections central values, their combinations and associated covariance matrices.

\section{Acknowledgments}

We thank the J-PARC staff for superb accelerator performance. We thank the CERN NA61/SHINE Collaboration for providing valuable particle production data. We acknowledge the support of MEXT, Japan; NSERC (Grant No. SAPPJ-2014-00031), NRC and CFI, Canada; CEA and CNRS/IN2P3, France; DFG, Germany; INFN, Italy; National Science Centre (NCN) and Ministry of Science and Higher Education, Poland; RSF (Grant \#19-12-00325) and Ministry of Science and Higher Education, Russia; MICINN and ERDF funds, Spain; SNSF and SERI, Switzerland; STFC, UK; and DOE, USA. We also thank CERN for the UA1/NOMAD magnet, DESY for the HERA-B magnet mover system, NII for SINET4, the WestGrid and SciNet consortia in Compute Canada, and GridPP in the United Kingdom. In addition, participation of individual researchers and institutions has been further supported by funds from ERC (FP7), ``la Caixa'' Foundation (ID 100010434, fellowship code LCF/BQ/IN17/11620050), the European Union’s Horizon 2020 Research and Innovation Programme under the Marie Sklodowska-Curie grant agreements no. 713673 and no. 754496, and H2020 Grants No. RISE-RISE-GA822070-JENNIFER2 2020 and RISE-GA872549-SK2HK; JSPS, Japan; Royal Society, UK; French ANR Grant No. ANR-19-CE31-0001; and the DOE Early Career program, USA, RFBR, project number 20-32-70196.

\clearpage
\bibliography{bibliography}

\end{document}